\documentstyle[epsfig,here,12pt,french,feynmf,twoside,fancyhdr]{book}

\input amssym.def       
\input amssym.tex

\def\double{\Bbb}

\def\ccc{{\double C}}     

\def\nnn{{\double N}}       
\def\zzz{{\double Z}}
\def\qqq{{\double Q}}
\def\rrr{{\double R}}     
\def\kkk{{\double K}}
\def\hhh{{\double H}}

\def\aa{{\cal A}}
\def\bb{{\cal B}}
\def\cc{{\cal C}}
\def\dd{{\cal D}}
\def\ee{{\cal E}}
        
\def\hh{{\cal H}}   
\def\ll{{\cal L}}
\def\mm{{\cal M}}    
\def\nn{{\cal N}}
\def\jj{{\cal J}}
\def\oo{{\cal O}}
\def\rr{{\cal R}}
\def\ss{{\cal S}}
\def\tt{{\cal T}} 
\def\uu{{\cal U}}
\def\ww{{\cal W}}
\def\xx{{\cal X}}

\def\t{\mathrm{Tr}} 

\def\M{\mathrm{M}}
\def\dix{\int \!\!\!\!\!\! -}
\def\lb{\left[} 
\def\rb{\right]}
\def\lp{\left(} 
\def\rp{\right)}
\def\la{\left\{} 
\def\ra{\right\}}

\def\ul{\underline}
\def\ov{\overline}
\def\ot{\otimes}
\def\op{\oplus}
\def\om{\ominus}
\def\bbb{\begin{equation}}
\def\eee{\end{equation}}
\def\bbbb{\begin{eqnarray}}
\def\eeee{\end{eqnarray}}

\def\pp{\pmatrix}
\def\n{\nonumber}

\newtheorem{thm}{Th\'eor\`eme}[section]
\newtheorem{lem}{Lemme}[section]
\newtheorem{dfi}{D\'efinition}[section]
\newtheorem{axi}{Axiome}
\newtheorem{pro}{Proposition}[section]
\newtheorem{cor}{Corollaire}[section]

\newcounter{ex}[section]
\def\theex{\arabic{chapter}.\arabic{section}.\arabic{ex}}
\def\exe{\bigskip\noindent {\bf Exemple} \refstepcounter{ex}\theex\\ }
\def\eexe{\hfill$\blacksquare$\\}
\def\demo{\noindent\ul{D\'emonstration}:\\ \noindent }
\def\edemo{\hfill$\square$\\}

\begin{document}

\thispagestyle{empty}

\begin{center}

\huge{\bf TH\`ESE}\\

\vskip 2.5truecm

\large{pr\'esent\'ee par}\\

\vskip 0.5truecm

\Large{\bf Thomas KRAJEWSKI}\\

\vskip 1.5truecm

\large{pour obtenir le grade de}\\

\vskip 0.5truecm

\Large{Docteur de l'Universit\'e de Provence}\\

\vskip 0.5truecm

\large{\underline{sp\'ecialit\'e:}}\\

\vskip 0.5truecm
\large{Physique des Particules, Physique Math\'ematique et Mod\'elisation}\\
\vskip 1.5truecm
\Large{\bf G\'EOM\'ETRIE NON COMMUTATIVE ET}\\
\vskip 0.5truecm
\Large{\bf INTERACTIONS FONDAMAENTALES}\\

\vskip 2truecm
\large{Soutenue le 10 D\'ecembre 1998, devant le Jury compos\'e de}

\end{center}

\begin{center}
{\large
\begin{tabular}{l}
A. Connes (Pr\'esident)\\
B. Iochum (Directeur de th\`ese)\\ 
V.F.R. Jones\\
D. Kastler\\
F. Lizzi (Rapporteur)\\
E. de Rafael\\
T. Sch\"ucker
\end{tabular}}
\end{center}


\newpage

\thispagestyle{empty}
${}$
\newpage

\thispagestyle{empty}
{\bf\Large Remerciements}

\bigskip

En g\'en\'eral, il est impossible d'\'ecrire une th\`ese isol\'e dans une forteresse en reconstituant jour apr\`es jour toutes les lois de la Physique, par exemple en regardant des pommes tomber ou bien en prenant un bain...

\par

Aussi, je dois beaucoup \`a toutes les personnes qui m'ont aid\'e un jour ou l'autre. Il me tient \`a c\oe ur de les remercier ici.
 
\par 

En particulier, je remercie chaleureusement mon directeur de th\`ese. Bruno Iochum,  pour la confiance dont il m'a t\'emoign\'e. Son souci de rigueur, dans les explications et les d\'emonstrations, m'a \'et\'e de la plus grande utilit\'e. Je dois aussi beaucoup \`a Thomas Sch\"ucker, qui n'a jamais m\'enag\'e son temps pour me donner des explications, ainsi qu'\`a Raimar Wulkenhaar avec qui j'ai eu le plaisir de collaborer. Tous trois m'ont aid\'e \`a transformer quelques pages obscures en une th\`ese que j'esp\`ere pr\'esentable.  

\par

Je tiens aussi \`a exprimer toute ma gratitude aux membres de Jury ainsi qu'\`a  Jos\'e Maria Gracia-Bond\'\i a qui a \'et\'e rapporteur mais n'a malheureusement pas pu \^etre pr\'esent lors de la soutenance. Malgr\'e un emploi du temps charg\'e, ils ont accept\'e d'\'evaluer mon travail et certains d'entre eux m'ont d'ores et d\'ej\`a rendu de grands services en me fournissant des explications, des critiques ou de nouvelles perspectives de travail.   

\par

Pendant mon s\'ejour au Centre de Physique Th\'eorique de Marseille, j'ai pu b\'en\'eficier d'excellentes conditions de travail, tant sur le plan materiel  que scientifique. En particulier, j'adresse mes remerciements au directeur du laboratoire, Pierre Chiappetta, ainsi qu'\`a l'ensemble du personnel admninistratif.

\par

De plus, l'ambiance stimulante du s\'eminaire de g\'eom\'etrie non commutative ainsi que du groupe "interactions fondamentales" m'a permis de progresser \'enorm\'ement au cours de ces derni\`eres ann\'ees, par des discussions avec (entre autres et en sus des personnes d\'ej\`a remerci\'ees) Rafaele Buffa, Robert Coquereaux, Max Hasler, Carl Herrmann, Ctirad Klim$\check{\mathrm{c}}$ik, Marc Knecht, Chris Korthals-Altes, Serge Lazzarini, Pierre Martinetti, Carlo Rovelli, Daniel Testard et Koumarane Valavane. 

\par

Une mention sp\'eciale pour ce dernier qui va, en compagnie d'Anna-Maria Kiss, m'aider \`a organiser le traditionnel pot de th\`ese.

\par

Enfin, tout cela n'aurait jamais \'et\'e possible sans le soutien de ma famille, \`a laquelle je suis infiniment redevable.

\newpage
\thispagestyle{empty}
${}$

\newpage

\tableofcontents

\newpage

{\renewcommand{\thechapter}{}\renewcommand{\chaptername}{}
\addtocounter{chapter}{-1}
\chapter{Introduction}\markboth{\sl INTRODUCTION}{\sl INTRODUCTION}}

Le d\'ebut de ce si\`ecle a vu l'\'emergence de deux nouvelles th\'eories qui ont compl\`etement boulvers\'e le monde de la physique: la th\'eorie de la relativit\'e g\'en\'erale et la m\'ecanique quantique. Ces deux r\'evolutions sont \`a l'origine de la grande majorit\'e des d\'eveloppements de la physique th\'eorique de notre si\`ecle, y compris les avanc\'ees les plus r\'ecentes.  

\par

La m\'ecanique quantique est une th\'eorie de l'infiniment petit qui cherche \`a d\'ecrire des syt\`emes physiques dont l'action est une quantit\'e de m\^eme ordre de grandeur que la constante de Planck $\hbar=1,05\; 10^{-34}$ J$\cdot$s. Cette constante a \'et\'e introduite par Planck dans son \'etude du corps noir, qui a \'et\'e amen\'e \`a faire l'hypoth\`ese suivante: l'action ne peut varier que par multiples entiers de $\hbar$.
 
\par

Ult\'erieurement, la m\'ecanique quantique a \'et\'e d\'evelopp\'ee sur la base de  l'hypoth\`ese de Planck, et s'applique \`a des syt\`emes physiques tr\`es g\'en\'eraux, allant de l'atome aux syst\`emes macroscopiques d\'ecrits par le biais de la physique statistique. Le point central de cette th\'eorie est le principe d'incertitude de Heisenberg, qui peut \^etre formul\'e comme suit: Il n'est pas possible de mesurer simultan\'ement la position et l'impulsion d'une particule car l'incertitude $\Delta x$ sur la mesure de sa position et l'incertitude $\Delta p$ sur son impulsion doivent satisfaire \`a la relation $\Delta x\cdot\Delta p\simeq \hbar$. Ceci a pour cons\'equence que la notion de trajectoire dans l'espace des phases n'a plus de sens et est une premi\`ere invitation \`a reconsid\'erer nos conceptions actuelles de la g\'eom\'etrie.

\par 

Le formalisme math\'ematique de la m\'ecanique quantique est la th\'eorie des espaces de Hilbert. L'\'etat du syst\`eme physique consid\'er\'e est un vecteur d'un certain espace de Hilbert et les observables sont des op\'erateurs hermitiens agissant sur cet espace de Hilbert, le spectre d'une observable correpondant aux diff\'erentes valeurs que cette observable peut prendre lors d'une mesure physique.  

\par

En cons\'equence, la position $x$ et l'impulsion $p$ d'une particule sont remplac\'ees, en m\'ecanique quantique, par des op\'erateurs $\hat{x}$ et $\hat{p}$ agissant sur un certain espace de Hilbert. Cet espace n'est autre que l'espace $L^{2}(\rrr)$ des fonctions de carr\'e sommable de la variable $x$ et les op\'erateurs $\hat{x}$ et $\hat{p}$ sont d\'efinis par
$$
\hat{x}\Psi(x)=x\Psi(x)\quad\mathrm{et}\quad\hat{p}\Psi(x)=-i\hbar\frac{d}{dx}
\Psi(x)
$$ 
pour tout $\Psi\in L^{2}(\rrr)$. Il est remarquable que le commutateur de ces deux op\'erateurs soit non nul,
$$
\lb\hat{x},\hat{p}\rb=i\hbar,
$$
et que cette r\`egle de commutation ait pour cons\'equence la relation d'incertitude de Heisenberg. Puisque $\hat{x}$ et $\hat{p}$ engendrent l'alg\`ebre des observables quantiques, cette alg\`ebre non commutative doit remplacer l'alg\`ebre des fonctions sur l'espace des phases classique. Par cons\'equent, la g\'eom\'etrie de l'espace des phases quantique fait appel aux coordonn\'ees non commutatives $\hat{x}$ et $\hat{p}$.

\par

Parall\`element, la th\'eorie de la relativit\'e g\'en\'erale, qui a obtenu des succ\`es remarquables, se base sur une conception g\'eom\'etrique de l'espace-temps qui est assimil\'e \`a une vari\'et\'e munie d'une m\'etrique $g_{\mu\nu}$. Contrairement \`a la plupart des autres th\'eories physiques, cette m\'etrique n'est pas donn\'ee \`a priori mais est une variable dynamique comme peut l'\^etre, par exemple, la coordonn\'ee ou l'impulsion d'une particule.
La dynamique de cette th\'eorie est gouvern\'ee par les \'equations d'Einstein,
$$
\rr_{\mu\nu}-\frac{1}{2}g_{\mu\nu}\rr=\frac{8\pi G}{c^{4}}T_{\mu\nu},
$$
qui relient le tenseur de Ricci $\rr_{\mu\nu}$ et la courbure scalaire $\rr$, qui d\'ependent de la m\'etrique $g_{\mu\nu}$, au contenu mat\'eriel de l'espace-temps d\'ecrit par le tenseur \'energie-impulsion $T_{\mu\nu}$. La constante $G$ est la constante de Newton $G=6,67\; 10^{-11}$ N$\cdot$m$^{2}\cdot$kg$^{-2}$ et $c=3\; 10^{8}$ km$\cdot$s$^{-1}$ d\'esigne la vitesse de la lumi\`ere dans le vide .

\par

Ces \'equations ont pour cons\'equence que toute concentration importante de mati\`ere ou d'\'energie en un point donn\'e entra\^\i ne une courbure de l'espace-temps. Au-del\`a d'un certain seuil, cette courbure est si importante que m\^eme la lumi\`ere ne peut plus quitter le voisinage de ce point. Il appara\^\i t donc une sorte de singularit\'e et on en peut plus obtenir la moindre information en provenance de cette r\'egion de l'espace-temps.

\par

Supposons maintenant que nous voulions localiser une particule dans l'espace-temps avec une pr\'ecision arbitraire. D'apr\`es le principe de Heisenberg, l'incertitude sur son impulsion devient alors arbitrairement grande et nous devons pour cela disposer de particules incidentes d'\'energie tr\`es \'elev\'ee. Suivant la discussion pr\'ec\'edente, cette importante concentration d'\'energie donne naissance \`a une singularit\'e, si bien qu'il est impossible de localiser une particule dans l'espace-temps avec une pr\'ecision sup\'erieure \`a un certain seuil \cite{dop}.

\par

Cette incertitude minimale sur la position d'une particule est \'egale \`a la longueur de Planck    
$$
l_{P}=\sqrt{\frac{G\hbar}{c^{3}}}\simeq 1,6\;10^{-33}\mathrm{cm}.
$$  
Si on ram\`ene cette longueur \`a une \'echelle d'\'energie, on trouve une \'energie qui est de l'ordre de $10^{19}$ GeV, ce qui est consid\'erable et \'evidemment hors de port\'ee des acc\'el\'erateurs de particules. Les consid\'erations d\'evelopp\'ees \`a une telle \'echelle d'\'energie sont donc uniquement de nature th\'eorique, m\^eme s'il est possible d'en d\'eduire quelques cons\'equences sur les th\'eories valables \`a l'\'echelle d'\'energie du mod\`ele standard qui est de l'ordre de la centaine de GeV. 

\par

Cela implique qu'il doit exister des relations d'incertitude pour les coordonn\'ees de l'espace-temps lui-m\^eme. En suivant l'analogie avec l'espace des phases et la m\'ecanique quantique, cela nous am\`ene tout naturellement \`a supposer que les coordonn\'ees de l'espace-temps sont elles aussi des \'el\'ements d'une alg\`ebre non commutative. Bien entendu, ces relations d'incertitude ne peuvent \^etre d\'ecel\'ees \`a l'\'echelle macroscopique et nous supposerons toujours que cette structure non commutative de l'alg\`ebre des coordonn\'ees n'est importante qu'\`a l'\'echelle de la longueur de Planck.  

\par

Les principales th\'eories qui d\'ecrivent les interactions fondamentales, que ce soient les th\'eories de Yang-Mills ou la relativit\'e g\'en\'erale, sont des objets de nature g\'eom\'etrique. En effet, leur formulation la plus naturelle et la plus \'el\'egante se fait en utilisant le langage des espace fibr\'es \cite{nakahara}. Par cons\'equent, si nous voulons construire une th\'eorie susceptible de d\'ecrire la physique \`a l'\'echelle de Planck, nous devons \'etendre ces concepts g\'eom\'etriques \`a des espaces dont les coordonn\'ees ne forment plus n\'ecessairement une alg\`ebre commutative. Cette nouvelle g\'eom\'etrie, qui se base sur une alg\`ebre de coordonn\'ees non commutative prend tout naturellement le nom de {\it g\'eom\'etrie non commutative}.  

\par

A priori, il existe autant de g\'eom\'etries non commutative que de possibilit\'es de choisir les relations d\'efinissant l'alg\`ebre des coordonn\'ees. De plus, pour une alg\`ebre donn\'ee, il y a en g\'en\'eral plusieurs g\'en\'eralisations possibles des concepts g\'eom\'etriques. Toutes ces constructions peuvent \^etre regroup\'ees selon que l'accent est mis sur tel ou tel aspect de la g\'eom\'etrie. Avant d'\'etudier plus en d\'etail notre sujet, donnons quelques exemples de g\'eom\'etries non commutative, en renyoyant \`a \cite{madore} pour une discussion plus d\'etaill\'ee. 

\par

L'exemple le plus connu de la communaut\'e des physiciens des particules est certainement la supersym\'etrie. En effet, la supersym\'etie est une forme tr\`es particuli\`ere de g\'eom\'etrie non commutative, puisqu'elle contient, outre les coordonn\'ees ordinaires, des variables de Grassmann satisfaisant \`a des relations du type $\theta_{i}\theta_{j}+\theta_{j}\theta_{i}=0$. Il est alors possible de d\'efinir l'analogue de la d\'erivation et de l'int\'egration par rapport aux coordonn\'ees $\theta_{i}$. L'utilisation de concepts g\'eom\'etriques dans le cadre de la supersym\'etrie se r\'ev\`ele alors extr\^emement utile et permet de simplifier de nombreux calculs.

\par

Un autre exemple de g\'eom\'etrie non commutative est form\'e par la th\'eorie des groupes quantiques. Dans cette th\'eorie, l'accent est mis sur la notion de sym\'etrie. En g\'eom\'etrie ordinaire, une sym\'etrie est d\'ecrite par l'action d'un groupe sur un certain espace. Puisque l'on doit travailler avec l'alg\`ebre des coordonn\'ees plut\^ot qu'avec les points, cette action se traduit par une coaction de l'alg\`ebre des fonctions sur le groupe sur l'alg\`ebre des coordonn\'ees. Dans certains cas, il est possible de rendre ces deux alg\`ebres non commutatives tout en pr\'eservant la coaction de l'une sur l'autre. L'alg\`ebre des fonctions sur un groupe poss\`ede une structure tr\`es riche, appel\'ee alg\`ebre de Hopf,  due \`a la loi de multiplication sur le groupe. Lorsque cette alg\`ebre est une d\'eformation d'une alg\`ebre de fonction sur un groupe ordinaire, elle prend le nom de groupe quantique \cite{chari}.  

\par

L'exemple le plus simple d'espace non commutatif admettant une sym\'etrie quantique est certainement le plan quantique. Celui-ci est d\'ecrit par deux coordonn\'ees $x$ et $y$ satisfaisant \`a la relation $xy=qyx$, o\`u $q$ est un nombre complexe non nul. Il est possible de d\'efinir une d\'eformation $SL_{q}(2)$ de l'alg\`ebre des fonctions sur le groupe $SL(2)$ qui coagit sur le plan quantique. On peut alors construire l'analogue des formes diff\'erentielles sur le plan quantique et ces formes sont assujetties \`a une condition de covariance par rapport \`a cette coaction. 

\par

Dans ce travail, nous \'etudierons essentiellement l'approche d\'evelopp\'ee par  A. Connes dans \cite{bible} ainsi que dans \cite{grav}, renvoyant \`a \cite{hou}, \`a \cite{landi} et \`a \cite{var} pour une introduction p\'edagogique.

\par

En se basant sur l'analogie avec la m\'ecanique quantique, cette th\'eorie fait appel \`a des alg\`ebres de coordonn\'ees qui sont toujours repr\'esent\'ees commes des sous-alg\`ebres de l'alg\`ebre des op\'erateurs d'un certain espace de Hilbert. Ainsi, la majorit\'e des op\'erations se font dans le cadre de l'espace de Hilbert et nous avons \`a notre disposition le puissant arsenal de l'analyse fonctionnelle, qui devient indispensable lorsqu'il s'agit de travailler avec des alg\`ebres de dimension infinie. Malheureusement le prix \`a payer est assez lourd car il nous est absolument impossible de travailler avec des alg\`ebres qui ne sont pas semi-simples, comme par exemple l'alg\`ebre de Grassmann de la supersym\'etrie ou certains groupes quantiques lorsque le param\`etre de d\'eformation est une racine de l'unit\'e.   

\par 

Plus pr\'ecis\'ement, la d\'emarche employ\'ee en g\'eom\'etrie non commutative est la suivante:
\begin{itemize}
\item
Partant d'un ensemble $X$ muni d'une certaine structure g\'eom\'etrique (structure d'espace topologique ou mesurable, ou de vari\'et\'e diff\'erentiable, \dots), on formule la th\'eorie sous-jacente \`a l'aide de sous-alg\`ebres ad\'equates de l'alg\`ebre des fonctions \`a valeurs complexes d\'efinies sur $X$.
\item
On \'etend ensuite tous les r\'esultats de la th\'eorie pr\'ec\'edente qui ne font pas usage de la commutativit\'e de l'alg\`ebre. 
\end{itemize}
Cette d\'emarche est illustr\'ee par le lien entre C$^{*}$-alg\`ebres et topologie g\'en\'erale (cf Appendice A). Dans un premier temps, on montre que l'alg\`ebre des fonctions continues sur un espace topologique localement compact est une C$^{*}$-alg\`ebre commutative et que toute C$^{*}$-alg\`ebre commutative est l'alg\`ebre des fonctions continues sur un espace localement compact. On d\'efinit alors la donn\'ee d'un espace topologique non commutatif comme \'etant la donn\'ee d'une C$^{*}$-alg\`ebre non commutative, et on \'etend la plupart des d\'efinitions et des th\'eor\`emes de topologie qui ne reposent pas sur la commutativit\'e de l'alg\`ebre des coordonn\'ees. De plus, les sym\'etries, qui sont d\'ecrites dans le cas commutatif par les hom\'eomorphismes, correspondent dans le cas non commutatif aux automorphismes de l'alg\`ebre $\aa$.

\par

Bien d'autres aspects de la g\'eom\'etrie et de la topologie peuvent \^etre \'etendus au cas non commutatif en utilisant ce proc\'ed\'e. Par exemple, il est possible de d\'efinir de mani\`ere purement alg\'ebrique la notion de fibr\'e vectoriel qui nous m\`ene \`a la K-th\'eorie (cf Appendice B). D'un autre cot\'e, s'il n'est pas possible pour l'instant de caract\'eriser ainsi l'alg\`ebre des fonctions lisses sur une vari\'et\'e, on peut toujours d\'efinir de mani\`ere axiomatique les formes diff\'erentielles et leur int\'egration. Cette th\'eorie se g\'en\'eralise au cas non commutatif par le biais de la notion de cycle, qui est \`a la base de la cohomologie cyclique (Appendice C). Sur le plan purement math\'ematique, cette construction atteint son apog\'ee avec la d\'efinition du couplage de la cohomologie cyclique avec la K-th\'eorie par l'interm\'ediaire du caract\`ere de Chern. 
  
\par

Les th\'eories de jauge peuvent aussi \^etre d\'efinies en utilisant la notion de module projectifs et de connexion. Afin de pouvoir d\'efinir l'analogue non commutatif de la relativit\'e g\'en\'erale ainsi que le couplage de cette th\'eorie avec des fermions, la notion qui int\'eresse le physicien des hautes \'energies est celle de vari\'et\'e \`a spin. Cette derni\`ere peut \^etre caract\'erise\'ee de mani\`ere purement alg\'ebrique \`a l'aide de la notion de {\it triplet spectral} $(\aa,\hh,\dd)$ \cite{grav}. Puisque nous reviendrons largement sur la d\'efinition de cette notion, contentons-nous de d\'efinir un tel triplet comme \'etant une alg\`ebre involutive $\aa$ jouant le r\^ole de l'alg\`ebre des coordonn\'ees, repr\'esent\'ee sur une espace de Hilbert $\hh$. $\dd$ est un op\'erateur agissant sur $\hh$ qui g\'en\'eralise l'op\'erateur de Dirac.

\par

A partir de ce triplet spectral, on d\'efinit un calcul diff\'erentiel qui permet la construction de th\'eories de Yang-Mills, en incluant la g\'en\'eralisation de configurations ayant une topologie non triviale. Dans ce formalisme, le r\^ole des transformations de jauge est jou\'e par les \'el\'ements unitaires de l'alg\`ebre $\aa$. Tout comme en th\'eorie des champs ordinaire, l'invariance sous ces transformations de jauge nous am\`ene \`a remplacer l'op\'erateur de Dirac usuel par un op\'erateur de Dirac covariant, qui contient un champ de jauge. La dynamique de la th\'eorie est alors gouvern\'ee par le {\it principe d'action spectrale}, qui stipule que l'action ne d\'epend que du spectre de l'op\'erateur de Dirac covariant.

\par

Lorsque le triplet spectral est le produit du triplet spectral relatif \`a la g\'eom\'etrie de l'espace-temps par un {\it triplet spectral fini} (i.e. tel que $\aa$ et $\hh$ soient de dimension finie) judicieusement choisi, la construction pr\'ec\'edente permet de retrouver le mod\`ele standard coupl\'e \`a la th\'eorie de la gravitation. Outre l'interpr\'etation g\'eom\'etrique du boson de Higgs comme une connexion associ\'ee \`a la composante discr\`ete de l'espace-temps, ce mod\`ele impose de nombreuses contraintes aux masses et aux constantes de couplage du mod\`ele standard. En particulier, ce mod\`ele permet de pr\'edire une masse du boson de Higgs de l'ordre de 200 GeV \cite{hear}. 

\par

L'objet de cette th\`ese est la construction et l'\'etude de diff\'erents mod\`eles physiques que l'on peut construire \`a l'aide de la g\'eom\'etrie non commutative. Cela inclut l'analyse de toutes les th\'eories de Yang-Mills-Higgs que l'on peut obtenir en utilisant un triplet spectral proche du mod\`ele standard. Cependant, nous nous sommes aussi efforc\'e d'\'etudier des triplets spectraux d'un type diff\'erent, comme par exemple le tore non commutatif. De fa\c con g\'en\'erale, nous avons consid\'er\'e ces objets comme servant de base \`a des constructions g\'eom\'etriques permettant de g\'en\'eraliser certains aspects de la th\'eorie des champs ordinaire, \'etant entendu que l'expression "th\'eorie des champs" d\'esigne un syst\`eme dynamique ayant un nombre infini de degr\'es de libert\'e et susceptible d'une certaine interpr\'etation g\'eom\'etrique.

\par

Ce travail est divis\'e en quatre chapitres d'une importance \'egale. 

\par

Dans un premier temps, nous rappelons les bases axiomatiques de la g\'eom\'etrie non commutative, ainsi que quelque constructions classiques que nous utiliserons par la suite, comme celle de l'alg\`ebre diff\'erentielle ou des th\'eories de jauge. Nous introduisons aussi le caract\`ere de Chern ainsi que la version locale du th\'eor\`eme de l'indice. En nous basant sur ces constructions, nous proposons une d\'efinition de l'action de Chern-Simons en g\'eom\'etrie non commutative et nous \'etudions ses propri\'et\'es sous les transformations de jauge. En utilisant la forme de Chern-Simons, nous donnons \'egalement une  nouvelle d\'emonstration d'un r\'esultat relatif \`a la minoration de nature topologique de l'action de Yang-Mills pour un triplet spectral de dimension 4.   

\par

Le second chapitre est consacr\'e \`a une \'etude d\'etaill\'e des triplets spectraux finis. Nous donnons la forme simplifi\'ee des axiomes dans ce cas ainsi que leur solution g\'en\'erale et nous  proposons un interpr\'etation diagrammatique de la solution pr\'ec\'edente. Ce chapitre inclut \'egalement l'\'etude d'objets associ\'es \`a ces triplets spectraux comme les th\'eories de jauge ou les distances.

\par

Ensuite, nous utilisons les r\'esultats pr\'ec\'edents pour \'etudier syst\'ematiquement tous les mod\`eles qui peuvent \^etre construits \`a l'aide du produit tensoriel de la g\'eom\'etrie de l'espace-temps usuel par un triplet spectral fini. Ces mod\`eles sont des mod\`eles de Yang-Mills-Higgs coupl\'es \`a la gravitation dont nous analysons les secteurs fermionique et bosonique en nous servant autant que possible de la m\'ethode diagrammatique. 
  
\par

Le dernier chapitre est consacr\'e \`a l'\'etude du tore non commutatif. Nous commen\c cons par un bref rappel de ses propri\'et\'es alg\'ebriques et nous v\'erifions que les axiomes de la g\'eom\'etrie non commutative sont satisfaits. Ensuite, nous \'etudions les th\'eories de jauge sur le tore non commutatif en utilisant les m\'ethodes du chapitre 1; en particulier nous v\'erifions l'invariance de jauge de l'action de Chern-Simons ainsi que la quantification de la borne topologique de l'action de Yang-Mills en dimension 4. Enfin, nous terminons cette \'etude par la quantification de la th\'eorie de Yang-Mills: nous \'etendons la sym\'etrie BRS, donnons les r\`egles de Feynman associ\'ees et terminons en donnant les principales caract\'eristiques de cette th\'eorie.  

\par

Trois appendices suivent le corps de cette th\`ese. Ceux-ci n'ont d'autre pr\'etention que de fournir au lecteur un receuil de d\'efinitions et de r\'esultats classiques en g\'eom\'etrie non commutative, lui \'evitant de la sorte des renvois trop nombreux \`a la litt\'erature existante.

\chapter{G\'eom\'etrie spectrale non commutative}

\section{Les axiomes de la g\'eom\'etrie non commutative}

\subsection{Les triplets spectraux}

La d\'emarche employ\'ee en g\'eom\'etrie non commutative peut \^etre r\'esum\'ee en deux \'etapes.
\begin{itemize}
\item
Partant d'un espace g\'eom\'etrique $X$ (c'est-\`a-dire que $X$ est un espace mesurable ou topologique, ou encore une vari\'et\'e diff\'erentiable), on formule la th\'eorie math\'ematique sous-jacente \`a l'aide de certaines sous-alg\`ebres de l'alg\`ebre des fonctions \`a valeurs complexes d\'efinies sur $X$.
\item
Ensuite, on g\'en\'eralise les d\'efinitions et th\'eor\`emes de la th\'eorie qui ne font pas appel \`a la commutativit\'e de l'alg\`ebre.
\end{itemize}

Dans le cas des espaces mesurables et des espaces topologiques, cette d\'emarche ne faisait intervenir que l'alg\`ebre des fonctions correspondantes (fonctions mesurables et fonctions continues), repr\'esent\'ee dans un espace de Hilbert $\hh$, ainsi que leurs g\'en\'eralisations non commutatives qui sont les alg\`ebres de von Neumann et les $C^{*}$-alg\`ebres (cf Appendice A).

\par

Par contre, pour construire l'analogue non commutatif du calcul diff\'erentiel et de la g\'eom\'etrie riemanienne, nous sommes amen\'es \`a introduire un autre objet qui est un op\'erateur $\dd$ sur $\hh$. Dans le cas d'une vari\'et\'e munie d'une structure riemannienne \`a spin, cet op\'erateur n'est autre que l'op\'erateur de Dirac.

\par

Par exemple, dans le cas g\'en\'eral non commutatif, nous consid\'erons les \'el\'ements de $\aa$ comme g\'en\'eralisant les fonctions et nous voulons pouvoir d\'efinir la diff\'erentielle d'un \'el\'ement $x\in\aa$, en conservant la plupart des propri\'et\'es et usuelles de la diff\'erentielle. De plus, il est n\'ecessaire que cette d\'efinition de la diff\'erenciation co\"\i ncide, dans le cas commutatif, avec la d\'efinition usuelle. 

\par

La solution de ce probl\`eme fait appel \`a l'op\'erateur de Dirac $\dd$ \cite{bible}. En effet, pour tout $x\in\aa$, nous d\'efinissons la d\'eriv\'ee exterieure de la 0-forme $\pi(x)\in \Omega^{0}_{\dd}(\aa)$, o\`u $\pi$ d\'esigne la repr\'esentation de $\aa$ dans l'alg\`ebre des op\'erateurs sur $\hh$, par
\bbb
d\pi(x)=\lb\dd,\pi(x)\rb,\n
\eee
Nous verrons ult\'erieurement que cette d\'efinition, ainsi que ses g\'en\'eralisations d'ordre superieur, v\'erifie toutes les propri\'et\'es de la diff\'erentielle (cf \S 1.3.1). 

\par

Bornons-nous pour le moment \`a \'etudier le cas commutatif. Nous choisissons une vari\'et\'e riemannienne $\mm$ de dimension $n$ munie d'une structure de spin et $\aa=C^{\infty}(\mm)$ n'est autre que l'alg\`ebre involutive des fonctions lisses d\'efinies sur $\mm$ et \`a valeurs complexes. L'espace de Hilbert $\hh$ est le compl\'et\'e $L^{2}(\mm,\ss)$ pour la norme $L^{2}$ de l'espace des spineurs de carr\'e int\'egrable, sur lequel $\aa$ est repr\'esent\'ee par simple multiplication.  

\par

Enfin, $\dd$ est l'op\'erateur de Dirac, donn\'e en coordonn\'ees locales par $\dd=i\gamma^{\mu}(\partial_{\mu}+\omega_{\mu})$, o\`u $\gamma^{\mu}$ d\'esigne les matrices de Dirac hermitiennes reli\'ees \`a la m\'etrique $g_{\mu\nu}$ par
\bbb
\gamma^{\mu}\gamma^{\nu}+\gamma^{\nu}\gamma^{\mu}=2g^{\mu\nu}.\n
\eee
Bien entendu, $\partial_{\mu}$ est la d\'erivation par rapport \`a la coordonn\'ee locale $x^{\mu}$ et nous utilisons la convention d'Einstein pour les indices grecs uniquement: la somme sur tout indice r\'ep\'et\'e en position haute et basse est sous-entendue. $\omega^{\mu}$ est la connexion spinorielle qui n'est autre que la connexion de Levi-Civita exprim\'ee en partie dans la base des "vielbeins". Cette connexion ne jouera aucun r\^ole avant le chapitre 3, aussi nous contenterons nous de noter qu'elle commute avec l'action des fonctions.

\par

Il est alors tr\`es facile de calculer le commutateur de $\dd$ avec l'action d'une fonction $f\in C^{\infty}(\mm)$,
\bbb
\lb\dd,\pi(f)\rb=\lb i\gamma^{\mu}(\partial_{\mu}+\omega_{\mu}),f\rb=
i\gamma^{\mu}\lp\partial_{\mu}f+f\partial_{\mu}-f\partial_{\mu}\rp=
i\gamma^{\mu}\partial_{\mu}f.\n
\eee
En identifiant $i\gamma^{\mu}$ et la forme diff\'erentielle $dx^{\mu}$, nous pouvons consid\'erer que $[\dd,\pi(f)]$ est analogue \`a la diff\'erentielle $df=\partial_{\mu}fdx^{\mu}$. Nous verrons que cette construction se g\'en\'eralise aux formes diff\'erentielles de tous ordres. 

\par

Le second outil indispensable en g\'eom\'etrie diff\'erentielle est la th\'eorie de l'int\'egration par rapport \`a la forme volume $\sqrt{g}\,d^{n}x$ d\'etermin\'ee par la structure riemannienne. Celle-ci se reformule \`a l'aide de la trace de Dixmier (cf Appendice A) de la mani\`ere suivante,
\bbb
\int_{\mm}\sqrt{g}d^{n}x f=\lp 2^{n-[n/2]}\pi^{n/2}\Gamma(n/2+1)\rp\
Tr_{\omega}\lp \pi(x)|\dd|^{-n}\rp,\label{ts'1}
\eee
o\`u $\t_{\omega}$ est la trace de Dixmier. Cela nous am\`ene \`a d\'efinir l'analogue de l'int\'egrale d'un \'el\'ement $a\in\aa$ par
\bbb
\dix ads^{n}=\lp 2^{n-[n/2]}\pi^{n/2}\Gamma(n/2+1)\rp\
Tr_{\omega}\lp \pi(a)|\dd|^{-n}\rp,\n
\eee
o\`u on a not\'e $ds=\sqrt{D^{2}}$.

\par

L'utilisation de la trace de Dixmier dans le cas g\'en\'eral nous am\`ene \`a imposer une condition de d\'ecroissance aux valeurs propres de $|\dd|^{-n}$ (voir l'axiome de dimension).

\par

Ainsi, il apparait que l'objet fondamental en g\'eom\'etrie non commutative est un triplet $(\aa,\hh,\dd)$, appel\'e triplet spectral \cite{bible}.

\begin{dfi}
Un triplet spectral est un triplet $(\aa,\hh,\dd)$ form\'e d'une alg\`ebre involutive $\aa$ munie d'une repr\'esentation involutive $\pi$ sur un espace de Hilbert $\hh$. $\dd$ est un op\'erateur non born\'e et hermitien \`a r\'esolvante compacte et tel que $\lb\dd,\pi(x)\rb$ soit born\'e pour tout $x\in\aa$.
\end{dfi}

Dans tout ce qui suit, l'alg\`ebre $\aa$ n'intervient que par sa repr\'esentation en tant qu'op\'erateurs sur $\hh$. Si cette repr\'esentation n'est pas injective, son noyau $\jj$ est un id\'eal bilat\`ere et l'alg\`ebre $\aa/\jj$ a une repr\'esentation fid\`ele. Puisque les th\'eories construites \`a partir de $\aa$ et de $\aa/\jj$ sont identiques, nous supposerons toujours $\pi$ fid\`ele.

\par

Les propri\'et\'es de $\dd$ sont motiv\'ees par l'analogie avec le cas classique. Dans ce cas, $\hh$ est l'espace de Hilbert obtenu en compl\'etant pour la norme $L^{2}$ l'espace des sections du fibr\'e spinoriel. Les nouveaux \'el\'ements introduits par cette compl\'etion ne sont pas d\'erivables, aussi l'op\'erateur de Dirac est un op\'erateur non born\'e, d\'efini uniquement sur un sous-espace dense de $\hh$. 

\par

$\dd$ est un op\'erateur \`a r\'esolvante compacte signifie que pour tout nombre complexe non r\'eel $\lambda$ l'op\'erateur $(\dd-\lambda)^{-1}$ est compact. De mani\`ere \'equivalente, $\dd$ a un noyau de dimension finie et son inverse, d\'efini sur l'orthogonal de son noyau, est un op\'erateur compact. En cons\'equence, la suite d\'ecroissante des valeurs propres de $|\dd|^{-1}$ tend vers 0 et son taux de d\'ecroissance permet de minorer la dimension de l'espace.  
\par

Enfin, puisque dans le cas commutatif la diff\'erentielle d'une fonction $f$, donn\'ee par
\bbb
df=\lb\dd,f\rb=i\gamma^{\mu}\partial_{\mu}f,\n
\eee
est un op\'erateur born\'e, nous supposerons que $\lb\dd,\pi(x)\rb$ est born\'e pour tout $x\in\aa$. Cette hypoth\`ese s'av\`erera n\'ecessaire pour pouvoir d\'efinir la th\'eorie de l'int\'egration des formes diff\'erentielles en utilisant la trace de Dixmier.

\par

Lorsque la dimension de l'espace est paire, la construction de l'analogue non commutatif des formes diff\'erentielles et de leur int\'egration fait intervenir de mani\`ere cruciale une $\zzz_{2}$-graduation $\gamma$, qui dans le cas commutatif n'est autre que la graduation $\gamma^{n+1}$ de l'alg\`ebre de Clifford. Aussi supposerons-nous que lorsque $n$ est pair, le triplet spectral $(\aa,\hh,\dd)$ est pair dans le sens suivant. 

\begin{dfi}
Un triplet spectral est pair s'il existe une involution hermitienne $\gamma$, appel\'ee chiralit\'e, qui anticommute avec $\dd$ et commute avec $\pi(x)$ pour tout $x\in\aa$.
\end{dfi}

Dans le cas commutatif, il est possible de caract\'eriser, lorsqu'elles existent, les structures spinorielles sur une vari\'et\'e compacte \`a l'aide des triplets spectraux. Cependant, nous devons supposer que ces derniers satisfont \`a d'autres hypoth\`eses que nous donnons dans la suite \cite{grav}.  


\subsection{Caract\'erisation des vari\'et\'es \`a spin}

Pour commencer, nous allons caract\'eriser les structures de spin sur une vari\'et\'e compacte $\mm$ de dimension $n$ \`a l'aide des triplets spectraux $(\aa,\hh,\dd)$, o\`u $\aa$ est l'alg\`ebre des fonctions $C^{\infty}$ sur $\mm$ \`a valeurs complexes.

\par

La premi\`ere condition que nous devons imposer \`a l'op\'erateur de Dirac est relative au taux de d\'ecroissance des valeurs propres de l'op\'erateur compact $ds=|\dd|^{-1}$. En effet, nous voulons pouvoir d\'efinir l'int\'egrale d'un \'el\'ement de $\aa$ \`a l'aide de la formule (\ref{ts'1}), ce qui implique que $|\dd|^{-n}=ds^{n}$ doit \^etre un infinit\'esimal d'ordre $1$. En prenant la racine $n$-i\`eme, on est amen\'e \`a supposer que $\dd$ satisfait \`a l'axiome suivant.

\begin{axi}[Dimension]
Pour une g\'eom\'etrie de dimension n, $ds=|\dd|^{-1}$ est un infinit\'esimal d'ordre 1/n.
\end{axi}

Nous donnons dans l'Appendice A la d\'efinition et les principales propri\'et\'es des infinit\'esimaux. Rappelons cependant que par d\'efinition, un op\'erateur compact est un infinit\'esimal d'ordre $\alpha>0$ si la suite d\'ecroissante de ses valeurs propres $(\lambda_{k})_{k\in\nn}$ satisfait \`a
\bbb
\lambda_{k}\mathop{=}\limits_{k\rightarrow+\infty}O\lp\frac{1}{k^{\alpha}}\rp.\n
\eee

Ainsi, $\dd$ est un op\'erateur non born\'e dont les valeurs propres croissent au moins aussi vite que la suite $k^{1/n}$ pour une vari\'et\'e de dimension n.
 
\par

Ensuite, il nous faut imposer \`a l'op\'erateur de Dirac d'\^etre un op\'erateur diff\'erentiel d'ordre un. Pour caract\'eriser alg\'ebriquement ces op\'erateurs, supposons que $\dd$ soit un op\'erateur diff\'erentiel et $f\in C^{\infty}(\mm)$ une fonction. Dans ce cas, il est facile de montrer, en utilisant un syst\`eme de coordonn\'ees locales, que $\dd$ est un op\'erateur diff\'erentiel d'ordre $k$ si et seulement si $\lb\dd,f\rb$ est un op\'erateur diff\'erentiel d'ordre $k-1$. Puisque les op\'erateurs diff\'erentiels d'ordre $0$ sont ceux qui commutent avec les fonctions, $\dd$ est un op\'erateur diff\'erentiel d'ordre un si et seulement si il v\'erifie la condition suivante.

\begin{axi}[Ordre un]
Pour toutes fonctions $a,b\in\aa$, on a $\lb\lb\dd,\pi(a)\rb,\pi(b)\rb=0$.
\end{axi}

En g\'en\'eral, l'op\'erateur de Dirac et son carr\'e sont des op\'erateurs non born\'es ce qui induit certaines complications, en particulier lorsqu'on est amen\'e \`a faire commuter $\dd^{2}$ avec des \'el\'ements de $\pi(\aa)$ et $[\dd,\pi(\aa)]$. Pour rem\'edier \`a cela, nous ferons l'hypoth\`ese suivante.

\begin{axi}[R\'egularit\'e]
Pour toute fonction $a\in\aa$, $\pi(a)$ et $\lb\dd,\pi(a)\rb$ appartiennent \`a l'intersection des domaines de toutes les puissances $\delta^{k}$ de la d\'erivation $\delta$ d\'efinie par $\delta(b)=\lb|\dd|,b\rb$ lorsque $b$ appartient \`a l'alg\`ebre engendr\'ee par $\pi(\aa)$ et $\lb\dd,\pi(\aa)\rb$. 
\end{axi}
  
\par

L'axiome suivant permet de caract\'eriser alg\'ebriquement la chiralit\'e $\gamma$ en dimension paire. Il convient de noter que c'est le seul axiome faisant explicitement r\'ef\'erence \`a l'int\'egralit\'e de la dimension. 

\begin{axi}[Orientabilit\'e]
Il existe un cycle de Hochschild $c\in Z_{n}(\aa)$ de dimension $n$ tel que $\pi(c)=1$ lorsque $n$ est impair et $\pi(c)=\gamma$ lorsque $n$ est pair.
\end{axi}

Tout comme $\gamma^{n+1}$, $\gamma$ n'est d\'efini que lorsque la dimension est paire. Aussi, en dimension impaire $\gamma$ est remplac\'e par 1 dans l'axiome pr\'ec\'edent. Pour un produit tensoriel g\'en\'eral $c=a_{0}\ot a_{1} \ot\dots\ot a_{n}$, nous d\'efinissons,
\bbb
\pi(c)=\pi(a_{0})[\dd,\pi(a_{1})]\dots[\dd,\pi(a_{n})],\n
\eee
qui s'\'etend par lin\'earit\'e \`a tous les \'el\'ements de $\aa^{\ot(n+1)}$.

\par

L'homologie de Hochschild est la th\'eorie duale de la cohomologie de Hochschild. Puisque, dans le cas de l'alg\`ebre des fonctions sur une vari\'et\'e, cette derni\`ere est reli\'ee \`a l'int\'egration des formes diff\'erentielles \cite{ihes},
l'homologie de Hochschild correspond aux formes diff\'erentielles (cf Appendice C).

\par

Pour une alg\`ebre $\aa$ quelconque, l'homologie de Hochschild est l'homologie du complexe
\bbb
0\mathop{\leftarrow}\limits^{b}
C_{0}(\aa)\mathop{\leftarrow}\limits^{b}\dots
\mathop{\leftarrow}\limits^{b}C_{n}(\aa)
\mathop{\leftarrow}\limits^{b}C_{n+1}(\aa)
\mathop{\leftarrow}\limits^{b}\dots\n
\eee
o\`u
$C_{n}(\aa)$ est le produit tensoriel de l'alg\`ebre $\aa$ par elle-m\^eme $n+1$ fois et $b:\;\aa^{\ot(n+1)}\rightarrow\aa^{\ot n}$ est d\'efini par
\bbbb
&b(a_{0}\ot a_{1}\ot...\ot a_{n})=
a_{0}a_{1}\ot a_{2}\ot...\ot a_{n}+(-1)^{n}a_{n}a_{0}\ot a_{1}\ot ...\ot a_{n-1}&\n\\
&+\mathop{\sum}\limits_{i=1}^{n-1} (-1)^{i} a_{0}\ot a_{1}\ot...\ot a_{i}a_{i+1}\ot...\ot a_{n}.\n&
\eeee
Par d\'efinition, l'espace $Z_{n}(\aa)$ des cycles de Hochschild de longueur $n$ est l'ensemble des $c\in C^{n}(\aa)$ qui v\'erifient $b(c)=0$.

\par

Pr\'ecisons le lien entre l'homologie de Hochschild et les formes diff\'erentielles lorsque l'alg\`ebre $\aa$ est commutative. Dans ce cas, on d\'efinit un cycle de Hochschild  de dimension $m$ par
\bbb
c=\mathop{\sum}\limits_{\sigma\in S_{m}}\epsilon(\sigma)a_{0}\ot a_{\sigma(1)}\ot\dots\ot a_{\sigma(m)}\n
\eee
pour tous $a_{0},\dots,a_{m}\in\aa$.

\par

En revenant au cas $\aa=C^{\infty}(\mm)$, on repr\'esente ce cycle comme op\'erateur sur $\hh$ \`a l'aide d'une formule analogue \`a celle relative aux formes diff\'erentielles,
\bbb
\pi(c)=\mathop{\sum}\limits_{\sigma\in S_{m}}\epsilon(\sigma)
\pi(a_{0})\lb\dd,\pi(a_{\sigma(1)})\rb\dots\lb\dd,\pi(a_{\sigma(m)})\rb.\n
\eee
Lorsque $\dd$ est l'op\'erateur de Dirac $i\gamma_{\mu}(\partial_{\mu}+\omega_{\mu})$, on a
\bbb
\pi(c)=\mathop{\sum}\limits_{\sigma\in S_{n}}\epsilon(\sigma)
a_{0}\partial_{\mu_{\sigma(1)}}a_{1}i\gamma^{\mu_{\sigma(1)}}\dots
\partial_{\mu_{\sigma(m)}}a i\gamma^{\mu_{\sigma(m)}}.\n
\eee
Si on identifie $i\gamma^{\mu}$ et $dx^{\mu}$ comme nous l'avons fait lorsque nous avons d\'efini la diff\'erentielle d'une fonction $f\in\aa$, $\pi(c)$ s'identifie \`a la forme diff\'erentielle
\bbb
a_{0}\wedge da_{1}\wedge\dots\wedge da_{m}.\n 
\eee

\par

Puisque les formes diff\'erentielles correspondent aux cycles de Hochschild, il est naturel de chercher \`a d\'efinir la d\'eriv\'ee ext\`erieure dans cadre. Celle-ci doit \^etre un op\'erateur de carr\'e nul qui transforme $Z_{n}(\aa)$ en $Z_{n+1}(\aa)$. Ces propri\'et\'es sont v\'erifi\'ees par l'op\'erateur $B:\;\aa^{\ot(n+1)}\rightarrow\aa^{\ot(n+2)}$ d\'efini par
\bbbb
&B(a_{0}\ot a_{1}\ot...\ot a_{m})=&\n\\
&\mathop{\sum}\limits_{i=0}^{m-1}(-1)^{mi}\,
1\ot a_{i}\ot...\ot a_{m}\ot a_{0}\ot...\ot a_{i-1}&\n\\
&-\mathop{\sum}\limits_{i=0}^{m-1}(-1)^{m(i-1)}\,
a_{i-1}\ot 1\ot a_{i}\ot...\ot a_{m}\ot a_{0}\ot...\ot a_{i-2},&\n
\eeee
pour tous $a_{0},...,a_{m}\in\aa$. En effet, on montre que $B^{2}=0$  et que $Bb+bB=0$, ce qui implique que $B(Z_{m}(\aa))\subset Z_{m+1}(\aa)$. De m\^eme on obtient un analogue du produit ext\'erieur en consid\'erant le produit "shuffle" \cite{loday}. 

\par

En utilisant la repr\'esentation \`a l'aide de commutateurs avec l'op\'erateur de Dirac, il est facile de montrer que $\pi(B(c))$ s'identifie \`a
\bbb
da_{0}\wedge da_{1}\wedge\dots\wedge da_{m},\n
\eee
ce qui prouve que $B$ est l'analogue de la d\'eriv\'ee ext\'erieure.

\par

Lorsque l'on applique ces r\'esulats au cycle donn\'e en coordonn\'ees locales par
\bbb
c=\frac{i^{-[n/2]}}{n!}\mathop{\sum}\limits_{\sigma\in S_{n}}\epsilon(\sigma)\,1\ot x^{\mu_{\sigma(1)}}\ot\dots\ot x^{\mu_{\sigma(n)}},\n
\eee
on trouve $\pi(c)=\gamma^{n+1}$ lorsque la dimension est paire et $\pi(c)=1$ dans le cas impair, ce qui prouve que l'axiome de r\'ealit\'e est satisfait lorsque $(\aa,\hh,\dd)$ est le triplet spectral associ\'e \`a une structure spinorielle sur $\mm$.

\par

De plus, cet exemple nous montre que $\gamma$ est analogue \`a la forme volume
\bbb
\frac{1}{n!}\,dx^{0}\wedge\dots\wedge dx^{n},\n
\eee
ce que fera jouer \`a $\gamma$ un r\^ole important dans la th\'eorie de l'int\'egration des formes diff\'erentielles non commutatives.

\par

L'axiome suivant permet de caract\'eriser les spineurs de classe $C^{\infty}$. Ils doivent former un module projectif fini sur $\aa$, qui n'est autre que la formulation alg\`ebrique de la notion de fibr\'e vectoriel (cf Appendice B). 

\begin{axi}[Finitude]
L'intersection des domaines de toutes les puissances $\dd^{k}$ de l'op\'erateur de Dirac est un module projectif fini sur $\aa$ sur lequel la relation suivante d\'efinit une structure hermitienne $(,)$
\bbb
\langle \pi(a)\xi,\zeta\rangle=\int a(\xi,\zeta)ds^{n},\label{cvs1}
\eee
pour tous $\xi,\zeta\in\hh$ et $a\in\aa$.
\end{axi}

Cet axiome fait appel \`a la notion de structure hermitienne sur un module projectif fini que nous d\'ecrirons en d\'etail lorsque nous \'etudierons les th\'eories de jauge (cf \S  1.2.3.).  

\par

En particulier, cela interdit aux valeurs propres de $|\dd|$ de cro\^\i tre plus vite que la suite $k^{1/n}$, car si tel \'etait le cas, les valeurs propres de $ds^{n}$ d\'ecro\^\i traient plus vite que la suite $1/n$, et le second membre de (\ref{cvs1}) serait identiquement nul. Ce r\'esultat est \`a rapprocher de l'in\'egalit\'e obtenue dans \cite{vafa} sur les valeurs propres de l'op\'erateur de Dirac usuel, dont on pourra trouver une d\'emonstration nouvelle, bas\'ee sur la dualit\'e de Poincar\'e, dans \cite{mosco'}.

\par

Il est important de remarquer que cet axiome ne peut \^etre satisfait si la chiralit\'e $\gamma=\pi(c)$ est l'image d'un bord. En effet, si $n>0$, l'application 
\bbb
(a_{0},\cdots,a_{n})\mapsto \t_{\omega}\lp\gamma \pi(a_{0})
\lb\dd,\pi(a_{1})\rb\cdots\lb\dd,\pi(a_{n})\rb|\dd|^{-n}\rp,\n
\eee
o\`u on pose $\gamma=1$ dans le cas impair, est un cocycle de Hochschild (cf \S 1.3.1 pour la d\'emonstration). En cons\'equence, si $c$ \'etait un bord, on aurait 
\bbb
\t_{\omega}\lp\gamma\pi(c)|\dd|^{-n}\rp=\t_{\omega}|\dd|^{-n}=0,\n
\eee
ce qui est en contradiction avec l'axiome de finitude. En dimension $0$ nous avons un r\'esultat similaire sur lequel nous reviendrons (cf. \S 2.1.1).

\par

En association avec l'axiome de r\'egularit\'e, cette condition permet de caract\'eriser les fonctions $C^{\infty}$ comme les \'el\'ements de l'alg\`ebre de von Neumann engendr\'ee par $\pi(\aa)$ satisfaisant \`a l'axiome de r\'egularit\'e. 

\par

La dualit\'e de Poincar\'e est une des principales propri\'et\'es des vari\'et\'es compactes. Celle-ci stipule que la forme bilin\'eaire d\'efinie sur les groupes de cohomologie de de Rham $H^{p}(\mm)$ et $H^{n-p}(\mm)$ par 
\bbb
(\omega,\xi)\mapsto\int_{\mm}\omega\wedge\xi,\n
\eee
o\`u $\omega$ et $\xi$ sont des repr\'esentants respectifs de $H^{p}(\mm)$ et $H^{n-p}(\mm)$, est non d\'eg\'en\'er\'ee. Cette forme bilin\'eaire est apel\'ee forme d'intersection et l'axiome suivant exprime la dualit\'e de Poincar\'e de mani\`ere purement alg\'ebnrique. 

\begin{axi}[Dualit\'e de Poincar\'e]
La matrice de la forme d'intersection $\cap:\; K_{*}(\aa)\times K_{*}(\aa)\rightarrow \zzz$ est non d\'eg\'en\'er\'ee.
\end{axi}

$K_{*}(\aa)$ est une notation collective pour tous les groupes $K_{n}(\aa)$. 
Rappelons que gr\^ace au th\'eor\`eme de periodicit\'e de Bott, $K_{n+2}(\aa)=K_{n}(\aa)$, ce qui permet de r\'eduire notre \'etude aux cas des groupes $K_{0}(\aa)$ et $K_{1}(\aa)$.  

\par

Il est important de noter que nous employons la notation $K_{0}(\aa)$ pour l'alg\`ebre $\aa=C^{\infty}(\mm)$, alors qu'elle n'est en principe d\'efinie que si $\aa$ est une $C^{*}$-alg\`ebre. Cependant, la K-theorie peut \^etre d\'efinie de mani\`ere purement alg\'ebrique, sans faire r\'ef\'erence aux propri\'et\'es topologiques de $C^{*}$-alg\`ebres et nous supposons donc que $K_{0}(\aa)$ d\'esigne la K-th\'eorie alg\'ebrique de $\aa$. En particulier, l'alg\`ebre $C^{\infty}(\mm)$ est une pr\'e $C^{*}$-alg\`ebre dense dans l'alg\`ebre $C^{0}(\mm)$ des fonctions continues sur $\mm$. Par cons\'equent, sa K-th\'eorie alg\'ebrique co\"\i ncide avec la K-th\'eorie de la $C^{*}$-alg\`ebre $C^{0}(\mm)$. De plus, la K-th\'eorie de la $C^{*}$-alg\`ebre $C^{0}(\mm)$ est identique \`a sa K-th\'eorie topologique et classifie les fibr\'es vectoriels complexes sur $\mm$ (cf Appendice A). 

\par

La forme bilin\'eaire sur le groupe $K_{0}(\aa)\op K_{1}(\aa)$ est d\'efinie gr\^ace au couplage avec l'indice de l'op\'erateur $\dd$. D\'ecrivons plus en d\'etail la d\'efinition de la forme d'intersection sur le groupe $K^{0}(\aa)$ en dimension paire.

\par

Les \'el\'ements de $K_{0}$(\aa) sont des classes d'\'equivalence de projecteurs qui correspondent \`a des modules projectifs finis sur $\mm$. Pour d\'efinir une forme bilin\'eaire sur $K_{0}(\aa)$, nous partons d'un couple de deux \'el\'ements de $K_{0}(\aa)$ d\'etermin\'es par les projecteurs $e_{1}$ et $e_{2}$, auxquels nous devons associer un nombre entier $\cap(e_{1},e_{2})$.

\par

Cette construction passe par deux \'etapes. Dans un premier temps, nous construisons un nouvel \'el\'ement de $K_{0}(\aa)$ not\'e $m_{*}(e_{1}\ot e_{2})$. La construction de cet \'el\'ement repose sur les propri\'et\'es math\'ematiques du groupe $K_{0}(\aa)$. En effet, l'association d'un groupe $K_{0}(\aa)$ \`a toute alg\`ebre $\aa$ est un foncteur covariant de la cat\'egorie des alg\`ebres vers la cat\'egorie des groupes. Cela implique que la multiplication $m$, qui est un morphisme de $\aa\ot\aa$ dans $\aa$ puisque $\aa$ est commutative, d\'efinit une application bilin\'eaire de $m_{*}$ $K_{0}(\aa)\times K_{0}(\aa)$ dans $K_{0}(\aa)$. Cela permet d'associer au couple $(e_{1},e_{2})$ un nouveau projecteur que nous notons $e=m_{*}(e_{1}\ot e_{2})$.

\par

Ensuite, nous associons \`a $e$ un entier par l'interm\'ediaire de l'op\'erateur de Dirac. Puisque nous sommes en dimension paire, d\'efinissons deux op\'erateurs $\dd^{+}$ et $\dd^{-}$ adjoints l'un de l'autre par
\bbb
\dd^{+}=\frac{1-\chi}{2}\,\dd\,\frac{1+\chi}{2}\n
\eee
et
\bbb
\dd^{-}=\frac{1+\chi}{2}\,\dd\,\frac{1-\chi}{2}.\n
\eee
Puisque le noyau de $\dd$ est de dimension finie, il en est de m\^eme pour les op\'erateurs $e\dd^{+}e$ et $e\dd^{-}e$. Nous d\'efinissons alors $\cap(e_{1},e_{2})$ comme \'etant l'indice de l'op\'erateur $e\dd^{+}e$. En d'autres termes, $\cap(e_{1},e_{2})$ est \'egal \`a la diff\'erence des dimensions du noyau de $e\dd^{+}e$ et de son adjoint
\bbb
\cap(e_{1},e_{2})=\dim\ker e\dd^{+}e -\dim\ker e\dd^{-}e.\n 
\eee

Le lien entre $K_{0}(\aa)$ et les groupes de cohomologie pairs $H^{2p}(\mm)$ est donn\'e par le caract\`ere de Chern \cite{gilkey}. Ce dernier associe \`a tout fibr\'e vectoriel $\ee$ un \'el\'ement du groupe de cohomologie paire
\bbb
H^{2*}(\mm)=H^{0}(\mm)\op\cdots\op H^{2p}(\mm),\n
\eee
avec $p=[n/2]$, qui est d\'efini par
\bbb
\t\exp\frac{F}{2i\pi}=1\op \frac{F}{2i\pi}\op\cdots\op\frac{1}{(n/2)!}\lp\frac{F}{2i\pi}\rp^{p}.\n
\eee
Dans la relation pr\'ec\'edente, $F$ est la courbure d'une connexion $\nabla$ sur $\ee$ et $\t(F)^{p}$ est une forme diff\'erentielle ferm\'ee de degr\'e $2p$ qui d\'efinit un \'el\'ement du groupe de cohomologie $H^{2p}(\mm)$.

\par

$\t\exp\frac{F}{2i\pi}$ ne d\'epend que de la classe de $\ee$ dans $K_{0}(\aa)$ et le caract\`ere de Chern est un isomorphisme d'anneaux entre $K_{0}(\aa)\ot_{\zzz}\rrr$ et $H^{2*}(\mm)$. Notons que $K_{0}(\aa)\ot_{\zzz}\rrr$ s'obtient \`a partir de $K_{0}(\aa)$ en rempla\c cant chaque facteur $\zzz$ par un facteur $\rrr$ et en supprimant les groupes finis du type $\zzz/p\zzz$ qui correspondent \`a la torsion.

\par

Cela \'etablit, dans le cas pair, le lien entre la dualit\'e de Poincar\'e formul\'ee \`a l'aide des formes diff\'erentielles et la version alg\'ebrique donn\'ee ci-dessus. Dans le cas impair, on utilise un morphisme similaire entre
$K_{i}(\aa)\ot K_{j}(\aa)$ et $K_{i+j}(\aa)$, o\`u $i$, $j$ et $i+j$ sont d\'efinis modulo 2. Ensuite on couple $K_{1}(\aa)$ \`a l'op\'erateur de Dirac \`a l'aide de la version impaire du th\'eor\`eme de l'indice 
.

\par

Le dernier axiome introduit l'op\'erateur antiunitaire $\jj$, encore appel\'e structure r\'eelle. Lorsque $\dd$ est l'op\'erateur de Dirac, $\jj$ n'est autre que l'op\'erateur de conjugaison de charge. Dans le cas  g\'en\'eral, on suppose qu'il v\'erifie les relations suivantes, dont la p\'eriodicit\'e modulo 8 trouve son origine dans la m\^eme p\'eriodicit\'e de l'alg\`ebre de Clifford.   

\begin{axi}[R\'ealit\'e]
Il existe une application antiunitaire $\jj$ de $\hh$ dans lui-m\^eme telle que $\jj\pi(a)\jj^{-1}=\pi(a)^{*}$ pour tout $a\in\aa$. De plus, $\jj$ satisfait aux conditions $\jj^{2}=\epsilon$, $\jj\dd=\epsilon'\dd\jj$ et $\jj\gamma=\epsilon''\gamma\jj$, o\`u $\epsilon,\epsilon'$ et $\epsilon''$ sont donn\'es, selon la valeur de n modulo 8, par le tableau suivant.
\begin{center}
\begin{tabular}{|c|c|c|c|c|c|c|c|c|} 
\hline
n mod 8&0&1&2&3&4&5&6&7\\
\hline
$\epsilon$&1&1&-1&-1&-1&-1&1&1\\
\hline
$\epsilon'$&1&-1&1&1&1&-1&1&1\\
\hline
$\epsilon''$&1&&-1&&1&&-1&\\
\hline
\end{tabular}
\end{center}
\end{axi}

Lorsque toutes ces conditions sont r\'ealis\'ees, on peut montrer le r\'esultat suivant \cite{grav}.

\begin{thm}
Soit $\aa=C^{\infty}(\mm)$ l'alg\`ebre des fonctions lisses sur une vari\'et\'e compacte de dimension $n$ et soit $\tt=(\aa,\hh,\dd)$ un triplet spectral satisfaisant aux axiomes pr\'ec\'edents.
\begin{enumerate}
\item
Il existe une unique m\'etrique riemannienne $g_{\tt}$ sur $\mm$ dont la distance g\'eod\'esique associ\'ee est donn\'ee, entre deux points $x,y\in\mm$, par
\bbb
d(x,y)=
\mathop{\sup}\limits_{f\in\aa,\;||\lb\dd,f\rb||\leq 1}|f(x)-f(y)|.\label{cvs2}
\eee
\item 
La m\'etrique $g_{\tt}$ ne d\'epend que de la classe d'\'equivalence unitaire du triplet spectral $\pi$ et les fibres de l'application qui \`a une classe d'\'equivalence unitaire de triplets spectraux associe la m\'etrique correspondante est une famille finie d'espaces affines $A_{\sigma}$  param\'et\'ee par les structures spinorielles $\sigma$ sur $\mm$. 
\item
La fonctionelle 
\bbb
\tt\mapsto\dix ds^{n-2}
\eee
est quadratique et positive sur chaque $A_{\sigma}$ o\`u elle atteint un minimum $\tt_{\sigma}$ unique \`a une \'equivalence unitaire pr\`es.
\item
$\tt_{\sigma}$ est le triplet spectral obtenu en repr\'esentant $\aa$ par multiplication sur l'espace de Hilbert des spineurs de carr\'e int\'egrable et $\dd$ est l'op\'erateur de Dirac associ\'e \`a la connexion de Levi-Civita de $g_{\tt}$.
\item
La valeur de la fonctionelle $\displaystyle\dix ds^{n-2}$ en $\tt_{\sigma}$ est proportionelle \`a l'action d'Einstein-Hilbert de la m\'etrique $g_{\tt}$,
\bbb
\dix ds^{n-2}=C(n)\,\int_{\mm}\,\sqrt{g}\,dx^{n}\,\rr\n
\eee
avec
\bbb
C(n)=\frac{2^{[n/2]}(n-2)}{12(4\pi)^{n/2}\Gamma(n/2+1)}.\n
\eee
\end{enumerate}
\end{thm}

\par

Ce th\'eor\`eme n\'ecessite deux explications suppl\'ementaires. Deux triplets spectraux $(\aa,\hh,\dd)$ et $(\aa',\hh',\dd')$ sont unitairement \'equivalents (cf \S 1.2.1) s'il existe une application unitaire $U$ de $\hh$ dans $\hh'$ et un isomorphisme d'alg\`ebre $\phi$ de $\aa$ dans $\aa'$ tel que $U\,\dd\,U^{-1}=\dd'$, $U\,\jj\,U^{-1}=\jj'$ et $U\,\pi(x)\,U^{-1}=\pi\circ\phi(x)$ pour tout $x\in\aa$. Bien entendu, dans le cas pair on suppose aussi que $U\,\gamma\,U^{-1}=\gamma$.

\par

Si $P$ est un op\'erateur pseudo-diff\'erentiel d'ordre $-n$ agissant sur les sections d'un fibr\'e vectoriel sur $\mm$, la trace de Dixmier v\'erifie
\bbb
\t_{\omega}(P)=\frac{2}{n}\,\ww(P),\n
\eee
o\`u $\ww$ est le r\'esidue de Wodzicki de $P$ (cf Appendice A). 
\par
Lorsque $P$ est un op\'erateur pseudo-diff\'erentiel d'ordre quelconque, cette relation d\'efinit $\displaystyle\dix P$. De ce fait on a
\bbb
\dix ds^{n-2}=\frac{2}{n}\,\ww(ds^{n-2}).\n
\eee 

\par

Pour le physicien int\'er\'ess\'e par la th\'eorie de la relativit\'e g\'en\'erale, le r\'esultat relatif \`a la distance g\'eod\'esique est particuli\`erement important. C'est pourquoi nous allons donnons une d\'emonstration simplifi\'ee de ce r\'esultat, n'utilisant que les propri\'et\'es \'el\'ementaires de la distance g\'eod\'esique.


\subsection{La distance g\'eod\'esique}

V\'erifions que la relation (\ref{cs2}) nous donne la distance g\'eod\'esique lorsque $\aa$ est l'alg\`ebre des fonction $C^{\infty}$ \`a valeurs complexes sur une vari\'et\'e compacte $\mm$ munie d'une structure spinorielle et $\dd=i\gamma^{\mu}(+\partial_{\mu}+\omega_{\mu})$ est l'op\'erateur de Dirac associ\'e \`a la connexion de Levi-Civita.

\par

Dans ce cas, il est \'evident que la donn\'ee de l'op\'erateur de Dirac permet de reconstruire la m\'etrique $g$, car le symbole principal de l'op\'erateur de Dirac est donn\'e par les matrices $\gamma^{\mu}$ en espace courbe, qui satisfont \`a
\bbb
\gamma^{\mu}\gamma^{\nu}+\gamma^{\nu}\gamma^{\mu}=2g^{\mu\nu}.\n
\eee
Cependant, cette relation est locale car elle fait appel \`a l'usage d'un syst\`eme de coordonn\'ees et ne peut de ce fait \^etre g\'en\'eralis\'ee au cas non commutatif. 

\par

La relation (\ref{cvs2}) du th\'eor\`eme pr\'ec\'edent permet de reconstruire la distance g\'eod\'esique \`a l'aide de l'op\'erateur de Dirac. Nous allons montrer le r\'esultat suivant. 

\begin{thm}
Soit $\lp\aa,\hh,\dd\rp$ un triplet spectral commutatif correspondant \`a une vari\'et\'e \`a spin $\mm$. La distance g\'eod\'esique $L(x,y)$ entre deux points $x$ et $y$ de $\mm$ peut s'exprimer \`a l'aide de la relation
\bbb
L(x,y)=
\mathop{\sup}\limits_{f\in\aa,\;||\lb\dd,f\rb||\leq 1}|f(x)-f(y)|.\label{dg1}
\eee  
\end{thm}

\demo

Dans un premier temps, nous allons montrer que l'on peut se ramener \`a chercher la borne sup\'erieure sur les fonctions \`a valeurs r\'eelles. Soit $\aa_{\rrr}$ la sous-alg\`ebre de $\aa=C^{\infty}(\mm)$ form\'ee des fonctions \`a valeurs r\'eelles. Puisque $\aa_{\rrr}\subset\aa$, il est clair que
\bbb
\mathop{\sup}\limits_{f\in\aa_{\rrr},\;||\lb\dd,f\rb||\leq 1}|f(x)-f(y)|
\leq\mathop{\sup}\limits_{f\in\aa,\;||\lb\dd,f\rb||\leq 1}|f(x)-f(y)|.
\eee
Pour montrer l'\'egalit\'e, fixons deux points $x$ et $y$ de $\mm$ et consid\'erons une fonction quelconque $f$ \`a valeurs complexes satisfaisant \`a $||[\dd,f]||\leq 1$. En rempla\c cant $f$ par la fonction $g$ d\'efinie par
\bbb
g(z)=e^{-i\theta}\lp f(z)-f(y)\rp
\eee
o\`u $\theta$ est un argument de $f(x)-f(y)$, il est clair que l'on a toujours
\bbb
|f(x)-f(y)|=|g(x)-g(y)|\;\;\;\;\mathrm{ainsi}\;\mathrm{que}\;\;\;\;||[\dd,g]||\leq 1.
\eee
Pour obtenir une fonction r\'eelle, d\'efinissons
\bbb
h=\frac{g+\ov{g}}{2}.
\eee
Puisque $g(x)-g(y)$ est r\'eel, on a
\bbb
|h(x)-h(y)|=|g(x)-g(y)|.
\eee
D'autre part
\bbb
||[\dd,h]||\leq\frac{1}{2}||[\dd,g]||+\frac{1}{2}||[\dd,\ov{g}]||\leq 1,
\eee
ce qui implique que
\bbb
|f(x)-f(y)|\leq \mathop{\sup}\limits_{u\in\aa_{\rrr},\;||\lb\dd,u\rb||\leq 1}|u(x)-u(y)|
\eee
et prouve ainsi l'\'egalit\'e annonc\'ee.
 
\par

Ensuite, nous allons majorer le second membre de (\ref{dg1}) par la distance g\'eod\'esique. Pour cela, consid\'erons une fonction r\'eelle $f$ telle que $||[\dd,f]||\leq 1$ et remarquons que $\lb\dd,f\rb=i\gamma^{\mu}\partial_{\mu}f$. On en d\'eduit que
\bbb
||\lb\dd,f\rb||^{2}=
\mathop{\sup}\limits_{x\in\mm}g^{\mu\nu}(x)\partial_{\mu}f(x)\partial_{\nu}f(x)
\leq 1.
\eee
Cette relation est facile \`a montrer en utilisant les propri\'et\'es des matrices de Dirac et la relation  $||T||^{2}=||TT^{*}|||$ valable pour tout op\'erateur born\'e $T$. 

\par

Soit $f$ une fonction satisfaisant \`a $||\lb\dd,f\rb||\leq 1$ et $c:\,[0,1] \rightarrow\mm$ une g\'eod\'esique reliant $x$ \`a $y$. On a
\bbb
f(y)-f(x)=\int_{0}^{1}dt\,\frac{dc^{\mu}}{dt}\partial_{\mu}f(c(t)).
\eee
Par Cauchy-Schwarz on obtient 
\bbb
|\frac{dc^{\mu}}{dt}\partial_{\mu}f(c(t))|\leq
\sqrt{g_{\mu\nu}\frac{dc^{\mu}}{dt}\frac{dc^{\nu}}{dt}}
\sqrt{g^{\rho\sigma}\partial_{\rho}f\partial_{\sigma}f},
\eee
ce qui prouve que
\bbb
|f(y)-f(x)|\leq \int_{0}^{1}dt\,\sqrt{g_{\mu\nu}\frac{dc^{\mu}}{dt}\frac{dc^{\nu}}{dt}}=L(x,y),
\eee
apr\`es avoir utilis\'e
\bbb
\mathop{\sup}\limits_{x\in\mm}g^{\rho\sigma}(x)\partial_{\rho}f(x)
\partial_{\sigma}f(x)
\leq 1.
\eee

Pour montrer l'\'egalit\'e, il nous suffit d'exhiber une fonction $f$ telle que
\bbb
|f(x)-f(y)|=L(x,y)\label{dg2}
\eee
satisfaisant \`a $||\lb\dd,f\rb||\leq 1$. Le choix le plus simple consiste \`a fixer $y$ et \`a consid\'erer la fonction $f$ d\'efinie par $f(z)=L(z,y)$ pour tout $z\in\mm$. $f$ satisfait \`a (\ref{dg2}) et il nous reste \`a montrer que $||[\dd,f]||\leq 1$.

\par

Puisque $L(z+dz,y)$ est la longueur du plus court chemin joignant $z+dz$ et $y$, on a, par l'in\'egalit\'e triangulaire,
\bbb
L(z+dz,y)\leq L(z,y)+L(z+dz,z),\label{dg0}
\eee

\begin{figure}[H]
\centering
\epsfig{file={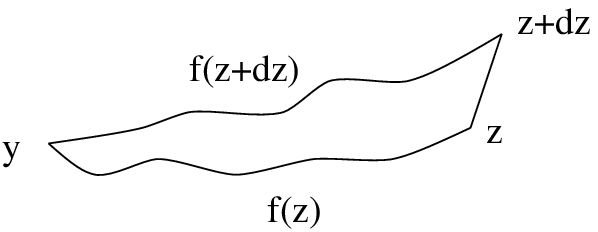},width=6cm}
\end{figure}

Lorsque le point $z=z+dz$ s'obtient \`a partir de $z$ par le flot d\'etermin\'e par le champ de vecteur $\partial^{\mu}$, on a, pour une transformation infinit\'esimale d\'etermin\'ee par $\epsilon$, $dz^{\mu}=\epsilon\partial^{\mu}f$. L'in\'egalit\'e (\ref{dg0}) s'\'ecrit donc
\bbb
f(z^{\mu}+\epsilon\partial^{\mu}f)\leq f(z)+\epsilon\sqrt{g_{\mu\nu}\partial^{\mu}f\partial^{\nu}f}+O(\epsilon^{2}).
\eee
En d\'eveloppant le premier membre au premier ordre en $\epsilon$, on obtient
\bbb
g_{\mu\nu}(z)\partial^{\mu}f(z)\partial^{\nu}f(z)\leq 1.
\eee
Puisque cela est vrai pour tout $z\in\mm$, on a $||\dd,f||\leq 1$.
\edemo

Cette formulation de la distance g\'eod\'esique s'\'etend sans difficult\'e au cas non commutatif. En effet, les points de $\mm$ sont les \'etats purs de l'alg\`ebre des fonctions sur $\mm$ (cf Appendice A), et la proposition suivante g\'en\'eralise la distance g\'eod\'esique \`a l'espace des \'etats purs d'une alg\`ebre non commutative.   

\begin{pro}
Soit $(\aa,\hh,\dd)$ un triplet spectral. La fonction 
\bbb
d(\phi,\psi)=
\mathop{\sup}\limits_{a\in\aa,\;||\lb\dd,\pi(a)\rb||\leq 1}|\phi(a)-\psi(a)|\n
\eee  
d\'efinit, lorsqu'elle est finie, une distance sur l'espace des \'etats purs de $\aa$.
\end{pro}

Nous verrons au cours du chapitre suivant des exemples de distances ainsi d\'efinies lorsque $\aa$ est une alg\`ebre de dimension finie.


\subsection{Les axiomes de la g\'eom\'etrie non commutative}

Les 7 axiomes pr\'ec\'edents permettent de caract\'eriser les triplets spectraux $(\aa,\hh,\dd)$ qui correspondent \`a la g\'eom\'etrie spinorielle sur une vari\'et\'e compacte. Pour \'etendre ces axiomes au cas d'une alg\`ebre non commutative, il faut les transformer de la mani\`ere suivante \cite{grav}.

\par

La principale modification, sur laquelle est bas\'ee celle de tous les autres axiomes, est l'introduction, dans l'axiome 7 (R\'eali\'t\'e) d'un op\'erateur $\jj$ qui est analogue \`a l'involution de Tomita-Takesaki (cf Appendice A). 

\bigskip

\noindent
{\bf Axiome 7' (R\'ealit\'e)}
\begin{it}
Il existe une application antiunitaire $\jj$ de $\hh$ dans lui-m\^eme telle que $\lb\jj\pi(a)\jj^{-1},\pi(b)\rb=0$ pour tous $a,b\in\aa$. De plus, $\jj$ satisfait aux conditions $\jj^{2}=\epsilon$, $\jj\dd=\epsilon'\dd\jj$ et $\jj\gamma=\epsilon''\gamma\jj$, o\`u $\epsilon,\epsilon'$ et $\epsilon''$ sont donn\'es, selon la valeur de n modulo 8, par le tableau suivant.
\begin{center}
\begin{tabular}{|c|c|c|c|c|c|c|c|c|} 
\hline
n mod 8&0&1&2&3&4&5&6&7\\
\hline
$\epsilon$&1&1&-1&-1&-1&-1&1&1\\
\hline
$\epsilon'$&1&-1&1&1&1&-1&1&1\\
\hline
$\epsilon''$&1&&-1&&1&&-1&\\
\hline
\end{tabular}
\end{center}
\end{it}
\bigskip

Puisque $\jj\pi(a)\jj^{-1}$ commute avec $\pi(b)$ pour tous $a,b\in\aa$, $\jj$ est tr\`es similaire \`a l'op\'erateur de Tomita-Takesaki qui permet de passer de l'alg\`ebre de von Neumann engendr\'ee par $\pi(\aa)$ \`a son commutant. Cependant, le commutant de $\pi(\aa)$ est en g\'en\'eral plus grand que $\jj\pi(\aa)\jj^{-1}$, ce qui implique que $\jj$ n'est pas \'egal \`a l'op\'erateur de Tomita-Takesaki en g\'en\'eral.

\par

En outre, l'op\'erateur $\jj$ permet de repr\'esenter sur $\hh$ l'alg\`ebre $\aa\ot\aa^{op}$ par
\bbb
\tilde{\pi}(\aa\ot\aa^{op})=\pi(a)\jj\pi(b^{*})\jj^{-1},\n
\eee
o\`u $\aa^{op}$ est une alg\`ebre identique \`a $\aa$ en tant q'espace vectoriel, mais dont le produit de $a$ par $b$ est d\'efini par $ba$ au lieu de $ab$. De fa\c con \'equivalente, $\pi$ et $\jj$ munissent $\hh$ d'une structure de bimodule sur $\aa$, qui nous sera utile lorsque nous construirons des mod\`eles en physique des particules.

\par

L'axiome 1 (dimension) ne fait appel qu'au spectre de l'op\'erateur de Dirac et reste inchang\'e dans le cas non commutatif.

\par

L'axiome 2 (condition d'ordre un) stipule que l'op\'erateur de Dirac est un op\'erateur diff\'erentiel du premier ordre. La g\'en\'eralisation de cette notion en g\'eom\'etrie non commutative requiert l'introduction de bimodules \cite{masson}. Cela nous m\`ene \`a la d\'efinition suivante.

\begin{dfi}
Soient $\aa$ et $\bb$ deux alg\`ebres et $\mm$ un $(\aa,\bb)$-bimodule. Une application lin\'eaire $D$ de $\mm$ dans lui-m\^eme est un op\'erateur du premier ordre si elle v\'erifie
\bbb
D(a\xi b)+aD(\xi)b=aD(\xi b)+D(a\xi) b,\n
\eee
pour tous $a,b\in\aa$ et $\xi\in\mm$.
\end{dfi}

L'axiome de r\'ealit\'e permet de munir $\hh$ d'une structure de bimodule sur $\aa$, aussi la g\'en\'eralisation de la condition d'ordre  un est donn\'ee par l'axiome suivant.

\bigskip

\noindent
{\bf Axiome 2' (Condition d'ordre un)}
\begin{it}
L'op\'erateur de Dirac v\'erifie 
\bbb
\lb\lb\dd,\pi(a)\rb,\jj\pi(b)\jj^{-1}\rb=0\n
\eee
pour tous $a,b\in\aa$.
\end{it}
\bigskip

L'axiome 3 (r\'egularit\'e) garde tous son sens lorsque $\aa$ est non commutative et ne subit aucune modification. 

\par

L'axiome 4 (orientabilit\'e) ne peut pas \^etre conserv\'e sous la m\^eme forme en g\'eom\'etrie non commutative. En effet, si on se place en dimension $n=0$, tout \'el\'ement de $\aa$ doit \^etre consid\'er\'e comme un cycle de dimension 0. Dans ce cas, l'axiome d'orientabilit\'e stipule que $\gamma=\pi(a)$ pour un $a\in\aa$. Par l'axiome de r\'ealit\'e on a $\jj\gamma=\gamma\jj$, d'o\`u $\gamma=\jj\pi(a)\jj^{-1}$. La condition d'ordre un nous donne $0=\lb\lb\dd,\pi(a)\rb,\jj\pi(a)\jj^{-1}\rb=4\dd$ puisque $\gamma^{2}=1$ et $\gamma\dd+\dd\gamma=0$, ce qui implique que le seul op\'erateur de Dirac compatible avec ces axiomes en dimension 0 est $\dd=0$.

\par

La n\'ecessaire modification de l'axiome 4 est obtenue en rempla\c cant l'homologie de Hochschild par l'homologie de Hochschild \`a valeurs dans un bimodule.

\begin{dfi}
Soit $\aa$ une alg\`ebre, $\mm$ bimodule sur $\aa$ et $C_{n}(\aa,\mm)$ l'espace vectoriel d\'efini par
\bbb
C_{n}(\aa,\mm)=\mm\ot\underbrace{\aa\ot\cdots\ot\aa}_{n\,fois}.\n
\eee
L'homologie de Hochschild de $\aa$ \`a valeurs dans $\mm$ est l'homologie du complexe
\bbb
0\mathop{\leftarrow}\limits^{b}
C_{0}(\aa,\mm)\mathop{\leftarrow}\limits^{b}\dots
\mathop{\leftarrow}\limits^{b}C_{n}(\aa,\mm)
\mathop{\leftarrow}\limits^{b}C_{n+1}(\aa,\mm)
\mathop{\leftarrow}\limits^{b}\dots\n
\eee
o\`u $b;\mm\ot \aa^{\ot n}\rightarrow \mm\ot\aa^{\ot (n-1)}$ est d\'efini par
\bbbb
&b(m\ot a_{1}\ot...\ot a_{n})=
ma_{1}\ot a_{2}\ot...\ot a_{n}+(-1)^{n}a_{n}m\ot a_{1}\ot ...\ot a_{n-1}&\n\\
&+\mathop{\sum}\limits_{i=1}^{n-1} (-1)^{i} m\ot a_{1}\ot...\ot a_{i}a_{i+1}\ot...\ot a_{n}.\n&
\eeee
\end{dfi}

Nous munissons $\aa\ot\aa^{op}$ d'une structure de bimodule par
\bbb
x\lp a\ot b\rp y=xay\ot b\n
\eee
pour $a,b,x,y\in\aa$ ce qui nous permet d'\'etendre l'axiome d'orientabilit\'e au cas non commutatif.

\bigskip

\noindent
{\bf Axiome 4' (Orientabilit\'e)}
\begin{it}
En dimension $n$, $\gamma$ est l'image d'un cycle de Hochschild de dimension $n$ \`a valeurs dans $\aa\ot\aa^{op}$.
\end{it}

\bigskip

Un \'el\'ement $c=\lp x\ot y\rp\ot a_{1}\ot\cdots\ot a_{n}\in C_{n}\lp\aa,\aa\ot\aa^{op}\rp$ est repr\'esent\'e en tant qu'op\'erateur sur 
$\hh$ par
\bbb
\pi(c)=\pi(x)\jj\pi(y)\jj^{-1}\lb\dd,\pi(a_{1})\rb\cdots\lb\dd,\pi(a_{n})\rb.\n
\eee

\par

L'axiome 5 (finitude) ne subit aucune modification et il est facile de montrer qu'il est incompatible avec la trivialit\'e du cycle repr\'esentant $\gamma$.

\par

Lorsque $\aa$ est non commutative, la multiplication n'est plus un morphisme d'alg\`ebre et ne peut pas induire l'application $m_{*}$ de $K_{*}(\aa)\times K_{*}(\aa)$ dans $K_{*}(\aa)$ dont nous avons besoin pour formuler alg\'ebriquement la dualit\'e de Poincar\'e.

\par

Partant de deux projections $e_{1}$ et $e_{2}$ d\'efinissant deux classes dans $K^{0}(\aa)$, nous leurs associons la nouvelle projection $e_{1}\ot e_{2}$ qui d\'efinit un \'el\'ement de $K^{0}(\aa\ot\aa^{op})$. Cet \'el\'ement est repr\'esent\'e par $\pi(e)=\pi(e_{1})\jj\pi(e_{2})\jj^{-1}$ et on d\'efinit
\bbb
\cap(e_{1},e_{2})=\dim\ker\lp\pi(e)\dd^{+}\pi(e)\rp 
-\dim\ker\lp \pi(e)\dd^{-}\pi(e)\rp.\n
\eee
Dans le cas impair, la m\^eme d\'emarche est employ\'ee en rempla\c cant les projections par des unitaires.

\par

Les axiomes pr\'ec\'edents sont appel\'es axiomes de la g\'eom\'etrie non commutative et nous m\`enent \`a la d\'efinition suivante.

\begin{dfi}
Une g\'eom\'etrie non commutative de dimension $n$ est un triplet spectral $(\aa,\hh,\dd)$ satisfaisant aux axiomes pr\'ec\'edents.
\end{dfi} 

Parfois, on dit aussi que les axiomes pr\'ec\'edents munissent le triplet spectral d'une structure r\'eelle.

\par

Dans tous ce qui suit, les triplets spectraux satisferons toujours, sauf mention explicite du contraire, les axiomes de la g\'eom\'etrie non commutative. Aussi, nous continuerons de les appeler triplets spectraux, ce qui sous-entendra toujours que les axiomes pr\'ec\'edents sont v\'erifi\'es.

\par

Il est int\'eressant de remarquer que m\^eme si $\aa$ est commutative, il peut \^etre n\'ecessaire de supposer que les triplets spectraux $(\aa,\hh,\dd)$ satisfont aux axiomes pr\'ec\'edents et non aux hypoth\`ese du th\'eor\`eme caract\'erisant les vari\'et\'e \`a spin (\,1.1.2). En effet, ces derni\`eres sont plus restrictives que les axiomes pr\'ec\'edents et ne peuvent pas permettre de construire des g\'eom\'etries int\'eressantes dans certains cas (cf \S\,2.1.5 ).

\par

Tout au long de cette revue des axiomes de la g\'eom\'etrie non commutative, nous n'avons donn\'e aucun exemple de triplet spectral non commutatif. Cette lacune sera combl\'ee au cours des chapitres suivants, dans lesquels nous \'etudierons les triplets spectraux finis, les produits de ces derniers avec les espaces ordinaires ainsi que les tores non commutatifs.

\section{Sym\'etries}

\subsection{Sym\'etries et automorphismes}

Pour d\'ecrire les sym\'etries d'un "espace non commutatif" associ\'e \`a une alg\`ebre de coordonn\'ees $\aa$, il est commode d'examiner d'abord le cas commutatif. En se basant sur le th\'eor\`eme de Gelfand-Na\"\i  marck (cf Appendice A), il est facile de montrer la proposition suivante.

\begin{pro}
Soit $\aa=C^{0}(X)$ une $C^{*}$-alg\`ebre commutative. Alors le groupe des automorphismes de $\aa$ est isomorphe au groupe des hom\'eomorphismes de l'espace topologique compact $X$.
\end{pro}

\demo
En effet, si $\phi$ est un hom\'eomorphisme de $X$, alors on d\'efinit un automorphisme $\Psi_{\phi}$ de $\aa$ par $\Psi_{\phi}(f)=f\circ\phi^{-1}$. Il est clair que l'application $\phi\mapsto\Psi_{\phi}$ est un morphisme du groupe des hom\'eomorphismes de $X$ vers le groupe des automorphismes de $\aa$. 

\par

Pour montrer que c'est un isomorphisme, nous d\'efinissons sa r\'eciproque comme suit. Soit $\psi$ un automorphisme de $\aa$ et $\chi$ un caract\`ere de $\aa$. Alors $\chi\circ\psi^{-1}$ est un autre caract\`ere de $\aa$ et l'application $\Phi_{\psi}\chi\mapsto\chi\circ\psi^{-1}$, d\'efinit, par application du th\'eor\`eme de Gelfand-Na\"\i marck, un hom\'eomorphisme de $X$.

\par

L'application $\psi\mapsto\Phi_{\psi}$ est un morphisme de groupe dont l'inverse est $\phi\mapsto\Psi_{\phi}$, ce qui prouve l'isomorphisme entre le groupe des hom\'eomorphismes de $X$ et le groupe des automorphismes de $\aa$.
\edemo

Ce r\'esultat nous am\`ene \`a d\'efinir les sym\'etries d'un "espace non commutatif" comme \'etant les automorphismes de son alg\`ebre de coordonn\'ees.
Ce faisant, nous ne tenons compte que de la structure de l'alg\`ebre des coordonn\'ees $\aa$ et nous ne prenons pas en compte la structure compl\`ete du triplet spectral $(\aa,\hh,\dd)$. Pour rem\'edier \`a cela, nous introduisons la notion d'\'equivalence de triplets spectraux.

\begin{dfi}
Deux triplets spectraux $(\aa,\hh,\dd)$ et $(\aa',\hh'\dd')$ sont \'equivalents s'il existe une application unitaire $U:\;\hh\rightarrow\hh'$ et un isomorphisme d'alg\`ebre $\phi:\;\aa\rightarrow\aa'$ tel que
\bbbb
\pi'\circ\phi&=&U\pi U^{-1},\n\\
\dd'&=&U\dd U^{-1},\n\\
\jj'&=&U\jj U^{-1},\n
\eeee
ainsi que
\bbb
\gamma'=U\gamma U^{-1}\n
\eee
dans le cas pair.
\end{dfi}

Une sym\'etrie d'un triplet spectral correspond alors \`a un automorphisme $\phi$ de l'alg\`ebre $\aa$ et un op\'erateur unitaire $U$ sur $\hh$ tels que les relations pr\'ec\'edentes soient v\'erifi\'ees avec $\aa=\aa'$, $\hh=\hh'$ et $\pi=\pi'$. On s'autorise cependant \`a modifier les op\'erateurs $\gamma$, $\jj$ et $\dd$.

\par

Par exemple dans le cas commutatif, si $\phi$ est un diff\'eomorphisme de la vari\'et\'e $\mm$, alors il induit un automorphisme de l'alg\`ebre $C^{\infty}(\mm)$ donn\'e par $f\mapsto f\circ\phi^{-1}$. L'op\'erateur unitaire $U_{\phi}$ agit sur un spineur $\Psi\in\hh$ par
\bbb
U_{\phi}\lp\Psi\rp(x)=\Psi\lp\phi^{-1}(x)\rp.\n
\eee

\subsection{Automorphismes int\'erieurs et fluctuations de la m\'etrique}

D\'es lors que l'alg\`ebre des coordonn\'ees est non commutative, il est tr\`es facile de construire une classe d'automorphismes appel\'es automorphismes int\'erieurs. En effet, pour tout unitaire $u\in\aa$, on d\'efinit un automorphisme int\'erieur par
\bbb
x\in\aa\mapsto uxu^{*}\in\aa.\n
\eee 
Cet automorphisme donne naissance \`a une sym\'etrie impl\'ement\'ee par l'op\'erateur unitaire $U=\pi(u)\jj\pi(u)\jj^{-1}$, puisque l'on a
\bbb
U\pi(x)U^{-1}=\pi(u)\pi(x)\pi(u^{-1})=\pi(uxu^{-1})\n
\eee
pour tout $x\in\aa$. Notons que nous avons utilis\'e de fa\c con essentielle l'axiome de r\'ealit\'e pour faire commuter $\pi(x)$ et $\pi(u)$ avec $\jj\pi(u^{-1})\jj^{-1}$.

\par

Sous cette transformation, $\gamma$ et $\jj$ restent inchang\'es et l'op\'erateur de Dirac se transforme en 
\bbbb
&\lp\pi(u)\jj\pi(u)\jj^{-1}\rp\dd\lp\pi(u)\jj\pi(u)\jj^{-1}\rp^{-1}=&\n\\
&\lp\jj\pi(u)\jj^{-1}\rp\dd\lp\jj\pi(u)\jj\rp^{-1}+
\lp\jj\pi(u)\jj^{-1}\rp\lb\dd,\pi(u)\rb\lp\jj\pi(u)\jj^{-1}\rp^{-1}&\n\\
&=\dd+\pi(u)\lb\dd,\pi(u^{-1})\rb+\jj\pi(u)\lb\dd,\pi(u^{-1})\rb\jj^{-1}.&\n
\eeee
Remarquons que pour montrer ces deux relations, il faut utiliser l'axiome de r\'ealit\'e et la condition d'ordre un.

\par

Conform\'ement aux habitudes de la th\'eorie des champs, cette transformation est compens\'ee par l'introduction de champs de jauge \`a partir desquels on forme un op\'erateur de Dirac covariant. Dans le cas le plus simple auquel nous  nous limitons ici, un champ de jauge n'est autre qu'une 1-forme hermitienne.

\begin{dfi}
L'espace $\Omega_{\dd}^{1}(\aa)$ des 1-formes est form\'e des op\'erateurs sur $\hh$ qui s'\'ecrivent sous la forme
\bbb
\mathop{\sum}\limits_{i}\pi(a_{i})\lb\dd,\pi(b_{i})\rb,\label{ai1}
\eee
avec $a_{i},b_{i}\in\aa$.
\end{dfi}

Par cons\'equent, un champ de jauge est un op\'erateur sur $\hh$ qui peut s'\'ecrire sous la forme (\ref{ai1}) et qui satisfait \`a $A=A^*$. 

\begin{pro}
Si $(\aa,\hh,\dd)$ est un triplet spectral, alors $(\aa,\hh,\dd_{A})$ est un triplet spectral de m\^eme dimension avec
\bbb
\dd_{A}=\dd+A+\jj A\jj^{-1},\n
\eee
o\`u $A$ est un champ de jauge.
\end{pro}

L'op\'erateur $\dd_{A}$ est appel\'e op\'erateur de Dirac covariant.

\par

Par d\'efinition, les transformations de jauge sont les transformations d\'etermin\'ees par l'op\'erateur unitaire $\pi(u)\jj\pi(u)\jj^{-1}$. Sous une telle transformation, l'op\'erateur $\dd_{A}$ devient
\bbb
\dd+\pi(u)A\pi(u^{-1})+\pi(u)\lb\dd,\pi(u^{-1})\rb+\jj\lp \pi(u)A\pi(u^{-1})+\pi(u)\lb\dd,\pi(u^{-1})\rb\rp\jj^{-1}=\dd_{A'}\n
\eee
avec
\bbb
A'=\pi(u)A\pi(u^{-1})+\pi(u)\lb\dd,\pi(u^{-1})\rb.\n
\eee
Par cons\'equent, la loi de transformation de $A$ est
\bbb
A\mapsto \pi(u)A\pi(u^{-1})+\pi(u)\lb\dd,\pi(u^{-1})\rb,\n
\eee
ce qui est tout \`a fait similaire \`a la loi de transformation habituelle
\bbb
A\mapsto uAu^{-1}+udu^{-1},\n
\eee
si nous identifions la diff\'erentielle de $u$ avec le commutateur $\lb\dd,\pi(u)\rb$.

\par

Puisque tout triplet spectral permet de d\'efinir une distance sur l'espace des \'etats purs de $\aa$, remplacer $\dd$ par $\dd_{A}$ induit une modification de la m\'etrique appel\'ee "fluctuation interne de la m\'etrique" \cite{grav}.

\par

Il est int\'eressant de remarquer que les fluctuations de la m\'etrique sont covariantes sous les transformations de jauge. En effet, si $\phi$ est un \'etat pur de $\aa$, alors $\phi_{u}$ d\'efini par
\bbb
\phi_{u}(x)=\phi(uxu^{-1})\n
\eee
pour tout $x\in\aa$ est aussi un \'etat pur sur $\aa$, appel\'e transform\'e de jauge de $\phi$.

\par

Pour tout champ de jauge $A$, notons $d_{A}(\phi,\psi)$ la distance d\'efinie sur l'espace des \'etats purs de $\aa$ par
\bbb
d_{A}(\phi,\psi)=
\mathop{\sup}\limits_{x\in\aa,\;||\lb\dd_{A},\pi(x)\rb||\leq 1}
|\phi(x)-\psi(x)|.\n
\label{ai4}
\eee  

\begin{pro}
Sous une transformation de jauge, la distance est covariante
\bbb
d_{\pi(u)A\pi(u^{-1})+\pi(u)\lb\dd,\pi(u^{-1})\rb}
(\phi,\psi)=d_{A}(\phi_{u},\psi_{u})
\eee 
\end{pro}

\demo
Pour montrer cela, posons $y=uxu^{-1}$ dans la d\'efinition de $d_{A}(\phi_{u},\psi_{u})$, on obtient
\bbb
d_{A}(\phi_{u},\psi_{u})=
\mathop{\sup}\limits_{y\in\aa,\;||\lb\dd_{A},\pi(u^{-1}yu)\rb||\leq 1}
|\phi(y)-\psi(y)|.\n
\eee
Ensuite, on utilise la relation 
\bbbb
\lb\dd_{A},\pi(u^{-1}yu) \rb&=&
\pi(u^{-1})\lb\pi(u)\dd_{A}\pi(u^{-1}),\pi(y)\rb\pi(u),\n\\
&=&\pi(u^{-1})\lb\dd_{\pi(u)A\pi(u^{-1})+\pi(u)\lb\dd,\pi(u^{-1})\rb},\pi(y)\rb\pi(u)\n
\eeee
qui m\`enent directement au r\'esultat annonc\'e.
\edemo

\par

Tout comme dans le cas commutatif, ces lois de transformation nous garantissent l'invariance de l'action fermionique. En effet, les \'el\'ements de $\hh$ jouent le r\^ole des spineurs en g\'eom\'etrie non commutative, et l'action fermionique est
\bbb
S_{fermionique}[\Psi]=\langle\Psi, \dd_{\aa}\Psi\rangle.\label{ai2}
\eee
Sous une transformation de jauge, on a
\bbb
\Psi\mapsto \pi(u)\jj\pi(u)\jj^{-1}\Psi,\label{ai3}
\eee 
ce qui prouve, compte tenu de la loi de transformation de l'op\'erateur de Dirac covariant, que $S_{fermionique}[\Psi]$ est invariant de jauge.

\par

Si la d\'efinition de l'action fermionique par la relation (\ref{ai2}) est une g\'en\'eralisation imm\'ediate de l'action de Dirac en th\'eorie des champs usuelle, la d\'efinition de l'action bosonique est nettement moins \'evidente.

\par

Tout d'abord, d\'efinissons les champs bosoniques comme \'etant l'ensemble des champs servant \`a param\'etrer l'op\'erateur de Dirac covariant $\dd_{A}$. Bien entendu, cela inclut le champ de jauge $A$ mais peut aussi contenir d'autres champs qui ne sont pas de m\^eme nature, comme par exemple la t\'etrade $e_{\alpha}^{\mu}$ ou la connexion de Levi-Civita $\omega_{\mu}$ qui entrent dans la d\'efinition de l'op\'erateur de Dirac dans le cas commutatif.

\par

Si on exclut ce type de champs, une action de type "Yang-Mills" peut \^etre construite pour le champ de jauge $A$. Cette action poss\`ede beaucoup de propri\'et\'es r\'eminiscentes de son analogue commutatif comme par exemple la positivit\'e ou l'invariance de jauge (cf \S\,1.3.3). 

\par

Motiv\'es par l'unification de la gravitation et des th\'eories de Yang-Mills, A. Chamseddine et A. Connes ont propos\'e de baser l'action bosonique sur le principe suivant \cite{spec}:

\bigskip

\noindent
{\bf Principe d'action spectrale}
\begin{it}
L'action bosonique ne d\'epend que du spectre de l'op\'erateur de Dirac covariant
\bbb
\dd_{A}=\dd+A+\jj A+\jj^{-1},\n
\eee
o\`u $A$ est un champ de jauge.
\end{it}

\bigskip

A partir de ce principe, l'action du mod\`ele standard coupl\'e \`a la gravitation de Weyl peut \^etre construite. Nous \'etudierons ce type de construction au cours du chapitre 3.

Terminons par l'\'etude de l'exemple le plus simple de g\'eom\'etrie non commutative qui puisse \^etre associ\'e \`a une vari\'et\'e spinorielle de dimension $n$. 

\exe
Soit $\mm$ une vari\'et\'e compacte munie d'une structure spinorielle et notons $\ss$ l'espace de Hilbert des spineurs de carr\'e int\'egrable sur $\mm$. Consid\'erons le triplet spectral $(\aa,\hh,\dd)$ d\'efini par
\bbbb
\aa&=&C^{\infty}(\mm)\ot M_{N}(\ccc),\n\\
\hh&=&\ss\ot M_{N}(\ccc),\n\\
\dd&=&i\gamma^{\mu}\lp\partial_{\mu}+\omega_{\mu}\rp\ot I,\n
\eeee
o\`u $M_{N}(\ccc)$ d\'esigne l'alg\`ebre des matrices $N\times N$ \`a coefficients complexes, $i\gamma^{\mu}\lp\partial_{\mu}+\omega_{\mu}\rp$ est l'op\'erateur de Dirac associ\'e \`a la g\'eom\'etrie spinorielle de $\mm$ et $I$ est l'identit\'e agissant sur l'espace vectoriel des matrices $N\times N$.

\par

$\aa$ agit sur $\hh$ par multiplication matricielle en chaque point et l'op\'erateur $\jj$ est \'egal \`a $C\ot*$, o\`u $C$ est la conjugaison de charge des spineurs et $*$ l'involution usuelle des matrices. De plus, lorsque la dimension $n$ de $\mm$ est paire, $\gamma$ est d\'efini par $\gamma=\gamma^{n+1}\ot I$.

\par

Le groupe des unitaires de l'alg\`ebre est le groupe des applications de $\mm$ dans le groupe $U(N)$ des unitaires de $M_{N}(\ccc)$. Lorsque ce groupe est repr\'esent\'e sur $\hh$ par (\ref{ai3}), il se trouve que son centre ne joue aucun r\^ole. Par cons\'equent, le groupe de jauge de cette th\'eorie est le groupe des applications de $\mm$ dans le groupe $SU(N)/Z_{N}$, o\`u $SU(N)$ est le sous-groupe des \'el\'ements de $U(N)$ de d\'eterminant 1 et $Z_{N}$ est le centre de $SU(N)$.

\par

Dans ce cas les fermions se transforment dans la repr\'esentation adjointe car on a
\bbb
\pi(u)\jj\pi(u)\jj^{-1}\Psi=u\Psi u^{-1},\n
\eee
pour tout $\Psi\in\hh$. L'op\'erateur de Dirac covariant est donn\'e par
\bbb
\dd_{A}=i\gamma^{\mu}\lp\partial_{\mu}+\omega_{\mu}+A_{\mu}^{a}T^{a}\rp,\n
\eee 
o\`u $A_{\mu}^{a}\in\rrr$ et o\`u $(T^{a})_{1\leq a\leq n^{2}-1}$ d\'esignent les g\'enerateurs de l'alg\`ebre de Lie de $SU(N)$ dans la repr\'esentation adjointe.

\par

L'espace des \'etats purs de $\aa$ est le produit $\mm\ot\ccc\mathrm{P}^{N-1}$, o\`u $\ccc\mathrm{P}^{N-1}$ est l'espace des droites de $\ccc^{N}$. Un couple $\phi=(x,\tilde{\xi})$ form\'e d'un point $x$ de $\mm$ et d'un \'el\'ement $\tilde{\xi}$ de $\ccc\mathrm{P}^{N-1}$ repr\'esent\'e par le vecteur unitaire $\xi$ de $\ccc^{N}$ agit sur $f\in\aa$ par
\bbb
\phi(f)=\t\lp \xi^{*}f(x)\xi\rp.
\eee 
La distance entre deux \'etats purs $(x_{1},\tilde{\xi}_{1})$ et $(x_{2},\tilde{\xi}_{2})$ est la longueur du plus court chemin joignant $x_{1}$ et $x_{2}$ tel que, relev\'e \`a $\mathrm{P}\ccc^{N}$ \`a l'aide de la connexion $A_{\mu}=A_{\mu}^{a}T^{a}$, il permette de transformer $\tilde{\xi}_{1}$ en $\tilde{\xi}_{2}$. 

\par

%

Nous renvoyons \`a \cite{grav} pour une discussion plus compl\`ete de cet exemple.
\eexe


\subsection{Connexions et th\'eories de jauge}

Nous allons maintenant \'etudier plus en profondeur la notion de champ de jauge en g\'eom\'etrie non commutative. En particulier, nous allons g\'en\'eraliser la th\'eorie pr\'ec\'edente aux modules projectifs finis qui sont les analogues non commutatifs des fibr\'es vectoriels (cf Appendice A).

\begin{dfi}
Soit $\lp\aa,\hh,\dd\rp$ un triplet spectral et $\ee$ un module projectif fini \`a droite sur $\aa$. Une connexion est une application lin\'eaire $\nabla$ de $\ee$ dans $\ee\ot_{\aa}\Omega_{\dd}^{1}(\aa)$ telle que
\bbb
\nabla(\xi a)=\nabla(\xi)a+\xi\ot da\n
\eee
pour tous $\xi\in\ee$ et $a\in\aa$.
\end{dfi}

Cette notion de connexion est l'analogue de la d\'eriv\'e covariante et $\ee$ est, lorsque l'alg\`ebre est commutative, l'espace des sections d'un fibr\'e vectoriel complexe. 

\par

La proposition suivante d\'etermine la structure de l'espace des connexions. 

\begin{pro}
Si $\nabla_{0}$ une connexion sur $\ee$, alors n'importe quelle connexion sur $\ee$ s'\'ecrit 
\bbb
\nabla=\nabla_{0}+A,\n
\eee
o\`u $A:\,\ee\rightarrow\ee\ot_{\aa}\Omega_{\dd}^{1}(\aa)$ est un morphisme de modules sur $\aa$.
\end{pro}

En d'autres termes, cela signifie que l'espace des connexions est un espace affine sur l'espace vectoriel des morphismes de $\ee$ dans $\ee\ot_{\aa}\Omega_{\dd}^{1}(\aa)$. 

\par

Il convient de remarquer que nous n'avons pas d\'efini les fibr\'es principaux car cette notion n'est pas disponible dans le formalisme que nous \'etudions. En effet, dans le cadre pr\'ec\'edent toutes les sym\'etries sont d\'ecrites par des groupes de Lie alors que la version non commutative des fibr\'es principaux fait appel \`a des sym\'etries plus g\'en\'erales d\'ecrites par des alg\`ebres de Hopf \cite{pflaum}.   

\par

En r\`egle g\'en\'erale, les groupes de jauge utiles en physique sont des groupes de Lie compacts. Pour introduire la notion de compacit\'e nous sommes amen\'es \`a d\'efinir la notion de structure hermitienne sur un module projectif.

\begin{dfi}
Soit $\ee$ un module projectif fini \`a droite sur une alg\`ebre involutive $\aa$. Une structure hermitienne sur $\ee$ est une application sesquilin\'eaire de $\ee\times\ee$ dans $\aa$ telle que
\begin{enumerate}
\item
$\langle\xi a,\zeta b\rangle=a^{*}\langle\xi,\zeta\rangle b$ pour tous $\xi,\zeta\in\ee$ et $a,b\in\aa$.
\item
$\langle\xi,\xi\rangle\geq0$ pour tout $\xi\in\ee$,
\item
$\ee$ est self-dual.
\end{enumerate} 
\end{dfi}

La d\'efinition pr\'ec\'edente fait appel \`a la notion de positivit\'e, d\'efinie comme suit pour toute alg\`ebre involutive. 

\begin{dfi}
Un \'el\'ement $x$ d'une alg\`ebre involutive $\aa$ est positif s'il existe $y\in\aa$ tel que $x=yy^{*}$. 
\end{dfi}

De m\^eme, la notion de self-dualit\'e est d\'efinie comme suit.

\begin{dfi}
Un module projectif fini \`a droite $\ee$ sur une alg\`ebre involutive $\aa$, muni d'une application sesquilin\'eaire $\langle,\rangle$ de $\ee\times\ee$ dans $\aa$ est self-dual si pour toute application $\aa$-lin\'eaire $\phi$ de $\ee$ dans $\aa$ il existe $\xi_{\phi}\in\ee$ tel que $\phi(\zeta)=\langle\xi_{\phi},\zeta\rangle$ pour tout $\zeta\in\ee$.
\end{dfi}

Les modules projectifs finis munis d'une telle structure nous permettent de d\'efinir les connexions hermitiennes. 

\begin{dfi}
Soit $(\aa,\hh,\dd)$ un triplet spectral et $\nabla$ un connexion sur un module projectif fini \`a droite sur $\aa$, muni d'une structure hermitienne. $\nabla:\;\ee\rightarrow\ee\ot\Omega_{\dd}^{1}(\aa)$ est une connexion hermitienne si elle v\'erifie
\bbb
\langle\xi,\nabla\zeta\rangle-\langle\nabla\xi,\zeta\rangle=
d\langle\xi,\zeta\rangle \label{ct1}
\eee
pour tous $\xi,\zeta\in\ee$.
\end{dfi}

La relation (\ref{ct1}) n\'ecessite quelques explications. En effet, si $x\in\ee$, alors $\nabla\xi$ est un \'el\'ement du produit tensoriel $\ee\ot_{\aa}\Omega_{\dd}^{1}(\aa)$ et peut se mettre sous la forme 
\bbb
\nabla\xi=\mathop{\sum}\limits_{i}\xi_{i}\ot\omega_{i}\n
\eee 
avec $\xi_{i}\in\ee$ et $\omega_{i}\in\Omega_{\dd}^{1}(\aa)$. On d\'efinit alors  $\langle\nabla\xi,\zeta\rangle$ par
\bbb
\langle\nabla\xi,\zeta\rangle=\mathop{\sum}\limits_{i}\omega_{i}^{*}
\langle\xi_{i},\zeta\rangle.\n
\eee
De la m\^eme mani\`ere, on \'ecrit $\nabla\zeta$ sous la forme
\bbb
\nabla\zeta=\mathop{\sum}\limits_{i}\zeta_{i}\ot\eta\n
\eee
avec $\zeta_{i}\in\ee$ et $\eta_{i}\in\Omega_{\dd}^{1}(\aa)$ ce qui permet de  d\'efinir $\langle\xi,\nabla\zeta\rangle$ par
\bbb
\langle\xi,\nabla\zeta\rangle=\mathop{\sum}\limits_{i}
\langle\xi,\zeta_{i}\rangle\eta_{i}.\n
\eee
Le signe $-$ apparaissant dans (\ref{ct1}) est du \`a la relation $d(x^{*})=-(dx)^{*}$, car nous avons d\'efini $dx$ par $dx=\lb\dd,x\rb$ pour tout $x\in\aa$.

\par

Ces connexions sont compatibles avec la structure hermitienne qui d\'efinit l'analogue d'une m\'etrique sur $\ee$, aussi sont-elles parfois appel\'ees "connexions compatibles avec la m\'etrique".

\par

Tout module projectif fini sur $\aa$ est de la forme $\ee=e\aa^{N}$ avec $e\in M_{N}(\aa)$ satisfaisant \`a
$e^{2}=e$ (cf Appendice B). Si de plus $e$ est hermitien, alors la structure hermitienne naturelle de $\aa^{N}$ donn\'ee par
\bbb
\langle \xi,\zeta\rangle=\mathop{\sum}\limits_{i=1}^{N}\xi^{*}_{i}\zeta_{i}\n
\eee
induit sur $\ee$ une structure hermitienne et toute structure hermitienne sur $\ee$ est de cette forme.

\par

Pour d\'eterminer les morphismes de module de $\ee$ dans $\Omega_{\dd}^{1}(\aa)$, nous commen\c cons par remarquer que $\ee\ot_{\aa}\Omega^{1}_{\dd}(\aa)$ et $e\lp\Omega_{\dd}^{1}(\aa)\rp^{N}$ sont canoniquement isomorphe puisque $\ee=e\aa^{N}$. Il s'en suit que les morphismes recherch\'es sont donn\'es par les \'el\'ements de $M_{N}(\aa)\ot_{\aa}\Omega_{\dd}^{1}(\aa)=M_{N}(\Omega^{1}_{\dd}(\aa))$ agissant sur les \'el\'ements de $e\aa^{N}$ par multiplication \`a droite suivi par la projection $e$.

\par

Nous pouvons toujours construire une connexion hermitienne  $\nabla_{0}$ sur $\ee$ de la mani\`ere suivante. Pour tout $e\xi\in\ee$, nous d\'efinissons $\nabla_{0}(e\xi)$ par $ed\xi$, o\`u le vecteur $d\xi\in\Omega_{\dd}^{1}(\aa)^{N}=\aa^{N}\ot_{\aa}\Omega^{1}_{\dd}(\aa)$ est obtenu en diff\'erenciant $\xi$ composantes par composantes. Cette connexion est appel\'ee connexion grassmanienne.

\par

Il est facile de v\'erifier que $\nabla_{0}$ est compatible avec la m\'etrique  de $e\aa^{N}$ induite par la m\'etrique canonique de $\aa^{N}$. En effet, si $\xi$ et $\zeta$ sont des \'el\'ements de $\aa^{N}$, on a
\bbbb
\langle e\zeta,\nabla_{0}(e\xi)\rangle-\langle\nabla_{0}(e\zeta),e\xi\rangle&=&
(e\zeta)^{*}ed(e\xi)-\lp d(e\zeta)\rp^{*}e\xi\n\\
&=&\zeta^{*}ede\xi+\zeta^{*}ed\xi+\zeta^{*}dee\xi+d\zeta e\xi\n\\
&=&d(\zeta^{*}e\xi)\n\\
&=&d\langle e\zeta,e\xi\rangle.\n
\eeee 

\par

Il s'en suit que toute connexion sur $\ee$ est donn\'ee par
\bbb
\nabla(e\xi)=ed\xi+eAe\xi\qquad\forall\,\xi\in\aa^{N},\n
\eee
o\`u $A\in M_{N}\lp\Omega^{1}_{\dd}(\aa)\rp$. Par analogie avec le cas classique, $A$ peut \^etre vu comme un champ de jauge qui est une 1-forme \`a valeurs matricielles.

\par

La connexion pr\'ec\'edente est compatible avec la m\'etrique si et seulement si
\bbb
\langle e\zeta, eAe\xi\rangle-\langle eAe\zeta,e\xi\rangle=0\qquad\forall\zeta,\xi\in\aa^{N},\n
\eee
ce qui est v\'erifi\'e si et seulement si la matrice de 1-formes $eAe$ est hermitienne.

\par

La proposition suivante r\'esume la discusssion pr\'ec\'edente.

\par

\begin{pro}
Tout module projectif fini muni d'une structure hermitienne est isomorphe \`a un module du type $e\aa^{N}$ o\`u $e\in M_{N}(\aa)$ est une projection hermitienne et dont la structure hermitienne est induite par la structure hermitienne canonique de $\aa^{N}$. Sur ce module, toutes les connexions hermitiennes  sont donn\'ees par
\bbb
\nabla(e\xi)=ed\xi+eAe\xi\;\;\;\;\forall\,\xi\in\aa^{N},
\eee
o\`u $A\in M_{N}\lp\Omega^{1}_{\dd}(\aa)\rp$ est une matrice de 1-formes hermitiennes.
\end{pro}

\par

Terminons en introduisant les transformations de jauge. Celles-ci sont obtenues en faisant agir le groupe des endomorphismes bijectifs de $\ee$ qui conservent la structure hermitienne, appel\'es endomorphismes unitaires, sur l'espace des connexions. 

\begin{pro}
Si $\nabla$ est une connexion hermitienne sur $\ee$ et u un \'el\'ement du groupe des endomorphismes unitaires $\uu(\ee)$, alors $(u\ot 1)\nabla u^{*}$ est \'egalement une connexion hermitienne.
\end{pro}

Pour simplifier nos notations, nous \'ecrirons $u\nabla u^{-1}$ \`a la place de 
$(u\ot 1)\nabla u^{-1}$.

\par

Cela d\'efinit l'action du groupe des endomorphismes unitaires de $\ee$ sur l'espace des connexions.

\begin{dfi}
L'action par conjugaison du groupe $\uu(\ee)$ sur les connexions hermitiennes est appel\'e une transformation de jauge. 
\end{dfi}

Lorsque le module hermitien est $\ee=e\aa^{N}$, une matrice $u\in M_{N}(\aa)$ d\'efinit un endomorphisme unitaire $eue$ de $\ee$ si et seulement si
\bbb
eueu^{*}e=eu^{*}eue=e.\n
\eee


\subsection{Application \`a la construction de triplets spectraux}

A partir d'un triplet spectral $(\aa,\hh,\dd)$ et d'un module projectif fini sur $\aa$ muni d'une connexion hermitienne il est possible de construire un nouveau triplet spectral \cite{grav}.  

\begin{pro}
Soit $\lp\aa,\hh,\dd\rp$ un triplet spectral de dimension n et $\nabla$ une connexion hermitienne sur un module projectif fini $\ee$ \`a droite sur $\aa$. Soit $\lp\tilde{\aa},\tilde{\hh},\tilde{\dd}\rp$ d\'efini par
\bbbb
\tilde{\aa}&=&End_{\aa}(\ee),\n\\
\tilde{\hh}&=&\ee\ot_{\aa}\hh\ot_{\aa}\ov{\ee},\n\\
\tilde{\dd}(\xi\ot_{\aa}\psi\ot_{\aa}\ov{\zeta})&=&
\nabla(\xi)\psi\ot_{\aa}\ov{\zeta}+
\xi\ot_{\aa}\dd(\psi)\ot_{\aa}\ov{\zeta}+
\xi\ot_{\aa}\psi\ov{(\nabla(\zeta))},\n
\eeee
pour tous $\xi,\zeta\in\ee$ et $\psi\in\hh$, et o\`u $End_{\aa}(\ee)$ d\'esigne l'alg\`ebre des endomorphismes de $\ee$. Alors $\lp\tilde{\aa},\tilde{\hh},\tilde{\dd}\rp$ est \'egalement un triplet spectral de dimension n.
\end{pro}

Notons que les op\'erateurs $\tilde{\jj}$ et $\tilde{\chi}$ associ\'es au triplet spectral  $\lp\tilde{\aa},\tilde{\hh},\tilde{\dd}\rp$ sont donn\'es par
\bbbb
\tilde{\jj}(\xi\ot_{\aa}\psi\ot_{\aa}\ov{\zeta})&=&
\zeta\ot_{\aa}\jj\psi\ot_{\aa}\ov{\xi}\n\\
\tilde{\chi}(\xi\ot_{\aa}\psi\ot_{\aa}\ov{\zeta})&=&
\xi\ot_{\aa}\chi\psi\ot_{\aa}\ov{\zeta},\n
\eeee
pour tous $\xi,\zeta\in\ee$ et $\psi\in\hh$. 

\par

Ce r\'esultat n\'ecessite quelques explications suppl\'ementaires. Tout d'abord, indiquons que $\ov{\ee}=\la\xi^{*}e|\xi\in\aa^{N}\ra$ est un module \`a gauche sur $\aa$ si $\ee=e\aa^{N}$. L'action \`a gauche de $a\in\aa$ sur $\xi^{*}e\in\ov{\ee}$ est donn\'ee par $a\xi^{*}e=(e\xi a^{*})^{*}.$

\par

Pour donner un sens \`a l'expression $\nabla(\xi)\Psi$ apparaissant dans la d\'efinition de l'op\'erateur de Dirac $\tilde{\dd}$, nous \'ecrivons d'abord $\nabla(\xi)$ sous la forme $\nabla(\xi)=\sum_{i}\xi_{i}\ot\omega_{i}$ 
avec $\xi_{i}\in\ee$ et $\omega_{i}\in\Omega_{\dd}^{1}(\aa)$, puis nous utilisons la repr\'esentation de $\Omega^{1}_{\dd}(\aa)$ comme op\'erateurs sur $\hh$  pour \'ecrire
\bbb
\nabla(\xi)\Psi=\mathop{\sum}\limits_{i}\xi_{i}\ot\omega_{i}\Psi.\n
\eee
De m\^eme, on d\'efinit
\bbb
\Psi\ov{\nabla(\zeta)}=\jj\eta_{i}\jj^{-1}\psi\ot\zeta_{i},\n
\eee
o\`u on a utilis\'e $\nabla(\zeta)=\mathop{\sum}\limits_{i}\zeta_{i}\ot\eta_{i}$ avec $\zeta_{i}\in\ee$ et $\eta_{i}\in\Omega_{\dd}^{1}(\aa)$.

\par

A partir de cette d\'efinition, nous pouvons construire l'action fermionique pour une module projectif g\'en\'eral. En effet,  nous consid\'erons $\tilde{\Psi}\in\tilde{\hh}$, et nous d\'efinissons l'action fermionique par
\bbb
S_{fermionique}[\tilde{\Psi}]=\langle\tilde{\Psi},\tilde{\dd}\tilde{\Psi}\rangle.
\eee 

Cette construction permet \'egalement de retrouver les fluctuations internes de la m\'etrique comme nous allons le voir sur l'exemple suivant.

\exe 
Soit $(\aa,\hh,\dd)$ un triplet spectral et $\ee=\aa$ le module hermitien trivial sur $\aa$. Les connexions hermitiennes sur $\ee$ sont donn\'ees par $\nabla\xi=d\xi+A\xi$ pour tout $\xi\in\aa$ et o\`u $A$ est une 1-forme hermitienne. 

\par

Il est facile de voir que la construction pr\'ec\'edente nous donne un nouveau triplet spectral $(\tilde{\aa},\tilde{\hh},\tilde{\dd})$ avec $\tilde{\aa}=\aa$ et $\tilde{\hh}=\hh$.

\par

Pour tous $\xi,\zeta\in\ee$ et $\Psi\in\hh$ nous avons
\bbbb
\nabla(\xi)\Psi&=&\lb\dd,\xi\rb\psi+A\xi\Psi,\n\\
\Psi\ov{\nabla(\zeta)}&=&\jj\lb\dd,\zeta\rb\jj^{-1}\Psi+\jj A\xi\jj^{-1}\Psi.
\eeee
On en d\'eduit que
\bbbb
\tilde{\dd}(\xi\jj\zeta\jj^{-1}\Psi)&=&\lb\dd,\xi\rb\jj\zeta\jj^{-1}\Psi+
A\xi\jj\zeta\jj^{-1}\Psi\n\\
&+&\xi\jj\zeta\jj^{-1}\dd\Psi+\xi\jj\lb\dd,\zeta\rb\jj^{-1}\Psi+
\xi\jj A\zeta\jj^{-1}\Psi.\label{ac1}
\eeee
En d\'eveloppant tous les commutateurs apparaissant au second membre de (\ref{ac1}) et en utilisant les relations de commutation de l'axiome de r\'ealit\'e, on montre que
\bbb
\tilde{\dd}\xi\jj\zeta\jj^{-1}\Psi=
\lp\dd+A+\jj A\jj^{-1}\rp\xi\jj\zeta\jj^{-1}\Psi
\eee
ce qui prouve que $\tilde{\dd}=\dd+A+\jj A\jj^{-1}$ puisque $\xi\jj\zeta\jj^{-1}\Psi$ est un \'el\'ement g\'en\'erique de $\tilde{\hh}=\hh$.
\eexe 

Ainsi on retrouve l'op\'erateur de Dirac correspondant aux fluctuations internes de la m\'etrique comme un cas particulier de la construction pr\'ec\'edente. Celle-ci g\'en\'eralise les fluctuations internes aux th\'eories de jauge d\'efinies sur un module projectif non trivial.


\section{Th\'eorie de Yang-Mills}

\subsection{L'alg\`ebre diff\'erentielle}

Au cours des sections pr\'ec\'edentes, nous avons d\'efini l'espace des formes de degr\'e 0 associ\'ees \`a un triplet spectral $(\aa,\hh,\dd)$ par $\Omega^{0}_{\dd}(\aa)=\pi(\aa)$ ainsi que l'espace des 1-formes
\bbb
\Omega_{\dd}^{1}(\aa)=\la\mathop{\sum}\limits_{i}\pi(a_{i})\lb\dd,\pi(b_{i})\rb
\;|\;a_{i},b_{i}\in\aa\ra,\n
\eee
la diff\'erentielle $d:\;\Omega_{\dd}^{0}(\aa)\rightarrow\Omega_{\dd}^{1}(\aa)$ \'etant d\'efinie par $d\pi(x)=\lb\dd,\pi(x)\rb$ pour tout $x\in\aa$.

\par

Pour d\'efinir les formes diff\'erentielles de degr\'e sup\'erieur de mani\`ere analogue, il est utile d'introduire l'alg\`ebre diff\'erentielle universelle sur $\aa$.

\begin{dfi}
Soit $\aa$ une alg\`ebre. On appelle alg\`ebre diff\'erentielle universelle sur $\aa$ l'alg\`ebre gradu\'ee $\Omega(\aa)=\op_{k\in\nnn}\,\Omega^{k}(\aa)$, o\`u $\Omega^{k}(\aa)$ est l'espace vectoriel engendr\'e par les symboles $a_{0}\delta a_{1}\dots\delta a_{k}$ pour tous $a_{0},a_{1},\dots,a_{k}\in\aa$ et les relations $\delta(ab)=\delta a\,b+a\,\delta b$ pour tous $a,b\in\aa$ ainsi que $\delta(1)=0$. La diff\'erentielle $d\,:\, \Omega^{k}(\aa)\rightarrow \Omega^{k+1}(\aa)$ est une application lin\'eaire d\'efinie par
\bbb
d\lp a_{0}\delta a_{1}\dots\delta a_{k}\rp=
\delta a_{0}\delta a_{1}\dots\delta a_{k}\n
\eee
pour tous $a_{0},a_{1},\dots,a_{k}\in\aa$.
\end{dfi}

Cette diff\'erentielle v\'erifie les principales propri\'et\'es de la diff\'erentielle usuelle:

\begin{pro}
La diff\'erentielle $d$ est nilpotente et satisfait \`a la r\`egle de Leibniz gradu\'ee  $d(\omega\xi)=d\omega\,\xi+(-1)^{p}\omega\,d\xi$ pour tous $\omega\in\Omega^{p}$ et $\xi\in\Omega^{q}$.
\end{pro}

Cependant, cette construction est purement alg\'ebrique et ne nous permet pas en tant que telle de retrouver les formes diff\'erentielles usuelles. En particulier, lorsque $\aa$ est une alg\`ebre de fonctions sur une vari\'et\'e, les formes diff\'erentielles ne satisfont aucune propri\'et\'e d'antisym\'etrisation, ce qui implique qu'il existe, en dimension $n$, des formes diff\'erentielles de degr\'e $n+1$.

\par

Pour tenir compte de l'information g\'eom\'etrique contenue dans tout le triplet spectral $(\aa,\hh,\dd)$, nous repr\'esentons ces formes diff\'erentielles \`a l'aide de commutateurs avec l'op\'erateur de Dirac. 
En d'autres termes, nous \'etendons la repr\'esentation $\pi$ \`a l'ensemble de l'alg\`ebre diff\'erentielle universelle par
\bbb
\pi(a_{0}\delta a_{1}\dots\delta a_{p})=\pi(a_{0})\lb\dd,\pi(a_{1})\rb\dots\lb\dd,\pi(a_{p})\rb\;\;\;\;\forall a_{0},\dots,a_{p}\in\aa.
\eee
Cependant, cette repr\'esentation n'est pas une repr\'esentation de la structure diff\'erentielle de $\Omega(\aa)$ et ne permet pas de d\'efinir une differentielle sur l'alg\`ebre $\pi(\Omega(\aa))$. En effet, si $\omega\in\Omega(\aa)$, alors on ne peut pas d\'efinir la diff\'erentielle de $\pi(\omega)$ par $d\pi(\omega)=\pi(d\omega)$ car il peut exister des \'el\'ements $\omega$ de $\Omega(\aa)$ tels que $\pi(\omega)=0$ et $\pi(d\omega)\neq0$.

\par

Pour rem\'edier \`a cela, on introduit l'id\'eal bilat\`ere de $\Omega(\aa)$ d\'efini par
\bbb
J=\ker\pi+d(\ker\pi).\n
\eee
Cet id\'eal est stable par $d$, ce qui permet de donner une  structure d'alg\`ebre diff\'erentielle au quotient $\pi(\omega)/\pi(J)$ \cite{bible}.

\begin{pro}
La repr\'esentation $\pi$ d\'efinit une repr\'esentation de l'alg\`ebre diff\'erentielle universelle sur le quotient, not\'e $\Omega_{\dd}(\aa)$ et donn\'ee par
\bbb
\pi\lp\Omega(\aa)\rp/\pi(J)=\pi(\Omega(\aa))/\pi\lp d(\ker\pi)\rp.\n
\eee
La diff\'erentielle d'un \'el\'ement repr\'esent\'e par sa classe $\pi(\omega)$ est d\'efinie par la classe de $\pi(d\omega)$. Elle est nilpotente et satisfait \`a la r\`egle de Leibniz gradu\'ee.
\end{pro}

\par

Dans la litt\'erature, l'id\'eal $J$ est souvent appel\'e "junk", et ses \'el\'ements sont nomm\'es "champs auxiliaires".

\par

Les \'el\'ements de l'alg\`ebre $\Omega_{\dd}(\aa)$ sont en fait des classes d'\'equivalence d'op\'erateurs. Pour pouvoir les manipuler ais\'ement dans un cas concret, il est commode d'introduire un produit scalaire  sur $\pi(\Omega^{n}(\aa))$ de mani\`ere \`a pouvoir choisir un et un seul repr\'esentant orthogonal \`a $\pi(J)$ \cite{kastlerschucker}. Ce produit scalaire sur $\pi\lp\Omega(\aa)\rp$ est construit \`a l'aide de la trace de Dixmier,
\bbb
\langle\pi(\omega),\pi(\eta)\rangle=\t_{\omega}\lp\pi(\omega)^{*}\pi(\eta)|\dd|^{-n}\rp,\n
\eee
si $\omega$ et $\eta$ sont des \'el\'ements de $\Omega(\aa)$ de m\^eme degr\'e, des formes de degr\'es diff\'erents \'etant par d\'efinition orthogonales.

\par
 
Nous rappelons que $\t_{\omega}$ est la trace de Dixmier et que puisque $(\aa,\hh,\dd)$ est un triplet spectral de dimension $n$, $\pi(\omega)$ et $\pi(\eta)$ sont des op\'erateurs born\'ees et $\pi(\omega)^{*}\pi(\eta)|\dd|^{-n}$ est un infinit\'esimal d'ordre un. 

\par 

Cependant, dans le cas g\'en\'eral, il n'est pas \'evident que cette application d\'efinisse un produit scalaire. En effet, il n'est pas possible de montrer que cette forme sesquilin\'eaire positive est d\'efinie positive car il peut arriver que les valeurs propres de $\pi(\omega)$ d\'ecroissent assez rapidement pour que la trace de Dixmier s'annule. Aussi travaillerons-nous directement avec les classes d'equivalence d'\'el\'ements de $\pi\lp\omega(\aa)\rp$. 

\par

Les propri\'et\'es usuelles des formes diff\'erentielles et de leur int\'egration sont r\'esum\'ees dans la notion de cycle (cf Appendice C). Pour faire le lien entre cette notion et la construction pr\'ec\'edente, nous utiliserons le r\'esultat suivant. 

\begin{pro}
La relation
\bbb
\pi(\omega)\mapsto
Tr_{\omega}\lp\gamma\pi(\omega)|\dd|^{-n}\rp\n
\label{ad3}
\eee
d\'efinit une trace gradu\'ee sur $\pi\lp\Omega^{n}(\aa)\rp$,
\end{pro}

\demo
Commen\c cons par remarquer que pour tout $\omega,\eta\in\Omega(\aa)$ on a
\bbb
\t_{\omega}\lp\pi(\omega)\pi(\eta)|\dd|^{-n}\rp=
\t_{\omega}\lp\pi(\eta)\pi(\omega)|\dd|^{-n}\rp.\label{ad2}
\eee
En effet, d'apr\`es l'axiome de r\'egularit\'e, les hypoth\`eses du th\'eor\`eme d\'emontr\'e dans \cite{cipriani} sont v\'erifi\'ees. La relation (\ref{ad2}) d\'ecoule imm\'ediatement de ce r\'esultat. De m\^eme, on pourra trouver une preuve de cette relation dans \cite{waltze}.  

\par

Soit $\omega\in\Omega^{p}(\aa)$ et $\eta\in\Omega^{q}(\aa)$ deux formes diff\'erentielles telles que $p+q=n$. D'apr\`es l'axiome d'orientabilit\'e, $\gamma=\pi(c)$, o\`u $c$ est un n-cycle de Hochschild \`a valeurs dans $\aa\ot\aa^{op}$. Le facteur \`a valeurs dans $\aa^{op}$ est repr\'esent\'e par $x\in\aa\mapsto\jj\pi(x)\jj^{-1}$, donc commute avec tous les \'el\'ements de $\pi\lp\Omega(\aa)\rp$. Par cons\'equent, en appliquant la relation (\ref{ad2}), on montre que
\bbb
\t_{\omega}\lp\gamma\pi(\omega)\pi(\eta)|\dd|^{-n}\rp=
(-1)^{p}\t_{\omega}\lp\gamma\pi(\eta)\pi(\omega)|\dd|^{-n}\rp,\n
\eee
puisque $\gamma\pi(\omega)=(-1)^{p}\pi(\omega)\gamma$ en dimension paire et $\gamma\pi(\omega)=\pi(\omega)\gamma$ en dimension impaire. La relation (\ref{ad3}) d\'efinit donc une trace gradu\'ee sur $\pi\lp\Omega^{n}(\aa)\rp$.

\edemo

Pour pouvoir construire un cycle \`a l'aide de l'alg\`ebre $\Omega_{\dd}(\aa)$, nous devons faire une hypoth\`ese suppl\'ementaire sur le triplet spectral $(\aa,\hh,\dd)$, qui correspond, dans le cas commutatif, \`a la condition de "vari\'et\'e sans bord".

\begin{dfi}[Condition de fermeture]
Un triplet spectral de dimension n satisfait \`a la condition de fermeture si
\bbb
Tr_{\omega}\lp\gamma\,\lb\dd,\pi(a_{1})\rb\dots\lb\dd,\pi(a_{n})\rb\,|\dd|^{-n}
\rp=0\label{ad8}
\eee
pour tous $a_{1},\dots,a_{p}\in\aa$.
\end{dfi}

Cela est \`a relier \`a la notion de cycle qui pr\'esente de mani\`ere purement alg\'ebrique les propri\'et\'es des formes diff\'erentielles et de leur int\'egration \cite{bible}. 

\begin{dfi}
Un cycle de dimension $n$ sur une alg\`ebre $\aa$ est un quadruplet $\lp\Omega,\pi,d,\int\rp$ o\`u
\begin{enumerate}
\item
$\Omega=\mathop{\op}\limits_{k=0}^{n}\;\Omega^{k}$ est une alg\`ebre gradu\'ee,
\item
$\pi$ est une repr\'esentation de $\aa$ dans $\Omega^{0}$,
\item
$d\,:\, \Omega^{k}\rightarrow \Omega^{k+1}$ est une application lin\'eaire satisfaisant \`a $d^{2}=0$. \`a $d\lp\Omega^{n}(\aa)\rp=0$ ainsi qu'\`a la r\`egle de Leibniz gradu\'ee $
d(\omega\xi)=d\omega\,\xi+(-1)^{p}\omega\,d\xi$ pour tous $\omega\in\Omega^{p}$ et $\xi\in\Omega^{q}$,
\item
$\int\,:\,\Omega^{n}\rightarrow\cc$ est une forme lin\'eaire telle que $\int\,d\omega=0\;\mathrm{si}\;\omega\in\Omega^{n-1}$, et $\int\,\omega\xi=(-1)^{pq}\int\,\xi\omega$, pour tous $\omega\in\Omega^{p}$ et $\xi\in\Omega^{q}$ tels que $p+q=n$.
\end{enumerate}
\end{dfi}

Lorsque la condition de fermeture est satisfaite, on montre le r\'esultat suivant. 

\begin{pro}
Si le triplet spectral $(\aa,\hh,\dd)$ satisfait \`a la condition de fermeture, alors l'alg\`ebre diff\'erentielle $\Omega_{\dd}(\aa)=\pi\lp\Omega(\aa)\rp/\pi(J)$ d\'efinit un cycle $\lp\Omega_{\dd}(\aa),d,\int\rp$ avec $\int:\;\Omega^{n}_{\dd}(\aa)\rightarrow\ccc$ donn\'e par
\bbb
\int\pi(\omega)=\frac{1}{(4\pi)^{n/2}\Gamma(n/2+1)}
\t_{\omega}\lp\gamma\pi(\omega)|\dd|^{-n}\rp\label{ad4}
\eee 
\end{pro}

\demo
Puisque les \'el\'ements de $\Omega_{\dd}^{n}(\aa)$ sont des classes d'\'equivalence d'\'el\'ements de $\pi(\Omega^{n}(\aa))$ nous devons montrer que la relation (\ref{ad4}) d\'efinit bien une application sur l'espace quotient $\Omega^{n}_{\dd}(\aa)$. Pour cela, remarquons que deux \'el\'ements $\pi(\omega)$ et $\pi(\eta)$ de $\pi\lp\Omega^{n}(\aa)\rp$ qui d\'efinissent la m\^eme classe dans $\Omega_{\dd}^{n}(\aa)$ diff\`erent par un \'el\'ement de $\pi(J)\cap\pi\lp\Omega^{n}(\aa)\rp$ qui peut s'\'ecrire
\bbb
\pi(\omega)-\pi(\eta)=\mathop{\sum}\limits_{i}\lb\dd,\pi(a_{1}^{i})\rb\dots
\lb\dd,\pi(a_{n}^{i})\rb\n
\eee
avec $a_{1}^{i},\dots a_{n}^{i}\in\aa$. Par application de la condition de fermeture (\ref{ad8}), on obtient 
\bbb
Tr_{\omega}\lp\gamma\pi(\omega)|\dd|^{-n}\rp=
Tr_{\omega}\lp\gamma\pi(\eta)|\dd|^{-n}\rp,\n
\eee
ce qui prouve que $\int$ passe au quotient. Il d\'ecoule alors des r\'esultats pr\'ec\'edents que $\int$ que $\lp\Omega_{\dd}(\aa),d,\int\rp$ est un cycle.

\edemo

Le facteur de normalisation de la relation (\ref{ad4}) est choisi de mani\`ere \`a retrouver l'int\'egration usuelle des formes diff\'erentielles. De fa\c con g\'en\'erale, nous notons
\bbb
\dix\gamma \pi(\omega)ds^{n}=\frac{1}{(4\pi)^{n/2}\Gamma(n/2+1)}
\t_{\omega}\lp\pi(\omega)|\dd|^{-n}\rp\n
\eee
pour tout \'el\'ement $\omega$ de $\Omega(\aa)$.

\par

On en d\'eduit ausit\^ot que le caract\`ere de ce cycle est un cocycle cyclique de dimension $n$.

\begin{pro}
$(\aa,\hh,\dd)$ satisfait \`a la condition de fermeture si et seulement si l'application
\bbb
\phi(a_{0},a_{1},\dots,a_{n})=
\dix\,\gamma a_{0}\lb\dd,a_{1}\rb\dots\lb\dd,a_{n}\rb\,ds^{n}
\eee
est un cocyle cyclique de dimension $n$.
\end{pro}

\demo
En effet, si la condition de fermeture est satisfaite, alors $\phi$ est le caract\`ere d'un cycle et est un cocycle cyclique. R\'eciproquement, si $\phi$ est un cocycle cyclique alors
\bbbb
\dix\,\gamma\lb\dd,a_{1}\rb\dots\lb\dd,a_{n}\rb\,ds^{n}&=&
\phi(1,a_{1},\dots,a_{n})\n\\
&=&(-1)^{n}\phi(a_{1},\dots,a_{n},1)\n\\
&=&0.\n
\eeee

\edemo

Nous verrons ult\'erieurement que la g\'en\'eralisation non commutative des th\'eories de Yang-Mills et de Chern-Simons nous am\`ene \`a coupler ce cocycle cylique \`a la K-th\'eorie pour obtenir des quantit\'es stables par d\'eformation. 

\par

Il est aussi inter\'essant de noter que la condition de fermeture rend possible l'int\'egration par parties.

\begin{pro}
Si le triplet spectral $(\aa,\hh,\dd)$ v\'erifie la condition de fermeture, alors la r\`egle d'int\'egration par parties, 
\bbb
\dix\gamma d\pi(\omega)\pi(\eta)ds^{n}=(-1)^{p+1}\dix\gamma \pi(\omega)d\pi(\eta)ds^{n}\n
\eee
pour tous $\omega\in\Omega^{p}(\aa)$ et $\eta\in\Omega^{n-p-1}(\aa)$, est valide.
\end{pro}

Ce r\'esultat, utile dans les calculs, passe au quotient et permet de d\'efinir l'int\'egration par parties sur $\Omega^{n}_{\dd}(\aa)$.

Enfin, terminons par l'\'etude du cas commutatif.

\exe
Soit $\aa=C^{\infty}(\mm)$ l'alg\`ebre des fonctions lisses sur une vari\'et\'e compacte sans bord $\mm$ de dimension $n$ munie d'une structure spinorielle, $\hh$ l'espace de Hilbert des spineurs de carr\'e sommable sur $\mm$ et $\dd=i\gamma^{\mu}(\partial_{\mu}+\omega_{\mu})$ l'op\'erateur de Dirac usuel. Nous allons montrer que le r\'esultat de la construction pr\'ec\'edente est identique au complexe de de Rham.

\par

Les fonctions $f\in\aa$ agissent sur $\hh$ par multiplication, ce qui entra\^\i  ne que $\lb\dd,f\rb=i\gamma^{\mu}\partial_{\mu}f$. Un \'el\'ement g\'en\'erique de $\pi(\Omega^{p}(\aa))$ s'\'ecrit sous la forme
\bbb
\mathop{\sum}\limits_{i}\gamma^{\mu_{1}}\dots\gamma^{\mu_{p}}
f_{0}^{i}\partial_{\mu_{1}}f_{1}^{i}\dots\partial_{\mu_{p}}f_{p}^{i},\label{ad5}
\eee
avec $f_{0},\dots,f_{p}\in\aa$.

\par

Il est \'evident que pour $p=0$, $\pi\lp\Omega^{0}(\aa)\rp=\Omega^{0}_{\dd}(\aa)=\aa$, ce qui est identique au 0-formes de de Rham. De m\^eme, lorsque $p=1$, on a $\pi\lp\Omega^{1}(\aa)\rp=\Omega^{1}_{\dd}(\aa)$, un \'el\'ement quelconque de cet ensemble pouvant \^etre assimil\'e \`a une 1-forme de de Rham $\omega_{\mu}dx^{\mu}$ si on identifie $i\gamma^{\mu}$ et $dx^{\mu}$ et si on pose
\bbb
\omega_{\mu}=\mathop{\sum}\limits_{i}
f_{0}^{i}\partial_{\mu}f_{1}^{i}\partial_{\mu}.\n
\eee
R\'eciproquement, toute 1-forme de de Rham peut se mettre sous la forme pr\'ec\'edente. 

\par

Lorsque $p=2$, si on veut encore pouvoir identifier $i\gamma^{\mu}$ et $dx^{\mu}$, nous devons faire appara\^\i tre le produit compl\`etement antisym\'etrique des matrices de Dirac dans (\ref{ad5}), et \'ecrire un \'el\'ement g\'en\'erique de $\pi(\Omega^{2}(\aa))$ sous la forme
\bbb
\mathop{\sum}\limits_{i}\frac{1}{2}\lp i\gamma_{\mu_{1}}i\gamma_{\mu_{2}}-i\gamma_{\mu_{2}}i\gamma_{\mu_{1}}\rp f_{0}^{i}\partial_{\mu_{1}}f_{1}^{i}\partial_{\mu_{2}}f_{2}^{i}
-\mathop{\sum}\limits_{i}g^{\mu_{1}\mu_{2}}
f_{0}^{i}\partial_{\mu_{1}}f_{1}^{i}\partial_{\mu_{2}}f_{2}^{i}
\label{ad6}
\eee
Montrons que le second terme de (\ref{ad6}) est un \'el\'ement de $\pi(J)\cap\pi\lp\Omega^{2}(\aa)\rp$. En effet, tous les \'el\'ements $\pi(J)\cap\pi\lp\Omega^{2}(\aa)\rp$ s'\'ecrivent comme
\bbb
\mathop{\sum}\limits_{i}i\gamma_{\mu_{1}}i\gamma_{\mu_{2}} \partial_{\mu_{1}}h_{1}^{i}\partial_{\mu_{2}}h_{2}^{i}
=-\mathop{\sum}\limits_{i}g^{\mu_{1}\mu_{2}}
\partial_{\mu_{1}}h_{1}^{i}\partial_{\mu_{2}}h_{2}^{i}\n
\eee 
o\`u les fonctions $h_{1}^{i}$ et $h_{2}^{i}$ satisfont \`a la condition
\bbb
\mathop{\sum}\limits_{i}h_{1}^{i}\partial_{\mu}h_{2}^{i}=0.\n
\eee
En utilisant l'identit\'e
\bbb
\mathop{\sum}\limits_{i}f_{0}^{i}f_{1}^{i}\partial_{\mu}f_{2}^{i}+
f_{0}^{i}\partial_{\mu}f_{1}^{i}f_{2}^{i}-
f_{0}^{i}\partial_{\mu}\lp f_{1}^{i}f_{2}^{i}\rp=0,
\eee
on montre que le second terme de (\ref{ad6}) est un \'el\'ement de $\pi(J)\cap\pi\lp\Omega^{2}(\aa)\rp$. Par cons\'equent, on peut identifier le premier membre de (\ref{ad6}) \`a la 2-forme $1/2\omega_{\mu_{1}\mu_{2}}dx^{\mu_{1}}dx^{\mu_{2}}$ avec 
\bbb
\omega_{\mu_{1}\mu_{2}}=\mathop{\sum}\limits_{i}\frac{1}{2}
f_{0}^{i}\lp\partial_{\mu_{1}}f_{1}^{i}\partial_{\mu_{2}}f_{2}^{i}
-\partial_{\mu_{2}}f_{1}^{i}\partial_{\mu_{1}}f_{2}^{i}\rp
\eee
Cela nous a permis de choisir dans l'espace quotient $\Omega^{2}_{\dd}(\aa)$ un unique repr\'esentant de chaque classe qui correspond au produit totalement antisym\'etris\'e des matrices de Dirac.

\par 

La d\'eriv\'e ext\'erieure d'une 1-forme $\omega_{\mu}dx^{\mu}$ est donn\'ee par
\bbb
d\omega=\frac{1}{2}\partial_{\mu}\omega_{\nu}dx^{\mu}dx^{\nu}.\n
\eee
Le produit ext\'erieur de $\omega$ par une autre 1-forme $\eta=\eta_{\nu}dx^{\nu}$ s'obtient en le multipliant en tant qu'op\'erateurs sur $\hh$ puis en antisym\'etrisant le r\'esultat obtenu. Le r\'esultat est
\bbb
\omega\wedge\eta=\lp
\omega_{\mu}\eta_{\nu}-\omega_{\nu}\eta_{\mu}\rp dx^{\mu}\wedge dx^{\nu}.\n
\eee

\par

En ce qui concerne la d\'etermination des formes de degr\'e $p>2$, la d\'emarche employ\'ee est en tout point identique: on obtient un isomorphisme d'alg\`ebres diff\'erentielles entre $\Omega^{n}_{\dd}(\aa)$, dont le repr\'esentant de chaque classe est un produit totalement antisym\'etrique de matrices de Dirac, et les formes de degr\'e $n$ de de Rham en identifiant $i\gamma^{\mu}$ et $dx^{\mu}$ \cite{bible}.

\par

En utilisant la relation $\gamma^{n+1}\gamma^{\sigma(1)}\dots\gamma^{\sigma(n)}=i^{n/2}\epsilon(\sigma)$ valable en dimension paire, on montre facilement que, en toute dimension,
\bbb
\dix(f_{0},f_{1},\dots,f_{n})=i^{\frac{n(n-1)}{2}}\int_{\mm} f_{0}\wedge df_{1}\wedge \dots \wedge df_{n}.\n
\eee
On retouve le complexe de de Rham et la th\'eorie usuelle de l'int\'egration des formes diff\'erentielles. 

\par

Avant de clore l'\'etude de cet exemple, il convient de faire deux remarques. Tout d'abord, dans le cas particulier consid\'er\'e, il est possible de d\'efinir sur $\pi\lp\Omega^{p}(\aa)\rp$ un produit scalaire par
\bbb
\langle\pi(\omega),\pi(\eta)\rangle=
\t_{\omega}\lp\pi(\omega)^{*}\pi(\eta)|\dd|^{-n}\rp\label{ad7}.
\eee
La v\'erification de tous les axiomes relatifs au produit scalaire est \'evidente car on a (cf Appendice A)
\bbb
\t_{\omega}\lp\pi(\omega)^{*}\pi(\eta)|\dd|^{-n})\rp=
\frac{1}{(4\pi)^{n/2}\Gamma(n/2+1)}\int_{\mm}\,\sqrt{g}\,dx^{n}\t\lp\pi(\omega)^{*}\pi(\eta)\rp,
\eee 
$\t$ d\'esignant la trace sur les matrices de Dirac. Le choix d'un repr\'esentant compl\`etement antisym\'etrique correspond \`a la d\'efinition de $\Omega_{\dd}(\aa)$ comme le supl\'ementaire orthogonal de $\pi(\jj)$ pour ce produit scalaire.

\par

En second lieu, il est utile de remarquer que notre d\'etermination des formes de degr\'e 0, 1 et 2 n'utilise que les propri\'et\'es de base des matrices de Dirac et la commutation $\lb\partial_{\mu},\partial_{\nu}\rb=0$ des d\'erivations de $\aa$, mais ne repose absolument pas sur la commutativit\'e de l'alg\`ebre. Nous verrons ult\'erieurement que ce r\'esultat s'applique encore au tore non commutatif (cf \S  4.1.4).  
\eexe
 

\subsection{Courbure et identit\'es de Bianchi}

Pour d\'efinir la courbure d'une connexion $\nabla$, il est n\'ecessaire de prolonger $\nabla$ en une application de $\ee\ot_{\aa}\Omega_{\dd}(\aa)$ dans $\ee\ot_{\aa}\Omega_{\dd}(\aa)$ \cite{bible}.

\begin{pro}
$\nabla$ s'\'etend de mani\`ere unique en une application lin\'eaire de $\ee\ot_{\aa}\Omega_{\dd}(\aa)$ dans $\ee\ot_{\aa}\Omega_{\dd}(\aa)$ satisfaisant \`a la relation
\bbb
\nabla(\xi\ot\omega)=\nabla(\xi)\omega+\xi\ot d\omega\n
\eee
pour tous $\xi\in\ee$ et $\omega\in\Omega_{\dd}(\aa)$.
\end{pro}

Le produit $\nabla(\xi)\omega$ est d\'efini de la mani\`ere suivante. Puisque $\nabla(\xi)$ est un \'el\'ement de $\ee\ot_{\aa}\Omega_{\dd}^{1}(\aa)$, on peut le mettre sous la forme $\nabla(\xi)=\sum_{i}\xi_{i}\ot\omega_{i}$ avec $\xi_{i}\in\ee$ et $\omega_{i}\in\Omega_{\dd}^{1}(\aa)$. Par d\'efinition, on prend
\bbb
\nabla(\xi)\omega=\mathop{\sum}\limits_{i}\xi_{i}\ot\omega_{i}\omega,
\eee
o\`u le dernier produit $\omega_{i}\omega$ est le produit usuel dans $\Omega_{\dd}(\aa)$. Il convient de remarquer que puisque $\Omega^{p}_{\dd}(\aa)$ est un ensemble quotient pour $p>1$, le produit $\omega_{i}\omega$ est un ensemble d'op\'erateurs sur $\hh$.

\par

A partir de cette d\'efinition de $\nabla$ sur tous les \'el\'ements de $\ee\ot_{\aa}\Omega_{\dd}(\aa)$, on v\'erifie ais\'ement la proposition suivante \cite{bible}. 

\begin{pro}
$\nabla:\;\ee\ot_{\aa}\Omega_{\dd}(\aa)\rightarrow\ee\ot_{\aa}\Omega_{\dd}(\aa)$ satisfait \`a la r\`egle de Leibnitz gradu\'ee
\bbb
\nabla(\eta\omega)=\nabla(\eta)\omega+(-1)^{p}\eta\nabla(\omega),\n
\eee
pour tous $\eta\in\ee\ot_{\aa}\Omega_{\dd}^{p}(\aa)$ et $\omega\in\Omega_{\dd}^{q}(\aa)$.
\end{pro}

Cela permet de d\'efinir toutes les puissances $\nabla^{n}$ de la connexion $\nabla$. En particulier, nous pouvons consid\'erer $\nabla^{2}$ comme une application de $\ee$ dans $\ee\ot_{\aa}\Omega^{2}_{\dd}(\aa)$. Cette application v\'erifie la propri\'et\'e suivante \cite{bible}.

\begin{dfi}
La restriction $F$ de $\nabla^{2}$ de $\ee$ dans $\ee\ot_{\aa}\Omega_{\dd}^{2}(\aa)$ est un morphisme de modules \`a droite sur $\aa$ appel\'e courbure.
\end{dfi}

A partir de cette d\'efinition, il est facile de montrer que sous une transformation de jauge, la courbure se transforme comme suit.

\begin{pro}
Sous une transformation de jauge d\'etermin\'ee par un endomorphisme unitaire $u\in\uu(\ee)$, la courbure $F$ se transforme en $uFu^{*}$.
\end{pro}

\demo
Sous une transformation de jauge, $\nabla$ devient $u\nabla u^{*}$, donc $\nabla^{2}$ devient $u\nabla^{2}u^{*}$. Puisque la multiplication par $u$ laisse $\ee\ot_{\aa}\Omega^{2}_{\dd}(\aa)$ stable, il est clair que $F$ devient $uFu^{*}$.
\edemo

Lorsque le module hermitien est \'ecrit sous la forme $\ee=e\aa^{N}$, il est possible d'obtenir une expression explicite de $F$.

\begin{pro}
Soit $e\in M_{N}(\aa)$ la projection hermitienne associ\'ee au module $\ee=e\aa^{N}$, $\nabla_{0}$ la connexion grassmanienne et $\nabla=\nabla_{0}+eAe$ une connexion hermitienne quelconque,  o\`u $A$ est un \'el\'ement hermitien de $M_{N}\lp\Omega^{1}_{\dd}(\aa)\rp$. La courbure de cette connexion est
\bbb
F=edede+ed(eAe)e+(eAe)^{2}.\label{ci1}
\eee
\end{pro}

\demo
Soit $\xi$ un \'el\'ement quelconque de $\aa^{N}$. Par d\'efinition de $\nabla$, on a
\bbbb
\nabla(e\xi)&=&ed(e\xi)+eAe\xi.\n\\
&=&ede\xi+ed\xi+eAe\xi\n\\
&=&ed\xi+B\xi,
\eeee
o\`u $B=ede+eAe$ est une matrice de 1-formes qui d\'efinit un homomorphisme entre les modules $e\aa^{N}$ et $e\lp\Omega_{\dd}^{1}(\aa)\rp^{N}$. En d\'esignant par $(\epsilon_{i})_{1\leq i\leq N}$ la base canonique de $\aa^{N}$, nous avons $B(\xi)=\sum_{i,j}e\epsilon_{i}B_{ij}\xi_{j}$, avec $B_{ij}\in\Omega^{2}_{\dd}(\aa)$ ainsi que $ed\xi=\sum_{i}e\epsilon_{i}d\xi_{i}$
pour tout $\xi=\sum_{i}\epsilon_{i}\xi_{i}$.

\par

En utilisant l'isomorphisme canonique entre $e\aa^{N}\ot_{\aa}\Omega_{\dd}^{1}(\aa)$ et $e\lp\Omega^{1}_{\dd}(\aa)\rp^{N}$, nous pouvons \'ecrire
\bbb
\nabla(e\xi)=\mathop{\sum}\limits_{i}e\epsilon_{i}\ot d\xi_{i}
+\mathop{\sum}\limits_{i,j}e\epsilon_{i}\ot B_{ij}\xi_{j}.\n
\eee
Sous cette forme, il est facile d'appliquer \`a nouveau $\nabla$ qui est d\'efini sur les produits tensoriels. On obtient
\bbbb
&\nabla^{2}(e\xi)=\mathop{\sum}\limits_{i}ed(e\epsilon_{i})\ot d\xi_{i}
Be\epsilon_{i}\ot d\xi_{i}+e\epsilon_{i}\ot d^{2}\xi_{i}&\n\\
&+\mathop{\sum}\limits_{i,j}ed(e\epsilon_{i})\ot B_{ij}\xi_{j}+
Be\epsilon_{i}\ot B_{ij}\xi_{j}+e\epsilon_{i}\ot d(B_{ij}\xi_{j}).&
\eeee
En utilisant les propri\'et\'es de la d\'eriv\'ee ext\'erieure $d$ et l'isomorphisme canonique entre $e\aa^{N}\ot_{\aa}\Omega_{\dd}^{2}(\aa)$ et $e\lp\Omega^{2}_{\dd}(\aa)\rp^{N}$, nous pouvons exprimer $\nabla^{2}(e\xi)$ \`a l'aide de $A$,
\bbbb
&\nabla^{2}(e\xi)=edeede\xi+edeeAe\xi+eAede\xi+eAeAe\xi&\n\\
&+ed(ede)\xi+ed(eAe)\xi-eded\xi-eAed\xi+eded\xi+eAed\xi.
\eeee
Puisque $e$ est une projection, on a $de=ede+dee$, ce qui implique que $edee=0$. L'expression pr\'ec\'edente se simplifie en
\bbb
\nabla^{2}(e\xi)=edede\xi+eAede\xi+eAeAe\xi+ed(eAe)\xi,\n
\eee
et ne contient plus de termes en $d\xi$, ce qui est normal car $\nabla^{2}$ est $\aa$-lin\'eaire. Pour obtenir le r\'esultat annonc\'e, il suffit de d\'evelopper $ed(eAe)$ en utilisant la r\`egle de Leibniz gradu\'ee et d'utiliser une fois encore la relation $de=ede+dee$. 
\edemo

Ce r\'esultat appelle deux remarques. Tout d'abord, il est sous-entendu dans la d\'emonstration pr\'ec\'edente que lorsque nous d\'erivons une matrice dont les coefficients sont des formes diff\'erentielles, nous d\'erivons en fait chacun de ces coefficients. De m\^eme, lorsque nous multiplions une matrice de $p$-formes par une matrice de $q$-formes, nous utilisons la d\'efinition usuelle du produit de matrice et nous multiplions les formes entre elles de mani\`ere \`a avoir une matrice de $(p+q)$-formes. Ces deux d\'efinitions donnent un sens aux expressions $d(eAe)$ et $eAeAe$ et justifient l'emploi des propri\'et\'es usuelles de $d$ ($d^{2}=0$ ainsi que la r\`egle de Leibniz gradu\'ee) dans la d\'emonstration pr\'ec\'edente.

\par

Ensuite, il est int\'eressant de remarquer que lorsque $e=1$, ce qui correspond \`a un module libre, la courbure se r\'eduit \`a sa forme usuelle $F=dA+A^{2}$. Dans le cas contraire la th\'eorie correspond \`a une th\'eorie de jauge sur un fibr\'e non trivial. La relation (\ref{ci1}) nous montre que m\^eme si $A=0$, la courbure $F$ est en g\'en\'eral non nulle lorsque $e$ est non trivial. 
 
\par

En g\'eom\'etrie diff\'erentielle usuelle, il exite une identit\'e appel\'ee identit\'e de Bianchi permettant de relier une connexion et sa courbure. Cela se g\'en\'eralise ais\'ement au cas non commutatif.

\begin{dfi}
La relation $\lb\nabla,F\rb=0$ est appel\'ee identit\'e de Bianchi.
\end{dfi}

Puisque $F$ est une matrice de 2-formes, l'identit\'e de Bianchi est une relation entre matrices de 3-formes, qui s'\'ecrit, lorsque la topologie est triviale, sous la forme usuelle $dF+[A,F]=0$.

\par

Pour nous convaincre de la pertinence de toutes ces d\'efinitions, examinons le cas particulier d'une th\'eorie de jauge avec un fibr\'e trivial sur une vari\'et\'e ordinaire.

\exe
Soit $\mm$ une vari\'et\'e compacte munie d'une structure de spin et $E$ un fibr\'e vectoriel complexe trivial de rang $N$ au dessus de $\mm$. Bien entendu, nous identifions l'alg\`ebre diff\'erentielle $\Omega_{\dd}(\aa)$ avec le complexe de de Rham en choisissant dans chaque classe le repr\'esentant totalement antisym\'etrique.

\par

En coordonn\'ees locales $x^{\mu}$, l'action d'une connexion $\nabla$ sur une section $\xi$ de $E$ s'\'ecrit
\bbb
\nabla(\xi)=D_{\mu}\xi\ot dx^{\mu},\n
\eee
o\`u $D_{\mu}=\partial_{\mu}+A_{\mu}$, o\`u $A_{\mu}$ est une matrice $N\times N$ antihermitienne, fonction des coordonn\'ees $x^{\mu}$. En utilisant le caract\`ere antisym\'etrique du produit de 1-formes on obtient
\bbb
\nabla^{2}(\xi)=1/2\,F_{\mu\nu}\xi\ot dx^{\mu}\wedge dx^{\nu},\n
\eee
o\`u
\bbb
F_{\mu\nu}=\lb D_{\mu},D_{\nu}\rb=\partial_{\mu}A_{\nu}-\partial_{\mu}A_{\nu}
+\lb A_{\mu},A_{\nu}\rb\n
\eee
est la courbure usuelle du champ de jauge $A_{\mu}$.

\par

Les transformations de jauge correspondent aux matrices unitaires fonctions des coordonn\'ees $x_{\mu}$. Par cons\'equent, le groupe de jauge est $U(N)$ et sous la transformation de jauge d\'etermin\'ee par l'unitaire $g$, la connexion $\nabla$ devient $g\nabla g^{-1}$ ce qui est \'equivalent \`a
\bbb
A_{\mu}\rightarrow gA_{\mu}g^{-1}+g\partial_{\mu}g^{-1}.\n
\eee

\par

La relation $\lb\nabla,\nabla^{2}\rb=0$ nous donne
\bbb
\lb D_{\lambda},F_{\mu\nu}\rb P(dx^{\lambda}\ot dx^{\mu}\wedge dx^{\nu})=0,\n
\eee
o\`u $P$ est le projecteur sur la partie totalement antisym\'etrique. Apr\`es avoir explicit\'e $P$, on obtient
\bbb
\frac{1}{3}\lb D_{\lambda},F_{\mu\nu}\rb dx^{\lambda}\wedge dx^{\mu}\wedge dx^{\nu}=0.\n
\eee
Apr\`es deux permutations cycliques des indices, on obtient
\bbb
D_{\lambda}F_{\mu\nu}+D_{\mu}F_{\nu\lambda}+D_{\nu}F_{\lambda\mu}=0,\n
\eee
ce qui est la forme usuelle des identit\'es de Bianchi.
\eexe


\subsection{Action de Yang-Mills}

Au cours de la section pr\'ec\'edente, nous avons vu que la courbure d'une connexion pouvait \^etre identifi\'ee de mani\`ere canonique \`a une matrice \`a coefficients dans $\Omega^{2}_{\dd}(\aa)$. Cela nous permet d'associer \`a toute connexion un nombre r\'eel, qui, dans le cas classique, n'est autre de l'action de Yang-Mills
\bbb
S_{YM}[A_{\mu}]=-\frac{1}{2}\int_{\mm}\t \lp F_{\mu\nu}F^{\mu\nu}\rp.\n
\eee 
Dans le cas g\'en\'eral, cette construction est plus compliqu\'ee car les \'el\'ements de $\Omega_{\dd}^{2}(\aa)$ sont des classes d'\'equivalences d'op\'erateurs. 

\par

Consid\'erons un triplet spectral $(\aa,\hh,\dd)$ de dimension $n$ et un module projectif fini $\ee$ \`a droite sur $\aa$ muni d'une structure hermitienne. Sans perte de g\'en\'eralit\'e, nous pouvons supposer que $\ee=e\aa^{N}$, o\`u $e\in M_{N}(\aa)$ est une projection hermitienne. La structure hermitienne sur $\ee$ est la structure hermitienne induite par celle de $\aa^{N}$ et nous d\'esignons par $\nabla$ une connexion hermitienne sur $\ee$ de courbure $F$.

\par

Notons $\tilde{\pi}$ la projection canonique de $\pi\lp\Omega(\aa)\rp$ dans $\Omega_{\dd}(\aa)$. $\tilde{\pi}$ s'\'etend canoniquement en un morphisme d'alg\`ebres gradu\'ees de $M_{N}(\aa)\ot_{\aa}\pi(\Omega(\aa))$ dans $M_{N}(\aa)\ot_{\aa}\Omega_{\dd}(\aa)$. Soit $G\in M_{N}(\aa)\ot_{\aa}\pi(\Omega^{2}(\aa))$ tel que $\tilde{\pi}(G)=F$. $G$ est une matrice $N\times N$ dont les \'el\'ements $G_{ij}$ appartiennent \`a l'alg\`ebre $\bb(\hh)$ des op\'erateurs born\'es sur $\hh$. $GG^{*}$ est \'egalement une telle matrice et on lui associe l'op\'erateur born\'e $\t(GG^{*})=\sum_{i,j} G_{ij}G_{ij}^{*}$. Puisque $\t(GG^{*})$ est born\'e, $\t(GG^{*})|\dd|^{-n}$ est un infinit\'esimal d'ordre un si bien que la quantit\'e
\bbb
\dix\t(GG^{*})ds^{n}=\frac{1}{(4\pi)^{n/2}\Gamma(1+n/2)}
\t_{\omega}\lp\t(GG^{*})|\dd|^{-n}\rp\n
\eee
est bien d\'efinie. Nous d\'efinissons alors l'action de Yang-Mills comme suit \cite{bible}.

\begin{dfi}
L'action de Yang-Mills est la fonctionnelle d\'efinie sur l'espace des connexions hermitiennes par
\bbb
S_{YM}[\nabla]=\mathop{\inf}\limits_{\tilde{\pi}(G)=F}\dix\t\lp GG^{*}\rp ds^{n},\n
\eee
o\`u $F$ est la courbure de $\nabla$ et la borne inf\'erieure est \`a prendre sur tous les repr\'esentants possibles de $F$. 
\end{dfi} 

La fonctionnelle pr\'ec\'edente satisfait au propri\'et\'es usuelles de l'action de Yang-Mills \cite{bible}.

\begin{pro}
L'action de Yang-Mills est une fonctionnelle positive et invariante sous les transformations de jauge.
\end{pro}

\demo
Bien que la d\'emonstration de ce r\'esultat se trouve dans \cite{bible}, nous en donnons ici une version diff\'erente. Puisque la positivit\'e ne fait aucun doute, \'etudions l'invariance. 

\par

Dans un premier temps, remarquons que sous une transformation de jauge d\'etermin\'ee par un endomorphisme unitaire $u$ du module $\ee$, les lois de transformation de la connexion et de sa courbure sont $\nabla\rightarrow u\nabla u^{*}$ et $F\rightarrow uFu^{*}$.
 
\par

Soit $G\in\pi(\Omega^{2}(\aa))$ un repr\'esentant de $F$. {\it Stricto sensu}, la loi de transformation de $G$ ne peut pas \^etre d\'efinie puisque $G$ n'est pas d\'efini univoquement \`a partir de $\nabla$. Cependant, on peut remarquer que $\tilde{\pi}(uGu^{*})=\tilde{\pi}(u)\tilde{\pi}(G)\tilde{\pi}(u^{*})=uFu^{*}$. Ainsi $uGu^{*}$ est un repr\'esentant de la courbure de $u\nabla u^{*}$, ce qui implique
\bbb
S_{YM}[u\nabla u^{*}]\leq \dix\t\lp uGG^{*}u^{*}\rp ds^{n}.\n
\eee 
En utilisant la propri\'et\'e de trace sur l'alg\`ebre $\pi(\Omega(\aa))$ de l'application $\pi(\omega)\mapsto\displaystyle\dix\pi(\omega)ds^{n}$, on a
\bbb
\dix\t \lp uGG^{*}u^{*}\rp ds^{n}=\dix\t \lp GG^{*}\rp ds^{n},\n
\eee
d'o\`u l'in\'egalit\'e $S_{YM}[u\nabla u^{*}]\leq S_{YM}[\nabla]$.

\par

En appliquant cette m\^eme in\'egalit\'e \`a la connexion $u^{*}\nabla u$ puis en changeant $u$ en $u^{*}$, on obtient $S_{YM}[\nabla]\leq S_{YM}[u\nabla u^{*}]$, d'o\`u l'\'egalit\'e et l'invariance de jauge de l'action de Yang-Mills.
\edemo

La d\'efinition de l'action de Yang-Mills est compliqu\'ee par le fait que nous travaillons avec des classes d'\'equivalence d'op\'erateurs et que nous n'avons pas, en g\'en\'eral, de possibilit\'e de choisir canoniquement un et un seul \'el\'ement dans chaque classe. 

\par

Dans le cas commutatif, l'application 
\bbb
\langle\pi(\omega),\pi(\eta)\rangle=
\dix\t\lp\pi(\omega)^{*}\pi(\eta)\rp ds^{n}\n
\eee
permet de d\'efinir un produit scalaire sur l'alg\`ebre gradu\'ee $\pi(\Omega(\aa))\ot_{\aa} M_{N}(\aa)$. A l'aide de ce produit tensoriel, on peut choisir dans $\pi(\Omega^{p}(\aa))$ un et un seul repr\'esentant de la classe de $F^{2}$ qui est orthogonal \`a tous les \'el\'ements de $\pi(J)\cap\pi(\Omega^{2}(\aa))$ (cf \S 1.3.1). Si nous notons encore $F$ le repr\'esentant qui correspond au produit totalement antisym\'etrique de matrices de Dirac, alors $F$ est orthogonal \`a tout \'el\'ement de $\pi(J)\cap\pi(\Omega^{2}(\aa))$ et tout repr\'esentant de la courbure s'\'ecrit sous la forme $G=F+\theta$, avec $\theta\in\pi(J)\cap\pi(\Omega^{2}(\aa))$. De plus on a
\bbb
\dix\t(GG^{*})ds^{n}=\dix\t(FF^{*})ds^{n}+\dix\t(\theta\theta^{*})ds^{n}
\label{aym1}
\eee 
par suite de l'orthogonalit\'e de $\theta$ et $F$. On en d\'eduit que la valeur minimale du second membre de (\ref{aym1}) est obtenue pour $\theta=0$, ce qui implique que
\bbb
S_{YM}[\nabla]=\dix\t(FF^{*})ds^{n}.\n
\eee
On retrouve ainsi la forme usuelle de l'action de Yang-Mills car la projection orthogonale correspond au choix du repr\'esentant r\'ealisant le minimum de la fonctionnelle de Yang-Mills. Ce r\'esultat s'\'etend au cas d'un triplet spectral obtenu par produit d'une g\'eom\'etrie commutative par une triplet spectral fini \cite{zyl}.


\subsection{Une in\'egalit\'e en dimension 4}

En dimension 4, l'action de Yang-Mills admet une borne inf\'erieure de nature topologique qui joue un r\^ole central en th\'eorie quantique des champs \cite{instantons}. Avant de g\'en\'eraliser cette in\'egalit\'e au cas non commutatif, commen\c cons par \'etudier une th\'eorie de Yang-Mills associ\'ee \`a un fibr\'e principal sur une vari\'et\'e compacte munie d'une m\'etrique euclidienne. En coordonn\'ees locales, la connexion peut s'\'ecrire comme une 1-forme $A=A_{\mu}dx^{\mu}$ \`a valeurs dans l'alg\`ebre de Lie du groupe de jauge, $SU(N)$ par exemple, et sa courbure est repr\'esent\'ee par la 2-forme $F=1/2\,F_{\mu\nu}dx^{\mu}\wedge dx^{\nu}$ avec $F_{\mu\nu}=\partial_{\mu}A_{\nu}-\partial_{\mu}A_{\nu}+\lb A_{\mu},A_{\nu}\rb$.    L'action de Yang-Mills est 
\bbb
S_{YM}=-\frac{1}{4}\int_{\mm}\sqrt{g}\,dx^{4} g^{\mu\rho}g^{\nu\sigma}\t(F_{\mu\nu}F_{\rho\sigma})=
-\frac{1}{2}\int_{\mm}\t(F\wedge *F)\n
\eee
le signe $-$ \'etant choisi de mani\'ere \`a obtenir une action positive car $F_{\mu\nu}$ est une matrice antihermitienne. Le dual de Hodge $*$ est d\'efini sur les 2-formes par
\bbb
*F=\frac{\sqrt{g}}{4}\epsilon_{\mu\nu\rho\sigma}F^{\mu\nu}dx^{\rho}dx^{\sigma}
\n
\eee
et satisfait \`a $*(*F)=F$.

\par

Il est facile de voir que l'action de Yang-Mills satisfait \`a l'in\'egalit\'e
\bbb
S_{YM}=-\frac{1}{2}\int_{\mm}\t(F\wedge *F)\geq
-\frac{1}{2}\int_{\mm}\t(F\wedge F)
\label{idq1}.
\eee
De plus, le second membre de (\ref{idq1}) s'\'ecrit localement comme l'int\'egrale d'une d\'eriv\'ee totale,
\bbb
\frac{1}{2}\int_{\mm}\t(F\wedge F)=\frac{1}{2}\int_{\mm}d\,\t\lp A\wedge dA+\frac{2}{3}A^{3}\rp,
\eee
et ne d\'epend que de la topologie du fibr\'e sous-jacent.

\par

Nous allons g\'en\'eraliser les r\'esultats pr\'ec\'edents \`a une th\'eorie de jauge d\'efinie sur un triplet spectral de dimension 4 \cite{bible}.

\begin{thm}
Soit $(\aa,\hh,\dd)$ un triplet spectral de dimension 4 satisfaisant \`a la condition de fermeture et soit $\nabla$ une connexion sur un module hermitien $\ee$ sur l'alg\`ebre $\aa$. Si $e\in M_{N}(\aa)$ est le projecteur correspondant \`a $\ee$, alors l'action de Yang-Mills satisfait \`a l'in\'egalit\'e
\bbb
S_{YM}[\nabla]\geq | \dix\gamma\t\lp\,edededede\rp ds^{4} |,
\eee
o\`u $\gamma$ est la chiralit\'e associ\'ee \`a $(\aa,\hh,\dd)$ et $de=[\dd,e]$.
\end{thm}

\demo

Sans perte de g\'en\'eralit\'e, nous pouvons supposer que $\ee=e\aa^{N}$ o\`u $e\in M_{N}(\aa)$ v\'erifie $e^{2}=e^{*}=e$. L'action de $\nabla$ sur $e\xi\in\ee$ s'\'ecrit $\nabla(e\xi)=ed(e\xi)+eAe\xi$, o\`u $A$ est un \'el\'ement hermitien de $M_{N}(\aa)\ot_{\aa}\Omega^{1}_{\dd}(\aa)$. La courbure de cette connexion est une matrice d'\'el\'ements de $\Omega^{2}_{\dd}(\aa)$ donn\'ee par $F=edede+ed(eAe)e+eAeAe$. Rappelons que $\Omega^{2}_{\dd}(\aa)$ est form\'e de classes d'\'equivalence d'op\'erateurs born\'es, aussi choisissons un repr\'esentant quelconque $F_{1}$ de $F$.

\par

Puisque $\gamma$ est une involution hermitienne de carr\'e $1$ qui commute avec
$F_{1}$, on a
\bbb
\dix( \gamma\pm 1)^{2 }\t\lp F_{1}^{*}F_{1} \rp ds^{4}\geq 0,\n
\eee
puis, en d\'eveloppant le premier membre, on obtient
\bbb
\pm\dix\gamma\t\lp F_{1}^{*}F_{1}\rp ds^{4}
\leq\dix\t\lp F_{1}^{*}F_{1}\rp ds^{4}\label{idq2}. 
\eee
Cette in\'egalit\'e est l'analogue non commutatif de (\ref{idq1}), il nous reste cependant \`a comparer le premier membre de (\ref{idq2}) avec l'action de Yang-Mills et il nous faut aussi mettre en \'evidence la signification topologique de son second membre.

\par

Commen\c cons par noter que le second membre de (\ref{idq2}) ne d\'epend pas du repr\'esentant $F_{1}$ de la courbure $F$. En effet, si $F_{2}$ est un autre repr\'esentant de $F$, alors $F_{1}-F_{2}=\theta$, o\`u $\theta$ est une matrice \`a coefficients dans $ \pi(J)\cap\pi(\Omega^{2}(\aa))$, d'o\`u
\bbb
F_{1}F_{1}^{*}=F_{2}F_{2}^{*}+\underbrace{F_{2}\theta^{*}+
\theta F_{2}^{*}+\theta\theta^{*}}_{\theta'}.\n
\eee
Puisque $J$ est un id\'eal stable par involution, $\t(\theta')$ est un \'el\'ement de $J\cap\pi(\Omega^{4}(\aa))$ et peut s'\'ecrire sous la forme
\bbb
\t(\theta')=\mathop{\sum}\limits_{i}
\lb\dd,a_{1}^{i}\rb\lb\dd,a_{2}^{i}\rb\lb\dd,a_{3}^{i}\rb\lb\dd,a_{4}^{i}\rb\n
\eee
avec $a_{p}^{i}\in\aa$. En utilisant la condition de fermeture, on obtient $\displaystyle\dix\gamma\t(\theta')ds^{4}=0$, ce qui montre que $\displaystyle\dix\gamma\t(F_{1}F_{1}^{*})ds^{4}=\displaystyle\dix\gamma\t(F_{2}F_{2}^{*})ds^{4}$. Notant $\dix\gamma\t(FF^{*})ds^{4}$ ce nombre, on a
\bbb
\dix\gamma\t\lp F_{1}^{*}F_{1}\rp ds^{4}\geq\dix\gamma\t\lp F^{*}F\rp ds^{4}.\n
\eee
Puisque cette in\'egalit\'e est vraie pour tout repr\'esentant $F_{1}$ de $F$, on a, par passage \`a la borne inferieure,
\bbb
S_{YM}[\nabla]\geq\dix\gamma\t\lp F^{*}F\rp ds^{4}\label{idq3}
\eee
Pour d\'emontrer le th\'eor\`eme pr\'ec\'edent, il nous reste \`a expliciter le second membre de (\ref{idq3}) et \`a montrer qu'il ne d\'epend de la projection $e$.

\par

Le lemme suivant introduit en g\'eom\'etrie non commutative l'analogue de la forme de Chern-Simons $AdA+2/3\,A^{3}$.

\begin{lem}
Soit $B\in M_{N}(\aa)\ot_{\aa}\Omega_{\dd}^{1}(\aa)$ une 1-forme hermitienne \`a valeurs dans $M_{N}(\aa)$ et soit $G=dB+B^{2}\in M_{N}(\aa)\ot_{\aa}\Omega_{\dd}^{1}(\aa)$ sa courbure. Alors on a 
\bbb
\dix\gamma\t(G_{1}^{2})ds^{4}=0,\n
\eee
pour tout repr\'esentant $G_{1}$ de la classe de $G$.
\end{lem}

\demo
Pour montrer ce lemme, introduisons la 3-forme \`a valeurs matricielles
\bbb
K=BdB+\frac{2}{3}B^{3}.
\eee
Par d\'erivation, on obtient dans $M_{N}(\aa)\ot_{\aa}\Omega_{\dd}^{4}(\aa)$
\bbb
dK=dBdB+\frac{2}{3}dBB^{2}-\frac{2}{3}BdBB+\frac{2}{3}B^{2}dB,\n
\eee
soit en introduisant la courbure $G=dB+B^{2}$,
\bbb
dK=\lp G-B^{2}\rp^{2}+\frac{2}{3} \lp G-B^{2}\rp B^{2}
-\frac{2}{3}B\lp G-B^{2}\rp B+\frac{2}{3}B^{2}\lp G-B^{2}\rp.\n
\eee
Si $G_{1}$ est un repr\'esentant de la classe de $G$, alors il d\'ecoule de la structure d'id\'eal de $J$ que
\bbb
\lp G_{1}-B^{2}\rp^{2}+\frac{2}{3} B^{2}\lp G_{1}-B^{2}\rp
-\frac{2}{3}B\lp G_{1}-B^{2}\rp B+\frac{2}{3}\lp G_{1}-B^{2}\rp B^{2}\n
\eee
est un repr\'esentant de la classe de $dK$. Puisque $dK$ est une d\'eriv\'ee, la condition de fermeture implique que l'int\'egrale de tous ses repr\'esentants est nulle. On en d\'eduit que
\bbbb
&\dix\gamma\t\lp G_{1}-B^{2}\rp^{2}ds^{4}+
\frac{2}{3}\dix\gamma\t\lp B^{2}( G_{1}-B^{2})\rp ds^{4}&\n\\
&-\frac{2}{3}\dix\gamma\t\lp B(G_{1}-B^{2}) B\rp ds^{4}
+\frac{2}{3}\dix\gamma\t \lp (G_{1}-B^{2})B^{2}\rp ds^{4}=0.&
\eeee
En utilisant les propri\'et\'es de trace gradu\'ee de $\dix\gamma\t$ sur $\pi(\omega^{4}_{\dd}(\aa))$, la relation pr\'ec\'edente se r\'eduit \`a
\bbb
\dix\gamma\t\lp G_{1}\rp^{2}ds^{4}=0,\n
\eee
ce qui prouve le lemme.
\edemo

\par

Pour d\'emontrer le r\'esultat annonc\'e, appliquons le lemme pr\'ec\'edent \`a  \bbb
B=ede-dee+eAe.\n
\eee
Pour calculer la courbure $G=dB+B^{2}$, d\'eterminons
\bbb
dB=d\lp ede-dee+eAe\rp=2dede+d(eAe)
\eee 
et
\bbbb
B^{2}&=&\lp  ede-dee+eAe\rp^{2},\n\\
&=&-ededee-deede-deeAe+eAede+eAeAe,
\eeee
o\`u on a utilis\'e les relations $de=ede+dee$ et  $edee=0$ obtenues en d\'erivant $e^{2}=e$. Regroupant tous les termes, on a
\bbb
G=dede-deeAe+eAede+d(eAe)+eAeAe,\n
\eee
ce qui nous donne
\bbbb
&eGe=edede+ed(eAe)e+eAeAe=F&\label{idq4}\\
&eG(1-e)=(1-e)Ge=0&\label{idq5}\\
&(1-e)G(1-e)=deede.&\label{idq6}
\eeee
Introduisons un repr\'esentant hermitien $F_{1}$ de la classe d'\'equivalence de la courbure $F=edede+ed(eAe)e+eAeAe$. 
Ainsi
\bbb
G_{1}=(1-e)\lb\dd,e\rb\lb\dd,e\rb+F_{1}\n
\eee
est un repr\'esentant hermitien de $G$ et on a
\bbbb
\lp G_{1}\rp^{2}&=&(1-e)dede(1-e)dede+\lp F_{1}\rp^{2}\n\\
&=&(1-e)dededede+\lp F_{1}\rp^{2}.
\eeee
En appliquant le lemme pr\'ec\'edent, on trouve
\bbb
\dix\gamma\t\lp(1-e)dededede\rp ds^{4}+
\dix\gamma\t\lp F_{1}F_{1}^{*}\rp ds^{4}=0.\n
\eee
D'apr\`es la condition de fermeture on a
\bbb
\dix\gamma\t\lp dededede\rp ds^{4}=0,\n
\eee
d'\`ou
\bbb
\dix\gamma\t\lp F_{1}F_{1}^{*}\rp ds^{4}=
\dix\gamma\t\lp edededede\rp ds^{4},\n
\eee
ce qui ach\`eve de d\'emontrer le r\'esultat annonc\'e.
\edemo

Nous avons montr\'e que l'action de Yang-Mills est born\'ee par un nombre qui ne d\'epend que du projecteur $e$ d\'efinissant le module projectif $\ee$. Cependant, pour pouvoir interpr\'eter ce nombre comme une quantit\'e topologique, il nous faut prouver que seule importe la classe d'\'equivalence du projecteur $e$ dans le groupe $K_{0}(\aa)$ 

\begin{pro}
Sous les hypoth\`eses du th\'eor\`eme pr\'ec\'edent, la quantit\'e 
\bbb
\dix\gamma\t\lp edededede\rp ds^{4}\n
\eee
ne d\'epend que de la classe de $e$ dans $K_{0}(\aa)$.
\end{pro}

\demo
Puisque le triplet $(\aa,\hh,\dd)$ satisfait \`a la condition de fermeture, $\Phi$ d\'efini par
\bbb
\Phi(a_{0},a_{1},\dots,a_{4})=\dix\gamma a_{0}\lb\dd,a_{1}\rb\dots\lb\dd,a_{4}\rb ds^{4}\n
\eee
pour tout $a_{0},\dots,a_{4}\in\aa$ est le caract\`ere du cycle associ\'e \`a l'alg\`ebre diff\'erentielle $\Omega_{\dd}(\aa)$ (cf \S 1.3.1). De plus, ce caract\`ere d\'etermine un cocycle cyclique sur $M_{N}(\aa)$ en faisant le produit "cup" avec la trace naturelle de $M_{N}(\aa)$ dans $\aa$ (cf Appendice C). Le cocycle obtenu est donn\'e par
\bbb
\tilde{\Phi}(a_{0},a_{1},\dots,a_{4})=\dix\gamma\t\lp a_{0}\lb\dd,a_{1}\rb\dots\lb\dd,a_{4}\rb\rp ds^{4}\n
\eee
Par cons\'equent, $\tilde{\Phi}$ est un cocycle cyclique, et $\tilde{\Phi}(e,e,e,e,e)$ est, \`a un facteur de normalisation pr\`es, l'accouplement entre ce cocycle cyclique et le groupe $K_{0}(\aa)$, ce qui implique que $\tilde{\Phi}(e,e,e,e,e)$ ne d\'epend que de la classe d'\'equivalence de $e$ dans $K_{0}(\aa)$.
\edemo

Ainsi, nous avons g\'en\'eralis\'e au cas non commutatif la minoration de l'action de Yang-Mills par une quantit\'e de nature topologique. En g\'eom\'etrie diff\'erentielle ordinaire, on peut montrer que ce minorant est toujours le produit d'une constante num\'erique par un nombre entier appel\'e "indice de Pontryagin". Ce r\'esultat ne s'\'etend pas en toute g\'en\'eralit\'e au cas non commutatif, mais nous montrerons, en appliquant le th\'eor\`eme de l'indice \cite{local}, qu'il reste valide sur le tore non commutatif. 

\par

L'in\'egalit\'e pr\'ec\'edente ne peut \^etre valable que pour un triplet spectral de dimension 4. En effet, en dimension $n\neq 4$, l'action de Yang-Mills se transforme en $\Lambda^{4-n}$ sous la transformation $\dd\rightarrow\Lambda\dd$ pour tout $\Lambda>0$, alors qu'une quantit\'e topologique est n\'ecessairement invariante sous cette transformation. 


\subsection{Action de Chern-Simons}

Une th\'eorie des champs topologique \cite{blau} est une th\'eorie des champs dont la  construction ne n\'ecessite que des informations de nature topologique sur la structure de l'espace temps. La quanfication d'une telle th\'eorie permet de d\'eterminer explicitement des invariants topologiques, et se r\'ev\`ele tr\`es utile sur le plan math\'ematique.

\par

Parmi toutes les th\'eories topologiques, celle de Chern-Simons est certainement la plus c\'el\`ebre car elle permet de d\'eterminer des invariants topologiques des vari\'et\'es de dimension 3, qui ne sont autres que les valeurs de la fonction de partition pour diff\'erents choix de l'espace-temps. De plus, les valeurs moyennes dans le vide de certaines observables (les lignes de Wilson) sont directement reli\'es aux invariants polynomiaux de la th\'eorie des noeuds \cite{witten}.

\par

Plus pr\'ecis\'ement, si $\mm$ est une vari\'et\'e compacte de dimension 3 et $G$ un groupe de Lie compact, que l'on peut choisir comme \'etant $SU(N)$ pour fixer les id\'ees, on d\'efinit pour toute 1-forme $A$ \`a valeurs dans l'alg\`ebre de Lie de $G$, l'action de Chern-Simons par
\bbb
S_{CS}[A]=\frac{k}{4\pi}\int\t\lp AdA+\frac{2}{3}A^{3}\rp,\n
\eee
o\`u $k\in\rrr$ est une constante de couplage. Il est remarquable que cette action ne fasse pas appel au choix d'une m\'etrique sur $\mm$, puisqu'on int\`egre une 3-forme en dimension 3.

\par

Sous une transformation de jauge d\'etermin\'ee par une application $g$ de $\mm$ dans $G$, le champ de jauge $A$ devient $A^{g}=gAg^{-1}+gdg^{-1}$ et il est facile de montrer, en utilisant les propri\'et\'es usuelles de l'int\'egration des formes differentielles, que l'action n'est pas invariante de jauge, 
\bbb
S_{CS}[A^{g}]=S_{CS}[A]+\frac{k}{12\pi}\int\t\lp\lp gdg^{-1}\rp^{3}\rp.\n
\eee
Si nous normalisons les g\'en\'erateurs $T^{a}$ de l'alg\`ebre de Lie de $G$ par la condition $\t(T^{a}T^{b})=-1/2\delta^{ab}$, alors on montre que
\bbb
\frac{1}{24\pi^{2}}\int\t\lp\lp gdg^{-1}\rp^{3}\rp=n \n
\eee
est un nombre entier usuellement appel\'e "degr\'e" de l'application $g$ de $\mm$ dans $SU(N)$.

\par

Par cons\'equent, si $k$ est aussi un entier, $e^{ikS_{CS}[A]}$ est invariant sous une transformation de jauge si bien que la fonction de partition
\bbb
Z[\mm]=\int [DA]e^{ikS_{CS}[A]}\n
\eee
est invariante de jauge et nous permet de d\'efinir une th\'eorie quantique des champs, qui apr\`es r\'egularisation, nous donne un invariant topologique.

\par

Avant de chercher \`a \'etendre ce r\'esultat en g\'eom\'etrie non commutative, il est important de noter que le d\'efaut d'invariance de jauge de l'action de Chern-Simons est du \`a la non connexit\'e du groupe des applications de $\mm$ dans $G$. En effet, si $g_{t}$ en un chemin diff\'erentiable de telles applications reliant $g_{0}$ et $g_{1}$, on montre que
\bbb
\frac{d}{dt}\;\frac{1}{24\pi^{2}}\int\t\lp\lp g_{t}dg^{-1}_{t}\rp^{3}\rp=0,\n
\eee
ce qui prouve que ce nombre ne d\'epend que de la composante connexe de $g$.

\par

En g\'eom\'etrie non commutative, le groupe des applications de $\mm$ dans $G$ est remplac\'e par le groupe des unitaires de $M_{N}(\aa)$, o\`u $\aa$ est l'alg\`ebre des coordonn\'ees et le d\'efaut de connexit\'e de ce groupe est mesur\'e par le $K_{1}(\aa)$ (cf Appendice B). Aussi allons-nous relier l'invariance de jauge de la fonctionnelle de Chern-Simons au couplage entre la cohomologie cyclique et le groupe $K_{1}(\aa)$.  

\par

Commen\c cons par d\'efinir l'action de Chern-Simons dans le cadre non commutatif.

\begin{dfi}
Soit $\lp\aa,\hh,\dd\rp$ un triplet spectral de dimension 3 satisfaisant \`a la condition de fermeture et soit $A\in M_{N}(\aa)\ot_{\aa}\Omega_{\dd}^{1}(\aa)$ une 1-forme hermitienne \`a valeurs dans $M_{N}(\aa)$. L'action de Chern-Simons $S_{CS}[A]$ est d\'efinie par
\bbb
S_{CS}[A]=\dix\t\lp K_{1}\rp ds^{3},\n
\eee
o\`u $K_{1}$ est un repr\'esentant quelconque de la classe d'\'equivalence de la forme de Chern-Simons
\bbb
K=AdA+\frac{2}{3}A^{3}.\n
\eee
\end{dfi}

Il est important de noter que gr\^ace \`a la condition de fermeture, cette d\'efinition ne d\'epend pas du repr\'esentant choisi de la classe de $K$. En effet si $K_{2}$ est un autre repr\'esentant de $K$, alors $K_{1}-K_{2}$ est une matrice \`a coefficients dans $\pi(J)\cap\pi\lp\Omega^{3}(\aa)\rp$ se met sous la forme
\bbb
\t\lp K_{1}\rp -\t\lp K_{2}\rp =\mathop{\sum}\limits_{i}
\lb\dd,a_{1}^{i}\rb\lb\dd,a_{2}^{i}\rb\lb\dd,a_{3}^{i}\rb,\n
\eee  
o\`u $a_{1}^{i}$, $a_{2}^{i}$ et $a_{3}^{i}$ sont des \'el\'ements de $\aa$. D'apr\`es la condition de fermeture, on a
\bbb
\dix \lb\dd,a_{1}^{i}\rb\lb\dd,a_{2}^{i}\rb\lb\dd,a_{3}^{i}\rb ds^{3}=0,\n 
\eee
ce qui implique que
\bbb
\dix\t\lp K_{1}\rp ds^{3}=\dix\t\lp K_{2}\rp ds^{3}.\n 
\eee
A priori, il n'est pas \'evident que cette action ait r\'eellement un caract\`ere topologique en g\'eom\'etrie non commutative car elle d\'epend du choix de l'op\'erateur de Dirac qui rec\`ele une information de nature m\'etrique; une th\'eorie topologique ne devant d\'ependre que de la structure de $C^{*}$-alg\`ebre sous-jacente. Toutefois, il est clair que cette action ne d\'epend que de la structure de cycle sur $\Omega_{\dd}(\aa)$ d\'efinie par le triplet spectral $(\aa,\hh,\dd)$.

\par

\'Etudions l'invariance de jauge de cette action. 

\begin{thm}
Soit $\lp\aa,\hh,\dd\rp$ un triplet spectral de dimension 3 satisfaisant \`a la condition de fermeture. Alors, sous une transformation de jauge d\'etermin\'ee par $u\in U_{N}(\aa)$, l'action de Chern-Simons $S_{CS}[A]$ devient
\bbb
S_{CS}[uAu+udu^{-1}]=S_{CS}[A]+\Gamma[u],
\eee
avec
\bbb
\Gamma[u]=-\frac{1}{3}\dix\t\lp udu^{-1}udu^{-1}udu^{-1}\rp ds^{3}.
\eee
\end{thm}

\demo

Introduisons la courbure $F=dA+A^{2}$ de $A$. On peut \'ecrire
\bbb
AdA+\frac{2}{3}A^{3}=AF-\frac{1}{3}A^{3},\n
\eee
ce qui implique que si $F_{1}$ est un repr\'esentant de $F$, alors l'action de Chern-Simons est
\bbb
S_{CS}[A]=\dix\t\lp AF_{1}-\frac{1}{3}A^{3}\rp ds^{3}.\n
\eee
Sous une transformation de jauge $A$ devient $uAu^{-1}+udu^{-1}$ et $F$ se transforme en $uFu^{-1}$, d'o\`u
\bbb
S_{CS}[uAu^{-1}+udu^{-1}]=\dix\t\lp \lp uAu^{-1}+udu^{-1}\rp uF_{1}u^{-1}
-1/3\lp uAu^{-1}+udu^{-1}\rp^{3}\rp ds^{3}.\n
\eee
En utilisant uniquement la propri\'et\'e de trace de l'application $\pi(\omega)\mapsto\dix\t(\pi(\omega))ds^{3}$ et la relation $du^{-1}u+u^{-1}du=0$, on obtient 
\bbb
\dix\t\lp \lp uAu^{-1}+udu^{-1}\rp uF_{1}u^{-1}\rp ds^{3}=\dix\t\lp AF\rp ds^{3}-
\dix\t \lp duFu^{-1}\rp ds^{3}\n
\eee 
ainsi que
\bbbb
&-1/3\dix\t\lp\lp uAu^{-1}+udu^{-1}\rp^{3}\rp ds^{3}=&\n\\
&-\frac{1}{3}\dix\t\lp A^{3}\rp ds^{3}-\frac{1}{3}\dix\t\lp\lp udu^{-1}\rp^{3}\rp ds^{3}-\dix\t\lp udu^{-1}uA^{2}u^{-1}\rp ds^{3}+\dix\t\lp duAdu^{-1}\rp ds^{3}.&
\eeee
Regroupant tous les termes, on obtient
\bbbb
S_{CS}[uAu+udu^{-1}]&=&S_{CS}[A]+\Gamma[u]\n\\
&+&\dix\t\lp udu^{-1}uF_{1}u^{-1}-udu^{-1}uA^{2}u^{-1}+du Adu^{-1}\rp ds^{3}.\n
\eeee
L'int\'egrant du second membre de cette \'equation est un repr\'esentant de
\bbbb
udu^{-1}uF_{1}u^{-1}-udu^{-1}uA^{2}u^{-1}+du Adu^{-1}&=&udu^{-1}udAu^{-1}+duAdu^{-1}\n\\
&=&-dudAu^{-1}+duAdu^{-1}\n\\
&=&-dud\lp Au^{-1}\rp\n\\
&=&-d\lp ud(Au^{-1})\rp.
\eeee
Puisque c'est le repr\'esentant d'une d\'eriv\'ee totale, par application de la condition de fermeture, on a
\bbb
\dix\t\lp udu^{-1}uF_{1}u^{-1}-udu^{-1}uA^{2}u^{-1}+du Adu^{-1}\rp ds^{3}=0,\n
\eee
 ce qui prouve que
\bbb
S_{CS}[uAu+udu^{-1}]=S_{CS}[A]+\Gamma[u].\n
\eee
\edemo

De plus, nous pouvons interpr\'eter $\Gamma[u]$ comme un couplage entre un cocycle cyclique et le groupe $K_{1}(\aa)$, ce qui assure sa stabilit\'e par d\'eformation (cf Appendice C).

\begin{pro}
Sous les hypoth\`eses du th\'eor\`eme pr\'ec\'edent, $\Gamma[u]$ ne d\'epend que de la classe de $u$ dans $K_{1}(\aa)=\pi_{0}(U_{\infty}(\aa))$.
\end{pro}

\demo

La condition de fermeture nous montre que $\Phi$ d\'efini par
\bbb
\Phi(a_{0},a_{1},a_{2},a_{3})=\dix\lp a_{0}\lb\dd,a_{1}\rb\lb\dd,a_{2}\rb\lb\dd,a_{3}\rb\rp ds^{3},\n
\eee
est un cocyle cyclique sur $\aa$ qui s'\'etend \`a $M_{N}(\aa)$ \`a l'aide du produit "cup" avec la trace pour donner le cocycle cyclique
\bbb
\tilde{\Phi}(a_{0},a_{1},a_{2},a_{3})=\dix\t\lp a_{0}\lb\dd,a_{1}\rb\lb\dd,a_{2}\rb\lb\dd,a_{3}\rb\rp ds^{3}.\n
\eee
Par cons\'equent $\tilde{\Phi}(u-1,u^{-1}-1,u-1,u^{-1}-1)$ ne d\'epend que de la classe de $u$ dans le groupe $K_{0}(\aa)$. 

\par

Pour terminer, il nous suffit de remarquer que $\Gamma[u]=1/3\tilde{\Phi}(u,u^{-1}-1,u-1,u^{-1}-1)$ car on a
\bbb
\lp udu^{-1}\rp^{3}=-udu^{-1}dudu^{-1}=-ud\lp u^{-1}-1\rp d\lp u-1\rp d\lp u^{-1}-1\rp,\n
\eee
et que par application de la condition de fermeture,
\bbb
\tilde{\Phi}(1,u^{-1},u,u^{-1})=\dix\t\lp 1,u^{-1},u-1,u^{-1}-1\rp ds^{3}=0,\n
\eee
ce qui prouve que
\bbb
\Gamma[u]=\frac{1}{3}\tilde{\Phi}(u-1,u^{-1}-1,u-1,u^{-1}-1).\n
\eee
A un facteur de normalisation, $\Gamma[u]$ est \'egal au couplage d'un cocycle cyclique avec un unitaire $u$. Par cons\'equent, $\Gamma[u]$ ne d\'epend que de $u$ dans $K^{1}(\aa)=\pi_{0}(U_{\infty}(\aa))$. 
\edemo

En cons\'equence, lorsque la transformation de jauge est situ\'ee dans la composante connexe de l'identit\'e, $\Gamma[u]=0$ et l'action de Chern-simons est invariante sous cette transformation.

\par

Toutefois, nous ne pouvons pas montrer en toute g\'en\'eralit\'e que $\Gamma[u]$ est proportionel \`a un entier. Cependant, $\Gamma[u]$ ne peut prendre qu'un nombre r\'estreint de valeurs lorsque $u$ parcourt tout le groupe des unitaires de $M_{N}(\aa)$.

\begin{cor}
Lorsque l'alg\`ebre $\aa$ est s\'eparable, $\Gamma[u]$ ne prend qu'un nombre d\'enombrable de valeurs.
\end{cor}

En effet, lorsque $\aa$ est s\'eparable, $K_{1}(\aa)$ est d\'enombrable; les $C^{*}$-alg\`ebres s\'eparables jouant le r\^ole des espaces topologiques m\'etrisables (cf Appendices A et B).

\section{La formule de l'indice}

\subsection{Le probl\`eme de l'indice}

En g\'eom\'etrie diff\'erentielle classique, on peut montrer qu'en dimension 4 la borne inf\'erieure de l'action de Yang-Mills est proportionelle \`a un nombre entier. De m\^eme, en dimension 3, les seules valeurs de la fonction $\Gamma[u]$ apparaissant dans l'invariance de jauge de l'action de Chern-Simons sont des multiples entiers d'une certaine constante num\'erique. 

\par

Pour \'etendre ces r\'esultats au cas non commutatif, il est utile de relier ces quantit\'es \`a l'indice d'un certain op\'erateur, en suivant une d\'emarche analogue \`a celle employ\'ee par J. Bellissard dans son \'etude de l'effet Hall quantique \cite{hall}. Commen\c cons par rappeler certaines notions concernant l'indice d'un op\'erateur \cite{gilkey}.  

\begin{dfi}
Soit $\hh$ un espace de Hilbert et $T$ un op\'erateur born\'e sur $\hh$. On dit que $T$ est un op\'erateur de Fredholm si $\ker T$ et $\ker T^{*}$ sont de dimension finie et on appelle indice de $T$ la diff\'erence $\dim\ker T- \dim\ker T^{*}$.
\end{dfi}

La caract\'erisation suivante des op\'erateurs de Fredholm se r\'ev\`ele tr\`es utile dans notre contexte.

\begin{pro}
Un op\'erateur born\'e $T$ est un op\'erateur de Fredholm si et seulement si il existe un op\'erateur born\'e $S$ tels que $TS-1$ et $ST-1$ soient des op\'erateurs compacts.
\end{pro}

Pour pouvoir d\'eterminer explicitement l'indice d'un op\'erateur, il est commode d'introduire pour tout espace de Hilbert $\hh$ et tout r\'eel $p\geq 1$ l'ensemble $L^{p}(\hh)$ form\'e des op\'erateurs compacts tels que la s\'erie $\sum_{n}(\lambda_{n})^{p}$ converge, o\`u $(\lambda_{n})_{n\in\nnn}$ d\'esigne la suite des valeurs propres d\'ecroissantes de $|T|=\sqrt{TT^{*}}$. Si $T\in L^{p}(\hh)$, pour toute base hilbertienne $(e_{n})_{n\in\nnn}$ la s\'erie $\sum_{n}\langle e_{n}, T^{p} e_{n}\rangle$
est convergente et sa somme, qui ne d\'epend pas de la base choisie, est not\'ee $\t(T^{p})$. On d\'emontre alors le r\'esultat suivant \cite{simon}.

\begin{pro}
$L^{P}(\hh)$ est un id\'eal bilat\`ere de l'alg\`ebre des op\'erateurs born\'es sur $\hh$ et on a, 
\bbb
L^{p}(\hh)L^{q}(\hh)\subset L^{r}(\hh)\n
\eee
pour tous nombres r\'eels $p$, $q$ et $r$ $\leq 1$ tels que $\frac{1}{p}+\frac{1}{q}=\frac{1}{r}$, ainsi que $L^{p}(\hh)\subset L^{q}(\hh)$ si $p\geq q$.
\end{pro}

Les id\'eaux $L^{p}(\hh)$ sont appel\'es id\'eaux de Schatten. Bien entendu, $L^{1}(\hh)$ est form\'e des op\'erateurs \`a trace. Il est important de remarquer que cette d\'efinition fait appel \`a la trace usuelle et non \`a la trace de Dixmier et il ne faut pas confondre les \'el\'ements de $L^{p}(\hh)$ avec les infinit\'esimaux d'ordre $p$. 

\par

Si nous supposons que les valeurs propres des op\'erateurs compacts $TS-1$ et  $ST-1$ d\'ecroissent suffisament rapidement pour que la trace de leur puissance p-i\`eme existe, nous obtenons une expression explicite de l'indice de $T$ \cite{ihes}.

\begin{pro}
Soient $p\geq 1$, $T$ et $S$ deux op\'erateurs born\'es tels que $TS-1\in L^{p}(\hh)$ et $ST-1\in L^{p}(\hh)$. Alors $T$ est un op\'erateur de Fredholm et pour tout entier $n\geq p$ on a
\bbb
\mathrm{Ind}(T)=\t\lp 1-ST\rp^{n}-\t\lp 1-TS\rp^{n}.\n
\eee
\end{pro}

Nous ne pouvons pas appliquer directement le r\'esultat pr\'ec\'edent \`a un triplet spectral $(\aa,\hh,\dd)$ car l'op\'erateur de Dirac est un op\'erateur non born\'e. En fait, la formulation du th\'eor\`eme de l'indice en g\'eom\'etrie non commutative fait appel \`a la notion de module de Fredholm.

\begin{dfi}
Soit $\aa$ une alg\`ebre involutive. Un module de Fredholm sur $\aa$ est un couple $(\hh,F)$, o\`u $\hh$ est un espace de Hilbert sur lequel $\aa$ a une repr\'esentation $\pi$ et $F$ est une isom\'etrie hermitienne telle que $\lb F,\pi(x)\rb$ soit compact pour tout $x\in\aa$.
\end{dfi} 

Pour pouvoir \^etre utilis\'ee dans notre contexte, cette d\'efinition doit \^etre affin\'ee en introduisant la parit\'e et la dimension.

\begin{dfi}
Un module de Fredholm $(\hh,F)$ sur une alg\`ebre involutive $\aa$ est pair s'il existe une involution hermitienne $\gamma$ telle que $\gamma F+F\gamma=0$, $\gamma\pi(x)=\pi(x)\gamma$ pour tout $x\in\aa$. Pour tout entier $p\geq 1$, $(\hh,F)$ est dit $p$-sommable si, pour tout $x\in\aa$, $[F,\pi(x)]\in L^{p}(\hh)$.  
\end{dfi}

L'introduction de la notion de module de Fredholm est motiv\'ee par le th\'eor\`eme suivant, qui forme la premi\`ere partie du th\'eor\`eme de l'indice en g\'eom\'etrie non commutative \cite{ihes}. 

\begin{thm}
Soit $\aa$ une alg\`ebre involutive et $(\hh,F)$ un module de Fedholm n+1-sommable sur $\aa$. En rempla\c cant $\hh$ par $\hh\ot\ccc^{N}$ et $F$ par $F\ot I_{N}$ nous d\'efinissons un module de Fredholm sur $M_{N}(\aa)$ encore not\'e $(\hh,F)$. De plus, si $n$ est pair nous supposons que $(\hh,F)$ est pair et nous notons $F^{+}=\frac{\gamma-1}{2}F\frac{\gamma+1}{2}$ et si $n$ est impair nous notons $P=\frac{1+F}{2}$. Alors on a, en omettnat la repr\'esentation $\pi$:

\begin{enumerate}
\item
si $n$ est pair, pour tout $e\in M_{N}(\aa)$ tel que $e^{2}=e^{*}=e$, $eF^{+}e$ est un op\'erateur de Fredholm et son indice est donn\'e par
\bbb
\mathrm{Ind}(eF^{+}e)=(-1)^{p}\t\lp \gamma e\lb F,e\rb^{2p}\rp\n
\eee
pour tout entier $p>(n+1)/2$.
\item
si $n$ est impair, alors pour tout unitaire $u\in M_{N}(\aa)$, $PuP$ est un op\'erateur de Fredholm sur $\hh'=P\hh$ et son indice est donn\'e par
\bbb
\mathrm{Ind}(PuP)=(-1)^{p+1}
\frac{1}{2^{2p-1}}\t\lp (u-1)[F,u^{-1}-1] \underbrace{[F,u-1][F,u^{-1}-1]\dots [F,u-1][F,u^{-1}-1]}_{2p-2\,termes} \rp\n
\eee
pour tout entier $p>\frac{n+1}{2}$.
\end{enumerate}
\end{thm}

\demo
La d\'emonstration de ce r\'esultat se trouve dans \cite{bible} lorsque $n$ est  pair, le cas impair \'etant laiss\'e en exercice. Nous nous occuperons de ce dernier, en nous inspirant des m\'ethodes d\'evelopp\'ees dans \cite{bible} et \cite{ihes}.

D\'efinissons deux op\'erateurs de $\hh'$ par $T=PuP$ et $S=Pu^{*}P$. Il v\'erifient
\bbb
ST-1_{\hh'}=Pu^{*}PuP-P=P[P,u^{*}][P,u]\n
\eee 
ainsi que
\bbb
TS-1_{\hh'}=PuPu^{*}P-P=P[P,u][P,u^{*}].\n
\eee
Puisque le module est $n+1$-sommable, par la proposition 1.4.2 $P[P,u^{*}][P,u]$ et $P[P,u][P,u^{*}]$ sont des \'el\'ements de $L^{\frac{n+1}{2}}(\hh)$. De plus, $\hh'$ est stable par ces op\'erateurs et leur action sur le suppl\'ementaire orthogonal de $\hh'$ dans $\hh$ est triviale. Par cons\'equent, on peut les consid\'erer comme des \'el\'ements de $L^{\frac{n+1}{2}}(\hh')$, et on a, par la proposition 1.4.3, 
\bbb
\mathrm{Ind}(T)=\t_{\hh'}\lp 1_{\hh'}-ST\rp^{p}-\t_{\hh'}\lp 1_{\hh'}-TS\rp^{p},\n
\eee
pour tout entier $p>\frac{n+1}{2}$. Cela s'\'ecrit aussi sous la forme
\bbb
\mathrm{Ind}(PuP)=(-1)^{p}\t\lp P[P,u^{*}][P,u]\rp^{p}-
(-1)^{p}\t\lp P[P,u][P,u^{*}]\rp^{p},\n
\eee
o\`u la trace est \`a prendre dans $\hh$.

\par

Puisque $P$ est un projecteur, on a $P[P,u]P=0$ et $P[P,u^{*}]P=0$ ce qui implique que $P[P,u^{*}][P,u]P=P[P,u^{*}][P,u]$ et que $P[P,u][P,u^{*}]P=P[P,u][P,u^{*}]$. On a donc 
\bbb
\mathrm{Ind}(PuP)=(-1)^{p}\t\lp PQ[P,u]\rp-(-1)^{p}\t\lp P[P,u]Q\rp\n
\eee 
avec
\bbb
Q=[P,u^{*}]\lp [P,u][P,u^{*}]\rp^{p-1}.\n
\eee
Le module de Fredholm \'etant $n+1$-sommable, on a $Q\in L^{\frac{n+1}{2p-1}}\subset L^{1}(\hh)$ ce qui prouve que $Q$ est un op\'erateur \`a trace, d`o\`u
\bbbb
\mathrm{Ind}(PuP)&=&(-1)^{p}\t\lp [P,u]PQ\rp-(-1)^{p}\t\lp P[P,u]Q\rp\n\\
&=&(-1)^{p}\lp 2\t\lp PuPQ\rp -\t\lp PuQ\rp-\t\lp uPQ\rp\rp.
\eeee 
De plus on a
\bbbb
P[P,u][P,u^{*}]&=&[P,u][P,u^{*}]-[P,u]P[P,u^{*}]\n\\
&=&[P,u][P,u^{*}]-[P,u][P,u^{*}]+[P,u][P,u^{*}]P\n\\
&=&[P,u][P,u^{*}]P,
\eeee
puisque $P^{2}=P$. On en d\'eduit que
\bbbb
PQ&=&P[P,u^{*}]\lp [P,u][P,u^{*}]\rp^{p-1}\n\\
&=&[P,u^{*}]\lp [P,u][P,u^{*}]\rp^{p-1}-[P,u^{*}]P\lp [P,u][P,u^{*}]\rp^{p-1}\n\\
&=&[P,u^{*}]\lp [P,u][P,u^{*}]\rp^{p-1}-[P,u^{*}]\lp [P,u][P,u^{*}]\rp^{p-1}P\n\\
&=&Q(1-P),
\eeee
ce qui implique que $\t\lp PuPQ\rp=0$ et $\t\lp uPQ\rp=\t\lp(1-P)uQ\rp$. On en d\'eduit que
\bbb
\mathrm{Ind}(PuP)=-(-1)^{p}\t\lp PuQ\rp-(-1)^{p}\t\lp (1-P)uQ\rp=
(-1)^{p+1}\t\lp uQ\rp.\n
\eee 
En utilisant la relation $PQ=Q(1-P)$, on montre que $\t(Q)=0$, ce qui implique que
\bbb
\mathrm{Ind}(PuP)=(-1)^{p+1}\t\lp (u-1)\underbrace{[P,u^{*}-1][P,u-1]\dots[P,u-1][P,u^{*}-1]}_{2p-1\,termes}\rp.\n
\eee 
Finalement, en utilisant $P=(1+F)/2$, il vient
\bbb
\mathrm{Ind}(PuP)=\frac{(-1)^{p+1}}{2^{2p-1}}\t\lp (u-1)\underbrace{[F,u^{*}-1][F,u-1]\dots[F,u-1][F,u^{*}-1]}_{2p-1}\rp,\n
\eee
ce qui est bien le r\'esultat annonc\'e.
\edemo

Ces relations sont tr\`es similaires aux relations apparaissant dans la d\'efinition du couplage entre la K-th\'eorie et la cohomologie cyclique p\'eriodique. En fait, on peut construire \`a l'aide d'un module de Fredholm une suite de cocycles cycliques \cite{bible}.

\begin{pro}
Soit $(\hh,F)$ un module de Fredholm $(n+1)$-sommable ayant la m\^eme parit\'e que $n$. Alors pour tout entier $m$, les relations suivantes d\'efinissent des $n+2m$-cocycles cycliques sur $\aa^{n+2m+1}$:
\bbbb
\tau_{n+2m}(a_{0},a_{1},\dots,a_{n+2m})&=&
\t'\lp\gamma a_{0}[F,a_{1}]\dots[F,a_{n+2m}]\rp
\;si\; n\;est\; pair,\n\\
\tau_{n+2m}(a_{0},a_{1},\dots,a_{n+2m})&=&
\t'\lp a_{0}[F,a_{1}]\dots[F,a_{n+2m}]\rp
\;si\; n\;est\; impair,\n
\eeee
o\`u $\t'(T)=\frac{1}{2}\t\lp F(FT+TF)\rp$.
\end{pro}

Nous utilisons $\t'$ au lieu de la trace ordinaire pour pouvoir d\'efinir $\tau_{n}$; lorsque $m>0$ on peut remplacer $\t'$ par $\t$ dans la d\'efinition de $\tau_{n+2m}$. Tous les termes de cette suite de cocyles sont reli\'es par l'op\'erateur de p\'eriodicit\'e $S$ (cf \S Appendice C).

\begin{pro}
Les cocycles pr\'ec\'edents satisfont \`a la relation
\bbb
\tau_{n+2m+2}=-\frac{2}{n+2m+2}S\tau_{n+2m},\n
\eee
o\`u $S:\; HC^{n+2m}(\aa)\rightarrow HC^{n+2m+2}$ est l'op\'erateur de p\'eriodicit\'e.
\end{pro}

Cela nous am\`ene naturellement \`a changer les normalisations de ces cocyles de mani\`ere \`a pouvoir definir une classe en cohomologie cyclique p\'eriodique.

\begin{dfi}
Le caract\`ere de Chern du module de Fredholm est la classe de cohomologie cyclique p\'eriodique d\'efinie par les cocyles suivants
\bbbb
\phi_{p}(a_{0},a_{1},\dots,a_{p})&=&
=(-1)^{\frac{p(p-1)}{2}}\Gamma(p/2+1)\tau_{p}(a_{0},a_{1},\dots,a_{p})
\;si\; p\;est\; pair,\n\\
\phi_{p}(a_{0},a_{1},\dots,a_{p})&=&
\sqrt{2i}(-1)^{\frac{p(p-1)}{2}}\Gamma(p/2+1)\tau_{p}(a_{0},a_{1},\dots,a_{p})
\;si\; p\;est\; impair,\n
\eeee
pour tout entier $p=n+2m$ et tous $a_{0},\dots,a_{p}\in\aa$.
\end{dfi}

On peut alors ais\'ement v\'erifier le r\'esulat suivant.

\begin{pro}
Lorsque $n$ est pair (resp. impair), l'indice de $eF^{+}e$ (resp. $PuP$) est obtenu en couplant le caract\`ere de Chern avec $K_{0}(\aa)$ (resp. $K_{1}(\aa)$).
\end{pro}

Nous allons maintenant appliquer ces r\'esultats aux triplets spectraux en suivant la proc\'edure donn\'ee dans \cite{bible} et \cite{ihes}. La premi\`ere \'etape consiste \`a associer \`a tout triplet spectral $(\aa,\hh,\dd)$ de dimension $n$ un module de Fredholm $(\hh',F)$ n+1-sommable et ayant la m\^eme parit\'e que $n$. Puisque la r\'esolvante de $\dd$ est compacte, le noyau de $\dd$ est de dimension finie et nous d\'esignons par $\hh_{1}$ le suppl\'ementaire orthogonal de $\ker\dd$ dans $\hh$.

\par

Dans le cas impair nous d\'efinissons $\hh'=\hh$ et sur $\hh_{1}$ nous pouvons \'ecrire $\dd=|\dd|F$, avec $F$ unitaire tel que $F^{2}=1$. Dans le cas pair nous posons $\hh'=\hh\op\ker\dd=\hh_{1}\op\ker\dd\op\ker\dd$. Nous d\'efinissons $F$ par la d\'ecomposition polaire $\dd=|\dd|F$ sur $\hh_{1}$ et $F$ \'echange simplement les deux facteurs $\ker\dd$, la chiralit\'e \'etant oppos\'ee sur le second facteur de mani\`ere \`a avoir $\gamma F+F\gamma=0$. On montre alors le r\'esultat suivant\cite{bible}.

\begin{pro}
Le module de Fredholm $(\hh',F)$ est $n+1$-sommable.
\end{pro}

Nous pouvons ainsi, au moins en principe, utiliser le caract\`ere de Chern pour relier certaines quantit\'es \`a l'indice d'un op\'erateur et montrer de la sorte leur int\'egralit\'e. Cependant, que ce soit  en dimension 3 avec l'action de Chern-Simons ou en dimension 4 avec la th\'eorie de Yang-Mills, les quantit\'es que nous avons rencontr\'ees faisaient intervenir l'op\'erateur de Dirac \`a la place de $F$ et la trace de Dixmier  \`a la place de la trace usuelle. Malheureusement, on ne peut obtenir de la sorte que la classe de cohomologie de Hochschild du caract\`ere de Chern \cite{bible}.

\begin{pro}
Soit $(\aa,\hh,\dd)$ un triplet spectral de dimension $n$. La relation suivante d\'efinit un cocycle de Hochschild
\bbb
\psi(a_{0},\dots,a_{n})=
\mu_{n}\t_{\omega}\lp\gamma\pi(a_{0})\lb\dd,\pi(a_{1})\rb\dots
\lb\dd,\pi(a_{n})\rb\rp,\n
\eee
avec 
\bbbb
\mu_{n}&=&(-1)^{\frac{n(n-1)}{2}}\Gamma(n/2+1)
\;si\; n\;est\; pair,\n\\
\mu_{n}&=&\sqrt{2i}(-1)^{\frac{n(n-1)}{2}}\Gamma(n/2+1)
\;si\; n\;est\; impair,\n.
\eeee
qui appartient \`a la classe de cohomologie de Hochschild du caract\`ere de Chern. 
\end{pro}
  
\edemo

Lorsque $\psi$ est un cocycle cyclique, son \'evaluation sur des projecteurs ou des unitaires ne d\'epend que de leur classe dans les groupes de K-th\'eorie, les nombres complexes ainsi obtenus sont donc relativement stables par d\'eformation. 

\par

Malheureusement, ce cocyle n'est pas suffisant pour d\'eterminer le caract\`ere de Chern puisqu'il est l'image de ce dernier par l'application $I$ qui est l'inclusion du complexe de cohomologie cyclique dans le complexe de Hochschild. Il nous manque donc d'\'eventuelles composantes du caract\`ere de Chern qui sont situ\'ees dans le noyau de $I$ (Appendice C). 

\par

Si $c$ est un cycle de Hochschild, alors le couplage $\Psi(c)$ est \'egal au caract\`ere de Chern \'evalu\'e en $c$. En dimension 1, il est clair que
\bbb
c=(u-1)\ot(u^{-1}-1)\n 
\eee 
est un cycle de Hoschild pour tout unitaire $u\in M_{N}(\aa)$. Puisque $\Psi$ est dans la classe de cohomologie de Hochschild du caract\`ere de Chern, il ne diff\`erent que par un bord et leur couplage avec $c$ est identique. En tenant compte de toutes les normalisations, on a 
\bbb
\mathrm{Ind}(PuP)=\frac{1}{2}\t_{\omega}\lp (u-1)[\dd,u^{-1}]|\dd|^{-1}\rp,\n
\eee 
ce qui nous donne une forme particuli\`erement simple du th\'eor\`eme de l'indice en dimension 1.  

\exe
Consid\'erons le triplet spectral de dimension 1 correspondant aux fonctions sur le cercle $S^{1}$ de longueur 1. Autrement dit, nous prenons $\aa=C^{\infty}(S^{1})$ qui agit par multiplication sur $\hh=L^{2}(S^{1})$ et $\dd=2\pi i\frac{d}{dt}$, o\`u $t$ est la coordonn\'ee usuelle sur $S^{1}$ obtenue en identifiant les fonctions de $C^{\infty}(S^{1})$ avec les fonctions ind\'efiniment d\'erivables de p\'eriode 1.   
\par

La trace de Dixmier est proportionnelle \`a l'int\'egrale (le coefficient de proportionalit\'e \'etant $1/\pi$ en dimension 1) et on a
\bbb
\t_{\omega}\lp f_{0}[\dd,f_{1}]|\dd|^{-1}\rp=2i\int_{0}^{1}f_{0}\frac{d}{dt}f_{1}.\n
\eee
Les modes de Fourier $e_{n}=e^{inx}$ forme une base Hilbertienne de $\hh$ dans laquelle l'op\'erateur $F$ est simplement donn\'e par $Fe_{n}=\mathrm{sign}(n)e_{n}$, ce qui implique que $Pe_{n}=e_{n}$ si $n\geq 0$ et $Pe_{n}=0$ si $n<0$. 

\par

V\'erifions le th\'eor\`eme de l'indice pour l'unitaire $u=e^{inx}$. L'op\'erateur $PuP$ transforme le vecteur $e_{k}$ avec $k\geq 0$ en $e_{n+k}$ si $n+k\geq 0$ et en 0 si $n+k<0$. Par cons\'equent, il est injectif si $n\leq 0$, et si $n>0$, son noyau est de dimension $n$ et engendr\'e par $(e_{0},e_{1},\dots,e_{n-1})$. De m\^eme, on etudie $Pu^{*}P$ en changeant $n$ en $-n$. On obtient
\bbb
\mathrm{Ind}(PuP)=\dim\ker PuP - \dim\ker Pu^{*}P=n.\n
\eee 
D'autre part, on a $\frac{d}{dt}u^{*}=-inu^{*}$, ce qui donne
\bbb
2i\int_{0}^{1}u\frac{d}{dt}u^{*}=2n,\n
\eee
d'o\`u, puisque la trace de Dixmier n'est autre que l'int\'egration,
\bbb
\mathrm{Ind}(PuP)=n=\frac{1}{2}\t_{\omega}\lp (u[\dd,u^{-1}]|\dd|^{-1}\rp,\n
\eee
ce qui est bien le th\'eor\`eme de l'indice en dimension 1.
\eexe

\subsection{Spectre de dimension}

Pour pouvoir \^etre utilis\'e en pratique, le th\'eor\`eme de l'indice doit exprimer l'indice d'un certain op\'erateur non pas \`a l'aide de la trace et de l'op\'erateur $F$, mais en fonction de la trace de Dixmier et de l'op\'erateur de Dirac. En r\`egle g\'en\'erale, il ne sera pas possible de n'utiliser que la trace de Dixmier. Nous aurons besoin de g\'en\'eraliser cette derni\`ere par l'
interm\'ediaire de l'analogue du r\'esidu de Wodzicki (cf Appendice A) pour un triplet spectral g\'en\'eral \cite{local}.  

\par

En vue de g\'en\'eraliser  le r\'esidu de Wodzicki, nous devons supposer que les fonctions $\zeta$ construites \`a partir de l'op\'erateur de Dirac peuvent \^etre continu\'ees analytiquement sauf peut-\^etre sur un sous-ensemble discret de $\ccc$. 

\begin{dfi}
Soit $(\aa,\hh,\dd)$ un triplet spectral de dimension $n$. On dit que $(\aa,\hh,\dd)$ a un spectre de dimension discret $\Sigma\in\ccc$ si les fonctions d\'efinies par
\bbb
\zeta_{b}(z)=\t(b|\dd|^{-z})\n
\eee
lorsque la partie r\'eelle de $z$ est assez grande et pour tout $b$ appartenant \`a l'alg\`ebre $\bb$ engendr\'ee par $\aa$ et $\lb\dd,\aa\rb$ et les commutateurs it\'er\'es $\delta^{n}(\pi(\aa))$ et $\delta^{n}(\lb\dd,\aa\rb)$ avec $|\dd|$, se prolongent analytiquement \`a $\ccc$ priv\'e de $\Sigma$. Nous supposons aussi que pour tout $z=s+it$ avec $s>0$ la fonction $t\mapsto\Gamma(z)\zeta_{b}(z)$ est \`a d\'ecroissance rapide.  
\end{dfi}

Par d\'efinition, le spectre de dimension est un sous-ensemble discret de $\ccc$ qui correspond aux p\^oles simples des fonctions $\zeta$ g\'en\'eralis\'ees dans le cas classique \cite{gilkey}. 

\par

Il existe de nombreux triplets spectraux ayant un spectre de dimension simple comme par exemple le tore non commutatif, qui correspond \`a un cas particulier de produit crois\'e d'une vari\'et\'e riemannienne par le groupe engendr\'e par une de ses isom\'etries. Lorsqu'on fait le produit crois\'e par le groupe form\'e de tous les diff\'eomorphismes, le triplet spectral ainsi obtenu a encore un spectre de dimension discret et correspond \`a "la g\'eom\'etrie invariante sous les diff\'eomorphismes" \cite{mosco}.

\par

On g\'en\'eralise alors le r\'esidu de Wodzicki de la mani\`ere suivante.

\begin{dfi}
Pour tout entier k et tout $b\in\bb$, nous d\'efinissons
\bbb
\tau_{k}(b)= \mathop{\mathrm{Res}}\limits_{z=0}z^{k}\zeta_{b}(2z).
\eee
\end{dfi}

Dans le cas commutatif, les p\^oles des fonctions $\zeta$ sont simples et nous avons un seul r\'esidu $\tau_{0}$ qui correspond au r\'esidu de Wodzicki. Puisque ce dernier est une trace sur l'alg\`ebre des op\'erateurs pseudodiff\'erentiels qui \'etend la trace de Dixmier, $\tau_{0}$ est \'egalement une trace.  

\par

Ce r\'esultat se g\'en\'eralise dans le cas non commutatif \cite{local}

\begin{pro}
Lorsque le spectre de dimension est simple, $\tau_{0}$ est une trace sur $\bb$. 
\end{pro}

Lorsque certains de ces p\^oles sont multiples, seul le r\'esidu d'ordre le plus \'elev\'e est une trace. 

\subsection{Th\'eor\`eme local de l'indice}

Nous allons exprimer le caract\`ere de Chern associ\'e au triplet spectral $(\aa,\hh,\dd)$ \`a l'aide des r\'esidus $\tau_{k}$ introduits au cours de la section pr\'ec\'edente. Plus pr\'ecis\'ement, nous allons donner un cocycle dans le bicomplexe $(b,B)$ (cf Appendice C) dont la classe de cohomologie cyclique p\'eriodique nous donne le caract\`ere de Chern.  

\par

Si $(\aa,\hh,\dd)$ est un triplet spectral pair de dimension $n$ ayant un spectre de dimension discret,  nous d\'efinissons pour tout entier pair $p$ compris entre 0 et $n$ des cochaines $\phi_{p}$ sur $\aa^{\ot(p+1)}$ par
\bbb
\phi_{0}(a_{0})=\tau_{-1}(\gamma a_{0})\n
\eee
avec $\tau_{-1}(b)=\mathop{\mathrm{Res}}\limits_{z=0}z^{-1}\t(b|\dd|^{-2z})$ pour tout $b\in\bb$ et 
\bbbb
&\phi_{p}(a_{0},a_{1},\dots,a_{p})=&\n\\
&\mathop{\sum}\limits_{|k|\leq n-p,\:0\leq q\leq |k|+n/2}
\frac{(-1)^{k}}{k_{1}!\dots k_{p}!}\alpha_{k}\sigma_{q}(|k|+p/2)
\tau_{q}(\gamma a_{0}(da_{1})^{k_{1}}\dots(da_{p})^{k_{p}}|\dd|^{-(2|k|+p)}),&\n
\eeee
o\`u $\sigma_{j}(m)$ est d\'efini par
\bbb
\frac{\Gamma(m+z)}{\Gamma(1+z)}=
\mathop{\sum}\limits_{j=0}^{m-1}\sigma_{j}(m)z^{j}\n
\eee
et
\bbb
\alpha_{k}=\frac{1}{(k_{1}+1)(k_{1}+k_{2}+1)\dots(k_{1}+k_{2}+\dots+k_{p}+p)}\n
\eee
et o\`u $k=(k_{1},\dots,k_{n})$ est un multi-indice $|k|=k_{1}+\dots+k_{n}$. De plus, nous d\'esignons par $\nabla$ la d\'erivation d\'efinie par $\nabla(T)=[\dd^{2},T]$ pour tout op\'erateur born\'e $T$ et $T^{k}=\nabla^{k}(T)$. On a alors le r\'esultat suivant \cite{local}.

\begin{thm}
Soit $(\aa,\hh,\dd)$ un triplet spectral de dimension $n$ paire admettant un spectre de dimension discret. Alors les formules pr\'ec\'edentes d\'efinissent, pour tout entier pair $p$ inf\'erieur ou \'egal \`a $n$, des cocycles cycliques $\phi_{p}$ dans le bicomplexe $(b,B)$ dont la classe de cohomologie est le caract\`ere de Chern du triplet spectral $(\aa,\hh,\dd)$. 
\end{thm}

Il est remarquable que ces cocycles ne font intervenir qu'un nombre fini de r\'esidus $\tau_{k}$ avec $k\leq n$, m\^eme si le spectre de dimension a des p\^oles d'ordre plus \'elev\'e. La d\'emonstration de ce fait utilise des m\'ethodes tr\`es similaires \`a celles du groupe de renormalisation (on utilise de fa\c con essentielle l'invariance sous la transformation $\dd\rightarrow\lambda\dd$, avec $\lambda>0$) \cite{local}. 

\begin{thm}
Soit $e\in M_{N}(\aa)$. Alors pour $F=\frac{\dd}{|\dd|}$ on a
\bbb
\mathrm{Ind}(eF^{+}e)=\mathop{\sum}\limits_{0\leq 2m\leq n}\frac{(-1)^{m}(2m)!}{m!}\phi_{m}(e,\dots,e).
\eee
\end{thm}

\demo
Ce r\'esultat n'a pas \'et\'e \'enonc\'e dans \cite{local} car il est une cons\'equence directe du r\'esultat pr\'ec\'edent. Nous allons cependant en donner la preuve car nous l'utiliserons ult\'erieurement.

\par

Les cocycles $\phi_{p}$ sont des cocycles dans le bicompelxe $(b,B)$, c'est-\`a-dire qu'il v\'erifient $b\phi_{p}+B\phi_{p}=0$ pour $p>0$ et $B\phi_{0}=0$. Pour passer du bicomplexe $(b,B)$ au complexe $(C_{\lambda}(\aa),b)$ de la cohomologie cyclique p\'eriodique (cf Appendice C), nous devons changer la normalisation des cochaines $\phi_{p}\rightarrow (-1)^{[p/2]}p!\phi_{p}$. 

Nous pouvons alors utiliser les formules habituelles pour coupler les cochaines $(-1)^{[p/2]}p!\phi_{p}$, avec $p=2m$, au groupe $K_{0}(\aa)$ pour obtenir (cf Appendice C) 
\bbb
\mathrm{Ind}(eF^{+}e)=\mathop{\sum}\limits_{0\leq 2m\leq n}\frac{(-1)^{m}(2m)!}{m!}\phi_{m}(e,\dots,e).\n
\eee
\edemo

Il faut remarquer que toute la construction pr\'ec\'edente suppose implicitement que le noyau de l'op\'erateur de Dirac  $\dd$ est nul, car sinon la notation $|\dd|^{z}$ n'a pas de sens pour $z$ complexe. Dans le cas g\'en\'eral, le noyau est un sous-espace de dimension finie et toutes les traces sont \`a prendre sur le suppl\'ementaire orthogonal de $\ker\dd$.  

\par

En dimension 4, ce th\'eor\`eme fait intervenir trois cochaines, $\phi_{0}$, $\phi_{2}$ et $\phi_{4}$ qui v\'erifient
\bbb
\phi_{0}(e)-2\phi_{2}(e,e,e)+12\phi_{4}(e,e,e,e,e)\in\zzz.
\eee
Dans le cas g\'en\'eral, seul $\phi_{4}$ peut \^etre reli\'e directement au minorant de l'action de Yang-Mills que nous avons \'etudi\'e (cf \S 1.3.4). Aussi nous ne pouvons pas prouver de la sorte la quantification de ce minorant en toute g\'en\'eralit\'e.

\par

Toutefois, nous montrerons qu'un tel r\'esultat peut ais\'ement \^etre obtenu dans le cas du tore non commutatif, en v\'erifiant que dans ce cas les cocycles $\phi_{0}$ et $\phi_{2}$ sont identiquement nuls.


\subsection{Th\'eor\`eme de l'indice en dimension impaire}

De la m\^eme mani\`ere, on peut obtenir une formule locale du th\'eor\`eme de l'indice pour un triplet spectral $(\aa,\hh,\dd)$ impair \cite{local}.

\par

Comme dans le cas pair, nous d\'efinissons, pour tout entier $p$ impair compris entre 1 et $n$, des cochaines $\phi_{p}$ sur $\aa^{\ot (p+1)}$ par
\bbbb
&\phi_{p}(a_{0},a_{1},\dots,a_{p})=&\n\\
&\mathop{\sum}\limits_{|k|\leq n-p,\:0\leq q\leq |k|+n/2}
\frac{(-1)^{k}}{k_{1}!\dots k_{p}!}\alpha_{k}\sigma_{q}(|k|+p/2)
\tau_{q}(a_{0}(da_{1})^{k_{1}}\dots(da_{p})^{k_{p}}|\dd|^{-(2|k|+p)}).&
\eeee
A part $\sigma_{j}(m)$ qui est d\'efini par
\bbb
\mathop{\Pi}\limits_{l=0}^{m-1}\lp z+\frac{2l+1}{2}\rp=\mathop{\sum}\limits_{j=0}^{m-1}\sigma_{m-j}(m)z^{j},\n
\eee
toutes les autres notations sont identiques \`a celles introduites au cours de la section pr\'ec\'edente.

\begin{thm}
Soit $(\aa,\hh,\dd)$ un triplet spectral de dimension $n$ impaire admettant un spectre de dimension discret. Alors les formules pr\'ec\'edentes d\'efinissent des cocycles cycliques dans le bicomplexe $(b,B)$ ayant la m\^eme classe que le caract\`ere de Chern. 
\end{thm}

En couplant ces cocycles au groupe $K_{1}(\aa)$, on obtient la version impaire du th\'eor\`eme de l'indice:

\begin{thm}
Soit $u\in M_{N}(\aa)$ un unitaire. Alors on a
\bbb
\mathrm{Ind}(PuP)=\frac{1}{\sqrt{2i\pi}}
\mathop{\sum}\limits_{0\leq 2m+1\leq n}(-1)^{m}m!
\phi_{m}(\underbrace{u,u^{-1},\dots,u,u^{-1}}_{2m\quad fois}),
\eee
avec $P=\frac{1}{2}\lp 1+F\rp$ et $F=\dd/|\dd|$.
\end{thm}

\exe
L'exemple le plus simple est celui d'un triplet spectral de dimension $1$. Dans ce cas, le seul cocycle non nul est
\bbb
\phi_{1}(a_{0},a_{1})=\sqrt{2i\pi}\mathop{\mathrm{Res}}\limits_{z=0}
\t\lp a_{0}[\dd,a_{1}]|\dd|^{-1-2z}\rp.\n
\eee
Ce cocycle correspond directement \`a la classe de Hochschild du caract\`ere de Chern. M\^eme si le spectre n'est pas simple, aucun des r\'esidus d'ordre plus \'elev\'e n'apparait dans la formule de l'indice.
\eexe

\exe
En dimension 3, il y a deux cocycles qui sont non nuls. Le premier est donn\'e par
\bbbb
\frac{1}{\sqrt{2i\pi}}\phi_{3}(a_{0},a_{1},a_{2},a_{3})&=&
\frac{1}{12}\tau_{0}\lp a_{0}[\dd,a_{1}][\dd,a_{2}][\dd,a_{3}]|\dd|^{-3}\rp\n\\
&-&\frac{1}{6}\tau_{1}\lp a_{0}[\dd,a_{1}][\dd,a_{2}][\dd,a_{3}]|\dd|^{-3}\rp\n
\eeee
et fait intervenir le r\'esidu d'ordre sup\'erieur. Le second cocycle est $\phi_{1}$, donn\'e par
\bbbb
\frac{1}{\sqrt{2i\pi}}\phi_{1}(a_{1},a_{2})&=&
\tau_{0}(a_{0}[ \dd,a_{1}]|\dd|^{-1})-\frac{1}{4}\tau_{0}\lp a_{1}\lb\dd^{2},[ \dd,a_{1}]\rb|\dd|^{-3}\rp\n\\
&-&
\frac{1}{2}\tau_{1}\lp a_{1}\lb\dd^{2},[\dd,a_{1}]\rb|\dd|^{-3}\rp
+\frac{1}{8}\tau_{0}\lp a_{1}\lb\dd^{2},\lb\dd^{2},[ \dd,a_{1}]\rb\rb|\dd|^{-5}\rp\n\\
&+& \frac{1}{3}\tau_{1}\lp a_{1}\lb\dd^{2},\lb\dd^{2},[ \dd,a_{1}]\rb\rb|\dd|^{-5}\rp
+\frac{1}{12}\tau_{2}\lp a_{1}\lb\dd^{2},\lb\dd^{2},[
\dd,a_{1}] \rb\rb|\dd|^{-5}\rp  .\n
\eeee
Lorsque le spectre de dimension est simple, les r\'esidus $\tau_{1}$ et $\tau_{2}$ sont nuls et le th\'eor\`eme de l'indice nous donne
\bbbb
\mathrm{Ind}(PuP)&=&\tau_{0}(u[\dd,u^{-1}]|\dd|^{-1})-\frac{1}{4}\tau_{0}\lp u\lb\dd^{2},[\dd,u^{-1}]\rb|\dd|^{-3}\rp\n\\
&+&\frac{1}{8}\tau_{0}\lp u\lb\dd^{2},\lb\dd^{2},[\dd,u^{-1}]\rb\rb|\dd|^{-5}\rp \n\\
&-&\frac{1}{12}\tau_{0}\lp u[ \dd,u^{-1}][ \dd,u][ \dd,u^{-1}]|\dd|^{-3}\rp .
\n
\eeee
Seul le dernier terme apparait dans l'\'etude de l'invariance de jauge de l'action de Chern-Simons, aussi le th\'eor\`eme de l'indice ne nous permet-il pas de prouver en toute g\'en\'eralit\'e son int\'egralit\'e. 
\eexe

Avant de clore cette partie conscacr\'ee au th\'eor\`eme de l'indice, signalons que le calcul de tous ces cocycles devient rapidement tr\`es p\'enible car il y a de plus en plus de termes. En particulier, lorsque le triplet spectral correspond \`a "la g\'eom\'etrie invariante sous les diff\'eomorphismes" \cite{mosco}, le classement de tous les termes obtenus fait appara\^\i tre une alg\`ebre de Hopf $\hh_{n}$.

\par

Cette alg\`ebre de Hopf, qui peut \^etre d\'ecrite de mani\`ere diagrammatique \`a l'aide d'arbres dont on a isol\'e un des sommets ("rooted trees"), semble devoir \^etre amen\'ee \`a jouer un r\^ole central en g\'eom\'etrie non commutative et en th\'eorie quantique des champs. En effet, cette alg\`ebre de Hopf apparait aussi dans la formule des for\^ets, qui permet de r\'esoudre le difficile probl\`eme des divergences enchevetr\'ees \cite{kreimer}.   
 
\par

Faute de temps, nous ne d\'ecrivons pas ces derniers d\'eveloppements de la g\'eom\'etrie non commutative et de ses relations avec la th\'eorie des champs. Nous renvoyons \`a l'article de Connes et Kreimer \cite{kreimer} pour un expos\'e complet ainsi qu'\`a \cite{raimart}, dans lequel nous proposons une modification du travail de Kreimer visant \`a tenir compte des divergences enchev\'etr\'ees de mani\`ere purement alg\'ebrique. 


\chapter{Classification des triplets spectraux finis}

\section{Les axiomes de la g\'eom\'etrie non commutative finie}

\subsection{D\'efinition et premi\`eres cons\'equences}

Parmi tous les exemples d'espaces topologiques, les plus simples sont les espaces finis, munis de la topologie discr\`ete. Du point de vue de la g\'eom\'etrie non commutative, nous devons remplacer un espace par l'alg\`ebre des fonctions \`a valeurs complexes d\'efinies sur cet espace. Dans le cas d'un ensemble fini \`a $N$ points, cette alg\`ebre n'est autre que l'alg\`ebre $\ccc^{N}$, ce qui nous am\`ene \`a consid\'erer des triplets spectraux construits avec une alg\`ebre $\aa$ de dimension finie. Plus g\'en\'eralement, nous supposerons que l'espace de Hilbert $\hh$ est de dimension finie, ce qui entraine que $\pi(\aa)$ est de dimension finie m\^eme si $\aa$ est de dimension infinie.

\par

D'autre part, il est clair que dans le cas classique un espace fini correspond \`a une "vari\'et\'e de dimension 0". Il semble donc naturel de supposer que le triplet spectral correspondant soit de dimension 0 au sens des axiomes pr\'ec\'edents.

\par

Cependant, il y a eu, au cours de l'histoire de la physique, de nombreuses tentatives d'approximation de l'espace infini et continu par un espace fini et discret. Pour d\'evelopper cette approche dans le cadre de la g\'eom\'etrie non commutative, il nous faut donc approximer une vari\'et\'e de dimension n par un triplet spectral dont l'alg\`ebre est de dimension finie. Au niveau alg\'ebrique cette d\'emarche fonctionne admirablement bien dans le cas de la sph\`ere, dont l'alg\`ebre des coordonn\'ees peut \^etre approxim\'ee par une suite croissante d'alg\`ebres de matrices qui forment la "sph\`ere non commutative".

\par

Bien entendu, il ne peut y avoir de "saut" de la dimension: les triplet spectraux qui r\'ealisent l'approximation doivent \^etre des triplets spectraux de dimension n. Nous allons voir que cette condition cr\'ee, dans le cadre des triplets spectraux finis, une obstruction fondamentale \`a une telle approximation. Enon\c cons d'abord le r\'esultat suivant.

\begin{pro}
Il n'existe pas de triplet spectral de dimension $d>0$ construit \`a l'aide d'un espace de Hilbert $\hh$ de dimension finie.
\end{pro}

\demo
La d\'emonstration de ce r\'esultat passe par deux \'etapes. Nous allons d'abord montrer que si un tel triplet spectral existait, son alg\`ebre serait une alg\`ebre de matrices. En effet, puisque l'alg\`ebre $\aa$ n'intervient que par l'interm\'ediaire de sa repr\'esentation $\pi$, nous pouvons, sans perte de g\'en\'eralit\'e, remplacer $\aa$ par $\aa/\jj$, o\`u $\jj$ est le noyau de $\pi$. L'alg\`ebre involutive $\aa/\jj$ admet une repr\'esentation fid\`ele sur un espace de Hilbert de dimension finie, elle est donc une somme directe d'alg\`ebres de matrices, comme cela d\'ecoule du r\'esultat suivant.  

\par

\begin{pro}
Toute alg\`ebre involutive sur le corps des nombres complexes qui admet une repr\'esentation involutive et fid\`ele sur un espace de Hilbert de dimension finie est somme directe d'alg\`ebres de matrices.
\end{pro}

\par

Nous supposerons donc que $\aa$ est une somme directe d'alg\`ebres de matrices \`a coefficients complexes,
\bbb
\aa=\mathop{\op}\limits_{i=1}^{N}M_{n_{i}}(\ccc).
\eee
D'apr\`es l'axiome d'{\it orientabilit\'e}, la chiralit\'e, que nous notons $\chi$ au lieu de $\gamma$ dans le cas fini, doit \^etre construite \`a l'aide d'un cycle de Hochschild de dimension $d$ \`a valeurs dans l'alg\`ebre $\aa\ot\aa^{op}$. Plus pr\'ecis\'ement, $\chi$ doit \^etre de la forme
\bbbb
\chi=\mathop{\sum}\limits_{i}\pi(x_{i})\jj\pi(x^{'}_{i})\jj^{-1}
\lb\dd,\pi(y^{1}_{i})\rb...\lb\dd,\pi(y^{d}_{i})\rb
\eeee
o\`u $x_{i}$, $x_{i}^{'}$ et $y_{i}^{j}$ sont des \'el\'ements de $\aa$ tels que
\bbbb
\mathop{\sum}\limits_{i}\pi(x_{i})\jj\pi(x^{'}_{i})\jj^{-1}
\ot\pi(y^{1}_{i})\ot...\ot\pi(y^{d}_{i})
\eeee
soit un cycle de Hochschild \`a valeurs dans $\aa\ot\aa^{op}$. Pour simplifier nos notations, nous r\'esumons la relation pr\'ec\'edente par $\chi=\pi(c)$, o\`u $c$ d\'esigne un cycle de Hochschild. Or il est connu que l'homologie de Hochschild des alg\`ebres de matrices est triviale pour des cycles de dimension $d>0$. Ce r\'esultat s'\'etend sans difficult\'e aux sommes directes d'alg\`ebres de matrices ainsi qu'\`a l'homologie de Hochschild \`a valeurs dans $\aa\ot\aa^{op}$, puisque la structure de bimodule n'agit que sur le facteur $\aa$. Prendre le produit tensoriel avec $\aa^{op}$ \'equivaut donc simplement \`a prendre un certain nombre de copies de l'homologie de Hochschild des alg\`ebres de matrices, qui est triviale en dimension $d>0$. Par cons\'equent, $c$ est en r\'ealit\'e un bord, ce qui m\`ene \`a une contradiction. En effet, l'application
\bbbb
(x,y_{0},....,y_{d})\in\aa^{d+2}\mapsto \t\lp\chi\jj\pi(x^{'})\jj^{-1}\pi(y_{0})
\lb\dd,\pi(y_{1})\rb...\lb\dd,\pi(y_{d})\rb \rp,
\eeee
o\`u $\t$ d\'esigne la trace ordinaire est un cocyle de Hochschild. La v\'erification de ce r\'esultat est en tout point analogue \`a la d\'emonstration que nous avons donn\'ee dans le chap\^\i tre pr\'ec\'edent, car le terme $\jj\pi(x^{'})\jj^{-1}$ est central et ne joue aucun r\^ole. Lorsque nous \'evaluons un cocycle de Hochschild sur un cobord, nous trouvons 0, et donc
\bbb
\mathrm{dim}\,\hh=\t(1)=\t(\chi^{2})=\t(\chi\pi(c))=0,
\eee
ce qui est \'evidemment inacceptable. Ainsi, les triplets spectraux finis d\'ecrivent des espaces non commutatifs finis,  et les axiomes correspondants sont obtenus en prenant la dimension \'egale \`a 0. 
\edemo

\par

Parall\`element, peut-on concevoir des triplets spectraux de dimension 0 avec une alg\`ebre et/ou un espace de Hilbert de dimension infinie? A priori, cela est possible et un tel triplet spectral pourrait d\'ecrire un espace non commutatif discret form\'e d'un nombre infini de "points". Cependant, cette notion est relativement peu naturelle car la plupart des axiomes ont une forme d\'eg\'en\'er\'ee lorsque $d=0$. Par exemple, si l'espace de Hilbert est de dimension infinie, l'op\'erateur de Dirac admet une infinit\'e de valeurs propres non nulles $\lambda_{n}$, que l'on range dans le sens des modules croissants. Dans ce cas, l'axiome de {\it dimension} nous impose, en dimension $d$, d'avoir $|\lambda_{n}|^{-1}=O(n^{-1/d})$. Lorsque $d\rightarrow 0$, cela signifie que la suite $|\lambda_{n}|$ doit cro\^\i tre plus vite que n'importe quelle suite $n^{\alpha}$ pour $n\rightarrow +\infty$. Cela nous am\`ene \`a choisir un op\'erateur de Dirac dont les valeurs propres ont une croissance de type exponentielle, ce qui n'est pas tr\`es naturel pour un op\'erateur devant jouer le r\^ole de la d\'erivation pour un espace discret, car cela exclut, par exemple, de construire un espace discret comme une somme infinie d'espace finis, avec un op\'erateur de Dirac qui est lui-m\^eme une somme directe de matrices, et donc un op\'erateur born\'e. Quoi qu'il en soit, nous ne consid\'ererons que des espace finis, et nous supposerons donc que l'alg\`ebre $\aa$ et l'espace de Hilbert $\hh$ sont de dimension finie. Tout ceci motive la d\'efinition suivante.

\begin{dfi}
Un triplet spectral fini est un triplet spectral $(\aa,\hh,\dd)$ de dimension 0 tel que $\aa$ et $\hh$ soient de dimension finie.
\end{dfi}

Dans ce cas, puisque $\hh$ est de dimension finie et que $\aa$ est une somme directe d'alg\`ebres de matrices, il est clair que la plupart des axiomes de la g\'eom\'etrie non commutative se simplifie grandement. En particulier, tous les axiomes relatifs \`a l'analyse fonctionnelle (dimension, r\'egularit\'e et finitude) sont v\'erifi\'es. De plus, les trois autres axiomes (premier ordre, orientabilit\'e, dualit\'e de Poincar\'e et r\'ealit\'e) prennent une forme si simple qu'il est possible d'en donner la solution g\'en\'erale explicitement. Supposons donc donn\'e un triplet spectral fini $(\aa,\hh,\dd)$ ainsi que, puisque la dimension est paire, un op\'erateur de chiralit\'e $\chi$. Bien entendu, nous supposons que l'alg\`ebre $\aa$ et sa repr\'esentation sont lin\'eaires sur $\ccc$. L'\'etude du cas r\'eel se faisant de la m\^eme mani\`ere mais n\'ecessitant des notations plus encombrantes, nous la renvoyons \`a la derni\`ere partie de ce chapitre.

\par

La plupart de r\'esultats que nous allons exposer dans la suite se trouvent dans \cite{cla} et \cite{cla'}.
   
\subsection{R\'ealit\'e}

Pour pouvoir formuler l'analogue de la dualit\'e de Poincar\'e en g\'eom\'etrie non commutative, il nous faut remplacer la structure de module de l'espace de Hilbert $\hh$ sur l'alg\`ebre $\aa$ par une structure de bimodule. En effet, dans le cas commutatif, peut importe que nous multipliions les spineurs \`a droite ou \`a gauche par des fonctions, car nous obtenons toujours le m\^eme r\'esultat. En g\'eom\'etrie non commutative, cela a une certaine importance et nous devrons supposer que les multiplications \`a droite et \`a gauche sont des op\'erations qui commutent entre elles. En d'autres termes, l'espace de Hilbert doit \^etre un bimodule sur l'alg\`ebre, ce qui s'\'etait d\'ej\`a av\'er\'e n\'ecessaire lors de la construction du mod\`ele standard en g\'eom\'etrie non commutative \cite{conneslott} pour pouvoir introduire les interactions fortes. 

\par

Cependant cette structure de bimodule n'est pas quelconque et nous supposons qu'il existe un op\'erateur $\jj$ permettant de relier l'action \`a droite \`a l'action \`a gauche qui est simplement donn\'ee par la repr\'esentation $\pi$. 

\medskip
\noindent
{\bf Axiome de r\'ealit\'e (dimension 0)} {\it Il existe un op\'erateur $\jj$ qui est une involution antiunitaire sur $\hh $(i.e. $\jj^{*}=\jj^{-1}=\jj$) qui commute avec $\chi$ 
et telle que, pour tout $x,y\in\aa$, on ait $\lb\pi(x),\jj\pi(y)\jj^{-1}\rb=0$.}
\medskip

Nous avons obtenu cet axiome \`a partir de l'axiome plus g\'en\'eral, valable en dimension $d$, simplement en faisant $d=0$. Il n'y a donc, au niveau de la structure r\'eelle, aucune subtilit\'e particuli\`ere de la dimension 0. Dans le cas du mod\`ele standard, cet op\'erateur n'est autre que l'op\'erateur antilin\'eaire qui \'echange particules et antiparticules, c'est pourquoi $\jj$ est appel\'e, en g\'en\'eral, la conjugaison de charge.

\par

En utilisant la conjugaison de charge, nous pouvons d\'efinir sur $\hh$ une structure de bimodule. L'action \`a gauche est donn\'ee par 
\bbb
(x,\Psi)\in\aa\times\hh\mapsto x\Psi=\pi(x)\Psi,
\eee
tandis que l'action \`a droite est
\bbb
(\Psi,y)\in\hh\times\aa\mapsto\jj\pi(y^{*})\jj^{-1}\Psi.
\eee
Cela d\'efinit sur $\hh$ une structure de  $(\aa,\aa)$-bimodule ou, de fa\c con \'equivalente, une structure de $(\aa\ot\aa^{op})$-module. Ceci se v\'erifie ais\'ement car, du fait de l'introduction de l'involution $*$, nous d\'efinissons bien une action \`a droite, qui commute avec l'action \`a gauche donn\'ee par la repr\'esentation $\pi$ du fait de la relation $\lb\pi(x),\jj\pi(y)\jj^{-1}\rb=0$ pour tout $x,y\in\aa$.

\par

Sur un plan plus math\'ematique, $\jj$ est tr\`es voisin de l'involution de Tomita-Takesaki (cf Appendice A) car cet op\'erateur  permet de relier, par conjugaison, l'alg\`ebre $\pi(\aa)$ et son commutant $\pi(\aa)'$, puisque, suite \`a l'axiome de r\'ealit\'e, on a $\jj\pi(\aa)\jj^{-1}\subset\pi(\aa)'$. Cependant, nous n'avons pas en g\'eneral l'\'egalit\'e car il n'y a pas toujours de vecteur cyclique et s\'eparateur. 

\par

Malgr\'e tout, dans les cas les plus simples, la conjugaison de charge  est exactement l'op\'erateur de Tomita-Takesaki. Par exemple, si nous choisissons $\aa=M_{n}(\ccc)$, agissant sur $\hh=M_{n}(\ccc)$ par simple multiplication \`a droite, nous pouvons prendre $\jj(\Psi)=\Psi^{*}$ pour tout $\Psi\in\hh$, car $\Psi$  est une matrice $n\times n$. Il est facile de v\'erifier que l'axiome de r\'ealit\'e est satisfait. Pour d\'emontrer que $\jj$ est l'op\'erateur de Tomita-Takesaki, nous devons construire un vecteur cyclique et s\'eparateur $\Psi_{0}$. Nous prenons $\Psi_{0}=I_{n}$ (matrice identit\'e $n\times n$), ce qui nous assure que $\Psi_{0}$ est cyclique puisque $\pi(\aa)\Psi_{0}=M_{n}(\ccc)=\hh$ et clairement s\'eparateur. Dans ce cas, l'op\'erateur qui transforme, pour tout $x\in\aa$, le vecteur $\Psi=x\Psi_{0}$ en le vecteur $x^{*}\Psi_{0}=\Psi^{*}$ d\'efinit un op\'erateur de Tomita-Takesaki qui n'est autre que la conjugaison de charge $\jj$. Ceci peut se g\'en\'eraliser \`a des triplets spectraux finis plus compliqu\'es, mais si nous prenons $\aa=M_{n}(\ccc)$ agissant par multiplication \`a gauche sur $M_{n}(\ccc)\ot\ccc^{m}$, o\`u les \'el\'ements de $\ccc^{m}$ sont laiss\'es invariants, il n'existe pas de vecteur cyclique, donc pas d'op\'erateur de Tomita-Takesaki. Cependant, nous pouvons construire une conjugaison de charge $\jj$, qui est \'egale \`a l'involution de Tomita-Takesaki sur les diff\'erentes composantes.

\par

Enfin, signalons que nous avons \'enonc\'e l'axiome de r\'ealit\'e dans le cas non commutatif. La version commutative de cet axiome stipule simplement qu'il existe une involution antilin\'eaire $\jj$ telle que $\ov{\pi(x)}=\jj\pi(x)\jj^{-1}$ pour $x\in\ccc^{N}$. Un tel op\'erateur peut facilement \^etre construit en choisissant simplement pour $\jj$ l'op\'erateur qui prend le conjugu\'e complexe des composantes des vecteurs de $\hh$, dans n'importe quelle base orthonormale. On peut consid\'erer un triplet spectral fini avec l'alg\`ebre commutative $\aa=\ccc^{N}$ soit comme un triplet spectral commutatif, c'est-\`a-dire que l'on suppose que $\jj$ satisfait \`a la d\'efinition pr\'ec\'edente, soit comme un triplet spectral non commutatif en imposant simplement l'axiome de r\'ealit\'e. Il est \'evident que la premi\`ere condition est nettement plus restrictive que la seconde, et nous verrons au cours de ce chapitre que l'axiome commutatif ne permet pas, pour un triplet spectral fini, de construire des g\'eom\'etries int\'eressantes.

\subsection{Orientabilit\'e}

Dans le cas g\'en\'eral, l'axiome d'orientabilit\'e nous assure de l'existence de l'analogue non commutatif de la notion de forme volume. Etant donn\'e que, dans le cas commutatif, les formes diff\'erentielles sont en tout point analogues aux cycles de Hochschild repr\'esent\'es comme op\'erateurs sur $\hh$, il est naturel d'imposer que la forme volume soit, en dimension $d$, un cycle de Hochschild d'ordre $d$. Le r\^ole de la forme volume  est jou\'e par l'op\'erateur $\chi$ et nous supposons donc, en dimension 0, que $\chi$ est un cycle d'ordre 0, c'est-\`a-dire que l'on a $\chi=\pi(x)$ pour un certain $x\in\aa$.

\par

Cependant, cette construction est trop restrictive et n'est pas v\'erifi\'ee par le mod\`ele standard. C'est pourquoi on suppose que $\chi$ est l'image d'un cycle \`a valeurs dans le bimodule $\mm=\aa\ot\aa^{op}$, dont la structure de bimodule est donn\'ee par
\bbb
y(x\ot x')y'=yxy'\ot x',
\eee
pour tout $y,y'\in\aa$ et $x\ot x'\in\aa\ot\aa^{op}$. Un \'el\'ement $x\ot y$ de $\mm$ est repr\'esent\'e comme op\'erateur sur $\hh$ par
\bbb
\pi(x\ot y)=\pi(x)\jj\pi(y)\jj^{-1}.
\eee
L'axiome d'orientabilit\'e se formule donc, en dimension 0, comme suit.

\medskip
\noindent
{\bf Axiome d'orientabilit\'e (dimension 0)} {\it L'op\'erateur $\chi$ peut \^etre \'ecrit sous la forme $\chi=\sum_{i}\pi(x_{i})\jj\pi(y_{i})\jj^{-1}$ avec $x_{i},y_{i}\in\aa$.}
\medskip

Dans le cas  du mod\`ele standard, $\chi$ est simplement un op\'erateur qui prend la valeur 1 pour les fermions droits et -1 pour les fermions gauches, c'est pourquoi on l'appelle, en g\'en\'eral, la chiralit\'e.

\par

Cet axiome permet, comme nous allons le voir plus en d\'etail, d'\'eliminer beaucoup de possibilit\'es. En particulier, nous ne pouvons pas construire un op\'erateur de Dirac non trivial avec une alg\`ebre simple $M_{n}(\ccc)$. De m\^eme, si nous consid\'erons un triplet spectral fini avec une alg\`ebre commutative $\aa=\ccc^{N}$, satisfaisant \`a l'axiome de r\'ealit\'e commutatif, alors on a $\chi=\pi(z)$ avec $z\in\aa$. Nous verrons \'egalement que cette derni\`ere relation interdit la construction d'un op\'erateur de Dirac non nul.

\subsection{Dualit\'e de Poincar\'e}

Pour une vari\'et\'e compacte de dimension $n$, il existe un isomorphisme, non canonique si on ne choisit pas une m\'etrique, entre les groupes de cohomologie d'ordre p et d'ordre $n-p$. L'existence de cet isomorphisme \'equivaut \`a dire que l'application bilin\'eaire
\bbb
(f,g)\mapsto\int_{V} f\wedge g,
\eee
est non d\'eg\'en\'er\'ee.

\par

Pour formuler cela en g\'eom\'etrie non commutative, on introduit en g\'en\'eral le caract\`ere de Chern qui d\'efinit un isomorphisme entre les groupes de cohomologie de de Rham et les groupes de K-th\'eorie. Ainsi, la forme bilin\'eaire pr\'ec\'edente est d\'efinie sur le groupe $K_{*}(\aa)$ et est appell\'ee forme d'intersection.

\par

Avant de donner une d\'efinition plus pr\'ecise de la forme d'intersection, rappellons rapidement quelques \'el\'ements de K-th\'eorie (cf Appendice B). Tout d'abord, gr\^ ace \`a la p\'eriodicit\'e de Bott, les seuls groupes qui intervienennt dans la pratique sont les groupes $K_{0}(\aa)$ et $K_{1}(\aa)$. Rappelons aussi, que pour les alg\`ebres de matrices , on a  $K_{0}(M_{n}(\ccc))=\zzz$ et  $K_{1}(M_{n}(\ccc))=0$. On en d\'eduit que si $\aa$ est somme directe de $N$ alg\`ebres de matrices, on a $K_{0}(\aa)=\zzz^{N}$ et $K_{1}(\aa)=0$. 
 
\par

Les g\'en\'erateurs du groupe ab\'elien $K_{0}(\aa)$ sont donn\'es par les projections hermitiennes minimales non nulles de $\aa$. Un projection est minimale si elle ne peut \^etre \'ecrite comme une somme directe de plusieures autres projections. Par cons\'equent, pour chaque facteur simple, nous d\'efinissons la projection $p_{i}$ dont toutes les composantes sont nulles \`a l'exception de la i-\`eme, qui est une projection de rang 1 dans $M_{n_{i}}(\ccc)$. Les projections $(p_{i})_{1\leq i\leq N}$ sont bien entendu lin\'eairement ind\'ependantes et minimales; elle forment donc un syst\`eme de g\'en\'erateurs de $K_{0}(\aa)$.

\par

Pour formuler la dualit\'e de Poincar\'e dans le cas des triplets spectraux finis, nous devons calculer la forme d'intersection $\cap$ dans ce cas particulier. Plus pr\'ecis\'ement, puisque $\cap$ est une forme $\zzz$-bilin\'eaire sur $K_{0}(\aa)\times K_{0}(\aa)$, nous devons d\'eterminer ses \'el\'ements de matrice $\cap_{ij}=\cap(p_{i},p_{j})$. Pour cela, partons de la d\'efinition g\'en\'erale de la forme d'intersection comme le couplage entre deux projections hermitiennes $e$ et $f$ par l'interm\'ediaire de l'indice de l'op\'erateur de Dirac,
\bbb
\cap(e,f)=\mathrm{Ind}\lb\;\pi(e)\jj\pi(f)\jj^{-1}\;\frac{1-\chi}{2}
\;\dd\;\frac{1+\chi}{2}\;\pi(e)\jj\pi(f)\jj^{-1}\rb.
\eee
En dimension finie, si $\oo$ est une application lin\'eaire d'un espace de dimension $p$ dans un espace de dimension $q$, son indice est
\bbbb
\mathrm{Ind}\;\oo&=&\mathrm{dim}\,\mathrm{ker}\,\oo
-\mathrm{dim}\,\mathrm{ker}\,\oo^{*}\\
&=&p-\mathrm{dim}\,\mathrm{Im}\,\oo
-\lp q-\mathrm{dim}\,\mathrm{Im}\,\oo^{*}\rp\\
&=&p-q.
\eeee
L'indice est donc simplement la diff\'erence de dimension entre les espaces de d\'epart et d'arriv\'ee. Par cons\'equent on a
\bbbb
&\cap(e,f)=\mathrm{dim}\:\lb\frac{1+\chi}{2}\;\pi(e)\jj\pi(f)\jj^{-1}\;\hh\rb&\n\\
&-\mathrm{dim}\:\lb\frac{1-\chi}{2}\;\pi(e)\jj\pi(f)\jj^{-1}\;\hh\rb.&
\eeee
Les op\'erateurs 
\bbb
\frac{1\pm\chi}{2}\;\pi(e)\jj\pi(f)\jj^{-1}
\eee
sont des projections et la dimension de leur image est simplement donn\'ee par leur trace. D'o\`u
\bbbb
\cap(e,f)&=&\t\lb\frac{1+\chi}{2}\;\pi(e)\jj\pi(f)\jj^{-1}\rb
-\t\lb\frac{1-\chi}{2}\;\pi(e)\jj\pi(f)\jj^{-1}\rb\\
&=&\t\lp\chi\pi(e)\jj\pi(f)\jj^{-1}\rp,
\eeee
et
\bbb
\cap_{ij}=\t\lp\chi\pi(p_{i}\jj\pi(p_{j})\jj^{-1})\rp,
\eee
ce qui nous donne la matrice de la forme d'intersection dans la base $(p_{i})_{1\leq i\leq N}$. Nous pouvons formuler la dualit\'e de Poincar\'e en dimension 0 comme suit.

\medskip
\noindent
{\bf Dualit\'e de Poincar\'e (dimension 0)}
{\it La matrice d\'efinie par $\cap_{ij}=\t\lb\chi\lp\pi(p_{i})\jj\pi(p_{j})\jj^{-1}\rp\rb$ a un d\'eterminant non nul, o\`u $p_{i}\in M_{n_{i}}(\ccc)$ sont des projections autoadjointes de rang minimal de $M_{n_{i}}(\ccc)$.}
\medskip

Cette d\'efinition ne d\'epend pas du choix des projections de base $(p_{i})_{1\leq i\leq N}$, car si $(q_{i})_{1\leq i\leq N}$ est un autre syst\`eme de g\'en\'erateurs de $K_{0}(\aa)$, alors il existe une matrice $M\in M_{N}(\zzz)$ de d\'eterminant 1, permettant de passer d'un syst\`eme \`a l'autre.

\subsection{Condition d'ordre un}

Au cours du chapitre consacr\'e aux axiomes de la g\'eom\'etrie non commutative, nous avons formul\'e la notion d'op\'erateur diff\'erentiel en termes purement alg\'ebriques. Rappelons qu'en g\'eom\'etrie diff\'erentielle ordinaire, l'op\'erateur de Dirac est un op\'erateur diff\'erentiel du premier ordre. Ceci signifie que, en coordonn\'ees locales, il ne contient que des d\'eriv\'ees du premier ordre par rapport \`a ces coordonn\'ees. Par cons\'equent, si on \'ecrit  $\dd$ sous la forme $i\gamma^{\mu}\partial_{\mu}$, alors pour toute fonction $f$, on a $\lb\dd,f\rb=i\gamma^{\mu}\partial_{\mu}f$, qui est un op\'erateur diff\'erentiel d'ordre 0 car il ne contient plus d'op\'eration de  d\'erivation.
De ce fait, $\lb\dd,f\rb$ est simplement un op\'erateur de multiplication par une certaine matrice et il satisfait \`a
\bbb
\lb\lb\dd,f\rb,g\rb=0\label{cdo1}
\eee
pour toute fonction $g$. Ceci caract\'erise de fa\c con purement alg\'ebrique les op\'erateurs diff\'erentiels d'ordre au plus 1, mais ne peut \^etre g\'en\'eralis\'e \`a la g\'eom\'etrie non commutative de mani\`ere directe. En effet, nous avons suppos\'e que l'op\'erateur $\lb\dd,f\rb$ commute avec la multiplication par la fonction $g$ ce qui n'est assur\'ement pas le cas si l'alg\`ebre des coordonn\'ees est non commutative. 

\par

Par exemple, consid\'erons l'alg\`ebre $\aa=M_{n}(\ccc)$ repr\'esent\'ee sur l'espace $\hh=\ccc^{n}$. Soit $\dd\in M_{n}(\ccc)$ une matrice hermitienne, l'op\'erateur $d$ d\'efini pour $x\in\aa$ par $d(x)=\lb\dd,x\rb$ est une d\'erivation de l'alg\`ebre $\aa$ et g\'en\'eralise de mani\`ere naturelle la notion de diff\'erenciation des fonctions. Cependant, la condition (\ref{cdo1}) n'est pas satisfaite, puisque
\bbb
\lb\lb\dd,x\rb,y\rb+\lb x,\lb\dd,y\rb\rb=\lb\dd,\lb x,y\rb\rb\neq 0,
\eee
en g\'en\'eral.

\par

La g\'en\'eralisation de (\ref{cdo1}) utilise la structure de bimodule de $\hh$. En fait, si $\mm$ et $\nn$ sont deux $(\aa,\bb)$-bimodules, o\`u $\aa$ et $\bb$ sont deux alg\`ebres, une application lin\'eaire $\dd$ de $\mm$ dans $\nn$ est un op\'erateur du premier ordre si elle v\'erifie
\bbb
\dd(x\Psi y)=x\dd(\Psi y)+\dd(x\Psi)y-x\dd(\Psi)y,
\eee
pour tout $\Psi\in\mm$, tout $x\in\aa$ et tout $y\in\bb$. Cette notion d'op\'erateur diff\'erentiel du premier ordre a \'et\'e d\'evelopp\'ee dans \cite{masson}, o\`u on pourra aussi trouver la g\'en\'eralisation de la notion de symbole d'un op\'erateur diff\'erentiel en g\'eom\'etrie non commutative.

\par

Appliquons cela \`a l'op\'erateur de Dirac. $\dd:\;\hh\rightarrow\;\hh$ est un op\'erateur du premier ordre pour la structure de bimodule d\'efinie sur $\hh$ par l'axiome de r\'ealit\'e si et seulement si, pour tout $\Psi\in\hh$ et tout $x,y\in\aa$ on a
\bbbb
&\dd(\pi(x)\Psi \jj\pi(y)\jj^{-1})=\pi(x)\;\dd\lp\jj\pi(y)\jj^{-1}\Psi \rp&\n\\
&+\jj\pi(y)\jj^{-1}\;\dd\lp\pi(x)\Psi\rp-\pi(x)\jj\pi(y)\jj^{-1}\;\dd(\Psi)&,
\eeee
Ceci peut \^etre reformul\'e de la mani\`ere suivante.

\medskip
\noindent
{\bf Condition d'ordre un}
{\it L'op\'erateur de Dirac satisfait \`a $\lb\lb\dd,\pi(x)\rb,\jj\pi(y)\jj^{-1}\rb=0$ pour tout $x,y\in\aa$}.
\medskip

Une premi\`ere cons\'equence de cet axiome est la nullit\'e de l'op\'erateur de Dirac pour un triplet spectral fini commutatif. Par commutatif, nous sous-entendons bien entendu le choix de l'alg\`ebre commutative $\aa=\ccc^{N}$, mais nous supposons aussi que l'axiome de r\'ealit\'e, dans sa version commutative, est satisfait, c'est-\`a-dire que $\jj\pi(x)\jj^{-1}=\ov{\pi(x)}$ pour tout $x\in\aa$. Dans ce cas, nous avons vu que l'axiome d'orientabilit\'e implique l'existence d'un $x\in\aa$ tel que $\chi=\pi(x)$. Etant donn\'e que $\chi\dd+\dd\chi=0$, on trouve en appliquant la condition du premier ordre avec $y=\ov{x}$,
\bbb
\lb\lb\dd,\pi(x)\rb,\pi(x)\rb=\lb\lb\dd,\chi\rb,\chi\rb=4\dd=0.
\eee
Par cons\'equent, l'op\'erateur de Dirac est identiquement nul et il n'y a pas de g\'eom\'etrie commutative int\'eressante en dimension 0, ce qui est tout-\`a-fait coh\'erent avec les principes g\'en\'eraux de la g\'eom\'etrie, car un espace de dimension 0 est un ensemble fini de points ind\'ependants, et il ne peut par cons\'equent y avoir de notions non triviales de distance ou de formes diff\'erentielles. Par contre, m\^eme avec une alg\`ebre commutative, on peut construire des exemples de g\'eom\'etries non commutatives finies assez riches si on impose l'axiome de r\'ealit\'e dans sa forme la plus g\'en\'erale.

\subsection{$S^{0}$-r\'ealit\'e}

Les quatre derniers axiomes d\'efinissent les triplets spectraux finis. Cependant, il appara\^\i t que le triplet spectral du mod\`ele standard satisfait \`a une condition suppl\'ementaire appell\'ee $S^{0}$-{\it r\'ealit\'e}. Cette condition peut \^etre consid\'er\'ee comme un cinqui\`eme axiome, qui impose une contrainte suppl\'ementaire \`a la conjugaison de charge $\jj$ \cite{reality}.

\begin{dfi}
Un triplet spectral est $S^{0}$-r\'eel s'il existe une involution hermitienne $\epsilon$ qui commute avec $\pi(\aa)$, $\chi$ et $\dd$ et anticommute avec $\jj$.
\end{dfi}

\'Etant donn\'e que les op\'erateurs $\epsilon$ et $\chi$ sont hermitiens, de carr\'e 1 et commutent entre eux, on peut les diagonaliser simultan\'ement dans une base orthogonale. Leurs valeurs propres sont $\pm1$ et l'espace de Hilbert s'\'ecrit comme une somme directe,
\bbb
\hh=\hh_{+}^{+}\op\hh_{+}^{-}\op\hh_{-}^{+}\op\hh_{-}^{-},
\eee
o\`u l'indice haut (resp. bas) correspond \`a la valeur propre de $\epsilon$ (resp. $\chi$). Dans chacun de ces sous-espaces , on peut trouver une base de telle mani\`ere que l'op\'erateur $\jj$ corresponde \`a la matrice suivante, compos\'ee par la conjugaison complexe $C$  des composantes des vecteurs,
\bbb
\jj=
\pp{0&0&I_{n_{+}}&0\cr0&0&0&I_{n_{-}}\cr I_{n_{+}}&0&0&0\cr0&I_{n_{-}}&0&0\cr}
\;C\, .
\eee
Dans la relation pr\'ec\'edente, les entiers $n_{+}$ (resp. $n_{-}$) d\'esignent les dimensions communes des sous-espaces $\hh_{+}^{+}$ et $\hh_{-}^{+}$ (resp. $\hh_{+}^{+}$ et $\hh_{-}^{+}$), l'\'egalit\'e des dimensions de ces sous-espaces provenant de la relation $\jj^{2}=1$.

\par

Dans cette m\^eme base, l'op\'erateur de Dirac s'\'ecrit
\bbb
\dd=\pp{0&M&0&0\cr M^{*}&0&0&0\cr0&0&0&\ov{M}\cr0&0&\ov{M}^{*}&0\cr},
\eee 
o\`u $M$ est une matrice complexe $n_{+}\times n_{-}$. La repr\'esentation $\pi$ est diagonale par blocs et on a
\bbb
\pi=\mathrm{diag}\lp\pi_{+}^{+},\pi_{+}^{-},\pi_{-}^{+},\pi_{-}^{-}\rp,
\eee
ce qui nous permet de r\'e\'ecrire l'axiome de r\'ealit\'e et la condition de premier ordre sous la forme
\bbbb
&\lb\pi_{+}^{+}(x),\ov{\pi}_{-}^{+}(y)\rb=0,&\\
&\lb\pi_{-}^{+}(x),\ov{\pi}_{-}^{-}(y)\rb=0,&\\
&\lp M\pi_{+}^{-}(x)-\pi_{+}^{+}(x)M\rp\ov{\pi}\, _{-}^{-}(y)=
\ov{\pi}\, _{+}^{+}(y)\lp M\pi_{+}^{-}(x)-\pi_{+}^{+}(x)M\rp.&
\eeee
En g\'en\'eral, l'axiome de $S^{0}$-r\'ealit\'e stipule simplement que la structure de bimodule est d\'ej\`a comprise dans la repr\'esentation $\pi$. En effet, gr\^ace \`a l'op\'erateur $\epsilon$, l'espace de Hilbert peut \^etre \'ecrit comme une somme directe de deux sous-espaces, $\hh_{+}^{+}\op\hh_{+}^{-}$ et $\hh_{-}^{+}\op\hh_{-}^{-}$ et l'op\'erateur antiunitaire $\jj$ ne fait qu'\'echanger ces sous-espaces.

\section{Solution g\'en\'erale}

\subsection{La matrice de multiplicit\'e}

Les axiomes que nous venons de donner sont si simples qu'il est possible de d\'eterminer explicitement tous les triplets spectraux finis. Nous allons commencer par \'etudier la structure de bimodule sous-jacente.

\par

Dans le cas g\'en\'eral, soit $\hh$ un espace de Hilbert de dimension finie muni d'une structure de $(\aa,\bb)$-bimodule sur deux alg\`ebres involutives $\aa$ et $\bb$ sur le corps $\ccc$.  Etant donn\'e que ces deux alg\`ebres involutives sont repr\'esent\'ees en tant qu'op\'erateurs sur $\hh$, elles peuvent \^etre \'ecrites comme sommes directes d'alg\`ebres de matrices \`a coefficients complexes,
\bbb
\aa=\mathop{\op}\limits_{i=1}^{P}M_{p_{i}}(\ccc),\;\;\;\;
\bb=\mathop{\op}\limits_{i=1}^{Q}M_{q_{i}}(\ccc).
\eee
A l'aide de cette structure de bimodule, nous d\'efinissons deux applications $\pi_{L}$ et $\pi_{R}$ de $\aa$ et $\bb$ dans l'alg\`ebre des op\'erateurs sur $\hh$ par
\bbb
\pi_{L}(x)\Psi=x\Psi\;\;\mathrm{et}\;\;\pi_{R}(y)\Psi=\Psi y^{*}
\eee
pour $x\in\aa$, $y\in\bb$ et $\Psi\in\hh$. Il est facile de voir que $\pi_{L}$ est une repr\'esentation de $\aa$,  que l'on d\'ecompose en somme directe de repr\'esentations irr\'eductibles. Dans une certaine base orthonormale de $\hh$, on a
\bbb
\pi(x)=\mathop{\op}\limits_{p=1}^{P}\;x_{p}\ot I_{m_{p}},
\eee
avec $x=(x_{p})_{1\leq p\leq P}$ et $x_{p}\in M_{n_{p}}(\ccc)$ et o\`u les multiplicit\'es $m_{i}$ sont des entiers positifs. Cela nous am\`ene \`a \'ecrire l'espace $\hh$ sous la forme d'une somme directe
\bbb
\hh=\mathop{\op}\limits_{p=1}^{P}\;\hh_{p},
\eee
o\`u $\hh_{p}$ est isomorphe \`a $\ccc^{n_{p}}\ot\ccc^{m_{p}}$.

\par

L'application $x\in\bb\mapsto\pi_{R}(x)$ satisfait \`a tous les axiomes relatifs aux repr\'esentations d'alg\`ebres, \`a l'exception de la lin\'earit\'e qui est remplac\'ee par $\pi_{R}(\lambda x)=\ov{\lambda}\pi_{R}(x)$ pour tout $x\in\bb$ et $\lambda\in\ccc$. Il d\'ecoule de la structure de bimodule que les applications $\pi_{L}(x)$ et $\pi_{R}(y)$ commutent pour tout $x\in\aa$ et $y\in\bb$, ce qui prouve  que $\hh_{p}$ est stable par $\pi_{R}(y)$. Nous pouvons alors d\'ecomposer la r\'estriction \`a $\hh_{p}$ de $\ov{\pi}_{R}$ en repr\'esentations irr\'eductibles. Compte tenu de la relation de commutation, on a
\bbb
\pi_{R}(y)=\mathop{\op}\limits_{q=1}^{Q}
I_{n_{p}}\ot I_{m_{pq}}\ot \ov{y}_{q},
\eee 
pour la restriction de $\pi_{L}(y)$ \`a $\hh_{p}$. Les multiplicit\'es $m_{pq}$ sont des entiers positifs ou nuls et, par convention, $I_{m}$ n'appara\^ \i t pas dans la d\'ecomposition si $m=0$. De fa\c con  g\'en\'erale, l'espace de Hilbert s'\'ecrit comme une somme directe de sous-espaces deux \`a deux orthogonaux,
\bbb
\hh=\mathop{\op}\limits_{p,q}\hh_{pq},\qquad \hh_{pq}=\ccc^{p}\ot\ccc^{m_{pq}}\ot\ccc^{q}.
\eee
De plus, dans une certaine base orthogonale de $\hh$ compatible avec cette d\'ecomposition, les applications $\pi_{L}$ et $\pi_{R}$ sont donn\'ees par
\bbbb
\pi_{L}(x)&=&\mathop{\op}\limits_{p,q}\;
x_{p}\ot I_{m_{pq}}\ot I_{n_{q}},\\
\pi_{R}(y)&=&\mathop{\op}\limits_{p,q}\;
I_{n_{p}}\ot I_{m_{pq}}\ot \ov{y}_{q}.
\eeee
La structure de bimodule de $\hh$ est donc donn\'ee, \`a une \'equivalence unitaire pr\`es, par une matrice $P\times Q$ \`a coeffcients entiers. R\'eciproquement, tout $(\aa,\bb)$-bimodule peut \^etre reconstruit \`a l'aide d'une matrice $P\times Q$ gr\^ace aux formules pr\'ec\'edentes. Bien entendu, nous supposons qu'une telle matrice ne contient pas de ligne ou de colonne identiquement nulle, car sinon un des facteurs ne serait pas repr\'esent\'e. Ceci peut \^etre r\'esum\'e de la mani\`ere suivante:

\begin{pro}
Soient $\aa$ et $\bb$ deux alg\`ebres sommes directes de respectivement p et q alg\`ebres de matrices. Alors, tout $(\aa,\bb)$-bimodule est d\'etermin\'e, \`a une \'equivalence unitaire pr\`es, par sa matrice de multiplicit\'e $m\in M_{p,q}(\nnn)$.
\end{pro}

Ayant caract\'eris\'e les bimodules \`a l'aide de matrices \`a coeffcients entiers et positifs, nous pouvons nous demander \`a quelles op\'erations sur les bimodules correspondent les op\'erations habituelles sur les matrices. Par exemple, si $\mm$ et $\nn$ sont des $(\aa,\bb)$-bimodules de matrices de multiplicit\'e $M$ et $N$, alors la somme directe $\mm\op\nn$ est un  $(\aa,\bb)$-bimodule dont la matrice de multiplicit\'e est $M+N$. De m\^eme, en \'echangeant les actions \`a droite et \`a gauche, $\mm$ est muni d'une structure de $(\bb,\aa)$-bimodule dont la matrice de multiplicit\'e est la transpos\'ee de $M$.

\par

L'analogue du produit de matrice est plus compliqu\'e et s'obtient en consid\'erant des produits tensoriels. 

\begin{pro}
Soient $\aa$, $\bb$ et $\cc$ trois alg\`ebres, $\mm$ un $(\aa,\bb)$-bimodule et $\nn$ un $(\bb,\cc)$-bimodule. Alors $\mm\ot_{\bb}\nn$ est un $(\aa,\cc)$-bimodule.
\end{pro}

Passant aux matrices de multiplicit\'e, on obtient ais\'ement le r\'esultat suivant.

\begin{pro}
Soient $\aa$, $\bb$ et $\cc$ trois alg\`ebres sommes directes d'alg\`ebres de matrices, $\mm$ un $(\aa,\bb)$-bimodule de matrice de multiplicit\'e m et $\nn$ un $(\bb,\cc)$-bimodule de matrice de multiplicit\'e n. Alors la matrice de multiplicit\'e de $\mm\ot_{\bb}\nn$ est le produit mn. 
\end{pro}

En dimension infinie, lorsque les alg\`ebres $\aa$, $\bb$ et $\cc$ sont des alg\`ebres de von Neumann, cette op\'eration joue un r\^ole important et est connue sous le nom de "fusion" ou "composition des correspondances" \cite{jones}.

\subsection{Op\'erateurs du premier ordre}

Nous avons vu que l'op\'erateur de Dirac est un op\'erateur du premier ordre pour la structure de bimodule donn\'ee par l'axiome de r\'ealit\'e. Avant d'\'etudier en d\'etail les propri\'et\'es de l'op\'erateur de Dirac, consid\'erons un $(\aa,\bb)$-bimodule $\hh$, o\`u $\aa$ et $\bb$ sont deux alg\`ebres involutives que nous supposons \^etre des sommes directes d'alg\`ebres de matrices.

\par

Un op\'erateur $\ll$ (resp. $\rr$) de $\hh$ dans lui-m\^eme est dit lin\'eaire \`a gauche (resp. lin\'eaire \`a droite) si pour tout $\Psi\in\hh$ et tout $x\in\aa$ on a $\ll(x\Psi)=x\ll(\Psi)$ (resp. $\rr(\Psi y)=\rr(\Psi)y$ pour $y\in\bb$). Dans ce cas, $\dd=\ll+\rr$ est un op\'erateur du premier ordre, car on a
\bbbb
\dd(x\Psi y)&=&\ll(x\Psi y)+\rr(x\Psi y),\n\\
&=&x\ll(\Psi y)+\rr(x\Psi)y,\n\\
&=&x\lb \ll(\Psi y)+\rr(\Psi y)\rb+\lb\rr(x\Psi)+\ll(x\Psi)\rb y\n\\
&&-\lb x\rr(\Psi y)+\ll(x\Psi )y\rb,
\eeee
ce qui se r\'e\'ecrit
\bbb
\dd(x\Psi y)=x\dd(\Psi y)+\dd(x\Psi)y-x\dd(\Psi)y, \label{op0}  
\eee
puisque
\bbb
x\rr(\Psi y)+\ll(x\Psi )y=x\rr(\Psi)y+x\ll(\Psi)y=x\dd(\Psi)y.
\eee
Cette d\'emonstration prouve que la somme d'un op\'erateur lin\'eaire \`a gauche et d'un op\'erateur lin\'eaire \`a droite, est un op\'erateur du premier ordre. De plus, nous n'avons utilis\'e que les propri\'etes alg\'ebriques des bimodules, ce qui montre que cette d\'emonstration reste valable en dimension infinie.

\par

Cependant, en dimension finie, on peut prouver la r\'eciproque.

\begin{pro}
Soient $\aa$ et $\bb$ deux alg\`ebres sommes directes d'alg\`ebres de matrices et $\hh$ un $(\aa,\bb)$-bimodule de dimension finie. Alors tout op\'erateur du premier ordre sur $\hh$ est somme directe d'un op\'erateur lin\'eaire \`a droite et d'un op\'erateur lin\'eaire \`a gauche.
\end{pro}

\demo
En dimension finie, nous avons montr\'e que la structure de bimodule est donn\'e par la matrice de multiplicit\'e. Avec les notations de la section pr\'ec\'edente, $\hh$ peut \^etre d\'ecompos\'e en somme directe,
\bbb
\hh=\mathop{\op}\limits_{p,q}\;\hh_{pq}.
\eee
Si on note $1_{p}$ (resp. $1_{q}$) l'\'el\'ement de $\aa$ (resp. de $\bb$) dont toutes les composantes sont nulles \`a l'exception de la $p$-i\`eme (resp $q$-i\`eme) qui est \'egale \`a la matrice unit\'e, alors l'op\'erateur de projection sur $\hh_{pq}$  est, avec les notations du paragraphe pr\'ec\'edent,  \'egal \`a $\pi_{L}(1_{p})\pi_{R}(1_{q})$.

\par

La condition du premier ordre (\ref{op0}) se r\'e\'ecrit
\bbb
\dd\pi_{L}(x)\pi_{R}(y)=\pi_{L}(x)\dd\pi_{R}(y)+\pi_{R}(y)\dd\pi_{L}(x)
-\pi_{L}(x)\pi_{R}(y)\dd,
\eee
ou de mani\`ere \'equivalente,
\bbb
\lb\lb\dd,\pi_{L}(x)\rb,\pi_{R}(y)\rb=0,
\eee
avec $x\in\aa$ et $y\in\bb$. Appliquons cela \`a $x=1_{r}$ et $y=1_{s}$ et multiplions \`a droite par $\pi_{L}(1_{r})\pi_{R}(1_{s})$ et \`a gauche par $\pi_{L}(1_{p})\pi_{R}(1_{q})$. Nous obtenons la relation
\bbb
\dd_{pq}^{rs}=\delta_{pr}\dd_{pq}^{rs}+\delta_{qs}\dd_{pq}^{rs} -\delta_{pr}\delta_{qs}\dd_{pq}^{rs},\label{op1}
\eee
o\`u nous avons introduit les \'el\'ements de matrice de l'op\'erateur de Dirac, d\'efinis par
\bbb
\dd_{pq}^{rs}=\pi_{L}(1_{p})\pi_{R}(1_{q})\;
\dd\;\pi_{L}(1_{r})\pi_{R}(1_{s}).
\eee

\par

La condition (\ref{op1}) nous montre que $\dd_{pq}^{rs}=0$ sauf si $p=r$ ou $q=s$.
Nous pouvons donc r\'eecrire $\dd$ sous la forme
\bbb
\dd=\mathop{\sum}\limits_{p=r,q\neq s}\,\dd_{pq}^{rs}
+\mathop{\sum}\limits_{p\neq r,q=s}\,\dd_{pq}^{rs}
+\mathop{\sum}\limits_{p=r,q=s}\,\dd_{pq}^{rs}.
\eee
En utilisant les d\'ecompostions de $\pi_{L}$ et de $\pi_{R}$ introduites au cours du paragraphe pr\'ec\'edent, la condition du premier ordre est \'equivalente aux relations
\bbb
\lp x_{p}\ot I_{pq}\ot I_{n_{q}}\rp\,\dd_{pq}^{ps}=
\dd_{pq}^{ps}\,\lp x_{p}\ot I_{ps}\ot I_{n_{s}}\rp
\eee 
si $p=r$ et $q\neq s$, ainsi que
\bbb
\lp I_{n_{p}}\ot I_{pq}\ot \ov{y}_{q}\rp\,\dd_{pq}^{qs}=
\dd_{pq}^{qs}\,\lp I_{n_{p}}\ot I_{pq}\ot \ov{y}_{q}\rp
\eee 
si $p\neq r$ et $q=s$.

\par

Cela prouve que l'op\'erateur
\bbb
\mathop{\sum}\limits_{p=r,q\neq s}\,\dd_{pq}^{rs}
\eee
commute avec $\pi_{L}$ et que
\bbb
\mathop{\sum}\limits_{p\neq r,q=s}\,\dd_{pq}^{rs}
\eee
commutent avec $\pi_{R}$.
Si $q=s$ et $p\neq r$, on obtient
\bbb
\lb\lb\dd_{pq}^{pq}, x_{p}\ot I_{pq}\ot I_{n_{q}}\rb,
I_{n_{p}}\ot I_{pq}\ot \ov{y}_{q}\rb=0.
\eee
La solution de cette \'equation est
\bbb
\dd_{pq}^{pq}=M_{pq}\ot I_{n_{q}}+I_{n_{p}}\ot N_{pq},
\eee
avec $M_{pq}\in M_{m_{pq}n_{p}}(\ccc)$ et $N\in M_{m_{pq}n_{q}}(\ccc)$ quelconques. En effet, si $(I_{n_{p}},E_{\alpha})_{1\leq\alpha\leq n_{p}^{2}-1}$ (resp. $(I_{n_{q}},F_{\beta})_{1\leq\beta\leq n_{q}^{2}-1}$) est une base de $M_{n_{p}}(\ccc)$ (resp. $M_{n_{q}}(\ccc)$), on peut \'ecrire
\bbb
\dd_{pq}^{pq}=M_{pq}\ot I_{n_{q}}+I_{n_{p}}\ot N_{pq}+
\mathop{\sum}\limits_{\alpha,\beta}E_{\alpha}\ot M_{\alpha\beta}\ot F_{\beta},
\eee
et on montre facilement, en raisonnant par l'absurde, que les matrices $M_{\alpha\beta}$ sont nulles. Cela prouve que 
\bbb
\mathop{\sum}\limits_{p=r,q=s}\,\dd_{pq}^{rs}=
\mathop{\sum}\limits_{p=r,q=s}\,M_{pq}\ot I_{n_{q}}+
\mathop{\sum}\limits_{p=r,q=s}\,I_{n_{p}}\ot N_{pq}.
\eee
est somme d'un op\'erateur qui commute avec $\pi_{L}$ et d'un op\'erateur qui commute avce $\pi_{R}$.

\edemo

Cette d\'emonstration se base uniquement sur la semi-simplicit\'e des alg\`ebres $\aa$ et $\bb$ et sur la finitude de la dimension du bimodule. Elle est donc appliquable dans un cadre tr\`es g\'en\'eral, mais il est facile de voir que si la dimension est infinie, il existe des op\'erateurs du premier ordre qui ne sont pas de cette forme. Par exemple, dans le cas d'une vari\'et\'e, un op\'erateur diff\'erentiel du premier ordre ne commute pas avec toutes les fonctions et ne peut donc \^etre \'ecrit sous cette forme.

\subsection{La conjugaison de charge}

\'Etant donn\'e un $(\aa,\aa)$-bimodule $\hh$ de matrice de multiplicit\'e $m$, il est int\'eressant de se demander dans quel cas on peut passer de l'action \`a gauche \`a l'action \`a droite par l'op\'erateur de conjugaison de charge introduit par l'axiome de r\'ealit\'e. De plus, nous pouvons nous demander, lorsqu'un tel op\'erateur existe, s'il est unique ou s'il y a plusieurs possibilit\'es. Dans ce chapitre, nous consid\'erons un $(\aa,\aa)$-bimodule $\hh$, de matrice de multiplicit\'e $m$ et nous utilisons les lettres $i$, $j$, $k$, ... pour rep\'erer les facteurs simples de $\aa$. La r\'eponse aux deux questions pr\'ec\'edentes est donn\'ee par la proposition suivante.

\begin{pro}
Un bimodule sur $\aa$ correspond \`a un triplet spectral fini si et seulement si sa matrice de multiplicit\'e est sym\'etrique. Dans ce cas, l'op\'erateur $\jj$  tel que $\pi_{R}=\jj\pi_{L}\jj^{-1}$ est unique \`a une conjugaison par un unitaire pr\`es.
\end{pro}

\demo
Supposons tout d'abord que $\jj$ existe. L'action de $\jj$ sur $\psi\in\hh$ peut \^etre \'ecrite, dans n'importe quelle base, sous la forme $\jj\psi=K\ov{\psi}$, o\`u $K$ est une matrice. $\jj$ est une involution antiunitaire si et seulement si $K\ov{K}=KK^{*}=1$. Puisque l'on doit avoir
\bbbb
&\pi_{L}(x)=\mathop{\op}\limits_{i,j}\;
x_{i}\ot I_{m_{ij}}\ot I_{n_{j}},&\\
&\pi_{R}(y)=\jj\pi_{L}(y)\jj^{-1}=\mathop{\op}\limits_{i,j}\;
I_{n_{i}}\ot I_{m_{ij}}\ot \ov{y}_{j},&
\eeee
pour tous $x,y\in\aa$, nous avons aussi
\bbb
K\lp\mathop{\op}\limits_{i,j}x_{i}\ot I_{m_{ij}} \ot I_{n_{j}}\rp=
\lp\mathop{\op}\limits_{i,j}I_{n_{i}}\ot I_{m_{ij}} \ot x_{j}\rp K,
\eee
pour tout $x,y\in\aa$. Par cons\'equent, $K\lp\op_{j}\hh_{ij}\rp\subset\op_{j}\hh_{ji}$, et puisque $K^{-1}=K^{*}$, $K(\hh_{ij})=\hh_{ji}$. C'est pourquoi les dimensions de $\hh_{ij}$ et $\hh_{ji}$ sont \'egales, ou de mani\`ere \'equivalente, la matrice de multiplicit\'e $m$ est sym\'etrique: $m_{ij}=m_{ji}$.

\par

Notons $K_{ij}\, :\,\hh_{ij}\rightarrow\hh_{ji}$ la restriction de $K$. Elle doit satisfaire aux relations
\bbbb
&K_{ij}\lp x_{i}\ot I_{m_{ij}}\ot I_{n_{j}}\rp= 
\lp I_{n_{i}}\ot I_{m_{ij}}\ot x_{j}\rp K_{ij},&\\
&K_{ij}K_{ij}^{*}=K_{ij}\ov{K_{ij}}=1.&
\eeee
Par cons\'equent, si $e_{i}^{a}\ot\psi_{ij}^{p}\ot e_{j}^{b}$ est une base de $\hh_{ij}=\ccc^{n_{i}}\ot\ccc^{m_{ij}}\ot\ccc^{n_{j}}$, on a
\bbbb
K_{ij}\lp e_{i}^{a}\ot\psi_{ij}^{p}\ot e_{j}^{b}\rp&=&
e_{j}^{b}\ot L_{ij}\psi_{ij}^{p}\ot e_{i}^{a}.\\
L_{ij}L_{ij}^{*}&=&1.
\eeee
Si $i\neq j$, les espaces $\hh_{ij}$ et $\hh_{ji}$ sont distincts et il est toujours  possible, dans une base orthonormale convenable $(\psi_{ij}^{p})_{1\leq p\leq m_{ij}}$, d'avoir $L_{ij}\psi_{ij}^{p}=\psi_{ji}^{p}$. Lorsque $i=j$, $L_{ii}$ est unitaire et nous pouvons trouver une base orthonormale telle que $L_{ii}\psi_{ii}^{p}=e^{i\phi_{i}^{p}}\psi_{ii}^{p}$, avec $\phi_{i}^{p}\in\rrr$. Cependant, puisque $\jj$ est antilin\'eaire,
\bbbb
\jj\lp e^{i\phi_{i}^{p}/2}\,e_{i}^{a}\ot\psi_{ii}^{p}\ot e_{i}^{b}\rp&=&
e^{-i\phi_{i}^{p}/2}\,
K\lp \ov{e}_{i}^{a}\ot\ov{\psi}_{ii}^{p}\ot \ov{e}_{i}^{b}\rp,\\
&=& e^{-i\phi_{i}^{p}/2}\,
\ov{e}_{i}^{a}\ot L_{ii}\ov{\psi}_{ii}^{p}\ot\ov{e}_{i}^{b},\\
&=& e^{+i\phi_{i}^{p}/2}\,
\ov{e}_{i}^{a}\ot\ov{\psi}_{ii}^{p}\ot\ov{e}_{i}^{b},
\eeee
ce qui nous permet de faire disparaitre les phases si on fait la substitution  $\psi_{ij}^{p}\rightarrow e^{i\phi_{i}^{p}/2}\psi_{ij}^{p}$. Par cons\'equent, l'action de $\jj$ est simplement
\bbb
\jj\lp e_{i}^{a}\ot\psi_{ij}^{p}\ot e_{j}^{b}\rp=
\ov{e}_{j}^{b}\ot\ov{\psi}_{ji}^{p}\ot\ov{e}_{i}^{a},
\eee
dans tous les cas.

\par

R\'eciproquement, si la matrice de multiplicit\'e $m$  est sym\'etrique, il est possible de d\'efinir $\jj$ de cette fa\c con et il est clair qu'un tel op\'erateur satisfait \`a tous les contraintes impos\'ees \`a la conjugaison de charge. 
\edemo

\subsection{La chiralit\'e}

D'apr\`es l'axiome d'orientabilit\'e, la chiralit\'e $\chi$ commute avec $\pi(\aa)$ et $\jj\pi(\aa)\jj^{-1}$, et est hermitienne, ce qui nous permet de l'\'ecrire sous la forme
\bbb
\chi=\mathop{\op}\limits_{ij}I_{n_{i}}\ot\chi_{ij}\ot I_{n_{j}}.
\eee
Puisque $\chi=\sum_{p}\pi(x^{p})\jj\pi(y^{p})\jj^{-1}$, avec $x^{p},y^{p}\in\aa$,  $\chi_{ij}$ est une matrice scalaire. De plus $\chi_{ij}=\pm1$ car $\chi^{2}=1$. $\lb\chi,\jj\rb=0$ entraine la relation de sym\'etrie $\chi_{ij}=\chi_{ji}$.

\par

Toute l'information contenue dans les matrices  $\chi_{ij}$ et $m_{ij}$ peut \^etre int\'egr\'ee dans une matrice de multiplicit\'e \`a coefficients non n\'ecessairement positifs $\mu\in M_{N}(\zzz)$ d\'efinie par 
\bbb
\mu_{ij}=\chi_{ij}m_{ij}.
\eee
Nous retrouvons $\chi_{ij}$ et $m_{ij}$ comme le signe et le module de $\mu_{ij}$. Par cons\'equent,
\bbb
\chi=\mathop{\op}\limits_{ij}\mathrm{sign}(\mu_{ij})I_{n_{i}}\ot I_{|\mu_{ij}|}\ot I_{n_{j}}.
\eee

\par

Occupons-nous maintenant de la dualit\'e de Poincar\'e. Comme nous nous sommes plac\'es dans la cas des alg\`ebres complexes, les projecteurs $p_{i}=1_{i}$ qui appara\^\i ssent dans la formule de d\'efinition de la matrice de la forme d'intersection sont de trace 1. Cette matrice est donn\'ee par
\bbbb
\cap_{ij}&=&\t\lb\chi\pi(p_{i})\jj\pi(p_{j})\jj^{-1}\rb,\\
&=&\mathrm{sign}(\mu_{ij})\t\lb p_{i}\ot I_{|\mu_{ij}|}\ot \ov{p}_{j}\rb,\\
&=&\mu_{ij}.
\eeee
Par cons\'equent, la dualit\'e de Poincar\'e se traduit simplement par la non nullit\'e du d\'eterminant de la matrice de multiplicit\'e $\mu$.

\par

R\'eciproquement, si nous partons d'une matrice sym\'etrique et non d\'eg\'en\'er\'ee $\mu\in M_{N}(\zzz)$ (que nous appellerons dor\'enavant matrice de multiplicit\'e au lieu de $m$), nous pouvons reconstruire la structure de bimodule \`a l'aide des valeurs absolues $|\mu_{ij}|$. La chiralit\'e $\chi$ est obtenue comme
\bbb
\chi=\mathop{\op}\limits_{ij}\mathrm{sign}(\mu_{ij})I_{n_{i}}\ot I_{|\mu_{ij}|}\ot I_{n_{j}}.
\eee
$\chi$ satisfait \`a toutes les conditions impos\'ees par les axiomes. Par construction, la v\'erification des relations de commutation  ainsi que de $\chi=\chi^{*}$ et $\chi^2=1$ est \'evidente. Montrons qu'il existe des \'el\'ements $x^{p}$ and $y^{p}$ de $\aa$ tels que $\chi=\sum_{p}\pi(x^{p})\jj\pi(y^{p})\jj^{-1}$. Pour cela, \'ecrivons la matrice de multiplicit\'e $\mu$ comme une somme de matrices de rang 1
\bbb
\chi_{ij}=\mathop{\sum}\limits_{p}\alpha_{i}^{p}\ov{\beta}_{j}^{p},
\eee       
avec $\alpha_{i}^{p}$ et $\beta_{j}^{p}$ des nombres complexes. On en d\'eduit que  $\chi=\sum_{p}\pi(x^{p})\jj\pi(y^{p})\jj^{-1}$, avec $x^{p}=(\alpha_{i}^{p}\,I_{n_{i}})_{1\leq i\leq N}$ et $y^{p}=(\beta_{j}^{p}\, I_{n_{j}})_{1\leq j\leq N}$.

\par

Cela nous permet d'\'enoncer le r\'esultat interm\'ediaire suivant.

\begin{pro}
Les \'el\'ements $(\aa,\jj,\chi)$ d'un triplet spectral fini sont d\'etermin\'es, \`a une transformation unitaire pr\`es, par une matrice sym\'etrique et non d\'egen\'er\'ee  $\mu\in M_{N}(\zzz)$. 
\end{pro}

\subsection{L'op\'erateur de Dirac}

Comme nous l'avons montr\'e, tout op\'erateur du premier ordre d'un $(\aa,\bb)$-bimodule de dimension finie, o\`u $\aa$ et $\bb$ d\'esignent des sommes directes d'alg\`ebres de matrices, est la somme d'un op\'erateur lin\'eaire \`a gauche et d'un op\'erateur lin\'eaire \`a droite. D'apr\`es la condition d'ordre un, l'op\'erateur de Dirac $\dd$ est un op\'erateur du premier ordre pour la structure de bimodule de $\hh$ donn\'ee par l'axiome de realit\'e. On peut donc lui appliquer le r\'esultat pr\'ec\'edent pour d\'ecomposer $\dd$ sous la forme
\bbb
\dd=\ll+\rr,
\eee
o\`u $\ll$ est lin\'eaire \`a gauche et $\rr$ lin\'eaire \`a droite. Gr\^ace \`a l'axiome d'orientabilit\'e. il s'av\`ere que cette d\'ecomposition est essentiellement unique;

\begin{pro}
Dans le cas d'un triplet spectral fini, la d\'ecomposition de l'op\'erateur de Dirac consid\'er\'e comme un op\'erateur du premier ordre est unique si on impose aux applications lin\'eaires \`a droite et \`a gauche correspondantes d'anticommuter avec $\chi$.
\end{pro}

\demo
En effet, supposons qu'il existe quatre op\'erateurs  $\ll_{1}$, $\rr_{1}$, $\ll_{2}$ et $\rr_{2}$, qui anticommutent avec $\chi$ et tels que
\bbb
\dd=\ll_{1}+\rr_{1}=\ll_{2}+\rr_{2},
\eee
v\'erifiant \'egalement
\bbb
\lb\ll_{1},\pi(x)\rb=\lb\rr_{1},\jj\pi(x)\jj^{-1}\rb=
\lb\ll_{2},\pi(x)\rb=\lb\rr_{2},\jj\pi(x)\jj^{-1}\rb=0
\eee
pour tout $x\in\aa$. On en d\'eduit que 
\bbb
\ll_{1}-\ll_{2}=\rr_{2}-\rr_{1}
\eee
commute avec $\pi(x)$ et $\jj\pi(x)\jj^{-1}$ pour tout $x\in\aa$. D'apr\`es l'axiome d'orientabilit\'e, la chiralit\'e peut \^etre \'ecrite sous la forme
$\chi=\sum_{p}\pi(x^{p})\jj\pi(y^{p})\jj^{-1}$, avec $x^{p},y^{p}\in\aa$, ce qui prouve que $\ll_{1}-\ll_{2}$ et $\rr_{2}-\rr_{1}$ commutent avec $\chi$. Par hypoth\`ese, ils anticommutent \'egalement avec $\chi$, et puisque $\chi^{2}=1$, on en d\'eduit que 
\bbb
\ll_{1}-\ll_{2}=\rr_{2}-\rr_{1}=0,
\eee
ce qui prouve l'unicit\'e de la d\'ecomposition. 
\edemo

Ce r\'esultat a plusieures cons\'equences. Tout d'abord, il nous permet de montrer que les op\'erateurs $\ll$ et $\rr$ sont hermitiens. En effet, puisque $\dd$ est hermitien, si $\ll$ et $\rr$ sont deux op\'erateurs convenables, alors $\ll^{*}$ et $\rr^{*}$ le seront aussi. Par l'unicit\'e de la d\'ecomposition, on d\'eduit que $\ll=\ll^{*}$ et $\rr=\rr^{*}$.

\par

De plus, puisque $\dd$ commute avec $\jj$, on a $\dd=\jj\dd\jj^{-1}=\jj\ll\jj^{-1}+\jj\rr\jj^{-1}$. Puisque $\ll$ commute avec $\pi(x)$, on a $\lb\jj\ll\jj^{-1},\jj\pi(x)\jj^{-1}\rb=\jj\lb\ll,\pi(x)\rb\jj^{-1}=0$, et de la m\^eme fa\c con, on prouve que
\bbb
\lb\jj\rr\jj^{-1},\pi(x)\rb=\jj\lb\rr,\jj^{-1}\pi(x)\jj\rb\jj^{-1}=
\jj\lb\rr,\jj\pi(x)\jj^{-1}\rb\jj^{-1}=0,
\eee
o\`u la relation $\jj^{2}=1$ a \'et\'e utilis\'ee. Par l'unicit\'e de la d\'ecompostion, on obtient
\bbb
\ll=\jj\rr\jj^{-1}\;\;\;\mathrm{et}\;\;\;\rr=\jj\ll\jj^{-1}.
\eee
Si nous notons $\Delta$ l'op\'erateur $\rr$, alors nous avons 
\bbb
\dd=\Delta+\jj\Delta\jj^{-1}.
\eee
En r\'esum\'e, nous avons prouv\'e le r\'esultat suivant.

\begin{pro}
L'op\'erateur de Dirac $\dd$ d'un triplet spectral fini s'\'ecrit de mani\`ere unique sous la forme
\bbb
\dd=\Delta+\jj\Delta\jj^{-1},
\eee
o\`u $\Delta$ est un op\'erateur qui anticommute avec $\chi$ et commute avec $\jj\pi(x)\jj^{-1}$ pour tout $x\in\aa$. De plus, $\Delta$ est hermitien.
\end{pro}

Cette d\'ecomposition de l'op\'erateur de Dirac a plusieurs cons\'equences que nous d\'evelopperons au cours de ce chapitre. Premi\`erement, puisque $\jj\Delta\jj^{-1}$ commute avec $\pi(x)$, nous pouvons remplacer $\lb\dd,\pi(x)\rb$ par $\lb\Delta,\pi(x)\rb$ pour tout $x\in\aa$. Cela permet en g\'en\'eral de simplifier la plupart des \'equations; par exemple dans le calcul des distances. Ensuite, nous allons montrer que $\Delta$ est une 1-forme, ce qui nous permettra de d\'eterminer explicitement toutes les 1-formes.

\par

Enfin, puisque l'op\'erateur $\Delta$ satisfait des relations de commutation tr\`es simples, nous pouvons le d\'eterminer explicitement en termes de matrices \`a coefficients complexes. Rappelons que $\hh=\mathop{\op}\limits_{ij}\hh_{ij}$ est la d\'ecomposition de l'espace de Hilbert, o\`u les indices $i$ et $j$ rep\`erent les facteurs simples de l'alg\`ebre. Nous d\'efinissons les \'el\'ements de matrice  de l'op\'erateur $\Delta$ par
\bbb
\Delta_{ij}^{kl}=\pi(1_{i})\jj\pi(1_{j})\jj^{-1}\,\Delta\,
\pi(1_{k})\jj\pi(1_{l})\jj^{-1},
\eee
o\`u $1_{i}$ d\'esigne la matrice unit\'e de $M_{n_{i}}(\ccc)$ consid\'er\'ee comme \'el\'ement de $\aa$. D'apr\`es le r\'esultat du paragraphe 2.2, cet \'el\'ement de matrice doit \^etre nul sauf si $j=l$. Dans ce dernier cas, il doit satisfaire \`a 
\bbb
\Delta_{ij}^{kl}(I_{n_{k}\ot_{|\mu_{kj}|}}\ot y_{j})=
(I_{n_{i}\ot_{|\mu_{ij}|}}\ot y_{j})\Delta_{ij}^{kl}
\eee
pour tout $y_{j}\in M_{n_{j}}(\ccc)$. Cela impose que l'on \'ecrive $\Delta_{ij}^{kl}$ sous la forme
\bbb
\Delta_{ij}^{kl}=\delta_{jl}\,M_{ik,j}\ot I_{n_{j}}\label{od}
\eee
o\`u les matrices $M_{ik,j}\in M_{\mu_{ij}n_{i}\times\mu_{kj}n_{k}}(\ccc)$ sont telles que $M_{ik,j}=M_{ki,j}^{*}$ si $\mu_{ij}\mu_{kj}<0$, et nulles dans tous les autres cas. Cela nous donne une param\'etrisation de l'op\'erateur $\Delta$
\`a l'aide de matrices complexes, dont on d\'eduit l'op\'erateur de Dirac \`a l'aide de la relation $\dd=\Delta+\jj\Delta\jj^{-1}$.

\par

Pour d\'eterminer l'op\'erateur de Dirac complet \`a l'aide de matrices, nous devons calculer l'action de $\jj$ sur $\Delta$. Pour cela, il est commode d'\'ecrire les matrices $M_{ik,j}$ comme sommes de produits tensoriels,
\bbb
M_{ik,j}=\mathop{\sum}\limits_p\, E_{ik}^{p}\ot M_{ik,j}^{p},
\eee
avec $E_{ik}^{p}\in M_{n_{i}\times n_{k}}(\ccc)$ et $M_{ik,j}^{p}\in M_{|\mu_{ij}|\times |\mu_{kj}|}(\ccc)$. L'action de $\jj$ sur un produit tensoriel est $\jj\lp e\ot\psi\ot f\rp=\ov{f}\ot\ov{\psi}\ot\ov{e}$, ce qui donne
\bbb
\lp\jj\Delta\jj^{-1}\rp_{ij}^{kl}=\delta_{ik}\;\mathop{\sum}\limits_p\,
I_{n_{i}}\ot\ov{M}_{jl,i}^{p}\ot\ov{E}_{jl}^{p}. 
\eee
Ainsi, en additionnant les \'el\'ements de matrice de $\Delta$ et de $\jj\Delta\jj^{-1}$, nous obtenons l'\'el\'ement de matrice correspondant de $\dd$,
\bbb
\dd_{ij}^{kl}=\delta_{jl}\;\mathop{\sum}\limits_p\, 
E_{ik}^{p}\ot M_{ik,j}^{p}\ot I_{n_{j}}+
\delta_{ik}\;\mathop{\sum}\limits_p\,
I_{n_{i}}\ot\ov{M}_{jl,i}^{p}\ot\ov{E}_{jl}^{p}
\eee

\par

En conclusion, nous pouvons r\'esumer cette \'etude comme suit.

\begin{pro}
Un triplet spectral fini est donn\'e, \`a une \'equivalence unitaire pr\`es, par une matrice sym\'etrique et non d\'eg\'en\'er\'ee $\mu\in M_{N}(\zzz)$, \`a partir de laquelle on construit la structure de bimodule et la chiralit\'e, ainsi que d'un op\'erateur $\Delta$ dont les \'el\'ements de matrice sont donn\'ees par la relation (\ref{od}).
\end{pro}

\subsection{$S^{0}$-r\'ealit\'e}

Bien que les r\'esultats obtenus soient valables dans le cas des triplets spectraux finis $S^{0}$-r\'eels, il est utile de d\'ecrire les propri\'et\'es particuli\`eres de ces derniers en utilisant le langage que nous venons de d\'evelopper.

\par

Rappelons qu'un triplet spectral $(\aa,\hh,\dd)$ est $S^{0}$-r\'eel si il existe une involution hermitienne qui commute avec $\dd$, $\chi$ et $\pi(x)$ pour tout $x\in\aa$ et anticommute avec $\jj$.

\par

En fait, nous avons vu que la strucuture de bimodule d'un triplet spectral $S^{0}$-r\'eel est obtenue \`a partir de deux bimodules. Le premier correspond \`a la valeur propre $+1$ de $\epsilon$ et nous notons $\nu$ sa matrice de multiplicit\'e. Le second bimodule, qui est associ\'e \`a la valeur propre $-1$ de $\epsilon$ s'obtient simplement, puisque dans cette d\'ecomposition on a
\bbb
\jj=\pp{0&I\cr I&0}C,
\eee
en \'echangeant les actions \`a droite et \`a gauche du bimodule associ\'e \`a la valeur propre $+1$. Par cons\'equent, la matrice de multiplici\'e totale s'\'ecrit sous la forme $\mu=\nu+\nu^{*}$. Bien entendu, puisque $\chi$ et $\jj$ commutent, les entiers $\nu_{ij}$ et $\nu_{ji}$ doivent toujours avoir m\^eme signe.

\par

R\'eciproquement, si la matrice de multiplicit\'e $\mu$ satisfait \`a ces conditions, il est facile de construire un op\'erateur $\epsilon$ convenable. 

\par

De m\^eme, puisque l'op\'erateur de Dirac commute avec $\epsilon$ et avec $\jj$, il se d\'ecompose en deux op\'erateurs du premier ordre connect\'es par $\jj$. Cette condition est suffisante pour que $\dd$ commute avec $\epsilon$ et $\jj$.

\par

En cons\'equence, nous pouvons caract\'eriser les triplets spectraux finis $S^{0}$-r\'eels de la mani\`ere suivante:

\begin{pro}
Un triplet spectral fini est $S^{0}$-r\'eel si et seulement si
\begin{itemize}
\item
sa matrice de multiplicit\'e peut s'\'ecrire sous la forme $\mu=\nu+\nu^{*}$, o\`u $\nu\in M_{N}(\zzz)$ est une matrice telle que $\nu_{ij}$ et $\nu_{ji}$ soient de m\^eme signe,
\item 
l'op\'erateur de Dirac est somme de deux op\'erateurs du premier ordre correspondant aux bimodules d\'etermin\'es par $\nu$ et $\nu^{*}$ et se d\'eduisant l'un de l'autre par conjugaison par $\jj$.
\end{itemize}
\end{pro}

\subsection{Une approche diagrammatique}

Pour classifier les triplets spectraux finis, il faut regrouper ces derniers dans un certain nombre de classes dont on peut facilement saisir les propri\'et\'es caract\'eristiques.

\par

Un premier pas en ce sens a \'et\'e franchi par l'introduction de la matrice de multiplicit\'e: celle-ci permet, pour une alg\`ebre fix\'ee, de reconstruire \`a une \'equivalence unitaire pr\`es, la repr\'esentation $\pi$ et les op\'erateurs $\chi$ et $\jj$. Nous pourrions donc, dans un premier temps, consid\'erer que deux triplets spectraux sont \'equivalents si et seulement si ils ont m\^eme matrice de multiplicit\'e. Cependant, cette classification ne permet pas de distinguer deux triplets spectraux ayant m\^eme matrice de multiplicit\'e mais des op\'erateurs de Dirac diff\'erents. Pour rem\'edier \`a ce probl\`eme, il est int\'eressant d'associer \`a chaque triplet spectral fini un diagramme.

\begin{dfi}
\`A un triplet spectral fini $\lp\aa,\hh,\dd\rp$ dont la matrice de multiplicit\'e est $\mu$, on associe un diagramme construit de la mani\`ere suivante:
\begin{itemize}
\item
Au point de coordonn\'ees $(i,j)$ du plan on associe un sommet de type $\op$ si $\mu_{ij}>0$ ou un sommet de type $\om$ si $\mu_{ij}<0$.
\item
On relie les sommets situ\'es aux points de coordonn\'ees $(i,j)$ et $(k,l)$ si et seulement si l'\'el\'ement de matrice de l'op\'erateur de Dirac
\bbb
\dd_{ij}^{kl}=\lp\pi(1_{i})\jj\pi(1_{j})\jj^{-1}\rp\,\dd\,
\lp\pi(1_{k})\jj\pi(1_{l})\jj^{-1}\rp
\eee 
est non nul.
\end{itemize}
\end{dfi}

Les sommets de ce diagramme permettent de  d\'eterminer quels sont les \'el\'ements non nuls de la matrice de multiplicit\'e alors que les liens nous aident \`a visualiser la structure de l'op\'erateur de Dirac. 

\par

Nous classifions les triplets spectraux finis selon leurs diagrammes. L'int\'er\^et de ce type de classification se trouve dans le fait que beaucoup des propri\'et\'es des mod\`eles en physique des particules que nous allons construire apparaissent \`a travers la structure de ces diagrammes. De plus, cela permet une manipulation tr\`es simple d'objets qui sont souvent des matrices de taille importante ($90\times 90$ pour le mod\`ele standard) tout en perdant le moins d'information possible.

\par

Ces diagrammes poss\`edent les propri\'et\'es suivantes:

\begin{pro}
Le diagramme associ\'e \`a un triplet spectral fini n'est form\'e que de liens verticaux et horizontaux entre sommets de type diff\'erents de telle mani\`ere que le diagramme soit sym\'etrique par rapport \`a la diagonale (\'echange des indices $i$ et $j$).
\end{pro}

\demo
Ces propri\'et\'es r\'esultent des relations satisfaites par $\dd$. Les liens sont soit verticaux soit horizontaux \`a cause de la condition d'ordre un et seuls des sommets de type diff\'erents sont reli\'es suite \`a la relation d'anticommutation $\chi\dd+\dd\chi=0$. La relation $\jj\dd=\dd\jj$ implique la sym\'etrie par rapport \`a la diagonale.
\edemo

Il convient de remarquer que, puisque $\dd$ est hermitien, $\dd_{ij}^{kl}$ est non nul si et seulement si $\dd_{kl}^{ij}$ est non nul, ce que nous symbolisons par un seul lien non orient\'e sur le diagramme. 

\par

Sur ce diagramme, les liens verticaux correspondent aux \'el\'ements de matrices de l'op\'erateur $\Delta$ alors que que les liens horizontaux sont relatifs \`a $\jj\Delta\jj^{-1}$. 

\par

Cette remarque nous permet de construire diagrammatiquement tous les triplet spectraux;

\begin{enumerate}
\item
On choisit une alg\`ebre
\bbb
\aa=\mathop{\op}\limits_{i=1}^{N}M_{n_{i}}(\ccc),
\eee
ainsi qu'une matrice de multiplicit\'e $\mu\in M_{N}(\zzz)$ qui soit sym\'etrique et non d\'eg\'en\'er\'ee.
\item
A l'aide de la matrice de multiplicit\'e, on construit les sommets du diagramme en associant \`a chaque point $(i,j)$ du plan un sommet de type $\op$ si $\mu_{ij}>0$ ou un sommet de type $\om$ si $\mu_{ij}<0$.
\item
Au sommet $(i,j)$ on associe l'espace de Hilbert $\hh_{ij}=\ccc^{n_{i}}\ot \ccc^{|\mu_{ij}|}\ot\ccc^{n_{j}}$.
\item
On relie tous les sommets de type diff\'erent par des liens verticaux.
\item
A chaque lien vertical entre $(i,j)$ et $(k,j)$ avec $i<k$ on associe l'\'el\'ement de matrice de $\Delta$ donn\'e par

\bbb
\Delta_{ij}^{kj}= M_{ik,j}\ot I_{n_{j}},
\eee
o\`u $M_{ik,j}\in M_{|\mu_{ij}|n_{i}\times |\mu_{kj}|n_{j}}(\ccc)$ est une matrice quelconque.
\item
Si $i>k$, on d\'efinit
\bbb
\Delta_{ij}^{kl}=\lp M_{ki,j}\rp^{*}\ot I_{n_{j}}.
\eee
\item
On termine en compl\'etant par
\bbb
\dd=\Delta+\jj\Delta\jj^{-1}.
\eee
\end{enumerate}

\section{Th\'eories de jauge}

\subsection{Sym\'etries et th\'eories de jauge}

Les sym\'etries d'un espace sont en correspondance directe avec les automorphisnes de son alg\`ebre des coordonn\'ees, ce qui entraine que, pour \'etudier les sym\'etries d'un "espace quantique" d\'etermin\'e par une alg\`ebre $\aa$, nous devons rechercher tous les automorphismes de $\aa$.

\par

Ensuite, ces sym\'etries doivent \^etre impl\'ement\'ees au niveau du triplet spectral $(\aa,\hh,\dd)$ en associant \`a chaque automorphisme $\phi$ une transformation unitaire $U_{\phi}$ de l'espace de Hilbert telle que
\bbb
U_{\phi}\pi U_{\phi}^{*}=\pi\,\circ\,\phi^{-1}.
\eee

\par

Commen\c cons par d\'eterminer les automorphismes de l'alg\`ebre associ\'ee \`a un triplet spectral fini. Dans le cas le plus simple, l'alg\`ebre est commutative et ses automorphismes sont connus.

\begin{pro}
Les automorphismes de l'alg\`ebre $\ccc^{N}$ sont les permutations des diff\'erents facteurs simples.
\end{pro}

Cela se d\'emontre facilement en cherchant les images des \'el\'ements unit\'es $1_{i}$ de chaque facteur simple. 

\par

Le groupe des automorphismes de cette alg\`ebre est donc isomorphe au groupe des pertmutaioms de $N$ \'el\'ements. A chaque permutation $\sigma\in S_{N}$ doit \^etre associ\'ee une transfomation unitaire $U_{\sigma}$ telle que
\bbb
U_{\sigma}\pi(1_{i}) U_{\sigma}^{*}=\pi(1_{\sigma(i)}).\label{st1}
\eee
Si on d\'ecompose $\hh$ en somme directe, $\hh=\op_{i}\hh_{i}$ avec $\hh_{i}=\pi_{1_{i}}\hh$, la relation (\ref{st1}) implique $U_{\sigma}\hh_{i}\subset\hh_{\sigma(i)}$. Un argument identique pour $\sigma^{-1}$ nous prouve que l'on doit en fait avoir $\hh_{i}=\hh_{\sigma(i)}$. En cons\'equence, une permutation $\sigma$ ne peut \^etre une sym\'etrie du triplet spectral que si les dimensions des sous-espaces que l'on veut permuter co\"\i ncident. 

\par

Sur le plan diagrammatique, cela correspond \`a effectuer cette permutation sur les lignes et les colonnes. La nouvelle matrice de multiplicit\'e s'obtient par la relation 
\bbb
\mu'=P_{\sigma^{-1}}\,\mu\,P_{\sigma},
\eee  
o\`u $P_{\sigma}$ est la matrice de permutation d\'efinie par $\sigma$.

\par

Si les sym\'etries d'un triplet spectral commutatif sont discr\`etes, il n'en est plus de m\^eme dans le cas non commutatif. Par exemple, si $\aa=M_{n}(\ccc)$, alors pour tout unitaire $u\in M_{n}(\ccc)$, l'application $x\mapsto uxu^{-1}$ est un automorphisme non trivial de $\aa$. Ces automorphismes sont appel\'es automorphismes int\'erieurs et ce sont les seuls:.

\begin{pro}
Tous les automorphismes de l'alg\`ebre $M_{n}(\ccc)$ sont int\'erieurs.
\end{pro}

En cons\'equence, le groupe des automorphismes de $M_{n}(\ccc)$ est isomorphe \`a
\bbb
PSU(n)=SU(n)/Z_{n},
\eee
o\`u $SU(n)$ d\'esigne le groupe des matrices unitaires de d\'eterminant 1 et $Z_{n}$ est son centre est form\'e des matrices du type $q\,I_{n}$, avec $q^{n}=1$.

\par

Lorsque $\aa$ est une somme directe de $N$ alg\`ebres de matrices, chaque unitaire $u$ d\'etermine encore un automorphisme non trivial de $\aa$. Notons  que dans ce cas tous les automorphismes ne sont pas n\'ecessairement int\'erieurs car il peut y avoir un sous groupe fini d'automorphismes qui correspondent simplement \`a des permutations de facteurs identiques.

\par

Pour \^etre une sym\'etrie du triplet spectral $(\aa,\hh,\dd)$, chaque unitaire $u$ doit \^etre repr\'esent\'e sur l'espace de Hilbert par un op\'erateur $U$ satisfaisant \`a
\bbb
U\pi(x)U^{-1}=\pi(uxu^{-1}).\label{st2}
\eee
pour tout $x\in\aa$.

\par

Au cours de la section 2.2.2, nous avons montr\'e que, en prenant 
\bbb
U=\pi(u)\jj\pi(u)\jj^{-1},
\eee
alors $U$ est solution de (\ref{st2}) et laisse invariant les op\'erateurs $\chi$ et $\jj$. Seul l'op\'erateur de Dirac est modifi\'e et devient
\bbb
\dd\rightarrow U\,\dd\,U^{-1}=\dd+u\lb\dd,u^{-1}\rb+\jj u\lb\dd,u^{-1}\rb\jj^{-1},
\eee
o\`u nous avons simplement not\'e $u$ l'op\'erateur $\pi(u)$.

\par

Cette transformation de l'op\'erateur de Dirac peut \^etre report\'ee sur un champ de jauge $A$. Ce dernier est d\'efini comme une 1-forme hermitienne, c'est-\`a-dire que c'est un op\'erateur hermitien qui peut s'\'ecrire sous la forme
\bbb
A=\mathop{\sum}\limits_{p}x_{p}\lb\dd,y_{p}\rb
\eee
avec $x_{p},y_{p}\in\aa$.

\par

Sous une transformation de jauge, $A$ devient 
\bbb
A\rightarrow uAu^{-1}+u\lb\dd,u^{-1}\rb,
\eee
de telle fa\c con que l'op\'erateur de Dirac covariant $\dd+A+\jj A\jj^{-1}$ se transforme en 
\bbb
\dd+A+\jj A\jj^{-1}\rightarrow 
U\lp\dd+A+\jj A\jj^{-1}\rp U^{-1}.
\eee
Cette loi de transformation nous assure que l'action fermionique d\'efinie par
\bbb
\langle\Psi,\lp\dd+A+\jj A\jj^{-1}\rp\Psi\rangle
\eee
est invariante de jauge lorsque le fermion $\Psi$ se transforme en
\bbb
\Psi\rightarrow\pi(u)\jj\pi(u)\jj^{-1}\Psi.
\eee

Nous renvoyons \`a la section 1.2.2 le lecteur int\'eress\'e par une \'etude d\'etaill\'ee de la sym\'etrie de jauge dans un cadre g\'en\'eral.


\subsection{D\'etermination des 1-formes}

Etudions en d\'etail la structure de l'espace des 1-formes $\Omega_{\dd}^{1}(\aa)$ associ\'ees \`a un triplet spectral fini $(\aa,\hh,\dd)$. Pour cela, commen\c cons par rappeler que l'op\'erateur de Dirac $\dd$ peut \^etre \'ecrit sous la forme $\dd=\Delta+\jj\Delta\jj^{-1}$, o\`u $\Delta$ et un op\'erateur hermitien qui anticommute avec $\chi$ et commute avec $\jj\pi(x)\jj^{-1}$ pour tout $x\in\aa$. 

\begin{pro}
L'op\'erateur $\Delta$ apparaissant dans la d\'ecomposition de $\dd$ est une 1-forme.
\end{pro}

\demo
Si $G$ est le groupe des unitaires de $\aa$, alors $G$ est un groupe de Lie compact donc admet une mesure de Haar. En utilisant cette mesure, on a
\bbb
\dd=\int_{G}\,du\,u\dd u^{*}-\int_{G}\,du\,u\lb\dd,u^{*}\rb.\label{df1}
\eee
Puisque la mesure $du$ est invariante par translation \`a gauche, le premier terme du second membre de \ref{df1} commute avec tous les unitaires donc avec tous les \'el\'ements de $\aa$ car n'importe quel $x\in\aa$ peut s'\'ecrire comme une somme de 4 unitaires. 

\par

Le second terme du deuxi\`eme membre de (\ref{df1}) commute avec $\jj\pi(x)\jj^{-1}$ pour tout $x\in\aa$ en vertu de la condition d'ordre un.
De plus, c'est une 1-forme car, en utilisant des sommes de Riemann, $\int_{G}\,du\,u\lb\dd,u^{*}\rb$ peut \^etre \'ecrit comme une limite d'expressions du type $\mathop{\sum}\limits_{p}x_{p}\lb\dd,y_{p}\rb$ avec $x_{p},y_{p}\in\aa$. Puisqu'on est en dimension finie, l'espace $\Omega_{\dd}^{1}(\aa)$ est un sous espace ferm\'e et toute limite d'une suite de 1-formes est une 1-forme.

\par

Enfin, chacun des deux termes du second membre de (\ref{df1}) anticommute avec la chiralit\'e, ce qui implique, par application de l'unicit\'e de la d\'ecomposition de l'op\'erateur de Dirac, que
\bbb
\Delta=-\int_{G}\,du\,u\lb\dd,u^{*}\rb.\label{df2}
\eee
Par cons\'equent, $\Delta$ est une 1-forme.

\edemo

Il est \'egalement possible de donner une d\'emonstration de ce r\'esultat en utilisant le diagramme associ\'e au triplet spectral $(\aa,\hh,\dd)$. En effet, les \'el\'ements de matrice de $\Delta$ correspondent aux liens verticaux entre les diff\'erentes lignes. Puisque on ne peut relier une ligne \`a elle-m\^eme par un lien vertical, on a n\'ecessairement $\pi(1_{i})\Delta\pi(1_{i})=0$, o\`u $1_{i}$ est l'\'el\'ement unit\'e du facteur $M_{n_{i}}(\ccc)$ apparaissant dans la d\'ecomposition de $\aa$. On en d\'eduit que
\bbbb
\Delta&=&\Delta-\mathop{\sum}\limits_{i=1}^{N}\pi(1_{i})\Delta\pi(1_{i})\\
&=&-\mathop{\sum}\limits_{i=1}^{N}\pi(1_{i})\lb\Delta,\pi(1_{i})\rb\\
&=&-\mathop{\sum}\limits_{i=1}^{N}\pi(1_{i})\lb\dd,\pi(1_{i})\rb
\eeee
ce qui prouve que $\Delta$ est une 1-forme.

\par

Cette d\'emonstration n'est cependant pas valable dans le cas des triplets spectraux r\'eels, alors que la d\'emonstration utilisant la mesure de Haar ne d\'epend pas du fait que l'alg\`ebre soit r\'eelle ou complexe. En fait, nous avons montr\'e l'existence et l'unicit\'e de la d\'ecomposition de l'op\'erateur de Dirac que pour les alg\`ebres complexes. Dans le cas r\'eel, la d\'emonstration pr\'ec\'edente nous permet de d\'efinir $\Delta$ par la relation (\ref{df2}). Le r\'esultat d'unicit\'e s'\'etend sans difficult\'e au cas r\'eel et on a $\dd=\Delta+\jj\Delta\jj^{-1}$ avec $\Delta\in\Omega_{\dd}^{1}(\aa)$. Nous renvoyons aux derni\`eres sections de ce chapitre pour une discussion de certaines des propri\'et\'es des alg\`ebres r\'eelles.

\par

L'\'etape suivante dans la d\'etermination des 1-formes consiste \`a relier $\Omega_{\dd}^{1}(\aa)$ et $\Delta$.

\begin{pro}
L'espace des 1-formes est \'egal au bimodule engendr\'e par $\Delta$.
\end{pro}

\demo
Puisque $\Omega_{\dd}^{1}(\aa)$ est un bimodule et que $\Delta$ est une 1-forme, il est clair que $\Omega_{\dd}^{1}(\aa)$ contient le bimodule engendr\'e par $\Delta$.

\par

Pour montrer la r\'eciproque, il suffit d'\'ecrire un \'el\'ement g\'en\'erique de $\Omega_{\dd}^{1}(\aa)$ sous la forme $A=\mathop{\sum}\limits_{p}x_{p}\lb\dd,y_{p}\rb$ avec $x_{p},y_{p}\in\aa$, et de remplacer $\dd$ par $\Delta$ dans l'expression pr\'ec\'edente puisque $\jj\Delta\jj^{-1}$ commute avec $\pi(x)$ pour tout $x\in\aa$. 

\edemo

L'expression de $\Delta$ \`a l'aide de ses \'el\'ements de matrices, donn\'ee par la relation (\ref{od}), est
\bbb
\Delta_{ij}^{kl}=\mathop{\sum}\limits_{p}\delta_{jl}\, E_{ik}^{p}\ot M_{ik,j}^{p}\ot I_{n_{j}}.
\eee
Nous rappelons que $\Delta_{ij}^{kl}=\pi(1_{i})\jj\pi(1_{j})\jj^{-1}\,\Delta\,
\pi(1_{k})\jj\pi(1_{l})\jj^{-1}$, que $\lp E_{ik}^{p}\rp_{1\leq p\leq n_{i}n_{k}}$ est une base de $M_{n_{i}\times n_{k}}(\ccc)$ et que les matrices $M_{ik,j}^{p}\in M_{|\mu_{ij}|\times |\mu_{kj}|}(\ccc)$ sont telles que $\Delta$ soit hermitien. 

\par

Nous choisissons la base  de matrices \'el\'ementaires
\bbb
\lp E_{ik}^{ab} \rp_{1\leq a\leq n_{i}\atop 1\leq b\leq n_{k}},
\eee
dont le seul \'el\'ement non nul est \'egal \`a 1 et est situ\'e \`a l'intersection de la ligne $a$ et de la colonne $b$. Par commodit\'e, nous noterons toujours $p$ la paire $(a,b)$ permettant de distinguer les matrices \'el\'ementaires.

\par

A partir des matrices $M_{ik,j}^{p}$, formons les vecteurs colonnes $M_{ik}^{p}$ d\'efinis par
\bbb
M_{ik}^{p}=\pp{M_{ik,1}^{p}\cr\vdots\cr M_{ik,N}^{p}}.
\eee
Lorsque $p$ varie, les $n_{i}n_{k}$ vecteurs $M_{ik}^{p}$ engendrent un espace vectoriel $M_{ik}$ dont nous notons $p_{ik}$ la dimension.

\par

$(M_{ik}^{p})_{1\leq p\leq n_{i}n_{k}}$ est un syst\`eme de g\'en\'erateurs de $M_{ik}$ dont nous pouvons toujours extraire une base. Sans perte de g\'en\'eralit\'e, suposons que cette base soit donn\'ee par les $p_{ik}$ premiers vecteurs, ce qui entraine que
\bbb
\Delta_{ij}^{kl}=\mathop{\sum}\limits_{p=1}^{p_{ik}}\delta_{jl}\, E_{ik}^{p}\ot M_{ik,j}^{p}\ot I_{n_{j}}
+
\mathop{\sum}\limits_{1\leq p\leq p_{ik}\atop
p_{ik}\leq q\leq n_{i}n_{k}}\delta_{jl}\,\lambda_{ik}^{pq} E_{ik}^{p}\ot M_{ik,j}^{p}\ot I_{n_{j}},\label{df3}
\eee 
avec $\lambda_{ik}^{pq}$.

\par

Par multiplication \`a droite et \`a gauche par des matrices \'el\'ementaires bien choisies $E_{ii}\in M_{n_{i}}(\ccc)$ et $E_{kk}\in M_{n_{k}}(\ccc)$, on peut toujours obtenir un \'el\'ement de matrice de $\pi(E_{ii})\Delta\pi(E_{kk})$ du type
\bbb
\lp\pi(E_{ii})\Delta\pi(E_{kk})\rp_{ij}^{kl}=\delta_{jl}\, E_{ik}^{p}\ot M_{ik,j}^{p}\ot I_{n_{j}}.
\eee
Ensuite, en multipliant \`a droite et \`a gauche la relation pr\'ec\'edente par des matrices quelconques et en prenant toutes les combinaisons lin\'eaire possibles, on obtient une 1-forme $\Omega_{ik}^{p}$ dont l'\'el\'ement de matrice est
\bbb
\lp\Omega_{ik}^{p}\rp_{ij}^{kl}=\delta_{jl}\, \omega_{ik}^{p}\ot M_{ik,j}^{p}\ot I_{n_{j}},
\eee
o\`u $\omega_{ik}^{p}\in M_{n_{i}\times n_{k}}(\ccc)$ est une matrice quelconque.

\par

Puis nous rep\`etons cette op\'eration pour tous les indices $p\in\la 1,2,\dots,p_{ik}\ra$ et tous les couples $(i,k)\in\la 1,2,\dots,N\ra$, ce qui nous am\`ene \`a une 1-forme $\Omega$ dont les \'el\'ements de matrice sont
\bbb
\Omega_{ij}^{kl}=\mathop{\sum}\limits_{p=1}^{p_{ik}}\delta_{jl}\, \omega_{ik}^{p}\ot M_{ik,j}^{p}\ot I_{n_{j}},
\eee 
o\`u les matrices $\omega_{ik}^{p}\in M_{n_{i}\times n_{k}}(\ccc)$ sont quelconques.

\par

Cette 1-forme est en fait la 1-forme la plus g\'en\'erale car elle contient \'egalement les contributions du second terme du second membre de (\ref{df3}). En cons\'equence, nous pouvons \'enoncer le r\'esultat suivant.

\begin{pro}
Un \'el\'ement g\'en\'erique $\Omega$ de l'espace des 1-formes $\Omega_{\dd}^{1}(\aa)$ est donn\'e par ses \'el\'ements de matrices
\bbb
\Omega_{ij}^{kl}=\mathop{\sum}\limits_{p=1}^{p_{ik}}\delta_{jl}\, \omega_{ik}^{p}\ot M_{ik,j}^{p}\ot I_{n_{j}},
\eee 
o\`u les matrices $\omega_{ik}^{p}\in M_{n_{i}\times n_{k}}(\ccc)$ sont quelconques
\end{pro}

Pour chaque couple $(i,k)$, l'espace des matrices $M_{n_{i}\times n_{k}}(\ccc)$ est r\'ep\'et\'e $p_{ik}$ fois, ce qui justifie la d\'enomination suivante.

\begin{dfi}
Les entiers $p_{ik}$ sont appel\'es les multiplicit\'es de l'espace des 1-formes.
\end{dfi}

Il ressort de l'analyse pr\'ec\'edente que la dimension de $\Omega_{\dd}^{1}(\aa)$ peut \^etre d\'etermin\'ee en fonction de $p_{ik}$.

\begin{pro}
La dimension de $\Omega_{\dd}^{1}(\aa)$ est
\bbb
\mathop{\sum}\limits_{i,k}p_{ik}n_{i}n_{k}\leq 
\mathop{\sum}\limits_{i,k}\lp n_{i}n_{k}\rp^{2}.
\eee
\end{pro}

La derni\`ere inegalit\'e r\'esulte de $p_{ik}\leq n_{i}n_{k}$, qui est une cons\'equence imm\'ediate de la d\'efinition des entiers $p_{ik}$.

\par

Enfin, terminons par deux remarques qui nous seront utiles par la suite. Tout d'abord, puisque l'op\'erateur $\Delta$ est hermitien, il est toujours possible de supposer que les matrices $M_{ij,k}^{p}$ satisfont \`a $\lp M_{ij,k}^{p}\rp^{*}=M_{ji,k}^{p}$.

\par

Ensuite, par application du proc\'ed\'e d'orthonormalisation de Gram-Schmidt \`a la base $\lp M_{ij}^{p}\rp_{1\leq p\leq p_{ij}}$, nous pouvons, sans perte de g\'en\'eralit\'e, supposer que les matrices $M_{ij,k}^{p}$ satisfont \`a
\bbb
\mathop{\sum}\limits_{k=1}^{N}n_{k}\t\lb\lp M_{ij,k}^{p}\rp^{*}M_{ij,k}^{q}\rb=
X\,\delta_{p,q},
\eee
o\`u $X>0$ est une constante de normalisation que nous sp\'ecifierons plus tard.


\subsection{Application \`a la construction de triplets spectraux}

Au cours du chap\^\i tre pr\'ec\'edent, nous avons vu comment construire un triplet spectral $\lp\tilde{\aa},\tilde{\hh},\tilde{\dd}\rp$ \`a l'aide d'un  autre triplet $\lp\aa,\hh,\dd\rp$, d'un module hermitien $\ee$ sur $\aa$ et d'une connexion $\nabla$ sur $\ee$, compatible avec la structure hermitienne. Rappelons bri\`evement que $\tilde{\aa}$, $\tilde{\hh}$ et $\tilde{\dd}$ sont d\'efinis par:
\bbbb
\tilde{\aa}&=&\mathrm{End}_{\aa}(\ee),\\
\tilde{\hh}&=&\ee\ot_{\aa}\hh\ot_{\aa}\ov{\ee},\\
\tilde{\dd}(\xi\ot_{\aa}\psi\ot_{\aa}\ov{\zeta})&=&
\nabla(\xi)\psi\ot_{\aa}\ov{\zeta}+
\xi\ot_{\aa}\dd(\psi)\ot_{\aa}\ov{\zeta}+
\xi\ot_{\aa}\psi\ov{(\nabla(\zeta))},
\eeee
pour tous $\xi,\zeta\in\ee$ et $\psi\in\hh$. De m\^eme, les op\'erateurs $\tilde{\jj}$ et $\tilde{\chi}$ sont d\'efinis \`a l'aide de $\jj$ et $\chi$ par
\bbbb
\tilde{\jj}(\xi\ot_{\aa}\psi\ot_{\aa}\ov{\zeta})&=&
\zeta\ot_{\aa}\jj(\psi)\ot_{\aa}\ov{\xi},\\
\tilde{\chi}(\xi\ot_{\aa}\psi\ot_{\aa}\ov{\zeta})&=&
\xi\ot_{\aa}\chi(\psi)\ot_{\aa}\ov{\zeta}.
\eeee
Le triplet $\lp\tilde{\aa},\tilde{\hh},\tilde{\dd}\rp$ ainsi construit correspond \`a une th\'eorie de jauge sur l'espace non commutatif d\'efini par 
$\lp\aa,\hh,\dd\rp$, donn\'e par le module hermitien $\ee$ et la connexion $\nabla$. 

\par

Nous allons montrer qu'une large classe de triplets spectraux finis peut \^etre obtenue par un tel proc\'ed\'e \`a partir de triplets spectraux commutatifs.

\begin{pro}
Tout triplet spectral fini dont l'espace des 1-formes n'a pas de multiplicit\'e ($p_{ij}\leq 1$) peut \^etre obtenu \`a l'aide d'une th\'eorie de jauge avec une alg\`ebre  commutative. 
\end{pro}

Avant de prouver ce r\'esultat,  nous devons \'etudier les th\'eories de jauge construites \`a l'aide d'un triplet spectral fini commutatif  $\lp\aa,\hh,\dd\rp$. L'alg\`ebre $\aa$ est $\ccc^{N}$, l'espace de Hilbert, la repr\'esentation de $\aa$ ainsi que les op\'erateurs $\jj$ et $\chi$ sont donn\'es, \`a une \'equivalence unitaire pr\`es, par la matrice de multiplicit\'e $m\in M_{N}(\zzz)$. L'op\'erateur de Dirac est univoquement reconstruit \`a partir des \'el\'ements de matrice $\Delta_{ij}=\pi(1_{i})\Delta\pi(1_{j})$ de l'op\'erateur $\Delta$, o\` u $1_{i}$ est l'\'el\'ement de $\ccc^{N}$ dont toutes les composantes sont nulles, sauf la i-\`eme, qui est \'egale \`a 1. D\'esormais, pour all\'eger nos notations, nous omettrons le symbole $\pi$ de la repr\'esentation et nous identifierons $x$ et $\pi(x)$.

\par

Les modules projectifs finis sur $\aa$ sont tous de la forme
\bbbb
\ee=\ccc^{n_{1}}\op\ccc^{n_{2}}\op\dots\op\ccc^{n_{N}},
\eeee
o\`u les $n_{i}$ sont des nombres entiers positifs et chaque facteur $\ccc$ de $\aa$ agit sur le facteur correspondant de $\ee$ par simple multiplication. Dans chacun des espaces $\ccc^{n_{i}}$ on choisit une base $(e_{i}^{a})_{1\leq a\leq n_{i}}$ et les vecteurs $e_{i}^{a}$, lorsque les deux indices $i$ et $a$ varient, forment une base de $\ee$ en tant qu'espace vectoriel sur $\ccc$. La structure hermitienne sur $\ee$ est d\'efinie en choisissant la base pr\'ec\'edente comme base orthonormale, 
\bbbb
\langle e_{i}^{a},e_{j}^{b}\rangle=\delta_{ab}\delta_{ij}\, 1_{i}.
\eeee
Pour continuer notre construction, nous devons trouver toutes les connexions sur $\ee$, compatibles avec la m\'etrique pr\'ec\'edente.

\begin{pro}
Les connexions hermitiennes sont donn\'ees par
\bbb
\nabla(e_{i}^{a})=-\mathop{\sum}\limits_{j}e_{i}^{a}\ot\Delta_{ij}
+\mathop{\sum}\limits_{j,b}A_{ij}^{ab}e_{j}^{b}\ot\Delta_{ji},
\eee
o\`u les matrices $A_{ij}\in M_{n_{i}\times n_{j}}(\ccc)$ satisfont \`a $A_{ij}^{*}=A_{ji}$.
\end{pro}

\demo
Pour montrer cela, il est pratique de se rappeler que l'espace des connexions est un espace affine sur l'espace des morphismes entre les modules \`a droite $\ee$ et $\ee\ot_{\aa}\Omega_{\dd}^{1}(\aa)$. Cela signifie que si nous avons une connexion $\nabla_{0}$, nous pouvons les obtenir toutes en rajoutant \`a $\nabla_{0}$ un morphisme. Puisque les \'el\'ements de matrices non nuls $\Delta_{ij}$ forment une base de $\Omega_{\dd}^{1}(\aa)$, les vecteurs $e_{i}^{a}\ot\delta_{ij}$ forment une base de $\ee\ot_{\aa}\Omega_{\dd}^{1}(\aa)$ et nous d\'efinissons une application lin\'eaire $\nabla_{0}$ de $\ee$ dans $\ee\ot_{\aa}\Omega_{\dd}^{1}(\aa)$ par
\bbbb
\nabla_{0}(e_{i}^{a})=-\mathop{\sum}\limits_{j}e_{i}^{a}\ot\Delta_{ij}.
\eeee
Par lin\'earit\'e, la v\'erification de la r\`egle de Leibniz se fait sur les \'el\'ements de base.
\bbbb
\nabla_{0}(e_{i}^{a}1_{i})&=&
-\delta_{ij}\mathop{\sum}\limits_{j}e_{i}^{a}\ot\Delta_{ij},\n
\eeee
ainsi que
\bbbb
\nabla_{0}(e_{i}^{a})1_{j}+e_{i}^{a}\ot_{\aa}d(1_{j})&=&
-\mathop{\sum}\limits_{k}e_{i}^{a}\ot\Delta_{ik}1_{j}
e_{i}^{a}\ot\delta_{ij}\\
&-&\delta_{ij}\mathop{\sum}\limits_{k}e_{i}^{a}\ot\Delta_{ik}\\
&=&-\delta_{ij}\mathop{\sum}\limits_{j}e_{i}^{a}\ot\Delta_{ij},
\eeee
nous prouvent que
\bbbb
\nabla_{0}(e_{i}^{a}1_{j})&=&
\nabla_{0}(e_{i}^{a})1_{j}+e_{i}^{a}\ot_{\aa}d(1_{j}),\\
\eeee
ce qui n'est autre que la r\`egle de Leibniz. De m\^eme, le morphisme $A$ de module le plus g\'en\'eral entre $\ee$ et $\ee\ot_{\aa}\Omega_{\dd}^{1}(\aa)$ doit v\'erifier $A(e_{i}^{a}1_{j})=a(e_{i}^{a})1_{j}$ et s'\'ecrit, dans la base pr\'ec\'edente
\bbbb
A(e_{i}^{a}1_{i})&=&
\mathop{\sum}\limits_{j,b}A_{ij}^{ab}e_{j}^{b}\ot\delta_{ji},
\eeee
o\`u  $A_{ij}^{ab}$ d\'esignent des nombres complexes quelconques. Par cons\'equent, la connexion la plus g\'en\'erale sur $\ee$ est d\'efinie par
\bbbb
\nabla(e_{i}^{a}1_{i})&=&
-\mathop{\sum}\limits_{j}e_{i}^{a}\ot\Delta_{ij}+
\mathop{\sum}\limits_{j,b}A_{ij}^{ab}e_{j}^{b}\ot\Delta_{ji},
\eeee
Reste \`a traduire la condition de compatibilit\'e avec la m\'etrique. Celle-ci recquiert que
\bbbb
&\langle e_{i}^{a},\nabla(e_{j}^{b})\rangle-
\langle\nabla(e_{i}^{a}),e_{j}^{b}\rangle=
d(\langle e_{i}^{a},e_{j}^{b}\rangle).&
\eeee
En utilisant les relations
\bbbb
\langle e_{i}^{a},\nabla(e_{j}^{b})\rangle&=&
-\mathop{\sum}\limits_{k}\langle e_{i}^{a},e_{j}^{b}\rangle\Delta_{jk}+
\mathop{\sum}\limits_{k,c}
A_{ik}^{ac}\langle e_{i}^{a},e_{k}^{c}\rangle\Delta_{kj} ,\n\\
\langle\nabla(e_{i}^{a}),e_{j}^{b}\rangle&=&
-\mathop{\sum}\limits_{k}\langle e_{i}^{a},e_{j}^{b}\rangle\lp\Delta_{jk}\rp^{*}+
\mathop{\sum}\limits_{k,c}
\ov{A}_{ik}^{ac}\langle e_{i}^{a},e_{k}^{c}\rangle\lp\Delta_{kj}\rp^{*},\n
\eeee
la condition de compatibilit\'e avec la m\'etrique se traduit simplement par $\ov{A}_{ij}^{ab}=A_{ji}^{bc}$.
\edemo

Nous pouvons maintenant achever notre construction du triplet spectral $\lp\tilde{\aa},\tilde{\hh},\tilde{\dd}\rp$. L'alg\`ebre des endomorphismes du module $\ee$ est la somme directe des alg\`ebres correspondant aux endomorphismes de chacun des espaces vectoriels $\ccc^{n_{i}}$. Par cons\'equent, on a
\bbbb
\tilde{\aa}=\mathop{\op}\limits_{i=1}^{N}M_{n_{i}}(\ccc).
\eeee
En utilisant la matrice de multiplicit\'e, on d\'ecompose l'espace de Hilbert 
\bbbb
&\hh=\mathop{\op}\limits_{ij}\hh_{ij}
\>\>\>\mathrm{avec}\quad\hh_{ij}=\ccc^{|m_{ij}|}&,
\eeee
ce qui nous permet d'obtenir
\bbbb
&\tilde{\hh}=\mathop{\op}\limits_{ij}\tilde{\hh}_{ij}
\>\>\>\mathrm{avec}\quad\tilde{\hh}_{ij}=\ee\ot_{\aa}\hh_{ij}\ot_{\aa}\ov{\ee}=
\ccc^{n_{i}}\ot\ccc^{|m_{ij}|}\ot\ccc^{n_{j}}.&
\eeee
Les op\'erateurs $\tilde{\jj}$ et $\tilde{\chi}$ sont donn\'es par
\bbbb
\tilde{\jj}(e_{i}^{a}\ot\psi_{ij}\ot\ov{e_{j}^{b}})&=&
e_{j}^{b}\ot\jj(\psi_{ij})\ot\ov{e_{i}^{a}},\\
\tilde{\chi}(e_{i}^{a}\ot\psi_{ij}\ot\ov{e_{j}^{b}})&=&
e_{i}^{a}\ot\chi(\psi_{ij})\ot\ov{e_{j}^{b}},
\eeee
o\`u $\psi_{ij}$ d\'esigne un vecteur g\'en\'erique de $\hh_{ij}$. Ces relations nous montrent que les deux triplets spectraux $\lp\aa,\hh,\dd\rp$ et $\lp\tilde{\aa},\tilde{\hh},\tilde{\dd}\rp$ ont la m\^eme matrice de multiplicit\'e.

\par

L'op\'erateur de Dirac $\tilde{\dd}$ s'exprime \`a l'aide de la connexion par 
\bbbb
&\tilde{\dd}(e_{i}^{a}\ot\psi_{ij}\ot\ov{e_{j}^{b}})=
\nabla(e_{i}^{a})\psi_{ij}\ot\ov{e_{j}^{b}}+
e_{i}^{a}\ot\dd(\psi_{ij})\ot\ov{e_{j}^{b}}+
e_{i}^{a}\ot\psi_{ij}\ov{(\nabla(e_{j}^{b})}&.
\eeee
D'apr\`es les r\'esultats pr\'ec\'edents, on a
\bbbb
\nabla(e_{i}^{a})\psi_{ij}&=&
-\mathop{\sum}\limits_{k}e_{i}^{a}\ot\Delta_{ik}\psi_{ij}+
\mathop{\sum}\limits_{k,c}A_{ik}^{ac}e_{k}^{c}\ot\Delta_{ki}\psi_{ij},\n\\
\psi_{ij}\ov{(\nabla(e_{j}^{b})}&=&
-\mathop{\sum}\limits_{k}\jj\Delta_{jk}\jj^{-1}\psi_{ij}\ot e_{j}^{b}+
\mathop{\sum}\limits_{k,c}A_{jk}^{bc}\jj\Delta_{kj}\jj^{-1}\psi_{ij}\ot e_{k}^{c}
.\n
\eeee
En utilisant la relation $\Delta_{ii}=0$ et le fait que $\jj^{-1}\psi_{ij}\in\hh_{ji}$, on montre que tous les termes apparaissant dans la d\'efinition de $\tilde{\dd}$ sont nuls \`a l'exception de ceux contenant les coefficients $A_{ik}^{ac}$ et $\ov{A}_{jk}^{bc}$. L'op\'erateur $\tilde{\dd}$ peut donc se mettre sous la forme
\bbbb
\tilde{\dd}(e_{i}^{a}\ot\psi_{ij}\ot\ov{e_{j}^{b}})&=&
\mathop{\sum}\limits_{k,c}A_{ik}^{ac}
\lp e_{k}^{c}\ot\Delta_{ki}\psi_{ij}\ot\ov{e_{j}^{b}}\rp\n\\ &+&
\mathop{\sum}\limits_{k,c}\ov{A}_{jk}^{bc}
\lp e_{i}^{a}\ot\jj\Delta_{kj}\jj^{-1}\psi_{ij}\ot\ov{e_{k}^{c}}\rp.
\eeee
Nous pouvons \'egalement \'ecrire l'op\'erateur de Dirac sous la forme $\tilde{\dd}=\tilde{\Delta}+\tilde{\jj}\tilde{\Delta}\tilde{\jj}^{-1}$, avec 
\bbbb
&\tilde{\Delta}(e_{i}^{a}\ot\psi_{ij}\ot\ov{e_{j}^{b}})=
\mathop{\sum}\limits_{k,c}A_{ik}^{ac}
\lp e_{k}^{c}\ot\Delta_{ki}\psi_{ij}\ot\ov{e_{j}^{b}}\rp
\eeee
Cependant, nous n'avons pas la solution la plus g\'en\'erale pour l'op\'erateur de Dirac, car nous avons vu au d\'ebut de ce chap\^ \i tre que celle contient une sommation sur un indice de multiplicit\'e qui est ici absent.

\par

Ainsi, nous avons r\'eussi \`a construire une large classe de triplets spectraux fini \`a l'aide de modules et de connexions sur des triplets commutatifs associ\'es. En d'autres termes, une grande majorit\'e des espaces discrets non commutatifs correspondent simplement \`a des th\'eories de jauge sur des espaces commutatifs. Ces th\'eories de jauge sont d\'ecrites par des modules projectifs, c'est-\`a-dire des fibr\'es vectoriels non commmutatifs, qui peuvent \^etre interpr\'et\'es comme suit. 

\par

La base de ce fibr\'e est un ensemble fini de N points, associ\'e \`a l'alg\`ebre des coordonn\'ees $\ccc^{N}$. Si nous rep\'erons chacun de ces points par un indice $i$, au dessus du point label\'e par $i$, se trouve une "fibre" form\'ee d'un ensemble fini de $n_{i}$  points, de telle mani\`ere que le module des sections du fibr\'e soit la somme directe des espaces vectoriels $\ccc^{n_{i}}$ attach\'es \`a chaque point. Une transformation de jauge est simplement un changement de base orthonormale dans chacun de ces espaces vectoriels; le groupe de jauge est donc le produit direct $U(n_{1})\times\dots\times U(n_{N})$. Enfin, la connexion va relier les fibres au dessus de chaque point en utilisant le calcul diff\'erentiel sur la base, tout en restant compatible avec la sym\'etrie de jauge.

\section{Distances associ\'ees aux triplets spectraux finis}

\subsection{G\'en\'eralit\'es}

La g\'eom\'etrie non commutative permet de g\'en\'eraliser la notion de distance g\'eod\'esique en rempla\c cant les points de l'espace par les \'etats purs de l'alg\`ebre des coordonn\'ees. Bien entendu, cette notion garde un sens dans le cas des triplets spectraux finis et permet de d\'efinir une distance sur l'ensemble des \'etats purs d'une somme directe d'alg\`ebres de matrices.

\par

Lorsque l'alg\`ebre est commutative, l'espace des \'etats purs est un ensemble fini sur lequel la distance est donn\'ee par la formule suivante.

\begin{pro}
Soit $(\aa,\hh,\dd)$ un triplet spectral fini avec l'alg\`ebre commutative  $\aa=\ccc^{N}$. Les \'etats purs de $\aa$ forment un ensemble fini de $N$ points sur lesquels la distance est donn\'ee par
\bbb
d_{ij}=\mathop{\sup}\limits_{(x_{1},\dots,x_{N})\in\ccc^{N}}
\la |x_{i}-x_{j}|\;\mathrm{tel}\;\mathrm{que}\;||\lb\Delta,x\rb||\leq 1\ra .
\eee
\end{pro}

\demo
Il suffit de remarquer que les \'etats purs de $\ccc^{N}$ sont donn\'es par les formes lin\'eaires $\phi_{i}$ d\'efnies par $\phi_{i}(x)=x_{i}$ pour tout $x=(x_{1},\cdots,x_{N})\in\ccc^{N}$. La construction g\'en\'erale de la distance reste alors valable.
\edemo

Bien entendu, cette distance peut \^etre infinie, comme c'est le cas si on prend $\Delta=0$. Il convient de noter que nous avons remplac\'e l'op\'erateur de Dirac $\dd$ par $\Delta$ car $\lb\jj\Delta\jj^{-1},\pi(x)\rb=0$ pour tout $x\in\aa$. Lorsque le nombre de points $N$ est fix\'e, la distance d\'epend uniquement de $\Delta$. Se pose alors la question de savoir si $\Delta$ \'etant donn\'e, il existe un triplet spectral fini dont l'op\'erateur de Dirac satisfait \`a $\dd=\Delta+\jj\Delta\jj^{-1}$.

\begin{pro}
Soit $\hh=\op_{i=1}^{N}\hh_{i}$ un espace de Hilbert de dimension finie et $\Delta$ un op\'erateur hermitien sur $\hh$ dont les \'el\'ements de matrice diagonaux $\Delta_{ii}$ sont nuls. Alors il existe un triplet spectral fini $(\aa,\hh,\dd)$ avec $\aa=\ccc^{N}$ et $\dd=\Delta+\jj\Delta\jj^{-1}$. 
\end{pro}

\demo
Remarquons que la condition $\Delta_{ii}=0$ est n\'ecessaire car nous avons vu que $\Delta_{ii}=\pi(1_{i})\Delta\pi(1_{i})=0$. 

\par

Pour construire le triplet spectral, nous allons utiliser la construction diagrammatique de l'op\'erateur $\dd$. Les sommets de ce diagramme sont les points du plan de coordonn\'ees $(i,j)$, o\`u $i$ et $j$ vont de 1 \`a N. Nous relions les points de coordonn\'es $(N,1)$ et $(N,j)$ si et seulement si $\Delta_{1j}$ est non nul. Nous affectons un signe $+$ au point $(N,1)$ et un signe $-$ aux points $(N,j)$ qui sont reli\'es \`a $(N,1)$ et d\'etruisons tous les autres sommets de cette ligne. Par sym\'etrie par rapport \`a la premi\`ere diagonale, nous construisons la colonne form\'ee des points de coordonn\'ees $(i,1)$ avec $i<N$. Nous poursuivons l'op\'eration pour la seconde ligne en affectant un signe $+$ au point $(N-1,2)$, un signe $-$ \`a tous les points $(N-1,j)$ pour $j>2$ qui sont tels que $\Delta_{2j}\neq 0$ et nous compl\'etons par sym\'etrie pour obtenir les \'el\'ements de la deuxi\`eme colonne $(i,N-1)$ tels que $i<N-2$. Ensuite nous  proc\'edons de la m\^eme mani\`ere jusqu'a la derni\`ere ligne. 

\par

A partir du diagramme ainsi construit, on reconstruit un triplet spectral fini convenant par les m\'ethodes que nous avons expos\'ees au cours de ce chapitre. Par construction, tous les axiomes sont satisfaits, \`a l'exception peut-\^etre de la dualit\'e de Poincar\'e. Cependant, celle-ci peut toujours \^etre obtenue en ajustant la dimension des espaces vectoriels apparaissant dans la d\'ecomposition de l'espace de Hilbert, quitte \`a compl\'eter l'op\'erateur de Dirac par des lignes et des colonnes nulles. 
\edemo

Avant de nous consacrer exclusivement au cas commutatif, \'etudions un exemple de distance bas\'ee sur une alg\`ebre non commutative. 

\exe
Consid\'erons l'alg\`ebre $\aa=M_{n}(\ccc)\op\ccc$ repr\'esent\'ee dans $\hh=\ccc^{n}\op\ccc$. L'op\'erateur $\Delta$ correspondant est 
\bbb
\Delta=\pp{0&m\cr m^{*}&0},
\eee
o\`u $m\in\ccc^{n}$ est un vecteur colonne. Les \'etats purs de $\aa=M_{n}(\ccc)\op\ccc$ sont donn\'es par la r\'eunion disjointe de $\ccc P^{n-1}$ et d'un point. D\'esignons par $(x,y)$ un \'el\'ement de $\aa$ et posons $z=x-yI_{n}$, o\`u $I_{n}$ est la matrice unit\'e $n\times n$. Avec ces notations, on a
\bbb
\lb\Delta,(x,y)\rb=\pp{0&-zm\cr m^{*}z&0},
\eee
Montrons que la condition $||\lb \Delta,(x,y)\rb||\leq 1$ est \'equivalente \`a
\bbb
\la 
\begin{array}{l}
\t\lp mm^{*}zz^{*}\rp\leq 1 \\
\t\lp mm^{*}z^{*}z\rp\leq 1.\label{g1}
\end{array}
\right.   
\eee
En effet, on a
\bbb
\pp{0&-zm\cr m^{*}z&0}\pp{0&zm^{*}\cr -mz&0}=\pp{zmm^{*}z^{*}&0\cr 0&m^{*}zz^{*}m},
\eee
ce qui prouve que $||\lb \Delta,(x,y)\rb||\leq 1$ si et seulement si $zmm^{*}z^{*}$ et $m^{*}zz^{*}m$. La derni\`ere de ces matrices \'etant simplement un nombre r\'eel, elle est de norme un si et seulement si $\t\lp mm^{*}z^{*}z\rp\leq 1$. Puisque $mm^{*}$ est une matrice de rang 1, il en est de m\^eme de $zmm^{*}z^{*}$, ce qui implique que tous ses d\'eterminants extraits d'ordre $\geq 2$ sont nuls. On en d\'eduit que son \'equation caract\'eristique est
\bbb
\lambda^{n}-\t(zmm^{*}z^{*})\lambda^{n-1}=0,
\eee
ce qui prouve que
\bbb
||zmm^{*}z^{*}||=\t(zmm^{*}z^{*})=\t(mm^{*}z^{*}z).
\eee

\par

Avant d'aborder cette \'etude, nous pouvons simplifier le probl\`eme de la mani\`ere suivante. Si nous multiplions l'op\'erateur $\Delta$ par une constante $\lambda>0$, il est clair que la distance est multipli\'ee par la constante $1/\lambda$. En appliquant cela, on peut toujours supposer que le vecteur $m$ est normalis\'e, quitte \`a multiplier la distance par $1/|m|$, avec $|m|=\sqrt{m^{*}m}$. Bien entendu, nous supposons que $m$ est non nul, car si tel \'etait le cas, il est clair que toutes les distances seraient infinies car (\ref{g1}) est v\'erifi\'ee pour toutes valeurs de $x$ et $y$. 
 
\par

De plus, en utilisant l'invariance de jauge de la distance (cf \S 1.2.2), nous pouvons supposer que $m=(1,0,\dots,0)$. En effet, nous avons montr\'e qu'il est \'equivalent de faire une transformation de jauge sur l'op\'erateur de Dirac ou de la faire sur les \'etats. En effectuant la transformation associ\'ee \`a l'unitaire $(u,1)$, o\`u $u\in M_{n}(\ccc)$ permet de transformer $m$ en $(1,0,\dots,0)$, on se ram\`ene au cas $m=(1,0,\dots,0)$. Dans ce cas, les contraintes donn\'ee par (\ref{g1}) se r\'eduisent \`a
\bbb
\mathop{\sum}\limits_{j=1}^{n}|z_{1j}|^{2}\quad\mathrm{et}\quad
\mathop{\sum}\limits_{i=1}^{n}|z_{i1}|^{2},\label{g2}
\eee
o\`u $z_{ij}$ d\'esigne l'\'el\'ement de la matrice $z$ situ\'e sur la ligne $i$ et la colonne $j$.

\par

Il est aussi utile d'introduire sur $\ccc^{N}$ le produit scalaire usuel
\bbb
\langle x,y\rangle=x^{*}y,
\eee
ce qui fait de $m$ un vecteur unitaire.

\par

Les distances entre \'etats purs de cette alg\`ebre sont de deux types. Nous pouvons soit calculer la distance entre deux \'etats purs de $M_{n}(\ccc)$, soit calculer la distance entre un \'etat pur de $M_{n}(\ccc)$ et l'unique \'etat pur de $\ccc$.

\par

Commen\c cons par ce dernier cas. Soit $\phi_{\xi}$ l'\'etat pur de $M_{n}(\ccc)$ d\'etermin\'e par le vecteur unitaire $\xi\in \ccc^{n}$. Son action sur $x\in M_{n}(\ccc)$ est $\phi_{\xi}=\t(\xi^{*}x\xi)$ et on a
\bbb
|\t(\xi^{*}x\xi)-y|=|\t(\xi^{*}z\xi)|,
\eee
compte tenu de $\t(\xi^{*}\xi)=1$. Nous devons donc calculer le maximum de $|\t(\xi^{*}z\xi)|=|\t(\xi\xi^{*}z)|$ avec les contraintes donn\'ees par(\ref{g2}).

\par

Puisque
\bbb
|\t(\xi^{*}z\xi)|=\mathop{\sum}\limits_{ij}\ov{\xi}_{i}z_{ij}\xi_{j},
\eee
il est clair que si $\xi_{i}\neq 0$ pour au moins un des indices $i\in\la 2,3,\dots,n\ra$, en prenant $z_{1i}=z_{i1}=0$ et $z_{ij}=L>0$ si $i$ et $j$ sont distincts de 1, les contraintes (\ref{g2}) sont toujours satisfaites. En faisant tendre $L$ vers l'infini, on obtient une distance infinie.

\par

Dans le cas contraire, on a
\bbb
\t(\xi^{*}z\xi)=\mathop{\sum}\limits_{i}\ov{\xi}_{i}z_{i1}\xi_{1}+
\mathop{\sum}\limits_{j}\ov{\xi}_{1}z_{1j}\xi_{j}.
\eee
En appliquant Cauchy-Schwarz \`a chacun des termes du second membre de cette \'equation, en d\'eduit que
\bbb
|\t(\xi^{*}z\xi)|\leq 1.
\eee
R\'eciproquement, en choisissant $z_{i1}=\xi_{i}$ et $z_{j1}=\ov{\xi}_{j}$ il est clair que les conditions (\ref{g2}) sont satisfaites et que 
\bbb
|\t(\xi^{*}z\xi)|= 1,
\eee
ce qui prouve que la distance est \'egale \`a 1.

\par

La distance entre deux \'etats purs de $M_{n}(\ccc)$ d\'etermin\'es par les vecteurs unitaires $\xi$ et $\zeta$ se calcule de mani\`ere identique. Soit $x\in M_{n}(\ccc)$, $y\in\ccc$ et $z=x-y I_{n}$. Commen\c cons par remarquer que
\bbb
|\t(\xi^{*}x\xi)-\t(\zeta^{*}x\zeta)|=|\t(z\xi\xi^{*})-\t(z\zeta\zeta^{*})|,
\eee
ce qui s'\'ecrit aussi
\bbb
|\t(\xi^{*}x\xi)-\t(\zeta^{*}x\zeta)|=
\mathop{\sum}\limits_{i,j}
z_{ij}(\ov{\xi}_{i}\ov{\xi}_{j}-\ov{\zeta}_{i}\zeta_{j}).
\eee

\par

Supposons dans un premier temps que $|\xi_{2}|^{2}-|\zeta_{2}|^{2}$ soit non nul. Dans ce cas il est clair que si on choisit $z_{12}=z_{21}=z_{11}=0$ et $z_{22}=L$, les conditions donn\'ees par (\ref{g2}) sont satisfaites et on a
\bbb
|\t(\xi^{*}x\xi)-\t(\zeta^{*}x\zeta)|=
\left| |\xi_{2}|^{2}-|\zeta_{2}|^{2} \right| L.
\eee
En faisant tendre $L$ vers l'infini, on trouve $d(\xi,\zeta)=+\infty$. 

\par

Si on a $|\xi_{2}|^{2}-|\zeta_{2}|^{2}=0$, alors 
\bbb
|\t(\xi^{*}x\xi)-\t(\zeta^{*}x\zeta)|=
+z_{21}(\ov{\xi}_{2}\xi_{1}-\ov{\zeta}_{2}\zeta_{1})
+z_{12}(\ov{\xi}_{1}\xi_{2}-\ov{\zeta}_{1}\zeta_{2}).
\eee
car on a aussi $|\xi_{1}|^{1}-|\zeta_{1}|^{1}=0$ puisque $|\xi_{1}|^{2}+|\xi_{2}|^{2}=1$ et $|\zeta_{1}|^{2}+|\zeta_{2}|^{2}=1$. Dans ce cas on a, compte-tenu de (\ref{g2}), 
\bbb
|\t(\xi^{*}z\xi)-\t(\zeta^{*}z\zeta)|^{2}
\leq 2\left| \ov{\xi}_{1}\xi_{2}-\ov{\zeta}_{1}\zeta_{2}\right|,
\eee 
qui est atteint en prenant
\bbb
z_{11}=0,\quad z_{12}=\ov{z_{21}}=e^{i\arg(\ov{\xi}_{1}\xi_{2}-\ov{\zeta}_{1}\zeta_{2})}.
\eee
Cette derni\`ere relation suppose que $\ov{\xi}_{1}\xi_{2}-\ov{\zeta}_{1}\zeta_{2}$ est non nul. Si tel est le cas, alors, compte tenu de $|\xi_{1}|^{2}=|\xi_{2}|^{2}$ et de  $|\zeta_{1}|^{2}=|\zeta_{2}|^{2}$, on a
\bbb
\begin{array}{|cc|}
\xi_{1}&\zeta_{1}\\
\xi_{2}&\zeta_{2}
\end{array}
=0,
\eee
ce qui prouve que les \'etats d\'efinis par les vecteurs $\xi$ et $\eta$ sont identiques. Par cons\'equent la distance les s\'eparant est nulle et on a 
\bbb
d(\xi,\zeta)=2\left| \ov{\xi}_{1}\ov{\xi}_{2}-\ov{\zeta}_{1}\zeta_{2}\right|
\eee 
dans les deux cas.

\par

Cette relation peut se r\'e\'ecrire sous une forme intrins\`eque de la mani\`ere suivante. Puisque  $|\xi_{1}|^{2}=|\xi_{2}|^{2}$ et $|\zeta_{1}|^{2}=|\zeta_{2}|^{2}$, on a
\bbb
\xi\xi^{*}-\zeta\zeta^{*}=\pp{0&\ov{\xi}_{2}\xi_{1}-\ov{\zeta}_{2}\zeta_{1}
\cr {\xi}_{1}\ov{\xi}_{2}-\ov{\zeta}_{1}\zeta_{2}&0},
\eee
ce qui montre que
\bbb
2\left| \ov{\xi}_{1}\xi_{2}-\ov{\zeta}_{1}\zeta_{2}\right|^{2}=
\t\lp \xi\xi^{*}-\zeta\zeta^{*}\rp^{2}.
\eee
En d\'eveloppant la trace, on obtient
\bbb
\left| \ov{\xi}_{1}\xi_{2}-\ov{\zeta}_{1}\zeta_{2}\right|^{2}=
2\lp 1-|\langle\xi,\zeta|^{2}\rp,
\eee
d'o\`u
\bbb
d(\xi,\zeta)=\sqrt{2-2|\langle\xi,\zeta\rangle|^{2}}
\eee

\par

Nous pouvons utiliser l'homog\'en\'e\"\i t\'e de la distance et l'invariance de jauge pour nous ramener au cas g\'en\'eral d'un vecteur $m$ quelconque. En conclusion, 
\bbb
\la 
\begin{array}{l}
d(1,\xi)=\frac{1}{|m|}
\quad\mathrm{si}\;\xi\;\mathrm{et}\;\mathrm{m}\;\mathrm{sont}\;\mathrm{proportionnels},\\
d(1,\xi)=+\infty\quad\mathrm{sinon},
\end{array}
\right.   
\eee
o\`u nous avons not\'e 1 l'unique \'etat pur de $\ccc$. La distance entre deux \'etats purs de $M_{n}(\ccc)$ est donn\'ee par
\bbb
\la 
\begin{array}{l}
d(\xi,\zeta)=\frac{1}{|m|}\sqrt{2-2|\langle\xi,\zeta|^{2}\rangle}
\quad\mathrm{si}\;|\langle m,\xi\rangle|=|\langle m,\zeta\rangle|,\\
d(\xi,\zeta)=+\infty\quad\mathrm{sinon},
\end{array}
\right.   
\eee
Terminons l'\'etude de cet exemple par une interpr\'etation g\'eom\'etrique de cette distance en dimension 2.

\par

Nous savons que l'espace des \'etats purs de $M_{2}(\ccc)$ est la droite projective complexe $P\ccc^{1}$ qui est isomorphe \`a la sph\`ere $S^{2}$ en tant que vari\'et\'e diff\'erentiable. Cet isomorphisme peut \^etre construuit \`a l'aide de la fibration de Hopf \cite{nakahara}. En effet, si $(z_{1},z_{2})$ est un vecteur unitaire de $\ccc^{2}$, d\'efnissons trois nombres r\'eels $\alpha,\beta,\gamma$ par
\bbbb
\alpha&=&2\Re\lp z_{1}\ov{z}_{2}\rp,\n\\
\beta&=&2\Im\lp z_{1}\ov{z}_{2}\rp,\n\\
\gamma&=&|z_{1}|^{2}-|z_{2}|^{2}.
\eeee
Il est ais\'e de v\'erifier que $\alpha^{2}+\beta^{2}+\gamma^{2}=1$, ce qui montre que les relations pr\'ec\'edentes d\'efinissent une application $\pi$ de $S^{3}$ (les vecteurs unitaires de $\ccc^{2}$) dans $S^{2}$. Cette application est surjective mais non injective car deux \'el\'ements unitaires de $\ccc^{2}$ ont m\^eme image par $\pi$ si et seulement si l'un se d\'eduit de l'autre par multiplication par une phase. Par cons\'equent ces deux vecteurs d\'eterminent le m\^eme \'etat pur et nous avons bien un isomorphisme entre l'espace des \'etats purs et la sphe\`re $S^{2}$.

\par

Notons $(\alpha,\beta,\gamma)$ et $(\delta,\epsilon,\eta)$ les points de la sph\`ere associ\'es \`a deux \'etats purs $\xi$ et $\zeta$. si nous choisissons $m=(1,0)$ pour simplifier nos notations, la distance pr\'ec\'edente d\'etermine un distance sur $S^{2}$ donn\'ee par
\bbb
\la 
\begin{array}{l}
d(\xi,\zeta)=\sqrt{(\alpha-\delta)^{2}+(\beta-\epsilon)^{2}}
\quad\mathrm{si}\quad\gamma=\eta,\\
d(\xi,\zeta)=+\infty\quad\mathrm{sinon},
\end{array}
\right.   
\eee
Cela signifie que notre distance n'est autre que la distance euclidenne r\'estreinte aux plan d'altidue constante, la distance entre des points situ\'es dans des plans d'altitudes diff\'erentes \'etant infinie.
\eexe

\par

Dans le cas que nous venons d'\'etudier, la distance est invariante de jauge en ce sens que si on prend un unitaire $u\in U(n)$, la distance entre les \'etats transform\'es de jauge $u\xi$ et $u\zeta$ est la m\^eme que celle entre $\xi$ et $\zeta$, puisque cette distance ne d\'epend que du produit scalaire entre $\xi$ et $\zeta$. Ceci d\'ecoule du fait que la contrainte $||\lb \Delta,(x,y)\rb||\leq 1$ ne fait pas intervenir le vecteur unitaire $\psi$ apparaissant dans la d\'ecomposition de l'op\'erateur de Dirac donn\'ee par $m=|m|\psi$. En effet, nous savons, par la propri\'et\'e de covariance de la distance, que la distance entre $u\xi$ et $u\zeta$ avec l'op\'erateur $\Delta$ est \'egale \`a la distance entre $\xi$ et $\zeta$ avec l'op\'erateur $u\Delta u^{-1}$. Or cette derni\`ere transformation sur l'op\'erateur de Dirac \'equivaut \`a changer $\psi$ en $u\psi$. \'Etant donn\'e que la contrainte ne d\'epend pas de $\psi$, les distances obtenues sont les m\^emes dans les deux cas.

\par

Dans le cas g\'en\'eral, cette propri\'et\'e n'est plus v\'erifi\'ee, car les contraintes obtenues avec un op\'erateur de Dirac et son transform\'e de jauge ne sont plus n\'ecessairement \'equivalentes. Cependant les distances sont toujours covariantes, et on peut d\'efinir une distance invariante de jauge sur les classes d'\'equivalence d'\'etats purs en utilisant la m\'ethode introduite au cours du chapitre pr\'ec\'edent. Deux \'etats sont \'equivalents si et seulement si ils peuvent \^etre d\'eduits l'un de l'autre par une transformation de jauge. Dans notre cas, deux \'etats purs sont \'equivalents si et seulement si ils correspondent au m\^eme facteur $M_{n_{i}}(\ccc)$ apparaissant dans la d\'ecomposition de $\aa$. Par cons\'equent, l'ensemble des classes d'\'equivalences est un ensemble fini \`a $N$ points, o\`u $N$ est le nombre de facteur simples apparaissant dans la d\'ecomposition de $\aa$ en alg\`ebres de matrices.

\subsection{Distances et chemins}

Nous allons \'etudier quelques propri\'et\'es de la distance associ\'ee \`a un triplet spectral fini, en essayant d'en donner une interpr\'etation g\'eom\'etrique \`a l'aide de diagrammes.

\par

Nous partons d'un entier $N$ et d'une matrice hermitienne $\Delta$ donn\'ee par ses $N^{2}$ blocs $\Delta_{ij}$. Nous supposons que $\Delta_{ii}=0$, ce qui est une condition  n\'ecessaire et suffisante pour identifier $\Delta$ avec l'op\'erateur apparaissant dans la d\'ecompostion de l'op\'erateur de Dirac d'un triplet spectral fini. Sur l'ensemble fini \`a $N$ \'el\'ements, nous d\'efinissons une distance par
\bbb
d_{ij}=\mathop{\sup}\limits_{(x_{1},\dots,x_{N})\in\ccc^{N}}
\la |x_{i}-x_{j}|\;\mathrm{tel}\;\mathrm{que}\;||\lb\Delta,x\rb||\leq 1\ra .
\eee
qui est l'analogue discret de la distance d\'efinie dans la section 2.1.3.

\par

Bien entendu, la condition $\Delta_{ii}=0$ ne joue aucun r\^ole dans la d\'efinition de $d_{ij}$, nous pouvons l'abandonner et \'etudier $d_{ij}$ ind\'ependement de tout triplet spectral fini.

\par

En vue de pouvoir interpr\`eter g\'eom\'etriquement cette distance, construisons pour chaque couple $(N,\Delta)$ un diagramme.

\begin{dfi}
Le diagramme associ\'e au couple $(N,\Delta)$, o\`u $\Delta$ est une matrice hermitienne form\'ee de $N^{2}$ blocs $\Delta_{ij}$ tels que $\Delta_{ii}=0$, est form\'e de $N$ sommets qui sont reli\'es si et seulement si l'\'el\'ement de 
matrice $\Delta_{ij}$ est non nul.
\end{dfi}

Ce diagramme permet de rep\'erer imm\'ediatement les \'el\'ements de matrice non nuls de $\Delta$. Par exemple, si $N=5$ et si $\Delta$ est donn\'e par
\bbb
\Delta=\pp{0&*&*&0&0\cr*&0&0&*&0\cr*&0&0&*&0\cr0&*&*&0&*\cr0&0&0&*&0\cr},
\eee
o\`u $*$ d\'esigne un \'el\'ement non nul, le diagramme associ\'e est

\begin{figure}[H]
\centering
\epsfig{file={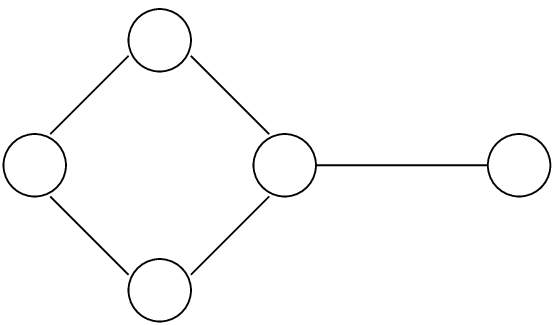},width=4.5cm}
\end{figure}

En g\'en\'eral, la matrice $\Delta$ ne contient pas n\'ecessairement beaucoup d'\'el\'ements de matrice nuls. Il est cependant parfois int\'eressant de pouvoir annuler certaines de ses lignes ou colonnes. Sur le plan diagrammatique, cela revient \`a isoler les sommets correspondants. Le r\'esultat suivant s'av\`ere utile.
 
\begin{pro}
Soit $\Delta'$ l'op\'erateur obtenu en annulant une ou plusieurs lignes de $\Delta$ ainsi que les colonnes correspondantes. Alors la distance calcul\'ee \`a partir de $\Delta'$ est sup\'erieure ou \'egale \`a celle calcul\'ee \`a partir de $\Delta$.
\end{pro}

\demo
Soit $e\in\ccc^{N}$ d\'efini par $e_{i}=0$ si la ligne et la colonne $i$ sont supprim\'es et $e_{i}=1$ sinon. $e$ est une projection commutant avec tout $x\in\aa$ et telle que $\Delta'=e\,\Delta\,e$. 
On en d\'eduit
\bbb
||\lb\Delta',x\rb||\leq ||e\lb\Delta,x\rb e||\leq ||\lb\Delta,x\rb||
\eee
pour tout $x\in\aa$. Par cons\'equent, la contrainte $||\lb\Delta,x\rb||\leq 1$ est plus forte que $||\lb\Delta',x\rb||\leq 1$, d'o\`u l'in\'egalit\'e annonc\'ee entre les distances.
\edemo

Nous allons utiliser ce r\'esultat pour donner une interpr\'etation g\'eom\'etrique de cette distance \`a l'aide de chemins sur le diagramme correspondant.

\begin{dfi}
Un chemin reliant les sommets i et j est un ensemble de n+1 sommets distincts $i=i_{0},i_{1},\dots,i_{n-1},i_{n}=j$ tels que $\Delta_{i_{k}i_{k+1}}\neq 0$ pour tout $k\in{0,1,\dots,n-1}$.
\end{dfi}

Cette notion de chemin nous permet de simplifier le calcul de la distance de la mani\`ere suivante.

\begin{pro}
La distance entre les points i et j ne d\'epend que des \'el\'ements de matrice situ\'es sur les chemins reliant i et j.
\end{pro}

\demo
Soit $\Delta'$ l'op\'erateur obtenu en enlevant toutes les lignes et les colonnes associ\'ees aux sommets qui ne sont pas situ\'es sur les chemin reliant $i$ et $j$ et $d'_{ij}$ la distance correspondante. La proposition pr\'ec\'edente montre que $d_{ij}\leq d'_{ij}$.

\par

Soit $I$ l'ensemble des sommets qui ne sont pas situ\'es sur les chemins reliant $i$ et $j$ et soit $x$ un \'el\'ement de $\ccc^{N}$ tel que, pour tout $k\in I$, on ait
\bbb
\la 
\begin{array}{l}
x_{k}=x_{i}\;\;
\mathrm{si}\;\Delta_{ki}\neq 0\\
x_{k}=x_{j}\;\;\mathrm{si}\;\Delta_{kj}\neq 0.
\end{array}
\right.   
\eee
Bien entendu, il n'est pas possible que $k$ soit reli\'e \`a $i$ et \`a $j$, car sinon il serait situ\'e sur un chemin reliant $i$ et $j$. Pour un tel $x$, il est facile de voir que $\lb\Delta,x\rb=\lb\Delta',x\rb$ car $\Delta_{kl}(x_{k}-x_{l})=0$ si ni $k$, ni $l$ ne sont reli\'es \`a $i$ ou $j$. On en d\'eduit que $d_{ij}'\leq d_{ij}$. D'o\`u l\'egalit\'e annonc\'ee.
 
\edemo

Cette derni\`ere propri\'et\'e de la distance permet de simplifier grandement beaucoup de calculs. Par exemple, si nous voulons calculer la distance associ\'ee \`a un diagramme du type suivant,

\begin{figure}[H]
\centering
\epsfig{file={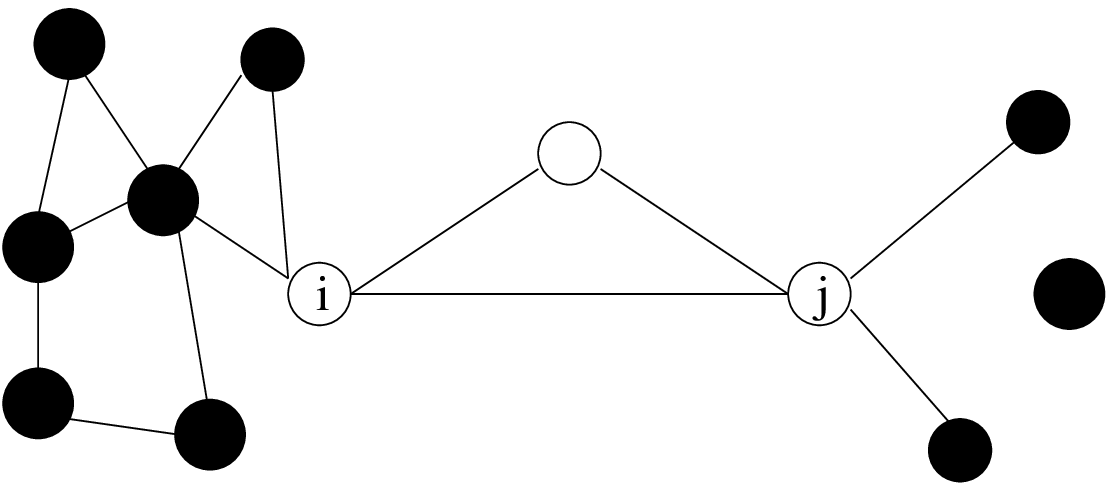},width=10cm}
\end{figure}
\noindent
nous pouvons \'eliminer tous les \'el\'ements de matrice de l'op\'erateur $\Delta$, en noir sur le dessin, qui correspondent \`a des ar\^etes situ\'es \`a gauche du sommet $i$ ou \`a droite du sommet $j$.

\par

Plus particuli\`erement, cette propri\'et\'e permet de ramener la d\'etermination des distances associ\'ees \`a un diagramme simplement connexe \`a un calcul effectu\'e dans le cadre d'un r\'eseau lin\'eaire. Un diagramme simplement connexe est  un diagramme tel que deux points quelconques peuvent \^etre reli\'es par un et un seul chemin. 

\par

En effet, la distance entre deux sommets $i$ et $j$ ne d\'epend que des \'el\'ements de matrice situ\'es sur les chemins reliant $i$ \`a $j$. Dans le cas d'un diagramme simplement connexe, il n'y a qu'un seul chemin joignant ces deux points, que nous d\'efinissons par ses sommets $i_{0}=i,i_{1},\dots,i_{n-1},i_{n}=j$. Comme nous pouvons \'eliminer tous les \'el\'ements de matrices qui n'appartiennent pas \`a ce chemin, nous pouvons remplacer $\Delta$ par la matrice tridiagonale suivante,
\bbb
\pp{0&\Delta_{i_{0}i_{1}}&0&\dots&\dots&0\cr
\Delta_{i_{1}i_{0}}&0&\Delta_{i_{1}i_{2}}&0&\dots&0\cr
0&\ddots&\ddots&\ddots&\ddots&\vdots\cr
\vdots&\ddots&\ddots&\ddots&\ddots&0\cr
0&\dots&0&\Delta_{i_{N-2}i_{N-3}}&0&\Delta_{i_{N-2}i_{N-1}}\cr
0&\dots&\dots&0&\Delta_{i_{N-1}i_{N-2}}&0},
\eee
qui correspond \`a une cha\^ \i ne lin\'eaire repr\'esent\'ee par le diagramme \`a N points suivant.
\begin{figure}[H]
\centering
\epsfig{file={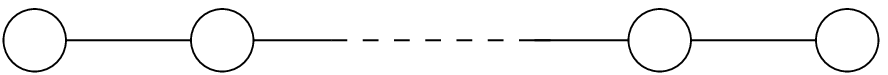},width=7cm}
\end{figure}

\par

Si on ne consid\`ere que 2 points, il est clair que la distance entre ces deux points est donn\'ee par 
\bbb
d_{12}=\frac{1}{||\Delta_{12}||},
\eee
o\`u $||\Delta_{12}||$ d\'esigne la norme de l'\'el\'ement de matrice $\Delta_{12}$.
Cela nous am\`ene \`a d\'efinir la longueur d'un chemin comme la somme des distances entre deux points cons\'ecutifs, consid\'er\'es comme des points isol\'es du diagramme. 

\begin{dfi}
La longueur d'un chemin $\gamma=i_{0},i_{1},\dots,i_{n_1},i_{n}$ reliant les points $i$ et $j$ est d\'efinie par
\bbb
L(\gamma)=\mathop{\sum}\limits_{k=1}^{n}\frac{1}{||\Delta_{i_{k-1}i_{k}}||}.
\eee
\end{dfi}

Dans le cas d'une vari\'et\'e, la distance g\'eod\'esique entre deux points est la longueur de la courbe la plus courte joignant ces deux points. Cela se transpose dans notre contexte de la mani\`ere suivante.

\begin{dfi}
La distance g\'eod\'esique $L_{ij}$ entre i et j est la longueur du plus court chemin reliant i et j.
\end{dfi}

Nous pouvons comparer la distance g\'eod\'esique avec la distance que nous avons d\'efinie au d\'ebut de ce chapitre.

\begin{pro}
La distance $d_{ij}$ est inf\'erieure ou \'egale \`a la distance g\'eod\'esique $L_{ij}$.
\end{pro}

\demo
Soit $\Delta'$ la matrice obtenue en supprimant toutes les lignes et le colonnes qui ne sont pas situ\'ees sur un chemin realisant le minimum de la distance g\'eod\'esique et $ d'_{ij}$ la distance correspondante. Alors, en appliquant un des r\'esultats pr\'ec\'edents, on obtient $d_{ij}\leq d'_{ij}$. Si $i=i_{0},i_{1},\dots,i_{n-1},i_{n}=j$ d\'esigne le chemin pr\'ec\'edent, on a, par l'in\'egalit\'e triangulaire,
\bbb
d'_{ij}\leq\mathop{\sum}\limits_{k=1}^{n}d'_{i_{k-1}i_{k}}.
\eee
Puisque la distance entre $i_{k-1}$ et $i_{k}$ est donn\'ee par $d_{i_{k-1}i_{k}}=\frac{1}{||\Delta_{i_{k-1}}i_{k}||}$, on obtient
\bbb
d_{ij}\leq\mathop{\sum}\limits_{k=1}^{n}\frac{1}{||\Delta_{i_{k-1}}i_{k}||},
\eee
ce qui prouve le r\'esultat annonc\'e.
\edemo

En g\'en\'eral, nous ne pouvons pas montrer l'\'egalit\'e entre la distance pr\'ec\'edente et la distance g\'eod\'esique. Nous verrons aux cours des sections suivantes deux exemples dans lesquel l'\'egalit\'e n'est pas v\'erifi\'ee.

\par

Lorsque deux points ne peuvent pas \^etre reli\'es par un chemin, il est clair que leur distance g\'eod\'esique est infinie. Pour montrer qu'il en est de m\^eme pour la distance que nous avons d\'efinie, nous avons besoin du r\'esultat suivant.

\begin{pro}
$x$ satisfait \`a $[\Delta,x]=0$ si et seulement si l'application $i\mapsto x_{i}$ est constante sur chaque composante connexe.
\end{pro}

\demo
La relation $[\Delta,x]=0$ \'equivaut \`a
\bbb
\Delta_{ij}(x_{i}-x_{j})=0.
\eee
Par cons\'equent, on doit avoir $x_{i}=x_{j}$ si et seulement si $i$ et $j$ peuvent \^etre reli\'es par un chemin.
\edemo

Cette relation peut s'interpr\'eter g\'eom\'etriquement de la mani\`ere suivante. En effet, si nous consid\'erons $d(x)=\lb\Delta,x\rb$ comme l'analogue discret de la diff\'erentielle, nous avons montr\'e que la diff\'erentielle d'une fonction est nulle si et seulement si cette fonction est constante sur chaque composante connexe.

\par 

Nous pouvons relier la finitude de la distance \`a la connexit\'e du diagramme.

\begin{pro}
La distance entre deux points est finie si et seulement si ils peuvent \^etre reli\'es par un chemin. 
\end{pro}

\demo
Il est \'evident que si $i$ et $j$ peuvent \^etre reli\'es par un chemin, alors la distance entre deux points est major\'ee par la distance g\'eod\'esique, ce qui entraine sa finitude.

\par

R\'eciproquement, supposons que $i$ et $j$ ne puissent pas \^etre reli\'es par un chemin. Alors, en prenant $x_{i}=t>0$, $x_{k}=x_{i}$ si $k$ est reli\'e \`a $i$ et $x_{k}=0$ sinon, on obtient $[\Delta,x]=0$ et $|x_{i}-x_{j}|=t$. En passant \`a la limite $t\rightarrow+\infty$, on obtient $d_{ij}=+\infty$.  
\edemo

Enfin, terminons par un r\'esultat qui s'av\`ere tr\`es utile lorsqu'on cherche \`a d\'eterminer explicitement une distance.

\begin{pro}
La distance $d_{ij}$ est donn\'ee par
\bbb
d_{ij}=\mathop{\sup}\limits_{(x_{1},\dots,x_{N})\in\ccc^{N}}
\la |x_{i}-x_{j}|\;\mathrm{tel}\;\mathrm{que}\;||\lb\Delta,x\rb||= 1\ra .
\eee
\end{pro}

\demo
Pour prouver cela, il suffit de remarquer que l'ensemble des $x\in\ccc^{N}$ tels que $[\Delta,x]\leq 1$ est un ensemble convexe et que les fonctions $x\mapsto|x_{i}-x_{j}|$ sont convexes. Une fonction convexe ne pouvant atteindre son maximum que sur le bord de son domaine de d\'efinition, il est clair que l'on peut remplacer la condition $[\Delta,x]\leq 1$ par $[\Delta,x]= 1$.
\edemo


\subsection{Un exemple simple}

Dans les deux sections suivantes, nous allons \'etudier en d\'etail deux exemples simples de distances.

\par

Le cas le plus simple correspond \`a l'espace \`a trois points, c'est \`a dire que nous prenons $N=3$, avec un op\'erateur $\Delta\in M_{3}(\rrr)$ donn\'e par
\bbb
\Delta=\pp{0&\Delta_{12}&\Delta_{13}\cr
\Delta_{12}&0&\Delta_{23}\cr
\Delta_{13}&\Delta_{23}&0},
\eee
Cela correspond au diagramme suivant.
\begin{figure}[H]
\centering
\epsfig{file={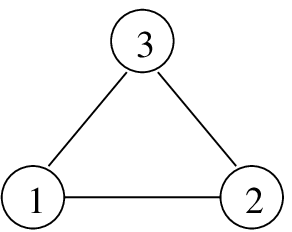},width=3cm,}
\end{figure}
La distance est d\'efinie par $d_{ij}=\mathop{\mathrm{sup}}\limits_{x\in\ccc^{3}}|x_{i}-x_{j}|$ o\`u $x=(x_{1},x_{2},x_{3})$ est tel que la norme de la matrice
\bbb
\lb \Delta,x\rb=\pp{0&\Delta_{12}(x_{2}-x_{1})&\Delta_{13}(x_{3}-x_{1})\cr
-\Delta_{12}(x_{2}-x_{1})&0&\Delta_{23}(x_{3}-x_{2})\cr
-\Delta_{13}(x_{3}-x_{1})&-\Delta_{23}(x_{3}-x_{2})&0}
\eee
soit 1.

\par

Gr\^ ace \`a une transformation unitaire, nous pouvons \'eliminer les phases des nombres complexes $x_{j}-x_{i}=e^{i\phi_{ij}}|x_{j}-x_{i}|$. En effet, pour toute matrice unitaire $U$, on a $||U\lb\Delta,x\rb U||=||\lb\Delta,x\rb||$. On choisit
\bbb
U=\pp{e^{i\alpha_{1}}&0&0\cr
0&e^{i\alpha_{2}}&0\cr
0&0&e^{i\alpha_{3}}}
\eee
telle que
\bbb
\alpha_{i}+\alpha_{j}=\phi_{ij},
\eee
ce qui est toujours possible car le syst\`eme suivant est non d\'eg\'en\'er\'e,
\bbb
\pp{0&1&1\cr 1&0&1\cr 1&1&0}\pp{\alpha_{1}\cr\alpha_{2}\cr\alpha_{3}}
=\pp{\phi_{23}\cr\phi_{13}\cr\phi_{12}}.
\eee

\par

\'Etant donn\'e que la matrice $U\lb\Delta,x\rb U^{*}$ est antisym\'etrique r\'eelle, sa norme est \'egale au module de sa plus grande valeur propre. Le polyn\^ome caract\'eristique de la matrice antisym\'etrique
\bbb
\pp{0&-r&q\cr r&0&-p\cr -q&p&0}
\eee
est 
\bbb
P(X)=X^{3}-(p^{2}+q^{2}+r^{2})X,
\eee 
ce qui nous m\`ene \`a
\bbb
||\lb\Delta,x\rb ||=
(\Delta_{12})^{2}|x_{1}-x_{2}|^{2}+(\Delta_{13})^{2}|x_{1}-x_{3}|^{2}+
(\Delta_{23})^{2}|x_{2}-x_{3}|^{2}.
\eee
Supposons que nous voulions calculer la distance entre les points 1 et 2. En posant $x=x_{1}-x_{2}$ et $y=x_{1}-x_{3}$, nous devons chercher la plus grande valeur de $|x|$ lorsque $x$ et $y$ satisfont \`a la contrainte
\bbb
(\Delta_{12})^{2}|x|^{2}+(\Delta_{13})^{2}|y|^{2}+(\Delta_{23})^{2}|x-y|^{2}=1.
\eee
Cela peut encore \^etre simplifi\'e en introduisant un r\'eel $\theta$ tel que $|x-y|^{2}=|x|^{2}+|y|^{2}+2|x||y|\cos\theta$, ce qui nous permet de supposer que $x$ et $y$ sont r\'eels, car il n'interviennent plus que par l'interm\'ediaire de leur modules. Finalement, nous devons chercher la plus grande valeur de $x$ lorsque
\bbb
(\Delta_{12})^{2}x^{2}+(\Delta_{13})^{2}y^{2}+(\Delta_{23})^{2}
\lp x^{2}+y^{2}+2xy\cos\theta\rp=1.
\eee
avec $x\geq0$, $y\geq0$ et $\theta\in[0,\pi]$.

\par

La contrainte donn\'ee par cette \'equation d\'efinit une ellipse $\ee_{\theta}$ dans le plan $xy$ pour chaque valeur de $\theta$. La distance entre les points 1 et 2 correspond au maximum de la coordonn\'ee $x$ sur l'ensemble des ellipses $\ee_{\theta}$, avec $\theta\in[0,\pi]$. Compte tenu de $\cos\theta\leq 1$, il n'est pas difficile de voir que toutes les ellipses $\ee_{\theta}$ sont situ\'ees \`a l'int\'erieur de l'aire limit\'ee par $\ee_{0}$. Le maximum de $x$ est donc atteint sur $\ee_{0}$, dont l'\'equation s'\'ecrit 
\bbb
(\Delta_{12})^{2}x^{2}+(\Delta_{13})^{2}y^{2}+(\Delta_{23})^{2}\lp x+y\rp^{2}=1.
\eee
La valeur maximale $d$ de $x$ lorsque $(x,y)\in\ee_{0}$, est \'egale \`a la valeur du param\`etre $h$ tel que la droite d'\'equation $x=h$ soit tangente \`a l'ellipse $\ee_{0}$.
\begin{figure}[H]
\centering
\epsfig{file={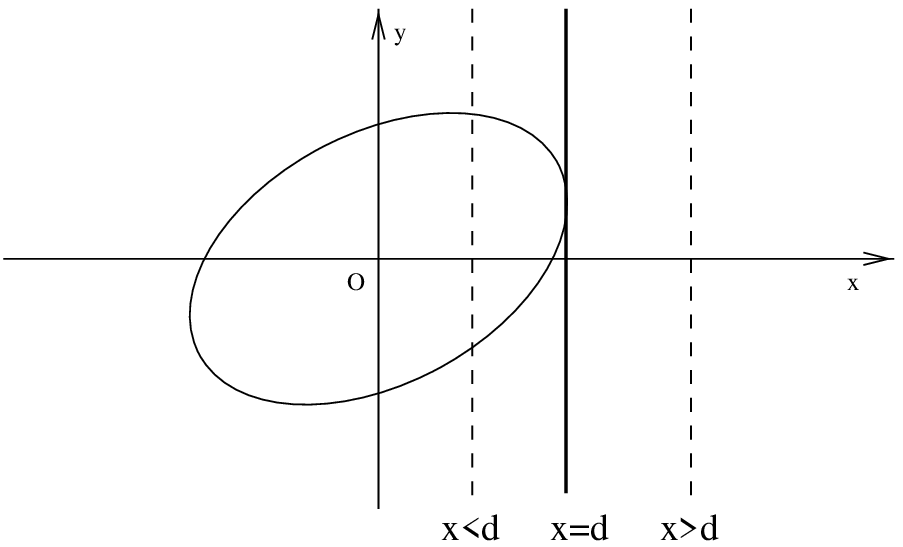},width=10cm,}
\end{figure}
Le changement $x=h$ dans l'\'equation de $\ee_{0}$ nous donne une \'equation du second degr\'e en $y$,
\bbb
\lp(\Delta_{13})^{2}+(\Delta_{23})^{2}\rp\,y^{2}+2h(\Delta_{23})^{2}\,y
+h^{2}\lp(\Delta_{12})^{2}+(\Delta_{23})^{2}\rp-1=0
\eee
La droite $x=h$ est tangente \`a l'ellipse pour les valeurs de $h$ telles que cette \'equation  admette une unique solution en $y$. Cela \'equivaut \`a annuler son discriminant r\'eduit,
\bbb
h^{2}\lb(\Delta_{23})^{4}-\lp(\Delta_{13})^{2}+(\Delta_{23})^{2}\rp
\lp(\Delta_{12})^{2}+(\Delta_{23})^{2}\rp\rb+(\Delta_{13})^{2}+(\Delta_{23})^{2}=0,
\eee 
ce qui nous donne
\bbb
d_{12}=\sqrt{\frac{(\Delta_{13})^{2}+(\Delta_{23})^{2}}
{(\Delta_{13})^{2}(\Delta_{23})^{2}+(\Delta_{23})^{2}(\Delta_{12})^{2}+
(\Delta_{12})^{2}(\Delta_{13})^{2}}}.\label{es1}
\eee
Par sym\'etrie, on obtient
\bbb
d_{13}=\sqrt{\frac{(\Delta_{12})^{2}+(\Delta_{23})^{2}}
{(\Delta_{13})^{2}(\Delta_{23})^{2}+(\Delta_{23})^{2}(\Delta_{12})^{2}+
(\Delta_{12})^{2}(\Delta_{13})^{2}}},
\eee
et
\bbb
d_{23}=\sqrt{\frac{(\Delta_{13})^{2}+(\Delta_{12})^{2}}
{(\Delta_{13})^{2}(\Delta_{23})^{2}+(\Delta_{23})^{2}(\Delta_{12})^{2}+
(\Delta_{12})^{2}(\Delta_{13})^{2}}}.
\eee

Il est int\'eressant d'\'etudier les cas limites obtenus en annulant certains des \'el\'ements de la matrice $\Delta$. Par exemple, si on a $\Delta_{13}=0$ avec $\Delta_{12}\neq0$ et $\Delta_{23}\neq0$ et, ce qui correspond au diagramme
\begin{figure}[H]
\centering
\epsfig{file={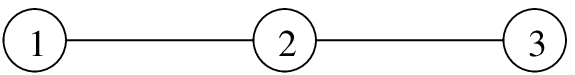},width=6cm,}
\end{figure}
on a
\bbb
d_{13}=\sqrt{\frac{1}{(\Delta_{12})^{2}}+\frac{1}{(\Delta_{23})^{2}}},\;\;
d_{12}=\frac{1}{|\Delta_{12}|},\;\;d_{23}=\frac{1}{|\Delta_{23}|}.
\eee
Si de plus on suppose que $\Delta_{13}=0$,  ce qui signifie que le point 1 n'est plus reli\'e \`a 2 et \`a 3, on obtient
\bbb
d_{12}=d_{13}=\infty,\;\;\;d_{23}=\frac{1}{|\Delta_{23}|},
\eee
ce qui correspond au diagramme suivant.
\begin{figure}[H]
\centering
\epsfig{file={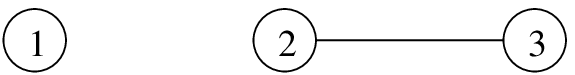},width=6cm,}
\end{figure}
Ces deux cas limites illustrent les propri\'etes g\'en\'erales des distances, relatives \`a la connexit\'e et \`a la simple connexit\'e du diagramme, que nous avons montr\'ees au cours du chap\^\i tre pr\'ec\'edent.

\par

Se pose le probl\`eme de la r\'eciproque: Etant donn\'e trois nombres $d_{12}$, $d_{13}$ et $d_{23}$ satisfaisant l'in\'egalit\'e triangulaire, existe-t-il un op\'erateur de Dirac du type pr\'ec\'edent permettant de reconstruire ces distances? En d'autres termes, est-ce que nous pouvons obtenir une distance quelconque sur un ensemble de trois points \`a l'aide de ce proc\'ed\'e? La r\'eponse est non, car il est facile de voir, en utilisant leurs expressions explicites en fonction de $\Delta$, que les distances satisfont toujours \`a l'in\'egalit\'e  
\bbb
d_{ij}^{2}\leq d_{ik}^{2}+d_{kj}^{2},\label{es2}
\eee
pour $i,j,k\in\la 1,2,3\ra$. Cette in\'egalit\'e est plus forte que l'in\'egalit\'e triangulaire, car
\bbb
d_{ik}^{2}+d_{kj}^{2}\leq\lp d_{ik}+d_{kj}\rp^{2}.
\eee
En particulier, cette in\'egalit\'e nous emp\^eche de mod\'eliser ainsi un r\'eseau unidimensionnel form\'e de trois points dont les distances mutuelles sont $d_{12}=d_{23}=d$ et $d_{13}=2d$, car on a toujours
\bbb
d_{13}\leq\sqrt{d_{12}^{2}+d_{23}^{2}}=\sqrt{2}\,d<2d,
\eee
ce qui explique, dans le cas particulier de trois points, les r\'esultats obtenus dans \cite{lizzi} et dans \cite{atzmon}. 

\par

Cela dit, si  on impose les in\'egalit\'es (\ref{es2}), nous allons voir qu'il est toujours possible de construire un op\'erateur de Dirac qui correspond \`a cette m\'etrique. En principe, nous avons \`a r\'esoudre un syst\`eme d'\'equations non lin\'eaires, car il faut inverser les formules donnant les distances en fonction des \'el\'ements de matrice de $\Delta$. Cependant, nous pouvons toujours r\'e\'ecrire (\ref{es1}) sous la forme
\bbb
\frac{1}{d_{12}^{2}}=(\Delta_{12})^{2}+
\frac{1}{\frac{1}{(\Delta_{13})^{2}}+\frac{1}{(\Delta_{23})^{2}}}.
\eee
En posant $R_{23}=1/|\Delta_{23}|^{2}$, $R_{13}=1/|\Delta_{13}|^{2}$ et $R_{12}=1/|\Delta_{12}|^{2}$, nous pouvons \'ecrire cette \'equation sous la forme
\bbb
\frac{1}{d_{12}^{2}}=\frac{1}{R_{12}}+
\frac{1}{R_{23}+R_{13}},
\eee
ce qui signifie que $d_{12}^{2}$ est la r\'esistance \'equivalente au montage en parall\`ele de la r\'esistance $R_{12}$ avec l'ensemble des r\'esistances $R_{13}$ et $R_{23}$ mont\'ees en s\'erie. Par permutations circulaires, nous obtenons des formules analogues pour  $1/d_{13}^{2}$ et  $1/d_{23}^{2}$.

\begin{figure}[H]
\centering
\epsfig{file={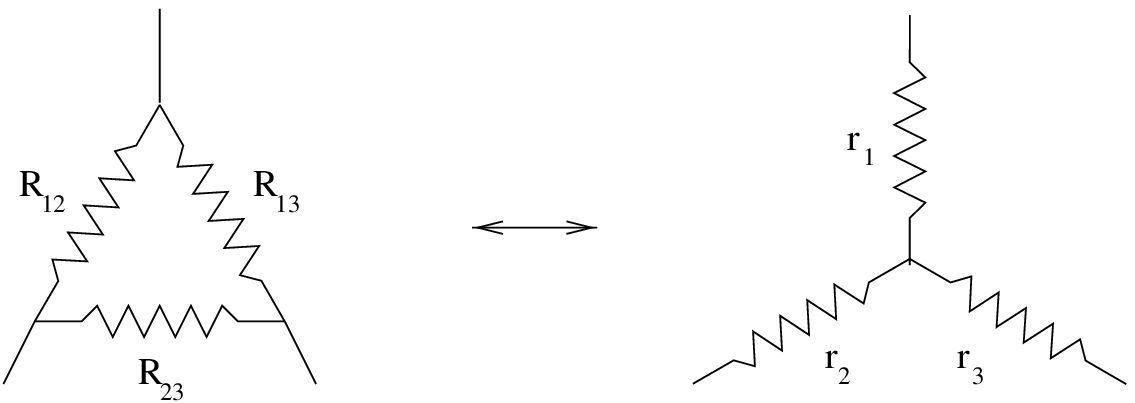},width=10cm,}
\end{figure}

Ainsi, passer des carr\'es des \'el\'ements de matrice de l'op\'erateur $\Delta$ aux inverses des carr\'es des distances consiste simplement \`a transformer le syst\`eme de r\'esistances "triangle" en un syt\`eme "\'etoile". Si nous posons $R_{ij}=1/|\Delta|_{ij}^{2}$, alors il ressort de l'\'etude pr\'ec\'edente que $d_{ij}^{2}=r_{i}+r_{j}$. Donnons nous trois nombres r\'eels strictement positifs $d_{12}$, $d_{13}$ et $d_{23}$ satisfaisant aux in\'egalit\'es (\ref{es2}). Nous construisons les r\'esistances "\'etoiles" comme solution du syst\`eme lin\'eaire
\bbb
\la 
\begin{array}{l}
r_{1}+r_{2}=d_{12}^{2}\\
r_{1}+r_{3}=d_{13}^{2}\\
r_{2}+r_{3}=d_{23}^{2}\\
\end{array}
\right. 
\eee
ce qui nous donne
\bbb
\la 
\begin{array}{l}
2r_{1}=d_{12}^{2}+d_{13}^{2}-d_{23}^{2}\\
2r_{2}=d_{12}^{2}+d_{23}^{2}-d_{13}^{2}\\
2r_{3}=d_{13}^{2}+d_{32}^{2}-d_{12}^{2}.\\
\end{array}
\right. 
\eee
Il est \`a noter que pour que ce syt\`eme d\'efinisse des r\'esistances il est capital que les in\'egalit\'es (\ref{es2}) soient satisfaites. Les formules de passage entre les syt\`emes  "triangle" et "\'etoile" sont classiques et peuvent \^etre trouv\'ees dans \cite{schaum}. Ainsi, connaissant les r\'esistances $r_{i}$ nous pouvons construire les r\'esistances $R_{ij}$ ainsi que les \'el\'ements de matrice de l'op\'erateur de Dirac, que nous choisirons positifs. Par exemple, nous avons
\bbb
R_{12}=\frac{r_{1}r_{2}+r_{1}r_{3}+r_{2}r_{3}}{r_{3}},
\eee 
d'o\`u il ressort que
\bbb
\Delta_{12}=\sqrt{\frac{2(d_{13}^{2}+d_{23}^{2}-d_{12}^{2})}
{2d_{12}^{2}d_{13}^{2}-d_{23}^{4}+
2d_{12}^{2}d_{23}^{2}-d_{13}^{4}+
2d_{13}^{2}d_{23}^{2}-d_{12}^{4}}}.
\eee
Par permutations circulaires, nous obtenons des formules analogues pour $\Delta_{13}$ et $\Delta_{23}$. Etant donn\'ee la correspondance que nous avons montr\'ee entre r\'esistances \'equivalentes et distances, il est clair qu'un tel op\'erateur de Dirac nous redonnera, apr\`es calcul par les m\'ethodes pr\'ec\'edemment expos\'ees, des distances \'egales aux nombres $d_{ij}$. Cela prouve que n'importe quel syst\`eme de distances sur un espace \`a trois points satisfaisant aux in\'egalit\'es (\ref{es2}) peut \^etre obtenu de la sorte.

\par

Bien que cette analogie entre distances et r\'esistances \'equivalentes soit s\'eduisante, elle n'est plus valable pour un syst\`eme de plus de trois points.

\subsection{Un exemple plus compliqu\'e}

Apr\`es avoir calcul\'e explicitement les distances sur un ensemble de trois points, il est naturel de chercher \`a \'etendre ces r\'esultats \`a des exemples plus compliqu\'es. En particulier, \`a partir de quatre points, nous rencontrons des difficult\'es nouvelles. 

\par

Consid\'erons donc un triplet spectral correspondant \`a un ensemble \`a quatre points, avec l'alg\`ebre commutative $\aa=\ccc^{4}$ repr\'esent\'ee de mani\`ere naturelle sur  $\hh=\ccc^{4}$. Nous choisissons un op\'erateur $\Delta\in M_{4}(\ccc)$, hermitien et dont tous les \'el\'ements situ\'es sur la diagonale sont nuls.

\par

Avec nos notations habituelles, la matrice $\lb\Delta,x\rb$ s'\'ecrit
\bbb
\lb\delta,x\rb=
\pp{0&\Delta_{12}(x_{2}-x_{1})&\Delta_{13}(x_{3}-x_{1})
&\Delta_{14}(x_{4}-x_{1})\cr
-\Delta_{12}^{*}(x_{2}-x_{1})&0&\Delta_{23}(x_{3}-x_{2})
&\Delta_{24}(x_{4}-x_{2})\cr
-\Delta_{13}^{*}(x_{3}-x_{1})&-\Delta_{23}^{*}(x_{3}-x_{2})&0
&\Delta_{34}(x_{4}-x_{3})\cr
-\Delta_{14}^{*}(x_{4}-x_{1})&-\Delta_{24}^{*}(x_{4}-x_{2})
&-\Delta_{34}^{*}(x_{4}-x_{3})&0}.
\eee
Dans le cas de trois points, nous avons traduit la condition $||\lb\Delta,x\rb||\leq 1$ \`a l'aide des valeurs propres de cette matrice, ce qui a n\'ecessit\'e l'\'elimination des phases pour nous ramener au cas plus simple d'une  matrice antisym\'etrique. Il est facile de voir que dans le cas de quatre points, une telle m\'ethode ne sera pas utilisable en toute g\'en\'eralit\'e, car si nous voulons absorber les phases par multiplication par des unitaires, nous ne diposons que de 4+4=8 param\`etres alors qu'il y a 6+6=12 phases qui apparaissent.

\par

Cependant, m\^eme si nous avions \'elimin\'e toutes ces phases, la contrainte $||\lb\Delta,x\rb||\leq 1$ est encore trop compliqu\'ee pour \^etre exploitable dans le cas g\'en\'eral. En effet, supposons que l'alg\`ebre soit r\'eelle, ainsi que la matrice $\Delta$, ce qui implique que $\lb\Delta,x\rb$  est une matrice antisym\'etrique. Le polyn\^ome caract\'eristique de la matrice antisym\'etrique
\bbb
\pp{0&E_{1}&E_{2}&E_{3}\cr
-E_{1}&0&-B_{3}&B_{2}\cr
-E_{2}&B_{3}&0&-B_{1}\cr
-E_{3}&-B_{2}&B_{1}&0},
\eee
est
\bbb
P(X)=X^{4}+\lp\vec{E}^{2}+\vec{B}^{2}\rp\, X^{2}+\lp\vec{E}\cdot\vec{B}\rp^{2}.
\eee 
Par cons\'equent, cette matrice est de norme plus petite que 1 si et seulement si
\bbb
\frac{\vec{E}^{2}+\vec{B}^{2}+
\sqrt{\lp\vec{E}^{2}+\vec{B}^{2}\rp^{2}-4\lp\vec{E}\cdot\vec{B}\rp^{2}}}{2}\leq 1,
\eee
ce qui \'equivaut \`a 
\bbb
\vec{E}^{2}+\vec{B}^{2}-\lp\vec{E}\cdot\vec{B}\rp^{2}\leq 1\quad\mathrm{et}\quad\vec{E}^{2}+\vec{B}^{2}\leq 2.\label{ep1}
\eee 
Lorsque nous appliquons cela \`a la matrice $\lb\Delta,x\rb$ nous voyons appara\^\i tre des termes quartiques en les variables $x_{i}$, ce qui rend la d\'etermination de la solution g\'en\'erale tr\`es difficile.

\par

Nous allons \'etudier le cas particulier du r\'eseau lin\'eaire \`a quatre points. Pour cela, nous revenons \`a l'alg\`ebre complexe $\aa=\ccc^{4}$ et nous choisissons un op\'erateur $\Delta$ tridiagonal,
\bbb
\Delta=\pp{0&\Delta_{12}&0&0\cr
\Delta_{12}^{*}&0&\Delta_{23}&0\cr
0&\Delta_{23}^{*}&0&\Delta_{34}\cr
0&0&\Delta_{34}^{*}&0},
\eee
ce qui correspond \`a un diagramme lin\'eaire du type

\begin{figure}[H]
\centering
\epsfig{file={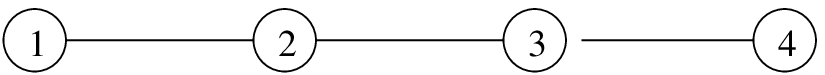},width=10cm,}
\end{figure}

\par

Introduisons les variables $X=x_{2}-x_{1}$, $Y=x_{3}-x_{2}$ et $Z=x_{3}-x_{4}$, ce qui nous permet d'\'ecrire $\lb\Delta,x\rb$ sous la forme
\bbb
\lb\Delta,x\rb=
\Delta=\pp{0&\Delta_{12}X&0&0\cr
-\Delta_{12}^{*}X&0&\Delta_{23}Y&0\cr
0&-\Delta_{23}^{*}Y&0&\Delta_{34}Z\cr
0&0&-\Delta_{34}^{*}Z&0}.
\eee
Nous s\'eparons les modules et les arguments de chacun des termes de cette matrice
\bbb
\lb\Delta,x\rb=
\pp{0&|\Delta_{12}||X|e^{i\theta_{12}}&0&0\cr
-|\Delta_{12}||X|e^{i\theta_{21}}&0&|\Delta_{23}||Y|e^{i\theta_{23}}&0\cr
0&-|\Delta_{23}||Y|e^{i\theta_{32}}&0&|\Delta_{34}||Z|e^{i\theta_{34}}\cr
0&0&-|\Delta_{34}||Z|e^{i\theta_{43}}&0}.
\eee
Il est facile de voir que l'on peut toujours \'eliminer ces phases par multiplication \`a droite et \`a gauche par des matrices unitaires et diagonales. La matrice $\lb\Delta,x\rb$ et la matrice antisym\'etrique r\'eelle
\bbb
\pp{0&|\Delta_{12}||X|&0&0\cr
-|\Delta_{12}||X|&0&|\Delta_{23}||Y|&0\cr
0&-|\Delta_{23}||Y|&0&|\Delta_{34}||Z|\cr
0&0&-|\Delta_{34}||Z|&0}
\eee
ont donc m\^eme norme et la condition $||\lb\Delta,x\rb||\leq1$ est \'equivalente \`a
\bbb
|\Delta_{12}|^{2}|X|^{2}+|\Delta_{23}|^{2}|Y|^{2}+|\Delta_{34}|^{2}|Z|^{2}-
|\Delta_{12}|^{2}|\Delta_{34}|^{2}|X|^{2}|Y|^{2}\leq 1\label{ep2}
\eee
compte tenu de (\ref{ep1}).

\par

Pour calculer la distance $d_{14}$, nous devons chercher le maximum de $|X+Y+Z|$ lorsque la contrainte (\ref{ep2}) est satisfaite. Puisque cette contrainte ne d\'epend que des modules des variables, il est clair que le maximum sera atteint lorsque ces variables sont des r\'eels positifs. Nous pouvons donc, sans perte de g\'en\'eralit\'e, supposer que $X$, $Y$ et $Z$ sont des r\'eels positifs. De plus, nous avons montr\'e que le maximum est atteint lorsqu'il y a \'egalit\'e, ce qui nous permet, en posant 
\bbb
\la 
\begin{array}{l}
x=X\\
y=Y\\
z=X+Y+Z\\
\end{array}
\right. 
\eee  
d'extraire $z$ sous la forme
\bbb
z(x,y)=x+y+
\frac{1}{|\Delta_{23}|}\sqrt{(1-\Delta_{12}^{2}x^{2})(1-\Delta_{34}^{2}y^{2})}.
\eee

\par

Puisque $d_{14}$ est le maximum de $X+Y+Z=z$, nous allons chercher les extrema de la fonction $z$. Ceux-ci sont atteints lorsque
\bbb
\la 
\begin{array}{l}
\frac{\partial z}{\partial x}=1-x\,\frac{\Delta_{12}^{2}}{|\Delta_{23}|}
\sqrt{\frac{1-\Delta_{34}^{2}y^{2}}{1-\Delta_{12}^{2}x^{2}}}=0\\
\\
\frac{\partial z}{\partial y}=1-y\,\frac{\Delta_{34}^{2}}{|\Delta_{23}|}
\sqrt{\frac{1-\Delta_{12}^{2}x^{2}}{1-\Delta_{34}^{2}y^{2}}}=0.
\end{array}
\right. 
\eee
Apr\`es des calculs assez longs mais tr\`es simples, on trouve que $z$ a un unique maximum, ce qui nous permet de donner la distance entre les points 1 et 4,
$$
d_{14}=\frac{|\Delta_{23}|}{|\Delta_{12}||\Delta_{34}|}
\lb
\sqrt{\frac{|\Delta_{23}|^{2}+|\Delta_{34}|^{2}}
{|\Delta_{23}|^{2}+|\Delta_{12}|^{2}}}
+
\sqrt{\frac{|\Delta_{23}|^{2}+|\Delta_{12}|^{2}}
{|\Delta_{23}|^{2}+|\Delta_{34}|^{2}}}\right.
$$
\bbb
+
\left.\frac{\left| |\Delta_{23}|^{4}-|\Delta_{12}|^{2}|\Delta_{34}|^{2}\right| }
{\sqrt{|\Delta_{23}|^{2}+|\Delta_{12}|^{2}}
\sqrt{|\Delta_{23}|^{2}+|\Delta_{34}|^{2}}}
\rb.
\eee

\par

Dans le cas particulier o\`u tous les  \'el\'ements de matrices sont \'egaux \`a un nombre r\'eel positif $1/L$, cette formule se simplifie consid\'erablement. Tenant compte des distances donn\'ees pour deux et trois points, on obtient
\bbb
d_{12}=L,\;\;d_{13}=\sqrt{2}\;L,\;\;d_{14}=2\,L.
\eee
En g\'en\'eral, pour une cha\^\i ne lin\'eaire de longueur quelconque dont tous les \'el\'ements de matrice sont des nombres r\'eels positifs \'egaux \`a $L$, la distance entre les points $1$ et $n$ est donn\'ee par \cite{atzmon}
\bbb
\la 
\begin{array}{lll}
d_{1n}=n\,L/2\;\;
&\mathrm{si}&\;n\;\mathrm{est}\;\mathrm{pair},\\
\\
d_{1n}=\sqrt{(n-1)(n+1)}\,L/2\;\;
&\mathrm{si}&\;n\;\mathrm{est}\;\mathrm{impair}.
\end{array}
\right.   
\eee
Dans tous les cas, cette distance  ne correspond pas \`a la distance naturelle $d_{ij}=|i-j|\,L$ que l'on peut d\'efinir sur un r\'eseau lin\'eaire. De ce fait, les triplets spectraux finis correspondant ne peuvent d\'ecrire un r\'eseau unidimensionnel. 

\par

Une solution \`a ce probl\`eme a \'et\'e apport\'ee dans \cite{dimakis}. Consid\'erons un triplet spectral fini avec l'alg\`ebre $\aa=\ccc^{N}$, repr\'esent\'ee sur $\hh=\ccc^{N}\op\ccc^{N}$ par
\bbb
x\in\ccc^{N}\mapsto\pp{x&0\cr0&x}\in M_{2N}(\ccc).
\eee  
L'op\'erateur de Dirac (ce que nous appelons $\Delta$, le v\'eritable op\'erateur de Dirac \'etant obtenu par sym\'etrisation par la conjugaison de charge) est
\bbb
\Delta=\pp{0&M\cr M^{*}&0},
\eee
o\`u la matrice $M\in M_{N}(\ccc)$ est d\'efinie par
\bbb
M=\pp{0&1/L&0&\dots&0\cr
      \vdots&\ddots&\ddots& &\vdots\cr 
      \vdots&&\ddots&&1/L\cr
        0&\dots&\dots&&0}.
\eee
Ce triplet spectral admet une chiralit\'e $\chi$ donn\'ee par
\bbb
\chi=\pp{1&0\cr 0&-1}
\eee
et il n'est pas difficile de montrer qu'il satisfait \`a tous les axiomes en utilisant la m\'ethode diagrammatique, comme nous l'avons d\'ej\`a fait au cours de ce chapitre.

\par

En utilisant ce triplet spectral, on montre \cite{dimakis} que la distance entre le point $i$ (correspondant \`a l'\'etat pur $\phi_{i}(x_{1,}\dots,x_{N})=x_{i})$ et le point $j$ (correspondant \`a l'\'etat pur $\phi_{j}(x_{1,}\dots,x_{N}=x_{j})$ est donn\'ee par $d_{ij}=|i-j|\, L$. Cette distance est la distance naturelle sur un r\'eseau lin\'eaire, et on peut \'etendre cette construction au cas d'un r\'eseau cyclique.

\par

Pour faire le lien avec les exemple que nous avons \'etudi\'es, il est commode d'effectuer un changement de base. D\'efinissons l'op\'erateur $U$ de $\hh=\ccc^{2}\ot\ccc$ dans lui-m\^eme par $U(x\ot y)=y\ot x$. Il est facile de voir que par conjugaison par $U$ la repr\'esentation de $x$ devient
\bbb
\pp{x_{1}\,I_{2}&0&0\cr
 0&\ddots&0\cr
0&0&x_{N}\, I_{2}}.
\eee
De m\^eme l'op\'erateur de Dirac s'\'ecrit, dans cette nouvelle base, sous la forme d'une matrice tridiagonale,
\bbb
\pp{0&N&0&\dots&0\cr
N^{*}&0&N&\ddots&\vdots\cr
0&\ddots&\ddots&\ddots&0\cr
\vdots&\ddots&N^{*}&0&N\cr
0&\dots&0&N^{*}&0},
\eee
avec
\bbb
N=\pp{0&1/L\cr 0&0}.
\eee
Ainsi, ce choix de l'op\'erateur de Dirac diff\`ere du notre en ce sens que les \'el\'ements de matrice de $\Delta$ sont des matrices $2\times 2$ alors que dans les exemples que nous avons \'etudi\'es, ces \'el\'ements de matrice \'etaient des nombres complexes.

\section{Alg\`ebres r\'eelles}

\subsection{La matrice de multiplicit\'e}

La m\'ethode que nous avons d\'evelopp\'ee au cours de ce chapitre pour \'etudier les triplets spectraux finis n'est appliquable que dans le cas d'une alg\`ebre complexe dont la repr\'esentation sur l'espace de Hilbert est lin\'eaire sur $\ccc$. Ceci est suffisant pour une \'etude de nature math\'ematique des g\'eom\'etries non commutatives finis, car aucune modification substancielle des r\'esultats que nous avons montr\'es n'est \`a pr\'evoir dans le cas r\'eel.

\par

Cependant, le triplet spectral fini correspondant au mod\`ele standard est construit avec une alg\`ebre r\'eelle. Pour pouvoir \'etudier les applications de la g\'eom\'etrie non commutative \`a la physique des particules, il est donc absolument n\'ecessaire d'\'etendre nos r\'esultats au cas des alg\`ebre r\'eelles.

\par

Lorsque l'alg\`ebre est r\'eelle, elle s'\'ecrit comme une somme directe d'alg\`ebres de matrices
\bbb
\aa=\mathop{\op}\limits_{i=1}^{N}M_{n_{i}}(\kkk),
\eee
dont les coefficients peuvent \^etre des r\'eels $(\kkk=\rrr)$, des complexes $(\kkk=\ccc)$ ou des quaternions $\kkk=\hhh$.

\par

Bien entendu, toute repr\'esentation de $\aa$ est semi-simple et ses repr\'esentations irr\'eductibles sont donn\'ees par les repr\'esentations irr\'eductibles des diff\'erents facteurs $M_{n_{i}}(\kkk)$.

\par

Les repr\'esentations irr\'eductibles de l'alg\`ebre $M_{n}(\kkk)$ sont 
\begin{itemize}
\item
la repr\'esentation fondamentale sur $\ccc^{n}$ lorsque $\kkk=\rrr$,
\item
la repr\'esentation fondamentale sur $\ccc^{n}$ et sa complexe conjugu\'ee lorsque $\kkk=\ccc$,
\item 
la repr\'esentation fondamentale sur $\ccc^{2n}$ lorsque $\kkk=\hhh$. 
\end{itemize}
Il convient de remarquer que si $\kkk=\rrr$, la repr\'esentation sur $\ccc^{n}$ n'est pas irr\'eductible car elle s'\'ecrit comme somme de deux repr\'esentations sur $\rrr^{n}$. Cependant, l'espace de Hilbert $\hh$ du triplet spectral  est toujours un espace vectoriel complexe, aussi nous n'effectuerons jamais cette d\'ecomposition et nous consid\'ererons cette repr\'esentation comme irr\'eductible.

\par

Dans le cas quaternionique, nous identifions les \'el\'ements $q$ de $\hhh$ avec les matrices $2\times 2$ complexes du type
\bbb
q=\pp{x&-\ov{y}\cr y&\ov{x}},\label{mm'1}
\eee
ce qui permet de d\'efinir la repr\'esentation fondamentale de $M_{n}(\kkk)$ sur $\ccc^{2n}$. De plus, la relation (\ref{mm'1}) entraine
\bbb
\ov{q}=u\,q\,u^{-1}
\eee
avec
\bbb
u=\pp{0&1\cr -1&0},
\eee
ce qui entraine que la repr\'esentation fondamentale de $M_{n}(\kkk)$ et sa complexe conjugu\'ee sont \'equivalentes.

\par

A partir de ces consid\'erations, nous pouvons d\'ecomposer toutes les repr\'esentations de $\aa$ sur $\hh$ qui apparaissent dans les sections 2.2.1-2.2.3 en repr\'esentations irr\'eductibles. Nous devons simplement remplacer les indices $i$, $j$, $\dots$  permettant de rep\'erer les facteurs simples par des indices qui d\'esignent les repr\'esentations irr\'eductibles de ces facteurs. Nous modifions donc la d\'efinition de la matrice multiplicit\'e comme suit.

\begin{dfi}
Les \'el\'ements de la matrice de multiplicit\'e d'un bimodule construit avec des alg\`ebres r\'eelles sont rep\'er\'es par les repr\'esentations irr\'eductibles de ces alg\`ebres. 
\end{dfi}

Avec cette d\'efinition de la matrice de multiplicit\'e, nos r\'esultats s'\'etendent au cas r\'eel.

\begin{pro}
Tous les r\'esulats des sections 2.2.1-2.2.3 restent valables dans le cas des alg\`ebres r\'eelles.
\end{pro}

Il est important de noter que cette extension n'est valide que pour les r\'esultats concernant la structure de bimodule et les op\'erateurs du premier ordre. La chiralit\'e et la dualit\'e de Poincar\'e subissent certaines modifications lors du passage aux alg\`ebres r\'eelles.


\subsection{La chiralit\'e}

La chiralit\'e est une involution hermitienne $\chi$ qui doit s'\'ecrire sous la forme
\bbb
\chi=\mathop{\sum}\limits_{p}\pi(x_{p})\jj\pi(y_{p})\jj^{-1}
\eee
avec $x_{p},y_{p}\in\aa$.
La d\'ecomposition par blocs de la chiralit\'e est toujours valable
\bbb
\chi=\mathop{\op}\limits_{ij}\chi_{ij}\, I_{n_{i}}\ot I_{m_{ij}}\ot I_{n_{j}}
\eee
avec $\chi_{ij}=\pm1$. Bien entendu, les indices $i$ et $j$ correspondent aux repr\'esentations irr\'eductibles de $\aa$ et $m_{ij}$ est la matrice de multiplicit\'e introduite dans la d\'efinition pr\'ec\'edente.

\par

Cependant, $\chi_{ij}$ et $\chi_{kl}$ ne sont pas ind\'ependants lorsque les repr\'esentations $k$ et $l$ sont \'egales ou conjugu\'ees \`a $i$ et $j$. En introduisant la matrice
\bbb
\mu_{ij}=\chi_{ij}\,m_{ij},
\eee
les contraintes impos\'ees \`a la chiralit\'e sont donn\'ees par la proposition suivante.

\begin{pro}
La matrice $\mu_{ij}$ satisfait \`a $\mu_{ij}\mu_{\ov{i}\ov{j}}\geq 0$.
\end{pro}

$\ov{i}$ d\'esigne la repr\'esentation complexe conjugu\'ee de celle associ\'ee \`a $i$. Si elle est r\'eelle ou quaternionique on a simplement $\ov{i}=i$. Ce r\'esultat se d\'emontre ais\'ement en \'ecrivant explicitement $\chi_{ij}$ en fonction de $x_{p}$ et $y_{p}$ et en utilisant $\ov{\chi}_{ij}=\chi_{ij}$.

\subsection{Dualit\'e de Poincar\'e}

La K-th\'eorie des $C^{*}$-alg\`ebres r\'eelles est un peu plus compliqu\'ee que dans le cas complexe \cite{schroeder}. Le th\'eor\`eme de periodicit\'e de Bott est toujours valide mais la periode est 8 au lieu de 2. On a donc $K_{p+8}(\aa)=K_{p}(\aa)$, o\`u $K_{p}(\aa)$ est d\'efini par it\'eration de la suspension (voir Appendice B). Bien entendu on a toujours
\bbb
K_{p}\lp\mathop{\op}\limits_{i=1}^{N}M_{n_{i}}(\kkk)\rp=
\mathop{\op}\limits_{i=1}^{N}K_{p}\lp M_{n_{i}}(\kkk)\rp
\eee
ainsi que $K_{p}(M_{n}(\kkk))=K_{p}(\kkk)$.

\par

On a $K_{0}(\kkk)=\zzz$ et les groupes d'ordre sup\'erieur contiennent de la torsion, i.e. un sous-groupe fini du type $\zzz_{2}$ dont nous ne tenons pas compte ici.  

\par

Dans ce cas, la dualit\'e de Poincar\'e ne fait intervenir que les projections hermitiennes $e\in\aa$ qui forment une base de $K_{0}(\aa)$. Par cons\'equent, la formulation de la dualit\'e de Poincar\'e est parfaitement similaire \`a celle que nous avons rencontr\'ee dans le cas complexe. Nous devons malg\'re tout noter deux diff\'erences importantes:
\begin{itemize}
\item
La seule projection non nulle $e\in\kkk$ est l'identit\'e, ce qui entraine $\t(e)=2$,
\item
La projection \'egale \`a l'identit\'e de $\ccc$ apparait dans la repr\'esentation fondamentale de $\M_{n}(\ccc)$ et dans sa complexe conjugu\'ee,
\end{itemize}  
En cons\'equence, la matrice de la forme d'intersection $\cap_{ij}$ n'est plus \'egale \`a $\mu_{ij}$.

\begin{pro}
La matrice $\cap_{ij}$ de la forme d'intersection s'obtient en multipliant les lignes et les colonnes de $\mu_{ij}$ associ\'ees aux quaternions par 2 et en additionnant les contributions d'une repr\'esentation et de sa complexe conjugu\'ee en une seule ligne ou colonne.
\end{pro}

Par cons\'equent, la dimension de $\cap_{ij}$ est donn\'ee par le nombre de facteurs simples de la d\'ecomposition de $\aa$ alors que la taille de $\mu_{ij}$ est donn\'ee par le nombre de repr\'esentations irr\'eductibles de $\aa$.

\chapter[Construction de mod\`eles en physique\dots]{Construction de mod\`eles en physique des particules}

\section{Le mod\`ele d'espace-temps}

\subsection{G\'en\'eralit\'es}

Depuis que les th\'eories de jauge sont devenues la pierre angulaire de la physique des particules, la construction de mod\`eles fait de plus en plus appel \`a des notions de g\'eom\'etrie et de topologie. En effet, la notion de sym\'etrie de jauge est d'essence purement g\'eom\'etrique et sa formulation la plus \'el\'egante repose sur la g\'eom\'etrie diff\'erentielle. De plus, cette approche g\'eom\'etrique permet de mieux comprendre les similarit\'es entre les th\'eories de Yang-Mills, comme l'\'electromagn\'etisme o\`u la chromodynamique, et la gravitation: toutes deux sont des th\'eories de jauge.

\par

Sur le plan math\'ematique, la g\'eom\'etrie non commutative permet, gr\^ace \`a un formalisme alg\'ebrique, de g\'en\'eraliser la plupart des notions de base de la g\'eom\'etrie diff\'erentielle en s'affranchissant de la commutativit\'e de l'alg\`ebre des coordonn\'ees. Il est donc naturel d'utiliser cette nouvelle approche de la g\'eom\'etrie diff\'erentielle en physique.

\par  
 
Puisque le mod\`ele standard, confirm\'e \`a l'heure actuelle par la plupart des exp\'eriences, est une th\'eorie de jauge avec brisure spontan\'ee de sym\'etrie, nous essayerons de comprendre comment on peut construire de tels mod\`eles, ainsi que leur couplage \`a la gravitation, \`a l'aide de la g\'eom\'etrie non commutative. Nous essayerons de d\'elimiter la classe de mod\`eles qu'il est possible d'obtenir et nous mettrons l'accent sur les contraintes que cela engendre.

\par

Pour cela, nous devons choisir un mod\`ele d'espace-temps, \`a partir duquel sera construite la th\'eorie. Pour d\'eterminer l'alg\`ebre $\aa$ \`a partir de laquelle on construit les triplets spectraux que nous utiliserons, il est utile de partir des sym\'etries que doit poss\'eder la th\'eorie.  Les th\'eories de jauge coupl\'ees \`a la gravitation sont invariantes sous deux types de transformations, qui doivent \^etre des automorphismes de l'alg\`ebre des coordonn\'ees. D'une part, la th\'eorie est invariante sous l'action des diff\'eomorphismes de l'espace-temps qui forment le groupe des automorphismes de l'alg\`ebre $C^{\infty}(\mm)$. D'autre part les particules \'el\'ementaires sont situ\'ees dans certaines repr\'esentations de groupes de Lie compacts et semi-simples. Ces derniers correspondent naturellement aux automorphismes d'une somme directe d'alg\`ebre de matrices. Pour obtenir simultan\'ement ces deux types de sym\'etries, il est n\'ecessaire de combiner l'alg\`ebre $C^{\infty}(\mm)$ avec une alg\`ebre de dimension finie $\aa_{F}$. Le choix le plus simple consiste \`a prendre pour $\aa$ un produit tensoriel $C^{\infty}(\mm)\ot\aa_{F}$, dont les automorphismes comprennent les diff\'eomorphismes de $\mm$ et les transformations de jauge locales qui sont donn\'ees par les \'el\'ements unitaires de $\aa$.

\par

Le mod\`ele retenu combine la g\'eom\'etrie de l'espace-temps ordinaire avec celle d'un espace discret, qui d\'ecrit les degr\'es de libert\'e internes de la th\'eorie. Plus pr\'ecis\'ement nous allons partir d'un triplet spectral qui est le produit du triplet spectral de la g\'eom\'etrie ordinaire par un triplet spectral fini. 

\par

Bien entendu, il est possible de consid\'erer une situation plus g\'en\'erale dans laquelle nous rempla\c cons ce produit de l'espace-temps par un espace discret, qui d\'ecrit un fibr\'e trivial, par un espace fibr\'e dont la topologie est non triviale; la base de ce fibr\'e est l'espace-temps alors que la fibre d\'ecrit les sym\'etries internes.
A partir d'un tel objet, il est possible de construire un triplet spectral qui g\'en\'eralise la notion de produit d'un triplet spectral par un triplet spectral fini  (cf \S \, 2.2.4).

\subsection{Produit tensoriel de triplets spectraux}

Commen\c cons par donner la d\'efinition d'un produit tensoriel de triplets spectraux.

\begin{pro}
Soit $(\aa_{1},\hh_{1},\dd_{1})$ et $(\aa_{2},\hh_{2},\dd_{2})$ deux triplets spectraux de dimension $n_{1}$ et $n_{2}$, o\`u $(\aa_{1},\hh_{1}\dd_{1})$ est suppos\'e pair. Alors $(\aa,\hh,\dd)$ d\'efini par
\bbbb
&\aa=\aa_{1}\ot\aa_{2},\qquad \hh=\hh_{1}\ot\hh_{2},&\n\\
&\dd=\dd_{1}\ot id_{2}+\gamma_{1}\ot\dd_{2},&
\eeee
est un triplet spectral de dimension $n=n_{1}+n_{2}$ appel\'e produit de $(\aa_{1},\hh_{1},\dd_{1})$ par $(\aa_{2},\hh_{2},\dd_{2})$.
\end{pro}

Bien entendu, la repr\'esentation $\pi$ de $\aa$ sur $\hh$ est le produit tensoriel $\pi_{1}\ot\pi_{2}$, de m\^eme que l'on a $\jj=\jj_{1}\ot\jj_{2}$.

\par

Pour que ce produit puisse \^etre d\'efini, il est n\'ecessaire que l'un au moins des triplets spectraux soit pair. Cela nous permet de d\'efinir l'op\'erateur de Dirac du produit comme \'etant $\dd=\dd_{1}\ot id_{2}+\gamma_{1}\ot\dd_{2}$, ce qui nous assure que nous obtenons  l'op\'erateur de Dirac usuel dans le cas du tore plat $T^{2+n}=T^{2}\times T^{n}$. 

\par

De plus, pous avons
\bbb
\dd^{2}=(\dd_{1})^{2}\ot id_{2}+id_{1}\ot(\dd)^{2},
\eee
ce qui garantit la validit\'e de la formule d'addition des dimensions: la dimension du produit est la somme des dimensions.

\begin{pro}
Lorsque les deux triplets spectraux sont pairs, leur produit est \'egalement pair.
\end{pro}

La chiralit\'e est d\'efinie comme \'etant le produit tensoriel des chiralit\'es, $\chi=\chi_{1}\ot\chi_{2}$. Il est clair que c'est une involution hermitienne satisfaisant \`a toutes les relations de commutation impos\'ees \`a la chiralit\'e. 

\par

Pour montrer qu'elle satisfait \`a l'axiome d'orientabilit\'e, il faut trouver un cycle de Hochschild $c$ de  dimension $n$ tel que $\pi(c)=\chi$, sachant qu'il existe deux cycles de Hochschild $c_{1}$ et $c_{2}$ de dimensions $n_{1}$ et $n_{2}$ tels que $\pi_{1}(c_{1})=\chi_{1}$ et $\pi_{2}(c_{2})=\chi_{2}$. Pour cela, introduisons la notion suivante.

\begin{dfi}
Soit $\sigma$ une permutation des $p+q$ indices $i_{1},\dots,i_{p},i_{p+1},\dots,i_{p+q}$. $\sigma$ est un $(p,q)$-battement si on a
\bbb
\sigma(i_{1})<\sigma(i_{2})<\dots<\sigma(i_{p})
\eee
et
\bbb
\sigma(i_{p+1})<\sigma(i_{2})<\dots<\sigma(i_{p+q})
\eee
pour tous les indices $i_{1},\dots,i_{p+q}$. Nous notons $S_{p,q}$ l'ensemble des $(p,q)$-battements.
\end{dfi}

L'introduction de cette notion se justifie par le r\'esultat suivant \cite{loday}.

\par

\begin{pro}
Soient $\aa$ et $\bb$ deux alg\`ebres, $a=a_{0}\ot a_{1}\dots\dots\ot a_{p}\in C^{p}(\aa)$ et  
$b=b_{0}\ot b_{p+1}\ot\dots\ot b_{p+q}\in C^{q}(\bb)$ deux cycles de Hochschild. Notons $c_{0}=a_{0}\ot b_{0}$, $c_{i}=a_{i}\ot 1$ si $i\in\la 1,\dots,p\ra$ et $c_{i}=1\ot b_{i}$ si $i\in\la p+1,\dots,p+q\ra$. Alors $sh(a,b)\in C^{p+q}(\aa\ot\bb)$ d\'efini par
\bbb
sh(a,b)=\mathop{\sum}\limits_{\sigma\in S_{p,q}}\epsilon(\sigma)
c_{0}\ot c_{\sigma(1)}\ot\dots\ot c_{\sigma(p+q)}
\eee
est un cycle de Hochschild, appel\'e "shuffle product" de $a$ par $b$.
\end{pro}

\par

Appliquons cette construction pour d\'efinir $c$ \`a partir des cycles $c_{1}$ et $c_{2}$. Pour cela, commen\c cons par remarquer que le r\'esultat pr\'ec\'edent n'est pas modifi\'e si les premiers termes $a_{0}$ et $b_{0}$ appartiennent aux alg\`ebres $\aa\ot\aa^{op}$ et $\bb\ot\bb^{op}$, puisque les facteurs $\aa^{op}$ et $\bb^{op}$ ne sont pas concern\'es par la multiplication par les \'el\'ements de $\aa$ et de $\bb$.

\par

D\'efinissons $c$ par
\bbb
c=\frac{(n_{1}+n_{2})!}{n_{1}!n_{2}!}sh(c_{1},c_{2}).
\eee 
Gr\^ace \`a la proposition pr\'ec\'edente, $c$ est un cycle de Hochschild. En utilisant les relations
\bbbb
\lb\dd,\pi(a_{1}\ot 1)\rb&=&\lb\dd_{1},\pi_{1}(a_{1})\rb\ot id_{1},\\
\lb\dd,\pi(1\ot a_{2})\rb&=&\chi_{1}\ot\lb\dd_{2},\pi_{2}(a_{2})\rb,
\eeee
pour tout $a_{i}\in\aa_{i}$, il est facile de v\'erifier que $\pi(c)=\chi_{1}\ot\chi_{2}$, ce qui justifie la d\'efinition de $\chi$ comme le produit tensoriel $\chi_{1}\ot\chi_{2}$.

\par

De m\^eme que la formule de K\" unneth permet d'exprimer l'alg\`ebre diff\'erentielle d'un produit de vari\'et\'es \`a l'aide des formes diff\'erentielles sur chaque facteur, le calcul diff\'erentiel associ\'e \`a un produit de triplets spectraux a \'et\'e d\'etermin\'e \`a l'aide des calculs diff\'erentiels associ\'es \`a chaque facteur \cite{tes}. Nous n'aurons besoin que du r\'esultat relatif aux 1-formes, qui ne n\'ecessite pas l'introduction des champs auxiliaires.

\begin{pro}
L'espace des 1-formes $\Omega_{\dd}^{1}(\aa)$ associ\'ees au produit s'exprime en fonction de 0-formes et des 1-formes de chaque facteur par
\bbb
\Omega_{\dd}^{1}(\aa)=
\Omega^{1}_{\dd_{1}}(\aa_{1})\ot\Omega^{0}_{\dd_{2}}(\aa_{2})
+\gamma_{1}\Omega_{\dd_{1}}^{0}(\aa_{1})\ot\Omega_{\dd_{2}}^{1}(\aa_{2}).
\eee
\end{pro}

\par

Consid\'erons le produit tensoriel du triplet spectral relatif \`a une vari\'et\'e compacte de dimension paire $n$  munie d'une strcture spinorielle, par un triplet spectral fini $(\aa_{F},\hh_{F},\dd_{F})$. Il en r\'esulte le triplet spectral suivant,
\bbbb
&\aa = C^{\infty}(\mm)\ot\aa_{F},\qquad \hh = L^{2}(\mm,\ss)\ot\hh_{F},&\\
&\dd = i\gamma^{\mu}(\partial_{\mu}+\omega_{\mu})\ot Id_{F}
+\gamma^{n+1}\ot \dd_{F},& 
\eeee  
o\`u $\gamma^{\mu}$ d\'esigne les matrices de Dirac en espace courbe et $\omega_{\mu}$ la connexion spinorielle.

\par

Il convient de remarquer que ce produit fait appel \`a la matrice $\gamma^{n+1}$ et ne peut pas \^etre d\'efini lorsque la dimension $n$ est impaire. Cependant, un produit de triplets spectraux peut \^etre d\'efini si un au moins des deux triplets spectraux est pair. Etant donn\'e qu'un triplet spectral fini est pair, il est tout aussi l\'egitime d'utiliser la chiralit\'e $\chi_{F}$ associ\'e au triplet spectral $(\aa_{F},\hh_{F},\dd_{F})$. Nous d\'efinissons un nouveau triplet spectral $\lp\aa',\hh',\dd'\rp$ par
\bbbb
&\aa' = \aa_{F}\ot C^{\infty}(\mm),\qquad\hh' = \hh_{F}\ot L^{2}(\mm,\ss),&\\
&\dd' = \dd_{F}\ot I+\chi_{F}\ot\gamma^{\mu}\lp\partial_{\mu}+\omega_{\mu}\rp.&
\eeee  
D\'esignons par $K:\,\hh_{F}\ot L^{2}(\mm,\ss)\rightarrow L^{2}(\mm,\ss)\ot\hh_{F}$ l'isomorphisme d\'efini par 
\bbb
K(\xi\ot\Psi)=\Psi\ot\xi
\eee
pour tous $\xi\in\hh_{F}$ et $\Psi\in\ss$ et notons $U$ la transformation unitaire de $\hh_{F}\ot L^{2}(\mm,\ss)$ dans $L^{2}(\mm,\ss)\ot\hh_{F}$ donn\'ee par
\bbb
U=\lp\frac{1+\chi}{2}\ot I+\frac{\chi-1}{2}\ot\gamma^{n+1}\rp K.
\eee
Il est facile de voir que $U$ est unitaire et que
\bbb
U\,\dd'\, U^{-1}=\dd.
\eee
Puisque l'action ne d\'epend que du spectre de l'op\'erateur de Dirac, il est facile de voir que $\dd$ et $\dd'$ nous donnent la m\^eme action. Etant donn\'e que la d\'efinition de $\dd'$ ne n\'ecessite pas d'hypoth\`ese sur la parit\'e de $n$, il serait plus logique d'utiliser $\dd'$ \`a la place de $\dd$. Cependant, nous sommes principalement int\'eress\'es par le cas $n=4$ et nous nous conformerons \`a l'usage en vigueur en prenant $\dd$.  

\par

Comme d'habitude, l'invariance de l'action fermionique $\langle\Psi,\dd\Psi\rangle$ sous les transformations de jauge $\Psi\mapsto u\jj u\jj^{-1}\Psi$ pour tout $\Psi$ et tout unitaire $u\in\aa$ nous am\`ene \`a remplacer l'op\'erateur de Dirac $\dd$ par l'op\'erateur de Dirac covariant $\dd_{A}$,
\bbb
\dd_{A}=\dd+A+\jj A\jj^{-1}.
\eee
A est une 1-forme hermitienne, et se transforme, sous une transformation de jauge en
\bbb
\pi(u)\,A\,\pi(u)^{-1}+\pi(u)\lb\dd,\pi(u^{-1})\rb,
\eee
de sorte que l'op\'erateur de Dirac covariant se transforme en
\bbb
\dd_{A}\mapsto \pi(u)\jj \pi(u)\jj^{-1}\,\dd_{A}\,\pi(u^{-1})\jj \pi(u^{-1})\jj^{-1},
\eee
ce qui assure l'invariance de jauge de l'action fermionique $\langle\Psi,\dd_{A}\Psi\rangle$. Pour simplifier nos notations, nous omettrons le symbole $\pi$ de la repr\'esentation, tout \'el\'ement de $x\in\aa$ \'etant identifi\'e \`a $\pi(x)$. Nous omettrons \'egalement tous les symboles relatifs au produit tensoriels: $\gamma^{n+1}\ot id$ est identifi\'e \`a $\gamma^{n+1}$, \dots

\par

$A$ est une 1-forme hermitienne du produit tensoriel, elle peut donc s'\'ecrire sous la forme
\bbb
A=i\gamma^{\mu}\lp A_{\mu}+\jj A_{\mu}\jj^{-1}\rp+
\gamma^{n+1}(H+\jj H\jj^{-1}),
\eee
o\`u $A_{\mu}$ est une 1-forme antihermitienne d'espace-temps \`a valeurs dans l'alg\`ebre $\aa_{F}$ et repr\'esent\'ee sur $\hh$ et $H$ est un champ scalaire hermitien \`a valeurs dans $\Omega_{\dd_{F}}^{1}(\aa_{F})$. L'op\'erateur de Dirac covariant se r\'e\'ecrit 
\bbb
\dd_{A_{\mu},H}=i\gamma^{\mu}\lp \partial_{\mu}+\omega_{\mu}+A_{\mu}+\jj A_{\mu}\jj^{-1}\rp+
\gamma^{n+1}(\dd_{F}+H+\jj H\jj^{-1}).
\eee

\par

Sous une transformation de jauge, le champ $A_{\mu}$ se transforme de mani\`ere habituelle
\bbb
A_{\mu}\mapsto uA_{\mu}u^{-1}+u\partial_{\mu}u^{-1}.
\eee
Le champ scalaire $H$ a une loi de transformation analogue 
\bbb
H\mapsto u H u^{-1}+u\lb\dd_{F},u^{-1}\rb,
\eee
faisant intervenir la d\'eriv\'ee ext\'erieure de l'espace discret.

\par
 
Cependant, $H$ n'intervient dans l'op\'erateur de Dirac covariant que par l'interm\'ediaire de la combinaison $\dd_{F}+H+\jj H\jj^{-1}$. En utilisant la d\'ecomposition de l'op\'erateur de Dirac d'un triplet spectral fini, $\dd_{F}=\Delta+\jj\Delta\jj^{-1}$, on peut \'ecrire $\dd_{F}+H+\jj H\jj^{-1}$ sous la forme $\Phi+\jj\Phi\jj^{-1}$, avec 
\bbb
\Phi=H+\Delta.
\eee
Puisque $\Delta$ est une 1-forme hermitienne, $\Phi$ est une 1-forme hermitienne et on peut \'ecrire l'op\'erateur de Dirac covariant sans utiliser $\dd_{F}$,
\bbb
\dd_{A_{\mu},\Phi}=i\gamma^{\mu}\lp \partial_{\mu}+\omega_{\mu}+A_{\mu}+\jj A_{\mu}\jj^{-1}\rp+
\gamma^{n+1}\lp\Phi+\jj\Phi\jj^{-1}\rp.
\eee

\par

Puisque 
\bbb
u\lb\dd_{F},u^{-1}\rb=u\lb\Delta,u^{-1}\rb
=u\Delta u^{-1}-\Delta,
\eee
le champ scalaire $\Phi$ se transforme de mani\`ere homog\`ene,
\bbb
\Phi\mapsto u\Phi u^{-1}. 
\eee
Gr\^ace \`a ce changement de variable, nous avons r\'eussi \`a \'eliminer le terme inhomog\`ene dans la loi de transformation du champ scalaire.

\bigskip

\noindent
{\bf Conclusion}
\begin{it}
Lorsque l'espace est le produit de l'espace-temps par un espace fini non commutatif, l'op\'erateur de Dirac covariant s'\'ecrit sous la forme
\bbb
\dd_{A_{\mu},\Phi}=i\gamma^{\mu}\lp \partial_{\mu}+\omega_{\mu}+\tilde{A}_{\mu}\rp+
\gamma^{n+1}\tilde{\Phi},
\eee
o\`u $\omega_{\mu}$ est la connexion spinorielle, $\tilde{A}_{\mu}=A_{\mu}+\jj A_{\mu}\jj^{-1}$ est un champ de jauge et $\tilde{\Phi}=\Phi+\jj\Phi\jj^{-1}$
un champ scalaire tel que $\Phi$ soit une 1-forme pour la g\'eom\'etrie de l'espace discret. Sous une transformation de jauge, on a
\bbb
A_{\mu}\mapsto uA_{\mu}u^{-1}+u\partial_{\mu} u^{-1},\;\;\;\Phi\mapsto u\Phi u^{-1}.
\eee
\end{it}

\subsection{Le carr\'e de l'op\'erateur de Dirac}

Pour pouvoir d\'eterminer l'action spectrale \`a l'aide du noyau de la chaleur, nous devons mettre le carr\'e de l'op\'erateur de Dirac covariant sous la forme 
\bbb
\lp\dd_{A_{\mu},\Phi}\rp^{2}=-\Delta+E,
\eee
o\`u $E$ est un endomorphisme du fibr\'e spinoriel et $\Delta$ un Laplacien g\'en\'eralis\'e \cite{gilkey}.

\par

Pour cela, \'ecrivons l'op\'erateur de Dirac de la mani\`ere suivante,
\bbb
\dd_{A_{\mu},\Phi}=i\gamma^{\mu}\nabla_{\mu}+\gamma^{n+1}\tilde{\Phi},
\eee
o\`u 
\bbb
\nabla_{\mu}=\partial_{\mu}+\omega_{\mu}+\tilde{A}_{\mu}
\eee
est la d\'eriv\'ee covariante des spineurs.

\par

En utilisant les relations $\gamma^{n+1}\gamma^{\mu}+\gamma^{\mu}\gamma^{n+1}=0$ et $(\gamma^{n+1})^{2}=1$, on obtient
\bbb
\lp\dd_{A_{\mu},\Phi}\rp^{2}=-\gamma^{\mu}\nabla_{\mu}\gamma^{\nu}\nabla_{\nu}+
i\gamma^{\mu}\gamma^{n+1}D_{\mu}\tilde{\Phi}+\tilde{\Phi}^{2},\label{co1}
\eee
o\`u la d\'eriv\'ee covariante du champ scalaire $\Phi$ est donn\'ee par
\bbb
D_{\mu}\tilde{\Phi}=
\partial_{\mu}\tilde{\Phi}+\lb\tilde{A_{\mu}},\tilde{\Phi}\rb.
\eee

\par

Le premier terme du second membre de (\ref{co1}) s'\'ecrit
\bbb
\gamma^{\mu}\nabla_{\mu}\gamma^{\nu}\nabla_{\nu}=
\gamma^{\mu}\gamma^{\nu}\nabla_{\mu}\nabla_{\nu}+
\gamma^{\mu}\lb\nabla_{\mu},\gamma_{\nu}\rb\nabla_{\nu}.
\eee
En introduisant
\bbb
\gamma^{\mu}\gamma^{\nu}=g^{\mu\nu}+\gamma^{\mu\nu},
\eee
on obtient
\bbb
\gamma^{\mu}\nabla_{\mu}\gamma^{\nu}\nabla_{\nu}=
g^{\mu\nu}\nabla_{\mu}\nabla_{\nu}+\frac{1}{2}\Omega_{\mu\nu}\gamma^{\mu\nu}
+\gamma^{\mu}\lb\nabla_{\mu},\gamma_{\nu}\rb\nabla_{\nu},
\eee
avec 
\bbb
\gamma^{\mu\nu}=\frac{1}{2}\lb\gamma^{\mu},\gamma^{\nu}\rb\;\;
\mathrm{et}\;\;\Omega_{\mu\nu}=\lb\nabla_{\mu},\nabla_{\nu}\rb.
\eee

\par

Pour d\'eterminer le commutateur $[\nabla_{\mu},\gamma_{\nu}]$, il faut exprimer $\gamma^{\mu}$ et $\omega_{\mu}$ \`a l'aide de la t\'etrade $e^{\mu}_{\;a}$,
\bbb
\gamma^{\mu}=e^{\mu}_{\;a}\gamma^{a}\;\;\mathrm{et}\;\;
\omega_{\mu}=\frac{1}{4}\omega_{ab\mu}\gamma^{ab},
\eee
o\`u $\omega_{ab\mu}$ est la connexion de Levi-Civita exprim\'ee en partie dans la base d\'etermin\'ee par la t\'etrade, $\gamma^{a}$ les matrices de Dirac euclidiennes satisfaisant \`a $\gamma^{a}\gamma^{b}+\gamma^{b}\gamma^{a}=\delta^{ab}$ et  $\gamma^{ab}=1/2[\gamma^{a},\gamma^{b}]$. Conform\'ement \`a l'usage, nous r\'eservons les indice grecs $\mu$, $\nu$, $\rho$, $\dots$ pour les coordonn\'ees locales et les indices latins $a$, $b$, $c$, $\dots$ pour les coordonn\'es plates obtenues \`a l'aide de la t\'etrade. Le commutateur $[\nabla_{\mu},\gamma_{\nu}]$ s'\'ecrit
\bbb
\lb\partial_{\mu}+\omega_{\mu}+\tilde{A}_{\mu},\gamma_{\nu}\rb=
\partial_{\mu}e^{\nu}_{\;c}\gamma^{c}+
\frac{1}{4}\omega_{ab\mu}e^{\nu}_{\;c}\lb\gamma^{ab},\gamma^{c}\rb.
\eee
En utilisant les propri\'et\'es des matrices de Dirac euclidiennes, il est facile de prouver que
\bbb
\lb\gamma^{ab},\gamma^{c}\rb=2\delta^{bc}\gamma^{a}-2\delta^{ac}\gamma^{b}.
\eee
On en d\'eduit que
\bbb
\lb\nabla_{\mu},\gamma^{\nu}\rb=
\lp\partial_{\mu}e^{\nu}_{\;a}+\omega_{ab\mu}e^{\nu}_{\;b}\rp\gamma^{a}.
\eee
La condition $\partial_{\mu}e^{\nu}_{\; a}+\omega_{ab\mu}e^{\nu}_{\; b}
+\Gamma^{\nu}_{\mu\lambda}e^{\lambda}_{\; a}=0$, qui exprime l'absence de torsion pour la connexion de Levi-Civita, nous permet d'\'ecrire
\bbb
\lb\nabla_{\mu},\gamma^{\nu}\rb=-\Gamma^{\nu}_{\mu\lambda}\gamma^{\lambda},
\eee
o\`u $\Gamma^{\lambda}_{\mu\nu}=1/2\,g^{\kappa\lambda}(\partial_{\mu}g_{\kappa\nu}+\partial_{\nu}g_{\kappa\mu}-\partial_{\kappa}g_{\mu\nu})$ d\'esigne les symboles de Christoffel.

\par

Rassemblant tous ces r\'esultats, on trouve
\bbb
\gamma^{\mu}\nabla_{\mu}\gamma^{\nu}\nabla_{\nu}=
g^{\mu\nu}\nabla_{\mu}\nabla_{\nu}-
g^{\mu\nu}\Gamma_{\mu\nu}^{\lambda}\nabla_{\lambda}
+\frac{1}{2}\gamma^{\mu\nu}\Omega_{\mu\nu}.
\eee
La d\'eriv\'ee covariante $\nabla_{\mu}$ d\'efinit une connexion $\nabla:\;\ee\rightarrow \ee\ot_{C^{\infty}(\mm)}\Omega^{1}(\mm)$ sur le fibr\'e spinoriel $\ee$ par
\bbb
\nabla(\Psi)=\nabla_{\mu}(\Psi)\ot dx^{\mu}
\eee
pour tout $\Psi\in\ee$. Par d\'efinition, l'op\'erateur
\bbb
\Delta=g^{\mu\nu}\nabla_{\mu}\nabla_{\nu}-
g^{\mu\nu}\Gamma_{\mu\nu}^{\lambda}\nabla_{\lambda}
\eee
est un laplacien g\'en\'eralis\'e associ\'e \`a la connexion $\nabla$.

\par

Pour terminer la d\'ecomposition de l'op\'erateur de Dirac, il nous reste \`a d\'eterminer l'endomorphisme $E$. D\'efinissons $E$ par
\bbb
E=-\frac{1}{2}\gamma^{\mu\nu}\Omega_{\mu\nu}
+i\gamma^{\mu}\gamma^{n+1}D_{\mu}\tilde{\Phi}+\tilde{\Phi}^{2}.
\eee
Il est clair que $E$ est un endomorphisme, puisque $\Omega_{\mu\nu}=[\nabla_{\mu},\nabla_{\nu}]$ en est un, et que
\bbb
\lp\dd_{A_{\mu},\Phi}\rp^{2}=-\Delta+E.
\eee 
Nous avons r\'eussi \`a d\'ecomposer le carr\'e de l'op\'erateur de Dirac en un laplacien g\'en\'eralis\'e et un endomorphisme du fibr\'e spinoriel. Puisque cette d\'ecomposition est unique, il n'y a aucune ambigu\"\i t\'e sur la d\'efinition de $E$.

\par

Afin de simplifier des calculs ult\'erieurs, \'ecrivons explicitement la courbure $\Omega_{\mu\nu}$
\bbb
\Omega_{\mu\nu}=
[\nabla_{\mu},\nabla_{\nu}]=\tilde{F}_{\mu\nu}+
\partial_{\mu}\omega_{\nu}-\partial_{\nu}\omega_{\mu}+
[\omega_{\mu},\omega_{\nu}],
\eee 
o\`u $\tilde{F}_{\mu\nu}=\partial_{\mu}\tilde{A}_{\nu}-\partial_{\nu}\tilde{A}_{\mu}+[\tilde{A}_{\mu},\tilde{A}_{\nu}]$ est la courbure du champ de jauge $\tilde{A}_{\mu}$. Notons que, gr\^ace \`a l'axiome de r\'ealit\'e, on a
\bbb
\tilde{F_{\mu\nu}}=
\partial_{\mu}\tilde{A}_{\nu}-\partial_{\nu}\tilde{A}_{\mu}
+[\tilde{A}_{\mu},\tilde{A}_{\nu}]
=F_{\mu\nu}+\jj F_{\mu\nu}\jj^{-1}.
\eee
En utilisant la relation
\bbb
\lb\gamma^{ab},\gamma^{cd}\rb=
2\lp\delta^{ad}\gamma^{bc}-\delta^{ac}\gamma^{bd}
+\delta^{bc}\gamma^{ad}-\delta^{bd}\gamma^{ac}\rp,
\eee
le terme gravitationnel s'\'ecrit
\bbb 
\partial_{\mu}\omega_{\nu}-\partial_{\nu}\omega_{\mu}+
[\omega_{\mu},\omega_{\nu}]=\frac{1}{4}
\lp\partial_{\mu}\omega_{ab\mu}-\partial_{\nu}\omega_{ab\nu}
+\omega_{ac\mu}\omega_{cd\nu}-\omega_{ac\nu}\omega_{cd\mu}\rp.\label{co2}
\eee
Puisque $\omega_{ab\mu}$ n'est autre que la connexion de Levi-Civita partiellement exprim\'ee dans la base d\'etrmin\'ee par la t\'etrade, le second membre de (\ref{co2}) est le tenseur de Riemann $R_{ab\mu\nu}$. On en d\'eduit
\bbb
\gamma^{\mu\nu}[\partial_{\mu}+\omega_{\mu},\partial_{\nu}+\omega_{\nu}]=
\frac{1}{4}\gamma^{\mu\nu}R_{ab\mu\nu}\gamma^{ab}=
\frac{1}{4}R_{\kappa\lambda\mu\nu}\gamma^{\mu\nu}\gamma^{\kappa\lambda}.
\eee
Utilisant les propri\'et\'es de sym\'etrie du tenseur de Riemann, on obtient
\bbb
R_{\kappa\lambda\mu\nu}\gamma^{\mu\nu}\gamma^{\kappa\lambda}=
2\rr,
\eee
o\`u $\rr$ est la courbure scalaire, que nous choisissons positive pour la sph\`ere. Ainsi, nous avons
\bbb
E=-\frac{1}{4}\rr-\frac{1}{2}\gamma^{\mu\nu}\tilde{F}_{\mu\nu}+
i\gamma^{\mu}\gamma^{n+1}D_{\mu}\tilde{\Phi}+\tilde{\Phi}^{2}.
\eee

\bigskip

\noindent
{\bf Conclusion}
\begin{it}
Nous avons \'ecrit le carr\'e de l'op\'erateur de Dirac $\dd_{A_{\mu.\Phi}}$ sous la forme $(\dd_{A_{\mu},\Phi})^{2}=-\Delta+E, $ o\`u $\Delta$ est un laplacien g\'en\'eralis\'e associ\'e \`a la connexion d\'efinie par la d\'eriv\'ee covariante $\nabla_{\mu}=\partial_{\mu}+\omega_{\mu}+\tilde{A_{\mu}}$ et $E$ est un endomorphisme du fibr\'e spinoriel donn\'e  par 
\bbb
E=-\frac{1}{4}\rr-\frac{1}{2}\gamma^{\mu\nu}\tilde{F}_{\mu\nu}+
i\gamma^{\mu}\gamma^{n+1}D_{\mu}\tilde{\Phi}+\tilde{\Phi}^{2}.
\eee
La courbure de la connexion $\nabla_{\mu}$ est donn\'ee par
\bbb
\Omega_{\mu\nu}=\frac{1}{4}R_{ab\mu\nu}\gamma^{ab}+\tilde{F}_{\mu\nu}.
\eee
\end{it}
 
\subsection{Action spectrale}

Suivant le principe d'action spectrale, l'action bosonique ne d\'epend que du spectre de l'op\'erateur de Dirac covariant. \'Etant donn\'e que $\dd_{A_{\mu},\Phi}$ contient les champs de Yang-Mills $A_{\mu}$ et scalaires $\Phi$ ainsi que la m\'etrique $g_{\mu\nu}$, l'action obtenue d\'ecrira un mod\`ele de Yang-Mills-Higgs coupl\'e \`a la gravitation.

\par

Si on consid\`ere que l'action fermionique est l'objet essentiel de la th\'eorie des champs, l'id\'ee d'une action bosonique qui ne d\'epend que du spectre de l'op\'erateur de Dirac est relativement naturelle. Par exemple, l'action effective obtenue en int\'egrant sur les champs de fermions nous donne un d\'eterminant fonctionnel
\bbb
\int[\dd\ov{\Psi}][\dd\Psi]
\mathrm{e}^{-\langle\ov{\Psi},\dd_{A_{\mu},\Phi}\Psi\rangle}
=\det\dd_{A_{\mu},\Phi}.
\eee
Apr\`es r\'egularisation, par une fonction $\zeta$ par exemple, ce d\'eterminant ne d\'epend que du spectre de l'op\'erateur de Dirac $\dd_{A_{\mu},\Phi}$. Ainsi, cette action effective satisfait au principe d'action spectrale.

\par

De fa\c con g\'en\'erale, nous \'ecrirons l'action bosonique sous la forme
\bbb
S=\t\,\mathrm{F}\lb\dd^{2}/\Lambda^{2}\rb,\label{as1}
\eee
o\`u, pour simplifier, nous notons $\dd$ l'op\'erateur de Dirac covariant $\dd_{A_{\mu},\Phi}$. $F$ est une fonction quelconque, mais suppos\'ee suffisament r\'eguli\`ere et \`a d\'ecroissance rapide \`a l'infini pour que la trace existe. De plus, pour obtenir une action positive, nous faisons l'hypoth\`ese que $F(x)\geq 0$ si $x\geq0$. $\Lambda$  est un cut-off homog\`ene \`a une masse de l'ordre de grandeur de la masse de Planck $m_{P}=10^{19}$ Gev. 

\par

Puisque la masse de Planck est tr\`es grande par rapport \`a l'echelle d'\'energie \'electrofaible $M_{W}=80\;GeV$, nous nous int\'eresserons uniquement au d\'eveloppement asymptotique de (\ref{as1}) lorsque $\Lambda\rightarrow\infty$. En dimension $n$, le d\'eveloppement asymptotique de cette fonctionnelle est donn\'e par
\bbb
\t F\lb\dd^{2}/\Lambda^{2}\rb=
\mathop{\sum}\limits_{0\leq k\leq n/2}
\Lambda^{n-2k}F_{2k}\int_{\mm}a_{2k}(\dd^{2})\,\sqrt{g}\,d^{n}x + O(1/\Lambda^{2}),
\eee
o\`u $a_{2k}(\dd^{2})$ d\'esigne les coefficients de Seeley-de Witt qui interviennent dans le d\'eveloppement limit\'e de
\bbb
\t\,\mathrm{e}^{-t\dd^{2}}=
\mathop{\sum}\limits_{0\leq k\leq n/2} t^{k-n/2}\int_{\mm}a_{2k}(\dd^{2})\,\sqrt{g}d^{n}x
+O(t)\label{as2}
\eee
quand $t\rightarrow 0$. Les nombres r\'eels $F_{2k}$ ne d\'ependent que de $F$ et sont donn\'es par
\bbb
F_{2k}=\frac{1}{\Gamma(n/2-k)}\int_{0}^{\infty}t^{n/2-k-1}F(t)dt
\;\;\;\mathrm{si}\;2k<n,\;\;\;F_{n}=F(0).
\eee

\par

Cette relation peut \^etre justifi\'ee en \'ecrivant $F$ sous la forme d'une transform\'ee de Laplace $F(s)=\int_{0}^{\infty}G(t)\mathrm{e}^{-st}$, ce qui donne
\bbb
\t\,F[ \dd^{2}/\Lambda^{2}]=\int_{0}^{\infty} \t\,\mathrm{e}^{-\dd^{2}/\Lambda^{2}}\,G(t)\,dt.
\eee
En ins\'erant le d\'eveloppement limit\'e (\ref{as2}), on obtient
\bbb
\t F\lb\dd^{2}/\Lambda^{2}\rb=
\mathop{\sum}\limits_{0\leq k\leq n/2}
\Lambda^{n-2k}F_{2k}
\int_{\mm}a_{2k}(\dd^{2})\,\sqrt{g}d^{n}x + O(1/\Lambda^{2}).
\eee
avec
\bbb
F_{2k}=\int_{0}^{\infty}t^{k-n/2}G(t)dt.
\eee
En utilisant
\bbb
\int_{0}^{\infty}s^{-p}F(s)\,ds=
\Gamma(p+1)\int_{0}^{\infty}\frac{G(t)}{t^{p+1}}\,dt,
\eee
valide pour $Re(p)>-1$, on montre que
\bbb
F_{2k}=\frac{1}{\Gamma(n/2-k)}\int_{0}^{\infty}t^{n/2-k-1}F(t)dt
\eee
si $2k<n$, ce qui justifie notre r\'esultat.

\par

Les trois premiers coefficients de Seeley-de Witt sont donn\'es \cite{gilkey}, \`a une d\'eriv\'ee totale pr\`es, en fonction de $E$ et $\Omega_{\mu\nu}$ par
\bbbb
a_{0}(\dd^{2})&=&\frac{1}{(4\pi)^{n/2}}\,2^{[n/2]}\t(1),\\
a_{2}(\dd^{2})&=&\frac{1}{(4\pi)^{n/2}}\lp\frac{1}{6}\rr\,2^{[n/2]}\t(1)
-\t E\rp,\\
a_{4}(\dd^{2})&=&\frac{1}{(4\pi)^{n/2}}\lp\frac{1}{72}\rr^{2}\,2^{[n/2]}\t(1)
-\frac{1}{180}R_{\mu\nu}R^{\mu\nu}\,2^{[n/2]}\t(1)\right.\n\\
&+&\left. 
\frac{1}{180}R_{\mu\nu\rho\sigma}R^{\mu\nu\rho\sigma}\,2^{[n/2]}\t(1)+\frac{1}{12}\t\Omega_{\mu\nu}\Omega^{\mu\nu} 
-\frac{1}{6}\rr\t E
+\frac{1}{2}\t E^{2}
\rp.
\eeee
o\`u $R_{\mu\nu\rho\sigma}$ est le tenseur de Riemann, $R_{\mu\nu}$ le tenseur de Ricci, $\rr$ la courbure scalaire et $\t(1)$ la dimension de l'espace $\hh_{F}$.

\par

Utilisant les expression explicites de $E$ et de $\Omega_{\mu\nu}$, nous obtenons
\bbbb
a_{0}(\dd^{2})&=&\frac{\t(1)}{(2\pi)^{\frac{n}{2}}},\\
a_{2}(\dd^{2})&=&-\frac{1}{12}\;\frac{\t(1)}{(2\pi)^{\frac{n}{2}}}\;\rr
-\frac{1}{(2\pi)^{\frac{n}{2}}}\;\t(\tilde{\Phi}^{2}),\\
a_{4}(\dd^{2})&=&\frac{1}{1440}\;\frac{\t(1)}{(2\pi)^{\frac{n}{2}}}\;
\lp 5\rr^{2}-8 R_{\mu\nu}R^{\mu\nu}
-7 R_{\kappa\lambda\mu\nu}R^{\kappa\lambda\mu\nu}\rp\n\\
&&+\frac{1}{2}\;\frac{1}{(2\pi)^{\frac{n}{2}}}\;
\lp \t(D_{\mu}\tilde{\Phi}D^{\mu}\tilde{\Phi})
+\t(\tilde{\Phi}^{4})\rp\n\\
&&-\frac{1}{6}\;\frac{1}{(2\pi)^{\frac{n}{2}}}\;
\t(\tilde{F}_{\mu\nu}\tilde{F}^{\mu\nu})
+\frac{1}{12}\;\frac{1}{(2\pi)^{\frac{n}{2}}}\;
\rr\t(\tilde{\Phi}^{2}).
\eeee

\par

En dimension $n>4$, l'action spectrale contient des termes en $a_{6}(\dd^{2})$, puisque ces derniers ne sont plus en facteur de puissances du cut-off qui tendent vers 0 quand $\Lambda$ tend vers l'infini. Physiquement, l'action ainsi obtenue n'est pas acceptable car $a_{6}(\dd^{2})$ contient des termes d'ordre trop \'elev\'es tels que $\t{F}^{3}$. Bien entendu, il est toujours possible de choisir une fonction $F$ (non positive) telle que $F_{2k}=0$ si $k>2$, mais cela n'am\'eliore pas les probl\`emes dus \`a la non renormalisabilit\'e de la th\'eorie en dimension $n>4$. Par cons\'equent, m\^eme si les relations que nous obtenons sont donn\'ees en dimension quelconque, nous ne devons pas perdre de vue que $\Lambda$ n'est un cut-off naturel qu'en dimension 4. 

\par

Enfin, notons que puisque la fonction $F$ est une fonction de cut-off qui doit limiter la trace aux valeurs propres inf\'erieures \`a $\Lambda$, le choix le plus naturel est de prendre pour $F$ la fonction caract\'eristique de l'intervalle $[0,1]$. Cependant, cette fonction n'est pas r\'eguli\`ere et nous ne pouvons pas appliquer la m\'ethode pr\'ec\'edente. Dans ce cas, il est n\'ecessaire de recourir \`a des m\'ethodes plus \'elabor\'ees \cite{asymp}.

\bigskip

\noindent
{\bf Conclusion}
\begin{it}
En utilisant le noyau de la chaleur, le d\'eveloppement asymptotique de l'action spectrale s'\'ecrit, en dimension 4, 
\bbbb
\t F\lb\dd^{2}/\Lambda^{2}\rb&=&
\Lambda^{4}\,F_{0}\int_{\mm}a_{0}(\dd^{2})\,\sqrt{g}\,d^{4}x+ 
\Lambda^{2}\,F_{2}\int_{\mm}a_{2}(\dd^{2})\,\sqrt{g}\,d^{4}x\n\\
&+&\Lambda^{0}\,F_{4}\int_{\mm}a_{4}(\dd^{2})\,\sqrt{g}\,d^{4}x+
O(1/\Lambda^{2})
\eeee
o\`u $F_{0}$, $F_{2}$ et $F_{4}$ sont des constantes reli\'ees \`a la fonction $F$, $a_{0}(\dd^{2})$, $a_{2}(\dd^{2})$ et $a_{4}(\dd^{2})$ sont les coefficients de Seeley-de Witt et $\Lambda$ est un cut-off de l'ordre de la masse de Planck. Cette action d\'ecrit un mod\`ele de Yang-Mills-Higgs coupl\'e \`a la gravitation $\rr^{2}$. En dimension $n>4$, l'action spectrale contient des termes d'ordre sup\'erieurs, non physiques.
\end{it}

\subsection{Termes gravitationnels}

Dans cette section, nous allons \'etudier quelques aspects sp\'ecifiques au secteur gravitationnel de la theorie en dimension 4. Les termes gravitationnels obtenus \`a l'aide de l'expansion de l'action spectrale sont de quatre types: un terme contenant une constante cosmologique, l'action d'Einstein-Hilbert, un couplage entre les champs scalaires et la courbure scalaire et des termes du second degr\'e en $R_{\mu\nu\rho\sigma}$.

\par

Commen\c cons par simplifier ces derniers en introduisant l'invariant de Gauss-Bonnet,
\bbb
\chi=
R_{\kappa\lambda\mu\nu}R^{\kappa\lambda\mu\nu}-4R_{\mu\nu}R^{\mu\nu}+\rr^{2},
\eee 
qui est une d\'eriv\'ee totale en dimension 4. Les termes en $\rr^{2}$ de l'action spectrale apparaissent par l'interm\'ediaire du coefficient $a_{4}(\dd^{2})$ et sont donn\'ees par
\bbb
\frac{1}{1440}\;\frac{\t(1)}{(2\pi)^{2}}\;
\lp 5\rr^{2}-8 R_{\mu\nu}R^{\mu\nu}
-7 R_{\kappa\lambda\mu\nu}R^{\kappa\lambda\mu\nu}\rp
\eee
En leur  retranchant un terme proportionel \`a $\chi$, on obtient
\bbb
 5\rr^{2}-8 R_{\mu\nu}R^{\mu\nu}
-7 R_{\kappa\lambda\mu\nu}R^{\kappa\lambda\mu\nu}-11\chi=
 -6\rr^{2}+36 R_{\mu\nu}R^{\mu\nu}
-18 R_{\kappa\lambda\mu\nu}R^{\kappa\lambda\mu\nu}.\label{tg1}
\eee
Le second membre de (\ref{tg1}) s'exprime \`a l'aide du carr\'e du tenseur de Weyl. Ce dernier est d\'efini, en dimension 4, par
\bbb
C_{\kappa\lambda\mu\nu}=R_{\kappa\lambda\mu\nu}
-\frac{1}{2}\lp g_{\mu\rho}R_{\nu\sigma}-g_{\mu\sigma}R_{\nu\rho}
+g_{\nu\sigma}R_{\mu\rho}-g_{\nu\rho}R_{\mu\sigma}\rp
+\frac{1}{6}\lp g_{\mu\rho}g_{\nu\sigma}-g_{\mu\sigma}g_{\nu\rho}\rp\rr.
\eee
Son carr\'e est donn\'e  par $C_{\kappa\lambda\mu\nu}C^{\kappa\lambda\mu\nu}=
R_{\kappa\lambda\mu\nu}R^{\kappa\lambda\mu\nu}
-2R_{\mu\nu}R^{\mu\nu}
+\frac{1}{3}\rr^{2}$, ce qui nous permet d'\'ecrire
\bbb
5\rr^{2}-8R_{\mu\nu}R^{\mu\nu}-7R_{\kappa\lambda\mu\nu}R^{\kappa\lambda\mu\nu}=
-18C_{\kappa\lambda\mu\nu}C^{\kappa\lambda\mu\nu}+11\chi.
\eee
A une d\'eriv\'e totale pr\`es, les termes en $\rr^{2}$ sont 
\bbb
-\frac{1}{80}\;\frac{\t(1)}{(2\pi)^{2}}\;
C_{\kappa\lambda\mu\nu}C^{\kappa\lambda\mu\nu}.
\eee

\par

L'action d'Einstein-Hilbert provient du coefficient $a_{2}(\dd^{2})$. En effet, outre un terme de masse pour les champs scalaires, celui-ci contient
\bbb
-\frac{1}{12}\;\frac{\t(1)}{(2\pi)^{2}}\;\rr.
\eee
L'action d'Einstein-Hilbert s'\'ecrit 
\bbb
S_{EH}=-\frac{1}{16\pi G}\int_{\mm}\,d^{4}x\,\sqrt{g}\,\rr,
\eee
o\`u la constante de Newton est donn\'ee, avant brisure spontan\'ee de sym\'etrie, par
\bbb
G=\frac{3}{2}\frac{2\pi}{\t(1)}\,F_{2}\,\Lambda^{2},\label{tg2}
\eee
En utilisant $G$, nous pouvons r\'e\'ecrire le terme de Weyl sous la forme
\bbb
S_{W}=\frac{3}{160\pi\mu^{2}G^{2}}\,\int_{\mm}\,d^{4}x\,\sqrt{g}\,
C_{\kappa\lambda\mu\nu}C^{\kappa\lambda\mu\nu},
\eee
o\`u $\mu$ est une quantit\'e homog\`ene \`a une masse donn\'e par
\bbb
\mu=\sqrt{\frac{2F_{2}}{F_{4}}}\Lambda.\label{tg3}
\eee
Nous verrons que $\mu$ pourra \^etre interpr\'et\'e comme la "masse" des champs scalaires avant la brisure spontann\'ee de sym\'etrie.

\par

Le coefficient $a_{0}(\dd^{2})$ g\'en\`ere un terme
\bbb
\frac{\t(1)}{(2\pi)^{2}}\,\int_{\mm}\,d^{4}x\,\sqrt{g}
\eee 
qui s'identifie, avant brisure spontan\'ee de sym\'etrie, \`a une constante cosmologique
\bbb
\Lambda_{c}=\frac{1}{12}\,\frac{\t(1)^{2}}{(2\pi)^{4}}\,\frac{F_{0}}{F_{2}}\,
\Lambda_{2},\label{tg4}
\eee
de telle sorte que l'action d'Einstein-Hilbert avec constante cosmologique soit
\bbb
S_{EHC}=\frac{1}{16\pi G}\int_{\mm}\,d^{4}x\,\sqrt{g}\,(-\rr+\Lambda_{c}).
\eee
Bien entendu, cette constante cosmologique peut \^etre annul\'ee en choisissant une fonction $F$ telle que $F_{0}=0$. Cependant, cela n\'ecessite l'emploi d'une fonction non positive \cite{rov}.

\par

Enfin, nous avons un terme de couplage entre les champs scalaires et la gravitation donn\'e par
\bbb
\frac{1}{12}\,\frac{1}{(2\pi^{2})}\,F_{4}\,\rr\,\t(\tilde{\Phi}^{2}).\label{tg6}
\eee
Nous verrons que, gr\^ace \`a la normalisation des champs scalaires, ce terme est simplement 
\bbb
\frac{1}{6}\,\rr\mathop{\sum}\limits_{scalaires}\t(\Phi^{2}),\label{tg5}
\eee
o\`u la somme est une somme sur tous les champs scalaires que contient la th\'eorie. Il est \`a noter que dans certains cas ce terme peut menacer la brisure spontan\'ee de sym\'etrie \cite{gef}. En effet, toutes les constantes sont donn\'ees avant brisure spontan\'ee de sym\'etrie. Lorsque l'on d\'eveloppe la th\'eorie au voisinage en fonction des champs scalaires physiques, il appara\^\i t un d\'ecalage dans les constantes $\Lambda_{c}$, $G$ et $\mu$.

\par

Terminons par quelques mots concernant l'unitarit\'e et la renormalisabilit\'e de cette th\'eorie \cite{buch}. Bien que la th\'eorie d'Einstein soit non renormalisable, l'incorporation de termes d'ordre sup\'erieurs rend cette th\'eorie renormalisable. En effet, si on incorpore \`a l'action d'Einstein-Hilbert les termes en $C_{\kappa\lambda\mu\nu}C^{\kappa\lambda\mu\nu}$ et en $\rr^{2}$, la th\'eorie est renormalisable. Cependant, nous avons besoin de ces deux termes simultan\'ement, alors que l'action spectrale ne nous fournit que le terme de Weyl. Nous n'obtenons donc pas une th\'eorie renormalisable au sens strict du terme, car les corrections quantiques vont g\'en\'erer des contretermes en $\rr^{2}$. De plus, l'introduction des termes d'ordre sup\'erieurs brise l'unitarit\'e.

\bigskip

\noindent
{\bf Conclusion}
\begin{it}
En dimension 4, le secteur gravitationnel de l'action spectrale contient un terme d'Einstein-Hilbert avec constante cosmologique
\bbb
S_{EHC}=\frac{1}{16\pi G}\int_{\mm}\,d^{4}x\,\sqrt{g}\,(-\rr+\Lambda_{c}),
\eee
un terme de Weyl
\bbb
S_{W}=\frac{3}{160\pi\mu^{2}G^{2}}\,\int_{\mm}\,d^{4}x\,\sqrt{g}\,
C_{\kappa\lambda\mu\nu}C^{\kappa\lambda\mu\nu},
\eee
ainsi qu'un couplage entre la gravitation et les champs scalaires donn\'e par (\ref{tg5}). Les constantes $G$, $\Lambda_{c}$ et $\mu$ sont donn\'ees, avant brisure spontann\'e de sy\'metrie, par les \'equations (\ref{tg2}), (\ref{tg3}) et (\ref{tg4}). Cette th\'eorie est non unitaire et n'est pas renormalisable au sens strict du terme, car les 
corrections quantiques engendrent un contreterme en $\rr^{2}$.
\end{it}

\section{Th\'eorie de jauge}

\subsection{Les fermions et leur repr\'esentation}

Les transformations de jauge sont donn\'ees par les \'el\'ements unitaires $u$ de $\aa$. $u$ peut \^etre consid\'er\'e comme une application de $\mm$ dans le groupe $G_{F}$ des \'el\'ements unitaires de $\aa_{F}$. Puisque $\aa_{F}$ est une somme directe d'alg\`ebres de matrices \`a coefficients dans les corps $\ccc$, $\rrr$ et $\hhh$, $G_{F}$ est un produit direct de groupes de Lie simples du type $U(n)$, $O(n)$ et $SP(n)$, qui sont les \'el\'ements unitaires des alg\`ebres de matrices $M_{n}(\ccc)$,  $M_{n}(\rrr)$ et $M_{n}(\hhh)$. Par cons\'equent, le groupe de jauge est un produit direct de groupes de Lie compacts qui correspondent aux quatres s\'eries infinies de la classification des alg\`ebres de Lie semi-simples, les cinq alg\`ebres de Lie exceptionnelles \'etant exclues.

\par

Pour \^etre une sym\'etrie de l'action spectrale, le groupe des unitaires doit \^etre repr\'esent\'e sur l'espace de Hilbert $\hh$. L'action de l'unitaire $u$ sur $\hh$ se fait par l'interm\'ediaire de la repr\'esentation $\tilde{\pi}$ d\'efinie par $\tilde{\pi}(u)=\pi(u)\jj\pi(u)\jj^{-1}$.

\par

Par d\'efinition, $\hh$ est le produit tensoriel de l'espace de Hilbert des spineurs de carr\'e int\'egrable sur $\mm$ par l'espace $\hh_{F}$ correspondant au triplet spectral fini. De m\^eme, $\pi$ et $\jj$ sont les produits tensoriels des op\'erateurs correspondants et il nous suffit, pour \'etudier les lois de transformation des \'el\'ements de $\hh$, de ne consid\'erer que la partie correspondant au triplet spectral fini $(\aa_{F},\hh_{F},\dd_{F})$. Pour all\'eger nos \'equations,  nous omettrons l'indice $F$ jusqu'a la fin de ce chapitre.

\par

Rappelons que si $\aa$ est une somme directe d'alg\`ebres de matrices, $\aa=\op_{i}M_{n_{i}}(\kkk)$, l'espace de Hilbert se d\'ecompose en $\hh=\op_{i,j}\hh_{ij}$, o\`u les indices $i$ et $j$ correspondent aux diff\'erentes repr\'esentations irr\'eductibles des facteurs $M_{n_{i}}(\kkk)$.

\par

A cette d\'ecomposition de $\hh$ sont associ\'ees les d\'ecompositions de $\pi$ et $\jj\pi\jj^{-1}$,
\bbbb
\pi(x)&=&\mathop{\op}\limits_{i,j}\;
x_{i}\ot I_{m_{ij}}\ot I_{n_{j}},\\
\jj\pi(x)\jj^{-1}&=&\mathop{\op}\limits_{i,j}\;
I_{n_{i}}\ot I_{m_{ij}}\ot \ov{x}_{j},
\eeee
pour tout $y\in\aa$. Un \'el\'ement $u$ du groupe $G=G_{1}\times\dots\times G_{N}$ des unitaires de $\aa$ est donc repr\'esent\'e par
\bbb
\tilde{\pi}(u)=\pi(u)\jj\pi(u)\jj^{-1}=\mathop{\op}\limits_{i,j}\;
u_{i}\ot I_{m_{ij}}\ot \ov{u}_{j},
\eee
Par cons\'equent, les vecteurs de $\hh_{ij}=\ccc^{n_{i}}\ot\ccc^{|\mu_{ij}|}\ot\ccc^{n_{j}}$ se transforment sous 
la repr\'esentation $n_{i}\ot \ov{n}_{j}$ avec la multiplicit\'e $|\mu_{ij}|$, o\`u $n_{i}$ d\'esigne la repr\'esentation fondamentale de $G_{i}$ ou sa complexe conjugu\'ee.

\par

Cette condition est particuli\`erement restrictive, car parmi toutes les repr\'esentations de $G$, nous ne pouvons construire que des produits tensoriels de deux repr\'esentations fondamentales de chacun des groupes simples $G_{i}$.
Cela impose de tr\`es s\'ev\`eres contraintes sur les th\'eories de grande unification \cite{gut}. Par exemple, le mod\`ele de grande unification bas\'e sur le groupe $SO(10)$ regroupe tous les fermions existant ainsi que le neutrino droit dans un multiplet de dimension 16. Ces fermions se transforment dans la repr\'esentation spinorielle de $SO(10)$ qui ne peut \^etre obtenue \`a partir de sa repr\'esentation fondamentale.

\par

Dans le but de construire de tels mod\`eles en utilisant les outils de la g\'eom\'etrie non commutative, une autre approche a \'et\'e d\'evelopp\'ee dans \cite{raimar}. Elle consiste \`a remplacer l'alg\`ebre associative $\aa$ par une alg\`ebre de Lie, repr\'esent\'ee sur un espace de Hilbert, \`a partir de laquelle on peut reconstruire un calcul diff\'erentiel ainsi que les th\'eories de jauge qui lui sont associ\'ees. 

\par

L'espace $\hh$ est \'equipp\'e d'une graduation $\chi$. Dans le cas du mod\`ele standard, cet op\'erateur prend la valeur $+1$ pour les fermions droits et la valeur $-1$ pour les fermions gauches. C'est pourquoi on l'interpr\`ete g\'en\'eralement comme un op\'erateur de chiralit\'e, les fermions de $\hh_{ij}$ \'etant consid\'er\'es comme des fermions chiraux, encore appel\'es fermions de Weyl, d'h\'elicit\'e droite si $\mu_{ij}>0$ et d'h\'elicit\'e gauche si $\mu_{ij}<0$.

\par

Cette interpr\'etation de $\chi$ est controvers\'ee car, au sens strict du terme, l'espace de Hilbert des spineurs contient d\'eja, pour chaque particule, des composantes droite et gauche. Il semble donc que la distinction entre particules droites et gauches dans l'espace $\hh$ relatif au triplet spectral fini soit superflue et conduise \`a un doublement
 du nombre de particules \cite{doubling}.

\par

Cependant, cette difficult\'e n'est pas propre \`a la g\'eom\'etrie non commutative. Elle apparait dans les th\'eories des champs contenant des fermions chiraux d\'efinis avec une signature euclidienne. En effet, avec une telle signature, l'action de Dirac est donn\'ee par $\Psi^{\dag}\dd\Psi$ alors que dans l'espace de Minkowski, elle est $\ov{\Psi}\dd\Psi$, avec $\ov{\Psi}=\Psi^{\dag}\gamma^{0}$. Dans le premier cas, des fermions chiraux, i.e. satisfaisant \`a $\gamma^{5}\Psi=\pm\Psi$, ont une action de Dirac nulle alors que ce n'est pas le cas en signature lorentzienne \`a cause de la pr\'esence de $\gamma^{0}$. Pour \'ecrire une action de Dirac non nulle dans l'euclidien pour des fermions de Weyl le doublement est n\'ecessaire et nous consid\'ererons que les fermions de $\hh_{ij}$ sont r\'eellement des fermions chiraux.

\par

Dans le triplet spectral du mod\`ele standard, l'op\'erateur $\jj$ est un op\'erateur antilin\'eaire qui \'echange particules et antiparticules. Par cons\'equent, dans le cas g\'en\'eral nous consid\'erons que $\jj$ joue le r\^ole de la conjugaison de charge. Nous interpr\'etons les \'el\'ements de $\hh_{ji}=\jj\hh_{ij}$ comme les antiparticules de $\hh_{ij}$. 

\par

En g\'en\'eral, seuls les vecteurs de $\hh_{ii}$ peuvent \^etre des particules de Majorana, qui se transforment n\'ecessairement dans la repr\'esentation adjointe de $G_{i}$. La condition d'absence de fermions de Majorana et donn\'e par l'axiome de $S^{0}$-r\'ealit\'e. En effet, celui-ci stipule qu'il existe un op\'erateur $\epsilon$, hermitien et de carr\'e 1, qui commute avec $\chi$, $\dd$ et $\pi$ et anticommute avec $\jj$. Cette derni\`ere relation permet de d\'ecomposer $\hh=\hh_{+}\op\hh_{-}$ suivant les valeurs propres de $\epsilon$. On a $\jj\hh_{+}=\hh_{-}$ et $\jj\hh_{-}=\hh_{+}$ ce qui nous am\`ene \`a interpr\'eter $\hh_{+}$ comme l'espace des particules et $\hh_{-}$ comme celui des antiparticules. Puisque $\jj$ \'echange ces deux espaces, aucune particule n'est sa propre antiparticule.

\bigskip

\noindent
{\bf Conclusion}
\begin{it}
Le groupe de jauge d'un mod\`ele de Yang-Mills-Higgs construit \`a l'aide de la g\'eom\'etrie non commutative est un produit direct de groupes de Lie compacts du type $G=O(n_{i})$, $U(n_{i})$ et $SP(n_{i})$. L'espace de Hilbert des fermions se d\'ecompose en 
\bbb
\hh=\mathop{\op}\limits_{i,j}\hh_{ij},
\eee
o\`u $\hh_{ij}$ contient des fermions chiraux se transformant sous le produit tensoriel $n_{i}\ot\ov{n}_{j}$, $n_{i}$ d\'esignant la repr\'esentation fondamentale de $G_{i}$ ou sa complexe conjugu\'ee. A partir de la matrice de multiplicit\'e du triplet spectral fini, on peut reconstruire la multiplicit\'e de la representation $n_{i}\ot\ov{n}_{j}$ donn\'ee par $|\mu_{ij}|$, ainsi que l'h\'elicit\'e des fermions de $\hh_{ij}$ donn\'ee par le signe de $\mu_{ij}$. Les \'el\'ements de $\hh_{ji}$ sont les antiparticules de ceux de $\hh_{ji}$ et la condition de $S^{0}$-r\'ealit\'e correspond \`a l'absence de fermions de Majorana.

\end{it}
 
\subsection{Constantes de couplage non ab\'eliennes}

En plus du secteur gravitationnel, l'action spectrale contient un terme de Yang-Mills qui apparait \`a travers le coefficient $a_{4}(\dd^{2})$. Plus pr\'ecis\'ement, l'action de Yang-Mills obtenue est, en dimension $n$.
\bbb
S_{YM}=
-\frac{1}{6}\;\frac{1}{(2\pi)^{n/2}}\;F_{4}\;\Lambda^{n-4}\;
\int_{\mm}d^{4}x\,\sqrt{g}\,\t(\tilde{F}_{\mu\nu}\tilde{F}^{\mu\nu}),
\eee  
avec $\tilde{F}_{\mu\nu}=\partial_{\mu}\tilde{A}_{\nu}-\partial_{\nu}\tilde{A}_{\mu}+[\tilde{A}_{\mu},\tilde{A}_{\nu}]$ et $\tilde{A}_{\mu}=A_{\mu}+\jj A_{\mu}\jj^{-1}$. 

\par

Par d\'efinition, $A_{\mu}$ est un \'el\'ement antihermitien de l'alg\`ebre $\aa$ repr\'esent\'e sur $\hh$. On peut toujours \'ecrire $A_{\mu}$ sous la forme
\bbb
A_{\mu}=\mathop{\op}\limits_{i,j}\;
(g_{i}A_{\mu}^{i}+B_{\mu}^{i})\ot I_{m_{ij}}\ot I_{n_{j}},\\
\eee
o\`u $A_{\mu}^{i}$ est un \'el\'ement antihermitien de $M_{n_{i}}(\kkk)$ tel que $\t(A_{\mu}^{i})=0$ et $B_{\mu}^{i}\in i\rrr$ un champ de jauge ab\'elien. $g_{i}$ est un nombre r\'eel strictement positif qui joue le r\^ole d'une constante de couplage pour le champ de jauge non ab\'elien $A_{\mu}^{i}$. Cela nous permet d'\'ecrire $F_{\mu\nu}$ et $\tilde{F}_{\mu\nu}$ sous la forme
\bbbb
F_{\mu\nu}&=&\mathop{\op}\limits_{i,j}\;
\lp g_{i}F_{\mu\nu}^{i}+G_{\mu\nu}^{i}\rp\ot I_{m_{ij}}\ot I_{n_{j}},\\
\jj F_{\mu\nu}\jj^{-1}&=&\mathop{\op}\limits_{i,j}\;
I_{n_{i}}\ot I_{m_{ij}}\ot\lp g_{i}\ov{F}_{\mu\nu}^{i}+\ov{G}_{\mu\nu}^{i}\rp,
\eeee
avec
\bbbb
F_{\mu\nu}^{i}&=&\partial_{\mu}A_{\mu}^{i}-\partial_{\nu}A_{\mu}^{i}
+g_{i}[A_{\mu}^{i},A_{\nu}^{i}],\\
G_{\mu\nu}^{i}&=&\partial_{\mu}B_{\mu}^{i}-\partial_{\nu}B_{\mu}^{i}.
\eeee
En rempla\c cant $F_{\mu\nu}$ et $\jj F_{\mu\nu}\jj^{-1}$ par leurs expressions dans $\t(\tilde{F}_{\mu\nu}\tilde{F}_{\mu\nu})$, on obtient
\bbb
\t(\tilde{F}_{\mu\nu}\tilde{F}^{\mu\nu})=\mathop{\sum}\limits_{ij}
2g_{i}^{2}|\mu_{ij}|n_{j}\;\t(F_{\mu\nu}^{i}F^{i\mu\nu})
+Q_{ij}\; G_{\mu\nu}^{i}G^{j\mu\nu}. \label{cc1}
\eee
La matrice sym\'etrique et positive $Q_{ij}$ est donn\'ee par
\bbb
Q_{ij}=2\lp\mathop{\sum}\limits_{k}m_{ik}n_{i}n_{k}\rp\delta_{ij}-
2|\mu_{ij}|n_{i}n_{j}.\label{cc3}
\eee
Le premier terme du second membre de (\ref{cc1}) d\'etermine l'action de Yang-Mills non ab\'elienne, alors que le second terme correspond aux champs ab\'eliens. Dans cette section, nous ne nous occupperons pas de ce dernier, car la discussion des champs de jauge ab\'eliens est plus compliqu\'ee et sera trait\'ee dans la partie suivante. L'action de Yang-Mills non ab\'elienne peut s'\'ecrire
\bbb
-\frac{1}{6}\;\frac{1}{(2\pi)^{n/2}}\;F_{4}\;\Lambda^{n-4}\;
\mathop{\sum}\limits_{i,j}
2g_{i}^{2}|\mu_{ij}|n_{j}\;\int_{\mm}d^{4}x\,\sqrt{g}\,
\t(F_{\mu\nu}^{i}F^{i\mu\nu}).\label{cc2}
\eee
Usuellement, les constantes de couplage non ab\'eliennes sont normalis\'ees de telle sorte que l'action de Yang-Mills soit
\bbb
-\mathop{\sum}\limits_{i}
\frac{1}{2}\;\int_{\mm}d^{4}x\,\sqrt{g}\,
\t(F_{\mu\nu}^{i}F^{i\mu\nu}).
\eee
La comparaison avec l'expression obtenue dans (\ref{cc2}) nous permet de d\'eterminer les constantes de couplages explicitement,
\bbb
g_{i}=(2\pi)^{n/4}
\sqrt{\frac{3/2}{F_{4}\Lambda^{n-4}\mathop{\sum}\limits_{j}|\mu_{ij}|n_{j}}}.
\eee
Il en ressort que les constantes de couplage non ab\'eliennes sont enti\`erement d\'etermin\'ees, \`a un facteur global $F_{4}$ par la matrice de multiplicit\'e $\mu_{ij}$. 

\par

Il convient de noter que ces relations font intervenir les constantes de couplage au niveau classique uniquement. A priori, il n'existe aucune identit\'e de type Ward-Takahashi ou Slavnov-Taylor \cite{zuber} qui nous assure que ces relations sont pr\'eserv\'ees au niveau quantique par le processus de renormalisation. 

\par

Dans le cas du mod\`ele standard, le point de vue adopt\'e consiste \`a consid\'erer la th\'eorie comme valide pour une valeur \'elev\'ee de $\Lambda$, de l'ordre de la masse de Planck, et \`a utiliser les \'equations du groupe de renormalisation pour d\'eterminer les constantes de couplage lorsque $\Lambda$ est de l'ordre de l'\'energie \'electrofaible \cite{hear}.

\bigskip

\noindent
{\bf Conclusion}
\begin{it}
Les constantes de couplage des champs de jauge non ab\'eliens sont donn\'ees explicitement en fonction de la matrice de multiplicit\'e par
\bbb
g_{i}=(2\pi)^{n/4}
\sqrt{\frac{3/2}{F_{4}\Lambda^{n-4}\mathop{\sum}\limits_{j}|\mu_{ij}|n_{j}}}.
\label{cc4}
\eee
\end{it}


\subsection{Champs de jauge ab\'eliens et conditions d'unimodularit\'e}

L'\'etude des champs de Yang-Mills ab\'eliens est l\'eg\`erement plus compliqu\'ee pour trois raisons. Tout d'abord, contrairement aux champs de non-ab\'eliens, les diverses composantes $G_{\mu\nu}^{i}$ ne sont pas n\'ecessairement orthogonales, c'est pourquoi l'action fait intervenir la matrice $Q_{ij}$. Ensuite, il apparait que certains unitaires de l'alg\`ebre, qui sont dans le noyau de l'application $u\mapsto\pi(u)\jj^{-1}\pi(u)\jj^{-1}$, n'ont pas de terme cin\'etique. De tels champs sont toujours ab\'eliens et nous devons les identifier. Enfin, il arrive que certains champs de jauge ab\'eliens doivent \^etre \'elimin\'es par une condition d'unimodularit\'e appliqu\'ee "\`a la main" pour des raisons physiques, comme par exemple l'annulation des anomalies dans le mod\`ele standard. Par cons\'equent, nous devons adapter le formalisme de l'action spectrale de mani\`ere \`a pouvoir traiter ces trois aspects du probl\`eme. 

\par

Commen\c cons par \'ecrire l'action correspondant aux champs $B_{\mu}^{i}$
\bbb
-\frac{1}{6}\;\frac{1}{(2\pi)^{n/2}}\;F_{4}\;\Lambda^{n-4}\;
\mathop{\sum}\limits_{i,j}
Q_{ij}\;\int_{\mm}d^{4}x\,\sqrt{g}\,
\t(G_{\mu\nu}^{i}G^{j\mu\nu}).\label{cj1}
\eee
o\`u $Q_{ij}$ est une matrice sym\'etrique et positive donn\'ee par (\ref{cc3}).

\par

Param\'etrons lin\'eairement les $N$ champs $B_{\mu}^{i}$ \`a l'aide de $N'\leq N$ champs $C_{\mu}^{i}$, $N$ d\'esignant le nombre de facteurs simples dans la d\'ecomposition de l'alg\`ebre qui donnent naissance \`a un champ de jauge ab\'elien,
\bbb
B_{\mu}^{i}=\mathop{\sum}\limits_{j=1}^{N'}P_{ij}C_{\mu}^{j},
\eee
o\`u $P_{ij}\in M_{N\times N'}(\rrr)$ est une matrice de rang $N'$. Nous avons choisi d'exprimer les champs $B_{\mu}^{i}$ \`a l'aide d'un nombre plus petit de champs $C_{\mu}^{i}$, cette r\'eduction du nombre de degr\'es de libert\'e nous permet d'\'eliminer certains des champs $B_{\mu}^{i}$ en leur imposant des contraintes lin\'eaires. 

\par

A l'aide des nouvelles variables $H_{\mu\nu}^{i}=\partial_{\mu}C_{i}-\partial_{\mu}C_{j}$, la partie ab\'elienne de l'action de Yang-Mills s'\'ecrit
\bbb
\frac{1}{6}\;\frac{1}{(2\pi)^{n/2}}\;F_{4}\;\Lambda^{n-4}\;
\mathop{\sum}\limits_{i,j,k,l}
P_{ki}P_{lj}Q_{kl}\;\int_{\mm}d^{4}x\,\sqrt{g}\,
H_{\mu\nu}^{i}H^{j\mu\nu}.
\eee
Celle-ci doit \^etre identifi\'ee \`a l'action de Yang-Mills usuelle donn\'ee par
\bbb
-\frac{1}{4}\;
\mathop{\sum}\limits_{i}
\int_{\mm}d^{4}x\,\sqrt{g}\,
H_{\mu\nu}^{i}H^{i\mu\nu}.
\eee
Cela impose que les matrices $P$ et $Q$ satisfassent \`a 
\bbb
P^{t}QP=\lambda\, I_{N'},\label{cj2}
\eee
avec
\bbb
\lambda=\frac{3}{2}\,\frac{(2\pi)^{n/2}}{\lambda^{n-4}\,F_{4}}.
\eee
Les charges des diff\'erents fermions peuvent \^etre d\'etermin\'ees en analysant la d\'eriv\'e covariante $\partial_{\mu}+\tilde{A}_{\mu}$, o\`u $\tilde{A}_{\mu}$ contient les champs $C_{\mu}^{i}$ ainsi que les \'el\'ements de matrice $P_{ij}$. Cela permet d'exprimer lin\'eairement les charges en fonction des coefficients $P_{ij}$.

\par

La relation (\ref{cj2}) ne peut \^etre satisfaite que si aucun des vecteurs colonne $(P_{ij})_{1\leq i\leq N}$ n'appartient au noyau de $Q$. Les \'el\'ements du noyau de $Q$ sont des champs  dont le terme cin\'etique est 
identiquement nul.

\par

Le terme cin\'etique des champs de jauge ab\'eliens est proportionnel \`a 
\bbb
\t(\tilde{G}_{\mu\nu}\tilde{G}^{\mu\nu})=\t\lp G_{\mu\nu}+\jj G_{\mu\nu}\jj^{-1}\rp^{2}.
\eee
Il ne peut s'annuler que pour des champs satisfaisant \`a 
\bbb
\tilde{B}_{\mu}=B_{\mu}+\jj B_{\mu}\jj^{-1}=0.
\eee
En utilisant les expressions explicites de $B_{\mu}^{j}$ et de $\jj B_{\mu}\jj^{-1}$, on voit que ceci \'equivaut aux relations 
\bbb
B_{\mu}^{i}=B_{\mu}^{j}\label{cg3}
\eee
pour $\mu_{ij}\neq 0$.

\par

En particulier, tout champ de jauge ab\'elien $B_{\mu}^{i}$ qui n'apparait que sur la diagonale de la matrice de multiplicit\'e (i.e. $\mu_{ij}=0$ si $i\neq j$), n'a pas de terme cin\'etique. De m\^eme, si le mod\`ele ne comprend que des alg\`ebres complexes avec des repr\'esentations complexes, l'\'equation (\ref{cg3}) admet une solution non triviale donn\'ee par $B_{\mu}^{i}=\lambda_{i}$. Un tel champ appartient n\'ecessairement au noyau de $Q$ et n'a pas de terme cin\'etique. Dans le cas o\`u le mod\`ele contient des quaternions ou des repr\'esentations r\'eelles, comme c'est le cas pour le mod\`ele standard, un tel champ n'est plus n\'ecessairement solution de (\ref{cg3}).

\par

Les champs qui satisfont \`a $B_{\mu}+\jj B_{\mu}\jj^{-1}=0$ correspondent, au niveau infinit\'esimal, aux \'el\'ements du groupe de jauge situ\'es dans le noyau de la repr\'esentation $u\mapsto \pi(u)\jj\pi(u)\jj^{-1}$. Par cons\'equent, il ne jouent absolument aucun r\^ole dans la th\'eorie et peuvent \^etre totalement ignor\'es.

\par

Terminons cette section par une discussion du nombre de possibilit\'es d'imposer des conditions d'unimodularit\'e pour r\'eduire de $N$ \`a $N'$ le nombre de champs de jauge ab\'eliens. La matrice $P_{ij}$ a $NN'$ coefficients et elle satisfait \`a $1/2\,N'(N'-1)$ conditions, puisqu'elle permet de r\'eduire une matrice sym\'etrique \`a une matrice scalaire $N'\times N'$. De plus, toute rotation dans l'espace de dimension $N'$ engendr\'e par les champs $C_{\mu}^{i}$ laissera l'action invariante et n'a pas de signification physique. Comme il y a $1/2\,N'(N'-1)$ telles rotations, le nombre de param\`etres ind\'ependants jouant un r\^ole physique dans la matrice $P$ est donn\'e par
\bbb 
NN'-\frac{N'(N'+1)}{2}-\frac{N'(N'-1)}{2}=N'(N-N').
\eee
Par exemple, dans le cas du mod\`ele standard le groupe des unitaires de l'alg\`ebre est $SU(2)\times U(1)\times U(3)$, alors que le groupe de jauge du mod\`ele standard est $SU(2)\times U(1)\times SU(3)$. On v\'erifie qu'il n'y a pas de solutions non triviales aux \'equations (\ref{cg3}), ce qui implique que nous devons \'eliminer un des facteurs $U(1)$. Puisque les quaternions ne contribuent pas au secteur ab\'elien, nous avons $N=2$ et $N'=1$, donc il n'y a qu'un seul degr\'e de libert\'e dans la mani\`ere d'\'eliminer le champ de jauge ab\'elien superflu.

\bigskip

\noindent
{\bf Conclusion}
\begin{it}
Le nombre de champs de jauge ab\'eliens est r\'eduit de $N$ \`a $N'$ par l'interm\'ediaire d'une matrice $N\times N'$. Cette matrice contient $N'(N-N')$ param\`etres jouant un r\^ole physique \`a partir desquels on peut exprimer les charges des diff\'erents fermions de la th\'eorie. 
\end{it}


\subsection{Anomalies}

Lorsque des fermions chiraux sont coupl\'es \`a des th\'eories de jauge ou \`a la gravitation, certaines des sym\'etries pr\'esentes au niveau classique de la th\'eorie ne sont plus des sym\'etries de la th\'eorie quantique (voir \cite{ber} et les r\'ef\'erences contenues \`a l'int\'erieur). Au cours de cette section, nous nous int\'eresserons essentielement \`a deux types d'anomalies: les anomalies de jauge et les anomalies mixtes gravitationnelles \cite{alv}.

\par

Ces derni\`eres apparaissent lorsque des fermions de Weyl sont coupl\'es \`a une th\'eorie de jauge et au champ gravitationnel. La condition d'absence d'anomalie mixte s'exprime \`a l'aide du contenu fermionique de la th\'eorie par
\bbb
\mathop{\sum}\limits_{fermions\,droits}\t(T^{a})-
\mathop{\sum}\limits_{fermions\,gauches}\t(T^{a})=0,
\eee
o\`u $T^{a}$ d\'esigne les g\'en\'erateurs de l'alg\`ebre de Lie du groupe de jauge.

\par

Dans le cadre des mod\`eles que nous construisons, cette relation peut se mettre sous la forme
\bbb
\t_{P}(\chi\tilde{A}_{\mu})=0,\label{a1}
\eee
o\`u $\t_{P}$ est la trace restreinte \`a l'espace des particules et $\tilde{A}_{\mu}$ un champ de jauge. 

\par

Pour simplifier notre discussion, supposons que le mod\`ele satisfasse \`a la condition de $S^{0}$-r\'ealit\'e, ce qui permet de s\'eparer les particules et les antiparticules. Dans ce cas, la matrice de multiplicit\'e se r\'e\'ecrit sous la forme $\mu_{ij}=\epsilon_{ij}+\epsilon_{ji}$, la composante $\epsilon_{ij}$ correspondant aux particules. Dans le cas g\'en\'eral, seuls les champs de jauge ab\'eliens contribuent \`a l'anomalie mixte et nous faisons l'hypoth\`ese suppl\'ementaire que ceux-ci sont donn\'es par un triplet spectral complexe. Dans le cas r\'eel, il faut distinguer la repr\'esentation fondamentale de $M_{n}(\ccc)$ de sa conjugu\'ee complexe, ce qui ne change pas le principe du calcul mais nous oblige \`a introduire des notations relativement lourdes. Avec ces hypoth\`eses, la relation (\ref{a1}) devient
\bbb
\mathop{\sum}\limits_{i}\lp\epsilon_{ij}-\epsilon_{ji}\rp n_{i}\,B_{\mu}^{j}=0,\label{a2}
\eee
o\`u $B_{\mu}$ d\'esigne le champ ab\'elien associ\'e \`a $A_{\mu}$ (cf \S 3.2.2).
La solution la plus simple de (\ref{a2}) consiste \`a choisir une matrice $\epsilon$ sym\'etrique. Dans le cas contraire, la relation (\ref{a2}) est une relation lin\'eaire qui doit \^etre impos\'ee aux champs de jauge ab\'eliens, par exemple par l'interm\'ediaire de la matrice $P$ que nous avons introduite au cours de la section pr\'ec\'edente. 

\par

Le second type d'anomalies que nous allons \'etudier sont les anomalies de jauge. Celles-ci peuvent appara\^\i tre lorsque des fermions chiraux sont coupl\'es \`a des champs de Yang-Mills et peuvent poser un probl\`eme particuli\`erement s\'erieux car elles menacent les identit\'es de Slavnov et mettent en p\'eril la renormalisabilit\'e de la th\'eorie. La condition d'absence de ces anomalies est
\bbb
\mathop{\sum}\limits_{fermions\,droits}\t\lp T^{a}\la T^{b},T^{c}\ra\rp-
\mathop{\sum}\limits_{fermions\,gauches}\t\lp T^{a}\la T^{b},T^{c}\ra\rp=0,\label{a3}
\eee
o\`u $\la T^{b},T^{c}\ra=T^{b}T^{c}+T^{c}T^{b}$. 

\par

Nous cherchons \`a \'eliminer ces anomalies en utilisant une condition d'unimodularit\'e, c'est-\`a-dire en imposant une ou plusieurs contraintes lin\'eaires aux champs ab\'eliens. Dans le cas du mod\`ele standard, l'absence d'anomalie de jauge est \'equivalente \`a une condition d'unimodularit\'e \cite{ano}. 

\par

En g\'eneral, la condition (\ref{a3}) est difficile \`a r\'esoudre car elle est non lin\'eaire. Cependant, pour nos mod\`eles, elle se reformule en
\bbb
\t_{P}[\chi(\tilde{A}_{\mu})^{3}]=0.\label{a4}
\eee
De cette expression, il ressort que les matrices orthogonales et symplectique ne contribuent pas \`a l'anomalie car si une telle matrice est antihermitienne, ses \'el\'ement diagonaux sont nuls. Ceci est \'egalement vrai pour sa puissance troisi\`eme et (\ref{a4}) est v\'erifi\'ee. 

\par

Les seules contributions \`a l'anomalie de jauge proviennent donc des matrices complexes. En supposant que le triplet spectral soit $S^{0}$-r\'eel et complexe, la relation (\ref{a4}) est \'equivalente au syst\`eme suivant.
\bbbb
&\mathop{\sum}\limits_{j}
n_{j}\lp\epsilon_{ij}-\epsilon_{ji}\rp=0
\;\;\mathrm{si}\;n_{i}\geq3,&\\
&\mathop{\sum}\limits_{j}
n_{j}\lp\epsilon_{ij}-\epsilon_{ji}\rp\lp B^{i}_{\mu}-B^{j}_{\mu}\rp=0
\;\;\mathrm{si}\;n_{i}\geq2,&\\
&\mathop{\sum}\limits_{ij}
n_{i}n_{j}\lp\epsilon_{ij}-\epsilon_{ji}\rp\lp B^{i}_{\mu}-B_{\mu}^{j}\rp^{3}=0.
&
\eeee
Ces \'equations sont obtenues en rempla\c cant $\tilde{A_{\mu}}$ par son expression en fonction des champs de jauge ab\'eliens $B_{\mu}^{i}$ et non ab\'eliens $A_{\mu}^{i}$ et en imposant l'absence d'anomalie pour les champs non ab\'eliens, car nous ne voulons \'eliminer que des champs ab\'eliens, 

\par

Une fois de plus, il apparait que si la matrice $\epsilon_{ij}$ est sym\'etrique, ces relations sont satisfaites et il n'y a pas d'anomalie.

\par

La premi\`ere de ces relations est une contrainte rigide impos\'ee \`a la matrice de multiplicit\'e. Si elle n'est pas satisfaite, on ne peut obtenir un mod\`ele sans anomalie par application d'une condition d'unimodularit\'e. Remarquons qu'elle ne fait intervenir que les matrices d'ordre $n>2$, car dans le cas $n=2$, on est ramen\'e au groupe $SU(2)$ qui ne contribue pas \`a l'anomalie puisque $\t[(A_{\mu})^{3}]=0$ dans ce cas.

\par

La seconde de ces relations est une contrainte lin\'eaire qu'il faut imposer aux champs ab\'eliens. Elle est donc \'equivalente \`a une condition d'unimodularit\'e.

\par

Enfin, la derni\`ere relation est une contrainte d'ordre 3 qui n'est pas, en g\'eneral, \'equivalente \`a une condition d'unimodularit\'e. Dans le cas particulier du mod\`ele standard, elle est donn\'ee par la relation
\bbb
\sum Y_{R}^{3}-\sum Y_{L}^{3}=0
\eee
entre les hypercharges $Y_{R}$ des fermions droits et $Y_{L}$ des fermions gauches.

\bigskip

\noindent
{\bf Conclusion}
\begin{it}
La condition d'absence d'anomalies mixtes gravitationnelles (\ref{a1}) peut toujours \^etre satisfaite par une condition d'unimodularit\'e en imposant les contraintes lin\'eaires (\ref{a2}) sur les champs ab\'eliens. Ceci n'est plus le cas pour les anomalies de jauge (\ref{a4}), car la condition d'absence de telles anomalies impose, outre une condition d'unimodularit\'e, des contraintes rigides sur la matrice de multiplicit\'e et une relation cubique sur les champs ab\'eliens. 
\end{it}


\section{Les champs scalaires}

\subsection{Les champs scalaires et leurs lois de transformation}

L'op\'erateur de Dirac covariant $\dd_{A_{\mu},\Phi}$, contient, outre les champs de jauge $A_{\mu}$ que nous venons d'\'etudier, un champ scalaire $\Phi$. Ce champ scalaire \`a la propri\'et\'e particuli\`ere d'\^etre \`a la fois une 0-forme pour la g\'eom\'etrie de l'espace-temps et une 1-forme pour la g\'eom\'etrie de l'espace interne. Plus pr\'ecis\'ement, c'est une fonction sur l'espace-temps \`a valeurs dans l'espace des 1-formes hermitiennes associ\'ees au triplet spectral fini. Puisque les d\'eriv\'ees par rapport au coordonn\'ees n'interviennent pas dans la discussion qui suit, la d\'ependance en $x$ de ce champ scalaire ne joue aucun r\^ole et nous nous concentrerons exclusivement sur la g\'eom\'etrie du triplet spectral fini et son interpr\'etation physique.

\par  

En g\'en\'eral, un \'el\'ement $\Phi\in\Omega_{\dd}^{1}(\aa)$ est une matrice de taille importante ($90\times 90$ dans le cas du mod\`ele standard) dont beaucoup d'\'el\'ements de matrice entre les diff\'erents sous-espaces sont nuls.
Pour interpr\'eter $\Phi$ comme un champ scalaire, nous devons d\'eterminer quels sont les \'el\'ements de matrice non nuls et les param\'etrer par un certain nombre de matrices qui seront les v\'eritables champs scalaires de la th\'eorie.

\par

Commen\c cons par rappeler quelques r\'esultats de la section 2.3.2 qui nous seront utiles. L'espace des 1-formes est donn\'e par
\bbb
\Omega_{\dd}^{1}(\aa)=\la\mathop{\sum}\limits_{ijkp}
P_{ik}^{*}\lp \omega_{ij}^{p}\ot M_{ij,k}^{p}\ot I_{n_{k}}\rp P_{jk}
\;|\;\omega_{ij}^{p}\in M_{n_{i}\times n_{j}}(\ccc)\ra,
\eee
o\`u $P_{ij}=\pi(1_{i})\jj\pi(1_{j})\jj^{-1}$, $1_{i}$ d\'esignant l'unit\'e de la sous-alg\`ebre $M_{n_{i}}(\kkk)$. Les matrices $M_{ij,k}^{p}$, satisfaisant \`a $\lp M_{ij,k}^{p}\rp^{*}=M_{ji,k}^{p}$, sont choisies telle que les vecteurs
\bbb
M_{ij}^{p}=\lp M_{ij,1}^{p},\dots,M_{ij,N}^{p}\rp
\eee   
forment une base d'un certain sous-espace. L'indice $p$, qui varie de 1 \`a $r_{ij}$, permet de distinguer les multiplicit\'es apparaissant dans les champs scalaires. De plus, il est toujours possible, par application du proc\'ed\'e de Gram-Schmidt, de supposer que ces matrices satisfont \`a
\bbb
\mathop{\sum}\limits_{k=1}^{N}\t\lb\lp M_{ij,k}^{p}\rp^{*}M_{ij,k}^{p'}\rb n_{k}
=X\delta_{p,p'},\label{cs3}
\eee
o\`u $X>0$ est un coefficient de normalisation.

\par

En toute rigueur, ces r\'esultats ne sont applicables que dans le cas d'un triplet spectral complexe. Lorsque la repr\'esentation est r\'eelle, la m\^eme param\'etrisation de l'espace des 1-formes est valide si les indices $i$, $j$ et $k$ distinguent les repr\'esentations irr\'eductibles de $\aa$. Bien entendu, dans ce cas les matrices $\omega_{ij}^{p}$ ne sont \`a coefficients complexes que si l'un au moins des indices $i,j$ correspond \`a une repr\'esentation d'un facteur complexe. Si les deux indices correspondent \`a des repr\'esentations de facteur r\'eels ou quaternioniques, les matrices sont r\'eelles ou quaternionique et dans le cas mixte o\`u un indice est quaternionique et l'autre r\'eel, la matrice est quaternionique. Enfin, si $i$ est un indice r\'eel ou quaternionique et $j$ un indice complexe, $\theta_{i\ov{j}}$ est le conjugu\'e complexe ou quaternionique de $\theta_{ij}$. 

\par

\'Etant donn\'e que $\Phi$ est une 1-forme hermitienne, on peut \'ecrire
\bbb
\Phi=\mathop{\sum}\limits_{ijkp}
P_{ik}^{*}\lp \Phi_{ij}^{p}\ot M_{ij,k}^{p}\ot I_{n_{k}}\rp P_{jk},\label{cs1}
\eee
o\`u $\Phi_{ij}^{p}\in M_{n_{i}\times n_{j}}(\kkk)$ satisfait \`a $\lp\Phi_{ij}^{p}\rp^{*}=\Phi_{ji}^{p}$. 

\par

Les champs $\Phi_{ij}^{p}$ sont identifi\'es comme les v\'eritables champs scalaires de la th\'eorie. Il portent trois indices, deux indices $i$ et $j$ correspondent aux repr\'esentation irr\'eductibles de l'alg\`ebre $\aa$, alors que le troisi\`eme est un indice de multiplicit\'e. Les deux indices $i$ et $j$ peuvent \^etre interpr\'et\'es diagrammatiquement: le champ $\Phi_{ij}^{p}$ correspond \`a l'ensemble des liens verticaux entre les sommets de la ligne $i$ et les sommets de la ligne $j$.

\par

L'op\'erateur de Dirac covariant contient \'egalement le champ $\jj\Phi\jj^{-1}$, qui s'exprime \`a l'aide des matrices $\Phi_{ij}^{p}$ par
\bbb
\jj\Phi\jj^{-1}=\mathop{\sum}\limits_{ijkp}
P_{ki}^{*}\lp I_{n_{k}}\ot \ov{M}_{ij,k}^{p}\ot \ov{\Phi}_{ij}^{p}\rp P_{kj},
\label{cs2}
\eee
Sur le plan diagrammatique, ses \'el\'ements de matrice correspondent aux diff\'erents liens horizontaux obtenus par sym\'etrie par rapport \`a la diagonale des liens associ\'es \`a $\Phi$.

\par

Sous la transformation de jauge d\'etermin\'ee par l'unitaire $u$, le champ $\Phi$ devient $u\Phi u^{-1}$. Si nous \'ecrivons $u$ \`a l'aide de ses composantes dans les facteurs simple du groupe de jauge, $u=(u_{1},\dots,u_{N})$, ce qui nous donne les lois de transformation des champs $\Phi_{ij}^{p}$
\bbb
\Phi_{ij}\mapsto u_{i}\,\Phi_{ij}^{p}\, u_{j}^{-1}.
\eee
Une fois encore, ces relations ne sont vraies en toute rigueur que dans le cas complexe. Dans le cas r\'eel, elles restent valides si on suppose que $u_{i}$ d\'esigne la composante de $u$ dans la repr\'esentation associ\'ee \`a l'indice $i$. Dans tous les cas, si nous notons $n_{i}$ cette repr\'esentation, le champ $\Phi_{ij}^{p}$ appartient \`a la repr\'esentation $n_{i}\ot\ov{n}_{j}$. 

\par

Rappelons que sur le diagramme associ\'e au triplet spectral fini, les champs $\Phi_{ij}^{p}$ correspondent aux liens verticaux entre les lignes $i$ et $j$. \'Etant donn\'e que nous ne pouvons relier un sommet \a lui-m\^eme, nous ne pouvons relier une ligne \`a elle-m\^eme par un lien vertical. Par cons\'equent, il n'y a jamais de champ scalaire $\Phi_{ij}^{p}$ avec $i=j$, qui se transformerait dans la repr\'esentation adjointe $n_{i}\ot\ov{n}_{i}$. En revanche, pour tous indices $i$ et $j$ distincts, on peut construire un mod\`ele contenant des champs scalaires $\Phi_{ij}^{}$ qui se transforment dans la repr\'esentation $n_{i}\ot\ov{n}_{j}$.

\bigskip

\noindent
{\bf Conclusion}
\begin{it}
La param\'etrisation du champ scalaire $\Phi$  par des matrices $\Phi_{ij}^{p}\in M_{n_{i}\times n_{j}}(\kkk)$ est donn\'ee par la relation (\ref{cs1}). Les champs $\Phi_{ij}^{p}$ se transforment sous les repr\'esentations $n_{i}\ot\ov{n}_{j}$, \`a l'exception de la repr\'esentation adjointe $n_{i}\ot\ov{n}_{i}$, l'indice $p$ \'etant un indice de multiplicit\'e qui distingue des champs ayant des lois de transformation identiques. Sur le diagramme, $\Phi_{ij}^{p}$ correspond aux liens verticaux entre les lignes $i$ et $j$, alors que $\jj\Phi\jj^{-1}$, donn\'e par (\ref{cs2}) est form\'e des liens sym\'etriques par rapport \`a la diagonale.
\end{it}


\subsection{Couplages de Yukawa}

Le couplage de Yukawa entre le champ fermionique $\Psi$ et le scalaire $\Phi$ est donn\'e par
\bbb
\langle\Psi,\lp\Phi+\jj\Phi\jj^{-1}\rp\Psi\rangle.\label{cy1}
\eee
L'espace de Hilbert $\hh$ \'etant une somme directe orthogonale $\hh=\mathop{\op}\limits_{i,j}\hh_{ij}$, o\`u $\hh_{ij}=P_{ij}\hh$. En introduisant la relation de fermeture pour les projecteurs $P_{ij}$ dans (\ref{cy1}), le couplage de Yukawa s'\'ecrit
\bbb
\mathop{\sum}\limits_{i,j,k,l}
\langle P_{ij}\Psi,P_{ij}\lp\Phi+\jj\Phi\jj^{-1}\rp P_{kl}\Psi\rangle
\eee
Par cons\'equent, on peut interpr\'eter 
\bbb
P_{ij}\lp\Phi+\jj\Phi\jj^{-1}\rp P_{kl}
\eee
comme un couplage de Yukawa entre les fermions de $\hh_{ij}$ et ceux de $\hh_{kl}$.

\par

En utilisant les relations  (\ref{cs1}) et (\ref{cs2}), on peut expliciter ce couplage en fonction des champs scalaires $\Phi_{ij}^{p}$
\bbbb
P_{ij}\lp\Phi+\jj\Phi\jj^{-1}\rp P_{kl}&=&
\delta_{k,l}\,\mathop{\sum}\limits_{p}
\Phi_{ij}^{p}\ot M_{ij,k}^{p}\ot I_{n_{k}}\n\\
&+&\delta_{i,j}\,\mathop{\sum}\limits_{p}
I_{n_{i}}\ot \ov{M}_{kl,i}^{p}\ot \ov{\Phi}_{kl}^{p}.\label{cy2}
\eeee
Sur le plan diagrammatique, ces termes sont les \'el\'ements de matrice de l'op\'erateur $\Phi+\jj\Phi\jj^{-1}$ entre les sous espaces $\hh_{ij}$ et $\hh_{kl}$ et correspondent aux liens entre les diff\'erents sommets. Un lien vertical entre les sommets de coordonn\'ees $(i,k)$ et $(j,k)$ est associ\'e \`a un couplage de Yukawa entre les fermions de $\hh_{ik}$ et de $\hh_{jk}$ et les champs scalaires $\Phi_{ij}^{p}$. De m\^eme, un lien horizontal entre $(k,i)$ et $(k,j)$ correspond \`a un couplage de Yukawa entre les fermions de $\hh_{ki}$ et de $\hh_{kj}$ et les champs scalaires $\ov{\Phi}_{ij}^{p}$.

\bigskip

\noindent
{\bf Conclusion}
\begin{it}
Les sommets du diagramme associ\'e au triplet spectral fini correspondent \`a des fermions chiraux entre lesquels les liens symbolisent des couplages de Yukawa avec le champ scalaire correspondant. Les expressions explicites de ces couplages sont donn\'ees par la relation (\ref{cy2}).
\end{it}


\subsection{Le potentiel scalaire}

le champ scalaire $\Phi$ n'apparait dans le carr\'e de l'op\'erateur de Dirac covariant que par l'interm\'ediaire de l'endomorphisme
\bbb
E=-\frac{1}{4}\rr-\frac{1}{2}\gamma^{\mu\nu}\tilde{F}_{\mu\nu}+
i\gamma^{\mu}\gamma^{n+1}D_{\mu}\tilde{\Phi}+\tilde{\Phi}^{2}
\eee
avec $\tilde{\Phi}=\Phi+\jj\Phi\jj^{-1}$. En utilisant les expressions explicites des coefficients $a_{2n}(\dd^{2})$ pour $n=1$ ,$2$ et $3$ \cite{gilkey}, il est facile de voir que les seuls termes qui vont contribuer  au potentiel scalaire sont $\t(E^{n})$, tous les autres termes impliquant le scalaire sont des couplages entre celui-ci et d'autres champs. Chacun de ces trois termes donnera naissance \`a une contribution au potentiel scalaire qui est proportionnelle aux mon\^omes $\t(\tilde{\Phi})$, $\t(\tilde{\Phi}^{2})$ et $\t(\tilde{\Phi}^{3})$. 

\par

De fa\c con g\'en\'erale, nous nous int\'eresserons aux mon\^omes du type $\t(\tilde{\Phi}^{n})$, ce qui nous permet de construire le potentiel scalaire jusqu'en dimension 6.
 
\par

En introduisant les \'el\'ements de matrice $\tilde{\Phi}_{ij}^{kl}$ de $\tilde{\Phi}$ d\'efinis par $\tilde{\Phi}_{ij}^{kl}=P_{ij}\tilde{\Phi}P_{kl}$,
on a
\bbb
\t(\tilde{\Phi}^{n})=\mathop{\sum}\limits_{i_{1},\dots,i_{n}\atop
j_{1},\dots,j_{n}}
\t\lp\tilde{\Phi}_{i_{1}j_{1}}^{i_{2}j_{2}}
\dots
\tilde{\Phi}_{i_{n-1}j_{n-1}}^{i_{n}j_{n}}
\tilde{\Phi}_{i_{n}j_{n}}^{i_{1}j_{1}}\rp.\label{ps1}
\eee
Les \'el\'ements de matrices $\tilde{\Phi}_{ij}^{kl}$ correspondent \`a des liens entre les sommets $(i,j)$ et $(k,l)$ et la suite $(i_{1},j_{1})\rightarrow (i_{2},j_{2})\rightarrow\dots\rightarrow(i_{n},j_{n})
\rightarrow (i_{1},j_{1})$ est une boucle de longueur $n$ sur le diagramme. Par cons\'equent, le second membre de (\ref{ps1}) s'interpr\`ete comme une somme sur toutes les boucles de longueur $n$ du diagramme.

\par

Bien entendu, cette somme contient des boucles du type $(i,k)\rightarrow (j,k)\rightarrow (i,k)$ obtenues en parcourant un chemin dans un sens puis dans l'autre. De m\^eme, nous devons pour le moment distinguer une boucle de celle parcourue dans le sens oppos\'e, ainsi que des boucles ne diff\'erant que par leur point de d\'epart. 

\par

Tous les liens du diagramme sont soit horizontaux, soit verticaux. Cela  entra\^\i ne que toute boucle de longueur $n$ est form\'ee de $r$ liens verticaux et de $s$ liens horizontaux avec $r+s=n$. Puisqu'il ne peut y avoir de lien vertical entre sommets d'une m\^ eme ligne ou de lien horizontal entre sommets d'une m\^eme colonne, les entiers $r$ et $s$ sont n\'ecesairement pairs.  

\par

Les \'el\'ements de matrice de $\tilde{\Phi}$ sont donn\'es par
\bbb
\tilde{\Phi}_{ij}^{kl}=\delta_{k,l}\,\mathop{\sum}\limits_{p}
\Phi_{ij}^{p}\ot M_{ij,k}^{p}\ot I_{n_{k}}\n\\
+\delta_{i,j}\,\mathop{\sum}\limits_{p}
I_{n_{i}}\ot \ov{M}_{kl,i}^{p}\ot \ov{\Phi}_{kl}^{p}.
\eee
A toute boucle $\gamma$ form\'ee de $r$ liens verticaux correspondant \`a la suite $i_{1}\rightarrow i_{2}\rightarrow \dots\rightarrow i_{r}$ et $s$ liens horizontaux associ\'es \`a $j_{1}\rightarrow j_{2}\rightarrow \dots\rightarrow j_{s}$ correspond un terme du potentiel scalaire donn\'e par
\bbb
\t\lp\Phi_{i_{1}i_{2}}^{p_{1}}\Phi_{i_{2}i_{3}}^{p_{2}}
\dots\Phi_{i_{r}i_{1}}^{p_{r}}\rp
\t\lp\ov{\Phi}_{j_{1}j_{2}}^{q_{1}}\ov{\Phi}_{j_{2}j_{3}}^{q_{2}}
\dots\ov{\Phi}_{j_{s}j_{1}}^{q_{s}}\rp
\lambda_{\gamma}^{p,q}.\label{ps3}
\eee
Le coefficient $\lambda_{\gamma}^{p,q}$ est un nombre complexe qui d\'epend des multiplicit\'es $p$ et $q$ et du d\'etail de la suite form\'ee par les sommets de $\gamma$. On peut l'\'ecrire comme une trace,
\bbb
\lambda_{\gamma}^{p,q}=\t M_{\gamma}^{p,q},
\eee
o\`u $M_{\gamma}^{p,q}$ est une matrice construite \`a l'aide de $\gamma$, en multipliant, dans l'ordre d\'etermin\'e par $\gamma$, les matrices 
\begin{itemize}
\item
$M_{i_{m},i_{m+1},j_{m}}^{p_{m}}$ pour chaque lien vertical entre $(i_{m},j_{m})$ et $(i_{m+1},j_{m})$,
\item
$\ov{M}_{j_{n},j_{n+1},i_{n}}^{q_{n}}$ pour chaque lien horizontal entre $(i_{n},j_{n})$ et $(i_{n},j_{n+1})$.
\end{itemize}

\par

Par exemple, s'il n'y a pas de multiplicit\'e, \`a la suite
\bbb
(i_{1},j_{1})\rightarrow
(i_{2},j_{1})\rightarrow
(i_{3},j_{1})\rightarrow
(i_{3},j_{2})\rightarrow
(i_{2},j_{2})\rightarrow
(i_{1},j_{2})\rightarrow
(i_{1},j_{1})
\eee
correspond le mon\^ ome
\bbb
\t\lp\Phi_{i_{1}i_{2}}\Phi_{i_{2}i_{3}}\Phi_{i_{3}i_{2}}\Phi_{i_{2}i_{1}}\rp
\t\lp\ov{\Phi}_{j_{1}j_{2}}\ov{\Phi}_{j_{2}j_{1}}\rp
\eee
affect\'e du coefficient
\bbb
\t\lp M_{i_{1}i_{2},j_{1}}M_{i_{2}i_{3},j_{1}}\ov{M}_{j_{1}j_{2},i_{3}}
M_{i_{3}i_{2},j_{2}}M_{i_{2}i_{1},j_{2}}\ov{M}_{j_{2}j_{1},i_{1}}\rp.
\eee
Pour simplifier nos notations, nous omettrons les indices de multiplicit\'e $p$ et $q$. Dans toute somme sur des champs scalaires, la somme sur $p$ et $q$ sera sous entendue.

\par

Etant donn\'e la loi de transformation des champs scalaires,
\bbb
\Phi_{ij}\mapsto u_{i}\Phi_{ij}u_{j}^{-1},\label{ps2},
\eee
le potentiel invariant de jauge le  plus g\'en\'eral est une fonction polynomiale des variables
\bbb
X_{i_{1},\dots,i_{n}}=\t\lp\Phi_{i_{1}i_{2}}\Phi_{i_{2}i_{3}}
\dots\Phi_{i_{n}i_{1}}\rp.
\eee 
Puisque le champ $\phi_{ii}$ n'existe pas, l'entier $n$ est pair.

\par

En utilisant l'action spectrale, nous ne pouvons pas construire le potentiel scalaire le plus g\'en\'eral compatible avec les lois de transformation (\ref{ps2}). En effet, le terme du potentiel scalaire donn\'e par (\ref{ps3}) est un de polyn\^ome de degr\'e au plus 2 en chacune des variables $X_{i_{1},\dots,i_{n}}$.

\par

Lorsque le potentiel est un polyn\^ome de degr\'e $\leq4$ en $\Phi$, cela n'impose aucune contrainte sur la forme des mon\^omes, car les mon\^omes les plus g\'en\'eraux invariants de jauge sont $1$, $\t( \Phi_{ij}\Phi_{ji})$, $\t( \Phi_{ij}\Phi_{jk}\Phi_{kl}\Phi_{li})$ et $\t( \Phi_{ij}\Phi_{ji})\t(\Phi_{kl}\Phi_{lk}) $, qui apparaissent tous dans le d\'eveloppement de l'action spectrale.

\par

En revanche, le mon\^ omes de degr\'e 6 les plus g\'en\'eraux sont $\t( \Phi^{6})$, $\t( \Phi^{4}) \t(\Phi^{2})$ et $\t( \Phi^{2})^{3}$, alors que seuls les deux premiers peuvent appara\^\i tre dans l'action spectrale.

\par

Afin de rendre plus facile la construction de mod\`eles de Yang-Mills-Higgs, il est utile de donner des r\`egles simplifi\'ees pour la construction des mon\^omes de degr\'e inf\'erieur ou \'egal \`a 4. En effet, le nombre de boucles de longueur donn\'ee devient rapidement tr\`es grand lorsque le nombre de sommets augmente, mais la plupart de ces boucles se d\'eduisent les une des autres par des transformations simples comme les changements d'orientation ou d'origine. De plus, la majorit\'e de ces boucles sont en fait obtenues en parcourant un certain chemin dans un sens puis dans le sens oppos\'e. 

\par

Commen\c cons par le terme de masse. Celui-ci est obtenu \`a l'aide des boucles de longueur 2 qui sont toutes du type suivant: 

\begin{figure}[H]
\centering
\epsfig{file={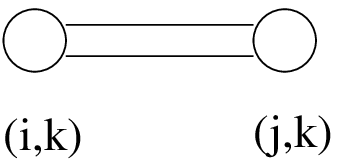},width=2.5cm}
\end{figure}

Ces boucles correspondent directement aux liens non orient\'es du diagramme. Chacun de ces liens apporte une contribution
\bbb
2n_{k}\,
\t(M_{ij,k}M^{*}_{ij,k})\,
\t(\Phi_{ij}\Phi^{*}_{ij})\label{ps4}
\eee
au terme de masse. Bien entendu, nous avons omis les indices de multiplicit\'e et (\ref{ps4}) contient une somme sur toutes les multiplicit\'es. 

\par

Si nous tenons compte de toutes les contributions au terme de masse pour le champ scalaire $\Phi_{ij}^{p}$, celui-ci s'\'ecrit
\bbb
\mathop{\sum}\limits_{k,p,q} 2n_{k}\,
\t\lp M_{ij,k}^{p}\lp M^{q}_{ij,k}\rp^{*}\rp\,
\t(\Phi_{ij}^{p}\lp\Phi^{q}_{ij}\rp^{*}).
\eee 
Compte tenu de la relation (\ref{cs3}) qui s'\'ecrit
\bbb
\mathop{\sum}\limits_{k=1}^{N}\t\lb\lp M_{ij,k}^{p}\rp^{*}M_{ij,k}^{p'}\rb n_{k}
=X\delta_{p,p'},
\eee
les termes de masse $\t(\tilde{\Phi}^{2})=2\t(\Phi^{2})$ sont donn\'es par
\bbb
\t(\tilde{\Phi}^{2})=4X\,\mathop{\sum}\limits_{scalaires}
\t(\Phi_{ij}^{p}\lp\Phi^{q}_{ij}\rp^{*}),
\eee
o\`u la somme sur tous les champs scalaires correspond \`a une somme sur tous les indices de multiplicit\'e $p$ et toutes les paires $(i,j)$ telles que le champ $\Phi_{ij}$ existe.

\par

Ce coefficent $X$ est d\'etermin\'e par la normalisation du terme cin\'etique des champs scalaires. D'apr\`es le principe d'action spectrale, celui-ci est, en dimension n, 
\bbb
\frac{1}{2}\;\frac{1}{(2\pi)^{\frac{n}{2}}}\;
 \t(\partial_{\mu}\tilde{\Phi}\partial^{\mu}\tilde{\Phi})\,F_{4}\,\Lambda^{n-4}.
\eee
Il doit \^etre identifi\'e au terme cin\'etique usuel des scalaires
\bbb
\mathop{\sum}\limits_{scalaires}\frac{1}{2}
\t(\partial_{\mu}\lb\Phi_{ij}^{p}\lp\partial^{\mu}\Phi_{ij}^{p}\rp^{*}\rb,
\eee
ce qui donne
\bbb
X=\frac{(2\pi)^{\frac{n}{2}}}{4\Lambda^{n-4}F_{4}}.\label{ps5}
\eee
Le terme de masse des champs scalaires est donn\'e par le coefficient $a_{2}(\dd^{2})$,
\bbb
-\frac{1}{(2\pi)^{\frac{n}{2}}}F_{2}\,\Lambda^{n-2}\,\t(\tilde{\Phi}^{2}).
\eee
Compte-tenu de la condition de normalisation (\ref{ps5}), cela donne
\bbb
-\mathop{\sum}\limits_{scalaires}\frac{1}{2}\mu^{2}
\t\lp \Phi_{ij}^{p}\lp\Phi_{ij}^{p}\rp^{*}\rp
\eee 
avec
\bbb
\mu=\sqrt{\frac{2F_{2}}{F_{4}}}\Lambda.\label{ps6}
\eee
Par cons\'equent, les champs scalaires ont tous des termes de masse identiques. Cependant, nous allons voir au cours de la section suivante qu'il y a toujours brisure spontan\'ee de sym\'etrie, ce terme n'est donc pas un v\'eritable terme de masse. La brisure spontan\'ee de sym\'etrie peut entrainer des modifications des masses des scalaires, comme l'illustre l'exemple trait\'e dans \cite{spe}. 

\par 

La condition de normalisation (\ref{ps5}) permet aussi de d\'eterminer le couplage entre les champs scalaires et le tenseur de courbure donn\'e par (\ref{tg6}) en dimension $n=4$,
\bbb
\frac{1}{12}\,\frac{1}{(2\pi)^{\frac{n}{2}}}\,F_{4}\Lambda^{n-4}
\,\rr\,\t(\tilde{\Phi}^{2}).
\eee
En utilisant l'expression explicite de $\t( \tilde{\Phi}^{2})$ et la condition de normalisation (\ref{ps5}), on obtient 
\bbb
\frac{1}{12}\,\rr\mathop{\sum}\limits_{scalaires}
\t\lp\Phi_{ij}^{p}\lp\Phi_{ij}^{p}\rp^{*}\rp,
\eee
ce qui d\'emontre la relation (\ref{tg5}). 

\par

Les couplages quartiques des champs scalaires sont donn\'es par les boucles de longueur 4. L'analyse de tous les diagrammes possibles permet de nous ramener \`a une sommation sur les 5 types de sous-diagrammes suivants. Leur contribution au potentiel scalaire inclut un facteur num\'erique correspondant aux diff\'erentes possibilit\'es de choix de l'orientation et de l'origine de la boucle.

\begin{figure}[H]
\centering
\epsfig{file={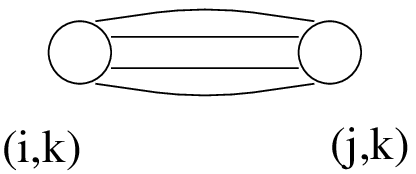},width=3cm}
\end{figure}

\bbb
2n_{k}\,
\t(M_{ij,k}M_{ij,k}^{*}M_{ij,k}M_{ij,k}^{*})\,
\t(\Phi_{ij}\Phi_{ij}^{*}\Phi_{ij}\Phi_{ij}^{*})
\eee

\begin{figure}[H]
\centering
\epsfig{file={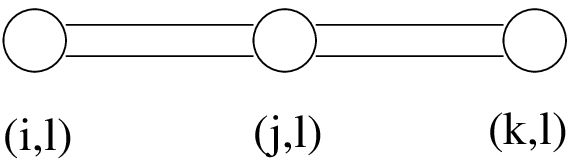},width=4.5cm}
\end{figure}

\bbb
4n_{k}\,
\t(M_{ij,l}M_{jk,l}M_{jk,l}^{*}M_{ij,l}^{*})\,
\t(\Phi_{ij}\Phi_{jk}\Phi_{jk}^{*}\Phi_{ij}^{*})
\eee

\begin{figure}[H]
\centering
\epsfig{file={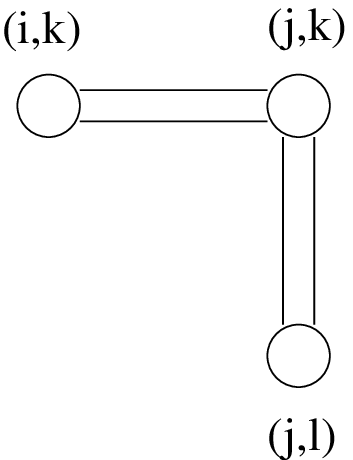},width=2.5cm}
\end{figure}

\bbb
4\,
\t(M_{ij,k}\ov{M}_{kl,j}\ov{M}_{kl,j}^{*}M_{ij,k}^{*})\,
\t(\Phi_{ij}\Phi_{ij}^{*})\,
\t(\Phi_{kl}\Phi_{kl}^{*})
\eee

\begin{figure}[H]
\centering
\epsfig{file={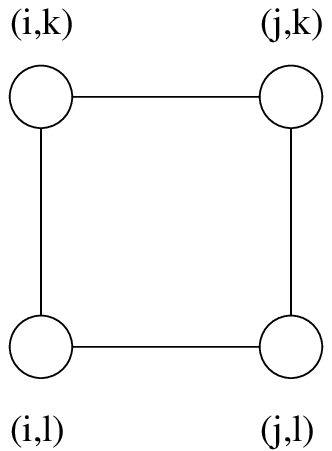},width=3cm}
\end{figure}

\bbb
8\,
\t(M_{ij,k}\ov{M}_{kl,j}M_{ij,l}^{*}\ov{M}_{kl,i}^{*})\,
\t(\Phi_{ij}\Phi_{ij}^{*})\,
\t(\Phi_{kl}\Phi_{kl}^{*})
\eee

\begin{figure}[H]
\centering
\epsfig{file={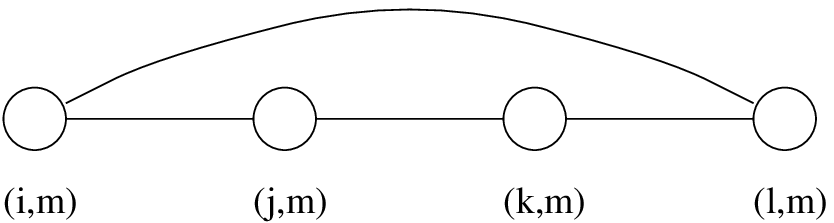},width=8cm}
\end{figure}

\bbb
8n_{k}\,
\t(M_{ij,m}M_{jk,m}M_{kl,m}M_{li,m})\,
\t(\Phi_{ij}\Phi_{jk}\Phi_{kl}\Phi_{li})
\eee

Il convient de remarquer que chacun de ces sous-diagrammes et leur sym\'etrique par rapport \`a la diagonale donnent les m\^emes termes. En effet, dans la cas g\'en\'eral, un diagramme et son sym\'etrique ont des contributions complexes conjugu\'ees l'une de l'autre. Puisque les matrices satisfont aux conditions de r\'ealit\'e
\bbb
\Phi_{ij}^{*}=\Phi_{ji}\;\;\;\mathrm{et}\;\;\;M_{ij,k}^{*}=M_{ji,k},
\eee
la contribution du diagramme sym\'etrique par rapport \`a la diagonale s'obtient simplement en changeant l'orientation. Pour les termes de degr\'e $4$, les r\`egles pr\'ec\'edentes sont ind\'ependantes de l'orientation ce qui entraine l'\'egalit\'e des contributions d'un diagramme et de son sym\'etrique.

\par

Les couplages quartiques des champs scalaires sont donn\'es par le coefficient $a_{4}(\dd^{2})$, ils sont \'egaux \`a
\bbb
\frac{1}{2}\frac{1}{(2\pi)^{n/2}}F_{4}\,\Lambda^{n-4}\t(\tilde{\Phi}^{4}),
\eee
ce qui entraine que nous devons multiplier tous les couplages pr\'ec\'edents par un facteur
\bbb
\lambda=\frac{1}{2}\frac{1}{(2\pi)^{n/2}}F_{4}\,\Lambda^{n-4}.\label{ps7}
\eee

\bigskip

\noindent
{\bf Conclusion}
\begin{it}
Les termes de degr\'e $n$ du potentiel scalaire s'\'ecrivent comme une somme sur toutes les boucles de longueur n du diagramme associ\'e au triplet spectral. Un potentiel de degr\'e $4$ s'\'ecrit
\bbbb
&V(\Phi_{ij})=\mathop{\sum}\limits_{ij}\;-\frac{1}{2}\mu^{2}\;
\t\lp\Phi_{ij}\Phi_{ij}^{*}\rp&\n\\
&+\mathop{\sum}\limits_{ijkl}\;\kappa_{ijkl}\;
\t\lp\Phi_{ij}\Phi_{ij}^{*}\rp\t\lp\Phi_{kl}\Phi_{kl}^{*}\rp\;
+\mathop{\sum}\limits_{ijkl}\;\lambda_{ijkl}\;
\t\lp\Phi_{ij}\Phi_{jk}\Phi_{kl}\Phi_{li}\rp.&\label{ps8}
\eeee
$\mu$ est un terme de masse identique pour tous les scalaires et donn\'e par (\ref{ps6}). Les coefficients $\kappa_{ijkl}$ et $\lambda_{ijkl}$ sont donn\'es par une somme sur tous les 5 types de sous-diagrammes pr\'ecedemment \'enum\'er\'es, \`a un facteur de proportionalit\'e donn\'e par (\ref{ps7}) pr\`es.
\end{it}


\subsection{Brisure spontan\'ee de sym\'etrie et masses}

Dans le potentiel scalaire donn\'e par (\ref{ps8}), nous avons un terme de masse n\'egatif alors que les constantes de couplages $\kappa_{ijkl}$ et $\lambda_{ijkl}$ sont positives, il y a donc brisure spontan\'ee de sym\'etrie. Cela apparait clairement si on \'ecrit le potentiel sous la forme
\bbb
V(\Phi)=-\frac{\mu'^{2}}{2}\t\lp\tilde{\Phi}^{2}\rp+
\frac{\lambda'}{4}\t\lp\tilde{\Phi}^{4}\rp.
\eee
En d\'esignant par $\varphi_{i}$ les valeurs propres de $\tilde{\Phi}$,
le potentiel scalaire se met sous la forme
\bbb
V(\Phi)=\mathop{\sum}\limits_{i}V(\varphi_{i}),
\eee
avec 
\bbb
V(\varphi)=-\frac{1}{2}\mu'^{2}\varphi^{2}+
\frac{\lambda'}{4}\varphi^{4}.
\eee
Le potentiel $V(\varphi)$ est caract\'eristique de la brisure de sym\'etrie car il atteint son minimum lorsque
\bbb
\varphi^{2}=v^{2}=\frac{6\mu'^{2}}{\lambda'}\neq 0.\label{bs1}
\eee
Par cons\'equent il y a brisure spontan\'ee de sym\'etrie et la valeur moyenne dans le vide est donn\'ee par le minimum qui, \`a priori, est atteint lorsque toutes les valeurs propres satisfont \`a (\ref{bs1}).
 
\par

Cependant, ces conditions ne peuvent pas \^etre satisfaites car les valeurs propres de $\tilde{\Phi}$ ne sont pas forc\'ement des variables ind\'ependantes. En effet, $\tilde{\Phi}$ est une matrice de taille importante dont beaucoup d'\'el\'ements sont identiquement nuls, son spectre ne peut donc satisfaire \`a ces relations en g\'en\'eral.

\par

Pour mener \`a bien une \'etude d\'etaill\'ee de la brisure de sym\'etrie, il faudrait
exprimer le potentiel \`a l'aide des champs $\Phi_{ij}$ qui sont les variables ind\'ependantes de la th\'eorie et \'etudier ses extrema. \'Etant donn\'e le nombre important de champs scalaires, ceci est impossible en pratique. Nous nous contenterons d'\'etudier quelques propri\'et\'es de la brisure de sym\'etrie, sachant qu'il existe au moins une valeur $V_{ij}$ des champs scalaires $\Phi_{ij}$ tel que le potentiel soit minimal.

\par

En rempla\c cant $\Phi_{ij}$ par $V_{ij}$ dans $\tilde{\Phi}$, on construit l'op\'erateur $\tilde{V}$. Le couplage de Yukawa entre les scalaires et le fermions est donn\'e par $\langle\Psi,\tilde{\Phi}\Psi\rangle$, ce qui entraine, apr\`es brisure de sym\'etrie, que $\tilde{V}$ est la matrice de masse des fermions.

\par

En nous basant sur cette interpr\'etation de $\tilde{V}$ comme la matrice de masse des fermions nous allons montrer une in\'egalit\'e g\'en\'erale entre masses des fermions et masses des bosons. 

\par

Pour cela, nous avons besoin du r\'esultat interm\'ediaire suivant. Soit $V\in M_{n}(\ccc)$ une matrice hermitienne dont la plus grande valeur propre en module est $m$. Alors pour toute matrice $A\in M_{n}(\ccc)$ on a
\bbb
\t\lp\lb A,V\rb^{*}\lb A,V\rb\rp
\leq 4m^{2}\,\t\lp A^{*}A\rp.\label{bs2}
\eee
Ce r\'esultat se montre en munissant l'espace vectoriel des matrices $M_{n}(\ccc)$ du produit scalaire $(A,B)\mapsto\t\lp A^{*}B\rp$. Pour ce produit scalaire, l'op\'erateur $T_{V}$ d\'efini par $T_{V}(A)=\lb V,A\rb$ est hermitien et on a $\t\lp\lb A,V\rb^{*}\lb A,V\rb\rp=\langle A,T_{V}^{2}A\rangle$.
Pour terminer, il nous suffit de remarquer que si $e_{i}$ est une base orthogonale de vecteur propres de $V$ tels que $Ve_{i}=m_{i}e_{i}$, les matrices $e_{i}e_{j}^{*}$ forment une base orthonormale de $M_{n}(\ccc)$ telle que
\bbb
T_{V}(e_{i}e_{j}^{*})=(m_{i}-m_{j})e_{i}e_{j}^{*}.
\eee
l'in\'egalit\'e (\ref{bs2}) r\'esulte alors de la majoration $(m_{i}-m_{j})^{2}\leq 4m^{2}$.

\par

Appliquons cela au terme de masse des bosons de jauge. Celui-ci est donn\'e par la d\'eriv\'ee covariante du champ scalaire $\tilde{\Phi}$ \'evalu\'ee en rempla\c cant  $\tilde{\Phi}$ par sa valeur moyenne dans le vide $\tilde{V}$. Le terme de masse des bosons de jauge est donn\'e par
\bbb
\frac{1}{2}\;\frac{1}{(2\pi)^{\frac{n}{2}}}\;
 \t \lp\lb \tilde{A}_{\mu},\tilde{V}\rb^{*}\lb \tilde{A}^{\mu},\tilde{V}\rb\rp.
\eee
Par l'in\'egalit\'e (\ref{bs2}), ceci est major\'e par
\bbb
2m_{f}^{2}\frac{1}{(2\pi)^{\frac{n}{2}}}\;
 \t A_{\mu}^{*}A^{\mu},
\eee
o\`u $m_{f}$ est la masse du fermion le plus lourd puisque $\tilde{V}$ est la matrice de masse des fermions. Lorsque les champs de jauge non-ab\'eliens $A_{\mu}^{i}$ et ab\'eliens $C_{\mu}^{i}$ sont correctement normalis\'es, on a
\bbb
\t\lp\tilde{A}_{\mu}^{*}\tilde{A^{\mu}}\rp=
\frac{3(2\pi)^{\frac{n}{2}}}{F_{4}\Lambda^{n-4}}
\lp\mathop{\sum}\limits_{i=1}^{N}\t\lp A_{\mu}^{i*}A^{i\mu}\rp
+\mathop{\sum}\limits_{i=1}^{N'}\frac{1}{2}C_{\mu}^{i*}C^{i\mu}\rp.
\eee
ce qui entraine
\bbb
\frac{1}{2}\,\frac{\Lambda^{n-4}F_{4}}{(2\pi)^{\frac{n}{2}}}\,
\t\lp\lb \tilde{A}_{\mu},\tilde{V}\rb^{*}\lb\tilde{A}^{\mu},\tilde{V}\rb\rp
\leq
6M_{f}^{2}\lp\mathop{\sum}\limits_{i=1}^{N}\t\lp A_{\mu}^{i*}A^{i\mu}\rp
+\mathop{\sum}\limits_{i=1}^{N'}\frac{1}{2}C_{\mu}^{i*}C^{i\mu}\rp.
\eee
Le premier membre de cette in\'equation (terme de masse) est une forme quadratique qui doit \^etre diagonalis\'ee dans une base qui pr\'eserve le second membre (terme cin\'etique). Les valeurs propres ainsi obtenues sont les carr\'es des masses des bosons, ce qui entraine que la plus grande de ces masses $m_{b}$ satisfait \`a
\bbb
m_{b}^{2}\leq 6m_{f}^{2}.\label{bs4}
\eee
Bien entendu, dans cette in\'egalit\'e $m_{b}^{2}$ et $m_{f}^{2}$ d\'ependent de $\Lambda$ implicitement car le potentiel scalaire en d\'epend, ce qui entraine une d\'ependance de la matrice $V$. Cette in\'egalit\'e n'est donc valable seulement pour des \'energies de l'ordre du cut-off $\Lambda\simeq 10^{19}$ Gev.  En utilisant le flot du groupe de renormalisation, il est toutefois possible d'\'etudier les valeurs de $m_{b}^{2}(\Lambda)$ et $m_{f}^{2}(\Lambda)$ dans le cas o\`u $\Lambda$ est de l'ordre de l'\'energie \'electrofaible.

\par

Il convient de noter que cette in\'egalit\'e ne fait intervenir que des arguments classiques. Aussi, si on consid\`ere l'action spectrale non pas comme une action effective mais comme une action classique qu'il convient de quantifier, rien ne nous assure que cette relation sera pr\'eserv\'ee par les corrections quantiques.

\par

En g\'en\'eral, la valeur moyenne $V$ dans le vide du champ $\Phi$ brise les sym\'etries non ab\'eliennes auxquelles il est coupl\'e. En effet, les sym\'etries qui subsistent apr\`es la brisure spontan\'ee de sym\'etrie sont donn\'ees par le stabilisateur de $\tilde{V}$. Par cons\'equent, ce sont les \'el\'ements du goupe de jauge qui satisfont \`a 
\bbb
u_{i}V_{ij}u_{j}^{-1}=V_{ij}.\label{bs5}
\eee
Dans le cas d'un triplet spectral dont la repr\'esentation est complexe, $u_{i}$ et $u_{j}$ sont ind\'ependants et la relation (\ref{bs5}) ne peut \^etre satisfaite, si $V_{ij}\neq 0$, par les \'el\'ements $u_{i}\in SU(n_{i})$. En cons\'equence, d\'es que les lignes $i$ et $j$ du diagramme sont reli\'es par la valeur moyenne dans le vide du champ scalaire, les sym\'etries non ab\'eliennes correspondantes sont bris\'ees.

\par

Si nous avons un triplet spectral dont la repr\'esentation est seulement lin\'eaire pour les r\'eels, la situation est plus d\'elicate car $u_{i}$ et $u_{j}$ ne sont pas n\'ecessairement ind\'ependants. Par exemple, il peut arriver que $u_{j}$ soit le complexe conjugu\'e de $u_{i}$. Dans ce cas la relation (\ref{bs5}) s'\'ecrit
\bbb
u_{i}V_{ij}u_{j}^{T}=V_{ij},
\eee
o\`u $u_{j}^{T}$ est la transpos\'ee de $u_{j}$. Cette relation est satisfaite, si $V_{ij}$ est proportionnel \`a l'unit\'e, par tous les \'el\'ements r\'eels $u_{i}\in SO(n_{i})$. Dans ce cas, la sym\'etrie de jauge $SO(n_{i})$ n'est pas bris\'ee et le r\'esultat pr\'ec\'edent n'est plus valable.

\par

Revenons \`a l'\'etude des sym\'etries ab\'eliennes et faisons \`a nouveau l'hypoth\`ese que la repr\'esentation est complexe. Dans ce cas, chaque $u_{i}$ contient un facteur ab\'elien $e^{i\theta_{i}}$ avec $\theta_{i}\in\rrr$. Ces sym\'etries ne sont pas bris\'ees si et seulement si
\bbb
V_{ij}e^{i(\theta_{i}-\theta_{j})}=V_{ij}.\label{bs3}
\eee
La relation (\ref{bs3}) \'equivaut \`a l'\'egalit\'e des phases $\theta_{i}$ et $\theta_{j}$ lorsque les lignes $i$ et $j$ sont reli\'es sur le diagrame d\'etermin\'e par $V$. Pour chaque groupe de lignes qui sont reli\'ees entre elles, il y a  une sym\'etrie $U(1)$ non bris\'ee et un champ de jauge ab\'elien de masse nulle.

\par

Bien entendu, ceci n'est valable que dans le cas complexe. Si les alg\`ebres sont r\'eelles, (\ref{bs5}) n'est plus \'equivalente \`a (\ref{bs3}) dans le cas ab\'elien. Cependant, on peut voir que pour le mod\`ele standard, cela permet encore de recenser les sym\'etries non bris\'ees \cite{spe}.

\par

Enfin, terminons par remarquer que si une ligne n'est reli\'ee \`a aucune autre ligne, la sym\'etrie de jauge associ\'ee est enti\`erement pr\'eserv\'ee par la brisure de sym\'etrie, car les champs de jauge correspondant \`a cette ligne ne sont pas coupl\'es aux champs scalaires.

\bigskip

\noindent
{\bf Conclusion}
\begin{it}
Le potentiel donn\'e par (\ref{ps8}) entraine toujours une brisure de sym\'etrie. Lorsque deux lignes du diagramme sont reli\'ees entre elles, c'est-\`a-dire que les champs de jauge correspondants sont coupl\'es au champs scalaires, les sym\'etries non ab\'eliennes associ\'ees sont bris\'ees, alors que pour chaque groupement de lignes uniquement reli\'ees entre elles, une sym\'etrie ab\'elienne persiste. De plus, dans le cas o\`u la sym\'etrie est bris\'ee la masse du boson le plus lourd $m_{b}$ et celle du fermion le plus lourd $m_{f}$ satisfont, au niveau classique, \`a l'in\'egalit\'e (\ref{bs4}). 
\end{it}  


\section{Le mod\`ele standard}

\subsection{Le triplet spectral}

Parmi tous les mod\`eles qu'il est possible de construire \`a l'aide de la g\'eom\'etrie non commutative se trouve le mod\`ele standard. Historiquement, celui-ci a \'et\'e construit dans le cadre du mod\`ele de Connes-Lott dans \cite{conneslott} puis am\'elior\'e dans \cite{reality} par l'incorporation des antiparticules. On pourra trouver une revue de cette construction dans \cite{sm}, \cite{his} et \cite{por}, ainsi qu'une analyse des contraintes que cette construction impose aux th\'eories de Yang-Mills-Higgs dans \cite{iochum}. Enfin, dans \cite{higgs} une interpr\'etation d\'etaill\'ee des contraintes num\'eriques sur les masses et les constantes de couplage est donn\'ee. 

\par

Contrairement \`a l'action spectrale, ces mod\`eles ne permettaient pas de coupler le mod\`ele standard \`a la gravit\'e car il ne contenaient qu'une th\'eorie de Yang-Mills sur un espace du type discret $\times$ continu, \'elabor\'ee en utilisant les m\'ethodes d\'ecrites dans le chapitre 1.

\par

Malgr\'e tout, ils pr\'esentaient d\'eja deux caract\'eristiques fondamentales des mod\`eles que nous venons d'\'etudier:
\begin{itemize}
\item
l'espace est le produit de l'espace-temps ordinaire par un espace fini non commutatif d\'ecrit par un triplet spectral fini,
\item
le boson de Higgs est un champ de jauge associ\'e \`a la g\'eom\'etrie de l'espace fini.
\end{itemize}  
La seule diff\'erence fondamentale entre le mod\`ele de Connes-Lott et ceux bas\'es sur l'action spectrale est la d\'efinition de l'action, car dans le premier cas elle ne peut inclure la th\'eorie de la gravitation.

\par

Le triplet spectral $(\aa,\hh,\dd)$ correspondant au mod\`ele standard doit \^etre choisi de telle mani\`ere que le groupe de jauge soit obtenu comme le groupe des unitaires de $\aa$. Puisque le groupe de jauge du mod\`ele standard est $SU(2)\times U(1) \times SU(3)$, l'alg\`ebre la plus adapt\'ee est
\bbb
\aa=\hhh\op\ccc\op M_{3}(\ccc),\label{ts1}
\eee
dont le groupe des unitaires est $SU(2)\times U(1)\times U(3)$. Il n'est pas possible d'obtenir exactement le groupe de jauge du mod\`ele standard car alors le seul choix possible de l'alg\`ebre est
\bbb
\aa'=\hhh\op M_{3}(\ccc),
\eee
qui nous oblige \`a avoir des leptons qui ne sont pas des singlets de couleur!

\par 

Si (\ref{ts1}) est le choix le plus \'economique, nous devrons tout de m\^eme \'eliminer un facteur $U(1)$ par une condition d'unimodularit\'e.

\par

L'espace de Hilbert est form\'e de toutes les particules et antiparticules du mod\`ele standard, consid\'er\'ees comme des fermions de Weyl. Pour $N_{f}=3$ g\'en\'erations de fermions, il y a en tout 90 particules qui se r\'epartissent dans les 4 espaces suivants
\bbb
\hh=\hh_{L}^{P}\op\hh_{R}^{P}\op\hh_{L}^{A}\op\hh_{L}^{A}.
\eee
L'espace $\hh_{L}^{P}=\ccc^{24}$ contient toutes les particules gauches, sa base est donn\'ee par
\bbb
\pp{u\cr d}_{L},\,
\pp{c\cr s}_{L},\,
\pp{t\cr b}_{L},\,
\pp{\nu_{e}\cr e}_{L},\,
\pp{\nu_{\mu}\cr \mu}_{L},\,
\pp{\nu_{\tau}\cr \tau}_{L},\label{ts2}
\eee
o\`u nous avons omis les indices de couleur pour les quarks.

\par

De m\^eme, $\hh_{R}^{P}=\ccc^{21}$ est l'espace de Hilbert de toutes les particules droites dont une base est 
\bbb
(u)_{R},\, (d)_{R},\, (c)_{R},\, (s)_{R},\, (t)_{R},
\, (b)_{R},\, (e)_{R},\, (\mu)_{R},\,(\tau)_{R}.\label{ts3}
\eee
Les espaces $\hh_{L}^{A}$ et $\hh_{R}^{A}$ contiennent les antiparticules de 
 $\hh_{L}^{P}$ et $\hh_{R}^{P}$.

\par

Puisque le mod\`ele standard ne contient aucune particule de Majorana, le triplet spectral est $S^{0}$-r\'eel. La chiralit\'e $\chi$ et la conjugaison de charge $\jj$ sont donn\'ees par
\bbb
\chi=
\pp{-I_{24}&0&0&0\cr0&I_{21}&0&0\cr 0&0&-I_{24}&0\cr0&0&0&I_{21}\cr}
\;\;\;\;
\jj=
\pp{0&0&I_{24}&0\cr0&0&0&I_{21}\cr I_{24}&0&0&0\cr0&I_{21}&0&0\cr}
\,C,
\eee
o\`u $C$ d\'esigne l'op\'eration de conjugaison complexe.

\par

La repr\'esentation $\pi$ est diagonale par blocs,
\bbb
\pi=\pi_{L}^{P}\op\pi_{R}^{P}\op\pi_{L}^{A}\op\pi_{L}^{A}.\label{ts4}
\eee
Dans les bases donn\'ees par (\ref{ts2}) et (\ref{ts3}) ainsi que celles correspondant \`a leur antiparticules, ces blocs sont donn\'es par
\bbbb
\pi_{L}^{P}(a)&=&\mathrm{diag}\lp a\ot I_{9},\,a\ot I_{3}\rp,\\
\pi_{R}^{P}(b)&=&\mathrm{diag}\lp b\,I_{9},\,\ov{b}\,I_{9},\,\ov{b}\,I_{3}\rp,\\
\pi_{L}^{A}(b,c)&=&\mathrm{diag}\lp I_{6
}\ot c,\,b\,I_{6}\rp,\\
\pi_{R}^{A}(b,c)&=&\mathrm{diag}\lp
I_{6}\ot c,\,b\,I_{3}\rp,
\eeee
o\`u $(a,b,c)\in\hhh\op\ccc\op M_{3}(\ccc)$.

\par

L'op\'erateur de Dirac est
\bbb
\dd=\pp{0&M&0&0\cr M^{*}&0&0&0\cr0&0&0&\ov{M}\cr0&0&\ov{M}^{*}&0\cr},\label{ts5}
\eee
o\`u $M$ est la matrice de masse des fermions. Celle-ci s'\'ecrit
\bbb
M=\pp{\pp{M_{u}\ot I_{3}&0\cr 0&M_{d}\ot I_{3}}&0\cr
0&\pp{0\cr M_{e}}},
\eee
avec 
\bbbb
M_{u}&=&\mathrm{diag}\lp m_{u},m_{c},m_{t}\rp,\\
M_{d}&=&V_{CKM}\,\mathrm{diag}\lp m_{d},m_{s},m_{b}\rp,\\
M_{e}&=&\mathrm{diag}\lp m_{e},m_{\mu},m_{\tau}\rp.
\eeee
Dans ce qui pr\'ec\`ede, $m_{p}$ d\'esigne la masse de la particule $p$ et $V_{CKM}$ est la matrice de Cabibbo-Kobayashi-Maskawa. Remarquons que les op\'erateurs $\dd$ et $\lambda\dd$ donnent naissance \`a la m\^eme th\'eorie lorsque $\lambda$ est un r\'eel non nul, compte-tenu de la condition de normalisation (\ref{ps5}). Ceci implique que le triplet spectral $(\aa,\hh,\dd)$ ne contient aucune \'echelle absolue de masse, cette derni\`ere \'etant introduite par l'interm\'ediaire du cut-off $\Lambda$.

\par

Ceci termine notre description du triplet spectral fini associ\'e au mod\`ele standard. Il est assez facile de v\'erifier qu'il satisfait \`a tous les axiomes relatifs au triplets spectraux finis. Ces axiomes peuvent \'egalement \^etre utilis\'es pour s\'electionner certaines extensions du mod\`ele standard. Par exemple, dans \cite{zgb}, un boson de jauge $U(1)$ suppl\'ementaire a \'et\'e introduit en choisissant une alg\`ebre du type
\bbb
\aa=\hhh\op\ccc\op\ccc\op M_{3}(\ccc).
\eee
Les axiomes de la g\'eom\'etrie non commutative permettent alors de r\'eduire drastiquement le nombre de solutions pour ce type d'extension.

\bigskip

\noindent

{\bf Conclusion}
\begin{it}
Le triplet spectral associ\'e au mod\`ele standard est bas\'e sur l'alg\`ebre
\bbb
\aa=\hhh\op\ccc\op M_{3}(\ccc).
\eee
L'espace de Hilbert est l'espace contenant tous les fermions et les antifermions
sur lequel $\aa$ est repr\'esent\'ee par (\ref{ts4}). L'op\'erateur de Dirac $\dd$ est construit \`a l'aide des matrices de masse et de m\'elange des fermions (\ref{ts5}).
\end{it}


\subsection{Matrice de multiplicit\'e et diagramme}

A partir de ce triplet spectral nous allons construire la matrice de multiplicit\'e. Le triplet spectral est r\'eel et nous avons $5$ repr\'esentations irr\'eductibles: les trois repr\'esentations fondamentales de $\hhh$, $\ccc$ et $M_{3}(\ccc)$ ainsi que les complexes conjugu\'ees des deux derni\`eres. Puisque la repr\'esentation complexe conjugu\'ee de $M_{3}(\ccc)$ n'intervient pas, nous avons seulement 4 repr\'esentations irr\'eductibles. En rep\`erant les entr\'ees de la matrice de multiplicit\'e par les repr\'esentations $\ccc$, $\hhh$, $\ov{\ccc}$ et $M_{3}(\ccc)$,
la matrice de multiplicit\'e est 
\bbb
\mu=3\,\pp{0&0&1&1\cr 0&0&-1&-1\cr 1&-1&0&1\cr 1&-1&1&0\cr}.\label{mm1}
\eee
Le facteur 3 correspond au nombre de g\'en\'erations de fermions et si on a $N_{g}$ g\'en\'erations il est remplac\'e par $N_{g}$.

\par

A partir des \'el\'ements de matrice de l'op\'erateur de Dirac, on construit le diagramme correspondant.

\begin{figure}[H]
\centering
\epsfig{file={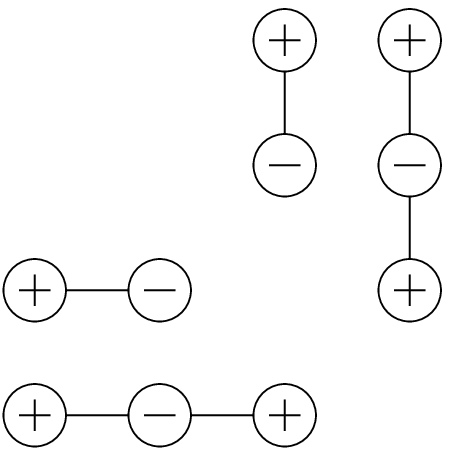},width=3cm}
\end{figure}

Les \'el\'ements de ce diagramme situ\'es au-dessus de la diagonale correspondent aux particules alors que l'autre moiti\'e est associ\'ee aux antiparticules. Puisque la matrice de multiplicit\'e  peut \^etre mise sous la forme
\bbb
\mu=\epsilon+\epsilon^{*}
\eee
avec
\bbb
\epsilon=3\,\pp{0&0&1&1\cr 0&0&-1&-1\cr 0&-0&0&1\cr 0&-0&0&0\cr},
\eee
et qu'il n'y a aucun lien entre particules et antiparticules, le triplet spectral du mod\`ele standard v\'erifie l'axiome de $S^{0}$-r\'ealit\'e.

\par

La premi\`ere ligne verticale du diagramme est associ\'ee \`a un couplage des champs scalaires avec le doublet d'isospin des leptons gauches et le singlet des leptons droits. La ligne suivante d\'ecrit le couplage du champ scalaire avec les quarks. Les deux sommets $\op$ correspondent au quarks droits $(u,c,t)$ et $(d,s,b)$ alors que le sommet $\om$ est associ\'e au doublet d'isospin des quarks gauches.

\par

Ce diagramme n'est pas le plus g\'en\'eral compatible avec les axiomes de la g\'eom\'etrie non commutative. En effet, on peut rajouter les liens suivants (en pointill\'es).

\begin{figure}[H]
\centering
\epsfig{file={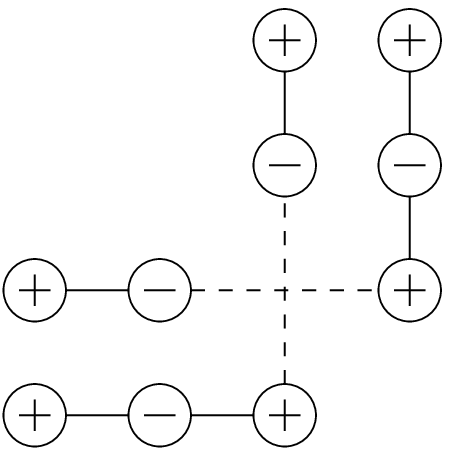},width=3cm}
\end{figure}

Ces liens pr\'eservent les axiomes de la g\'eom\'etrie non commutative, puisqu'ils sont verticaux ou horizontaux et ne relient que des sommets de types diff\'erents. Cependant, ils m\'elangent particules et antiparticules, ce qui entraine que la $S^{0}$-r\'ealit\'e n'est plus v\'erifi\'ee.

\par

Ils correspondent \`a un couplage de Yukawa entre les champs scalaires, les quarks $(u,c,t)$ droits et les antileptons gauches ainsi qu'a son sym\'etris\'e par conjugaispon de charge. Ce couplage a \'et\'e \'etudi\'e et interpr\`et\'e dans \cite{lep} en termes de leptoquarks.

\bigskip

\noindent
{\bf Conclusion}
\begin{it}
La matrice de multiplicit\'e, donn\'ee par (\ref{mm1}) et le diagramme associ\'e au triplet spectral du mod\`ele standard ont \'et\'e construits. Cependant, ce diagramme n'est pas le diagramme le plus g\'en\'eral compatible avec les axiomes de la g\'eom\'etrie non commutative et la matrice de multiplicit\'e du mod\`ele standard, car il est possible d'y ajouter un couplage du scalaire de Higgs avec les  leptons et les antiquarks ainsi qu'avec les   antileptons et les quarks.
\end{it}


\subsection{Dualit\'e de Poincar\'e et neutrino massif}

Dans cette section, nous allons v\'erifier en d\'etail la validit\'e de la dualit\'e de Poincar\'e dans le cas du mod\`ele standard. A partir de la matrice de multiplicit\'e
\bbb
\mu=3\,\pp{0&0&1&1\cr 0&0&-1&-1\cr 1&-1&0&1\cr 1&-1&1&0\cr},
\eee
nous pouvons construire la matrice de la forme d'intersection $\cap$ en utilisant les r\`egles \'enonc\'es au cours de la section 2.6.3. Celle-ci est obtenue \`a partir de $\mu$ en additionant les contributions des deux lignes (et colonnes) correspondant \`a l'alg\`ebre $\ccc$ en une seule ligne (et colonne) et en multipliant la ligne et la colonne associ\'ee au quaternions par 2. Nous obtenons la matrice suivante,
\bbb
\cap=6\pp{1&-1&1\cr -1&0&-1\cr 1&-1&0},
\eee 
dont le d\'eterminant est non nul. Ainsi, la dualit\'e de Poincar\'e est satisfaite dans le cas du mod\`ele standard.

\par

Incorporons au mod\`ele standard un neutrino droit. Ce dernier ne doit \^etre sujet \`a aucune interaction et doit \^etre un singlet sous toutes les transformations de jauge. Aussi, la seule possibilit\'e est de le mettre sur les termes diagonaux de la matrice de multiplicit\'e qui correspondent aux deux facteurs $\ccc$. Ajoutons donc \`a un de ces deux termes la quantit\'e $\epsilon n$, o\`u $n\leq3$ est le nombre de g\'en\'erations dans lequel nous voulons introduire ce neutrino et $\epsilon$ un facteur qui vaut 2 si ce neutrino droit est distinct de son antiparticule, et 1 si c'est un neutrino de Majorana.

\par

Dans ce cas, la matrice de la forme d'intersection est donn\'ee par
\bbb
\cap=\pp{6+n\epsilon&-6&6\cr -6&0&-6\cr 6&-6&0},
\eee 
dont le d\'eterminant vaut
\bbb
\det\cap=6(-n\epsilon+6).
\eee
Il s'annule si et si seulement si $n=3$ et $\epsilon=2$. Par cons\'equent, ajouter un neutrino droit distinct de son antiparticule dans les trois g\'en\'erations est incompatible avec la dualit\'e de Poincar\'e. Par contre ajouter un tel neutrino dans une ou deux g\'en\'erations ou ajouter un neutrino droit de Majorana dans un nombre quelconque de g\'en\'erations n'est pas en contradiction avec la dualit\'e de Poincar\'e.

\par

Il y a des indications exp\'erimentales d'oscillations de neutrinos. L'explication la plus simple est l'existence d'une matrice de masse pour les neutrinos comme pour les quarks. Il est important de noter qu'en g\'eom\'etrie non commutative des masses de Majorana pour les neutrinos sont exclues car l'op\'erateur de Dirac doit anticommuter avec la chiralit\'e. Pour les masses de Dirac, on a besoin de neutrinos droits, ce qui est en contradiction avec le dualit\'e de Poincar\'e.

\bigskip

\noindent
{\bf Conclusion}
\begin{it}
Dans le cas du mod\`ele standard, la dualit\'e de Poincar\'e est satisfaite pourvu qu'il n'y ait pas de neutrino droit distinct de son antiparticule dans les trois g\'en\'erations. Etant donn\'e que les masses de Majorana sont exclues, cela nous emp\^eche de construire un mod\`ele comportant des neutrinos massifs dans les trois g\'en\'erations. 
\end{it}

\subsection{Les bosons de jauge}

Le groupe des unitaires de l'alg\`ebre du mod\`ele standard est $SU(2)\times U(1)\times U(3)$, il y a donc un champ $U(1)$ en exc\'es qui doit \^etre \'elimin\'e par une condition d'unimodularit\'e. Cette condition d'unimodularit\'e est obtenue en imposant la condition $\t(A_{\mu})=0$ sur le champ de jauge \cite{grav}.

\par

Si nous notons $B_{\mu}$ le champ ab\'elien provenant du facteur $\ccc$ et $B_{\mu}'$ celui de $M_{3}(\ccc)$, la condition d'unimodularit\'e impose
\bbb
B_{\mu}'=\frac{1}{3}B_{\mu}.
\eee
Cette condition est \'equivalente \`a l'absence d'anomalies de jauge et gravitationnelle et permet de s\'electionner les bonnes hypercharges.

\par

Les constantes de couplage non ab\'eliennes sont donn\'ees par la relation (\ref{cc4})
\bbb
g_{i}=(2\pi)^{n/4}
\sqrt{\frac{3/2}{F_{4}\Lambda^{n-4}\mathop{\sum}\limits_{j}|\mu_{ij}|n_{j}}}.
\eee
Dans le cas du champ \'electrofaible $SU(2)$ on obtient, en dimension 4,
\bbb
g_{SU(2)}=\pi\sqrt{\frac{F_{4}}{2}}\label{bj1}
\eee
et dans le cas de $SU(3)$,
\bbb
g_{SU(3)}=\pi\sqrt{\frac{F_{4}}{2}}.\label{bj2}
\eee
Ces relations nous donnent des constantes de couplage $\sqrt{2}$ fois plus petites que celles obtenues dans \cite{spec} ou dans \cite{hear}. Ceci s'explique par le fait que dans ces derniers articles, la trace est prise sur l'espace de Hilbert des particules uniquement, alors que nous avons pris la trace sur tout l'espace de Hilbert, qui contient particules et antiparticules. Cela revient simplement \`a changer la fonction $F$ en $F/2$.
 
\par

En d\'eterminant explicitement la constante de couplage ab\'elienne $g_{U(1)}$, on montre \cite{spec} que 
\bbb
\frac{5}{3}g_{U(1)}^{2}=g_{SU(2)}^{2}=g_{SU(3)}^{2}.
\eee
De plus, en utilisant les \'equations du groupe de renormalisation, on peut donner ces constantes de couplages \`a l'\'echelle \'electrofaible \cite{hear}.

\bigskip

\noindent
{\bf Conclusion}
\begin{it}
Dans le cas du mod\`ele standard, le nombre de champs de jauge ab\'eliens est r\'eduit de $2$ \`a $1$ en imposant la condition d'unimodularit\'e $\t(A_{\mu})=0$. Les constantes de couplage non ab\'eliennes sont donn\'ees par les relations (\ref{bj1}) et (\ref{bj2}). 
\end{it}


\subsection{Le potentiel scalaire}

Le potentiel scalaire peut \^etre calcul\'e en utilisant les m\'ethodes que nous avons expos\'ees. Pour cela, rappelons le diagrame associ\'e au modele standard.

\begin{figure}[H]
\centering
\epsfig{file=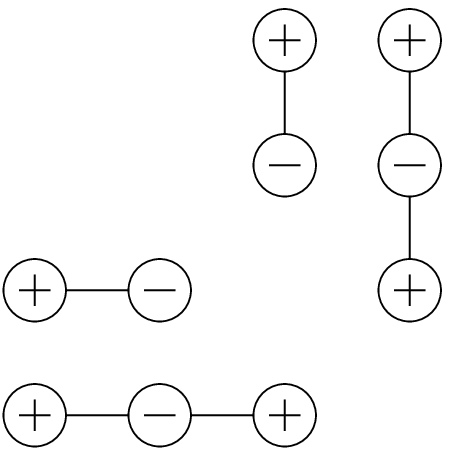},width=3cm}
\end{figure}

Les sommets reli\'es par les lignes horizontales correspondent \`a l'espace des antiparticules. Puisque le triplet spectral est $S^{0}$-r\'eel, ces contributions sont identiques \`a celles de l'espace des particules et nous nous occuperons essentiellement de ces derni\`eres.

\par

La derni\`ere ligne de ce diagramme, qui correpond aux champs de jauge $SU(3)$, n'est reli\'ee \`a aucune autre ligne. Les champs correspondant ne sont pas coupl\'es aux champs scalaires et restent sans masse. 

\par

Au lien entre la repr\'esentation fondamentale de $\hhh$ et la repr\'esentation fondamentale de $\ccc$ est associ\'e un seul doublet de champs scalaires complexes,
\bbb
\Phi=\pp{\phi_{1}\cr \phi_{2}}.
\eee
L'autre lien relie la repr\'esentation fondamentale de $\hhh$ \`a la complexe conjugu\'ee de $\ccc$, le champ scalaire correspondant est le conjugu\'e quaternionique de $\Phi$ donn\'e par
\bbb
i\sigma_{2}\ov{\Phi}=\pp{-\ov{\phi}_{2}\cr\ov{\phi}_{1}}.
\eee
Enfin, les trois matrices $M_{u}$, $M_{d}$ et $M_{e}$ jouent le r\^ole des matrices $M_{ij,k}$. Notons qu'il n'y a pas de multiplicit\'e car il n'y a qu'un seul doublet de champs scalaires complexes.

\par

Bien entendu, le terme de masse est donn\'e par
\bbb
-\frac{1}{2}\lp\phi_{1}^{2}+\phi_{2}^{2}\rp,
\eee
avec 
\bbb
\mu=\sqrt{\frac{2F_{2}}{F_{4}}}\Lambda.
\eee
Notons que ce r\'esultat est absolument identique \`a celui obtenu dans \cite{por} car la condition de normalisation, qui porte sur un terme quadratique, ne joue aucun r\^ole ici.
\par

Le calcul des termes d'ordre 4 se fait en recherchant toutes les boucles de longueur 4 sur le diagrammme. Celles-ci peuvent \^etre obtenues \`a partir des sous-diagrammes de longueur 1 et 2 en appliquant les r\`egles expos\'ees dans la section 3.3.3.

\par

La contribution des sous-diagrammes de longueur 1 est
\bbb
2\times 2\times\lp 
\t\lp
M_{e}M_{e}^{*}M_{e}M_{e}^{*} 
+3M_{u}M_{u}^{*}M_{u}M_{u}^{*}
+3M_{d}M_{d}^{*}M_{d}M_{d}^{*}
\rp \rp\times 
\t\lp\Phi\Phi^{*}\Phi\Phi^{*}\rp
\eee  
Le premier facteur 2 vient de la contribution des antiparticules alors que le second facteur est donn\'e par les r\`egles expos\'ees dans la section 3.3.3.

\par

La contribution de l'unique sous-diagramme de longueur 2 est proportionelle \`a
\bbb
\t\lb\Phi\lp i\sigma_{2}\ov{\Phi}\rp^{*}\Phi^{*}i\sigma_{2}\ov{\Phi}\rb.
\eee
Celle-ci est nulle car on a toujours
\bbb
\Phi^{*}i\sigma_{2}\ov{\Phi}=0.
\eee
\par
En incorporant les facteurs dus \`a la g\'eom\'etrie de l'espace-temps et la condition de normalisation (\ref{ps5}), on obtient le terme quadratique
\bbb
\lambda\lp(\phi_{1})^{2}+(\phi_{2})^{2}\rp^{2},
\eee
avec 
\bbb
\lambda=\frac{\pi^{2}}{2F_{4}}\frac{Q}{L^{2}}.\label{ps'1}
\eee
$Q$ et $L$ sont d\'efinis par
\bbbb
L&=&\t\lp M_{e}M_{e}^{*}+3M_{u}M_{u}^{*}+3M_{d}M_{d}^{*}\rp\\
Q&=&\t\lb \lp M_{e}M_{e}^{*}\rp^{2}+3\lp M_{u}M_{u}^{*}\rp^{*}+3\lp M_{d}M_{d}^{*}\rp^{*}\rb.
\eeee
Nous obtenons un couplage qui est deux fois plus faible que celui obtenu dans \cite{hear}. Encore une fois, ceci est du \`a la normalisation du terme cin\'etique. 

\bigskip

\noindent
{\bf Conclusion}
\begin{it}
Dans le cas du mod\`ele standard, le potentiel de Higgs peut \^etre calcul\'e \`a l'aide d'une somme sur toutes les boucles de longueur 4. Le r\'esultat donne un couplage qui est donn\'e par (\ref{ps'1})

\end{it}


\chapter{Le tore non commutatif}

\section{Le triplet spectral}

\subsection{L'alg\`ebre}

Dans ce dernier chapitre, nous allons construire et \'etudier en d\'etail un triplet spectral $(\aa,\hh,\dd)$ construit \`a l'aide d'une alg\`ebre non commutative qui est une d\'eformation de l'alg\`ebre des fonctions sur le tore.
Ce triplet spectral est relativement simple \`a d\'efinir et nous permet, pour tout entier $n>0$, de construire une g\'eom\'etrie non commutative de dimension $n$ qui n'est pas simplement obtenue en consid\'erant une alg\`ebre de fonctions \`a valeurs matricielles.

\par

De mani\`ere g\'en\'erale, on pourra trouver beaucoup d'informations concernant le tore non commutatif dans la revue de M. Rieffel \cite{rieffel}. Cependant, la plupart des travaux sur le tore non commutatifs sont ant\'erieurs \`a la notion de triplet spectral, aussi reformulerons-nous certains r\'esultats assez anciens en utilisant ce langage. 

\par

Pour commencer, rappelons que l'alg\`ebre des fonctions lisses sur le tore de dimension $n$ peut \^etre identifi\'ee \`a l'alg\`ebre des fonctions ind\'efiniment d\'erivables sur $\rrr^{n}$ satisfaisant \`a la condition de p\'eriodicit\'e $f(x+e_{i})=f(x)$ pour tout $x\in\rrr^{n}$ et tout vecteur $e_{i}$ de la base canonique de $\rrr^{n}$. 

\par

Cela revient \`a consid\'erer que le tore $T^{n}$ est le quotient de $\rrr^{n}$ par la relation d'\'equivalence
\bbb
x\rr y\;\mathrm{si}\;\mathrm{et}\;\;\mathrm{seulement}\;\mathrm{si}\;x-y=\mathop{\sum}\limits_{i=1}^{n}p_{i}e_{i},
\eee
avec $p_{i}\in\zzz$. Bien entendu, il est toujours possible de remplacer la base canonique ~par n'importe quelle autre base de $\rrr^{n}$ dans la relation pr\'ec\'edente. Les deux alg\`ebres de fonctions ainsi obtenues sont isomorphes et seule une structure suppl\'ementaire (structure de vari\'et\'e complexe ou  Riemannienne) permet de les distinguer.

\par

De telles fonctions peuvent toujours \^etre d\'evelopp\'ees en s\'erie de Fourier,
\bbb
f(x)=\mathop{\sum}\limits_{p\in\zzz^{n}}f_{p}e^{2\pi ip\cdot x}\,\forall x\in\rrr^{n},
\eee
avec
\bbb
f_{n}=\frac{1}{(2i\pi)^{n}}\int_{0}^{1}dx_{1}\dots\int_{0}^{1}dx_{n}f(x)e^{2i\pi p\cdot x},
\eee
o\`u $p=(p_{1},\dots,p_{n})\in\zzz^{n}$ est un multiindice et $p\cdot x=p_{1}x_{1}+\dots+p_{n}x_{n}$.

\par

Il est clair qu'une telle fonction est ind\'efinimemt d\'erivable si et seulement si la suite $f_{p}$ est \`a d\'ecroissance rapide, c'est-\`a-dire que
$\lim p_{1}^{k_{1}}\dots p_{n}^{k_{n}}f_{p}=0$ lorsque $|p_{1}|+\dots+|p_{n}|$ tend vers l'inifini, et cela pour $k=(k_{1},\dots,k_{n})\in\zzz^{n}$.

\par

De fa\c con plus abstraite, nous pouvons consid\'erer l'alg\`ebre des fonctions lisses sur le tore comme l'alg\`ebre involutive des s\'eries formelles
\bbb
f=\mathop{\sum}\limits_{p\in\zzz^{n}}f_{p}U^{p},
\eee
o\`u $(f_{p})_{p\in\zzz^{n}}$ est une suite \`a d\'ecroissance rapide et $U^{p}$ sont des unitaires v\'erifiant la relation $U^{p}U^{q}=U^{p+q}$. Ces unitaires s'identifient \`a la base des modes de Fourier $(e^{ip.x})_{p\in\zzz^{n}}$ et on peut toujours consid\'erer que $U^{p}=U_{1}^{p_{1}}\dots U_{n}^{p_{n}}$, o\`u $(U_{i})_{1\leq i\leq n}$ est un syst\`eme de g\'en\'erateurs unitaires satisfaisant \`a $U_{i}U_{j}=U_{j}U_{i}$.

\par

Il est alors facile de d\'efinir le tore non commutatif en ins\'erant dans la derni\`ere relation un facteur de phase.

\begin{dfi}
Soit $\theta\in M_{n}(\rrr)$ une matrice antisym\'etrique. On d\'efinit l'alg\`ebre $\aa_{\theta}$ comme \'etant l'alg\`ebre involutive des s\'eries formelles \`a d\'ecroissance rapide
\bbb
\mathop{\sum}\limits_{p\in\zzz^{n}}f_{p}U_{1}^{p_{1}}\dots U_{n}^{p_{n}},
\eee 
o\`u $(U_{i})_{1\leq i\leq n}$ est un syst\`eme d'unitaires qui v\'erifient
\bbb
U_{i}U_{j}=e^{2i\pi\theta_{ij}}U_{j}U_{i}.
\eee 
\end{dfi}

$\aa_{\theta}$ est consid\'er\'ee comme l'alg\`ebre des "fonctions" sur le tore non commutatif de dimension $n$. Si $n=1$, on retrouve bien entendu l'alg\`ebre des fonctions sur le cercle.

\par

Nous n'avons pas encore introduit de norme sur cette alg\`ebre pour nous assurer de la convergence des s\'eries consid\'er\'ees. Bien que cela puisse \^etre fait en introduisant une norme type "convergence uniforme", nous nous restreignons dans cette section \`a des consid\'erations purement alg\'ebriques.

\par

Il est commode d'introduire, pour tout $p\in\zzz^{n}$, la notation $U^{p}=(U_{1})^{p_{1}}\dots (U_{n})^{p_{n}}$. Il est facile de montrer que l'on a 
\bbb
U^{p}U^{q}=e^{2i\pi\theta(p,q)}U^{q}U^{p}, 
\eee
avec $\theta(p,q)=\theta_{\mu\nu}p^{\mu}q^{\nu}$. Par convention, toute sommation sur un indice grec $\lambda,\mu,\nu,\rho,\sigma,\dots$ r\'ep\'et\'e en position basse et haute est sous-entendue. Lorsque le contraire n'est pas explicitement sp\'ecifi\'e, nous utilisons la m\'etrique euclidienne.

\par

De plus, on montre que $U^{p}U^{q}=e^{2i\pi\chi(p,q)}U^{p+q}$, o\`u $\chi$ est la matrice triangulaire superieure obtenue en ne gardant que les \'el\'ements de $\theta$ situ\'es au dessus de la diagonale. En vue de simplifier la relation pr\'ec\'edente, d\'efinissons $V^{p}=e^{i\pi\chi(p,p)}U^{p}$.
On montre alors que
\bbbb
V^{p}V^{q}&=&e^{-i\pi(\chi(p,p)+\chi(q,q))}U^{p}U^{q}\n\\
&=&e^{i\pi(\chi(p,p)+\chi(q,q)+2\chi(p,q)-\chi(p+q,p+q))}V^{p+q}\n\\
&=&e^{i\pi(\chi(p,q)-\chi(q,p))}V^{p+q}\n\\
&=&e^{i\pi\theta(p,q)}V^{p+q}.
\eeee 
Quitte \`a multiplier les unitaires $U^{p}$ par une phase, nous obtenons la d\'escription suivante de $\aa_{\theta}$.

\begin{pro}
L'alg\`ebre $\aa_{\theta}$ est l'alg\`ebre involutive des s\'eries formelles
\bbb
\mathop{\sum}\limits_{p\in\zzz^{n}}f_{p}U^{p},
\eee 
o\`u $(f_{p})_{p\in\zzz^{n}}$ est une suite \`a d\'ecroissance rapide de nombres complexes et $(U^{p})_{p\in\zzz^{n}}$ un syst\`eme d'unitaires satisfaisant \`a 
\bbb
U^{p}U^{q}=e^{i\pi\theta(p,q)}U^{p+q}.
\eee
\end{pro}
En particulier cette relation nous montre que $U^{-p}$ est l'inverse de $U^{p}$.

\par

Il est utile de noter $\theta(p)$ le vecteur de $\rrr^{n}$ d\'efini par $\lp\theta(p)\rp_{\mu}=\theta_{\mu\nu}p^{\nu}$. En notant $\cdot$ le produit scalaire de vecteurs de $\rrr^{n}$, nous avons $\theta(p,q)=\theta(p)\cdot q=-p\cdot \theta(q)$. Nous pouvons alors \'enoncer le r\'esultat suivant.

\begin{pro}
Le centre de $\aa_{\theta}$ est form\'e des s\'eries formelles qui ne contiennent que les mon\^omes $U^{p}$ tels que $\theta(p)\in\zzz^{n}$. Lorsqu'il est non trivial, il est isomorphe \`a l'alg\`ebre des fonctions sur un tore de dimension $\leq n$.
\end{pro}

\demo
Soit $f=\sum_{p}f_{p}U^{p}\in\aa_{\theta}$ un \'el\'ement central. Puisqu'il doit commuter avec $U^{q}$ pour tout $q\in\zzz^{n}$, on doit avoir $U^{q}fU^{-q}=f$, ce qui \'equivaut \`a $f_{p}e^{2i\pi\theta(p).q}=f_{p}\;\forall q\in\zzz^{N}$. En cons\'equence, si $f_{p}\neq 0$ on doit avoir $\theta(p)\cdot q\in\zzz^{n}$ pour tout $q$, ce qui \'equivaut \`a $\theta(p)\in\zzz^{n}$ et d\'emontre la premi\`ere assertion.

\par

Soit $H$ le sous-ensemble de $\zzz^{n}$ form\'e des vecteurs $p$ tels que $\theta(p)\in\zzz^{n}$. Lorsque le centre de $\aa_{\theta}$ est non trivial, $H$ est un sous-groupe non trivial de $\zzz^{n}$ et on peut trouver un syst\`eme $e_{1},\dots,e_{n'}$ de g\'en\'erateurs de $H$ lin\'eairement ind\'ependants avec $n'\leq n$. Par cons\'equent, le centre de $\aa_{\theta}$ s'identifie avec l'alg\`ebre engendr\'e par les unitaires $U^{e_{1}},\dots,U^{e_{n'}}$, qui n'est autre que l'alg\`ebre des fonctions lisses sur un tore de dimension $n'$. 
\edemo

Lorsque le centre de $\aa_{\theta}$ est trivial, on dit que $\aa_{\theta}$ est non d\'eg\'en\'er\'ee. Dans ce cas, beaucoup de d\'emonstrations se simplifient consid\'erablement.

\par

De mani\`ere g\'en\'erale, nous appelons trace sur une alg\`ebre $\aa$ toute forme lin\'eaire $\tau$ sur $\aa$ satisfaisant \`a $\tau(ab)=\tau(ba)$ pour tout $a,b\in\aa$. De plus, si $\aa$ est une alg\`ebre involutive, une trace est dite positive si $\tau(aa^{*})\geq 0$ pour tout $a\in\aa$ et fid\`ele si elle v\'erifie $\tau(aa^{*})>0$ lorsque $aa^{*}\neq 0$.

\begin{pro}
La forme lin\'eaire $\tau$ d\'efinie sur $\aa_{\theta}$ par 
\bbb
\tau(U^{p})=\delta(p),
\eee
o\`u $\delta(p)$ est le symbole de Kronecker qui vaut 0 si $p\neq 0$ et 1 si $p=0$, est une trace positive et fid\`ele appel\'e trace canonique. De plus, toute autre trace s'annule sur les unitaires $U^{p}$ qui ne sont pas centraux. 
\end{pro}

\demo
Par lin\'earit\'e, il suffit de v\'erifier que $\tau(U^{p}U^{q})=\tau(U^{q}U^{p})$ ce qui est \'evident. De plus, si $f=\sum_{p}f_{p}U^{p}$, alors
\bbb
\tau(ff^{*})=\mathop{\sum}\limits_{p\in\zzz^{n}}|f_{p}|^{2},
\eee 
qui est positif et ne s'annule que si $f=0$. Notons que puisque la suite $f_{p}$ est \`a d\'ecroissance rapide, la s\'erie pr\'ec\'edente est convergente. Enfin, si $\tau^{'}$ est une trace quelconque sur $\aa$, on doit avoir
\bbb
\tau'(U^{p})=\tau'(U^{q}U^{p}U^{-q})=e^{2i\pi\theta(p,q)}\tau'(U^{p})\;
\forall q\in\zzz^{n},
\eee
ce qui implique que $\tau'(U^{p})=0$ si $U^{p}$ n'est pas central. 
\edemo
  
Outre le cas non d\'eg\'en\'er\'e, il est aussi int\'eressant d'\'etudier le cas o\`u le centre est un tore de dimension maximale $n'=n$, que nous appellerons "cas rationnel". 

\begin{pro}
Le centre de $\aa_{\theta}$ est un tore de dimension $n'=n$ si et seulement si les coefficients de la matrice $\theta$ sont tous rationnels.
\end{pro}

\demo
Si les nombres $\theta_{\mu\nu}$ sont rationnels, alors tous les vecteurs $(\epsilon_{i})_{1\leq i\leq n}$ de la base canonique de $\zzz^{n}$ v\'erifient $k\theta(\epsilon_{i})\in\zzz^{n}$ pour $k\in\zzz$ bien choisi, aussi le centre est-il de dimension $n$.

\par

R\'eciproquement, si le centre est un tore de dimension $n$, alors le sous-groupe H (voir la d\'emonstration concernant le centre de $\aa_{\theta}$) est engendr\'e par $n$ vecteurs lin\'eairement ind\'ependants $(e_{\mu})_{1\leq \mu\leq n}$. Tout vecteur de $\zzz^{n}$ s'\'ecrit donc comme combinaison lin\'eaire \`a coefficients rationnels des vecteurs  $(e_{\mu})_{1\leq \mu\leq n}$ ce qui nous montre que $\theta_{\mu\nu}=\theta(\epsilon_{\mu}).\epsilon_{\nu}$ est rationnel.
\edemo

Lorsque $\theta$ est rationnel, comment peut-on interpr\'eter les \'el\'ements de $\aa_{\theta}$ qui ne sont pas centraux? Pour r\'epondre \`a cette question, il est commode d'\'etudier d'abord le cas de la dimension 2. Dans ce cas, l'alg\`ebre $\aa^{\theta}$ est l'alg\`ebre involutive engendr\'ee par deux unitaires $U_{1}$ et $U_{2}$ satisfaisant \`a la relation 
\bbb
U_{1}U_{2}=qU_{2}U_{1},\;\;\;q=e^{2i\pi\frac{M}{N}},
\eee
o\`u $M$ et $N$ sont deux entiers positifs premiers entre eux avec $M\in\la 0,1,\dots,N-1\ra$. Le centre de $\aa_{\theta}$ est l'alg\`ebre involutive engendr\'ee par $(U_{1})^{N}$ et $(U_{2})^{N}$ qui est isomorphe \`a l'alg\`ebre des fonctions lisses sur le tore de dimension 2. Nous identifions $(U_{1})^{N}$ et $(U_{2})^{N}$ aux fonctions $x=(x_{1},x_{2})\mapsto e^{2i\pi x_{1}}$ et $x=(x_{1},x_{2})\mapsto e^{2i\pi x_{2}}$ d\'efinies sur le tore de longueur 1. Pour interpr\`eter $U_{1}$ et $U_{2}$ dans ce formalisme, il est utile d'introduire deux matrices unitaires  $P$ et $Q$ de taille $N$ d\'efinies par
\bbb
P=\pp{1&& &\cr
      &q &      &\cr
       & &\ddots&\cr
      & &     &q^{N-1}}\:\:\:\:\:
Q=\pp{&&&1\cr 1&&&\cr &\ddots&&\cr &&1&}^{N},
\eee
o\`u, par convention nous n'avons repr\'esent\'e que les \'el\'ements de matrice non nuls. Ces deux matrices v\'erifient $PQ=qQP$ ainsi que $P^{N}=Q^{N}=1$ et engendrent toute l'alg\`ebre des matrices $N\times N$ complexes. 

\par

Il est facile de repr\'esenter $\aa_{\theta}$ en identifiant $U_{1}$ et $U_{2}$ aux fonctions $U_{1}(x)=e^{\frac{2i\pi x_{1}}{N}}P$ et $U_{2}(x)=e^{\frac{2i\pi x_{2}}{N}}Q$. Cependant, ces fonctions ne sont pas de p\'eriode 1 et ne peuvent pas \^etre consid\'er\'ees comme des fonctions d\'efinies sur le tore car on a
\bbbb
&U_{1}(x_{1}+1,x_{2})=qU_{1}(x_{1},x_{2})\;\;
U_{1}(x_{1},x_{2}+1)=U_{1}(x_{1},x_{2})&\n\\
&U_{2}(x_{1},x_{2}+1)=qU_{2}(x_{1},x_{2})\;\;
U_{2}(x_{1}+1,x_{2})=U_{2}(x_{1},x_{2})&\n.
\eeee 
Au voisinage de chaque point, on peut consid\'erer que $\aa_{\theta}$ est une alg\`ebre de fonctions \`a valeurs dans $M_{N}(\ccc)$, mais cette propri\'et\'e n'est vraie que localement. Cependant, $\aa_{\theta}$ est un fibr\'e vectoriel sur le tore usuel, dont la fibre est l'alg\`ebre de matrice $M_{N}(\ccc)$, et dont les lois de transformations sont pr\'ecis\'ees par les quatres relations pr\'ec\'edentes. En utilisant la relation $PQ=qQP$, on peut r\'e\'ecrire les lois de transformations non triviales sous la forme
\bbb
U_{1}(x_{1}+1,x_{2})=Q^{-1}U_{1}(x_{1},x_{2})Q\;\;\;\;
U_{2}(x_{1},x_{2}+1)=PU_{2}(x_{1},x_{2})P^{-1}.
\eee
Par cons\'equent, les fonctions de transition du fibr\'e sont simplement obtenues par conjugaison avec des matrices unitaires et constantes.

\par

Lorsque nous \'etudierons les th\'eories de jauge, il nous sera utile de pouvoir interpr\'eter ce fibr\'e comme l'alg\`ebre des endomorphismes d'un fibr\'e vectoriel de fibre $\ccc^{N}$ sur le tore usuel \cite{royal}.

\begin{pro}
Soient $a$ et $b$ deux entiers tels que $aM+bN=1$. Alors $\aa_{\theta}$ est l'alg\`ebre des endomorphismes du fibr\'e vectoriel de fibre $\ccc^{N}$ sur $T^{2}$ dont les sections v\'erifient 
\bbb
\psi(1,x_{2})=\psi(0,x_{2})\; et\; \psi(x_{1},1)=U(x_{1})\psi(x_{1},0)\;\forall(x_{1},x_{2})\in T^{2},
\eee
avec 
\bbb
U(x_{1})=\pp{&1&&\cr &&\ddots&\cr &&&1\cr e^{-2i\pi ax_{1}}&&&}.
\eee
\end{pro}
   
En vue de g\'en\'eraliser cette description du cas rationnel \`a la dimension sup\'erieure, nous avons besoin du lemme suivant, dont on pourra trouver la d\'emonstration dans \cite{lang}.

\begin{lem}
Soit $A$ une matrice antisym\'etrique $n\times n$ \`a coefficients entiers. Alors il existe une matrice $S\in SL_{n}(\zzz)$ (groupe des matrices \`a coefficients entiers dont le d\'eterminant est 1) telle que $S^{t}AS$ soit une matrice diagonale par blocs constitu\'ee uniquement de 0 et de $n'\leq [n/2]$ blocs du type
\bbb
\pp{0&\lambda_{i}\cr
-\lambda_{i}&0},
\eee
o\`u la suite $(\lambda_{i})_{1\leq i\leq n'}$ est une suite d'entiers telle que $\lambda_{i}$ divise $\lambda_{i+1}$. 
\end{lem}

En nous basant sur ce lemme, nous allons montrer le r\'esultat suivant.

\begin{pro}
Dans le cas rationnel, l'alg\`ebre $\aa_{\theta}$ est engendr\'ee par un syst\`eme de $N$ unitaires $V_{m}$ tels que les seules relations de commutations non triviales soient
\bbb
V_{2m-1}V_{2m}=e^{2i\pi\frac{M_{m}}{N_{m}}}V_{2m}V_{2m-1},
\eee
pour tout entier $m$ tel que $2m\leq n'$, o\`u $n'\leq n$ est un entier positif.
\end{pro}

\demo
Puisque $\theta$ est une matrice \`a coefficients rationnels, il existe un entier $N$ tels que $N\theta\in M_{n}(\zzz)$. En appliquant le lemme, on peut trouver une matrice $S\in SL_{n}(\zzz)$ telle que $NS^{t}\theta S=N\theta'$ soit diagonale par blocs form\'ee de blocs $2\times 2$ et de z\'eros.

\par

D\'efinissons les unitaires $V_{i}$ par
\bbb
V_{i}=U_{1}^{S_{i1}}U_{2}^{S_{i2}}\dots U_{n}^{S_{in}},
\eee
ce qui implique que $V_{i}V_{j}=e^{2i\pi\theta^{'}_{ij}}V_{j}V_{i}$ que les relations de commutations non triviales sont celles anonc\'ees.

\par

Enfin, puisque la matrice $S$ appartient \`a $SL_{n}(\zzz)$, il est clair qu'elle est inversible et que son inverse  est \'egalement une matrice \`a coefficients entiers. On peut donc inverser les relations pr\'ec\'edentes,
\bbb
U_{i}=V_{1}^{S^{-1}_{i1}}V_{2}^{S^{-1}_{i2}}\dots V_{n}^{S^{-1}_{in}},
\eee
ce qui prouve que $\aa_{\theta}$ est \'egalement engendr\'ee par le syst\`eme $(V_{i})_{1\leq i\leq n}$.
\edemo

Cela permet de d\'ecrire $\aa_{\theta}$ lorsque $\theta$ est rationnel par un fibr\'e sur le tore de dimension $n$ dont la fibre est une alg\`ebre de matrice dont la dimension est le produit des d\'enominateurs $N=N_{1}\dots N_{n}$ apparaissant dans la forme canonique de $\theta$. Les seules fonctions de transition non triviales de ce fibr\'e sont analogue \`a celles de la dimension 2.

\par   

Avant de clore cette \'etude purement alg\'ebrique des propri\'et\'es de $\aa_{\theta}$, il convient de faire deux remarques concernant l'origine g\'eom\'etrique de cette d\'eformation de l'alg\`ebre des fonctions sur le tore. 
\par

Tout d'abord, il appara\^\i t que $\aa_{\theta}$ est intimement li\'ee \`a la quantification par d\'eformation sur le tore usuel. En effet, la matrice antisym\'etrique $\theta$ peut \^etre consid\'er\'ee, lorsque son d\'eterminant est non nul, comme une forme symplectique sur le tore usuel. Le produit que nous avons d\'efini correspond alors \`a la d\'eformation du produit usuel associ\'ee \`a cette forme symplectique \cite{rieffel}.  

\par

En second lieu, il est inter\'essant de noter que cette alg\`ebre peut \^etre obtenue, en dimension 2, \`a l'aide d'un produit crois\'e \cite{pacific}. Consid\'erons le produit crois\'e de l'alg\`ebre $\aa=C^{\infty}(S^{1})$ des fonctions lisses sur le cercle par le groupe $\zzz$ agissant par une rotation d'angle $-\theta$. Dans ce cas, le produit crois\'e est l'alg\`ebre $\aa^{'}_{\theta}$ engendr\'ee par les couples $(f,m)$, o\`u $f$ est une fonction sur le cercle et $m\in\zzz$. La structure d'espace vectoriel sur $\aa^{'}_{\theta}$ est d\'etermin\'ee par celle de $\cc^{\infty}(S^{1})$ et le produit est
\bbb
(f,m).(g,n)=(f.g\,\circ\,(R_{m\theta})^{-1}, m+n),
\eee
o\`u $R_{m\theta}$ est la rotation d'angle $m\theta$. Si nous notons $U$ la fonction $x\mapsto e^{2i\pi x}$ d\'efinie sur le cercle, alors $U_{1}=(U,0)$ et $U_{2}=(U,1)$ sont deux g\'en\'erateurs unitaires de $\aa^{'}_{\theta}$ satisfaisant \`a
\bbb
U_{1}U_{2}=e^{2i\pi\theta}U_{2}U_{1},
\eee
ce qui d\'etermine un isomorphisme entre $\aa^{'}_{\theta}$ et $\aa_{\theta}$.


\subsection{L'op\'erateur de Dirac et l'espace de Hilbert}

La construction de l'espace de Hilbert $\hh$ associ\'e \`a $\aa_{\theta}$ repose essentiellement sur la construction GNS (voir Appendice A) pour l'\'etat $\phi$ d\'etermin\'e par la trace $\tau$. Soit $\hh$ la compl\'etion de l'espace vectoriel $\aa_{\theta}\ot\ccc^{2^{[n/2]}}$ muni du produit scalaire
\bbb
\langle(a_{1},\dots,a_{2^{[n/2]}}), (b_{1},\dots,b_{2^{[n/2]}})\rangle=
\mathop{\sum}\limits_{k=1}^{2^{[n/2]}}\tau\lp (a_{k})^{*}b_{k}\rp.
\eee
$\aa_{\theta}$ agit sur $\hh$ par multiplication \`a gauche et il est facile de voir, en utilisant les propri\'et\'es de la trace, que cette action $\pi$ est une repr\'esentation d'alg\`ebre involutive par des op\'erateurs born\'ees.  

\par

La compl\'etion de l'espace pr\'ec\'edent revient simplement \`a remplacer les suites \`a d\'ecroissance rapide de $\aa_{\theta}$ par les suites de carr\'e sommable, les vecteurs $U^{p}\ot e_{i}$ formant une base hilbertienne de $\hh$ si $(e_{i})_{1\leq i\leq 2^{[n/2]}}$ est une base de $\ccc^{[n/2]}$. 

\par

Nous d\'efinissons sur $\hh$ une structure de bimodule en utilisant l'op\'erateur $J=C*$, o\`u $C$ est la conjugaison de charge usuelle agissant sur les matrices de Dirac en dimension $n$ et $*$ est l'involution de $\aa_{\theta}$.

\par

Avant de d\'efinir l'op\'erateur de Dirac, commen\c cons par donner l'analogue de la d\'erivation pour $\aa_{\theta}$. Si $\partial_{i}$ est la d\'erivation qui v\'erifie $\partial_{i}(U_{j})=2i\pi\delta_{ij}$, son action sur les mon\^omes est $\partial_{i}U^{p}=2i\pi p_{i}U^{p}$. Ces d\'erivations sont parfaitement analogues aux d\'erivations par rapport aux coordonn\'ees usuelles sur le tore. Cependant, il convient de remarquer que dans le cas non commutatif, si $a$ est un \'el\'ement de $\aa_{\theta}$ qui n'est pas central et $\partial$ une d\'erivation, alors $a\partial$ n'est pas n\'ecessairement une d\'erivation. Aussi est-il difficile de d\'efinir sur le tore non commutatif des d\'erivations qui soient analogues aux champs de vecteurs non constants, ce qui est \`a mettre en parall\`ele avec la difficult\'e que l'on rencontre pour construire des automorphismes de $\aa_{\theta}$ non triviaux. De plus, cela implique qu'il n'est pas possible de construire l'analogue non commutatif du rep\`ere mobile et cela nous emp\^eche de d\'evelopper la th\'eorie de la relativit\'e g\'en\'erale sur le tore non commutatif. Ceci est en accord avec le fait que l'alg\`ebre $\aa_{\theta}$ n'admet pas suffisament d'automorphismes dans le cas non d\'eg\'en\'er\'e \cite{rieffel}.  

\par

En notant $(\gamma^{\mu})_{1\leq \mu\leq n}$ les matrices de Dirac hermitiennes satisfaisant \`a
\bbb
\gamma^{\mu}\gamma^{\nu}+\gamma^{\nu}\gamma^{\mu}=2g^{\mu\nu},\label{odh2}
\eee
o\`u $g^{\mu\nu}$ est la m\'etrique euclidienne, nous d\'efinissons l'op\'erateur de Dirac par $\dd=i\gamma^{\mu}\partial_{\mu}$. Bien entendu, cet op\'erateur est non born\'e car les d\'erivations ne sont pas d\'efinies sur tout $\hh$ mais seulement sur les suites \`a d\'ecroissance rapide. Le second membre de (\ref{odh2}) peut toujours \^etre remplac\'e par une matrice sym\'etrique d\'efinie positive, dont les valeurs propres correspondent aux inverses des "rayons" des diff\'erents cercles constituant le tore. Pour simplifier nos notations nous supposerons toujours que ces nombres sont \'egaux \`a 1, mais l'extension au cas g\'en\'eral est \'evidente.

\par

Puisque $\dd$ est form\'e de d\'erivations et que la structure de bimodule est donn\'ee par la multiplication \`a gauche et \`a droite, il est facile de v\'erifier que la condition du premier ordre est satisfaite. De plus, si, lorsque $n$ est pair on pose $\gamma=\gamma^{n+1}$, o\`u $\gamma^{n+1}$ est la chiralit\'e des matrices de Dirac en dimension $n$, on v\'erifie ais\'ement que l'axiome de r\'ealit\'e est aussi satisfait.


\subsection{V\'erification des axiomes}

Nous avons d\'ej\`a montr\'e que les axiomes de r\'ealit\'e et la condition du premier ordre sont satisfaites. Nous allons maintenant v\'erifier tous les autres axiomes, \`a l'exception de la dualit\'e de Poincar\'e car la v\'erification de cette derni\`ere n\'ecessite une description d\'etaill\'ee des modules projectifs sur $\aa_{\theta}$ qui nous entrainerait trop loin \cite{canadian}.

\par

Le premier axiome que nous allons v\'erifier est l'axiome de la dimension. Pour cela remarquons que $\dd^{2}=\Delta I_{2^{n/2}}$  est analogue au Laplacien et que ses vecteurs propres sont donn\'es par les modes de Fourier $\Delta U^{p}=4\pi^{2}|p|^{2}U_{p}$. Lorsque $\Lambda\rightarrow\infty$, le nombre de valeurs propres de $\Delta$ qui sont sup\`erieures \`a $\Lambda^{2}$ est \'egal au nombre d'\'el\'ements de $\zzz^{n}$ satisfaisant \`a $|p|\leq \frac{\Lambda}{2\pi}$, ce qui \'equivaut au volume de la boule de rayon $\frac{\Lambda}{2\pi}$ qui est $V_{n}\frac{\Lambda^{n}}{(2\pi)^{n}}$, o\`u $V_{n}$ est le volume de la boule unit\'e en dimension $n$. Par cons\'equent, si on note $(\lambda_{k})_{k\in\nnn}$ la suite d\'ecroissante des valeurs propres de $ds=|\dd|^{-n}$, on a
\bbb
\lambda_{k}\mathop{\simeq}\limits_{k\rightarrow+\infty}
\frac{1}{\Gamma(n/2+1)(4\pi)^{n/2}}\frac{1}{k},
\eee
ce qui prouve que l'axiome de dimension est satisfait. 

\par

De plus, on d\'eduit de la relation pr\'ec\'edente que 
\bbb
\mathop{\lim}\limits_{N\rightarrow+\infty}
\frac{1}{\log N}\mathop{\sum}\limits_{k=1}^{N}\lambda_{k}
=\frac{1}{\Gamma(n/2+1)(4\pi)^{n/2}},
\eee 
ce qui revient \`a dire que $\t_{\omega}(ds^{n})$ est ind\'ependant de $\omega$ et est \'egal \`a $\lp(4\pi)^{(n/2)}\Gamma(n/2+1)\rp^{-1}$. Enfin, lorsque nous modifions l'op\'erateur de Dirac de mani\`ere \`a introduire une m\'etrique non triviale $g_{\mu\nu}$ de d\'eterminant $g$, il est facile de v\'erifier que l'on obtient
\bbb
\t_{\omega}(ds^{n})=\frac{\sqrt{g}}{\Gamma(n/2+1)(4\pi)^{n/2}}.\label{vda1}
\eee

\par

Pour tout $a\in\aa$, $a$ et $[\dd,a]=i\gamma^{\mu}\partial_{\mu}a$ sont des op\'erateurs de multiplication par des \'el\'ements de $\aa$, ils appartiennent donc aux domaines des d\'erivations $\delta^{n}=[|\dd|^{n},.]$ puisque le domaine de toutes les puissances $|\dd|^{n}$ est form\'e des \'el\'ements de $\aa\ot\ccc^{2^{[n/2]}}$.

\par

Pour montrer que l'axiome d'orientabilit\'e est satisfait, nous devons exhiber un cycle de Hochschild $c$ tel que $\pi(c)=\gamma$. Pour cela, introduisons 
\bbb
c=\mathop{\sum}\limits_{\sigma\in S_{n}}\epsilon(\sigma)
U_{\sigma(n)}^{-1}\dots U^{-1}_{\sigma(1)}\ot U_{\sigma(1)}\ot\dots\ot U_{\sigma(n)}.
\eee
$c$ est un cycle de Hochschild car
\bbbb
&bc=\mathop{\sum}\limits_{i=1}^{N-1}(-1)^{i}
\mathop{\sum}\limits_{\sigma\in S_{n}}\epsilon(\sigma)
U_{\sigma(n)}^{-1}\dots U^{-1}_{\sigma(1)}\ot U_{\sigma^{1}}\ot\dots\ot U_{\sigma(i)}U_{\sigma(i+1)}\ot\dots\ot U_{\sigma(n)}&\n\\
&+\mathop{\sum}\limits_{\sigma\in S_{n}} \epsilon(\sigma)
U_{\sigma(n)}^{-1}\ot\dots\ot U_{\sigma(2)}^{-1}\ot U_{\sigma(2)}\ot \dots \ot U_{\sigma(n)}&\n\\
&
+\mathop{\sum}\limits_{\sigma\in S_{n}} \epsilon(\sigma)
(-1)^{N}
U_{\sigma(n-1)}^{-1}\ot\dots\ot U_{\sigma(1)}^{-1}\ot U_{\sigma(1)}\ot \dots \ot U_{\sigma(n-1)}&\n\\
&=0&
\eeee
car le premier terme du second membre de cette \'equation est identiquement nul (changer $\sigma$ en $\sigma\tau_{i}$, o\`u $\tau_{i}$ est la permutation qui \'echange $i$ et $i+1$) alors que la somme des deux autres termes est aussi nulle (changer $\sigma$ en $\sigma\tau$, o\`u $\tau$ est la permutation circulaire).  

\par

Il est clair que
\bbbb
\pi(c)&=&\mathop{\sum}\limits_{\sigma\in S_{n}}
(2i\pi)^{n}\epsilon(\sigma)\gamma^{\sigma(1)}\dots\gamma^{\sigma(n)}\n\\
&=&(2i\pi)^{n}n!\gamma^{1}\dots\gamma^{n}.
\eeee
Lorsque la dimension $n$ est paire, la chiralit\'e $\gamma^{n+1}$ est d\'efinie par
\bbb
\gamma^{n+1}=i^{n/2}\gamma^{1}\dots\gamma^{n},
\eee 
ce qui montre que $c'=\frac{i^{n/2}}{(2i\pi)^{n}n!}c$ est un cycle de Hochschild tel que $\pi(c')=\gamma^{n+1}$ et prouve de la sorte la validit\'e de l'axiome d'orientabilit\'e. De m\^eme, lorsque $n=m+1$ est impair les matrices de Dirac en dimension $n$ sont obtenues en adjoignant aux matrices de Dirac de la dimension $m$, la chiralit\'e $\gamma^{m+1}$. En posant $c'=\frac{i^{[n/2]}}{(2i\pi)^{n}n!}$, on obtient $\pi(c')=1$, ce qui v\'erifie l'axiome d'orientabilit\'e en dimension impaire. 

\par

Afin de v\'erifier l'axiome de finitude, nous devons d\'eterminer la trace de Dixmier de mani\`ere explicite pour tous les op\'erateurs de la forme $\pi(a)|\dd|^{-n}$. 

\begin{pro}
Pour tout $a\in\aa_{\theta}$, on a
\bbb
\t_{\omega}(\pi(a)|\dd|^{-n})=\frac{1}{2^{n-[n/2]}\pi^{(n/2)}\Gamma(n/2+1)}
\tau(a),
\eee
o\`u $\tau$ est la trace canonique de $\aa_{\theta}$.
\end{pro}
\demo
Puisque le tore satisfait \`a l'axiome de r\'egularit\'e, il est clair que l'application $a\mapsto\t_{\omega}(\pi(a)|\dd|^{-n})$ est une trace sur $\aa_{\theta}$ \cite{cipriani}. Ainsi, elle s'annule sur tous les \'el\'ements non centraux et est proportionelle \`a $\tau$ sur ces \'el\'ements. Le centre de $\aa_{\theta}$ est isomorphe \`a l'alg\`ebre des fonctions lisses sur un tore ordinaire et $\dd$ se r\'eduit dans ce cas \`a l'op\'erateur de Dirac usuel. Aussi la trace de Dixmier est-elle dans ce cas proportionnelle \`a l'int\'egrale usuelle et s'annule sur tous les unitaires centraux $U^{p}$ tels que $p\neq 0$. Ainsi, $a\mapsto\t_{\omega}(\pi(a)|\dd|^{-n})$ est proportionelle \`a la trace canonique $\tau$, le coefficient de proportionnalit\'e \'etant d\'etermin\'e par (\ref{vda1}).  
\edemo

L'intersection $\hh^{\infty}$ des domaines des puissances de $\dd$ est form\'e des \'el\'ements de $\hh$ qui sont des suites \`a d\'ecroissance rapide.  Par cons\'equent, $\hh^{\infty}=\aa^{\theta}\ot\ccc^{2^{[n/2]}}$ est un module trivial sur $\aa_{\theta}$. De plus la relation (\ref{cvs1}) intervenant dans l'axiome de finitude s'\'ecrit
\bbb
\mathop{\sum}\limits_{i=1}^{2^{[n/2]}}\tau(\xi^{*}a\zeta_{i})=
\t_{\omega}\lp(\xi,a\zeta)|\dd|^{-n}\rp=
\pi^{(n/2)}2^{n-[n/2]}\Gamma(n/2+1)\tau\lp(\xi,\zeta)a\rp
\eee
pour tous $\xi,\zeta\in\hh^{\infty}$ et $a\in\aa$. Puisque $\tau$ est fid\`ele, cela \'equivaut \`a 
\bbb
(\xi,\zeta)=\mathop{\sum}\limits_{i=1}^{2^{[n/2]}}\xi^{*}\zeta_{i}
\eee
qui d\'efinit bien une forme hermitienne sur $\hh^{\infty}$.

\par

Nous avons v\'erifi\'e que tous les axiomes de la g\'eom\'etrie non commutative sont satisfaits pour le tore non commutatif, \`a l'exception de la dualit\'e de Poincar\'e, que nous admettons sans preuve.
 

\subsection{Calcul diff\'erentiel}

Afin de pouvoir \'etudier les th\'eories de jauge sur le tore non commutatif, nous devons d\'eterminer le calcul diff\'erentiel associ\'e au triplet spectral $(\aa,\hh,\dd)$. Pour cela, commen\c cons par \'etudier la repr\'esentation $\pi$ de l'alg\`ebre diff\'erentielle universelle donn\'ee par
\bbb
a_{0}\delta  a_{1}\dots\delta a_{p}\mapsto
\pi(a_{0})\lb\dd,a_{1}\rb\dots\lb\dd,\pi(a_{p})\rb=
(i)^{n}\gamma^{\mu_{1}}\dots\gamma^{\mu_{p}} a_{0}\partial_{\mu_{1}}a_{1}\dots\partial_{\mu_{p}}a_{p}, 
\eee
pour tout $a_{0},a_{1},\dots,a_{p}\in\aa$. R\'eciproquement, tout produit de $p$ matrices de Dirac $\gamma^{\mu_{1}}\dots\gamma^{\mu_{p}}$ est \'el\'ement de $\pi(\Omega^{p}(\aa))$ puisque
\bbb
\pi\lp \delta U_{1}\dots\delta U_{p}\rp= (-2\pi)^{p}\gamma^{\mu_{1}}\dots\gamma^{\mu_{p}},
\eee 
ce qui montre que
\bbb
\pi(\Omega^{p}(\aa))=\la \omega_{\mu_{1}\dots\mu_{p}}\gamma^{\mu_{1}}\dots\gamma^{\mu_{p}}\;\mathrm{avec}\;\omega_{\mu_{1}\dots\mu_{p}}\in\aa_{\theta}\ra. 
\eee
On montre alors par r\'ecurrence sur $p$ qu'un \'el\'ement quelconque $\omega$ de $\pi(\Omega^{p}(\aa))$ s'\'ecrit de mani\`ere unique sous la forme
\bbb
\omega=\mathop{\sum}\limits_{0\leq 2k\leq \inf(p,n)}
\mathop{\sum}\limits_{\mu_{1} < \dots < \mu_{\inf(p,n)-2k}}
\omega_{\mu_{1}\dots\mu_{\inf(p,n)-2k}}\gamma^{\mu_{1}\dots\mu_{\inf(p,n)-2k}},
\eee
avec $\omega_{\mu_{1}\dots\mu_{k}}\in\aa_{\theta}$ compl\`etement antisym\'etrique et 
\bbb
\gamma^{\mu_{1}\dots\mu_{k}}=\frac{1}{k!}
\mathop{\sum}\limits_{\sigma\in S_{k}}\epsilon(\sigma)
\gamma^{\mu_{\sigma(1)}}\dots\gamma^{\mu_{\sigma(k)}}.
\eee

Cette d\'ecomposition nous permet de d\'eterminer $\pi(\jj)$ explicitement. 

\begin{lem}
Les \'el\'ements de $\pi(J)\cap\pi(\Omega^{p}(\aa))$ sont les \'el\'ements de $\pi(\Omega^{p}(\aa))$ qui se d\'eveloppent sur un nombre $k<p$ de matrices de Dirac. 
\end{lem}

\demo
Commen\c cons par montrer que $1\in\pi(\Omega^{2}(\aa))$. En effet, si $U=U_{1}$, on a
\bbb
U\lb\dd, U^{-1}\rb+\lb\dd,U\rb U^{-1}=0
\eee
ce qui implique que
\bbb
1=\frac{1}{8\pi^{2}}\lp
\lb\dd,U\rb\lb\dd, U^{-1}\rb+\lb\dd,U^{-1}\rb\lb\dd,U\rb\rp.
\eee
Puisque $J$ est un id\'eal bilat\`ere, on conclut en multipliant $1$ par  n'importe quel \'el\'ement de $\pi(\Omega^{p-2}(\aa))$, qui est une somme de produits arbitraires d'au plus $p-2$ matrices de Dirac.

\par

R\'eciproquement, si $c=\sum_{i}[\dd,\pi(a_{0}^{i})]\dots[\dd,\pi(a_{p}^{i})]$ avec 
$\sum_{i}\pi(a_{0}^{i})[\dd,\pi(a_{1}^{i})]\dots[\dd,\pi(a_{p})^{i}]=0$, on a
\bbbb
c&=&\sum_{i}\partial_{\mu_{0}}a_{0}^{i}\dots\partial_{\mu_{p}}a_{p}^{i}
\gamma^{\mu_{0}}\dots\gamma^{\mu_{p}}\n\\
&=&\sum_{i}\lp\partial_{\mu_{0}}
(a_{0}^{i}\partial_{\mu_{1}}a_{1}^{i}\dots\partial_{\mu_{p}}a_{p}^{i})-
a_{0}^{i}\partial_{\mu_{0}}
(\partial_{\mu_{1}}a_{1}^{i}\dots\partial_{\mu_{p}}a_{p}^{i})\rp
\gamma^{\mu_{0}}\dots\gamma^{\mu_{p}}\n\\
&=&-\sum_{i}
a_{0}^{i}\partial_{\mu_{0}}(
\partial_{\mu_{1}}a_{1}^{i}\dots\partial_{\mu_{p}}a_{p}^{i})
\gamma^{\mu_{0}}\dots\gamma^{\mu_{p}}.
\eeee
La r\`egle de commutation des d\'erivations $[\partial_{\mu},\partial_{\nu}]=0$ permet alors de remplacer le produit $\gamma^{\mu}\gamma^{\nu}$ par sa partie sym\'etrique, ce qui diminue de 2 le nombre de matrices de Dirac apparaissant dans l'expression de $c$.
\edemo

On en d\'eduit ais\'ement que si $p>n$ alors $\pi(\Omega^{p}(\aa))\subset \pi(J)$, ce qui implique que $\Omega_{\dd}^{p}(\aa)=0$. De plus, tout \'el\'ement de $\Omega^{p}_{\dd}(\aa)$ a, pour $p\leq n$, un unique repr\'esentant form\'e de sommes de produits totalement antisym\'etris\'es de $p$ matrices de Dirac. 

\par

Par convention, nous choisissons dans chaque classe le repr\'esentant totalement antisym\'etrique, ce qui nous m\`ene \`a l'identification de $\Omega^{p}_{\dd}(\aa)$, pour $0\leq p\leq n$ avec l'ensemble des \'el\'ements de la forme $\omega_{\mu_{1},\dots,\mu_{p}}\gamma^{\mu_{1}}\dots\gamma^{\mu_{p}}$ o\`u $\omega_{\mu_{1},\dots,\mu_{p}}\in\aa_{\theta}$ est totalement antisym\'etrique en $\mu_{1},\dots,\mu_{p}$, alors que $\Omega^{p}_{\dd}(\aa)={0}$ pour $p>n$.

\par

Par analogie avec le cas classique, nous notons $dx^{\mu}$ la 1-forme $i\gamma^{\mu}$ et nous \'ecrivons le produit de matrices de Dirac sous la forme
\bbb
dx^{\mu_{1}}\wedge\dots\wedge dx^{\mu_{p}}
=\gamma^{\mu_{1}}\dots\gamma^{\mu_{p}},
\eee 
avec $\mu_{1}<\dots<\mu_{p}$. On peut alors mettre n'importe quel \'el\'ement $\omega$ de $\Omega^{p}_{\dd}(\aa)$ sous la forme
\bbb
\omega=\frac{1}{p!}\omega_{\mu_{1}\dots\mu_{p}}
dx^{\mu_{1}}\wedge\dots\wedge dx^{\mu_{p}},
\eee
avec $\omega_{\mu_{1}\dots\mu_{p}}\in\aa$ totalement antisym\'etrique, ce qui nous permet d'obtenir une formulation du calcul diff\'erentiel qui est formellement analogue \`a celle du cas commutatif.

\begin{pro}
Si $\omega=\frac{1}{p!}\omega_{\mu_{1}\dots\mu_{p}}dx^{\mu_{1}}\wedge\dots\wedge dx^{\mu_{p}}\in\Omega^{p}_{\dd}(\aa)$ et $\eta=\frac{1}{q!}\eta_{\nu_{1}\dots\nu_{q}}dx^{\nu_{1}}\wedge\dots\wedge dx^{\nu_{q}}\in\Omega^{q}_{\dd}(\aa)$ sont deux formes differentielles alors la d\'eriv\'ee exterieure est donn\'ee par
\bbb
d\omega=\frac{1}{p!}\partial_{\nu}\omega_{\mu_{1}\dots\mu_{p}}dx^{\nu}\wedge dx^{\mu_{1}}\wedge\dots\wedge dx^{\mu_{p}}, 
\eee
et le produit par
\bbb
\omega\eta=\frac{1}{p!q!}\omega_{\mu_{1}\dots\mu_{p}}\eta_{\nu_{1}\dots\nu_{q}}
dx^{\mu_{1}}\wedge\dots\wedge dx^{\mu_{p}}\wedge dx^{\nu_{1}}\wedge\dots\wedge dx^{\nu_{q}}.
\eee
\end{pro}

Indiquons simplement le fil directeur de la d\'emonstration, qui se fait en trois \'etapes.  Nous devons d'abord exprimer $\omega$ et $\eta$ \`a l'aide de matrices de Dirac. Nous effectuons alors les op\'erations en questions (d\'erivation et produit) et nous antisym\'etrisons enti\`erement le r\'esultat. Puis nous \'ecrivons le r\'esultat en utilisant \`a nouveau les notations de l'alg\`ebre ext\'erieure.  

\par

Il est important de noter que $dx^{\mu}$ n'est pas la diff\'erentielle d'une quelconque coordonn\'ee $x^{\mu}$ puisque nous n'avons jamais d\'efini $x^{\mu}$! Seuls les symboles $dx^{\mu_{1}}\wedge\dots\wedge dx^{\mu_{p}}$ ont un sens: ils forment une base du $\aa$-module libre $\Omega_{\dd}(\aa)$. En particulier, les "produits" $dx^{\mu_{1}}\wedge\dots\wedge dx^{\mu_{p}}$ lorsque $1\leq\mu_{1}< \dots< \mu_{p}\leq n$ forment une base de $\Omega^{p}_{\dd}(\aa)$, ce qui prouve que sa dimension est $C^{p}_{n}=\frac{n!}{p!(n-p)!}$.

\par

Cela nous montre que $\Omega^{p}_{\dd}(\aa)$ et $\Omega^{n-p}_{\dd}(\aa)$ ont m\^eme dimension. En compl\`ete analogie avec le cas commutatif, nous d\'efinissons un ismorphisme de $\aa$-modules entre $\Omega^{p}_{\dd}(\aa)$ et $\Omega^{n-p}_{\dd}(\aa)$ par
\bbb
*(dx^{\mu_{1}}\wedge\dots\wedge dx^{\mu_{p}})=
\frac{1}{(n-p)!}\epsilon^{\mu_{1}\dots\mu_{p}}_{\quad\quad\quad\nu_{1}\dots\nu_{n-p}}dx^{\nu_{1}}\wedge\dots\wedge dx^{\nu_{n-p}}
\eee
o\`u $\epsilon_{}$ est le tenseur compl\`etement antisym\'etrique associ\'e \`a la m\'etrique euclidienne. Puisqu'il est absolument identique \`a celui d\'efini dans le cas commutatif, il est clair que nous avons $*^{2}\omega=(-1)^{p(n-p)}\omega$ pour tout $\omega\in\Omega^{p}_{\dd}(\aa)$.

\par

La trace canonique $\tau$ sur $\aa_{\theta}$ est parfaitement analogue \`a l'int\'egrale sur le tore usuel de volume 1, puisque
\bbb
\tau\lp\mathop{\sum}\limits_{p\in\zzz^{n}}a_{p}U^{p}\rp=a_{0}.
\eee 
Aussi la noterons-nous dor\'enavant comme une integrale: $\tau(a)=\int a$ pour tout $a\in\aa_{\theta}$. Etant donn\'e que la d\'eriv\'ee d'un \'el\'ement de $\aa_{\theta}$ ne contient plus de terme constant, l'int\'egrale d'une d\'eriv\'ee est nulle. 

\begin{lem}
Pour tout $a\in\aa_{\theta}$, on a
\bbb
\int\partial_{\mu}a=0.
\eee
\end{lem}

Ce r\'esultat \'el\'ementaire s'av\`ere d'une grande utilit\'e car il permet d'int\'egrer par parties au niveau des \'el\'ements de l'alg\`ebre,
\bbb
\int a\partial_{\mu}(b)=-\int\partial_{\mu}(a)b
\eee
pour tous $a,b\in\aa_{\theta}$.

\par

Rappelons que la trace de Dixmier est reli\'ee \`a la trace canonique $\displaystyle\int$ par
\bbb
\t_{\omega}\lp \pi(a)|\dd|^{-n}\rp=\frac{2^{[n/2]-n}}{\pi^{n/2}\Gamma(n/2+1)}
\int a
\eee 
pour tout $a\in\aa$. Par lin\'earit\'e, cela se g\'en\'eralise aux formes de degr\'e sup\'erieur en incluant une trace sur les matrices de Dirac. Si $\omega\in\pi(\Omega^{p}(\aa))$, alors on peut l'\'ecrire sous la forme $\omega_{\mu_{1}\dots\mu_{p}}\gamma^{\mu_{1}}\dots\gamma^{\mu_{p}}$ et on a
\bbbb
\t_{\omega}\lp\pi(\omega)|\dd|^{-n}\rp&=&
\frac{1}{2^{[n/2]}}\t\lp\gamma^{\mu_{1}}\dots\gamma^{\mu_{p}}\rp
\t_{\omega}\lp\pi(\omega_{\mu_{1}\dots\mu_{p}})|\dd|^{-n}\rp\n\\
&=&\frac{1}{2^{n}\pi^{n/2}\Gamma(n/2+1)}
\t\lp\gamma^{\mu_{1}}\dots\gamma^{\mu_{p}}\rp\int\omega_{\mu_{1}\dots\mu_{p}}.
\eeee
Cela prouve que l'application 
\bbb
\langle \pi(\omega),\pi(\eta)\rangle=
\t_{\omega}\lp\pi(\omega)\pi(\eta)^{*}|\dd|^{-n}\rp
\eee
defini un  produit scalaire sur $\pi(\Omega^{p}(\aa))$. En effet, puisque $\t_{\omega}$ se r\'eduit \`a la trace usuelle sur les matrices de Dirac et \`a la trace canonique sur $\aa_{\theta}$ qui sont deux traces fid\`eles, il est clair que l'on ne peut avoir $\langle\pi(\omega),\pi(\omega)\rangle=0$ sans avoir $\pi(\omega)=0$. Puisque nous savons d\'ej\`a (cf \S 1.2.1) que cette application est une forme sesquilin\'eaire hermitienne positive, nous avons montr\'e que c'est un produit scalaire.

\par

De plus, le choix d'un repr\'esentant totalement antisym\'etrique dans chaque classe correspond au choix du repr\'esentant orthogonal \`a tous les \'el\'ements de $\pi(\jj)\cap\pi(\Omega^{p}(\aa))$. En effet, le repr\'esentant totalement antisym\'etrique d'un \'el\'ement de $\Omega^{p}_{\dd}(\aa)$ peut toujours se mettre sous la forme $\omega\gamma^{\mu_{1}}\dots\gamma^{\mu_{p}}$, avec $1\leq\mu_{1}<\dots<\mu_{p}\leq n$ et tout \'el\'ement de $\pi(\jj)\cap\pi(\Omega^{p}(\aa))$ se d\'eveloppe sur une somme de produits de $k$ matrices de Dirac du type $\gamma^{\nu_{1}}\dots\gamma^{\nu_{p}}$, o\`u $k<p$ est un entier ayant m\^eme parit\'e que $p$. Puisque $k<p$, il existe dans le produit $\gamma^{\mu_{1}}\dots\gamma^{\mu_{p}}$ une matrice $\gamma^{\mu}$ qui ne soit pas dans le produit $\gamma^{\nu_{1}}\dots\gamma^{\nu_{k}}$. Quitte \`a changer le signe du produit, nous pouvons supposer que cette matrice est $\gamma^{\mu_{p}}$. On a alors
\bbbb
\t\lp (\gamma^{\mu_{1}}\dots\gamma^{\mu_{p}})^{*}
\gamma^{\nu_{1}}\dots\gamma^{nu_{k}}\rp
&=&
\t\lp \gamma^{\mu_{p}}\dots\gamma^{\mu_{1}}
\gamma^{\nu_{1}}\dots\gamma^{nu_{k}}\rp\n\\
&=&
(-1)^{p-1}\t\lp \gamma^{\mu_{p-1}}\dots\gamma^{\mu_{1}}\gamma^{\mu_{p}}
\gamma^{\nu_{1}}\dots\gamma^{\nu_{k}}\rp\n\\
&=&
(-1)^{p+k-1}\t\lp \gamma^{\mu_{p-1}}\dots\gamma^{\mu_{1}}
\gamma^{\nu_{1}}\dots\gamma^{\nu_{k}}\gamma^{\mu_{p}}\rp\n\\
&=&
-\t\lp \gamma^{\mu_{p}}\dots\gamma^{\mu_{1}}
\gamma^{\nu_{1}}\dots\gamma^{\nu_{k}}\rp.
\eeee 
Cela prouve que $\t\lp \gamma^{\mu_{p}}\dots\gamma^{\mu_{1}}
\gamma^{\nu_{1}}\dots\gamma^{\nu_{k}}\rp=0$, ce qui implique que le repr\'esentant totalement antisym\'etrique est orthogonal \`a $\pi(\jj)\cap\pi(\Omega^{p}(\aa))$.

Enfin, montrons que le tore non commutatif est une "vari\'et\'e sans bord", ce qui nous permettra d'appliquer tous les r\'esultats relatifs aux th\'eories de Yang-Mills et de Chern-Simons

\begin{pro}
Le triplet spectral $(\aa,\hh,\dd)$ du tore non commutatif satisfait \`a la condition de fermeture. 
\end{pro}

\demo
Soit $\gamma$ la chiralit\'e de $(\aa,\hh,\dd)$ ($\gamma=\gamma^{n+1}$ si la dimension est paire et $\gamma=1$ si la dimension est impaire). Par d\'efinition de la condition de fermeture, nous devons montrer que
\bbb
\t_{\omega}\lp[\dd,\pi(a_{0})]\dots[\dd,\pi(a_{n})]|\dd^{n}|\rp=0
\eee
pour tous $a_{0},\dots,a_{n}\in\aa$.

\par

Commen\c cons par \'etudier le cas de la dimension paire. Nous savons que
\bbb
\gamma^{n+1}\gamma^{\mu_{1}}\dots\gamma^{\mu_{n}}=
(-i)^{n/2}\epsilon^{\mu_{1}\dots\mu_{n}},
\eee
o\`u $\epsilon^{\mu_{1}\dots\mu_{n}}$ est le tenseur totalement antisym\'etrique tel que $\epsilon^{12\dots n}=1$. On en d\'eduit que
\bbb
\t\lp\gamma^{n+1}\gamma^{\mu_{1}}\dots\gamma^{\mu_{n}}\rp=
(-2i)^{n/2}\epsilon^{\mu_{1}\dots\mu_{n}}.
\eee
En dimension impaire, nous avons aussi
\bbb
\t\lp\gamma^{n+1}\gamma^{\mu_{1}}\dots\gamma^{\mu_{n}}\rp=
(-2i)^{[n/2]}\epsilon^{\mu_{1}\dots\mu_{n}}.
\eee
En effet, si $\gamma^{n+1}$ apparait un nombre paire de fois dans la relation pr\'ec\'edente, nous pouvons l'\'eliminer en utilisant $(\gamma^{n+1})^{2}=1$. Il nous reste un produit d'un nombre impair de matrices de Dirac en dimension paire $n-1$ et la relation  pr\'ec\'edente est v\'erifi\'ee puisque ces deux membres sont nuls. S'il y a un nombre impair de matrices \'egales \`a $\gamma^{n+1}$ et si ce nombre est plus grand que $1$, alors les deux membres de l'\'equation pr\'ec\'edente sont identiquement nuls. Enfin, si une seule de ces matrices est \'egale \`a $\gamma^{n+1}$ nous sommes ramen\'es \`a l'\'etude du cas pair. 

\par

En cons\'equence, nous avons
\bbbb
&\t_{\omega}\lp[\dd,\pi(a_{0})]\dots[\dd,\pi(a_{n})]|\dd^{n}|\rp&\n\\
&=\frac{(-2i)^{[n/2]}}{2^{n}\pi^{n/2}\Gamma(n/2+1)}
\epsilon^{\mu_{1}\dots\mu_{n}}\int \partial_{\mu_{1}}a_{1}\dots\partial_{\mu_{p}}a_{p}&\n\\
&=
\frac{(-2i)^{[n/2]}}{2^{n}\pi^{n/2}\Gamma(n/2+1)}
\epsilon^{\mu_{1}\dots\mu_{n}}\int
\lp\partial_{\mu_{1}}( a_{1}\partial_{\mu_{2}}a_{2}\dots\partial_{\mu_{p}}a_{p})-
a_{1}\partial_{\mu_{1}}(\partial_{\mu_{2}}a_{2}\dots\partial_{\mu_{p}}a_{p})
\rp&\n\\
&=0,&
\eeee
o\`u nous avons utilis\'e la relation $\int\partial_{\mu_{1}}=0$ ainsi que la commutation des d\'erivations.
\edemo

De plus, avec les notations que nous avons introduites, toute l'alg\`ebre diff\'erentielle sur $\aa_{\theta}$ est formellement identique \`a l'alg\`ebre des formes diff\'erentielles usuelles. 

\par

Enfin, il d\'ecoule de l'\'etude pr\'ec\'edente et de la normalisation de $\displaystyle\dix$ que celle-ci co\"\i ncide avec l'int\'egrale $\int$.

\begin{pro}
Pour tout $a\in\aa$, on a
\bbb
\dix a ds^{n}=\int a
\eee
ainsi que
\bbb
\dix\lp\gamma a_{0}da_{1}\dots da_{n}\rp ds^{n}=
\epsilon^{\mu_{1}\dots\mu_{n}}\dix\lp a_{0}\partial_{\mu_{1}}a_{1}\dots\partial_{\mu_{n}}a_{n}\rp.
\eee
\end{pro}
Par cons\'equent, l'utilisation de la notation $\displaystyle\dix$ ne pr\'esente pas d'utilit\'e dans le cas du tore et nous la remplacerons par $\int$ et, le cas \'ech\'eant, par une contraction avec le tenseur totalement antisym\'etrique .

\subsection{Les sym\'etries et le tenseur \'energie-impulsion}

Dans le cas g\'en\'eral, les sym\'etries d'un espace non commutatif sont d\'efinies comme les automorphismes de l'alg\`ebre des coordonn\'ees. Pour \'etudier les sym\'etries du tore non commutatif, il nous faut donc \'etudier les automorphismes de l'alg\`ebre $\aa_{\theta}$. Nous ne d\'eterminerons pas tous les automorphismes de $\aa_{\theta}$, mais nous allons simplement en donner quelques-uns et \'etudier leur signification en th\'eorie des champs que l'on peut d\'evelopper sur le tore non commutatif.

\par

Puisque $\aa_{\theta}$ est en g\'en\'eral une alg\`ebre non commutative, il est clair que l'on peut construire des automorphismes int\'erieurs. Ces derniers correspondent aux sym\'etries de jauge sur lesquelles nous reviendrons dans les deux derni\`eres parties de cette th\`ese.

\par

L'\'etude du cas commutatif sugg\`ere l'existence de sym\'etries de translation. En effet, le tore de dimension $n$ peut \^etre identifi\'e au groupe $(S^{1})^{n}$ qui agit sur lui-m\^eme par translation. Cette action translate toutes les fonctions sur le tore, aussi d\'efinissons-nous l'action de $\alpha\in(S^{1})^{n}$ (consid\'er\'e comme le quotient $\rrr^{n}/\zzz^{n}$) sur les mon\^omes de base de $\aa_{\theta}$ par
\bbb
T_{\alpha}(U^{p})=e^{2i\pi\alpha.p}U^{p}.
\eee
Le produit dans $\aa_{\theta}$ est donn\'e par $U^{p}U^{q}=e^{i\pi\theta(p,q)}U^{p+q}$ ce qui implique que $T_{\alpha}$ est un automorphisme de $\aa_{\theta}$ car
\bbb
T_{\alpha}(U^{p})T_{\alpha}(U^{q})=e^{2i\pi\alpha.(p+q)}U^{p}U^{q}=
T_{\alpha}(U^{p}U^{q}),
\eee
ainsi que $T_{\alpha}T_{-\alpha}=T_{-\alpha}T_{\alpha}=1$ pour tout $\alpha\in (S^{1})^{n}$.

\par

Au niveau infinit\'esimal, cette transformation correspond aux d\'erivations $\partial_{\mu}$ que nous avons introduites lors de la construction de l'op\'erateur de Dirac. En effet, lorsque $\delta\alpha$ tend vers 0 on a
\bbb
T_{\delta\alpha}(U^{p})=e^{2i\pi\delta\alpha.p}U^{p}=
U^{p}+(\delta\alpha)^{\mu}U^{p}+O(\delta\alpha^{2}).
\eee
Dans le cas classique, une sym\'etrie continue d'une th\'eorie de champs donne naissance \`a un courant conserv\'e. Pour \'etudier ce qui se passe sur le tore non commutatif, nous devons d'abors d\'efinir ce que nous entendons par champ sur le tore non commutatif.

\par

Un champ scalaire sur le tore non commutatif est d\'efini comme un \'el\'ement $\Phi$ de $\aa_{\theta}$, ce qui est l'analogue d'un champ scalaire usuel qui est simplement une fonction \`a valeurs r\'eelles ou complexes d\'efinie sur l'espace-temps. Bien entendu, cela se g\'en\'eralise \`a des champs scalaires portant un indice suppl\'ementaire qui sont des matrices $(\Phi_{i})_{1\leq i\leq N}\in\lp\aa^{\theta}\rp^{N}$ ou \`a des \'el\'ements d'un module projectif fini \`a droite sur $\aa$, ce qui repr\'esente des champs ayant une topologie non triviale.

\par

Consid\'erons une fonction $\ll$ de $\aa_{\theta}\times(\aa_{\theta})^{n}$, appel\'ee lagrangien, et associons \`a tout champs scalaire $\Phi$ un nombre complexe $S[\Phi]$, appel\'e action de $\Phi$, par
\bbb
S[\Phi]=\int\ll(\Phi,\partial_{\mu}\Phi).
\eee
Bien entendu, ces d\'efinitions sont identiques aux d\'efinitions classiques de la th\'eorie des champs et nous allons d\'eterminer les points critiques de la fonctionelle $S$.

\par

Pour les lagrangiens du type 
\bbb
\ll(\Phi,\partial_{\mu}\Phi)=\frac{1}{2}g^{\mu\nu}\partial_{\mu}\Phi\partial_{\nu}\Phi-V(\Phi),
\eee
o\`u $g^{\mu\nu}$ est une matrice positive et $V(\Phi)$ un  polyn\^ome en $\Phi$, nous pouvons d\'eterminer les \'equations d'Euler associ\'ees \`a la minimisation de l'action. Si nous ne faisons pas l'hypoth\`ese pr\'ec\'edente sur la forme de $\ll$, le probl\`eme est nettement plus difficile \`a traiter et nous \'etudierons ult\'erieurement le probl\`eme des \'equations de Yang-Mills (cf \S 4.2.2).

\par

$\Phi$ est un extremum de l'action si et seulement si, pour tout $\delta\Phi\in\aa_{\theta}$, on a
\bbb
S[\Phi+t\delta\Phi]-S[\Phi]=O(t^{2})\label{motion}
\eee 
lorsque $t\rightarrow 0$. On a
\bbbb
&\frac{1}{2}\int g^{\mu\nu}\partial_{\mu}(\Phi+t\delta\Phi)\partial_{\nu}(\Phi+t\delta\Phi)-
\frac{1}{2}\int g^{\mu\nu}\partial_{\mu}\Phi\partial_{\nu}\Phi
=&\n\\
&t\int g^{\mu\nu}\partial_{\mu}\Phi\partial_{\nu}(\delta\Phi)+o(t^{2})=&\n\\
&-t\int g^{\mu\nu}\partial_{\mu}\partial_{\nu}\Phi\delta\Phi+o(t^{2}),&
\eeee
apr\`es int\'egration par parties. Puisque $V$ est un polyn\^ome que l'on \'ecrit sous la forme
\bbb
V(\Phi)=\mathop{\sum}\limits_{k=0}^{k=p}a_{k}\Phi^{k},
\eee
on a
\bbb
V(\Phi+t\delta\Phi)=t\mathop{\sum}\limits_{k=0}^{k=p}
\mathop{\sum}\limits_{l=0}^{k-1}
a_{k}\Phi^{l}\delta\Phi\Phi^{k-1-l}
+O(t^{2}).
\eee
Utilisant les propri\'et\'es de la trace, on a
\bbb
\int
V(\Phi+t\delta\Phi)=t\int V'(\Phi)\delta\Phi+O(t^{2}).
\eee
On en d\'eduit que
\bbb
S[\Phi+t\delta\Phi]-S[\Phi]= -\int  (g^{\mu\nu}\partial_{\mu}\partial_{\nu}\Phi+V'(\Phi))\delta\Phi + o(t^{2}).
\eee 
Puisque (\ref{motion}) doit \^etre vraie quelque soit $\delta\Phi$, on obtient 
\bbb
g^{\mu\nu}\partial_{\mu}\partial_{\nu}\Phi+V'(\Phi)=0
\eee
en utilisant la fid\'elit\'e de la trace $\int$.

\par

Cette \'equation est tout-\`a-fait similaire \`a l'\'equation classique d'un champ scalaire dans un potentiel. Cependant, nous avons eu besoin de la forme particuli\`ere du lagrangien pour obtenir ces \'equations. Dans le cas g\'en\'eral, il n'est pas possible d'obtenir des \'equations du mouvement aussi simples, du fait de la non commutativit\'e de $\delta\Phi$: nous devons tenir compte des commutateurs non triviaux qui apparaissent lorsque nous mettons $\delta\Phi$ \`a gauche, comme on peut s'en convaincre en \'etudiant le lagrangien plus g\'en\'eral,
\bbb
\ll(\Phi,\partial_{\mu}\Phi)=\frac{1}{2}g^{\mu\nu}(\Phi)\partial_{\mu}\Phi
\partial_{\nu}\Phi,
\eee  
correspondant au mod\`ele $\sigma$ non lin\'eaire.

\par

Nous pouvons maintenant discuter les effets de l'invariance par translation sur le Lagrangien pr\'ec\'edent. Dans le cas classique, le th\'eor\`eme de Noether nous montre que l'invariance par translation est \`a l'origine de la conservation du tenseur \'energie-impulsion $\partial_{\mu}T^{\mu\nu}=0$ avec $T_{\mu\nu}=\partial_{\mu}\Phi\partial_{\nu}\Phi-g^{\mu\nu}\ll(\Phi,\partial_{\lambda}\Phi)$. En effet, on v\'erifie que cette \'equation est satisfaite en utilisant les \'equations du mouvement. Dans le cas du tore non commutatif, le m\^eme tenseur \'energie-impulsion n'est pas conserv\'e, car on a $\partial_{\mu}T^{\mu\nu}\neq 0$ avec
\bbb
T^{\mu\nu}=\partial^{\mu}\Phi\partial^{\nu}\Phi
-g^{\mu\nu}\lp\frac{1}{2}\partial_{\rho}\Phi\partial^{\rho}\Phi+V(\Phi)\rp.
\eee
Ce tenseur-\'energie impulsion n'est pas non plus sym\'etrique. Si on le sym\'etrise, on obtient un nouveau tenseur $\tilde{T}^{\mu\nu}$, donn\'e par
\bbb
\tilde{T}^{\mu\nu}=
\frac{1}{2}\lp\partial^{\mu}\Phi\partial^{\nu}\Phi+
\partial^{\nu}\Phi\partial^{\mu}\Phi\rp
-g^{\mu\nu}\lp\frac{1}{2}\partial_{\rho}\Phi\partial^{\rho}\Phi+V(\Phi)\rp.
\eee
On montre, en utilisant les \'equations du mouvement, que ce tenseur satisfait \`a la relation
\bbb
\partial_{\mu}\tilde{T}^{\mu\nu}=
\partial^{\nu}V(\Phi)-\frac{1}{2}\lp\partial^{\nu}\Phi V'(\Phi)+
V'(\Phi)\partial^{\nu}\Phi\rp.
\eee
En g\'en\'eral, ce tenseur \'energie-impulsion n'est pas conserv\'e, mais lorsque la th\'eorie est libre $V(\Phi)=\frac{m^{2}}{2}\Phi^{2}$, nous avons $\partial_{\mu}\tilde{T}^{\mu\nu}=0$.

\par

Nous ne pouvons pas en g\'en\'eral d\'eterminer un tenseur \'energie-impulsion qui soit conserv\'e car sa d\'efinition implique l'existence d'une m\'etrique non trviale qui permette, dans le cas classique d'une vari\'et\'e riemannienne, de d\'efinir $T^{\mu\nu}$ ind\'ependament de l'invariance par translation par 
\bbb
T^{\mu\nu}(x)=\frac{\delta}{\delta g_{\mu\nu}(x)}\int_{\mm}
\sqrt{g}d^{n}x\ll(\Phi,\partial\Phi).
\eee
Puisque nous n'avons pas de telle m\'etrique sur le tore non commutatif, nous ne pouvons pas appliquer cette construction g\'en\'erale de $T_{\mu\nu}$.

\par

Toutefois, il est possible de construire \`a partir du tenseur \'energie-impulsion des charges conserv\'ees en adjoignant au tore non commutatif une coordonn\'ee suppl\'ementaire qui correspond au temps. Cela revient \`a consid\'erer l'alg\`ebre $C^{\infty}(\rrr)\ot\aa_{\theta}$, o\`u $C^{\infty}(\rrr)$ est l'alg\`ebre des fonctions lisses sur $\rrr$, qui sont suppos\'ees d\'ecro\^\i tre suffisament vite \`a l'infini pour assurer la convergence des int\'egrales. Cela revient \`a dire que nous consid\'erons des champs $\Phi$ qui sont des fonctions de $t$ et qui se d\'eveloppent sur la base des modes de Fourier de $\aa_{\theta}$ par
\bbb
\Phi(t)=\mathop{\sum}\limits_{p\in\zzz^{n}}\Phi_{p}(t)U^{p},\;\;\;\forall
\;t\in\rrr.
\eee 
Si nous notons $\partial_{0}$ la d\'eriv\'ee par rapport au temps, la forme des \'equations du mouvement reste inchang\'ee et nous d\'efinissons
\bbb
P^{\mu}=\int T^{\mu 0}.
\eee
Bien entendu $\int$ est la trace canonique sur $\aa_{\theta}$, qui est l'analogue de l'int\'egration sur les variables d'espace, ce qui implique que $P^{\mu}$ d\'epend \`a priori du temps $t$. Tout comme dans le cas classique, nous allons montrer que $P^{\mu}$ est une charge conserv\'ee.
 
\begin{pro}
Lorsque les \'equations du mouvement sont satisfaites, on a
\bbb
\frac{d}{dt} P^{\mu}=0.
\eee
\end{pro}

\demo
La d\'emonstration de ce r\'esultat est purement alg\'ebrique car nous ne pouvons pas utiliser des concepts tels que "le flux \`a travers une surface ferm\'ee" dans le cas du tore non commutatif. Contentons-nous de calculer la d\'eriv\'ee par rapport au temps de $P^{\mu}$ dans le cas d'une m\'etrique euclidienne, le cas g\'en\'eral d'une m\'etrique quelconque, mais constante se d\'eduisant du cas euclidien par lin\'earit\'e. En notant $\dot{\Phi}$ et $\ddot{\Phi}$ les d\'eriv\'ees par rapport au temps de $\Phi$, on a
\bbb
T^{0\mu}=\int\lp\dot{\Phi}\partial^{\mu}\Phi-\frac{g^{0\mu}}{2}\dot{\Phi}
\dot{\Phi}-
\frac{g^{0\mu}}{2}\mathop{\sum}\limits_{i=1}^{n}
\partial_{i}\Phi\partial_{i}\Phi
+\frac{1}{2}V(\Phi)\rp .
\eee 
Lorsque $\mu=i\neq 0$ est une variable d'espace, on a simplement
\bbb
T^{0i}=\int\dot{\Phi}\partial_{i}\Phi,
\eee
d'o\`u 
\bbb
\frac{d}{dt} P^{i}=\int\lp\ddot{\Phi}\partial_{i}\Phi
+\dot{\Phi}\partial_{i}\dot{\Phi}\rp.
\eee 
En utilisant l'equation du mouvement, $\partial_{\mu}\Phi\partial^{\mu}\Phi+V'(\Phi)=0$, on \'elimine $\ddot{\Phi}$
\bbb
\frac{d}{dt} P^{i}=\int\lp -\Delta(\Phi)\partial_{i}\Phi-
V'(\Phi)\partial_{i}\Phi
+\dot{\Phi}\partial_{i}\dot{\Phi}\rp,
\eee
o\`u $\Delta=\sum_{i}\partial_{i}\partial^{i}$ est le Laplacien.
Puisque
\bbb
\int\lp -
V'(\Phi)\partial_{i}\Phi
+\dot{\Phi}\partial_{i}\dot{\Phi}\rp
=\int\lp -
\partial_{i}V(\Phi)
+\frac{1}{2}\partial_{i}(\dot{\Phi}\dot{\Phi})\rp=0,
\eee
il nous reste \`a  montrer que
\bbb
\int-\Delta(\Phi)\partial_{i}\Phi=0.
\eee
En d\'eveloppant $\Phi$ sur les mon\^omes
\bbb
\Phi=\mathop{\sum}\limits_{p\in\zzz^{n}}\Phi_{p}U^{p},
\eee
on obtient
\bbb
\int-\Delta(\Phi)\partial_{i}\Phi=
\mathop{\sum}\limits_{p\in\zzz^{n}}
-(2\pi)^{2}p_{i}p^{2}\Phi_{p}\Phi_{-p}=
0
\eee
en changeant $p$ en $-p$.

\par

Lorsque $\mu=0$, on obtient 
\bbb
T^{00}=\frac{1}{2}\dot{\Phi}\dot{\Phi}-
\frac{1}{2}\mathop{\sum}\limits_{i=1}^{n}
\partial_{i}\Phi\partial_{i}\Phi
+V(\Phi).
\eee
On en d\'eduit
\bbb
\frac{d}{dt}\int T^{00}=0,
\eee
apr\`es avoir \'elimin\'e $\ddot{\Phi}$ \`a l'aide des \'equations du mouvement. 
\edemo

Le second type de sym\'etrie que nous allons discuter ici sont les sym\'etries discr\`etes. Celles-ci sont l'analogue non commutatif des transformations modulaires du tore usuel, d\'efinies par l'action du groupe $SL_{2}(\zzz)$ sur le tore de dimension deux que nous identifions \`a $\rrr^{2}/\zzz^{2}$. 

\par

Ces transformations s'\'etendent en g\'eom\'etrie non commutative de la mani\`ere suivante. A toute matrice \`a coefficients entiers de la forme,
\bbb
\pp{a&b\cr c&d},\:\:\:\mathrm{avec}\;ad-bc=1,
\eee
on associe l'application de $\aa_{\theta}$ dans elle-m\^eme donn\'ee par
\bbb
U_{1}\mapsto V_{1}=U_{1}^{a}U_{2}^{b}\quad\mathrm{et}\quad U_{2}\mapsto V_{2}=U_{1}^{c}U_{2}^{d}.
\eee
En utilisant la relation $ad-bc=1$, il est facile de voir que $V_{1}$  et $V_{2}$ satisfont les m\^emes relations de commutation que $U_{1}$ et $U_{2}$. De plus, cette matrice est inversible dans $SL_{2}(\zzz)$, ce qui prouve que l'application pr\'ec\'edente est un automorphisme de $\aa_{\theta}$ et correspond \`a une sym\'etrie du tore non commutatif.

\par

En dimension sup\'erieure, toutes les transformations de $SL_{n}(\zzz)$ ne d\'efinissent pas des automorphismes de $\aa_{\theta}$. En effet, si $M$ est une matrice de $SL_{n}(\zzz)$, alors, par analogie avec la dimension 2, nous d\'efinissons une application de $\aa_{\theta}$ dans elle m\^eme par 
\bbb
U^{p}\mapsto V^{p}=U^{Mp}\;\;\;\forall\;p\in\zzz^{n}.
\eee

\begin{pro}
L'application $U^{p}\mapsto V^{p}$ est un automorphisme si et seulement si $M^{t}\theta M-\theta\in M_{n}(\zzz)$.
\end{pro}

\demo
Il suffit de remarquer que $V^{p}V^{q}=U^{Mp}U^{Mq}=e^{2i\pi\theta(Mp,Mq)}U^{p+q}$, ce qui implique que $V^{p}$ et $U^{p}$ satisfont aux m\^emes relations de commutation si et seulement si $e^{2i\pi\theta(Mp,Mq)}=e^{2i\pi\theta(p,q)}$. Cela \'equivaut \`a dire que $\theta(Mp,Mq)-\theta(p,q)\in\zzz$ quelque soient $p$ et $q$, ce qui prouve la proposition pr\'ec\'edente.
\edemo

Cette d\'emonstration nous montre aussi qu'il y a un isomorphisme entre les alg\`ebres $\aa_{\theta}$ et $\aa_{M\theta}$. La r\'eciproque est une question non r\'esolue en dimension $n\geq 3$ \cite{rieffel}: A quelles conditions sur $\theta$ et $\theta'$ les alg\`ebres $\aa_{\theta}$ et $\aa_{\theta'}$ (ou leurs fermetures normiques) sont-elles isomorphes? Certaines avanc\'ees concernant la Morita-\'equivalence de ces alg\`ebres peuvent \^etre trouv\'ees dans \cite{schwarz}.

\par

Enfin, signalons que ces trois types d'autmorphismes (automorphismes int\'erieurs, translations et automorphismes discrets) ainsi que leurs produits forment l'ensemble des automorphismes de $\aa_{\theta}$ dans un certain nombre de cas \cite{rieffel}. Cette remarque exclut la possibilit\'e d'avoir suffisamment d'automorphismes pour d\'evelopper une th\'eorie de la gravitation, ces automorphismes devant jouer le r\^ole des diff\'eomorphismes du tore usuel.

\section{Th\'eorie de Yang-Mills}

\subsection{Les champs de jauge}

Nous allons maintenant aborder la construction des th\'eories de jauge en utilisant le formalisme d\'evelopp\'e au cours du chapitre pr\'ecedent. Pour avoir des relations plus famili\`eres, nous utilisons la notation $dx^{\mu_{1}}\wedge\dots\wedge dx^{\mu_{p}}$ pour d\'esigner les \'el\'ements de base de l'espace des formes diff\'erentielles.

\par

Avant de commencer notre \'etude, signalons que la plupart des propri\'et\'es de la th\'eorie de Yang-Mills sur le tore non commutatif en dimension 2 ont \'et\'e formul\'ees dans \cite{connesrieffel} et dans \cite{spera} en dimension sup\'erieure. Cependant, cette formulation se base sur une toute autre d\'efinition de la th\'eorie de Yang-Mills qui ne peut \^etre g\'en\'eralis\'ee pour des triplets spectraux quelconques. Nous allons reprendre cette \'etude \`a l'aide du formalisme d\'evelopp\'e au cours du chapitre 1, quitte \`a red\'emontrer des r\'esultats connus. 

\par

Consid\'erons un module projectif fini $\ee$ \`a droite sur $\aa=\aa_{\theta}$. Sans perte de g\'en\'eralit\'e, nous supposons que $\ee$ est de la forme $e\aa^{N}$, o\`u $e\in M_{N}(\aa)$ satisfait \`a $e^{2}=e=e^{*}$. De plus, la structure hermitienne canonique de $\aa^{N}$ induit sur $\ee$ une structure hermitienne.  

\par

Puisque $\Omega^{1}_{\dd}(\aa)$ est form\'e des \'el\'ements de la forme $\omega_{\mu}dx^{\mu}$ avec $\omega_{\mu}\in\aa$ et $dx^{\mu}=i\gamma^{\mu}$, l'action d'une connexion $\nabla$ sur un \'el\'ement quelconque $e\xi$ de $\ee$, avec $\xi\in\aa^{N}$, s'\'ecrit
\bbb
\nabla(e\xi)=\lp e\partial_{\mu}(e\xi)+eA_{\mu}e\xi\rp\ot dx^{\mu},
\eee
o\`u $A_{\mu}\in M_{N}(\aa)$. Cette connexion est hermitienne si et seulement si la matrice $A_{\mu}$ est antihermitienne, ce que nous supposons toujours \^etre le cas.

\par

En utilisant la relation
\bbb
F=edede+ed(eAe)e+eAeAe
\eee
qui nous donne la courbure d'une connexion comme une matrice de 2-formes (cf \S1.3.2), nous pouvons, apr\`es avoir identifi\'e les 2-formes avec les produits totalement antisym\'etriques, \'ecrire $F$ sous la forme habituelle
\bbb
F=\frac{1}{2}F_{\mu\nu}dx^{\mu}\wedge dx^{\nu},
\eee
avec
\bbb
F_{\mu\nu}=e\partial_{\mu}e\partial_{\nu}e-e\partial_{\nu}e\partial_{\mu}e
+e\partial_{\mu}(eA_{\nu}e)e-e\partial_{\nu}(eA_{\mu}e)e
+[eA_{\mu}e,eA_{\nu}e].
\eee
Gr\^ace aux propri\'et\'es des matrices de Dirac et \`a la commutation des d\'erivations, la courbure peut aussi s'\'ecrire $F_{\mu\nu}=[\nabla_{\mu},\nabla_{\nu}]$, o\`u la d\'eriv\'ee covariante $\nabla_{\mu}:\;\ee\mapsto\ee$ est d\'efinie par la relation
\bbb
\nabla(e\xi)=\nabla_{\mu}(e\xi)\ot dx^{\mu}
\eee
pour tout $\xi\in\aa^{N}$.

\par

Lorsque $\aa$ est commutative (ce qui \'equivaut \`a $\theta\in M_{n}(\zzz)$), il n'existe pas de projecteur $e\in\aa$. En effet, un tel \'el\'ement est une fonction lisse sur le tore ordinaire qui ne prend que les valeurs $0$ et 1, et qui ne peut \^etre que constante en vertu de la connexit\'e du tore. Toutefois lorsque $\aa$ est non commutative, il est possible de d\'efinir des projecteurs non triviaux dans $\aa$ appel\'es projecteurs de Power-Rieffel \cite{pacific}.

\par

Soient $V_{1}$ et $V_{2}$ deux \'el\'ements unitaires (par exemple des mon\^omes $U^{p}$ et $U^{q}$) de $\aa$ satisfaisant \`a la relation $U_{1}V_{2}=e^{2i\pi\lambda}V_{2}V_{1}$ avec $\lambda\in ]0,1[$. Pour toute fonction lisse $f$  du cercle dans $\rrr$ dont les coefficients de Fourier sont not\'es $a_{k}$, nous d\'efinissons $f(U)$, pour $U\in\aa$, par la relation
\bbb
f(U)=\mathop{\sum}\limits_{k\in\zzz} a_{k}U^{k}.
\eee
Soit alors
\bbb
e=f(V_{1})V_{2}+g(V_{1})+(f(V_{1})V_{2})^{*}
\eee 
o\`u $f$ et $g$ sont deux fonctions lisses sur le cercle. Il est facile de montrer qu'il existe une infinit\'e de fonctions $f$ et $g$ telles que $e$ soit un projecteur. En particulier \cite{cras}, on peut construire une projection $e$ telle que
\bbb
\int e=\lambda
\eee
et
\bbb
\frac{1}{2i\pi}\int e(\partial_{1}e\partial_{2}e-\partial_{2}e\partial_{1}e)=1,
\eee
o\`u nous avons not\'e $\partial_{1}$ et $\partial_{2}$ les deux d\'erivations d\'efinies par $\partial_{i}V_{j}=2i\pi\delta_{ij}V_{i}$ pour $i,j\in \la1,2\ra$. Nous utiliserons cette projection ult\'erieurement lorsque nous \'etudierons les propri\'et\'es topologiques des champs de Yang-Mills (cf \S 4.2.3).

\par

Lorsque la matrice $\theta$  est rationnelle, nous avons vu (cf \S 4.1.1) que $\aa_{\theta}$ correspond \`a un fibr\'e sur le tore usuel dont la fibre est une alg\`ebre de matrices. Les fonctions de transition de ce fibr\'e sont simplement donn\'ees par la conjugaison par des matrices unitaires et constantes qui d\'eterminent les lois de transformation des champs de jauge $A_{\mu}$.

\par

Ainsi, lorsque $\theta\in M_{n}(\qqq)$ la th\'eorie de jauge bas\'ee sur le module trivial $\ee=\aa$ est une th\'eorie de jauge sur le tore ordinaire, avec un fibr\'e non trivial ayant des fonctions de transition constantes. L'\'etude de ce type de configurations sur le tore, avec des fonctions de transition constantes ou non, a \'et\'e initi\'ee dans \cite{torons} et est d\'evelopp\'ee dans \cite{vanbaal} et dans \cite{policarpov}. Nous n'\'etudierons pas plus avant le cas rationnel mais nous garderons pr\'esent \`a l'esprit que lorsque $\theta$ est rationnel la th\'eorie de jauge est une th\'eorie de jauge conventionelle sur le tore commutatif.

\subsection{Equations du mouvement et identit\'es de Bianchi}

Dans le cas du tore non commutatif, l'application
\bbb
\langle\pi(\omega),\pi(\eta)\rangle
=\t_{\omega}\lp \pi(\omega)^{*}\pi(\eta)|\dd|^{-n}\rp
\eee
est un produit scalaire sur $\pi(\Omega^{p}(\aa))$ et le choix du repr\'esentant totalement antisym\'etrique correspond \`a une projection orthogonale sur le suppl\'ementaire orthogonal de $\pi(\jj)$ (cf \S 4.1.4). 

\par

Dans ce cas, l'action de Yang-Mills est donn\'e par le carr\'e de la norme de la courbure, et on a
\bbb
S_{YM}[eA_{\mu}e]=-\frac{1}{2}
\int \t\lp F_{\mu\nu}F^{\mu\nu}\rp 
\eee
car
\bbb
F=\frac{1}{2}F_{\mu\nu}dx^{\mu}\wedge dx^{\nu}=\frac{1}{4}F_{\mu\nu}[i\gamma^{\mu},i\gamma^{\nu}]
\eee
et
\bbb
\t\lp[i\gamma^{\mu},i\gamma^{\nu}][i\gamma^{\rho},i\gamma^{\sigma}]\rp=
2^{[n/2]}\,4\lp g^{\mu\sigma}g^{\nu\rho}-g^{\mu\rho}g^{\nu\sigma}\rp.
\eee
Notons que le signe - dans l'action de Yang-Mills nous assure sa positivit\'e ,car $F_{\mu\nu}$ est antihermitien.

\par

Puisque nous connaissons explicitement la courbure en fonction du champ de jauge $eA_{\mu}e$, il est possible de d\'eterminer les \'equations du mouvement.
\begin{pro}
$eA_{\mu}e$ est un point stationnaire de l'action de Yang-Mills si et seulement si
\bbb
e\partial_{\mu}F^{\mu\nu}e+\lb eA_{\mu}e,F^{\mu\nu}\rb=0,
\eee
avec
\bbb
F_{\mu\nu}=e\partial_{\mu}e\partial_{\nu}e-e\partial_{\nu}e\partial_{\mu}e
+e\partial_{\mu}(eA_{\nu}e)e-e\partial_{\nu}(eA_{\mu}e)e
+[eA_{\mu}e,eA_{\nu}e].
\eee
\end{pro}

\demo
La connexion d\'efinie par $eA_{\mu}e$ est un extremum de l'action si et seulement si 
\bbb
\frac{d}{dt}S_{YM}[eA_{\mu}e+te\delta A_{\mu}e]|_{t=0}=0
\eee
pour tout $\delta A_{\mu}$. Au premier ordre en t, on a
\bbb
S_{YM}[eA_{\mu}e+te\delta A_{\mu}e]-S_{YM}[eA_{\mu}e]=t\delta S=
-t\int \t\lp F_{\mu\nu}\delta F^{\mu\nu}\rp+ o(t^{2})
\eee
avec
\bbbb
&\delta F_{\mu\nu}=
+e\partial_{\mu}(e\delta A_{\nu}e)e-e\partial_{\nu}(e\delta A_{\mu}e)e
[e\delta A_{\mu}e,eA_{\nu}e]+[eA_{\mu}e,e\delta A_{\nu}e]&\n\\
&=e\lb\partial_{\mu}+eA_{\mu}e,e\delta A_{\nu}e\rb e-
e\lb\partial_{\nu}+eA_{\nu}e,e\delta A_{\mu}e\rb e.& 
\eeee
En posant $D_{\mu}(a)=[\partial_{\mu}+eA_{\mu}e,a]$ pour $a\in M_{N}(\aa)$, on a
\bbb
\delta F_{\mu\nu}=eD_{\mu}(e\delta A_{\nu}e)e-eD_{\nu}(e\delta A_{\mu}e)e,
\eee
ce qui donne, puisque $\int$ est une trace v\'erifiant la r\`egle d'int\'egration par parties,
\bbb
\delta S=2\int \t\lp eD_{\mu}F^{\mu\nu}e\delta A_{\nu}\rp.
\eee
Puisque $\delta S=0$ doit \^etre vrai quelque soit $\delta A_{\nu}$ et que la trace est fid\`ele, on obtient
\bbb
eD_{\mu}F^{\mu\nu}e=0,
\eee
ce qui est bien le r\'esultat annonc\'e.
\edemo

Dans le cas g\'en\'eral d'un triplet spectral quelconque, il est clair que la m\^eme d\'emarche ne peut \^etre employ\'ee pour d\'eterminer les \'equations du mouvement car notre d\'emonstration repose sur les trois particularit\'es suivantes du tore non commutatif:
\begin{enumerate}
\item
l'expression explicite de l'action en fonction du champ de jauge $A$,
\item
la r\`egle d'int\'egration par parties,
\item
la fid\'elit\'e de la trace d\'efinie par l'int\'egration des fonctions.
\end{enumerate}

Ce r\'esultat nous permet de montrer que les connexions \`a courbure constante, lorsqu'elle existent \cite{canadian}, sont des solutions des \'equations de Yang-Mills.

\begin{pro}
Si le module $\ee$ admet une connexion \`a courbure constante, i.e. $F_{\mu\nu}=f_{\mu\nu}e$ avec $f_{\mu\nu}\in\ccc$, alors cette connexion est solution des \'equations du mouvement et on a
\bbb
f_{\mu\nu}=\frac{\int\t(e\partial_{\mu}e\partial_{\nu}e-
e\partial_{\nu}e\partial_{\mu}e)}{\int\t e}
\eee
qui ne d\'epend que de la classe de $\ee$ dans le groupe $K_{0}(\aa)$.
\end{pro}

\demo
Si $eA_{\mu}e$ est une connexion \`a courbure constante, alors $e\partial_{\mu}F^{\mu\nu}e=f^{\mu\nu}e\partial_{\mu}ee$=0 car $e^{2}=e$  donne par d\'erivation $e(\partial_{\mu}e)e=0$. De plus, $[eA_{\mu}e,F^{\mu\nu}]=f^{\mu\nu}[eA_{\mu}e,e]=0$, ce qui prouve que $eA_{\mu}e$ satisfait aux \'equations du mouvement.

\par

Puisque la courbure est $F_{\mu\nu}=f_{\mu\nu}e$, on a
\bbbb
&f_{\mu\nu}e=F_{\mu\nu}&\n\\
&=e\partial_{\mu}e\partial_{\nu}e-e\partial_{\nu}e\partial_{\mu}e
+e\partial_{\mu}(eA_{\nu}e)e-e\partial_{\nu}(eA_{\mu}e)e
+[eA_{\mu}e,eA_{\nu}e].&
\eeee
En utilisant les propri\'et\'es de $\int\t$ (int\'egration par parties et propri\'et\'e de trace), on obtient
\bbb
f_{\mu\nu}\int\t e=\int\t(e\partial_{\mu}e\partial_{\nu}e-
e\partial_{\nu}e\partial_{\mu}e),  
\eee
d'o\`u
\bbb
f_{\mu\nu}=\frac{\int\t(e\partial_{\mu}e\partial_{\nu}e-
e\partial_{\nu}e\partial_{\mu}e)}{\int\t e}.
\eee
Enfin, les applications $\phi_{0}$ et $\phi_{2}$ d\'efinies par
\bbbb
\phi_{0}(a_{0})&=&\int\t(a_{0})\n\\
\phi_{2}(a_{0},a_{1},a_{2})&=&
\int\t\lp a_{0}(\partial_{1}a_{1}\partial_{2}a_{2}-\partial_{2}a_{1}\partial_{1}a_{2})\rp
\eeee
pour tous $a_{0},a_{1},a_{2}\in\aa$ sont des cocycles cycliques. En cons\'equence, $\phi_{0}(e)$ et $\phi_{2}(e,e,e)$ ne d\'ependent que de la classe du projecteur $e$ dans le groupe $K_{0}(\aa)$.
\edemo

En fait, la derni\`ere partie de ce r\'esultat s'\'etend sans difficult\'e \`a des connexions qui ne sont pas n\'ecessairement \`a courbure constante:

\begin{pro}
Si $\ee=e\aa^{N}$ est un module ayant une connexion dont la courbure est $F_{\mu\nu}$, alors
\bbb
\int\t F_{\mu\nu}=\int\t\lp e(\partial_{1}e\partial_{2}e-\partial_{2}e\partial_{1}e)\rp
\eee
ne d\'epend que de la classe de $e$ dans $K_{0}(\aa)$.
\end{pro}

Par cons\'equent, nous pouvons interpr\'eter $\int\t F_{\mu\nu}$ comme une quantit\'e topologique, car elle ne d\'epend pas du choix de la connexion $A_{\mu}$ et est stable par d\'eformation..

\par

Dans le cas commutatif, il est connu \cite{policarpov} que les quantit\'es
\bbb
\frac{1}{2i\pi}\int\t F_{\mu\nu}
\eee
sont des nombres entiers. L'exemple suivant nous montre que ce n'est pas le cas en g\'eom\'etrie non commutative, m\^eme si ces nombres apparaissent comme des couplages entre la $K$-th\'eorie et la cohomologie cyclique.

\exe
Consid\'erons le tore non commutatif de dimension 4 d\'efini par la matrice
\bbb
\theta=\pp{0&\theta'&0&0\cr
-\theta'&0&0&0\cr
0&0&0&-\theta''\cr
0&0&\theta''&0},
\eee
avec $\theta',\theta''\in ]0,1[$.

\par

Dans l'alg\`ebre engendr\'ee par les unitaires $U_{1}$ et $U_{2}$ (resp. $U_{3}$ et $U_{4}$), nous d\'efinissons, gr\^ace \`a la construction de Power-Rieffel, un projecteur $e'$ (resp. $e''$) satisfaisant aux relations $\int e'=\theta'$ et $\int e'(\partial_{1}e'\partial_{2}e'-\partial_{2}e'\partial_{1}e')=2i\pi$
(resp.  $\int e''=\theta''$ et $\int e''(\partial_{3}e''\partial_{4}e''-\partial_{4}e''\partial_{3}e'')=2i\pi$).

\par

Puisque $e'$ et $e''$ commutent, $e=e'e''$ est un projecteur dans $\aa_{\theta}$ qui v\'erifie
\bbb
\frac{1}{2i\pi}\int\t F_{\mu\nu}=\frac{1}{2i\pi}\int\t\lp
e(\partial_{\mu}e\partial_{\nu}e-\partial_{\nu}e\partial_{\mu}e)\rp
=\theta_{\mu\nu},
\eee
qui n'est pas entier en g\'en\'eral. Bien entendu, en prenant $e=e'$ ou $e=e''$, $\frac{1}{2i\pi}\int\t F_{\mu\nu}$ est entier.

\par
 
Cette construction se g\'en\'eralise en dimension sup\'erieure pour des matrices $\theta$ qui sont diagonales par blocs. Aussi $\frac{1}{2i\pi}\int\t F_{\mu\nu}$ n'est-il pas n\'ecessairement entier lorsque $n\geq 4$, contrairement au cas commutatif.

\par

Toutefois, nous allons montrer (cf \S 4.2.4) qu'en dimension 4,
\bbb
\frac{1}{32\pi^{2}}\epsilon^{\mu\nu\rho\sigma}\int\t(F_{\mu\nu}F_{\rho\sigma})
\eee
est toujours entier.
\eexe

\par

Terminons par l'\'etude des identit\'es de Bianchi.

\begin{pro}
Sur le tore non commutatif, les identit\'es de Bianchi s'\'ecrivent
\bbb
[\nabla,F]=\lp e\partial_{\mu}(F_{\rho\sigma})+[eA_{\mu}e,F_{\rho\sigma}]\rp dx^{\mu}\wedge dx^{\rho}\wedge dx^{\sigma}=0.
\eee 
\end{pro}

\demo
Rappelons que l'identit\'e de Bianchi est d\'efinie comme \'etant la relation
$[\nabla, F]=0$ entre 3-formes \`a valeurs matricielles.

\par

En utilisant les notations habituelles, l'action de $F$ sur $e\xi\in e\aa^{N}$ s'\'ecrit
\bbb
F(e\xi)=F_{\rho\sigma}\xi dx^{\rho}\wedge dx^{\sigma}.
\eee
Puisque $\nabla(e\xi)=\lp e\partial_{\mu}(e\xi)+eA_{\mu}e\xi\rp\ot dx^{\mu}$, on en d\'eduit que
\bbb
[\nabla,F]e\xi=\lp e\partial_{\mu}(F_{\rho\sigma})e+[eA_{\nu}e,F_{\rho\sigma}]\rp\xi\ot dx^{\mu}\wedge dx^{\rho}\wedge dx^{\sigma}=0,
\eee
ce qui est bien le r\'esultat anonc\'e.
\edemo

En dimension 4, $\epsilon_{\mu\nu\rho\sigma}dx^{\nu}\wedge dx^{\rho}\wedge dx^{\sigma}$ forment une base de l'espace des 3-formes pour $\mu=1,\dots,4$, ce qui nous permet d'\'ecrire l'identit\'e de Bianchi de la mani\`ere suivante,
\bbb
\epsilon_{\mu\nu\rho\sigma}\lp e\partial_{\nu}F_{\rho\sigma}e
+[eA_{\nu}e,F_{\rho\sigma} ]\rp=0.
\eee
Cela donne, en posant $\tilde{F}^{\mu\nu}=1/2 \epsilon^{\mu\nu\rho\sigma}F_{\rho\sigma}$,
\bbb
e\partial_{\mu}\tilde{F}^{\mu\nu}e+\lb eA_{\mu}e,\tilde{F}^{\mu\nu}\rb=0.
\eee
La comparaison de ces relations avec les \'equations du mouvement,
\bbb
e\partial_{\mu}F^{\mu\nu}e+\lb eA_{\mu}e,F^{\mu\nu}\rb=0,
\eee
nous montre que toute connexion self-duale, i.e. telle que $\tilde{F}_{\mu\nu}=F_{\mu\nu}$, est automatiquement solution des \'equations du mouvement, en compl\`ete analogie avec le cas classique. 

\subsection{Le spectre de dimension du tore non commutatif}

En vue d'appliquer le th\'eor\`eme local de l'indice, nous devons \'etudier le spectre de dimension du tore non commutatif. Rappelons d'abord le r\'esultat suivant \cite{local}.

\begin{pro}
Le spectre de dimension d'une vari\'et\'e compacte de dimension n munie d'une structure de spin est simple et inclus dans $\la 0,1,\dots,n \ra$.
\end{pro} 

En particulier, cela implique que le spectre de dimension du tore usuel est inclus dans  $\la 0,1,\dots,n \ra$. Montrons que ce r\'esultat reste valide dans le cas non commutatif.

\begin{pro}
Le spectre de dimension du tore non commutatif de dimension n, muni de l'op\'erateur de Dirac $i\gamma^{\mu}\partial_{\mu}$, est simple et inclus dans $\la 0,1,\dots,n \ra$.
\end{pro} 

\demo
Par d\'efinition du spectre de dimension, nous devons montrer que la fonction
\bbb
\zeta_{b_{1},\dots,b_{r}}^{m_{1},\dots,m_{r}}(z)=
\t\lp \delta^{m_{1}}(b_{1})\dots\delta^{m_{r}}(b_{r})|\dd|^{-2z}\rp,
\eee
avec $b_{i}\in\bb$, $\bb$ d\'esignant l'alg\`ebre engendr\'ee par $\pi(\aa)$ et $[\dd,\pi(\aa)]$, et o\`u $\delta(b)=[|\dd|,b]$, qui est holomorphe pour $\Re(z)$ assez grand, se prolonge analytiquement sur le plan complexe priv\'e de $\la 0,1,\dots,n \ra$.

\par

Malheureusement, cette d\'emonstration est assez longue et nous nous contentons d'en donner les principales \'etapes. 

\begin{enumerate}

\item
Puisque $[\dd,\pi(a)]=i\gamma^{\mu}\partial_{\mu}a$ pour tout $a$ et que $|\dd|$ est un op\'erateur scalaire (il est diagonal dans l'espace $\ccc^{2^{[n/2]}}$ sur lequel les matrices de Dirac sont repr\'esent\'ees), on s\'epare la trace sur les matrices de Dirac pour se ramener au cas $b_{i}\in\aa$.

\item 
On commence par \'etudier le cas o\`u tous les $b_{i}$ sont des mon\^omes de base $U^{i}$. A un facteur de phase pr\`es, c'est la m\^eme chose que dans le cas commutatif. En utilisant \cite{gilkey}
\bbb
|\dd|^{-2z}=\frac{1}{\Gamma(z)}\int_{0}^{\infty} t^{z-1}e^{-t\dd^{2}}dt,
\eee
on ram\`ene le probleme \`a la d\'etermination d'un d\'eveloppement asymptotique de
\bbb
\frac{1}{\Gamma(z)}\int_{0}^{\infty} t^{z-1}
\t\lp \delta^{m_{1}}(U^{p_{1}})\dots\delta^{m_{r}}(U^{p_{r}}) e^{-t\dd^{2}}\rp dt
\eee
lorsque $t\rightarrow 0$.

\item
On montre que les coefficients du d\'eveloppement asymptotique pr\'ec\'edent ont une croissance au plus polynomiale en $p_{1},\dots,p_{r}$, ce qui nous permet de donner un d\'eveloppement asymptotique de
\bbb
\frac{1}{\Gamma(z)}\int_{0}^{\infty} t^{z-1}
\t\lp \delta^{m_{1}}(b_{1})\dots\delta^{m_{r}}(b_{r}) e^{-t\dd^{2}}\rp dt.
\eee

\item
On conclut en utilisant la m\^eme m\'ethode que dans \cite{gilkey} pour prolonger analytiquement les fonctions pr\'ec\'edentes.

\end{enumerate}

\edemo

Cela nous permet d'appliquer la formule locale de l'indice pour montrer l'int\'egralit\'e de la borne inf\'erieure de l'action de Yang-Mills ainsi que l'invariance de jauge de l'action de Chern-Simons \`a un multiple entier de $2i\pi$ pr\`es.

\subsection{Une application du th\'eor\`eme de l'indice en dimension 4}

Lorsqu'un triplet spectral de dimension 4 satisfait \`a la condition de fermeture, nous avons montr\'e (cf \S 1.3.4 ) que l'action de Yang-Mills admet une borne inf\'erieure donn\'ee par
\bbb
\dix\gamma \t\lp edededede\rp=\epsilon^{\mu\nu\rho\sigma}\int\t\lp e\partial_{\mu}e\partial_{\nu}e\partial_{\rho}e\partial_{\sigma}e\rp.
\eee
Dans le cas commutatif, cette quantit\'e est un nombre entier que multiplie une constante num\'erique. En appliquant la formule locale de l'indice, nous allons montrer que ce r\'esultat d'int\'egralit\'e est encore valide pour le tore non commutatif de dimension 4.

\begin{pro}
\bbb
\dix\gamma \t\lp edededede\rp ds^{4}=8\pi^{2}n.
\eee
avec $n\in\zzz$
\end{pro}

\demo

Puisque le tore non commutatif de dimension 4 a un spectre de dimension simple et contenu dans $\la 0,1,2,3,4\ra$, nous pouvons lui appliquer le th\'eor\`eme de l'indice (cf \S 1.4.3), ce qui montre que
\bbb
\phi_{0}(e)-2\phi_{2}(e,e,e)+12\phi_{4}(e,e,e,e,e)
\eee
est un entier si $e\in M_{N}(\aa)$ est une projection hermitienne. Bien entendu, nous appliquons le th\'eor\`eme de l'indice au triplet spectral $(M_{N}(\aa),\hh\ot\ccc^{N},\dd\ot I_{N})$, qui a m\^eme spectre de dimension que $(\aa,\hh,\dd)$.

\par

Etant donn\'e que le spectre de dimension est simple, les cochaines $\phi_{0}$, $\phi_{2}$ et $\phi_{4}$ sont d\'efinies par
\bbbb
&\phi_{0}=\mathop{\mathrm{Res}}\limits_{z=0}\frac{1}{z}
\t\lp \gamma a_{0}|\dd|^{-2z}\rp
&\\
&\phi_{2}(a_{0},a_{1},a_{2})=&\n\\
&\mathop{\sum}\limits_{0\leq k_{1}+k_{2}\leq 2}
\frac{(-1)^{k_{1}+k_{2}}(k_{1}+k_{2})!}{(k_{1}+1)(k_{1}+k_{2}+2)}
\mathop{\mathrm{Res}}\limits_{z=0}\t\lp\gamma a_{0}(da_{1})^{k_{1}}(da_{2})^{k_{2}}
|\dd|^{-2(k_{1}+k_{2}+1+z)}\rp&
\\
&\phi_{4}(a_{0},a_{1},a_{2},a_{3},a_{4})=
\frac{1}{24}
\mathop{\mathrm{Res}}\limits_{z=0}\t\lp\gamma a_{0}da_{1}da_{2}da_{3}da_{4}|\dd|^{-4-2z}\rp,&
\eeee
o\`u $(da)^{k}=\nabla^{k}(da)$ avec $\nabla(da)=[\dd^{2},da]$.

\par

Pour \'evaluer les traces pr\'ec\'edentes, nous nous pla\c cons dans la base $(U_{p}\ot \epsilon^{i}\ot e^{a})$ form\'ee des produits tensoriels des mon\^omes de Fourier $U^{p}$, d'une base $(\epsilon^{i})_{1\leq i\leq 4}$ de l'espace $\ccc^{4}$ sur lequel les matrices de Dirac agissent et d'une base orthonormale $(e^{a})_{1\leq a\leq N}$ de l'espace $\ccc^{N}$ muni de sa structure hermitienne canonique. Puisque $\gamma=\gamma^{5}$ et  $[\dd,\pi(a)]=i\gamma^{\mu}\partial_{\mu}a$ pour tout $a\in\aa$, nous allons s\'eparer la trace sur les vecteurs de base de $\ccc^{2^{[n/2]}}$ de la trace sur les autre vecteurs. Par un abus de notations, nous noterons de la m\^eme mani\`ere toutes les traces. De plus nous notons $\Delta=\dd^{2}$ l'analogue du laplacien, qui n'agit que sur les vecteurs $U^{p}$. Enfin, tout op\'erateur de multiplication par un \'el\'ement de $M_{N}(\aa)$ n'agit que sur $U^{p}$ et $e^{a}$, son action sur $\epsilon^{i}$ \'etant triviale.

\par

Puisque $\gamma=\gamma^{5}$ et $\t(\gamma^{5})=0$, on a pour $\Re(z)$ assez grand
\bbb
\t\lp\gamma a_{0}|\dd|^{-2z}\rp=\t(\gamma^{5})\t\lp a_{0}\Delta^{-z}\rp=0
\eee 
ce qui implique que $\phi_{0}(e)=0$. De m\^eme, pour $\Re(z)$ assez grand, 
\bbbb
&
\mathop{\sum}\limits_{k_{1}+k_{2}\leq 2}
\frac{(-1)^{k_{1}+k_{2}}(k_{1}+k_{2})!}{(k_{1}+1)(k_{1}+k_{2}+2)}
\t\lp\gamma^{5}\gamma^{\mu}\gamma^{\nu}\rp 
\t\lp a_{0}(\partial_{\mu}a_{1})^{k_{2}}(\partial_{\nu}a_{2})^{k_{2}}
|\Delta|^{-(k_{1}+k_{2}+1+z)}\rp&\n\\
&=0&\n
\eeee
car $\t\lp\gamma^{5}\gamma^{\mu}\gamma^{\nu}\rp=0$. On en d\'eduit que $\phi_{2}(e,e,e)=0$. Enfin, pour $\Re(z)$ assez grand, 
\bbbb
&\t\lp\gamma a_{0}da_{1}da_{2}da_{3}da_{4}|\dd|^{-4-2z}\rp=&\n\\
&\t\lp\gamma^{5}\gamma^{\mu}\gamma^{\nu}\gamma^{\rho}\gamma^{\sigma}\rp
\t\lp a_{0}\partial_{\mu}a_{1}\partial_{\nu}a_{2}\partial_{\rho}a_{3}\partial_{\sigma}a_{4}\Delta^{-2-z}\rp=&\n\\
&-4\epsilon^{\mu\nu\rho\sigma}\t\lp a_{0}\partial_{\mu}a_{1}\partial_{\nu}a_{2}\partial_{\rho}a_{3}\partial_{\sigma}a_{4}\Delta^{-2-z}\rp&,
\eeee
puisque $\gamma^{5}=-\gamma^{1}\gamma^{2}\gamma^{3}\gamma^{4}$. Pour \'evaluer $\t\lp a_{0}\partial_{\mu}a_{1}\partial_{\nu}a_{2}\partial_{\rho}a_{3}\partial_{\sigma}a_{4}\Delta^{-2-z}\rp$, nous nous pla\c cons dans la base $U^{p}\ot e^{a}$. Cette base est orthonormale pour le produit scalaire
\bbb
\langle U^{p}\ot e^{a},U^{q}\ot e^{b}\rangle=\int\lp U^{p*}U^{q}\rp\t\lp e^{a*}e^{b}\rp,
\eee
ce qui implique que si $A\in M_{N}(\aa)$ est une matrice agissant par multiplication sur $\aa^{N}$ et $\Delta$ est le Laplacien v\'erifiant $\Delta U^{p}=4\pi^{2}p^{2}U^{p}$, on a
\bbbb
\langle U^{p}\ot e^{a}, A\Delta^{-2-z}(U^{p}\ot e^{a})\rangle&=&\n\\
(4\pi^{2}p^{2})^{-2-z}\int\lp U^{p*}\t(e^{a*}Ae^{a})U^{p}\rp&=&\n\\
(4\pi^{2}p^{2})^{-2-z}\int\t(e^{a*}Ae^{a}),
\eeee
o\`u $\t(A)$ d\'esigne la trace de la matrice $A$ \`a valeurs dans $\aa$. On en d\'eduit que
\bbbb
&\t\lp a_{0}\partial_{\mu}a_{1}\partial_{\nu}a_{2}\partial_{\rho}a_{3}\partial_{\sigma}a_{4}\Delta^{-2-z}\rp=&\n\\
&\mathop{\sum}\limits_{p,a}
(4\pi^{2}p^{2})^{-2-z}\int\t\lp e^{a*}a_{0}\partial_{\mu}a_{1}\partial_{\nu}a_{2}\partial_{\rho}a_{3}\partial_{\sigma}a_{4}e^{a}\rp=&\n\\
&\t(\Delta^{-2-z})\int\t\lp a_{0}\partial_{\mu}a_{1}\partial_{\nu}a_{2}
\partial_{\rho}a_{3}\partial_{\sigma}a_{4}\rp.&\n
\eeee
$\t(\Delta^{-2-z})=\sum_{p}(4\pi^{2}p^{2})^{-2-z}$ est la fonction $\zeta$ usuelle associ\'ee au laplacien sur le tore euclidien. Cette fonction se prolonge analytiquement sur $\ccc$ priv\'e de l'origine et on a \cite{gilkey}
\bbb
\mathop{\mathrm{Res}}\limits_{z=0}\t(\Delta^{-2-z})=\frac{1}{16\pi^{2}}.
\eee
On en d\'eduit que
\bbb
\phi_{4}(a_{0},a_{1},a_{2},a_{3},a_{4})=
\frac{1}{96\pi^{2}}\epsilon^{\mu\nu\rho\sigma}
\int\t\lp a_{0}\partial_{\mu}a_{1}\partial_{\nu}a_{2}\partial_{\rho}a_{3}\partial_{\sigma}a_{4}\rp.
\eee
Puisque $\phi_{0}$ et $\phi_{2}$ sont nuls, le th\'eor\`eme de l'indice nous montre que $12\phi_{4}(e,e,e,e,e)\in\zzz$, ce qui prouve que
\bbb
\frac{1}{8\pi^{2}}\epsilon^{\mu\nu\rho\sigma}
\int\t\lp e\partial_{\mu}e\partial_{\nu}e\partial_{\rho}e\partial_{\sigma}e\rp=n
\eee
est un entier.

\par

Etant donn\'e que 
\bbb
\dix\gamma\lp edededede\rp ds^{4}=\epsilon^{\mu\nu\rho\sigma}
\int\t\lp e\partial_{\mu}e\partial_{\nu}e\partial_{\rho}e\partial_{\sigma}e\rp,
\eee
on a
\bbb
\dix\gamma\lp edededede\rp ds^{4}=8\pi^{2}n.
\eee
\edemo

Cela nous montre que l'action de Yang-Mills est minor\'ee par $8\pi^{2}|n|$, 
\bbb
\int\t\lp F_{\mu\nu}F^{\mu\nu*}\rp\geq 16\pi^{2}|n|,
\eee
avec $n$ entier. Ce r\'esulat est tout-\`a-fait similaire \`a celui que l'on obtient dans le cas commutatif.  

\subsection{Invariance de jauge de l'action de Chern-Simons}

De mani\`ere tout-\`a-fait analogue, nous allons \'etudier l'action de Chern-Simons sur le tore non commutatif de dimension 3. Commen\c cons par rappeler que pour un triplet spectral de dimension 3 satisfaisant \`a la condition de fermeture, l'action de Chern-Simons est (cf \S 1.3.5)
\bbb
S_{CS}[A]=\frac{k}{4\pi}\dix\t\lp K_{1}\rp ds^{3},
\eee
o\`u $A$ est une 1-forme hermitienne \`a valeurs matricielles et $K_{1}$ est un repr\'esentant quelconque de la forme de Chern-Simons
\bbb
K=AdA+\frac{2}{3}A^{3}.
\eee
Notons que nous avons introduit un facteur $k/4\pi$ suppl\'ementaire par rapport \`a la d\'efinition donn\'ee dans 1.3.5, o\`u $k$ est un nombre r\'eel. Ce coefficient est une constante de couplage, dont l'int\'er\^et appara\^\i tra ult\'erieurement. 

\par

Dans le cas du tore non commutatif, cette d\'efinition se simplifie car on peut construire explicitement la forme $K$.

\begin{pro}
L'action de Chern-Simons est donn\'ee par
\bbb
S_{CS}[A_{\mu}]=\frac{k}{4\pi}\epsilon^{\lambda\mu\nu}
\int\t\lp A_{\lambda}\partial_{\mu}A_{\nu}+\frac{2}{3}A_{\lambda}A_{\mu}A_{\nu}\rp.
\eee
\end{pro}

\demo
Avec les notation usuelles, on a
\bbb
A= A_{\mu}dx^{\mu},
\eee
d'o\`u
\bbb
K=\lp A_{\lambda}\partial_{\mu}A_{\nu}+\frac{2}{3}A_{\lambda}A_{\mu}A_{\nu}\rp
dx^{\lambda}\wedge dx^{\mu}\wedge dx^{\nu},
\eee
apr\`es avoir identifi\'e $K$ avec son repr\'esentant totalement antisym\'etrique.

\par

Les matrices de Dirac en dimension 3 n'\'etant autres que les matrices de Pauli,
on a 
\bbb
dx^{\mu}\wedge dx^{\nu}\wedge dx^{\nu}=\frac{1}{6}\mathop{\sum}\limits_{\sigma\in S_{3}}\epsilon(\sigma)
i\gamma^{\sigma(\lambda)}i\gamma^{\sigma(\mu)}i\gamma^{\sigma(\nu)}
=\epsilon^{\lambda\mu\nu}.
\eee
La normalisation de $\displaystyle\dix$ \'etant choisie de mani\`ere \`a \'eliminer tous les facteurs num\'eriques, on a
\bbb
\frac{k}{4\pi}\dix K ds^{3}=\frac{k}{4\pi}\epsilon^{\lambda\mu\nu}
\int\t\lp A_{\lambda}\partial_{\mu}A_{\nu}+\frac{2}{3}A_{\lambda}A_{\mu}A_{\nu}\rp.
\eee
\edemo

Le comportement de cette action sous une transformation de jauge est analogue au cas classique.

\begin{pro}
Sous une transformation de jauge $A_{\mu}\rightarrow uA_{\mu}u^{-1}+u\partial_{\mu}u^{-1} $ avec $u\in M_{N}(\aa)$ unitaire, l'action devient
\bbb
S_{CS}[uA_{\mu}u^{-1}+u\partial_{\mu}u^{-1}]
=S_{CS}[A_{\mu}]+\frac{k}{12\pi}
\epsilon^{\lambda\mu\nu}\int\t\lp u\partial_{\lambda}u^{-1} u\partial_{\mu}u^{-1} u\partial_{\nu}u^{-1}\rp
\eee
avec
\bbb
\epsilon^{\lambda\mu\nu}\int\t\lp u\partial_{\lambda}u^{-1} u\partial_{\mu}u^{-1} u\partial_{\nu}u^{-1}\rp=24\pi^{2}n,
\eee
o\`u $n$ est un entier.
\end{pro}

\demo
Pour montrer la premi\`ere partie de ce r\'esultat, il suffit de remarquer que le tore non commutatif de dimension 3 satisfait \`a la condition de fermeture.
Les r\'esultats g\'en\'eraux sur l'action de Chern-Simons (cf \S 1.3.5) nous permettent alors d'\'ecrire
\bbb
S_{CS}[uAu^{-1}+udu^{-1}]=S_{CS}[A]+\frac{k}{12\pi}\dix\lp udu^{-1}\rp^{3}ds^{3}.
\eee
En choisissant le repr\'esentant totalement antisym\'etrique de $(udu^{-1})^{3}$, on obtient
\bbb
\frac{k}{12\pi}\dix\lp udu^{-1}\rp^{3}ds^{3}=\frac{k}{12\pi}
\epsilon^{\lambda\mu\nu}\int\t\lp u\partial_{\lambda}u^{-1} u\partial_{\mu}u^{-1} u\partial_{\nu}u^{-1}\rp,
\eee
ce qui est bien le r\'esultat annonc\'e.

\par

Pour montrer la quantification de cette quantit\'e, nous allons employer le th\'eor\`eme de l'indice en dimension 3. Puisque le spectre du tore non commutatif est simple, le th\'eor\`eme de l'indice nous montre que
\bbbb
n&=&\tau_{0}(u[\dd,u^{-1}]|\dd|^{-1})-\frac{1}{4}\tau_{0}\lp u\lb\dd^{2},[\dd,u^{-1}]\rb|\dd|^{-3}\rp\n\\
&+&\frac{1}{8}\tau_{0}\lp u\lb\dd^{2},\lb\dd^{2},[\dd,u^{-1}]\rb\rb|\dd|^{-5}\rp \n\\
&-&\frac{1}{12}\tau_{0}\lp u[ \dd,u^{-1}][ \dd,u][ \dd,u^{-1}]|\dd|^{-3}\rp 
\eeee
est un entier car c'est l'indice d'un op\'erateur de Fredholm. $\tau_{0}$ est le r\'esidu de la trace ordinaire d\'efini par
\bbb
\tau_{0}(P)=\mathop{\mathrm{Res}}\limits_{z=0}\t\lp P|\dd|^{-2z}\rp
\eee
pour tout $P$ appartenant \`a l'alg\`ebre engendr\'ee par $\pi(\aa)$, $[\dd,\pi(\aa)]$ et $|\dd|$.

\par

Pour calculer ces r\'esidus, nous allons employer essentiellement les m\^emes m\'ethodes que dans la section pr\'ec\'edente. Bien entendu, nous travaillons avec le triplet spectral $(M_{N}(\aa), \hh\ot\ccc^{N},\dd\ot I_{N})$ obtenu en consid\'erant une alg\`ebre de matrices \`a valeur dans le tore non commutatif.  Introduisons la base $U^{p}\ot \epsilon^{i}\ot e^{a}$ form\'ee des mon\^omes de Fourier $U^{p}$, de la base canonique de l'espace $\ccc^{2}$ sur lequel les matrices de Pauli agissent et de la base canonique $\ccc^{N}$. Cette base est orthonormale pour le produit scalaire
\bbb
\langle U^{q}\ot \epsilon^{j}\ot e^{b},U^{p}\ot \epsilon^{i}\ot e^{a}\rangle
=\t\lp \epsilon^{i*}\epsilon^{j}\rp\t\lp e^{a*}e^{b}\rp\int\lp U^{p*}U^{q}\rp.
\eee

\par

En dimension 3 les matrices de Dirac sont simplement les matrices de Pauli que nous notons toujours $\gamma^{\mu}$ pour $\mu=1,2,3$. Pour tout unitaire $u\in M_{N}(\aa)$ on a $du^{-1}=i\gamma^{\mu}\partial_{\mu}u^{-1}$, o\`u $\partial_{\mu}u^{-1}$ est la matrice obtenue en faisant agir la d\'erivation $\partial_{\mu}$ sur chacun des coefficients de $u^{-1}$. Pour $\Re(z)$ assez grand, $\t\lp u[\dd,u^{-1}]|\dd|^{-2z}\rp$ est bien d\'efini et on a
\bbb
\t\lp u[\dd,u^{-1}]|\dd|^{-2z}\rp=i\t\lp\gamma^{\mu}\rp
\mathop{\sum}\limits_{p}\int\lp U^{p*}\t\lp u\partial_{\mu}u^{-1}\rp U^{p}\rp \lp 4\pi^{2}p^{2}\rp^{-z}=0
\eee
car $\t(\gamma^{\mu})=0$. Par cons\'equent, cette fonction se prolonge analytiquement en la fonction nulle sur tout le plan complexe et son r\'esidu est nul. Donc a
\bbb
\tau_{0}\lp u[\dd,u^{-1}]|\dd|^{-1}\rp =0.
\eee
De la m\^eme mani\`ere, on montre que
\bbbb
&\tau_{0}\lp u\lb\dd^{2},[\dd,u^{-1}]\rb|\dd|^{-3}\rp=0,&\n\\
&\tau_{0}\lp u\lb\dd^{2},\lb\dd^{2},[\dd,u^{-1}]\rb\rb|\dd|^{-5}\rp=0.&
\eeee
En effet, pour $\Re(z)$ assez grand, les traces d\'efinissant ces r\'esidus sont bien d\'efinies, et contiennent en facteur un terme du type $\t(\gamma^{\mu})$ car $\dd^{2}$ est un op\'erateur scalaire dans l'espace sur lequel agissent les matrices de Pauli. Ces traces sont donc identiquement nulles pour $\Re(z)$ assez grand et se prolongent analytiquement sur le plan complexe tout entier.

\par

Reste \`a \'evaluer
\bbb
\tau_{0}\lp u[ \dd,u^{-1}][ \dd,u][ \dd,u^{-1}]|\dd|^{-3}\rp
=\mathop{\mathrm{Res}}\limits_{z=0}\t
 \lp u[\dd,u^{-1}][\dd,u][\dd,u^{-1}]|\dd|^{-3-2z}\rp.
\eee 
Lorsque $\Re(z)$ est assez grand, cette trace est bien d\'efinie et on a, en utilisant la base pr\'ec\'edente,
\bbbb
&\t\lp u[\dd,u^{-1}][\dd,u][\dd,u^{-1}]|\dd|^{-3-2z}\rp
=&\n\\
&\mathop{\sum}\limits_{p\in\zzz^{3}}\t\lp i\gamma^{\lambda}i\gamma^{\mu}i\gamma^{\nu}\rp
\int\lp U^{p*}\t\lp u\partial_{\lambda}u^{-1}\partial_{\mu}u\partial_{\nu}u^{-1}\rp U^{p}\rp
|4\pi^{2}p^{2}|^{-3/2-z}&\n\\
&=2\epsilon^{\lambda\mu\nu}\int\t\lp u\partial_{\lambda}u^{-1}\partial_{\mu}u\partial_{\nu}u^{-1}\rp
\t\lp\Delta^{-3/2-z}\rp,&
\eeee
o\`u $\Delta$ est le laplacien usuel sur le tore commutatif de dimension 3 muni de la m\'etrique euclidienne. Ses valeurs propres sont $4\pi^{2}p^{2}$ et la trace est \'evalu\'ee dans la base de modes de Fourier.

\par

La fonction $z\mapsto \t\lp\Delta^{-z}\rp$ est holomorphe sur $\ccc$ priv\'e de $3/2$ et son r\'esidu est donn\'e par \cite{gilkey}
\bbb
\mathop{\mathrm{Res}}\limits_{z=0}\t\lp\Delta^{-3/2-z}\rp=\frac{1}{4\pi^{2}}.
\eee
On en d\'eduit que
\bbb
\tau_{0}\lp u[ \dd,u^{-1}][ \dd,u][ \dd,u^{-1}]|\dd|^{-3}\rp
=\frac{1}{2\pi^{2}}\epsilon^{\lambda\mu\nu}\int\t\lp u\partial_{\lambda}u^{-1}\partial_{\mu}u\partial_{\nu}u^{-1}\rp.
\eee
\'Etant donn\'e que ce terme est le seul terme non trivial apparaissant dans le th\'eor\`eme de l'indice,
\bbb
\frac{1}{12}\tau_{0}\lp u[ \dd,u^{-1}][ \dd,u][ \dd,u^{-1}]|\dd|^{-3}\rp
\eee
est un entier, ce qui prouve que
\bbb
\epsilon^{\lambda\mu\nu}\int\t\lp u\partial_{\lambda}u^{-1}\partial_{\mu}u\partial_{\nu}u^{-1}\rp=
24\pi^{2}n
\eee
avec $n$ entier.
\edemo

Par cons\'equent, si $k$ est un entier la fonction $e^{iS_{CS}[A_{\mu}]}$ est invariante de jauge car sous une transformation de jauge 
\bbb
iS_{CS}[A_{\mu}]\rightarrow iS_{CS}[A_{\mu}]+2ik\pi.
\eee
Cela implique que la fonction de partition du tore non commutatif
\bbb
Z(\aa_{\theta})=\int [\dd A_{\mu}]\,e^{iS_{CS}[A_{\mu}]}
\eee
est bien d\'efinie apr\`es fixation de la jauge.

\par

Pour \'evaluer cette int\'egrale fonctionnelle, on se place en g\'en\'eral dans l'approximation semi-classique: on \'ecrit le champ $A_{\mu}$ sous la forme $A_{\mu}^{0}+B_{\mu}$, o\`u $A_{\mu}^{0}$ est un champ satisfaisant aux \'equations du mouvement. On d\'eveloppe alors l'action au second ordre en $B_{\mu}$, en incluant les termes de fixation de jauge, et on r\'egularise l'int\'egrale fonctionnelle sur $B_{\mu}$  en utilisant l'invariant $\eta$ \cite{witten}.

\par

Bien que nous n'allons pas effectuer ce calcul ici, cherchons les points critiques de l'action de Chern-Simons.

\begin{pro}
Les points critiques de l'action de Chern-Simons sont les connexions \`a courbure nulle.
\end{pro}

\demo
Par d\'efinition, le champ de jauge  $A_{\mu}$ est un point critique si et seulement si
\bbb
\frac{d}{dt}S_{CS}[A_{\mu}+t\delta A_{\mu}]|_{t=0}=0
\eee
pour toute 1-forme hermitienne $\delta A_{\mu}$ \`a valeurs matricielles.
Pour exprimer cette condition, nous faisons un d\'eveloppement limit\'e au premier ordre en $t$ de
\bbb
S_{CS}[A_{\mu}+t\delta A_{\mu}]|_{t=0}.
\eee
On a
\bbbb
&\epsilon^{\lambda\mu\nu}\lp A_{\lambda}+t\delta A_{\lambda}\rp\partial_{\mu}\lp A_{\nu}+t\delta A_{\nu}\rp=&\n\\
&\epsilon^{\lambda\mu\nu}A_{\lambda}\partial_{\mu}A_{\nu}+
t\lp A_{\lambda}\partial_{\mu}\delta A_{\nu}
+\delta A_{\lambda}\partial_{\mu}A_{\nu}
\rp+O(t^{2})&
\eeee
ainsi que
\bbbb
&\epsilon^{\lambda\mu\nu}\lp  A_{\lambda}+t\delta A_{\lambda}\rp
\lp  A_{\mu}+t\delta A_{\mu}\rp
\lp  A_{\nu}+t\delta A_{\nu}\rp=&\n\\
&\epsilon^{\lambda\mu\nu}A_{\lambda}A_{\mu}A_{\nu}+
t\epsilon^{\lambda\mu\nu}
\lp \delta A_{\lambda}A_{\mu}A_{\nu}+A_{\lambda}\delta A_{\mu}A_{\nu}+A_{\lambda}A_{\mu}\delta A_{\nu}\rp+O(t^{2}).
&
\eeee
En utilisant les propri\'et\'es de trace de $\int\t$, on obtient
\bbb
S_{CS}[A_{\mu}+t\delta A_{\mu}]=S_{CS}[A_{\mu}]+
t\epsilon^{\lambda\mu\nu}\int\t\lp \delta A_{\lambda}F_{\mu\nu}\rp,
\eee
avec 
\bbb
F_{\mu\nu}=\partial_{\mu}A_{\nu}-\partial_{\nu}A_{\mu}+[A_{\mu},A_{\nu}].
\eee
Puisque la trace est fid\`ele, le coefficient en $t$ s'annule pour tout $\delta A_{\lambda}$ si et seulement si $F_{\mu\nu}=0$, ce qui prouve le r\'esultat annonc\'e.
\edemo

Cela ach\`eve notre discussion des propri\'et\'es classiques des th\'eories de jauge sur le tore non commutatif.

\section{Sym\'etrie BRS et r\`egles de Feynman}
\subsection{G\'en\'eralit\'es}

Nous allons terminer en abordant l'\'etude de la quantification de la th\'eorie de Yang-Mills sur le tore non commutatif. Notre approche est essentiellement perturbative et nous d\'evelopperons tous les outils usuels de la th\'eorie des champs: fixation de jauge, sym\'etrie BRS et diagrammes de Feynman. Nous adoptons pour la th\'eorie de Yang-Mills sur le tore non commutatif la m\^eme d\'emarche que pour une th\'eorie des champs ordinaire, ceci \'etant justifi\'e par le fait que si $\theta$ est une matrice \`a coefficients rationnels, nous obtenons une th\'eorie de Yang-Mills ordinaire.

\par

Nous nous limitons \`a la th\'eorie la plus simple qui consiste \`a prendre le module projectif trivial $\ee=\aa$, m\^eme si la discusssion se g\'en\'eralise 
facilement \`a $\ee=\aa^{N}$. La connexion la plus g\'en\'erale sur ce module est donn\'ee par
\bbb
\nabla(\xi)=\lp\partial_{\mu}\xi+gA_{\mu}\xi\rp\ot dx^{\mu}
\eee
pour tout $\xi\in\aa$. $g>0$ est  une constante de couplage et cette connexion est suppos\'ee compatible avec la m\'etrique, ce qui \'equivaut \`a dire que $\lp A_{\mu}\rp^{*}=-A_{\mu}$.

\par

Les transformations de jauge sont donn\'ees par les unitaires $u\in\aa$ qui agissent sur $\nabla$ par $u\nabla u^{-1}$, ce qui nous donne la loi de transformation du champ de jauge
\bbb
A_{\mu}\mapsto uA_{\mu}u^{-1}+\frac{1}{g}u\partial_{\mu}u^{-1}.
\eee

\par

Pour d\'evelopper la th\'eorie quantique des champs, nous nous pla\c cons dans l'espace des impulsions. Par cons\'equent, nous d\'eveloppons tous les champs sur les modes de Fourier $U^{p}$,
\bbb
A_{\mu}=\mathop{\sum}\limits_{p\in\zzz^{n}}A_{\mu}^{p}U^{p}
\eee  
avec $\ov{A}_{\mu}^{p}=-A_{\mu}^{-p}$ car $\lp A_{\mu}\rp^{*}=-A_{\mu}$ et $\lp U^{p}\rp^{*} =U^{-p}$.

\par

Pour simplifier nos notations, nous multiplions $\theta$ par $\pi$ ce qui implique que la r\`egle de multiplication s'\'ecrit simplement $U^{p}U^{q}=e^{i\theta(p,q)}U^{p+q}$. Puisque $\partial_{\mu}U^{p}=2i\pi p_{\mu}U^{p}$ et $[U^{p},U^{q}]=2i\sin\theta(p,q)$, la courbure 
\bbb
F_{\mu\nu}=\frac{1}{g}\lb\partial_{\mu}+gA_{\mu}, \partial_{\nu}+gA_{\nu}\rb=
\partial_{\mu}A_{\nu}-\partial_{\nu}A_{\mu}+g\lb A_{\mu},A_{\nu}\rb
\eee
se d\'eveloppe en s\'erie
\bbb
F_{\mu\nu}=\mathop{\sum}\limits_{p\in\zzz^{n}}F_{\mu\nu}^{p}U^{p}
\eee
avec
\bbb
F_{\mu\nu}^{p}=2i\pi p_{\mu}A_{\nu}^{p}-2i\pi p_{\nu}A_{\mu}^{p}+
g\mathop{\sum}\limits_{q+r=p}2i\sin\theta(q,r)A_{\mu}^{q}A_{\nu}^{r}.
\eee

Nous avons vu (cf \S 4.2.2) que l'action de Yang-Mills est donn\'ee par
\bbb
S_{YM}=-\frac{1}{2}\int F_{\mu\nu}F^{\mu\nu}.
\eee
Afin de retrouver l'\'electrodynamique sur le tore lorsque $\theta=0$, nous changeons la normalisation de l'action de Yang-Mills que nous prenons dor\'enavant comme \'etant \'egale \`a
\bbb
S_{YM}[A_{\mu}]=-\frac{1}{4}\int F_{\mu\nu}F^{\mu\nu}.
\eee
Bien entendu, cette action est invariante sous les transformations de jauge d\'etermin\'ees par les unitaires de $\aa$.

\par

Tout comme en th\'eorie de Yang-Mills ordinaire, nous consid\'erons un syst\`eme dynamique dont les variables de configuration sont les champs $A_{\mu}$, ou de fa\c con \'equivalente ses composantes $A_{\mu}^{p}$ dans la base de Fourier, et dont la dynamique est gouvern\'ee par l'action de Yang-Mills.  Pour \^etre plus rigoureux,  nous devrions consid\'erer l'alg\`ebre des fonctions sur le tore non commutatif de dimension $n-1$ auxquelles on adjoint une d\'ependance temporelle. Nous obtenons ainsi une th\'eorie de Yang-Mills non commutative form\'ee avec l'alg\`ebre $\aa_{\theta}\ot C^{\infty}(\rrr)$, o\`u $C^{\infty}(\rrr)$ est l'alg\`ebre des fonctions lisses sur $\rrr$ et nous consid\'erons les composantes de $A_{\mu}$ comme des fonctions d\'ependant du temps. L'action de Yang-Mills permet alors de d\'eterminer les moments conjugu\'es de $A_{\mu}^{p}$ et nous quantifions la th\'eorie en rempla\c cant les variables $A_{\mu}^{p}$ et leurs moments conjugu\'es par des op\'erateurs dont les r\`egles de commutation sont dict\'ees par les crochets de Poisson de la th\'eorie classique.

\par

A cause de la sym\'etrie de jauge, ce proc\'ed\'e est tr\`es complexe \`a mettre en oeuvre. C'est pourquoi nous adoptons un proc\'ed\'e de quantification qui ne repose pas sur une th\'eorie hamiltonienne, mais fait appel \`a la notion d'int\'egrale de chemin. Dans notre contexte, cette derni\`ere est simplement donn\'e par une "somme sur toutes les variables de configurations" qui sont les modes de Fourier $A_{\mu}^{p}$. Nous \'ecrivons une telle somme comme un produit d'un grand nombre d'int\'egrales simples sur les nombres complexes $A_{\mu}^{p}$ et l'objet principal de notre \'etude est la fonctionelle g\'en\'eratrice
\bbb
Z[J_{\mu}]=\int [\dd A_{\mu}] e^{-S_{YM}[A_{\mu}]+\int J^{\mu}A_{\mu}},
\eee  
o\`u $J^{\mu}\in A_{\theta}$ est un \'el\'ement antihermitien. Nous supposons aussi que  $Z[J_{\mu}]$ est normalis\'ee par $Z[0]=0$ et nous cherchons \`a d\'eterminer un d\'eveloppement perturbatif de $Z[J_{\mu}]$ en compl\`ete analogie avec le cas commutatif.

\par

Il est tr\`es important de remarquer qu'au-del\`a des apparentes similitudes avec la th\'eorie classique, cet objet est tr\`es diff\'erent, car nous ne pouvons pas interpr\'eter $A_{\mu}^{p}$ comme les modes de Fourier de champs de jauges d\'efinis au sens usuel. En effet, l'action que nous avons pour $A_{\mu}^{p}$ et les sym\'etries de jauge associ\'ees ne peuvent pas \^etre, lorsque $\theta$ est non d\'eg\'en\'er\'ee, interpr\'et\'ees en th\'eorie des champs usuelle sans perdre le caract\`ere local de la th\'eorie. 

\par

Il est \'egalement remarquable que $A_{\mu}^{0}$ ne soit pas une variable dynamique. En effet, $A_{\mu}^{0}$ est la composante de $A_{\mu}$ sur l'unit\'e de $\aa_{\theta}$, c'est donc un \'el\'ement central qui annule les d\'erivations. Par cons\'equent, la courbure $F_{\mu\nu}$ ne d\'epend pas des variables $A_{\mu}^{0}$ et la mesure $[\dd A_{\mu}]$ est \`a consid\'erer comme un produit de toutes les mesures $dA_{\mu}^{p}$ sauf $dA_{\mu}^{0}$. En cons\'equence, les modes z\'ero des champs de jauge disparaissent compl\`etement de la th\'eorie, ce dont nous discuterons ult\'erieurement lorsque nous aborderons l'\'etude des divergences infrarouges (cf \S 4.3.5).    

\par

La premi\`ere difficult\'ee majeure que nous rencontrons est la fixation de jauge. En effet, \`a cause de la sym\'etrie de jauge, cette int\'egrale fonctionnelle ne peut pas avoir de sens si nous n'avons pas un proc\'ed\'e nous pemettant de ne tenir compte qu'une seule fois des configurations qui sont \'equivalentes sous les transformations de jauge. 
   
\subsection{La fixation de jauge}

Conmmen\c cons par s\'eparer dans l'action de Yang-Mills les termes quadratiques, cubiques et quartiques.
\bbbb
&-\frac{1}{4}\int F_{\mu\nu}F^{\mu\nu}=&\n\\
&-\frac{1}{4}\int \lp\partial_{\mu}A_{\nu}-\partial_{\mu}A_{\nu}+g[A_{\mu},A_{\nu}]\rp
\lp\partial^{\mu}A^{\nu}-\partial^{\mu}A^{\nu}+g[A^{\mu},A^{\nu}]\rp&\n\\
&=
-\frac{1}{4}\int\lp 2\partial_{\mu}A_{\nu}\partial^{\mu}A^{\nu}-2\partial_{\mu}A_{\nu}\partial^{\nu}A^{\mu}+4g\partial_{\mu}A_{\nu}[A_{\mu},A_{\nu}]+
g^{2}[A_{\mu},A_{\nu}][A^{\mu},A^{\nu}]\rp&\n\\
&=
-\frac{1}{4}\int\lp -2A_{\nu}\partial^{\mu}\partial_{\mu}A^{\nu}+2A_{\nu}\partial_{\mu}\partial^{\nu}A^{\mu}+4g\partial_{\mu}A_{\nu}[A_{\mu},A_{\nu}]+
g^{2}[A_{\mu},A_{\nu}][A^{\mu},A^{\nu}]\rp&
\eeee
apr\`es int\'egration par parties et utilisation de la propri\'et\'e de trace de $\int$.
En d\'eveloppant toutes ces expressions \`a l'aide des modes de Fourier, on en d\'eduit que
\bbb
S_{YM}[A_{\mu}]=S_{YM}^{(2)}[A_{\mu}]+S_{YM}^{(3)}[A_{\mu}]+S_{YM}^{(4)}[A_{\mu}],
\eee
avec
\bbbb
S_{YM}^{(2)}[A_{\mu}]&=&\frac{(2i\pi)^{2}}{2}
\lp p^{2}g_{\mu\nu}-p^{\mu}p^{\nu}\rp\ A_{\mu}^{p}A_{\nu}^{-p}\n\\
S_{YM}^{(3)}[A_{\mu}]&=&2\pi g\mathop{\sum}\limits_{p,q,r}p^{\mu}\sin\theta(q,r)
\delta(p+q+r)A_{\mu}^{p}A_{\nu}^{q}A_{\rho}^{r}
\n\\
S_{YM}^{(4)}[A_{\mu}]&=&g^{2}\mathop{\sum}\limits_{p,q,r,s}g^{\mu\rho}g^{\nu\sigma}
\sin\theta(p,q)\sin\theta(r,s)
\delta(p+q+r+s)A_{\mu}^{p}A_{\nu}^{q}A_{\rho}^{r}A_{\sigma}^{s}.
\n
\eeee

Seules les parties cubiques et quartiques d\'ependent du param\`etre de d\'eformation $\theta$, la partie quadratique est similaire \`a celle que nous rencontrons en \'electrodynamique. Par cons\'equent, nous rencontrons ici le m\^eme probl\`eme que pour l'\'electrodynamique ou les th\'eories de jauge en g\'en\'eral: l'op\'erateur $p^{2}g_{\mu\nu}-p^{\mu}p^{\nu}$  n'est pas inversible car $\lp p^{2}g_{\mu\nu}-p^{\mu}p^{\nu}\rp p_{\nu}=0$. 

\par

Pour rendre ce propagateur inversible, nous devons "fixer la jauge" en suivant toutes les \'etapes de la th\'eorie des champs usuelle. Nous introduisons la condition de jauge de Lorentz par la relation $\partial_{\mu}A^{\mu}=0$. Cela veut dire que la somme sur tous les champs $A_{\mu}$ apparaissant dans l'int\'egrale fonctionelle  ne devra se faire que sur le repr\'esentant de chaque orbite sous l'action du groupe de jauge qui satisfait \`a la condition $\partial_{\mu}A^{\mu}=0$.

\par

Ce proc\'ed\'e nous permet d'isoler dans l'int\'egrale fonctionelle un facteur infini \' egal au volume du groupe de jauge. Pour cela, rappelons que si $f$ est une fonction r\'eelle ayant un seul z\'ero simple, on a
\bbb
\int_{-\infty}^{\infty}dx f'(x)\delta\lp f(x)\rp=1.
\eee 
Cela se g\'en\'eralise aux fonctions d\'efinies sur $\rrr^{n}$ en rempla\c cant la d\'eriv\'ee par le jacobien. Nous utilisons cette m\^eme identit\'ee en dimension infinie. Si $u$ est un unitaire de $\aa$ et $A_{\mu}$ un champ de jauge, nous notons $A_{\mu}^{u}=uA_{\mu}u^{-1}+\frac{1}{g}u\partial^{\mu}u^{-1}$. Nous avons alors
\bbb
\int du \Delta_{FP}\lp A_{\mu}^{u}\rp\delta\lp f\lp A_{\mu}^{u}\rp\rp=1,
\eee
avec $f(A_{\mu})=\partial_{\mu}A^{\mu}$ et $ \Delta_{FP}\lp A_{\mu}^{u}\rp$ est le jacobien de la transformation $u\mapsto  f\lp A_{\mu}^{u}\rp$, appel\'e d\'eterminant de Faddeev-Popov. $du$ d\'esigne la mesure de Haar sur le groupe des unitaires de $\aa$. Tout comme dans le cas classique, ce groupe n'est pas localement compact et cette mesure n'existe pas. Cependant, cette approche est heuristique et nous nous contentons de montrer que la th\'eorie des champs sur le tore non commutatif peut \^etre d\'evelopp\'ee de la m\^eme mani\`ere que la th\'eorie des champs habituelle.

\par

La fonction de partition s'\'ecrit donc
\bbb
Z=\int [\dd A_{\mu}]\int du \Delta_{FP}\lp A_{\mu}^{u}\rp\delta\lp f\lp A_{\mu}^{u}\rp \rp
e^{-S_{YM}[A_{\mu}]}.
\eee
En utilisant l'invariance sous la transformation de jauge $A_{\mu}^{u}\mapsto A_{\mu}$, celle-ci se r\'e\'ecrit
\bbb
Z=\int du\int [\dd A_{\mu}] \Delta_{FP}\lp A_{\mu}\rp\delta\lp f\lp A_{\mu}\rp\rp 
e^{-S_{YM}[A_{\mu}]},
\eee  
ce qui nous permet d'isoler le volume $\int du$ du groupe de jauge. Bien entendu, il peut y avoir un probl\`eme d'ambig\"uit\'e de Gribov: il peut exister plusieurs solutions non \'equivalentes de jauge de cette contrainte. Tout comme dans le cas classique, nous admettons que ces solutions n'interviennent pas au niveau de la th\'eorie des perturbations et nous n'en tenons pas compte.

\par

Pour \'evaluer la fonction de partition \`a l'aide de la th\'eorie des perturbation, nous devons transformer l'\'ecriture de la fonction $\delta$ et du d\'eterminant de Faddeev-Popov. 

\par

Afin d'\'eliminer la fonction $\delta$, introduisons un champ auxiliaire $B$ qui est un \'el\'ement antihermitien de $\aa$. Nous rempla\c cons la condition de jauge de Lorentz $\partial_{\mu}A^{\mu}=0$ par la condition plus g\'en\'erale $g\partial_{\mu}A^{\mu}=B$. On montre alors que la d\'ependance du d\'eterminant de Faddeev-Popov en $B$ correspond simplement \`a un changement de jauge et apr\`es normalisation par le facteur $\int du$, la fonction de partition s'\'ecrit simplement
\bbb
Z=\int [\dd A_{\mu}] \Delta_{FP}\lp A_{\mu}^{u}\rp\delta \lp g\partial_{\mu}A^{\mu}-B\rp 
e^{-S_{YM}[A_{\mu}]}.
\eee
Nous simplifions encore $Z$ en int\'egrant sur le champ auxiliaire $B$ avec un poids gaussien $e^{\frac{1}{2\alpha} g^{2}\int B^{2}}$, o\`u $\alpha$ est un r\'eel positif. Notre fonction de partition est donc
\bbb
Z=\int [\dd A_{\mu}][\dd B] \Delta_{FP}\lp A_{\mu}^{u}\rp\delta \lp g\partial_{\mu}A^{\mu}-B\rp 
e^{-S_{YM}[A_{\mu}]+\frac{1}{2\alpha} g^{2}\int B^{2}},
\eee
ce qui donne, apr\`es avoir effectu\'e l'int\'egrale sur $B$,
\bbb 
Z=\int [\dd A_{\mu}][\dd B] \Delta_{FP}\lp A_{\mu}^{u}\rp\delta \lp g\partial_{\mu}A^{\mu}-B\rp 
e^{-S_{YM}[A_{\mu}]-S_{GF}[A_{\mu}]},
\eee
o\`u le terme de fixation de jauge est donn\'e par
\bbb
S_{GF}[A_{\mu}]=-\frac{1}{2\alpha}\int \lp\partial_{\mu}A^{\mu}\rp^{2}.
\eee
Le d\'eterminant de Faddeev-Popov peut aussi se mettre sous la forme d'une int\'egrale fonctionnelle. En effet, c'est le jacobien de la transformation $u\mapsto \partial_{\mu}A^{u\,\mu}$, que l'on calcule au vosinage de l'identit\'e. Si nous posons $u=e^{\omega}$, alors  au premier ordre en $\omega$, on a
\bbb
uA_{\mu}u^{-1}+\frac{1}{g}u\partial_{\mu}u^{-1}=
\lb\omega, A_{\mu}\rb-\frac{1}{g}\partial_{\mu}\omega.
\eee
d'o\`u 
\bbb
f(A)=g\partial_{\mu}\lp uA_{\mu}u^{-1}+\frac{1}{g}u\partial_{\mu}u^{-1}\rp=
g\partial^{\mu}\lb\omega, A_{\mu}\rb-\partial^{\mu}\partial_{\mu}\omega,
\eee  
Si nous d\'eveloppons $f(A_{\mu})$ et $\omega$ dans la base de Fourier, on obtient
\bbb
f^{p}(A_{\mu})=\mathop{\sum}\limits_{q+r=p}g2i\pi p^{\mu}2i\sin\theta(q,r)\omega^{q}A_{\mu}^{r}
+4\pi^{2}p^{2}\omega^{p}.
\eee
On en d\'eduit que le d\'eterminant de Faddeev-Popov est donn\'e par
\bbb
\Delta_{FP}[A_{\mu}]=\det M_{p,q}
\eee 
avec
\bbb
M_{p,q}=\frac{\partial f^{p}}{\partial\omega^{q}}=g2i\pi p^{\mu}2i\sin\theta(q,p-q)A_{\mu}^{p-q}
-4\pi^{2}p^{2}\delta(p-q).
\eee
Pour terminer, nous allons r\'e\'ecrire ce d\'eterminant en utilisant une int\'egrale sur des variables de Grassmann en suivant la m\'ethode d\'evelopp\'ee lors de la quantification des th\'eories de Yang-Mills non ab\'eliennes,.

\par

Remarquons que si $C$ et $\ov{C}$ sont deux variables de Grassmann et $M$ un nombre complexe, alors on a
\bbb
\int d\ov{C}dC e^{-\ov{C}MC}=M.
\eee
Ce r\'esultat se g\'en\'eralise en dimension $n$ de la mani\`ere suivante. Si $C_{i}$ et $\ov{C}_{j}$ sont un ensemble de $2n$ variables deux \`a deux anticommutantes et $M_{ij}$ une matrice complexe, on a
\bbb
\int \mathop{\Pi}\limits_{i}d\ov{C_{i}}dC_{i} e^{-\mathop{\sum}\limits_{i,j}\ov{C}_{i}M_{ij}C_{j}}=\det M.
\eee
En dimension infinie, nous \'ecrivons le d\'eterminant pr\'ec\'edent \`a l'aide d'une int\'egrale fonctionelle
\bbb
\int \mathop{\Pi}\limits_{p\neq 0}d\ov{C}_{p}dC_{p} e^{-\mathop{\sum}\limits_{p,q}\ov{C}_{p}M_{pq}C_{q}}=\Delta_{FP}[A_{\mu}],
\eee
o\`u $C_{p}$ et $\ov{C}_{q}$ sont des variables de Grassmann, appel\'es fant\^omes de Faddeev-Popov, satisfaisant aux relations d'anticommutation
\bbbb
C_{p}C_{q}+C_{q}C_{p}&=&0\n\\
C_{p}\ov{C}_{q}+\ov{C}_{q}C_{p}&=&0\n\\
\ov{C}_{p}\ov{C}_{q}+\ov{C}_{q}\ov{C}_{p}&=&0.
\eeee
Bien entendu, il n'y a pas de mode z\'ero pour les fant\^omes de Faddeev-Popov car ceux-ci n'apparaissent pas dans $M_{p,q}$ qui n'a pas de ligne $p=0$ ou de colonne $q=0$.

\par
 
Pour simplifier notre \'ecriture, introduisons 
\bbb
C=\mathop{\sum}\limits_{p\neq 0}C_{p}U^{p}
\eee
et
\bbb
\ov{C}=\mathop{\sum}\limits_{p\neq 0}\ov{C}_{p}U^{p}
\eee
et d\'efinissons les op\'erations $\int$, $\partial_{\mu}$ et la multiplication par $A_{\mu}$ comme \'etant lin\'eaires par rapport \`a $C_{p}$ et $\ov{C}_{q}$.
En d\'efinissant l'action de Faddeev-Popov comme \'etant
\bbb
S_{FP}[A_{\mu},\ov{C},C]=\mathop{\sum}\limits_{p,q}\ov{C}_{p}M_{pq}C_{q}=
\int \ov{C}\partial^{\mu}\lp\partial_{\mu}C+g[A_{\mu},C]\rp,
\eee
on a
\bbb
\Delta_{FP}[A_{\mu}]=\int [\dd \ov{C},C]e^{-S_{FP}[A_{\mu},\ov{C},C]},
\eee
ce qui nous permet de r\'e\'ecrire la fonction de partition sous la forme
\bbb
Z=\int [\dd A_{\mu}][\dd \ov{C}][\dd C] e^{-S[A_{\mu},\ov{C},C]}
\eee
o\`u $S[A_{\mu},\ov{C},C]$ est l'action de Yang-Mills contenant les termes de fixation de jauge et l'action de Faddeev-Popov
\bbb
S[A_{\mu},\ov{C},C]=S_{YM}[A_{\mu}]+S_{GF}[A_{\mu}]+S_{FP}[A_{\mu},\ov{C},C].\label{fdg1}
\eee
Il est important de noter que la partie quadratique en $A_{\mu}$ de cette action
s'\'ecrit
\bbb
S^{(2)}[A_{\mu}]=\frac{1}{2}\int A_{\nu}\partial_{\mu}\partial^{\mu}A^{\nu}
-\lp1-\frac{1}{\alpha}\rp A_{\mu}\partial^{\mu}\partial^{\nu}A_{\nu}
\eee
puisque
\bbb
S_{GF}[A_{\mu}]=-\frac{1}{2\alpha}\int \lp\partial_{\mu}A^{\mu}\rp^{2}
=\frac{1}{2\alpha}\int A_{\mu}\partial^{\mu}\partial^{\nu}A_{\nu}
\eee
apr\`es int\'egration par parties.

\par

En composantes, cela s'\'ecrit
\bbb
S_{YM}^{(2)}[A_{\mu}]=\frac{(2i\pi)^{2}}{2}
\lp p^{2}g^{\mu\nu}-\lp 1-\frac{1}{\alpha}\rp p^{\mu}p^{\nu}\rp A_{\mu}^{p}A_{\nu}^{-p}.
\eee
L'op\'erateur $p^{2}g^{\mu\nu}-\lp1-\frac{1}{\alpha}\rp p^{\mu}p^{\nu}$ est inversible et son inverse est
\bbb
\Delta_{\mu\nu}(\alpha)=\frac{g_{\mu\nu}-(1-\alpha)p_{\mu}p_{\nu}}{p^{2}}.
\eee
Ainsi, la fixation de jauge nous a permis de construire une th\'eorie ayant un propagateur bien d\'efini $\Delta_{\mu\nu}$, ce qui est indispensable pour d\'evelopper la th\'eorie perturbative.

\par

Bien entendu, toutes les int\'egrales fonctionelles que nous avons \'ecrites ainsi que la g\'en\'eralisation du d\'eterminant et de la mesure de Haar en dimension infinie n'ont pas de signification math\'ematique bien claire. Notre seul but a \'et\'e de montrer que les outils habituels de la th\'eorie quantique des champs pouvaient \^etre utilis\'es sur le tore non commutatif sans changement profond.


\subsection{Sym\'etrie BRS}

L'action donn\'ee par la relation (\ref{fdg1}) n'est plus invariante de jauge car nous avons incorpor\'e le terme de fixation de jauge $S_{GF}[A_{\mu}]$. En compl\`ete analogie avec la sym\'etrie BRS de la th\'eorie des champs de Yang-Mills, nous allons montrer que cette action admet une sym\'etrie r\'esiduelle qui s'av\`ere \^etre de la plus haute importance lorsque l'on \'etudie la renormalisation de cette th\'eorie.

\par

La transformation $BRS$ est donn\'ee par son  g\'en\'erateur $s$ dont  l'op\'eration est  d\'efinie par
\bbb
s(A_{\mu})=\frac{1}{g}\partial_{\mu}C+[A_{\mu},C]
\eee
sur le champ de jauge. Bien entendu, cette relation doit \^etre consid\'er\'ee comme une \'ecriture compacte de la relation entre composantes, qui seule a un sens pr\'ecis
\bbb
s(A_{\mu}^{p})=\frac{2i\pi p_{\mu}}{g}C^{p}+
\mathop{\sum}\limits_{q+r=p}2i\sin\theta(q,r)A_{\mu}^{q}C^{r}.
\eee
Nous d\'efinissons l'action de $s$ sur le fant\^ome de Faddeev-Popov par
\bbb
s(C)=-\frac{1}{2}\la C,C\ra=-C^{2}.
\eee
Cette relation est formellement analogue \`a l'action de la transformation BRS usuelle, mais elle doit \^etre explicit\'ee en composantes pour acqu\'erir une r\'eelle signification. Puisque le champ $C$ se d\'eveloppe en $C=\sum_{p}C_{p}U^{p}$ o\`u $C_{p}$ est une suite de variables de Grassmann, on a
\bbb
\frac{1}{2}\la C,C\ra=\mathop{\sum}\limits_{q,r} C_{q}C_{r}U^{q}U^{r}
=\mathop{\sum}\limits_{q,r} i\sin\theta(q,r)C_{q}C_{r}U^{q+r}.
\eee
On en d\'eduit que
\bbb
s(C_{p})=-\mathop{\sum}\limits_{q+r=p} i\sin\theta(q,r)C_{q}C_{r}.
\eee

\par

La d\'efinition de $s(\ov{C})$ est plus d\'elicate et recquiert l'introduction d'un champ auxilaire $B$. Rempla\c cons le terme de fixation de jauge $S_{GF}[A_{\mu}]=-\frac{1}{2\alpha}\int \lp\partial_{\mu}A^{\mu}\rp^{2}$ par
\bbb
S_{GF}[A_{\mu},B]=\int\lp\frac{\alpha g^{2}}{2}B^{2}+g\partial_{\mu}BA^{\mu}\rp.
\eee
Ce terme est \'equivalent, au niveau classique, au terme pr\'ec\'edent car l'\'equation du mouvement pour le champ $B$ nous donne simplement $B=\frac{1}{\alpha g}\partial_{\mu}A^{\mu}$ ce qui implique
\bbb
S_{GF}[A_{\mu},B]=\frac{1}{2\alpha}\int\lp\partial_{\mu}A^{\mu}\rp^{2}+
\frac{1}{\alpha}\int\lp\partial_{\mu}\partial_{\nu}A^{\nu}A^{\mu}\rp^{2}
=S_{GF}[A_{\mu}]
\eee
apr\`es int\'egration par parties, ce qui nous permet d'interpr\'eter $B$ comme un multiplicateur de Lagrange pour la condition de jauge de Lorentz.

\par

De m\^eme, ces deux actions sont \'equivalentes au niveau quantique car on peut r\'eecrire le terme de fixation de jauge \`a l'aide d'une int\'egration par parties
\bbbb
S_{GF}[A_{\mu},B]&=&\int\lp\frac{\alpha g^{2}}{2}B^{2}-gB\partial_{\mu}A^{\mu}\rp\n\\
&=&\frac{\alpha g^{2}}{2}\int\lp B-\frac{1}{\alpha g}\partial_{\mu}A^{\mu}\rp^{2}-
\frac{1}{2\alpha}\int\lp\partial_{\mu}A^{\mu}\rp^{2}.
\eeee
Par cons\'equent, l'int\'egrale fonctionelle sur $B$ est une int\'egrale gaussienne qui nous donne, \`a un facteur de normalisation ind\'ependant des champs pr\`es,
\bbb
\int[\dd B]e^{-S_{GF}[A_{\mu},B]}=e^{-S_{GF[A_{\mu}]}}
\eee 

\par

On peut alors achever la d\'efinition de $s$ en posant $s(\ov{C})=B$ et $s(B)=0$, ce qui entraine que cette transformation est nilpotente. En effet, on a $s(C)=-C^{2}$ donc
\bbb
s^{2}(C)=-s(C)C+Cs(C)=C^{3}-C^{3}=0.
\eee
De m\^eme, 
\bbbb
s^{2}(A_{\mu})&=&s\lp\frac{1}{g}\partial_{\mu}C+[A_{\mu},C]\rp\n\\
&=&\frac{1}{g}\partial_{\mu}(s(C))+s(A_{\mu})C+A_{\mu}s(C)-s(C)A_{\mu}+Cs(A_{\mu})\n\\
&=&-\frac{1}{g}\partial_{\mu}(C^{2})+\frac{1}{g}\partial_{\mu}CC+[A_{\mu},C]C-A_{\mu}C^{2}+C^{2}A_{\mu}+\frac{1}{g}C\partial_{\mu}C+C[A_{\mu},C]\n\\
&=&0.
\eeee
Il est important de remarquer que nous avons utilis\'e le fait que $s$ est une transformation impaire
\bbb
s(PQ)=s(P)Q\pm Ps(Q),
\eee 
o\`u on a le signe $-$ si $P$ est un fant\^ome ou un antifant\^ome.

\par

Finalement, nous allons montrer que l'action totale, incluant le d\'eterminant de Faddeev-Popov et le terme de fixation de jauge, est invariante sous la transformation BRS. Commen\c cons par \'etudier la loi de transformation de la courbure. Puisque
\bbb
s(A_{\mu})=\frac{1}{g}\partial_{\mu}C+[A_{\mu},C]
\eee
on a
\bbb
s(\partial_{\mu}A_{\nu})=\frac{1}{g}\partial_{\mu}\partial_{\nu}C+[\partial_{\mu}A_{\nu},C]+[A_{\nu},\partial_{\mu}C]
\eee
et
\bbbb
s\lp g[A_{\mu},A_{\nu}]\rp&=&g\lb s(A_{\mu}),A_{\nu}\rb+g\lb A_{\mu},s(A_{\nu})\rb\n\\
&=&\lb \partial_{\mu}C,A_{\nu}\rb+\lb A_{\mu},\partial_{\nu}C\rb\n\\
&+&g\lb[A_{\mu},C],A_{\nu}\rb+\lb A_{\mu},[A_{\nu},C]\rb.
\eeee
On en d\'eduit que 
\bbb
s(F_{\mu\nu})=[F_{\mu\nu},C],
\eee
ce qui prouve que l'action de Yang-Mills est invariante sous cette transformation car
\bbbb
s\lp F_{\mu\nu}F^{\mu\nu}\rp&=&\int\lp s(F_{\mu\nu})F^{\mu\nu}+F_{\mu\nu}s(F^{\mu\nu})\rp\n\\
&=&\int\lp [F_{\mu\nu},C]F^{\mu\nu}+F_{\mu\nu}[F^{\mu\nu},C]\rp\n\\
&=&\int\lp[F_{\mu\nu}F^{\mu\nu},C]\rp\\
&=&0\n
\eeee
en utilisant la propri\'et\'e de trace de $\int$.

\par

La loi de transformation de l'action de Faddeev-Popov se d\'etermine de la mani\`ere suivante,
\bbbb
&s\lp \ov{C}\partial_{\mu}\partial^{\mu}C+g\ov{C}\partial_{\mu}[A^{\mu},C]\rp=&\n\\
&B\lp\partial_{\mu}\partial^{C}+g\partial_{\mu}[A^{\mu},C]\rp&\n\\ 
&+\ov{C}\partial_{\mu}\partial^{\mu}C^{2}-\ov{C}\partial_{\mu}\lp C\rp\partial^{\mu}C-\ov{C}\partial_{\mu}\lp\partial^{\mu}CC\rp&\n\\
&-g\ov{C}\partial_{\mu}\lp [A^{\mu},C]C-A^{\mu}C^{2}+C[A^{\mu},C]+C^{2}A^{\mu}\rp. &
\eeee
On en d\'eduit que
\bbb
s\lp S_{FP}[A_{\mu},\ov{C},C]\rp=
\int\lp B(\partial_{\mu}\partial^{C}+g\partial_{\mu}[A^{\mu},C])\rp,
\eee
car les termes des troisi\`eme et quatri\`eme lignes de la relation pr\'ec\'edente sont identiquement nuls.

\par 

Pour terminer, il nous suffit de d\'eterminer la loi de transformation du terme de fixation de jauge $S_{GF}[A_{\mu},B]$:
\bbbb
s\lp\frac{\alpha g^{2}}{2}B^{2}-gB\partial_{\mu}A^{\mu}\rp
&=&-gBs\lp \partial_{\mu}A^{\mu}\rp\n\\
&=&-B\lp\partial_{\mu}\partial^{\mu}C+g\partial_{\mu}[A^{\mu},C]\rp
\eeee
Ceci est exactement le terme oppos\'e \`a celui que nous avons rencontr\'e dans l'\'etude de la transformation du terme de Faddeev-Popov. On en d\'eduit que
\bbb
s\lp S_{GF}[A_{\mu,B}]+S_{FP}[A_{\mu},C,\ov{C}]\rp=0,
\eee
ce qui prouve l'invariance de l'action totale sous la transformation BRS. 

\par

En conclusion, il semble important de noter que nous avons pu d\'efinir la transformation BRS et v\'erifier l'invariance de l'action totale, incluant les termes de fixation de jauge et le d\'eterminant de Faddeev-Popov sur le tore non commutatif. Il est remarquable que toute cette construction se place sur un plan purement alg\'ebrique, car nous n'avons jamais \'et\'e amen\'e \`a utiliser la notion de point, que ce soit lors du proc\'ed\'e de fixation de jauge ou dans l'\'etude de la sym\'etrie BRS.

\subsection{R\`egles de Feynman}

Pour construire la th\'eorie quantique des champs perturbative sur le tore non commutatif, nous devons d\'evelopper un proc\'ed\'e nous permettant de calculer \`a chaque ordre de la th\'eorie des perturbations les valeurs moyennes de certaines observables. Plus pr\'ecis\'ement, si $\oo$ est une fonction de $A_{\mu}$ invariante de jauge, nous voulons calculer sa valeur moyenne d\'efinie par l'int\'egrale fonctionnelle
\bbb
\langle \oo(A_{\mu})\rangle=\int[\dd\ov{C}C][\dd A_{\mu}]
\oo(A_{\mu})e^{-S[A,C,\ov{C}]},
\eee 
o\`u nous supposons l'int\'egrale fonctionelle normalis\'ee par la condition $\langle1\rangle=1$. Il est souvent utile de calculer la valeur moyenne de fonctions du type $\oo(A_{\mu},C,\ov{C})$ qui d\'ependent aussi des fant\^omes de Faddeev-Popov.

\par

Dans le cas du tore non commutatif, les variables sur lesquelles est prise la moyenne sont les nombres r\'eels $A_{\mu}^{p}$ ainsi que les variables de Grassmann $C_{q}$ et $\ov{C}_{r}$. Nous pouvons aussi, pour simplifier, supposer que la fonction $\oo(A_{\mu},C,\ov{C})$ se d\'eveloppe en s\'erie sur les mon\^omes $A_{\mu_{1}}^{p_{1}}\dots A_{\mu_{i}}^{p_{i}}C_{q_{1}}\dots C_{q_{j}}\ov{C}_{1}\dots\ov{C}_{r_{k}}$, ce qui ram\`ene le calcul de sa valeur moyenne \`a celui de l'int\'egrale fonctionelle
\bbb
\int[\dd\ov{C}\dd C][\dd A_{\mu}]
\lp A_{\mu_{1}}^{p_{1}}\dots A_{\mu_{i}}^{p_{i}}C_{q_{1}}\dots C_{q_{j}}\ov{C}_{1}\dots \ov{C}_{r_{k}}\rp e^{-S[A,C,\ov{C}]},
\eee
qui repr\'esente, en th\'eorie des champs usuelle, la fonction de Green dans l'espace des impulsions.

\par

Pour pouvoir calculer ces valeurs moyennes \`a tous les ordres de la th\'eorie des perturbations, nous allons suivre la m\'ethode usuelle de la th\'eorie des champs en couplant les champs $A_{\mu}$, $C$ et $\ov{C}$ aux sources $J_{\mu}$, $\ov{\eta}$ et $\eta$. $J^{\mu}$ est un \'el\'ement antihermitien de $\aa$ qui se couple \`a $A_{\mu}$ par
\bbb
\int J^{\mu}A_{\mu}=\mathop{\sum}\limits_{p}J^{\mu -p}A_{\mu}^{p}.
\eee
De m\^eme, $\eta$ et $\ov{\eta}$ sont des sources pour les antifant\^omes et les fant\^omes qui se d\'eveloppent en s\'erie \`a l'aide de variables de Grassmann
\bbb
\eta=\mathop{\sum}\limits_{p}\eta_{p}U^{p}\;\;\;\;
\ov{\eta}=\mathop{\sum}\limits_{p}\ov{\eta}_{p}U^{p},
\eee
o\`u $\eta_{p}$ et $\ov{\eta}_{p}$ sont des variables de Grassmann qui anticommutent entre elles ainsi qu'avec toutes les autres variables de Grassmann.
Le couplage avec $C$ et $\ov{C}$ se fait selon
\bbb
\int \ov{\eta}C=\mathop{\sum}\limits_{p}\ov{\eta}_{-p}C_{p}
\eee
et
\bbb
\int \ov{C}\eta=\mathop{\sum}\limits_{p}\ov{C}_{p}\eta_{-p}.
\eee
En introduisant la fonctionelle g\'en\'eratrice
\bbb
Z[J_{\mu},\eta,\ov{\eta}]=
\int[\dd\ov{C}\dd C][\dd A_{\mu}] e^{-S[A_{\mu},C,\ov{C}]
+\int J_{\mu}A^{\mu}+\int \ov{\eta}C+\ov{C}\eta},
\eee
que nous supposons normalis\'ee par $Z[0]=1$, on peut, par simple d\'erivation par rapport aux composantes des sources $\jj_{\mu}^{-p}$, $\eta_{-q}$ (d\'erivation \`a gauche) et $\ov{\eta}_{-r}$ (d\'erivation \`a droite) calculer la valeur moyenne de n'importe quelle observable. En effet, ces op\'erations de d\'erivations sont respectivement \'equivalentes \`a l'insertion dans l'int\'egrale fonctionelle d\'efinissant la fonctionnelle g\'en\'eratrice des termes $A_{\mu}^{p}$, $C_{q}$ et $\ov{C}_{q}$.

\par

Nous s\'eparons alors l'action en une partie quadratique,
\bbb
S_{quad}[A_{\mu},C,\ov{C}]=\frac{(2i\pi)^{2}}{2}
\mathop{\sum}\limits_{p}\lp (p^{2}g_{\mu\nu}-(1-\alpha)p^{\mu}p^{\nu}) A_{\mu}^{p}A_{\nu}^{-p}+p^{2}\ov{C}_{-p}C_{p}\rp
\eee
qui d\'ecrit une th\'eorie libre et un partie non quadratique associ\'ee aux interactions
\bbbb
&S_{int}[A_{\mu},C,\ov{C}]=
2\pi g\mathop{\sum}\limits_{p,q,r}p^{\mu}\sin\theta(q,r)
\delta(p+q+r)A_{\mu}^{p}A_{\nu}^{q}A_{\rho}^{r}&\n\\
&+2\pi g\mathop{\sum}\limits_{p,q,r}p^{\mu}\sin\theta(q,r)
\delta(p+q+r)\ov{C}^{p}A_{\mu}^{q}C^{r}&\n\\
&+g^{2}\mathop{\sum}\limits_{p,q,r,s}g^{\mu\rho}g^{\nu\sigma}
\sin\theta(p,q)\sin\theta(r,s)
\delta(p+q+r+s)A_{\mu}^{p}A_{\nu}^{q}A_{\rho}^{r}A_{\sigma}^{s}.&
\n
\eeee
Cela nous permet d'\'ecrire la fonctionnelle g\'en\'eratrice $Z[J_{\mu},C,\ov{C}]$ \`a l'aide de d\'eriv\'ees de la fonctionelle g\'en\'eratrice des champs libres
\bbb
Z_{0}[J_{\mu},C,\ov{C}]=\int [\dd \ov{C},C][\dd A_{\mu}] e^{-S_{quad}[A_{\mu},C,\ov{C}]},
\eee
car on a,
\bbb
Z[J_{\mu},C,\ov{C}]=e^{-S_{int}[\frac{\partial}{\partial J_{\mu}},\frac{\partial}{\partial C},\frac{\partial}{\partial\ov{C}}]}
Z_{0}[J_{\mu},C,\ov{C}].
\eee
Cette relation est formelle et nous permet, tout comme en th\'eorie des champs usuelle, d'\'ecrire n'importe quelle fonction de Green tout ordre de la th\'eorie des perturbations \`a l'aide de quelques r\`egles simples appel\'es r\`egles de Feynman.

\par

Ces r\`egles permettent de construire des diagrammes associ\'ees \`a chaque fonction de Green et chacun de ces diagrammes correspond \`a une somme sur toutes les impulsions. Nous ne d\'ecrivons pas ici la m\'ethode permettant d'effectuer 
ces calculs et nous nous bornons \`a donner les r\`egles de Feymann pour la th\'eorie de Yang-Mills sur le tore non commutatif, leur d\'erivation \`a partir des expressions de $S_{quad}[A_{\mu},C,\ov{C}]$ et de $S_{int}[A_{\mu},C,\ov{C}]$ \'etant similaire \`a la d\'emarche employ\'ee en th\'eorie des champs usuelle.

\par

Les propagateurs s'obtiennent en inversant les op\'erateurs apparaissant dans la d\'efinition de l'action quadratique. Puisque l'inverse de $p^{2}g^{\mu\nu}-(1-\frac{1}{\alpha})p^{\mu}p^{\nu}$ est
\bbb
\frac{1}{p^{2}}\lp g_{\mu\nu}-(1-\alpha)\frac{p_{\mu}p_{\nu}}{p^{2}}\rp,  
\eee
nous associons au diagramme repr\'esentant le propagateur du champ $A_{\mu}$
$$
\epsfig{file=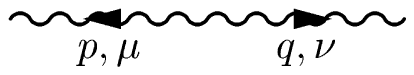}
$$
la quantit\'e
\bbb
-\frac{4\pi^{2}}{p^{2}}\lp g_{\mu\nu}-(1-\alpha)\frac{p_{\mu}p_{\nu}}{p^{2}}\rp\delta(p+q).
\eee
De m\^eme, au propagateur des fant\^omes
$$
\epsfig{file=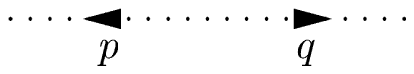}
$$
est associ\'e
\bbb
-\frac{4\pi^{2}}{p^{2}}\delta(p+q).
\eee
Ces propagateurs ne d\'ependent pas du param\`etre de d\'eformation $\theta$ et sont absolument identiques aux propagateurs usuels.

\par

L'interaction \`a trois champs de jauge est donn\'ee par le diagramme
$$
\epsfig{file=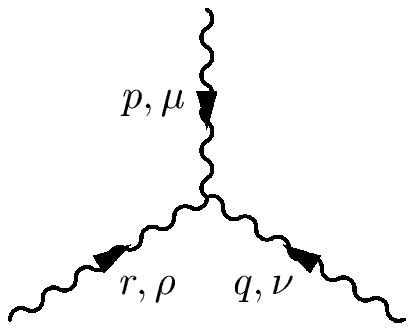}
$$
auquel est associ\'e
\bbb
-4\pi g\lp
(p-r)_{\nu}g_{\mu\rho}+(q-p)_{\rho}g_{\mu\nu}+(r-q)_{\mu}g_{\nu\rho}\rp
\sin\theta(p,q)\delta(p+q+r).
\eee

L'interaction de quatre bosons de jauge
$$
\epsfig{file=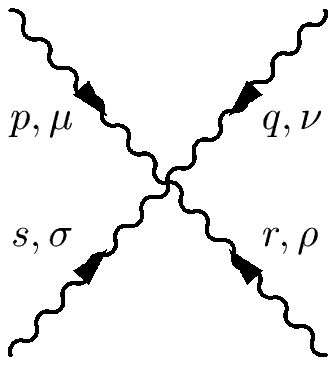}
$$
est donn\'ee par
\bbbb
&-16\pi^{2}g^{2}\lp
(g_{\mu\rho}g_{\nu\sigma}-g_{\mu\sigma}g_{\nu\rho})
\sin\theta(p,q)\sin\theta(r,s)\right.&\n\\
&+(g_{\mu\sigma}g_{\nu\rho}-g_{\mu\nu}g_{\rho\sigma})
\sin\theta(p,r)\sin\theta(s,q)\n\\
&\left.(g_{\mu\nu}g_{\rho\sigma}-g_{\mu\rho}g_{\nu\sigma})
\sin\theta(p,s)\sin\theta(q,r)
\rp\delta(p+q+r+s).&\n
\eeee

Enfin, l'interaction des bosons de jauge et des fant\^omes 
$$
\epsfig{file=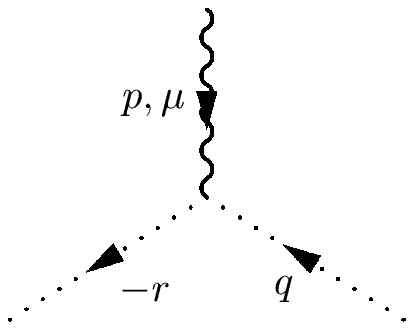}
$$
nous donne
\bbb
4\pi g r_{\mu}\sin\theta(p,q)\delta(p+q+r).
\eee

Si les propagateurs de la th\'eorie de Yang-Mills sur le tore non commutatif sont ceux de la th\'eorie de Yang-Mills usuelle, il n'en va pas de m\^eme pour les interactions. Celles-ci sont totalement nouvelles et leur d\'ependance en les impulsions est non polyn\^omiale, ce qui signifie que nous avons une th\'eorie des   champs non locale. En effet, nous avons \'ecrit le lagrangien dans l'espace de Fourier, mais il est impossible d'\'ecrire son \'equivalent \`a l'aide de fonctions sur le tore usuel sans avoir des interactions entre les valeurs des champs en des endroits diff\'erents de l'espace. Cette non localit\'e des interactions est le seul fait v\'eritablement nouveau de cette th\'eorie des champs; il est \`a relier directement \`a l'impossibilit\'e de d\'efinir la notion de point lorsque l'alg\`ebre est non commutative.

\par

Toutefois, lorsque $\theta$ est rationnel, nous savons que nous obtenons une th\'eorie de jauge $U(N)$ avec une topologie non triviale. Dans ce cas, la th\'eorie est \'evidemment locale car la fonction $\sin\theta(p,q)$ est p\'eriodique en $p$ et en $q$, ce qui implique que $\sin\theta(p,q)$ ne prend qu'un nombre fini de valeurs. Il est alors possible d'\'ecrire les r\`egles de Feynman \`a l'aide des impulsions sur le tore usuel, qui sont les indices $p$ tels que $\theta(p,q)\in 2i\pi\zzz$ pour tout $q\in\zzz^{n}$, et d'un indice ne prenant qu'un nombre fini de valeurs et jouant le r\^ole de l'indice de l'alg\`ebre de Lie de $SU(N)$.

\subsection{Vers une th\'eorie quantique}

Avec ces r\`egles de Feynman, nous disposons d'un proc\'ed\'e permettant de calculer \`a n'importe quel ordre de la th\'eorie des perturbations les fonctions de Green et donc les valeurs moyennes d'observables quelconques. Bien entendu, toutes les r\`egles de la th\'eorie des champs usuelle s'appliquent. Pour calculer \`a un ordre donn\'e une fonction de Green, il faut sommer toutes les contributions des diagrammes de Feynman dont les pattes externes correspondent \`a cette fonction de Green.

\par

Puisque nous sommes sur l'analogue non commutatif du tore, \`a chaque diagramme de Feynman contenant au moins une boucle est associ\'ee une somme sur toutes les impulsions $p\in\zzz^{n}$, qui remplace l'int\'egrale sur les impulsions que nous rencontrons en th\'eorie des champs usuelle.

\par

Faute de temps, nous allons nous limiter \`a donner les principaux r\'esultats de cette th\'eorie quantique, en renvoyant le lecteur int\'eress\'e \`a \cite{torus} pour un \'etude compl\`ete et une d\'emonstration des r\'esultats que nous allons \'enoncer.

\par

Tout d'abord, il s'av\`ere que les s\'eries que nous rencontrons dans le calcul des diagrammes de Feynman sont tout aussi divergentes qu'en th\'eorie des champs usuelle, du moins \`a une boucle. En effet, lorsque $\theta$ est rationnel, cette th\'eorie est une th\'eorie de Yang-Mills avec un groupe de jauge $SU(N)$ qui m\`ene \`a des s\'eries divergentes dans le calcul de certains diagrammes de Feynman. Le calcul explicite de quelques exemples simples, comme la correction \`a une boucle du propagateur du champ de jauge, nous montre que m\^eme si $\theta$ est non d\'eg\'en\'er\'e, nous obtenons le m\^eme type de divergence. Des r\'esultats similaires ont \'et\'e obtenus dans \cite{jose} et cela nous montre que m\^eme si on modifie profond\'ement la structure de l'espace-temps en introduisant des relations de commutation non triviales entre les coordonn\'ees, cela ne suffit pas \`a r\'egulariser les divergences ultraviolettes de la th\'eorie.

\par

Cependant, en utilisant l'in\'egalit\'e $|\sin\theta(p,q)|\leq 1$ et le th\'eor\`eme de Weinberg sur la convergence des graphes de Feynman \cite{zuber}, on montre ais\'ement que tout diagramme de Feynman convergent par comptage de puissance  dans la th\'eorie conventionelle est convergent sur le tore non commutatif. Se pose alors naturellement le probl\`eme de la renormalisation de cette th\'eorie ainsi que de l'influence du param\`etre de d\'eformation $\theta$ sur les diverses \'equations rencontr\'ees en th\'eorie de la renormalisation.

\par

Comme conjectur\'e dans \cite{M}, il s'av\`ere que cette th\'eorie est renormalisable. La preuve de la renormalisabilt\'e repose essentiellement sur la sym\'etrie BRS et l'utilisation d'un proc\'ed\'e de r\'egularisation pr\'eservant cette derni\`ere. Puisque nous ne pouvons pas utiliser la r\'egularisation dimensionnelle, nous avons recours \`a la m\'ethode des "hautes d\'eriv\'ees covariantes" \cite{faddeev}, qui nous permet de r\'egulariser tous les diagrammes ayant au moins deux boucles. Les diagrammes restant sont r\'egularis\'es par la m\'ethode de Pauli-Villars et nous construisons ainsi, ordre par ordre, les contretermes de la th\'eorie en utilisant la sym\'etrie BRS.
Il est remarquable que tout ceci est formellement analogue aux m\'ethodes de la th\'eorie des champs usuelle; m\^eme si notre objet d'\'etude est tr\`es diff\'erent, nous proc\'edons comme si c'\'etait un champ de Yang-Mills ordinaire.

\par

Toutefois, ce proc\'ed\'e de r\'egularisation n'est pas adapt\'e au calcul pratique des contretermes. Dans \cite{torus}, nous d\'eterminons les contretermes \`a une boucle dans le sch\'ema de renormalisation minimal MS en utilisant une fonction $\zeta$ g\'en\'eralis\'ee. Ces contretermes sont ind\'ependants du param\`etre de d\'eformation $\theta$ et identiques \`a ceux rencontr\'es en th\'eorie des champs usuelle.

\par

Il est remarquable que cette th\'eorie soit absolument d\'epourvue de divergences infrarouges. En effet, sur un espace compact ces derni\`eres sont essentiellement dues au mode z\'ero du champ qui nous donne une divergence du type $1/p$ pour $p=0$ dans toute s\'erie repr\'esentant un diagramme de Feynman.
Puisque ce mode z\'ero se d\'ecouple compl\`etement de la th\'eorie, il ne se propage pas et disparait compl\`etement. En cons\'equence, les sommes se font uniquement sur les impulsions non nulles et il n'y a aucune divergence infrarouge.  Sur le plan physique, ceci peut \^etre ais\'ement interpr\'et\'e dans le cas rationnel de la mani\`ere suivante. Puisque les divergences infrarouges correspondent \`a des divergences \`a grande \'echelle sur l'espace, celles-ci sont sensibles \`a la topologie nontriviale du fibr\'e d\'ecrivant la th\'eorie de jauge $SU(N)$ associ\'ee, alors que les divergences ultraviolettes apparaissent \`a petite \'echelle et ne peuvent pas distinguer deux th\'eories qui sont localement identiques.

{\renewcommand{\thechapter}{}\renewcommand{\chaptername}{}
\addtocounter{chapter}{5}
\chapter{Conclusion}\markboth{\sl CONCLUSION}{\sl CONCLUSION}}

En suivant les principes g\'en\'eraux de la g\'eom\'etrie non commutative \'edict\'es par A. Connes, il est possible de g\'en\'eraliser une grande partie des concepts g\'eom\'etriques rencontr\'es en physique th\'eorique. Cela inclut, par exemple, la notion de sym\'etrie de jauge et la th\'eorie de Yang-Mills, ainsi que les diverses propri\'et\'es topologiques de ces objets.

\par

Au cours de cette th\`ese, nous avons utilis\'e ces nouvelles id\'ees g\'eom\'etriques de deux mani\`eres tr\`es diff\'erentes.  

\par

D'un cot\'e, nous avons construit des g\'eom\'etries non commutatives tr\`es simples en prenant le produit tensoriel de la g\'eom\'etrie de l'espace-temps ordinaire par un triplet spectral fini. La g\'eom\'etrie r\'esultante d\'ecrit simplement le produit de l'espace-temps par un espace discret, qui est assimil\'e \`a l'espace interne de la th\'eorie, et nous permet d'interpr\'eter le boson de Higgs comme une connexion sur cet espace discret.

\par

En appliquant la construction que nous avons d\'ecrite, nous avons obtenu une th\'eorie de Yang-Mills-Higgs coupl\'ee \`a la gravitation. Notre point de d\'epart est un triplet spectral $(\aa,\hh,\dd)$, qui ne contient, outre la structure g\'eom\'etrique de l'espace-temps, que des informations relatives au secteur fermionique de la th\'eorie.  En effet, le triplet spectral fini d\'ecrivant l'espace interne du mod\`ele en question se compose de
\begin{itemize}
\item
une alg\`ebre $\aa$, qui est associ\'ee \`a un unique groupe de jauge \`a des facteurs ab\'eliens pr\`es;
\item
l'espace de Hilbert $\hh$ qui est simplement un espace vectoriel regroupant tous les fermions;
\item
l'op\'erateur de Dirac qui d\'ecrit la structure des couplages de Yukawa;
\item
un op\'erateur de chiralit\'e $\chi$ et un op\'erateur de conjugaison de charge $\jj$ qui permettent de distinguer fermions droits et fermions gauches ainsi que leurs antiparticules; 
\item
une repr\'esentation $\pi$ de $\aa$ sur $\hh$ qui est associ\'ee \`a la repr\'esentation fermionique du groupe de jauge. 
\end{itemize}

Ces 6 objets $\aa$, $\hh$, $\dd$, $\chi$, $\jj$ et $\pi$ sont fortement contraints par les axiomes de la g\'eom\'etrie non commutative. Par cons\'equent, le secteur fermionique des th\'eories de Yang-Mills-Higgs  que nous construisons, qui est en correspondance directe  avec les 6 objets pr\'ec\'edents, est lui aussi fortement contraint.

\par

De plus, lorsque ce dernier est fix\'e, le secteur bosonique est enti\`erement d\'etermin\'e. Les champs de jauge et leurs repr\'esentations apparaissent directement comme des \'el\'ements de $\aa$ repr\'esent\'es par $\pi$ alors que les champs scalaires sont d\'etermin\'es par le calcul, car ce sont des connexions associ\'ees \`a la th\'eorie de jauge sur la composante discr\`ete de l'espace-temps.

\par

Ainsi nous reconstruisons des mod\`eles de Yang-Mills-Higgs et nous mettons l'accent sur les contraintes impos\'ees \`a ces mod\`eles. Bien entendu, le mod\`ele standard entre dans la classe des mod\`eles que nous pouvons construire, et nous utilisons ainsi la g\'eom\'etrie non commutative pour r\'eduire l'arbitraire sur les param\`etres entrant dans sa description.

\par

Bien que l'\'etude de ce type d'exemple soit maintenant des plus classiques, il reste un certain nombre de points obscurs en relation avec le triplet spectral du mod\`ele standard.

\par

Tout d'abord, puisque nous avons mis l'accent sur la r\'eduction de l'arbitraire inh\'erent \`a la construction du mod\`ele standard, nous devons appliquer cette logique jusqu'au bout et essayer de r\'eduire encore la libert\'e de choix que nous avons en g\'eom\'etrie non commutative. 

\par

En tout premier lieu, ceci concerne la condition d'unimodularit\'e. En effet, la construction du mod\`ele standard en g\'eom\'etrie non commutative introduit un facteur $U(1)$  additionnel dans le groupe de jauge. La r\'eduction de ce dernier se fait par l'interm\'ediaire d'une condition d'unimodularit\'e, qui reste pour l'instant introduite {\it ad hoc}, m\^eme si elle est \'equivalente \`a l'absence d'anomalie. 

\par

En second lieu, il est \'evident que l'op\'erateur de Dirac contient lui aussi un arbitraire important car toutes les masses et les \'el\'ements de la matrice de Cabibbo-Kobayashi-Maskawa sont ind\'etermin\'es. La d\'etermination de contraintes v\'erifiables exp\'erimentalement sur ces \'el\'ements de matrice ne peut se faire \`a l'aide des axiomes pr\'ec\'edents et fait n\'ecessairement appel \`a une th\'eorie ext\'erieure. Dans \cite{grav}, il a \'et\'e sugg\'er\'e que le mod\`ele standard pouvait receler une sym\'etrie quantique. Un groupe quantique aux racines de l'unit\'e devrait jouer, pour l'espace fini, le r\^ole du groupe spinoriel et la covariance de l'op\'erateur de Dirac par rapport \`a ce groupe quantique pourrait \^etre \`a l'origine de contraintes sur les masses des fermions.     
 
\par

D'apr\`es le principe d'action spectrale, ce sont les valeurs propres de l'op\'erateur de Dirac qui sont les v\'eritables variables dynamiques de la th\'eorie. Cela peut \^etre consid\'er\'e comme le point de d\'epart d'une  nouvelle formulation de la th\'eorie de la gravitation. Par exemple, dans \cite{rov}, les crochets de Poisson des valeurs propres de l'op\'erateur de Dirac, consid\'er\'ees comme des fonctionelles d\'ependant de la t\'etrade ont \'et\'e calcul\'es.    
 
\par

Si cette approche a le m\'erite de nous fournir des contraintes qui pourront \^etre v\'erifi\'ees exp\'erimentalement, nous nous sommes born\'es \`a utiliser des triplets spectraux obtenus comme produits tensoriels de la g\'eom\'etrie ordinaire de l'espace-temps par des triplets spectraux finis. 

\par

D'un autre cot\'e, nous pouvons aussi utiliser la g\'eom\'etrie non commutative pour construire des objets r\'eellement nouveaux en th\'eorie des champs, mais sans v\'eritable signification en ce qui concerne la ph\'enom\'enologie des particules \'el\'ementaires. 
 
\par

Par exemple, nous avons construit l'action de Chern-Simons pour tout triplet spectral de dimension 3 satisfaisant \`a la condition de fermeture. Cette construction peut \^etre consid\'er\'ee comme un exemple de g\'en\'eralisation de th\'eorie des champs ordinaire que l'on peut obtenir \`a l'aide de la g\'eom\'etrie non commutative. 

\par

Dans le m\^eme ordre d'id\'ee, l'application des m\'ethodes g\'en\'erales de la g\'eom\'etrie non commutative au tore non commutatif permet de construire des th\'eories de jauge ayant des propri\'et\'es tr\`es similaires \`a celles des th\'eories de jauge ordinaires. Cependant, elles sont d'essence radicalement diff\'erente car nous ne pouvons plus les associer \`a une vari\'et\'e.

\par

A ce titre, l'exemple du tore non commutatif est particuli\`erement frappant. Bien que la th\'eorie de jauge sur ce dernier corresponde, dans l'espace des moments, \`a une th\'eorie ayant des interactions non locales, il est remarquable qu'on puisse lui appliquer les m\'ethodes usuelles de la th\'eorie quantique des champs quasiment sans aucun changement.

\par

Rappelons que nous avons v\'erifi\'e la sym\'etrie BRS et donn\'e les diagrammes de Feynman permettant le d\'eveloppement perturbatif de cette th\'eorie. Il appara\^\i t clairement que les propagateurs des bosons de jauge et des fant\^omes de Faddeev-Popov sont identiques \`a ceux que l'on obtient en th\'eorie des champs ordinaire. Les interactions des bosons de jauge entre eux ainsi qu'avec les fant\^omes sont non locales car elles ont une d\'ependance non polyn\^omiale en les impulsions.

\par

Nous disposons alors de tous les outils permettant de d\'evelopper la th\'eorie perturbative sur le tore non commutatif. Il s'av\`ere que tout diagramme de cette th\'eorie converge si et seulement si il converge en th\'eorie des champs usuelle. Par cons\'equent, la th\'eorie de Yang-Mills sur le tore non commutatif est tout aussi divergente dans l'ultraviolet que la th\'eorie de Yang-Mills usuelle (du moins \`a une boucle) et on peut montrer, en utilisant les m\'ethodes usuelles de la th\'eorie des champs, que c'est une th\'eorie renormalisable. Par contre, la th\'eorie de Yang-Mills est parfaitement finie dans l'infrarouge car le mode z\'ero se d\'ecouple totalement.   

\par

Bien que ce travail soit le premier exemple de th\'eorie quantique des champs d\'evelopp\'ee en g\'eom\'etrie non commutative qui ne corresponde pas directement \`a une th\'eorie d\'ej\`a connue dans le cadre de la physique des particules, il est clair que cet exemple est tr\`es particulier et ne saurait \^etre consid\'er\'e comme g\'en\'erique.

\par

Il semble donc n\'ecessaire de construire d'autres g\'eom\'etries non commutatives, par exemple en utilisant la notion de produit crois\'e qui nous permet d'obtenir le tore non commutatif comme cas particulier, et de d\'evelopper sur ces "espaces non commutatifs" une th\'eorie quantique des champs comme la th\'eorie de Yang-Mills ou celle de Chern-Simons.

\appendix

\chapter{Alg\`ebres d'op\'erateurs}

Cette appendice r\'esume les concepts essentiels de la th\'eorie des alg\`ebres d'op\'erateurs dont nous avons besoin dans cette th\`ese. Il est \'evidemment tr\`es incomplet et ne vise qu'\`a rappeler bri\`evement les notions indispensables \`a la compr\'ehension des pages pr\'ec\'edentes. Nous renvoyons le lecteur int\'er\'ess\'e par un expos\'e complet \`a \cite{ola} ou \cite{bible}.

\section{Alg\`ebres involutives et \'etats}

Les alg\`ebres employ\'ees dans ce qui pr\'ec\`ede sont destin\'ees \`a remplacer l'alg\`ebre des fonctions \`a valeurs complexes d\'efinies sur un espace $\xx$. Tout comme en m\'ecanique quantique, une fonction est remplac\'ee par un op\'erateur et nous devons \^etre capables d'identifier les analogues des  fonctions r\'eelles. Cela nous am\`ene \`a introduire le concept suivant:

\begin{dfi}
Une alg\`ebre involutive est une alg\`ebre associative $\aa$ munie d'une application antilin\'eaire $*$ de $\aa$ dans $\aa$ telle que $(x^{*})^{*}=x$ et $(xy)^{*}=y^{*}x^{*}$ pour tous $x,y\in\aa$.
\end{dfi}

Dans une telle alg\`ebre, il est possible de d\'efinir la positivit\'e ou l'unitarit\'e.

\begin{dfi}
Un \'el\'ement $x$ d'une alg\`ebre involutive $\aa$ est dit
\begin{enumerate}
\item
positif s'il existe $y\in\aa$ tel que $x=yy^{*}$;
\item
unitaire si $xx^{*}=x^{*}x=1$ (d\'efinition valable lorsque $\aa$ est unitaire ).
\end{enumerate}
\end{dfi}

L'analogie avec la m\'ecanique quantique nous am\`ene \`a concevoir $\aa$ comme une alg\`ebre d'observables. En m\'ecanique quantique nous voulons associer un nombre \`a chacune de ces observables qui n'est autre que la valeur moyenne de cette observable. Dans le cas g\'en\'eral, on introduit la notion d'\'etat. 

\begin{dfi}
Un \'etat sur une alg\`ebre involutive $\aa$ est une forme lin\'eaire $\phi$ sur $\aa$ telle que $\phi(1)=1$ et $\phi(xx^{*})\geq 0$ pour tout $x\in\aa$.
\end{dfi}

Un \'etat doit \^etre consid\'er\'e comme l'analogue de la matrice densit\'e $\rho$. En g\'en\'eral, $\rho$ correspond \`a un \'etat de m\'elange mais on peut identifier une classe de matrices densit\'e correspondant \`a des \'etats purs du syst\`eme en consid\'eration. 
 
\begin{dfi}
Un \'etat est pur s'il ne peut \^etre \'ecrit comme combinaison lin\'eaire convexe de deux \'etats. 
\end{dfi}

Avant de passer \`a la suite, \'etudions deux exemples simples.

\par

L'exemple le plus simple est l'alg\`ebre de matrices $\aa=M_{N}(\ccc)$. Toute forme lin\'eaire $\phi$ sur $\aa$ peut toujours se mettre sous la forme
\bbb
\phi(x)=\t\lp\rho x\rp\qquad\forall x\in\aa,
\eee
o\`u $\rho$ est une matrice $N\times N$.

\par

Il est alors facile de v\'erifier que $\phi$ est un \'etat si et seulement si $\rho$ est une matrice positive de trace 1. Pour que cet \'etat soit pur, il faut et suffit que $\rho$ soit une matrice de rang 1. Un tel projecteur est donn\'e par un vecteur unitaire de $\ccc^{N}$ et deux tels vecteurs d\'efinissent le m\^eme \'etat si et seulement si ils se d\'eduisent l'un de l'autre par multiplication par une phase. 

\par

Les \'etats de l'alg\`ebre des fonctions continues sur un espace topologique compact $\xx$ sont donn\'es par les mesures de probabilit\'e sur $X$. La correspondance entre une mesure $\mu$ et un \'etat $\phi$ est 
\bbb
\phi(f)=\int_{\xx}fd\mu\qquad f\in\aa.
\eee
$\phi$ est un \'etat pur si et seulement si $\mu$ est une mesure de Dirac; les \'etats purs de $\aa$ sont donc en correspondance directe avec les points de $\xx$. 

\par

A partir d'une alg\`ebre involutive $\aa$ et d'un \'etat $\phi$ sur $\aa$, il est toujours possible de construire un espace de Hilbert $\hh$ sur lequel $\aa$ admet une repr\'esentation involutive. Cette construction est appel\'e construction GNS du nom de ses auteurs Gelfand, Na\"\i marck et Segal. 

\begin{pro}
Soit $\aa$ une alg\`ebre involutive, $\phi$ un \'etat sur $\aa$ et $\jj$ l'ensemble des $x\in\aa$ tels que $\phi(xx^{*})=0$. Alors $\jj$ est un id\'eal \`a gauche de $\aa$ et le compl\'et\'e $\hh$ de $\aa/\jj$ est un espace de Hilbert dont le produit scalaire est d\'efini par $\langle x,y\rangle=\phi(y^{*}x)$. De plus, l'application qui \`a $a\in\aa$ associe l'op\'erateur $\pi_{\phi}(a)$ sur $\hh$ d\'efini par $\pi_{\phi}(a)[x]=ax$ pour tout $x\in\hh$ est une repr\'esentatiuon involutive de $\aa$ sur $\hh$.
\end{pro}

A partir de cette construction, on peut caract\'eriser l'irr\'eductibilit\'e de la repr\'esentation.   

\begin{dfi}
La repr\'esentation $\pi_{\phi}$ est irr\'eductible si et seulement si $\phi$ est un \'etat pur.
\end{dfi} 

Nous allons consid\'er\'e des classes plus particuli\`eres d'alg\`ebres involutives qui sont les alg\`ebres de von Neumann, les C$^{*}$-alg\`ebres et l'alg\`ebre des op\'erateurs compacts. Elles correspondent aux g\'en\'eralisations non commutatives des fonctions mesurables, des fonctions continues et des infinit\'esimaux.

\section{Alg\`ebres de von Neumann}

Commen\c cons par donner la d\'efinition d'une alg\`ebre de von Neumann.

\begin{dfi}
Soit $\hh$ un espace de Hilbert. Une alg\`ebre de von Neumann $\mm$ est une sous-alg\`ebre involutive de l'alg\`ebre des op\'erateurs born\'es sur $\hh$ qui est stable par limite faible. 
\end{dfi}

On dit que la suite $(T_{n})_{n\in\nnn}$ d'op\'erateurs born\'es converge faiblement vers l'op\'erateur born\'e $T$ si et seulement si
\bbb
\mathop{\lim}\limits_{n\rightarrow+\infty}\langle x,T_{n}y\rangle=\langle x,Ty\rangle
\eee 
pour tous vecteurs $x$ et $y$ de $\hh$. 

Le th\'eor\`eme suivant permet de caract\'eriser alg\'ebriquement les alg\`ebres de von Neumann. 

\begin{thm}
Une sous-alg\`ebre involutive de l'alg\`ebre des op\'erateurs born\'es sur un espace de Hilbert est une alg\`ebre de von Neumann si et seulement si elle co\"\i ncide avec son bicommutant.
\end{thm}

Le commutant d'un ensemble $\bb$ d'op\'erateurs born\'es est form\'e des op\'erateurs born\'es qui commutent avec tout \'el\'ement de $\bb$. Son bicommutant est le commutant de son commutant. En g\'en\'eral, on note $\bb'$ le commutant de $\bb$ et $\bb''$ son bicommutant.

\par

Bien que la th\'eorie des alg\`ebres de Von Neuman forme un chapitre fondamental de la physique math\'ematique, nous passons sous silence certains de ses aspects les plus fondamentaux, comme la classification des facteurs, et nous nous bornons \`a \'enoncer le th\'eor\`eme de Tomita-Takesaki. Ce r\'esultat est essentiel pour la compr\'ehension des axiomes de la g\'eom\'etrie non commutative. 

\begin{dfi}
Un vecteur $\xi$ de $\hh$ est dit s\'eparateur si les vecteurs $\mm\xi$ obtenus par l'action de $\mm$ sont denses dans $\hh$. Un vecteur $\xi$ est cyclique si les vecteurs $\mm'\xi$ sont denses dans $\hh$.
\end{dfi}

Avec cette d\'efinition, nous pouvons \'enoncer le

\begin{thm}
Si $\mm$ est une alg\`ebre de von Neumann admettant un vecteur cyclique et s\'eparateur $\xi$, alors l'op\'erateur antilin\'eaire de domaine $\mm\xi$ d\'efini par $x\xi\mapsto x^{*}\xi$ pour tout $x\in\mm$ est fermable. Sa fermeture $S$ admet une d\'ecomposition polaire $S=\jj\Delta^{1/2}$, o\`u $\jj$ est une involution antiunitaire et $\Delta$ un op\'erateur positif. De plus, on a
\bbb
\jj\mm\jj^{-1}=\mm'\;\;\;et\;\;\;
\Delta^{it}x\Delta^{-it}\in\mm
\eee
pour tous $x\in\mm$ et $t\in\rrr$.
\end{thm}

Ainsi, $\jj$ permet de d\'efinir un isomorphisme entre une alg\`ebre de Von Neuman et son commutant. C'est sous cette forme qu'il appara\^\i t dans la formulation des axiomes de la g\'eom\'etrie non commutative.

\section{C$^{*}$-alg\`ebres}

Pour pouvoir d\'efinir une version non commutative de la topologie, nous devons coder l'information relative \`a la structure d'espace topologique sur un espace $\xx$ \`a l'aide d'une certaine sous-alg\`ebre de l'alg\`ebre des fonctions sur $\xx$. 

\par

Cela nous am\`ene \`a introduire la notion de C$^{*}$-alg\`ebre.

\begin{dfi}
Une $C^{*}$-alg\`ebre est une alg\`ebre de Banach involutive dont la norme v\'erifie $||x||^{2}=||xx^{*}||$ pour tout $x\in\aa$.
\end{dfi}

La norme d'une telle alg\`ebre est enti\`erement d\'etermin\'ee par sa structure alg\'ebrique. 

\begin{pro}
Dans une $C^{*}$-alg\`ebre $\aa$, la norme est donn\'ee par
\bbb
||x||^{2}=\mathop{\sup}\limits_{\lambda\in\ccc}\la |\lambda|\;|\; xx^{*}-\lambda\;non\;inversible\ra
\eee
pour tout $x\in\aa$.
\end{pro}

Un premier exemple de C$^{*}$-alg\`ebre est donn\'e par l'alg\`ebre des op\'erateurs born\'es sur un espace de Hilbert. 

\begin{dfi}
L'alg\`ebre $\bb(\hh)$ des op\'erateurs born\'es sur un espace de Hilbert $\hh$ est une $C^{*}$-alg\`ebre dont la norme est donn\'ee par
\bbb
||T||=\mathop{\sup}\limits_{\Psi\in\hh,\,||\Psi||=1}||T\Psi||.
\eee
\end{dfi}

En g\'en\'eral, nous partons d'une alg\`ebre involutive engendr\'ee par certaines relations de commutation, comme par exemple l'alg\`ebre apparaissant dans la d\'efinition du tore non commutatif. A partir d'une telle alg\`ebre, il est toujours possible de d\'efinir une C$^{*}$-alg\`ebre. 

\begin{pro}
Soit $\aa$ une alg\`ebre involutive et $(\pi_{i})_{i\in I}$ une famille de representations de $\aa$ sur un espace de Hilbert $\hh$ telle que pour tout $x\in\aa$, la quantit\'e $||x||$ d\'efinie par
\bbb
||x||=\mathop{\sup}\limits_{i\in I}||\pi_{i}(x)||
\eee
soit finie. Alors l'application $x\in\aa\mapsto||x||\in\rrr^{+}$ d\'efinit une norme de $C^{*}$-alg\`ebre sur $\aa/\jj$, o\`u $\jj$ est l'intersection des noyaux des repr\'esentations $\pi_{i}$.
\end{pro}

Un autre exemple important de C$^{*}$-alg\`ebre est fourni par l'alg\`ebre des fonctions continues sur un espace compact. 

\begin{pro}
Soit $\xx$ un espace topologique compact et $C(\xx)$ l'alg\`ebre des fonctions continues sur $X$ \`a valeurs complexes. $C(\xx)$ est une $C^{*}$-alg\`ebre dont la norme est d\'efinie par
\bbb
||f||=\mathop{\sup}\limits_{x\in \xx}|f(x)|
\eee
pour toute fonction $f\in C(\xx)$.
\end{pro}

Cet exemple est le prototype de la C$^{*}$-alg\`ebre commutative. En fait, nous allons voir qu'il n'y en a pas d'autre. 

\begin{dfi}
Soit $\aa$ une $C^{*}$-alg\`ebre et $\chi$ une forme lin\'eaire sur $\aa$. On dit que $\chi$ est un caract\`ere de $\aa$ si $\chi$ est un morphisme d'alg\`ebres involutives de $\aa$ dans $\ccc$.
\end{dfi}

Cette d\'efinition nous permet d'\'enoncer le r\'esulat suivant, du \`a Gelfand et Na\"\i marck.

\begin{thm}
Soit $\aa$ une $C^{*}$-alg\`ebre commutative admettant une unit\'e et $\xx$ l'ensemble des caract\`eres de $\aa$. Alors $\xx$ est un espace topologique compact pour la topologie d\'efinie par la convergence simple et l'application qui \`a $x\in\aa$ associe la fonction $f_{x}$ de $\xx$ dans $\ccc$ d\'efinie par $f_{x}(\chi)=\chi(x)$ pour tout $\chi\in \xx$ est un isomorphisme entre les $C^{*}$-alg\`ebres $\aa$ et $C(\xx)$.
\end{thm}

Ce r\'esultat s'\'etend aux C$^{*}$-alg\`ebres commutatives n'admettant pas d'unit\'e en rempla\c cant l'espace compact $\xx$ par un espace localement compact et l'alg\`ebre $C(\xx)$ par l'alg\`ebre $C^{0}(\xx)$ des fonctions continues d\'ecroissant \`a l'infini. Cependant, nous nous limiterons toujours \`a l'\'etude des alg\`ebres avec unit\'e.  

\par

Ce r\'esultat est fondamental et il pose la premi\`ere pierre de la g\'eom\'etrie non commutative car il nous permet de d\'efinir la donn\'ee d'un espace topologique non commutatif comme \'etant celle d'une C$^{*}$-alg\`ebre non commutative. 

\par

Dans le cas commutatif, on peut caract\'eriser certaines propri\'et\'es de l'espace topologique sous-jacent \`a l'aide de la C$^{*}$-alg\`ebre des fonctions continues:

\begin{pro}
La $C^{*}$-alg\`ebre $C(\xx)$ est s\'eparable si et seulement si l'espace topologique $\xx$ est m\'etrisable.
\end{pro}

Nous reviendrons sur les cons\'equences de ce r\'esultat lors de l'\'etude du couplage de la K-th\'eorie avec la cohomologie cyclique.

\section{Op\'erateurs compacts et trace de Dixmier}

La g\'eom\'etrie non commutative permet de d\'efinir de fa\c con satisfaisante la notion d'infiniment petit qui est si utile en physique. En r\`egle g\'en\'erale, un infiniment petit est une quantit\'e qui doit \^etre inf\'erieure \`a tout nombre r\'eel strictement positif. Bien entendu, un infinit\'esimal ne peut \^etre un nombre r\'eel sans \^etre identiquement nul, ce qui ne pr\'esente aucun int\'er\^et. L'analyse non standard r\'esoud le probl\`eme en introduisant des nombres r\'eels "non standard" qui apparaissent uniquement comme des interm\'ediaires de calcul. Cependant, une telle approche n'est pas enti\`erement satisfaisante car il est impossible d'exhiber un seul nombre "non standard".    

\par

Toutefois, si on remplace les nombres par des op\'erateurs agissant sur un certain espace de Hilbert, il est naturel de d\'efinir les infiniment petits comme les op\'erateurs compacts.

\begin{dfi}
Soient $\hh$ un espace de Hilbert et $T$ un op\'erateur sur $\hh$. Par d\'efinition, T est compact si l'adh\'erence de l'image de la boule unit\'e de $\hh$ par T est une partie compacte de $\hh$.
\end{dfi}

Ces op\'erateurs forment un ensemble stable par les op\'erations usuelles, ce qui est indispensable dans les calculs pratiques. 

\begin{pro}
Les op\'erateurs compacts forment un id\'eal de l'alg\`ebre des op\'erateurs born\'es.
\end{pro}

On peut caract\'eriser les op\'erateurs compacts de plusieures mani\`eres diff\'erentes.

\begin{thm}
Soient $\hh$ un espace de Hilbert et $T$ un op\'erateur sur $\hh$. Les propositions suivantes sont \'equivalentes:
\begin{enumerate}
\item
T est un op\'erateur compact,
\item
la suite d\'ecroissante $(\lambda_{n})_{n\in\nnn}$ des valeurs propres de $\sqrt{T^{*}T}$ tend vers 0,
\item
pour tout $\epsilon>0$, il existe un sous-espace E de $\hh$ de dimension finie tel que la restriction de T au suppl\'ementaire orthogonal de E  soit de norme inf\'erieure \`a $\epsilon$,
\item
T est limite uniforme d'une suite d'op\'erateurs de rang fini et la distance entre $T$ et le sous-espace des op\'erateurs de rang $n$ est donn\'e par $\lambda_{n}$.
\end{enumerate}
\end{thm}

Ce th\'eor\`eme justifie l'analogie entre op\'erateurs compacts et infinit\'esimaux: alors qu'un infinit\'esimal doit verifier $|x|< \epsilon$ pour tout $\epsilon>0$, un op\'erateur compact v\'erifie $||T||< \epsilon$ sauf peut-\^etre sur un sous-espace de dimension finie. 

\par

En \'etudiant la suite des valeurs propres d'un op\'erateur compact, on peut d\'efinir des infinit\'esimaux de tout ordre.

\begin{dfi}
Soient $\hh$ un espace de Hilbert, $T$ un op\'erateur compact sur $\hh$ et $\alpha$ un nombre r\'eel strictement positif. On dit que $T$ est un infinit\'esimal d'ordre $\alpha$ si la suite $(\lambda_{n})_{n\in\nnn}$ des valeurs propres de $|T|=\sqrt{T^{*}T}$ satisfait \`a $\lambda_{n}=O(n^{-\alpha})$.
\end{dfi}

Il est possible de montrer que les infinit\'esimaux ob\'eissent aux r\`egles de calcul usuelles. Par exemple, si $T_{1}$ et $T_{2}$ sont des infinit\'esimaux d'ordres $\alpha_{1}$ et $\alpha_{2}$, leur produit $T_{1}T_{2}$ est un infinit\'esimal d'odre $\alpha_{1}+\alpha_{2}$.

\par

Puisqu'un op\'erateur g\'en\'eralise une fonction, un op\'erateur compact doit \^etre consid\'er\'e comme une fonction \`a valeurs infinit\'esimales. On veut pouvoir int\'egrer une telle fonction pour obtenir un nombre fini. A priori, le meilleur candidat pour l'int\'egration est simplement la trace, les op\'erateurs compacts ayant une trace sont d\'efinis commme int\'egrables et leur trace est leur int\'egrale. Toutefois, cette d\'efinition n'est pas la bonne car elle ne nous permet pas de retrouver l'int\'egrale usuelle dans le cas commutatif.  

\par

Pour cela, nous allons d\'efinir, sur l'id\'eal des infinit\'esimaux d'ordre 1, une fonctionnelle qui poss\`ede les propri\'et\'es de la trace et se r\'eduit \`a l'int\'egrale usuelle dans le cas commutatif.   

\par

Cette construction fait appel \`a la trace de Dixmier et n\'ecessite quelques pr\'eliminaires techniques. 

\par

Sur l'espace des suites born\'es de nombres r\'eels, soit $\lim_{\omega}$ une forme lin\'eaire satisfaisant aux conditions suivantes:

\begin{enumerate}
\item
$\lim_{\omega}(a_{n})$ est la limite de la suite $(\alpha_{n})_{n\in\nnn}$ si elle converge,
\item
$\lim_{\omega}(a_{n})\geq 0$ si $\alpha_{n}\geq 0$,
\item
$\lim_{\omega}(\alpha_{1},\alpha_{1},\alpha_{2},\alpha_{2},\alpha_{3},\alpha_{3},\dots)=\lim_{\omega}(\alpha_{n})$.
\end{enumerate}

On peut alors utiliser une telle forme lin\'eaire pour associer \`a tout infinit\'esimal d'ordre 1 un nombre complexe.

\begin{dfi}
Soient $T\geq 0$ un infinit\'esimal d'ordre 1, $(\lambda_{n})_{n\in\nnn^{*}}$ la suite  d\'ecroissante de ses valeurs propres et $\lim_{\omega}$ une forme lin\'eaire sur l'espace des suites born\'ees de nombre r\'eels satisfaisant aux propri\'et\'es pr\'ec\'edentes. On d\'efinit la trace de Dixmier par
\bbb
\t_{\omega}(T)=\lim_{\omega}\lp\lp
\frac{\lambda_{1}+\dots+\lambda_{N}}{\log N}\rp_{N}\rp.
\eee
On \'etend $\t_{\omega}$ par lin\'earit\'e aux infinit\'esimaux d'ordre 1 non positifs.
\end{dfi}

Lorsque la suite $(\lambda_{n})_{n\in\nn}$ converge, on a
\bbb
\t_{\omega}(T)=\mathop{\lim}\limits_{N\rightarrow +\infty}\frac{\lambda_{1}+\dots+\lambda_{N}}{\log N}.
\eee
En r\`egle g\'en\'erale, la trace de Dixmier v\'erifie les propri\'et\'es suivantes.

\begin{pro}\hfill
\begin{enumerate}
\item
Soit $T$ un infinit\'esimal d'ordre 1 et $S$ un op\'erateur born\'e. Alors $\t_{\omega}(ST)=\t_{\omega}(TS)$.
\item
Si $T\geq 0$ alors $\t_{\omega}(T)\geq 0$.
\item
La trace de Dixmier ne d\'epend que de la structure d'espace vectoriel topologique de $\hh$ et non du produit scalaire sous-jacent.
\item
La trace de Dixmier s'annule sur les infinit\'esimaux d'ordre $>1$.
\end{enumerate}
\end{pro}

Dans la plupart des cas, cette fonctionnelle d\'epend de la d\'efinition de $\lim_{\omega}$. Cependant, on peut parfois calculer $\t_{\omega}$ \`a l'aide d'une fonction $\zeta$ g\'en\'eralis\'ee d\'efinie par $\zeta(s)=\t(P^{s})$. 

\begin{pro}
Soient $T\geq 0$ un infinit\'esimal d'ordre 1 et $s>1$ un r\'eel. Alors $\t(T^{s})$ est bien d\'efinie et si
\bbb
\mathop{\lim}\limits_{s\rightarrow 1^{+}}(s-1)\t(T^{s}),
\eee
existe, elle est \'egale \`a $\t_{\omega}(T)$
\end{pro}

Terminons en \'etudiant le cas particulier des op\'erateurs pseudodiff\'erentiels qui nous montre comment la trace de Dixmier s'ins\`ere dans un contexte g\'eom\'etrique.  

\par

Commen\c cons par donner la d\'efinition d'un op\'erateur pseudodiff\'erentiel sur $\rrr^{n}$. Soit $\ss$ l'espace de Schwartz des fonctions \`a d\'ecroissance rapide ainsi que leurs d\'eriv\'ees sur $\rrr^{n}$. Toute fonction $f\in\ss$ peut s'\'ecrire \`a l'aide de sa transform\'ee de Fourier
\bbb
f(x)=\int_{\rrr^{n}}\,d^{n}k\,\hat{f}(k)e^{ikx},\quad\forall x\in\rrr^{n},
\eee   
avec
\bbb
\hat{f}(k)=\frac{1}{(2\pi)^{n}}
\int_{\rrr^{n}}\,d^{n}x\,\hat{f}(k)e^{-ikx},\quad\forall k\in\rrr^{n}.
\eee
L'application transform\'ee de Fourier $f\mapsto \hat{f}$ d\'efinit un isomorphisme de $\ss$ sur lui-m\^eme. 

\begin{dfi}
Un op\'erateur pseudodiff\'erentiel d'ordre $d\in\zzz$ sur $\rrr^{n}$ est un op\'erateur lin\'eaire $P$ qui \`a $f\in\ss$ associe la fonction $P\cdot f$ d\'efinie par
\bbb
P\cdot f(x)=\int_{\rrr^{n}}\,d^{n}k\,\sigma(x,k)\hat{f}(k)e^{ikx},\quad\forall x\in\rrr^{n},
\eee
o\`u $\sigma$ est une fonction sur $\rrr^{n}\times\rrr^{n}$ appel\'ee symbole de $P$ et satisfaisant \`a
\bbb
|\partial_{x}^{\alpha}\partial_{k}^{\beta}\sigma(x,k)|\leq C_{\alpha,\beta}(1+|k|)^{d-\beta},
\eee 
pour tous $\alpha,\beta\in\nnn^{n}$ et o\`u $C_{\alpha,\beta}$ est une constante positive.
\end{dfi}

Pour toute fonction $f$ de $n$ variables et tout $\alpha\in\nnn^{n}$, nous avons utilis\'e la notation
\bbb
\partial_{x}^{\alpha}f=\frac{\partial^{\alpha_{1}+\dots+\alpha_{n}}f}
{\partial x^{\alpha_{1}}\dots\partial x^{\alpha^{n}}}.
\eee

\par

Puisque $\ss$ est dense dans $L^{2}(\rrr^{n})$, on peut consid\'erer que $P$ est un op\'erateur sur l'espace de Hilbert $L^{2}(\rrr^{n})$ de domaine dense.

\par

Avant d'appliquer aux op\'erateurs diff\'erentiels le formalisme pr\'ec\'edent, il convient de remarquer que les op\'erateurs pseudodiff\'erentiels sont en g\'en\'eral d\'efinis entre les espaces des sections de deux fibr\'es vectoriels sur une vari\'et\'e \cite{gilkey}. Nous nous somme content\'es de donner une d\'efinition simplifi\'ee valable pour des fibr\'es triviaux de rang 1 au-dessus de $\rrr^{n}$. Cette d\'efinition peut \^etre g\'en\'eralis\'ee \`a des fibr\'es de rang quelconque au-dessus de $\rrr^{n}$ en prenant un symbole \`a valeurs matricielles. On \'etend alors la d\'efinition d'un op\'erateur pseudodiff\'erentiel sur une vari\'et\'e quelconque en utilisant des coordonn\'ees locales.

\par

Un op\'erateur pseudodiff\'erentiel $P$ est dit classique si son symbole admet un d\'eveloppement asymptotique du type
\bbb
\sigma(x,k)\simeq\mathop{\sum}\limits_{i\in\nnn}a_{d-i}(x,k),
\eee
o\`u $a_{d-i}(x,k)$ est une fonction homog\`ene de degr\'e $d-i$ de $k$.

\par

Nous pouvons maintenant introduire le r\'esidu de Wodzicki.

\begin{dfi}
Soit $\mm$ une vari\'et\'e compacte de dimension $n$ munie d'une m\'etrique riemannienne, $E$ un fibr\'e vectoriel complexe sur $\mm$ et $A$ un op\'erateur pseudodiff\'erentiel classique agissant sur les sections de $E$. Le r\'esidu de Wodzicki de $A$ est d\'efini par
\bbb
\ww(A)=\mathop{\mathrm{Res}}\limits_{s=0}\t(A\Delta^{-s}),
\eee
o\`u $\Delta$ est un laplacien g\'en\'eralis\'e.
\end{dfi}

Cette d\'efinition utilise la notion de laplacien g\'en\'eralis\'e: 

\begin{dfi}
Soit $\mm$ une vari\'et\'e compacte de dimension $n$ munie d'une m\'etrique riemannienne $g$, $E$ un fibr\'e vectoriel complexe sur $\mm$ et $\nabla:\,E\rightarrow E\ot_{C^{\infty}(\mm)}\Omega^{1}(\mm)$ une connexion sur $E$. Une application $\Delta$ de l'espace des sections de $E$ dans lui-m\^eme est un laplacien g\'en\'eralis\'e associ\'e \`a $\nabla$ si $\Delta$ s'\'ecrit en coordonn\'es locales
\bbb
\Delta=-g^{\mu\nu}\nabla_{\mu}\nabla_{\nu}+g^{\mu\nu}\Gamma_{\mu\nu}^{\lambda}
\nabla_{\lambda}
\eee   
o\`u l'action de $\nabla$ sur une section $\Psi$ est donn\'ee par $\nabla(\Psi)=\nabla_{\mu}\Psi\ot dx^{\mu}$ et o\`u $\Gamma_{\mu\nu}^{\lambda}$ d\'esigne la connexion de Levi-Civita.
\end{dfi}

Les op\'erateurs pseudodiff\'erentiels classiques forment une alg\`ebre et on montre le r\'esultat suivant (voir \cite{kassel} pour une revue).

\begin{thm}
L'application $A\mapsto\mathrm{Res}(A)$ est une trace sur l'alg\`ebre des op\'erateurs pseudodiff\'erentiels classiques. Lorsque $\mm$ est une vari\'et\'e connexe de dimension $n>1$, cette trace est unique \`a un facteur de normalisation pr\`es.
\end{thm}

Cette trace peut \^etre reli\'ee \`a la trace de Dixmier \cite{action}. 

\begin{thm}
Soit $A$ un op\'erateur pseudodiff\'erentiel d'ordre $-n$. Alors la trace de Dixmier $\t_{\omega}(A)$ est bien d\'efinie et on a
\bbb
\t_{\omega}(A)=\frac{2}{n}\ww(A).
\eee
\end{thm}

En g\'en\'eral, le r\'esidu ne se calcule pas en utilisant les fonctions $\zeta$ g\'en\'eralis\'ees, mais en utilisant son expression int\'egrale \`a l'aide du $a_{-n}(x,k)$ \cite{kassel}. On en d\'eduit ais\'ement que la trace de Dixmier combin\'ee avec l'op\'erateur de Dirac nous permet de retrouver l'int\'egrale usuelle:

\begin{pro}
Soit $\mm$ une vari\'et\'e riemanienne compacte de dimension n munie d'une structure de spin et soit $\dd$ l'op\'erateur de Dirac associ\'e. Pour toute fonction $f\in C^{\infty}(\mm)$, $f|\dd|^{-n}$ est un op\'erateur pseudodiff\'erentiel d'ordre $-n$ et on a
\bbb
\t_{\omega}(f|\dd|^{-n})=\lambda_{n}
\int_{\mm}\;f\,\sqrt{g}\,d^{n}x,
\eee 
avec
\bbb
\lambda_{n}=\frac{2^{[n/2]-n/2}}{(2\pi)^{n/2}\Gamma(n/2+1)}.
\eee
\end{pro}

Cela ach\`eve notre discussion de la trace de Dixmier et de ses applications en g\'eom\'etrie non commutative. Nous renvoyons \`a \cite{kastler} et \cite{kalau} pour les applications de cette th\'eorie \`a la relativit\'e g\'en\'erale.

\chapter{K-th\'eorie}

En g\'eom\'etrie usuelle, la K-th\'eorie se propose de classifier les fibr\'es vectoriels au-dessus d'un espace topologique $\xx$ fix\'e. Pour cela, on associe \`a chaque espace une s\'erie de groupes ab\'eliens et \`a chaque fibr\'e au-dessus de cet espace un \'el\'ement de ces groupes. Bien que de nature tr\`es diff\'erente, ces groupes sont assez similaires au groupes d'homotopie qui sont bien connus des physiciens. Alors que ces derniers n'ont pas de g\'en\'eralisation imm\'ediate en g\'eom\'etrie non commutative, la K-th\'eorie peut \^etre formul\'ee de mani\`ere purement alg\'ebrique et joue ainsi un r\^ole tr\`es important dans notre contexte. Nous allons donner les d\'efinitions de base de cette formulation alg\'ebrique dans le cas des C$^{*}$-alg\`ebres, en renvoyant \`a \cite{wegge} pour une construction d\'etaill\'ee.

\section{Fibr\'es vectoriels en g\'eom\'etrie non commutative}

Commen\c cons par rappeler la d\'efinition d'un fibr\'e vectoriel en g\'eom\'etrie ordinaire.

\begin{dfi}
Soit $\xx$ un espace topologique. Un fibr\'e vectoriel sur $\xx$, appel\'e la base du fibr\'e, est un triplet $(\ee,\xx,\pi)$, o\`u $\ee$ est un espace topologique appel\'e espace total et $\pi:\;\ee\rightarrow \xx$ une application continue et surjective telle que $\pi^{-1}(x)$ soit muni d'une structure d'espace vectoriel de dimension finie pour tout $x\in \xx$. De plus, ce fibr\'e est dit localement trivial s'il existe un recouvrement ouvert $(U_{i})_{i\in I}$ tel que $\pi^{-1}(U_{i})$ soit isomorphe \`a $U_{i}\times F$, o\`u $F$ est un espace vectoriel de dimension finie.
\end{dfi}

Par la suite, nous ne consid\'ererons que des fibr\'es vectoriels localement triviaux, aussi sous-entendrons nous l'expression "localement trivial".

\par

Afin de g\'en\'eraliser cette notion en g\'eom\'etrie non commutative, nous devons donner quelques d\'efinitions purement alg\'ebriques \cite{algebra}. 

\begin{dfi}
Soit $\aa$ une alg\`ebre et $\ee$ un $\aa$-module \`a droite. On dit que $\ee$ est un $\aa$-module \`a droite de type fini et projectif si
\begin{enumerate}
\item
$\ee$ est engendr\'e  par un nombre fini d'\'el\'ements en tant que $\aa$-module \`a droite,
\item
il existe un autre $\aa$-module \`a droite $\ee^{'}$ telle que la somme directe $\ee\op\ee^{'}$ soit isomorphe au $\aa$-module \`a droite trivial $\aa^{N}$.
\end{enumerate}
\end{dfi}

On montre alors le r\'esultat suivant, que nous avons utilis\'e lors de notre \'etude des th\'eories de Yang-Mills. 

\begin{pro}
Tout module projectif fini \`a droite sur $\aa$ est isomorphe \`a un module du type $e\aa^{N}$, o\`u $e\in M_{N}(\aa)$ est une projection.
\end{pro}

L'introduction de la notion de module projectif se justifie par le th\'eor\`eme de Serre-Swan, qui permet de relier modules projectifs et fibr\'es vectoriels: 

\begin{thm}
Si $\xx$ est un espace topologique compact et si $\aa=C^{0}(\xx)$ est l'alg\`ebre des fonctions continues sur $\xx$ \`a valeurs complexes, alors les modules des sections des fibr\'es vectoriels sur $\xx$ sont les modules projectifs de type fini sur l'alg\`ebre $\aa$. 
\end{thm}

Il est utile de remarquer que l'on construit explicitement cet isomorphisme en utilisant une partition de l'unit\'e telle qu'au-dessus de chacun des ouverts la composant, le fibr\'e est trivial. 

Ainsi, les modules projectifs sur une alg\`ebre g\'en\'eralisent les espaces de sections de fibr\'es vectoriels en g\'eom\'etrie non commutative.

\section{Le groupe $K_{0}(\aa)$}

Pour construire les groupes de K-th\'eorie dans ce contexte alg\'ebrique, commen\c cons par introduire une relation d'\'equivalence sur l'ensemble des modules projectifs sur une alg\`ebre donn\'ee. 

\begin{dfi}
Deux modules projectifs finis sur une alg\`ebre $\aa$ sont \'equivalents s'ils 
sont isomorphes, eventuellement apr\`es addition d'un module trivial $\aa^{N}$.
\end{dfi}

En utilisant la partition de l'unit\'e apparaissant dans la d\'emonstration du th\'eor\`eme de Serre-Swan, il est clair que les fibr\'es vectoriels triviaux sont en correspondance biunivoque avec les modules triviaux $\aa^{N}$. Cette relation d'\'equivalence est donc la m\^eme que celle introduite en g\'eom\'etrie usuelle.  

\begin{pro}
Muni de la somme directe, les classes d'\'equivalence de modules projectifs finis forment un semi-groupe ab\'elien $K_{0}^{+}(\aa)$.
\end{pro}

A partir de ce semi-groupe on construit un groupe \`a l'aide du proc\'ed\'e employ\'e lors du passage des entiers naturels $\nnn$ aux entiers relatifs $\zzz$. 

\begin{dfi}
$K_{0}(\aa)$ est le groupe  ab\'elien $\lp K_{0}^{+}(\aa)\times K_{0}^{+}(\aa)\rp/\simeq$, o\`u $\simeq$ est la relation d'\'equivalence d\'efinie sur $K_{0}^{+}(\aa)\times K_{0}^{+}(\aa)$ par
\bbb
(x,y)\simeq(x',y')\;\;\mathrm{si}\;\mathrm{et}\;\mathrm{seulement}\;\mathrm{si}\;\;x+y'=y+x'.
\eee
\end{dfi}

Ainsi, nous avons associ\'e \`a chaque alg\`ebre un groupe ab\'elien $K_{0}(\aa)$. Par la suite, nous appliquerons cette construction \`a des C$^{*}$-alg\`ebres qui g\'en\'eralisent l'alg\`ebre des fonctions continues sur un espace compact. 

\section{Groupes d'ordre sup\'erieurs et p\'eriodicit\'e de Bott}

De la m\^eme mani\`ere, il est possible de construire de mani\`ere purement alg\'ebrique les groupes d'ordre sup\'erieurs. Bien que cette construction soit relativement ardue, elle se simplifie consid\'erablement dans le cas des C$^{*}$-alg\`ebres.

\par

La premi\`ere \'etape consiste \`a introduire la suspension. 

\begin{dfi}
Soit $\aa$ une $C^{*}$-alg\`ebre et $C^{0}(\rrr)$ la $C^{*}$-alg\`ebre des fonctions continues sur $\rrr$ qui tendent vers 0 en $\pm\infty$. Par d\'efinition, la suspension de $\aa$ est $S\aa=C^{0}(\rrr)\ot\aa$.
\end{dfi}

Bien entendu, il est possible d'it\'erer ce proc\'ed\'e et de d\'efinir $S^{n}(\aa)$ par $S^{n}(\aa)=S\lp S^{n-1}(\aa)\rp$.

\begin{dfi}
Soit $\aa$ une C$^{*}$-alg\`ebre. Les groupes de K-th\'eorie d'ordre sup\'erieurs sont d\'efinis par $K_{n}(\aa)=K_{0}(S^{n}\aa)$.
\end{dfi}

En fait, il s'av\`ere que ces groupes sont isomorphes \`a $K_{0}(\aa)$ si $n$ est pair et \`a $K_{1}(\aa)$ si $n$ est impair. De plus $K_{0}(\aa)$ et $K_{1}(\aa)$ admettent une description simple en termes de groupes d'homotopie. 
\par

Pour tout entier $N$, notons $\uu_{N}(\aa)$ le groupe des \'el\'ements unitaires de $M_{N}(\aa)$. L'application
\bbb
u\in\uu_{N}(\aa)\mapsto\pp{u&0\cr 0&1}\in \uu_{N+1}(\aa)
\eee 
d\'efinit une inclusion de $U_{N}(\aa)$ dans $\uu_{N+1}(\aa)$. A l'aide de cette inclusion on d\'efinit
\bbb
\uu_{\infty}(\aa)=\mathop{\cup}\limits_{N\in\nnn}U_{N}(\aa).
\eee
On montre alors le

\begin{thm}
Avec les notations pr\'ec\'edentes, on a
\bbb
K_{0}(\aa)=\pi_{1}(\uu_{\infty}(\aa))\;\;\; et\;\;\;K_{1}(\aa)=\pi_{0}(\uu_{\infty}(\aa)).
\eee
\end{thm}

Ainsi, le groupe $K_{1}(\aa)$ mesure le d\'efaut de connexit\'e du groupe $\uu_{\infty}(\aa)$.

\section{Calcul fonctionnel holomorphe}

La th\'eorie pr\'ec\'edente s'applique remarquablement bien aux C$^{*}$-alg\`ebres. Cependant, ces derni\`eres correspondent aux alg\`ebres de fonctions continues alors que nous sommes int\'eress\'es, en particulier lors de l'\'etude du couplage avec la cohomologie cyclique, par des alg\`ebres de fonctions lisses. Pour passer de l'un \`a l'autre, on utilise la notion de pr\'e C$^{*}$-alg\`ebre, qui se base sur le calcul fonctionnel holomorphe \cite{bible}.   

\begin{dfi}
Soient $\aa$ une $C^{*}$-alg\`ebre avec unit\'e et a un \'el\'ement de $\aa$. Le spectre de $a$ est l'ensemble des $\lambda\in\ccc$ tel que $a-\lambda 1$ ne soit pas inversible.
\end{dfi}

Les formules de Cauchy permettent de d\'efinir l'image de $x$ par une fonction holomorphe sur le spectre de $x$.  

\begin{dfi}
Soient $\aa$ une $C^{*}$-alg\`ebre, a un \'el\'ement de $\aa$ et $f$ une fonction \`a valeurs complexes et holomorphe sur le spectre de a. On d\'efinit $f(a)$ par
\bbb
f(a)=\frac{1}{2i\pi}\int_{\Gamma}\frac{f(z)}{z-a}dz,
\eee
o\`u $\Gamma$ est n'importe quelle courbe ferm\'ee simple orient\'ee dans le sens positif et entourant le spectre de a.
\end{dfi}

On introduit alors la notion de pr\'e C$^{*}$-alg\`ebre.

\begin{dfi}
Une pr\'e $C^{*}$-alg\`ebre est une sous-alg\`ebre dense d'une $C^{*}$-alg\`ebre qui est stable par calcul fonctionnel holomorphe. 
\end{dfi}

Les alg\`ebres que nous rencontrons dans notre \'etude des triplets spectraux $(\aa,\hh,\dd)$ sont munies d'une structure de pr\'e C$^{*}$-alg\`ebre qui permet de construire une C$^{*}$-alg\`ebre repr\'esent\'ee sur $\hh$.

\begin{pro}
Soit $\aa$ une pr\'e $C^{*}$-alg\`ebre. Alors, toute repr\'esentation involutive de $\aa$ dans un espace de Hilbert est continue et s'\'etend de mani\`ere unique en une repr\'esentation de $C^{*}$-alg\`ebre.
\end{pro}

Enfin le r\'esulat suivant nous autorise \`a travailler avec des  pr\'e C$^{*}$-alg\`ebres.

\begin{pro}
Soient $\aa$ une $C^{*}$-alg\`ebre et $\bb$ une sous-alg\`ebre dense de $\aa$, stable par calcul fonctionnel holomorphe. Alors l'inclusion de $\bb$ dans $\aa$ induit un isomorphisme entre $K_{0}(\aa)$ et $K_{0}(\bb)$.
\end{pro}

Cela justifie le fait que nous travaillons avec des alg\`ebres de fonctions lisses plut\^ot qu'avec des  C$^{*}$-alg\`ebres qui sont d'un maniement plus d\'elicat.

\chapter{Cohomologie cyclique}

Enfin, nous terminons en donnant les quelques rudiments de cohomologie cyclique indispensables \`a la compr\'ehension de certains de nos r\'esultats concernant l'action de Yang-Mills ou celle de Chern-Simons. Le livre de r\'ef\'erence sur le sujet est incontestablement \cite{loday}, mais la lecture de certains passages de \cite{bible} se r\'ev\`ele indispensable pour l'\'etude du couplage avec la K-th\'eorie.

\section{Cohomologie cyclique}

La d\'efinition suivante caract\'erise de mani\`ere purement alg\'ebrique les propri\'et\'es des formes diff\'erentielles et de leur int\'egration.

\begin{dfi}
Un cycle de dimension $n$ sur une alg\`ebre $\aa$ est un quadruplet $\lp\Omega,\pi,d,\int\rp$ o\`u
\begin{enumerate}
\item
$\Omega=\mathop{\op}\limits_{k=0}^{n}\;\Omega^{k}$ est une alg\`ebre gradu\'ee,
\item
$\pi$ est une repr\'esentation de $\aa$ dans $\Omega^{0}$,
\item
$d\,:\, \Omega^{k}\rightarrow \Omega^{k+1}$ est une application lin\'eaire satisfaisant \`a $d^{2}=0$, \`a $d(\Omega^{n})=0$ ainsi qu'\`a la r\`egle de Leibniz gradu\'ee $
d(\omega\xi)=d\omega\,\xi+(-1)^{p}\omega\,d\xi$ pour tous $\omega\in\Omega^{p}$ et $\xi\in\Omega^{q}$,
\item
$\int\,:\,\Omega^{n}\rightarrow\ccc$ est une forme lin\'eaire telle que $\int\,d\omega=0\;\mathrm{si}\;\omega\in\Omega^{n-1}$, et $\int\,\omega\xi=(-1)^{pq}\int\,\xi\omega$, pour tous $\omega\in\Omega^{p}$ et $\xi\in\Omega^{q}$ tels que $p+q=n$.
\end{enumerate}
\end{dfi}

Cela nous m\`ene \`a la d\'efinition du caract\`ere d'un cycle.

\begin{dfi}
Le caract\`ere d'un cycle de dimension $n$ est l'application multilin\'eaire d\'efinie  sur $\aa^{n+1}$ par
\bbb
\tau(a_{0},a_{1},\dots,a_{n})=\int\,\pi(a_{0})\,d(\pi(a_{1}))\dots d(\pi(a_{n})).
\eee
\end{dfi}

En fait, il s'av\`ere que le caract\`ere d'un cycle v\'erifie certaines propri\'et\'es qui sont \`a la base de la cohomologie cyclique. 

\begin{pro}
Une application $(n+1)$-lin\'eaire $\tau$ sur $\aa$ est le caract\`ere d'un cycle sur $\aa$ si et seulement si
\bbb
\tau(a_{0},a_{1},\dots,a_{n})=(-1)^{n}\tau(a_{n},a_{0},\dots,a_{n-1})
\eee
pour tous $a_{0},a_{1},\dots,a_{n}\in\aa$ et
\bbb
\mathop{\sum}\limits_{k=0}^{n}
(-1)^{k}\tau(a_{0},a_{1},\dots,a_{k}a_{k+1},\dots,a_{n+1})
+(-1)^{n+1}\tau(a_{n+1}a_{0},a_{1},\dots,a_{n})=0
\eee
pour tous $a_{0},a_{1},\dots,a_{n+1}\in\aa$.
\end{pro}

Avant de d\'efinir la cohomologie cyclique dans un cadre plus abstrait, rappelons la d\'efinition d'un bimodule. 

\begin{dfi}
Soient $\aa$ et $\bb$ deux alg\`ebres et $\mm$ un espace vectoriel. On dit que $\mm$ est un $(\aa,\bb)$-bimodule si $\aa$ agit \`a gauche et $\bb$ \`a droite sur $\mm$ et si ces deux actions commutent. 
\end{dfi}

La proposition suivante est facile \`a v\'erifier.

\begin{pro}
Soient $\aa$ un alg\`ebre, $\mm$ un bimodule sur $\aa$ et $C^{n}(\aa,\mm)$ l'espace vectoriel des applications  multilin\'eaires de $\aa^{n}$ dans $\mm$. L'application $b:\;C^{n}(\aa,\mm)\rightarrow C^{n+1}(\aa,\mm)$ d\'efinie par
\bbbb
&bT(a_{1},\dots,a_{n+1})=a^{1}T(a_{2},\dots,a_{n+1})+&\n\\
&\mathop{\sum}\limits_{i=1}^{n}
(-1)^{i}T(a_{1},\dots,a_{i}a_{i+1},\dots,a_{n+1})+
(-1)^{n+1}T(a_{1},\dots,a_{n})a_{n+1}&
\eeee
pour tous $T\in C^{n}(\aa,\mm)$ et $a_{1},\dots,a_{n+1}\in\aa$ v\'erifie 
\bbb
b^{2}=0.
\eee
\end{pro}

Nous d\'efinissons alors la cohomologie de Hochschild de la mani\`ere suivante.

\begin{dfi}
La cohomologie du complexe
\bbb
C^{0}(\aa,\mm)\mathop{\longrightarrow}\limits^{b}C^{1}(\aa,\mm)
\mathop{\longrightarrow}\limits^{b}\dots\mathop{\longrightarrow}\limits^{b}
C^{n}(\aa,\mm)\mathop{\longrightarrow}\limits^{b}C^{n+1}(\aa,\mm)
\mathop{\longrightarrow}\limits^{b}\dots
\eee
est la cohomologie de Hochschild \`a valeurs dans le bimodule $\mm$.
\end{dfi}

De m\^eme, nous avons une version duale de la proposition pr\'ec\'edente:

\begin{pro}
Soit $\aa$ une alg\`ebre, $\mm$ un bimodule sur $\aa$ et $C_{n}(\aa,\mm)$ le produit tensoriel $\mm\ot\aa^{\ot n}$, L'application $b:\;C_{n}(\aa,\mm)\rightarrow C_{n-1}(\aa,\mm)$ d\'efinie par
\bbbb
&b(m\ot a_{1}\ot\dots,a_{n})=ma_{1}\ot a_{2}\ot\dots\ot a_{n}+&\n\\
&\mathop{\sum}\limits_{i=1}^{n-1}
(-1)^{i}m\ot a_{1}\ot\dots\ot a_{i}a_{i+1}\ot\dots\ot a_{n}+
(-1)^{n}a_{n}m\ot a_{1}\ot\dots\ot a_{n}\ot a_{n-1}&
\eeee
pour tous $m\in\mm$ et $a_{1},\dots,a_{n}\in\aa$ v\'erifie 
\bbb
b^{2}=0.
\eee
\end{pro}

Cela nous m\`ene \`a la d\'efinition de l'homologie de Hochschild.

\begin{dfi}
L'homologie du complexe
\bbb
\dots\mathop{\longrightarrow}\limits^{b}
C_{n}(\aa,\mm)\mathop{\longrightarrow}\limits^{b}C_{n-1}(\aa,\mm)
\mathop{\longrightarrow}\limits^{b}\dots
\mathop{\longrightarrow}\limits^{b}C_{0}(\aa,\mm)
\mathop{\longrightarrow}\limits^{b}\la 0\ra
\eee
est appel\'ee homologie de Hochschild \`a valeurs dans le bimodule $\mm$.
\end{dfi}

Ces d\'efinitions utilisant des bimodules sont tr\`es g\'en\'erales. Nous avons d\'ej\`a rencontr\'e l'homologie de Hochschild \`a valeurs dans un bimodule lors de la formulation de l'axiome d'orientabilit\'e. En particulier, il est facile de v\'erifier que l'homologie de Hochschild d'une alg\`ebre de matrices est \'egale \`a celle de $\ccc$ qui est triviale, ce qui est une obstruction \`a l'existence de triplets spectraux finis de dimension $n>0$. 

\par

Consid\'erons l'espace vectoriel $\mm$ form\'e de toutes les applications lin\'eaires $\phi$ de $\aa$ dans $\ccc$ muni de la structure de bimodule d\'efinie par
\bbb
a\phi b(x)=\phi(bxa)
\eee
pour tous $a,b,x\in\aa$. Nous identifions une applications lin\'eaire $\phi$ de $\aa$ dans $\mm$ \`a une forme $n+1$ lin\'eaire sur $\aa$ gr\^ace \`a la relation
\bbb
\phi(a_{0},a_{1},\dots,a_{n})=\phi(a_{0})(a_{1},\dots,a_{n})
\eee
pour tous $a_{0},\dots,a_{n}\in\aa$. 

\par

Le cas particulier de la cohomologie cyclique \`a valeurs dans ce bimodule est celui que nous avons consid\'er\'e dans cette th\`ese. Il nous m\`ene \`a la d\'efinition de la cohomologie cyclique. 

\begin{dfi}
Soit $C_{\lambda}^{n}(\aa)$ l'espace de formes multilin\'eaires $\phi:\;\aa^{n}\rightarrow\ccc$ satisfaisant \`a la condition
\bbb
\phi(a_{0},a_{1},\dots,a_{n})=(-1)^{n}\phi(a_{1},\dots,a_{n},a_{0}).
\eee
La cohomologie du complexe
\bbb
C^{0}_{\lambda}(\aa)\mathop{\longrightarrow}\limits^{b}C^{1}_{\lambda}(\aa)
\mathop{\longrightarrow}\limits^{b}\dots\mathop{\longrightarrow}\limits^{b}
C^{n}_{\lambda}(\aa)\mathop{\longrightarrow}\limits^{b}C^{n+1}_{\lambda}(\aa)
\mathop{\longrightarrow}\limits^{b}\dots
\eee
est la cohomologie cyclique de $\aa$. Les \'el\'ements de  $C_{\lambda}^{n}(\aa)$ situ\'es dans le noyau de $b$ sont appel\'es cocycles cycliques. 
\end{dfi}

En g\'en\'eral, on note $HC^{n}(\aa)$ les groupes de cohomologie cyclique. Par d\'efinition, on a
\bbb
HC^{n}(\aa)=Z^{n}_{\lambda}/B_{\lambda}^{n},
\eee
o\`u $Z_{\lambda}^{n}(\aa)$ (resp. $B_{\lambda}^{n}$) sont form\'es des \'el\'ements de $C_{\lambda}^{n}(\aa)$ situ\'es dans le noyau de $b$ (resp. dans l'image de $b$). De m\^eme, on note $H^{n}(\aa,\aa^{*})$ les groupes de cohomologie de Hochschild correspondant au bimodule $\mm=\aa^{*}$ form\'e des formes lin\'eaires sur $\aa$.

\par

Le complexe de la cohomologie cyclique est donc un sous-complexe du complexe de la cohomologie de Hochschild pr\'ec\'edente. L'\'etude des relations entre ces deux complexes est \`a l'origine de l'op\'erateur de p\'eriodicit\'e. 

\section{La cohomologie cyclique p\'eriodique}

Si $\aa$ et $\bb$ sont deux alg\`ebres et si $\lp\Omega_{\aa},\pi_{\aa},d_{\aa},\int_{\aa}\rp$ et 
$\lp\Omega_{\bb},\pi_{\bb},d_{\bb},\int_{\bb}\rp$ sont des cycles de dimension $n_{\aa}$ et $n_{\bb}$ sur $\aa$ et $\bb$, on d\'efinit de man\`i\`ere naturelle un cycle $\lp\Omega,\pi,d,\int\rp$ de dimension $n=n_{\aa}+n_{\bb}$ sur $\aa\ot\bb$ par
\begin{enumerate}
\item
$\Omega^{k}=\mathop{\op}\limits_{i+j=k}\Omega_{\aa}^{i}\ot\Omega_{\bb}^{j}$ et $\pi=\pi_{\aa}\ot\pi_{\bb}$,
\item
$d=d_{\aa}\ot 1+(-1)^{i}1\ot d_{\bb}$ sur $\Omega_{\aa}^{i}\ot\Omega_{\bb}^{j}$,
\item
$\int=\int_{\aa}\ot\int_{\bb}$ en tant que forme lin\'eaires sur $\Omega^{n}=\Omega_{\aa}^{n_{\aa}}\ot\Omega_{\bb}^{n_{\bb}}$.
\end{enumerate}

Cette notion correspond aux formules usuelles relatives aux formes difff\'erentielles sur un produit de deux  vari\'et\'es.  Puisqu'il y a une relation biunivoque entre caract\`eres de cycles et cocycles cycliques, on obtient le r\'esultat suivant.

\begin{pro}
Soit $\phi_{a}$ le caract\`ere du cycle $\lp\Omega_{\aa},\pi_{\aa},d_{\aa},\int_{\aa}\rp$ et $\phi_{\bb}$ celui de $\lp\Omega_{\bb},\pi_{\bb},d_{\bb},\int_{\bb}\rp$. Alors le caract\`ere du cycle
$\lp\Omega,\pi,d,\int\rp$ est un cocycle cyclique de dimension $n_{\aa}+n_{\bb}$ sur $\aa\ot\bb$ not\'e $\phi_{\aa}\#\phi_{\bb}$.
\end{pro}

Ce produit, appel\'e "cup product" en anglais, joue un r\^ole important en cohomologie cyclique. Par exemple, il peut \^etre utilis\'e pour \'etendre un cocycle cyclique $\phi$ de dimension $n$ sur une alg\`ebre $\aa$ \`a un cocycle cyclique $\tilde{\phi}$ sur $M_{N}(\aa)$. En effet, la trace est un cocycle cyclique de dimension 0 sur $M_{N}(\ccc)$, ce qui prouve que $\tilde{\phi}=\t\#\phi$ est un cocycle cyclique de dimension $n$ sur $M_{N}(\aa)=M_{N}(\ccc)\ot\aa$. Sur les produits tensoriels $\mu_{0}\ot a_{0},\dots,\mu_{n}\ot a_{n}$ de $M_{N}(\ccc)\ot\aa$, la valeur de $\tilde{\phi}$ est
\bbb
\tilde{\phi}\lp\mu_{0}\ot a_{0},\dots,\mu_{n}\ot a_{n}\rp=\t\lp\mu_{1}\dots\mu_{n}\rp\phi\lp a_{0},\dots,a_{n}\rp,\label{cup}
\eee 
ce que nous utilisons lors de l'\'etude des th\'eories de jauge.

\par

Abordons maintenant la d\'efinition de la cohomologie cyclique p\'eriodique. La proposition suivante, facile \`a v\'erifier, nous donne la cohomologie cyclique de l'alg\`ebre $\ccc$.

\begin{pro}
La cohomologie cyclique de $\ccc$ est donn\'ee par $H^{n}(\ccc)=\ccc$ si $n$ est pair et $H^{n}(\ccc)=\la 0\ra$ si $n$ est impair.
\end{pro}
 
Notons $\sigma$ le g\'en\'erateur de $HC^{2}(\ccc)$ normalis\'e par $\sigma(1,1,1)$. Remarquons que dans \cite{ihes}, la normalisation est $\sigma(1,1,1)=2i\pi$, ce qui entra\^\i ne l'apparition de facteurs $2i\pi$ suppl\'ementaires. Nous ici adoptons les conventions de \cite{bible}.

\begin{dfi}
L'op\'erateur de p\'eriodicit\'e $S:\;HC^{n}(\aa)\rightarrow HC^{n+2}(\aa)$ est d\'efini par
\bbb
S(\phi)=\sigma\#\phi
\eee
pour tout cocycle cyclique $\phi$ de dimension $n$.
\end{dfi}
 
Bien entendu, il faut v\'erifier la coh\'erence de cette d\'efinition, en particulier il faut s'assurer que l'image d'un bord est encore un bord. Nous renvoyons \`a \cite{bible} pour une construction d\'etaill\'ee. 
 
\par

L'op\'erateur de p\'eriodicit\'e permet de pr\'eciser la relation entre la cohomologie de Hochschild et la cohomologie cyclique. En effet, toute classe de cohomologie cyclique d\'etermine une classe de cohomologie de Hochschild, ce qui d\'efinit une inclusion $I:\;HC(\aa)\rightarrow H(\aa,\aa^{*})$ de la cohomologie cyclique dans la cohomologie de Hochschild. En g\'en\'eral, cette derni\`ere est plus facile \`a calculer et il est utile de caract\'eriser l'information perdue lors du passage de la cohomologie cyclique \`a la cohomologie de Hochschild. 

\begin{pro}
Le noyau de $I:\;HC^{n+1}(\aa)\rightarrow H^{n+1}(\aa,\aa^{*})$ est \'egal \`a l'image de $S:\;HC^{n-1}(\aa)\rightarrow HC^{n+1}(\aa)$.
\end{pro}

Les \'el\'ements que nous avons perdus lors du passage \`a la cohomologie de Hochschild sont donc exactement les images des cocycles cycliques de dimension inf\'erieure.

\par

L'op\'erateur de p\'eriodicit\'e permet aussi de d\'efinir la cohomologie cyclique p\'eriodique. En effet, on peut toujours inclure $HC^{n}(\aa)$ dans $HC^{n+2}(\aa)$ en l'identifiant avec $S\lp HC^{n}(\aa)\rp$. 

\begin{dfi}
Soit
\bbb
HC^{paire}(\aa)=\mathop{\cup}\limits_{n\in\nnn}HC^{2n}(\aa)
\eee
et
\bbb
HC^{impaire}(\aa)=\mathop{\cup}\limits_{n\in\nnn}HC^{2n+1}(\aa).
\eee
Les groupes de cohomologie $HC^{paire}(\aa)$ et $HC^{impaire}(\aa)$ forment la cohomologie cyclique p\'eriodique.
\end{dfi}

Cette derni\`ere joue un r\^ole fondamental en g\'eom\'etrie non commutative car c'est elle qui se couple \`a la K-th\'eorie. De plus, elle admet une d\'escription simplifi\'ee \`a l'aide d'un bicomplexe. 

\par

Pour d\'efinir ce bicomplexe, introduisons un nouvel op\'erateur $B:\; H^{n}(\aa,\aa^{*})\rightarrow HC^{n-1}(\aa)$. Pour toute forme $n$-lin\'eaire $\phi$ de $\aa^{\ot n}$ dans $\aa$, nous d\'efinissons
\bbb
B\phi(a_{0},a_{1},\dots,a_{n-1})=AB_{0}\phi(a_{0},a_{1},\dots,a_{n-1})
\eee
avec
\bbbb
B_{0}\phi(a_{0},a_{1},\dots,a_{n-1})&=&
\phi(1,a_{0},a_{1},\dots,a_{n-1})\n\\
&-&(-1)^{n}\phi(a_{0},a_{1},\dots,a_{n-1},1),
\eeee
et $A$ est d\'efini sur toute application multilin\'eaire $\Psi$ de $\aa^{\ot(n+1)}$ dans $\aa$ par
\bbb
A\Psi(a_{0},\dots,a_{n})=
\mathop{\sum}\limits_{j=0}^{n}
(-1)^{nj}\psi(a_{j},a_{j+1},\dots,a_{n},a_{0},\dots,a_{j-1}).
\eee
La proposition suivante relie $B$ \`a l'op\'erateur $b$ apparaisant dans la d\'efinition de la cohomologie de Hochschild. 

\begin{pro}
$B$ v\'erifie $B^{2}=0$ et $bB+Bb=0$ sur l'espace $C^{n}(\aa,\aa^{*})$ form\'e de toutes les application multilin\'eaires de $\aa^{\ot n}$ dans $\aa^{*}$.
\end{pro}

Par cons\'equent, les deux diff\'erentielles $b$ et $B$ peuvent servir \`a d\'efinir un bicomplexe. 

\begin{thm}
Si $m$ et $n$ sont deux entiers relatifs, d\'efinissons $C^{n,m}(\aa)$ comme \'etant l'espace des formes $(n-m)$ lin\'eaires de $\aa$ dans $\aa^{*}$ si $n-m\geq 0$ et par  $C^{n,m}(\aa)=0$ si $n<m$. Soient $d_{1}$ et $d_{2}$ les deux applications d\'efinies pour tous $m,n\in\zzz$ par
\bbb
d_{1}:\;C^{n,m}(\aa)\;\rightarrow\;C^{n+1,m}(\aa),\quad d_{1}=(n-m+1)b
\eee
\bbb
d_{2}:\;C^{n,m}(\aa)\;\rightarrow\;C^{n,m+1}(\aa),\quad d_{2}=\frac{1}{n-m}B,
\eee
$d_{2}$ \'etant nulle si $m=n$. Ces deux applications v\'erifient $(d_{1})^{2}=0$, $(d_{2})^{2}=0$ et $d_{1}d_{2}+d_{2}d_{1}=0$ et d\'efinissent un bicomplexe dont la cohomologie totale est \'egale \`a la cohomologie cyclique p\'eriodique.
\end{thm}

En d'autres termes, on peut reconstruire la cohomologie cyclique p\'eriodique \`a l'aide d'un bicomplexe, ce qui s'av\`ere fondamental dans la d\'emonstration de la formule locale de l'indice en g\'eom\'etrie non commutative.

\par

Il est important de noter que les diff\'erentielles de ce bicomplexe sont proportionelles \`a $b$ et $B$ mais non identiques \`a ces derniers. Cela a pour cons\'equence l'apparition d'un facteur de normalisation lors du passage de la cohomologie cyclique p\'eriodique au bicomplexe d\'etermin\'e par $(b.B)$.    

\section{Couplage avec la K-th\'eorie}

La cohomologie cyclique p\'eriodique se couple \`a la K-th\'eorie. Ce couplage g\'en\'eralise, dans le cadre de la g\'eom\'etrie non commutative, les quantit\'es topologiques apparaissant en physique th\'eorique, comme par exemple le nombre d'instantons.  

\par

Donnons les principaux r\'esultats relatifs \`a ce couplage, en commen\c cant par rappeler un r\'esultat que nous avons d\'ej\`a rencontr\'e (\ref{cup}).

\begin{pro}
Si $\tau$ est un cocyle cyclique sur $\aa$, alors $\tilde{\tau}$ d\'efini sur $M_{N}(\aa)=\aa\ot M_{N}(\ccc)$ par
\bbb
\tilde{\tau}(a_{0}\ot m_{0},a_{1}\ot m_{1},\dots,a_{N}\ot m_{N})=
\tau(a_{0},a_{1},\dots,a_{N})\;\t(m_{0}m_{1}\dots m_{N}) 
\eee
est un cocyle cyclique.
\end{pro}

Cela permet d'\'etendre tout cocycle sur $\aa$ \`a un cocycle sur $M_{N}(\aa)$, ce qui est indispensable si on veut appliquer les r\'esultats suivants aux th\'eories de jauge non ab\'eliennes. 
 
\par

Dans le cas pair, le couplage entre la cohomologie cyclique p\'eriodique et le groupe $K_{0}(\aa)$ est donn\'e par le

\begin{thm}
Soit $\tau$ un cocycle cyclique de degr\'e $2m$ sur $\aa$ et $e\in M_{N}(\aa)$ une projection hermitienne. La quantit\'e
\bbb
\frac{1}{m!}\tilde{\tau}(e,e,\dots,e)
\eee
ne d\'epend que de la classe de $\tau$ dans $HC^{paire}(\aa)$ et de celle de $e$ dans $K_{0}(\aa)$.
\end{thm}

De m\^eme, dans le cas impair, $HC^{impaire}(\aa)$ se couple \`a $K_{1}(\aa)$.

\begin{thm}
Soit $\tau$ un cocycle cyclique de degr\'e $2m+1$ sur $\aa$ et $u\in M_{N}(\aa)$ un unitaire. Alors
\bbb
\frac{1}{\sqrt{2i}\,2^{n}\Gamma(n/2+1)}
\tilde{\tau}(u-1,u^{-1}-1,\dots,u-1,u^{-1}-1)
\eee
ne d\'epend que de la classe de $\tau$ dans $HC^{impaire}(\aa)$ et de celle de $u$ dans $K_{1}(\aa)$.
\end{thm}

Lorsque $\tau$ est fix\'e, ces quantit\'es ne d\'ependent que de la classe de $e$ ou de $u$ en K-th\'eorie, ce qui prouve qu'elles sont stables par d\'eformation. 

\par

Lorsque $\aa$ est une C$^{*}$-alg\`ebre s\'eparable, ce qui correspond dans le cas commutatif aux espace topologiques m\'etrisables, on a le r\'esulat suivant:
 
\begin{cor}
Lorsque l'alg\`ebre $\aa$ est s\'eparable, les fonctions $e\mapsto\tilde{\tau}(e,e,\dots,e)$ et $\tilde{\tau}(u-1,u^{-1}-1,\dots,u-1,u^{-1}-1)$ ne prennent qu'un nombre d\'enombrable de valeurs.
\end{cor}

Toutefois, cela ne montre pas que ces couplages sont entiers. En g\'en\'eral, l'integralit\'e r\'esulte uniquement du couplage avec le caracat\`ere de Chern.




\end{document}